\documentclass[preprintnumbers, amsmath, amssymb, twocolumn, superscriptaddress, 
nofootinbib]{
revtex4}
\pdfoutput=1
\usepackage{graphicx}
\usepackage{epsfig}
\usepackage{dcolumn}
\usepackage{bm}
\usepackage{amsfonts}
\usepackage{subfigure}
\usepackage{color}

\def\bx{{\bf x}}
\def\bk{{\bf k}}
\def\tpk{{\tilde{\phi}_k}}

\newcommand{\ud}{\mathrm{d}}

\newcommand{\be}{\begin{equation}}
\newcommand{\ee}{\end{equation}}
\newcommand{\nn}{\nonumber}
\newcommand{\bea}{\begin{eqnarray}}
\newcommand{\eea}{\end{eqnarray}}
 \def\gsim{ \lower .75ex \hbox{$\sim$} \llap{\raise .27ex \hbox{$>$}} }
 \def\lsim{ \lower .75ex \hbox{$\sim$} \llap{\raise .27ex \hbox{$<$}} }

\begin{document}

%---------------------------------------
%\title{ Torsional, teleparallel and $f(T)$ formulations of gravity and their
%cosmological
%applications }
\title{$f(T)$ teleparallel gravity and cosmology}

\author {Yi-Fu Cai}
\email{yifucai@physics.mcgill.ca}
\affiliation{CAS Key Laboratory for Research in Galaxies and Cosmology, Department of
Astronomy,
University of Science and Technology of China, Chinese Academy of Sciences, Hefei, Anhui
230026,
China}
\affiliation{Department of Physics, McGill University, Montr\'eal, QC, H3A 2T8, Canada}

\author{Salvatore Capozziello}
\email{capozzie@na.infn.it}
\affiliation{Dipartimento di Fisica, Universit\'a di Napoli
\textquotedblleft{Federico II}\textquotedblright, Napoli, Italy,}
\affiliation{Gran Sasso Science Institute (INFN), Via F. Crispi 7, I-67100, L' Aquila,
Italy,}
\affiliation{INFN Sez. di Napoli, Compl. Univ. di Monte S. Angelo, Edificio G, Via
Cinthia, I-80126,
 Napoli, Italy,}

\author{Mariafelicia De Laurentis}
\email{mfdelaurentis@tspu.edu.ru}
\affiliation{Institut f\"ur Theoretische Physik, Goethe-Universit\"at,
  Max-von-Laue-Str. 1, 60438 Frankfurt, Germany,}
\affiliation{Frankfurt Institute for Advanced Studies,
  Ruth-Moufang-Str. 1, 60438 Frankfurt, Germany,}
\affiliation{Tomsk State Pedagogical University, ul. Kievskaya, 60, 634061 Tomsk, Russia}
\affiliation{Lab.Theor.Cosmology, Tomsk State University of Control Systems and
Radioelectronics (TUSUR), 634050 Tomsk, Russia,}
\affiliation{INFN Sez. di Napoli, Compl. Univ. di Monte S. Angelo, Edificio G, Via
Cinthia, I-80126,
 Napoli, Italy,}

\author{Emmanuel N. Saridakis}
\email{Emmanuel\_Saridakis@baylor.edu}
\affiliation{Instituto de F\'{\i}sica, Pontificia Universidad de Cat\'olica de
Valpara\'{\i}so,
Casilla 4950, Valpara\'{\i}so, Chile,}
\affiliation{CASPER, Physics Department, Baylor University, Waco, TX 76798-7310, USA,}
\affiliation{Physics Division, National Technical University of Athens, 15780 Zografou
Campus,
Athens, Greece.}

%---------------------------------------
\begin{abstract}
Over the past decades, the role of torsion in  gravity has been extensively investigated
along the main direction of bringing gravity closer to its gauge formulation and
incorporating spin in a geometric description. Here we review various torsional
constructions, from teleparallel, to Einstein-Cartan,  and metric-affine gauge
theories, resulting in extending torsional gravity in the paradigm of $f(T)$
gravity, where $f(T)$ is an arbitrary function of the torsion scalar. Based on
this theory, we further review the corresponding cosmological and astrophysical
applications. In particular, we study cosmological solutions arising from $f(T)$
gravity, both at the background and perturbation levels, in different eras along the
cosmic expansion. The $f(T)$ gravity construction can provide a theoretical
interpretation
of the late-time universe acceleration, alternative to a cosmological constant, and it
can
easily accommodate with the regular thermal expanding history including the radiation and
cold dark matter dominated phases. Furthermore, if one traces back to very early times,
for a certain class of $f(T)$ models, a sufficiently long period of inflation can be
achieved and hence can be investigated by cosmic microwave background observations, or
alternatively, the Big Bang singularity can be avoided at even earlier moments due to the
appearance of  non-singular bounces. Various observational constraints, especially the
bounds coming  from the large-scale structure data in the case of $f(T)$ cosmology, as
well as the behavior of gravitational waves, are described in detail. Moreover, the
spherically symmetric and black hole solutions of the theory are reviewed. Additionally,
we  discuss various extensions of the $f(T)$ paradigm. Finally, we consider the relation
with other modified gravitational theories, such as those based on curvature, like $f(R)$
gravity, trying to enlighten the subject of which formulation, or combination of
formulations, might be more suitable for quantization ventures and cosmological
applications.\\
\end{abstract}

\maketitle

%---------------------------------------
\tableofcontents{}

%\pagebreak
%\pagebreak

%---------------------------------------
\section{Introduction}

Einstein's General Relativity (GR) is based on the  assumption that space and time
constitute  a single  structure assigned on  Riemann's manifolds. Such manifolds are
dynamical  structures that reproduce the Minkowski space-time in absence of gravity. In
this sense, gravity is conceived as the curvature of space-time.

Any relativistic theory of gravity, included GR, has to match some requirements to be
 self-consistent. First of all, it has to reproduce the Newtonian theory in the
weak-field
 limit, hence it has to explain the  dynamics related of planets and the galactic
self-gravitating structures. Moreover, it has to pass  observational
tests in the  Solar System  \cite{Will93}. At cosmological scales, any theory of gravity
should reproduce the cosmological parameters as the expansion rate, the density
parameter,
etc., in a self-consistent way. Observations and experiments probe baryonic matter,
radiation, neutrinos  and an attractive overall interaction, acting at all scales:  This 
is the
gravity. In particular,  GR is based on some main assumptions that are:
\begin{quote}
The ``{\it Relativity Principle}'' - there is no preferred inertial frames.
\end{quote}
\begin{quote}
The ``{\it Equivalence Principle}'' - inertial effects are locally indistinguishable from
gravitational effects  which means that any gravitational field can be locally canceled.
\end{quote}
\begin{quote}
The ``{\it General Covariance Principle}'' - field equations must be ``covariant'' in
form,
and  invariant in form under  space-time diffeomorphisms.
\end{quote}
\begin{quote}
The ``{\it Causality Principle}'' - each point of space-time must admit a universally
valid notion
of past, present and future.
\end{quote}
\begin{quote}
The ``{\it Lorentz Covariance}'' -  experimental results are independent of the
orientation or the boost velocity of the laboratory through space.
\end{quote}

\noindent According to these assumptions, Einstein postulated that the
gravitational field is described in terms of the metric tensor field $ds^2 =
g_{\mu\nu}dx^{\mu}dx^{\nu}$, with the same signature of Minkowski metric in four
dimensions. The line element $ds^2$ is the  covariant scalar related to the space-time 
measurements.
 The metric coefficients are the gravitational potentials, and the
space-time, according to the former Riemann's intuition, is curved by
the distribution of matter-energy sources, in other words, the distribution of
astronomical bodies gives ``shape'' to space-time.

The above considerations  suggest  that the space-time structure has to be determined by
either one or both of the two following fields: a Lorentzian metric $g$ and a linear
connection $\Gamma$, assumed by Einstein to be torsionless. The metric $g$ fixes the {\it
causal structure} of space-time, that is the light cones and the way to make measurements
(clocks and rods); the connection $\Gamma$ fixes the free-fall and  the locally
inertial
observers (that is the {\it geodesic structure} of space-time). They must satisfy some
compatibility relations, such as the fact that photons have to follow null geodesics of
$\Gamma$, and hence $\Gamma$ and $g$ are {\it a priori} independent  but can be
constrained {\it a posteriori} by some physical properties. This fact means  that
$\Gamma$ has not to be necessarily the Levi-Civita connection of $g$, derived from a 
non-linear 
combination of $g$ and its derivatives  \cite{palatini,
palatini1}.

Furthermore, a physical quantity is Lorentz covariant if it transforms under the
Lorentz group. In particular, the {\it Local Lorentz Covariance} follows from GR and
refers to Lorentz covariance defined locally in any infinitesimal region of spacetime.
This property allows that GR results are fully coherent with Special Relativity as soon as
the effects of gravitational field can be neglected. In this sense, GR is a fully Lorentz
invariant theory since any physical quantity and equation has to be preserved under
Lorentz transformations. 

However, GR presents some  shortcomings at ultraviolet and infrared scales. From the
theoretical point of view, we have the non-renormalizability, the presence of
singularities, and the lack of a self-consistent theory of quantum gravity. From
the observational and experimental points of view GR is no longer capable of addressing
Galactic, extra-galactic and cosmic dynamics, unless   exotic forms of matter-energy
(interacting only at gravitational level) are  considered. These  elusive components
are addressed as {\it dark matter} and {\it dark energy} and constitute up to the $95\%$
of the total cosmological bulk of matter-energy  \cite{lamb}.

Alternatively, instead of changing the source side of the Einstein field equations,
one can ask for a {\it geometrical view} to fit  the missing matter-energy of the
observed Universe. In this case, the dark side could be addressed by including further
geometric invariants into the standard Hilbert-Einstein Action of GR. These effective
Lagrangians can be derived by any quantization scheme on curved space-times
\cite{Capozziello:2011et}. However, no final probe discriminating between dark matter and
extended gravity has been found up to now.

Another conceptual issue that has to be discussed in view of the present review paper is
the following. In GR it is assumed that the metric $g$ of space-time is the
fundamental object to describe gravity. The connection $\Gamma^\lambda_{\mu\nu}=\Big\{
\begin{array}{c} \lambda \\ \mu\nu
\end{array} \Big\}_g$
is derived from  $g$, which is thus the only object with dynamics. This implies that $g$
determines  the causal structure (light cones), the measurements (rods and clocks) and
the
free fall of test particles (geodesic structure). Geometrically speaking, the space-time
is given by  a couple $\{{\cal M},g\}$ constituted by a Riemannian manifold and a metric.
From a physical point of view, the metric formulation of gravitational theories assumes
the strict validity of the Equivalence Principle.

The above scheme can be enlarged assuming a {\it metric-affine} formulation of gravity.
In the Palatini formalism a  connection $\Gamma$ and a metric $g$ are   independent
objects. Space-time is a triple $\{{\cal M}, g, \Gamma\}$, where the metric determines
the
causal structure while $\Gamma$ determines the geodesic structure. In such a formalism,
$\Gamma^\lambda_{\mu\nu}=\Big\{
\begin{array}{c} \lambda \\
\mu\nu \end{array} \Big\}_g$
are differential equations. The $\Gamma$ is the Levi-Civita connection of $g$ as the
outcome of the field equations, however this fact is strictly valid only in GR and not in
any theory of gravity. In this approach, the connection (i.e. the ``force'') is the
fundamental field representing ``gravity''. The metric $g$ assumes an ``ancillary''
role needed to define lengths, distances, areas, volumes and clocks. Hence, there is no
fundamental reason to assume $g$ as the potential for $\Gamma$.  Furthermore, there is no
fundamental reason, apart from simplicity, to assume that space-time is torsionless.

%%%%%%%%%%%%%%%%%%%%%%%%%%%
%\subsection{The role of torsion and spin}
%%%%%%%%%%%%%%%%%%%%%%%%%%%%%%%%

Torsion appears in literature in quite different forms. Generally, spin is  considered to
be the source of torsion, but there are several other possibilities in which torsion
emerges in different contexts. In some cases a phenomenological counterpart is absent, in
some other cases torsion arises from sources without spin as a gradient of a scalar
field.
Nevertheless, some classification schemes for torsion are possible. For example, one  can
be  based on the possibility to construct torsion tensors from the product of  covariant
bivectors and  vectors, and their respective space-time properties. Another approach  can
be  obtained by starting from the decomposition of torsion into three irreducible pieces.
Their space-time properties, again, lead to a complete classification. The
classifications
can be made in {\bf U}$_4$, a four dimensional space-time where the torsion tensors have
some peculiar properties. However, other classification schemes are possible.

The issue to enlarge the GR with torsion  is felt strongly today, since several
questions strictly depend on whether the space-time connection is symmetric or not. GR
is essentially a classical theory which does not take into account quantum effects.
However, these effects must be considered in any theory which deals with gravity at a
fundamental level. Passing from {\bf V}$_4$ (Riemannian 4-dimensional manifolds) to {\bf
U}$_4$ manifolds is the first straightforward generalization which tries to include the
spin fields of matter into the same geometrical scheme of GR. The paradigm is that the
mass-energy is the source of curvature while the spin is the source of torsion. The
Einstein-Cartan-Sciama-Kibble (ECSK) theory is one of the most serious attempts in this
direction  \cite{Hehl:1976kj}. However, this mere inclusion of spin-matter fields does
not
exhaust the role of torsion which seems to have important roles in any fundamental theory.

For instance, a torsion field appears in (super)string theory, if we consider the
fundamental string modes; we need, at least, a scalar mode and two tensor modes: a
symmetric and an antisymmetric one. The latter one, in the low-energy limit for the
effective string action, gives the effects of a torsion field
\cite{GSW,Hammond:1994ds,DeSabbata:1991th,deSabbata:1992cp,Murase:1993xd}. Furthermore,
several attempts of unification between gravity and electromagnetism have to
take into account torsion in four and in higher dimensional theories such as the
Kaluza-Klein ones \cite{Kubyshin:1993wm,German:1993bq,Oh:1989wv}. Additionally, any
theory of gravity considering twistors needs the inclusion of torsion
\cite{Howe:1997pn,Howe:1996kj,deSabbata:1991nk} while supergravity is the natural arena
where torsion, curvature and matter fields are treated under the same standard
\cite{Papadopoulos:1994kj,Hull:1993ct}.

Besides, several people agree with the line of thinking that torsion could have played
some  specific role in the dynamics of the early universe and, by the way, it could have
yielded  macroscopically observable effects today. In fact, the presence of torsion
naturally gives  repulsive contributions to the energy-momentum tensor so that
cosmological models become singularity-free
\cite{SIV,Gonner:1984rw,Canale1984,Minkowski:1986kv,DeRitis:1988kb,Assad:1990ta,
Fennelly:1988dx,Buchbinder:1985ym,Chatterjee:1993rc,Wolf1995}. This feature, essentially,
depends on spin alignments of primordial particles which can be considered as the source
of torsion \cite{Dobado:1997wj}.

If the universe undergoes one or several phase transitions, torsion could give rise to
topological defects (e.g. torsion walls
\cite{Figueiredo:1991ze,GarciadeAndrade:1997xw,Chandia:1997hu,Letelier:1995ze,
Letelier:1995wt,Kuhne:1997rt,Ross:1989hy,Fabbri:2012yg,Fabbri:2014vda,Fabbri:2014wda,
Vignolo:2014wva, Fabbri:2015jaa}) which today can act as intrinsic
angular momenta  for cosmic  structures as galaxies. Moreover, the presence of torsion in
an effective energy-momentum tensor alters the spectrum of cosmological perturbations
giving characteristic lengths for large-scale structures \cite{ZZI}.

All these arguments, and several more, do not allow to neglect torsion in any
comprehensive theory of gravity which takes into account the non-gravitational
counterpart
of the fundamental interactions.

However, in most articles, a clear distinction is not made amongst the different kinds of
torsion. Usually torsion is related to the spin density of matter, but very often there
are examples where  it cannot be derived from spin matter and acquires interpretations
quite different from the models with spinning  fluids and particles. It can be shown that
there are many independent torsion tensors with  different properties.

In order to clarify these points,  classification schemes of torsion tensors can be
based,
as said above, on the geometrical properties of vectors and bivectors that can be used to
decompose them. Torsion tensors can be constructed as the tensor product of a simple
covariant bivector and a contravariant vector. Such objects are  well understood in GR
and
they can be easily classified  \cite{Misner:1974qy}. We call these tensors {\em
elementary torsions}.

Another classification follows from the decomposition, at one point of a {\bf U}$_4$
space-time, of the torsion tensors into three irreducible tensors with respect to the
Lorentz group. Again one can  use vectors and bivectors to identify their geometrical
properties. It follows that the elements of the second classification are generally
expressed as a ``combination of elementary torsion tensors'', while the ``elementary
torsion tensors'' are generally non-irreducible.

One of the main results of these classifications is that in many theories, such as the
ECSK theory, torsion is related to its sources by an algebraic equation; it follows that
the nature of the sources is clarified too. This feature leads to recognize which tensors
can be  generated by the spin and which not, and which do not even have a physical origin.

Other classifications have been proposed, too. The first one was given in
\cite{Davis:1978am,Atkins:1978ak} and it is based on the properties of the Riemann and
Ricci tensors, as defined in a {\bf U}$_4$ space-time, compared with the Weyl and Ricci
tensors as defined in a {\bf V}$_4$ space-time. The second one, given in
\cite{Moritsch:1993eg}, deals with the algebraic classification of space-times with
torsion following from the application of the Becchi-Rouet-Stora-Tyutin (BRST)
operator.

The classifications of the torsion tensors show how the different sources of torsion
can influence the physical phenomena. It is well known  \cite{Hehl:1976kj} that, the ECSK
theory can be recast in a nonminimal coupling version of ordinary GR, where the
energy-momentum tensor is modified according to  the torsion sources.

As we will discuss in detail along this review article, torsion theories may be relevant
in cosmology. The reason behind that is that the kinematical quantities such as shear,
vorticity, acceleration, expansion and their evolution equations, are modified by the
presence of torsion.

The plan of this Review is the following: Starting from the  definitions and
classifications of  torsion quantities, we will introduce the Einstein-Cartan theory and
the various realizations of torsion tensors. In particular, we will discuss the torsion
contributions to shear, expansion, vorticity and acceleration in Sec.
\ref{Seciontorsiontensors}. Sec. \ref{SecionPoincare} is devoted to  another important
gauge theory of gravity with  torsion, namely Poincar\'{e} Gauge Gravity, where the
Poincar\'{e} invariance plays a fundamental role. The formalism of teleparallel gravity
is
introduced in Sec. \ref{SecionTeleparallelgrav}, giving the field equations and
discussing
the equivalence with GR. The $f(T)$ extension is discussed in Sec. \ref{SecionfT} where
the main features of this approach are considered. In particular, we present the
equations
of motion, we discuss the issues of Lorentz violation, and we analyze the perturbations
in
such a theory and the implications for thermodynamics. $f(T)$ cosmology is considered in
Sec. \ref{SecionfTcosmology}. Specifically, we deal with early and late time cosmology,
cosmography and observational constraints from the $f(T)$ point of view. Sec.
\ref{SecionGravwaves} is devoted to some astrophysical applications like gravitational
waves in $f(T)$ gravity. In Sec.  \ref{sectionBlackHoles} we examine the black holes and
wormhole solutions. Extensions of $f(T)$ gravity are considered in Sec.
\ref{SecionextensionsfT}, where we discuss the role of further scalar fields, additional
couplings, or higher-order torsion invariants. Torsion and curvature gravity  are
compared
and discussed in Sec. \ref{Secioncomparison}. Finally, Sec. \ref{Seciontconclusions} is
devoted to the conclusions and discussion of the presented material. Lastly, in Appendix
\ref{SecionConventions} we give the adopted conventions throughout the Review, while in
App.
\ref{fTTCoefficients} we provide the coefficients of the stability equation of
$f(T,\cal{T})$ gravity.

\section{Torsion tensors}
\label{Seciontorsiontensors}

In this section, we will give general definitions of torsion and associated quantities
which,
below, will be specified in the particular {\bf U}$_{4}$ space-times. We shall use,
essentially, the notation of \cite{Hehl:1976kj}.

%%%%%%%%%%%%%%%%%%%%%%%%%%%%%%%%%%
\subsection{Tetrads and bivectors}
%%%%%%%%%%%%%%%%%%%%%%%%%%%%%%%%

Let us introduce the tetrad fields, using the conventions summarized in Appendix
\ref{SecionConventions}. They are defined at each point of the manifold as
a base of orthonormal vectors $e_{A}^{\mu}$, where $A,B,C\dots=0,1,2,3$ label the tangent
space-time coordinates, and $\mu,\nu,\rho\dots=0,1,2,3$ are the space-time coordinates
\cite{Capozziello:2001mq}.

Correspondingly, a cotetrad $e_{\mu}^{A}$ is defined such that
\begin{align}
\label{delta1}
  e_{A}^{\mu}e_{\nu}^{A} &= \delta^{\mu}_{\nu}\,, \\
\label{delta2}
 e_{A}^{\mu}e_{\mu}^{B} &= \delta^{B}_{A}\,.
\end{align}
The tetrad metric is
\begin{equation}\label{metrad}
 \eta_{AB}=\eta^{AB}= diag(1,-1,-1,-1),
\end{equation}
and then the space-time metric can be reconstructed in the following way
\begin{equation}\label{metrica}
 g_{\mu\nu} =\eta_{AB}e_{\mu}^{A}e_{\nu}^{B}.
\end{equation}
In the construction of the torsion tensors it will be useful to consider expressions of
the simple bivectors. These are given by the skew-symmetric tensor product of two
vectors.
Generally a bivector $B^{\mu\nu}$ is simple, if and only if it satisfies the equation
\begin{equation}
\label{simbiv}
  B^{[\mu\nu}B^{\rho]\sigma}=0.
\end{equation}
By the tetrad vectors in a $N$-dimensional manifold, one can construct the $N(N-1)/2$
simple
bivectors \begin{equation}\label{simple}
 F_{AB}^{\mu\nu}= e_{A}^{[\mu}e_{B}^{\nu]},
\end{equation}
and any bivector $B^{\mu\nu}$ is expressed as
\begin{equation}
\label{biv} B^{\mu\nu} = B^{AB}e_{A}^{\mu}e_{B}^{\nu},
\end{equation}
with $B^{AB}=-B^{BA}$.

%%%%%%%%%%%%%%%%%
\subsection{Definitions of torsion}
%%%%%%%%%%%%%%%%%%%%%%%%

The torsion tensor $T_{\mu\nu}^{\phantom{\mu\nu}\rho}$ it the antisymmetric part of the
affine
connection
coefficients $\Gamma_{\mu\nu}^{\rho}$, that is
\begin{eqnarray}
\label{t1}
 T_{\mu\nu}^{\phantom{\mu\nu}\rho} = \frac{1}{2} \left( \Gamma_{\mu\nu}^{\rho}
-\Gamma_{\nu\mu}^{\rho} \right)
\equiv\Gamma_{
[\mu\nu]}^{\rho}\,. \end{eqnarray}

In GR,  it is postulated that $T_{\mu\nu}^{\phantom{\mu\nu}\rho}=0$. It is a general
convention to
call {\bf U}$_4$ a $4$-dimensional space-time manifold endowed with metric and torsion.
The manifolds with metric and without torsion are labeled as {\bf V}$_4$ (see
\cite{schouten}).

Often in the calculations, torsion occurs in linear combinations as
in the {\it contortion tensor}, defined as
\begin{eqnarray}
\label{t3}
 K_{\mu\nu}^{\phantom{\mu\nu}\rho} = -T_{\mu\nu}^{\phantom{\mu\nu}\rho}
-T^{\rho}_{\phantom{\rho}\mu\nu}
+T^{\phantom{\mu}\rho}_{\nu\phantom{\rho}{\mu}} =
-K^{\phantom{\mu}\rho}_{\mu\phantom{\rho}\nu} \,,
\end{eqnarray}
and in the {\it modified torsion tensor}
\begin{eqnarray}
\label{t4}
 \hat{T}_{\mu\nu}^{\phantom{\mu\nu}\rho} = T_{\mu\nu}^{\phantom{\mu\nu}\rho}
+2\delta_{[\mu}^{\ \ \rho} T_{\nu]} ~,
\end{eqnarray}
where $T_\mu\equiv T_{\mu\nu}^{\phantom{\mu\nu}\nu}$. According to these definitions, it
follows that
the affine connection can be written as
\begin{eqnarray} \label{t5}
 \Gamma_{\mu\nu}^{\rho} = \left\{^{\rho}_{\mu\nu} \right\}
-K_{\mu\nu}^{\phantom{\mu\nu}\rho}\,,
\end{eqnarray}
where $\left\{^{\rho}_{\mu\nu}\right\}$ are the usual Christoffel symbols of the
symmetric
connection. The presence of torsion in the affine connection implies that the covariant
derivatives of a scalar field $\phi$ do not commute, that is
\begin{equation}\label{scalarder}
 \tilde
\nabla_{[\mu}\tilde\nabla_{\nu]}\phi=-T_{\mu\nu}^{\phantom{\mu\nu}\rho}\tilde\nabla_{\rho}
\phi ~,
\end{equation}
and for a vector $v^\mu$ and a covector $w_\mu$, one has the following relations
\begin{align}
\label{doppiader1}
 (\tilde\nabla_{\mu}\tilde\nabla_{\nu} -\tilde\nabla_{\nu}\tilde\nabla_{\mu}) v^{\rho} &=
R_{\mu\nu\sigma}^{\phantom{
abd}\rho}v^{\sigma}
 -2T_{\mu\nu}^{\phantom{\mu\nu}\sigma}\tilde\nabla_{\sigma}v^\rho ~, \\
\label{doppiader2}
 (\tilde\nabla_{\mu}\tilde\nabla_{\nu} -\tilde\nabla_{\nu}\tilde\nabla_{\mu}) w_{\sigma}
&=
R_{\mu\nu\rho}^{\phantom{
\mu\nu\rho}\sigma}w_{\sigma}
 -2T_{\mu\nu}^{\phantom{\mu\nu}\sigma}\tilde\nabla_{\sigma}w_{\rho} ~,
\end{align}
where the Riemann tensor is defined as
\begin{equation}
\label{Riemann}
 R_{\mu\nu\rho}^{\phantom{\mu\nu\rho}\sigma} =\partial_{\mu}\Gamma_{\nu\rho}^{\sigma}
-\partial_{\nu}\Gamma_{\mu\rho}^{\sigma}
+\Gamma_{\mu \lambda}^{\sigma}
\Gamma_{\nu\rho}^{\lambda} -\Gamma_{\nu\lambda}^{\sigma}\Gamma_{\mu\rho}^{\lambda} ~.
\end{equation}
The contribution of torsion to the Riemann tensor can be explicitly given by
\begin{align}
\label{riexpanded}
 R_{\mu\nu\rho}^{\phantom{\mu\nu\sigma}\sigma} =&
R_{\mu\nu\rho}^{\phantom{\mu\nu\sigma}\sigma}(\{\})
-\nabla_{\mu}K_{\nu\rho}^{\phantom{\nu]\rho}\sigma}
+\nabla_{\nu}K_{\mu\rho}^{\phantom{\mu\rho}\sigma}
\nonumber\\
 &+ K_{\mu \lambda}^{\phantom{\mu\lambda}\sigma}K_{\nu\rho}^{\phantom{\nu\rho}\lambda}
-K_{\nu \lambda}^{\phantom{\nu \lambda}\sigma}K_{\mu\rho}^{\phantom{\mu\rho}\lambda} ~,
\end{align}
where $R_{\mu\nu\rho}^{\phantom{\mu\nu\rho}\sigma}(\{\})$ is the tensor generated by the
Christoffel symbols. The symbols $\tilde\nabla$ and $\nabla$ have been introduced to 
indicate the
covariant derivative with and without torsion respectively. From (\ref{riexpanded})
the expressions for the Ricci tensor and the curvature scalar are respectively found to be
\begin{equation}
\label{ricci}
 \!\!\!
 R_{\mu\nu} = R_{\mu\nu}(\{\}) -2\nabla_{\mu}T_{\rho}
+\nabla_{\nu}K_{\mu\rho}^{\phantom{\mu\rho}\nu}
+K_{\mu \lambda}^{\phantom{\mu \lambda}\nu}K_
{\nu\rho}^{\phantom{\nu\rho}\lambda} -2T_\lambda K_{\mu\rho}^{\phantom{\mu\rho}\lambda}
\end{equation}
and
\begin{equation}
\label{curvscalar}
 R =R(\{\}) -4\nabla_{\mu}T^{\mu} +K_{\rho \lambda\nu}K^{\nu \rho\lambda} -4T_\mu T^\mu ~.
\end{equation}

\begin{figure}[ht]
\centering
\includegraphics[scale=0.5]{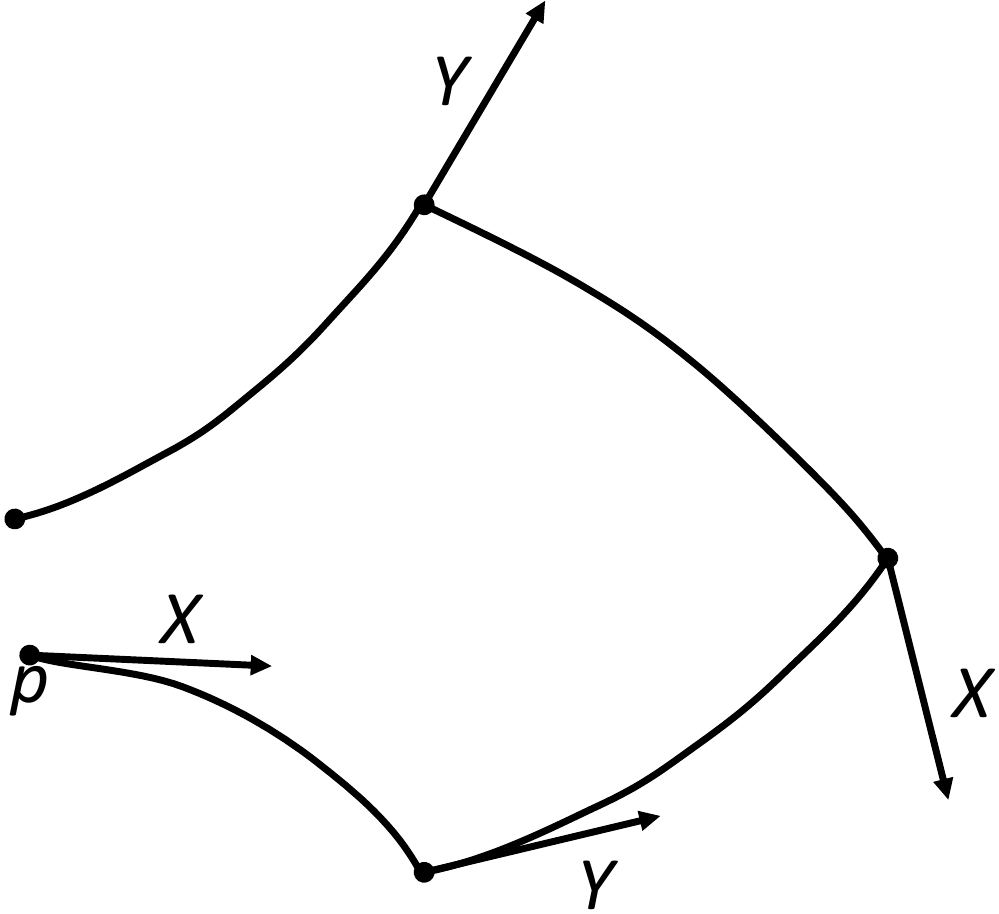}
\caption[]{\it A pictorial view of the breaking of parallelograms induced by torsion.}
\label{Quadrato1}
\end{figure}

We close this subsection by pointing out the genuine geometrical meaning of
torsion, namely that in a space with torsion the parallelograms break
\cite{Trautman:1973wy}. In particular, while in curved spaces considering two bits of
geodesics and displacing one along the other it will form an infinitesimal parallelogram
(and hence parallel-transporting a field from initial to final points  across both paths
one obtains a difference, determined by curvature), in twisted spaces the above procedure
of displacing one geodesic bit along the other leads to a gap between the
extremities, namely the infinitesimal parallelogram breaks. This
implies that performing the parallel transportation of a vector field in a space with
torsion, an intrinsic length, related to torsion, appears as it can be seen in
Fig.\ref{Quadrato1}.
%

%%%%%%%%%%%%%%%%%%%%%%%%%%%%%%%
\subsection{Decomposition of  torsion in  U$_4$}\label{gendec}
%%%%%%%%%%%%%%%%%%%%%%%%%%%%

An important property of torsion is that it can be decomposed with respect to the Lorentz
group
into three irreducible tensors, i.e. it can be written as
\begin{equation}
\label{decomposition}
 T_{\mu\nu}^{\phantom{\mu\nu}\rho} = {}^T T_{\mu\nu}^{\phantom{\mu\nu}\rho} + {}^A
T_{\mu\nu}^{\phantom{\mu\nu}\rho} +
{}^V
T_{\mu\nu}^{\phantom{\mu\nu}\rho} ~.
\end{equation}
Torsion has $24$ components, of which ${}^T T_{\mu\nu}$ has $16$ components, ${}^A
T_{\mu\nu}$
has $4$ and ${}^V T_{\mu\nu}$ has the remaining $4$ components  \cite{McCrea:1992wa,
Tsamparlis:1981xm,
Hehl:1994ue,Wanas:2001jy,Pereira:2001xf}.

One has
\begin{equation}\label{vector}
 {}^V T_{\mu\nu}^{\phantom{\mu\nu}\rho}=\frac{1}{3}(T_{\mu}\delta_{\nu}^{\rho}
-T_{\nu}\delta_{\mu}^{\rho}) ~,
\end{equation}
where $T_\mu =T_{\mu\nu}^{\phantom{\mu\nu}\nu}$,
\begin{equation}\label{axial}
 {}^A T_{\mu\nu}^{\phantom{\mu\nu}\rho}=g^{\rho\sigma}T_{[\mu \nu \sigma]} ~,
\end{equation}
which is called the axial (or totally anti-symmetric) torsion, and
\begin{equation}
\label{tensorTTT}
 {}^T T_{\mu\nu}^{\phantom{\mu\nu}\rho}= T_{\mu\nu}^{\phantom{\mu\nu}\rho} - {}^A
T_{\mu\nu}^{\phantom{\mu\nu}\rho} - {}^V
T_{\mu\nu}^{\phantom{\mu\nu}\rho} ~,
\end{equation}
which is the traceless non totally anti-symmetric part of torsion. For the sake of
brevity, in the following, we will refer respectively to the tensor (\ref{vector}) as a
V-torsion, to the tensor (\ref{axial}) as an A-torsion and to the tensor
(\ref{tensorTTT})
as a T-torsion. Finally, the dual operation (see \cite{McCrea:1992wa, Hehl:1994ue})
defined as
\begin{equation}
\label{dual}
 {}^{\star}T_{\mu\nu}^{\phantom{\mu\nu}\rho} = \frac{1}{2}
\epsilon^{\sigma\lambda}_{\phantom{\sigma\lambda}\mu\nu}
T_{\sigma\lambda}^{\phantom{\sigma\lambda}\rho} ~,
\end{equation}
has the relevant property that it associates an A-torsion tensor to a V-torsion tensor
and vice versa. Then it associates a V-torsion to a T-torsion.

%%%%%%%%%%%%%%%%%%%%%
\subsection{The classification of torsion tensors}
\label{bo}
%%%%%%%%%%%%%%%%%%%%%%%%%%%%

\subsubsection{Elementary torsion tensors}\label{element}

It can be observed that a tensor with all  properties of torsion can be constructed as
the tensor product of a bivector $F_{\mu\nu}$ with a vector $\Sigma^{\rho}$.

It is well known that any generic bivector in a four dimensional manifold can be always
reduced into the sum of two simple bivectors with a particular choice of the coordinates
(see e.g. \cite{Misner:1974qy}). In analogy with the electromagnetic case, we can call
the bivector with the timelike vector, the {\it electric} term and the one with two
spacelike vectors, the {\it magnetic} term, and label them respectively with $E_{\mu\nu}$
and
$B_{\mu\nu}$. Then we can introduce the concept of elementary torsion tensor given as the
tensor product of a {\it simple} bivector with a vector.

We say that a bivector $A_{\mu\nu}$ and a vector $V^{\rho}$ are orthogonal if
$V^\mu A_{\mu\nu}=0$. And
we consider only the cases where any four-vector $\Sigma^\rho$ is either orthogonal to a
simple bivector or is one of its components. All the other possible cases are
combinations
of these two cases. Then the $24$ elementary torsion tensors can be classified according
to the space-time properties of their bivectors and the corresponding vectors.

At this point an important remark is necessary. Any generic torsion tensor can be
decomposed in terms of these elementary parts. Let us practically construct the
elementary
torsion tensors by the vectors of a tetrad. In general, we have
\begin{equation}\label{elementa}
 T_{\ \ \ C\mu\nu}^{(el) AB\rho}=e^{A}_{[\mu}e^{B}_{\nu]}e_{C}^{\rho}\,,
\end{equation}
and then any torsion tensor can be expressed as
\begin{equation}\label{elementa1}
 T_{\mu\nu}^{\phantom{\mu\nu}\rho}=
 T_{AB}^{\phantom{AB}C}e^{A}_{[\mu}e^{B}_{\nu]}e_{C}^{\rho}\,,
\end{equation}
where the coefficients have to be
\begin{equation}\label{elementa2}
 T_{AB}^{\phantom{AB}C}=T_{\mu\nu}^{\phantom{\mu\nu}\rho}\,
 e_{A}^{[\mu}e_{B}^{\nu]}e^{C}_{\rho}\,.
\end{equation}

The classification of elementary torsion tensors in which $\Sigma^{\mu}$ does not lie on
the plane defined by the bivector is then the following:

a) If $E_{\mu\nu}$ is a bivector obtained from the antisymmetric tensor product of a
timelike
covector and a spacelike covector, $\Sigma^\mu$ must be any spacelike vector  orthogonal
to
$E_{\mu\nu}$. The pure electric case is represented just by one family of tensors. It
will be
labeled with the symbol $Es$.

b) In the pure magnetic case, one has that $\Sigma^\mu$ can be either a spacelike vector,
a
timelike vector or a null vector, leading to three family of tensors. These three
families
will be labeled respectively as $ Bs$, $ Bt$ and $Bn$.

c) In the null case, it turns out that there are two possibilities for $\Sigma^\mu$, i.e.
it can be either a null vector or a spacelike vector. The labels will be $Nn$ and $Ns$
respectively.

Regarding the case in which the vector $\Sigma^{\mu}$ lies on the plane described by the
bivector, it can be noted that if $B\equiv C$ in (\ref{elementa}) then we have V-torsions.

Finally, let us mention that the previous discussion changes if a null tetrad, defined by
$l^\mu = e_{0}
^{\mu}-e_{1}^{\mu}$, $n^\mu = e_{0}^{\mu}+e_{1}^{\mu}$,
$m^{\mu}=e_{2}^{\mu}-ie_{3}^{\mu}$ and
${m^*}^{\mu}=e_{2}^{\mu}
+ie_{3}^{\mu}$, is considered. In this case, it follows that the elementary torsion, for
instance $T_{\mu\nu}^{\phantom{\mu\nu}\rho}=m_{[\mu}l_{\nu]}l^{\rho}$,
bears all properties of a T-torsion.

%%%%%%%%%%%%%%%%%%%%
\subsubsection{Irreducible tensors in four dimensions}\label{irreduc}

To classify the torsion tensors, according to their irreducible properties, let us first
consider the V-torsion. It follows, from Eq.  (\ref{vector}), that the V-torsion is
characterized by a covector
\begin{equation}\label{vector2}
 T_\mu= T_{\mu\nu}^{\phantom{\mu\nu}\nu}.
\end{equation}
$T_\mu$ can be either time-like, space-like or light-like. So we have three different
possible types of V-torsion, which can be labeled respectively by  the symbols $Vt$, $Vs$
and $V\ell$. It can be noted that the V-torsion is expressed as a combination
of elementary torsion tensors.

From Eqs. (\ref{vector}) and (\ref{delta1}) it follows that
\begin{equation}\label{vector3}
 {}^V T_{\mu\nu}^{\phantom{\mu\nu}\rho}=\frac{2}{3}T_{[\mu}e^{A}_{\nu]}e_{A}^{\rho} ~.
\end{equation}
The A-torsion can be expressed by the equation
\begin{equation}\label{hodge}
 {}^A T_{\mu\nu\rho}= \epsilon_{\mu\nu \rho\lambda}\sigma^\lambda ~.
\end{equation}
Its properties can be characterized by the space-time properties of the vector
$\sigma^\lambda$. As for the V-torsion, we label the A-torsion with $At$, $As$ or $A\ell$
depending on whether the vector $\sigma^\lambda$ is time-like, space-like or light-like.

The statement given in subsection \ref{gendec} can be proved here by direct calculation.
In fact, we have that $ \epsilon^{\sigma\lambda}_{\phantom{\sigma\lambda}\mu
\nu}T_{[\sigma}\delta_{\lambda]}^{\rho} =
\epsilon^{\sigma\rho}_{\phantom{\sigma\lambda}\mu \nu} T_\sigma $,
which is an A-torsion, and that $ \epsilon^{\lambda \tau}_{\phantom{\sigma\lambda}\mu
\nu}
\epsilon_{\sigma \lambda \tau}^{\phantom{\sigma  \lambda \tau}\rho}T^\sigma =
T_{[\mu}\delta_{\nu]}^{\rho} $, which is a V-torsion.

Finally, the T-torsion tensors can be constructed through a combination of elementary
torsion tensors of the forms
\begin{equation}\label{ttor1}
 {}^T T_{\mu\nu}^{\phantom{\mu\nu}\rho} = V_{[\mu}e^{A}_{\nu]} C^{\phantom{A}B}_{A}
e^{\rho}_{B} \,,
\end{equation}
and
\begin{equation}\label{ttor2}
 {}^T T_{\mu\nu}^{\phantom{\mu\nu}\rho} = \epsilon^{\lambda \tau}_{\phantom{\lambda
\tau}\mu\nu} V_{[\lambda }e^{A}_{\tau]}
C^{\phantom{A}B}_{A}
e^{\rho}_{B} \,,
\end{equation}
where $C^{\phantom{A}B}_{A}$ is an arbitrary matrix. By the null-trace conditions
\begin{equation}\label{condit1}
 V_{[\mu}e^{A}_{\nu]}C^{\phantom{A}B}_{A}e^{\nu}_{B} =0 \,,
\end{equation}
and
\begin{equation}\label{condit2}
 \epsilon^{\lambda \tau}_{\phantom{\lambda \tau}\mu\nu} V_{[\lambda}e^{A}_{\tau]}
C^{\phantom{A}B}_{A} e^{\nu}_{B} =0 \,,
\end{equation}
on (\ref{ttor1}) and (\ref{ttor2}), we obtain 7 constraints on the matrix
$C^{\phantom{A}B}_A$, by
fixing $V_\mu$. As a consequence, $C^{\phantom{A}B}_A$ has 9 independent components. In
order to get the 16 components of the T-torsion from the expressions (\ref{ttor1}) and
(\ref{ttor2}), we have to impose a further condition. If $V^{2}\equiv V^{\mu}V_{\mu}\neq
0$,
we can impose that
\begin{equation}\label{condit3}
 C^{\phantom{A}B}_{A} e^{A}_{\mu} e^{\nu}_{B} V^\mu V_\nu =0 \,,
\end{equation}
which reduces one of the constraints following from (\ref{condit1}) to
\begin{equation}\label{trace}
 C^{\phantom{A}A}_{A}=0\, .
\end{equation}
If $V$ is a null vector, the constraint (\ref{condit3}) follows from (\ref{condit1}) and
the expression (\ref{trace}) has to be imposed as a supplementary constraint on the
matrix $C^{\phantom{A}B}_A$.

From the previous discussion, it follows that the T-torsion tensors can be classified
according to the nature of the vector $V_\mu$ which can be time-like, space-like, or
null.
We label the T-torsion with $Tt$, $Ts$ or $T\ell$ depending on whether the $4$-vector
$V^\mu$ is time-like, space-like or light-like, respectively.

\subsection{The Einstein-Cartan-Sciama-Kibble equations}\label{ecks}

The introduction of torsion as an extension of the gravitational field theories has some
relevant consequences. The closest theory to GR is the  Einstein-Cartan-Sciama-Kibble
(ECSK) theory. It is described by
\begin{eqnarray} \label{t6}
 L =\sqrt{-g} \left( \frac{R}{16 \pi G} +{\cal L}_{m} \right) ~,
\end{eqnarray}
which is the Lagrangian density of GR depending on the metric tensor $g_{\mu\nu}$ and on
the
connection
$\Gamma_{\mu\nu}^{\rho}$, where $R$ is the curvature scalar (\ref{curvscalar}) and ${\cal
L}_{m}$
the
Lagrangian function of matter fields, which yields
\begin{eqnarray} \label{t7}
 \mbox{{\bf t}}^{\mu\nu} =\frac{\delta {\cal L}_{m}}{\delta g_{\mu\nu}} ~,
 \end{eqnarray}
which is the symmetric stress--energy tensor, while
\begin{eqnarray} \label{t8}
 \tau_{\rho}^{\phantom{\rho}\nu\mu}=\frac{\delta {\cal L}_{m}}{\delta
K_{\mu\nu}^{\phantom{\mu\nu}\rho}}
\end{eqnarray}
is the source of torsion. In many instances, it can be identified with a spin density.
However, as will become clear from the following sections, there are many considerations
in which the source of the torsion field (\ref{t8}) is not spin but gravity itself.

From the variation of (\ref{t6}) and introducing the canonical energy-momentum tensor
\begin{equation}\label{canonical}
 {\bf \Sigma^{\mu\nu}}= {\bf t^{\mu\nu}} + {}^*\tilde\nabla_{\rho}(\tau^{\mu\nu\rho}
-\tau^{\nu \rho \mu}
+\tau^{c\mu\nu}) \, ,
\end{equation}
where we have used the abridged notation ${}^*\tilde\nabla_{\rho}:=\tilde\nabla_{\rho}
+2T_{\rho\sigma}^{\phantom{c\sigma}\sigma}
$, the following field equations are derived  \cite{Hehl:1976kj}:
\begin{equation}\label{t9}
 G^{\mu\nu}=8 \pi G {\bf\Sigma}^{\mu\nu}\, ,
\end{equation}
and
\begin{equation} \label{t10}
 T_{\mu\nu}^{\phantom{\mu\nu}\rho}=8 \pi G\tau_{\mu\nu}^{\phantom{\mu\nu}\rho}\,,
\end{equation}
where we have set the light speed as $c=1$.

Eq. (\ref{t9}) generalizes the Einstein equations in a $U_4$. Unlike (\ref{t9}), Eq.
(\ref{t10}) is algebraic and thus it is always possible to cast (\ref{t9}) in
a pure Einstein one, by substituting the torsion terms with their sources. It results in
defining an effective energy-momentum tensor as the source of the Riemannian part of the
Einstein tensor \cite{Hehl:1976kj}. Doing so one obtains
\begin{equation}
\label{t11}
 G^{\mu\nu}(\{\})=8 \pi G\tilde{\mbox{{\bf t}}}^{\mu\nu} ~,
\end{equation}
where $G^{\mu\nu}(\{\})$ is the Riemannian part of the Einstein tensor. The effective
energy-momentum tensor is
\begin{align}
\label{t12}
 \tilde{\mbox{{\bf t}}}^{\mu\nu} =& \mbox{{\bf t}}^{\mu\nu} +8 \pi G\bigg[
-4\tau^{\mu\rho}_{\phantom{\mu\rho}[\sigma}
\tau^{\nu\sigma}_
{\phantom{\nu\sigma}\rho]} -2\tau^{\mu\rho\sigma}\tau^{\nu}_{\rho\sigma}
+\tau^{\rho \sigma \mu}\tau_{\rho\sigma}^{\phantom{\rho\sigma}\nu}
\nonumber\\
 & \ \ \ \ \ \ \ \ \ \ \ \ \ \ \, +
\frac{1}{2}g^{\mu\nu}(4\tau^{\phantom{\lambda}\rho}_{\lambda\phantom{\rho}[\sigma}\tau^{
\lambda\sigma} _{ \phantom{ \lambda\sigma}\rho] }
+\tau^{\lambda \rho\sigma}\tau_{ \lambda \rho\sigma}) \bigg] ~.
\end{align}
The tensor ${\bf t}^{\mu\nu}$ can be of the form
\begin{eqnarray}\label{t13}
 \mbox{{\bf t}}^{\mu\nu} = (p+\rho)u^{\mu}u^{\nu}-pg^{\mu\nu} ~,
\end{eqnarray}
if standard perfect-fluid matter is considered. But when spin fluids are considered, one
has to
define a different stress-energy tensor in which the spin contributions are taken into
account as
in  \cite{Ritis:1983vb,deRitis:1985uw, drls, Ray:1982qr,Ray:1982qq, Obukhov:1987yu}.

%%%%%%%%%%%%%%%%%%%%%%%%%%%%%%
\subsection{Realizations of torsion tensors}\label{examples}
%%%%%%%%%%%%%%%%%%%%%%%%%%%%%%%%%

Let us now show how some torsion tensors, frequently found in literature, can be
classified according to the irreducible tensors classification given above.
\begin{enumerate}
\item
Scalar fields $\phi$ produce torsion only in non-minimally coupled theories with a $\xi
\phi^2 R$ term in the Lagrangian density, or in a $R^2$ theory in a {\bf U}$_4$ (where
the Ricci scalar is coupled to itself). As a result, the torsion is related to the
gradient of the field. For example, in homogeneous cosmologies, one obtains a $Vt$
tensor.
In a Schwarzschild solution one deals with a $Vs$ tensor. See for example
\cite{Soleng:1988wa, German:1986uw, cosimo,SIV,DeSabbata:1994nq}.

\item
According to  \cite{Tsamparlis:1981xm} and  \cite{goenner}, it turns out that the only
torsion tensors compatible with a  Friedmann-Robertson-Walker (FRW) universe are of class
$ Vt$ and $ At$. A cosmological solution with a torsion tensor of class $At$ is discussed
also in \cite{Minkowski:1986kv}.

\item
Examples of torsion of class $V \ell$ and $A\ell$ are found in
\cite{Davis:1978am,Atkins:1978ak} to describe null electromagnetic plane waves.

\item
$As$ and $Vs$ tensors introduce anisotropies in a space-time, since the spacelike vector
yields a privileged direction.

\item
The spin of classical Dirac particles is the source of an $As$ torsion for massive
particles and of an $An$ torsion for a massless neutrino  \cite{Hehl:1976kj}. The $At$
torsion is generated by tachyon Dirac particles.

\item
An example of T-torsion tensors can be found in simple supergravity where torsion is
given in terms of the Rarita-Schwinger spinors (see, for example,
 \cite{Freedman:1976xh,Freedman:1976py}). They contribute also to
torsion in the Weyssenhoff spin fluids (see below and discussion in  \cite{bauerle}).

\item
The Lanczos tensor was considered in  \cite{ Davis:1978am,Atkins:1978ak} as a candidate
of torsion in a non-ECSK theory. It is a sort of Weyl tensor potential and it bears all
the characteristics of a traceless torsion tensor. Then its properties depend on the
symmetries of space-time.

\item
The influence of an $At$ torsion on cosmological perturbations is discussed in
\cite{ZZI}.

\item
The helicity flip of fermions can be induced by an $Al$ torsion
 \cite{Capozziello:1999nt,Hammond:1996ua}.

\item
The same kind of torsion can induce a geometrical contribution to the Berry phase of
Dirac particles  \cite{berry}.\\

\noindent
Finally, the next group of examples is related to elementary torsion tensors found in the
literature:

\item
The torsion tensors related to spin, usually found in the literature, are generated by
the Weyssenhoff spinning particle and the classical Dirac particle. In the first case
the torsion tensor is a $Bs$ tensor, while in the second case one has an $As$ tensor.
Spin fluids {\it \`a la} Weyssenhoff can be found in
\cite{Hehl:1976kj,Ritis:1983vb,deRitis:1985uw,Ray:1982qr,Ray:1982qq,Obukhov:1987yu,
Trautman:1973wy, tafel}, and have been discussed by many other authors.

\item
Cosmological models with a $Bs$ torsion have been studied in  \cite{tafel}.
\end{enumerate}

%%%%%%%%%%%%%%%
\subsection{Torsion contributions to the energy-momentum tensor}\label{emtens}

After straightforward calculation, one obtains that the
contribution of the antisymmetric and vector parts of torsion to
the energy-momentum tensor are respectively proportional to the
following expressions:
\begin{eqnarray}
\label{emtens1}
 {}^A{\bf t}_{\mu}^{\nu} = 2\sigma^\nu\sigma_{\mu}
+\delta_{\mu}^{\nu}\sigma^\rho\sigma_\rho
\end{eqnarray}
and
\begin{eqnarray}
\label{emtens2}
 {}^V{\bf t}_\mu^\nu = \frac{8}{3}T^\nu T_\mu -\frac{4}{3}\delta^\nu_\mu T^\rho T_\rho \,.
\end{eqnarray}

The contribution of the T-torsion, when expressed from (\ref{ttor1}) reads
\begin{align}
\label{ttorsionem1}
 {}^{T_{1}}{\bf t}^{\mu\nu} =& -C^{\rho\sigma}C_{(\rho\sigma)}V^{\mu}V^{\nu}
-V^{\rho}C_{\rho\sigma}V^{(\mu}C^{\nu)\sigma} \nonumber\\
 & +\frac{1}{2}V^{\rho}V_{\rho}(C_{\tau}^{\phantom{\tau}\mu}C^{\tau \nu}
-C_{\phantom{\mu}\tau}^{\mu}C^{\nu \tau})
-\frac{1}{2}V^{\rho}
V^{\sigma}C_{\rho}^{\phantom{\rho}\mu}C_{\sigma}^{\phantom{\sigma}\nu} \nonumber\\
 & +\frac{1}{2}g^{\mu\nu}(C^{\rho\sigma}C_{(\rho\sigma)}V^{\tau}V_{\tau}
-\frac{1}{2}V^{\rho}C_{\rho\sigma}V^{\tau}C_{\tau}^{\phantom{\tau}\sigma}) ~,
\end{align}
otherwise when the T-torsion is expressed by (\ref{ttor2}) it writes as
\begin{align}
\label{ttorsionem2}
 {}^{T_{2}} {\bf t}^{\mu\nu} =& C^{\rho\sigma}C_{\rho\sigma}V^{\mu}V^{\nu}
+V^{\sigma}V_{\sigma}(C_{\tau}^{\phantom{\tau}\mu}C^{\tau \nu}
+C_{\phantom{\mu}\tau}^{\mu}C^{\nu \tau})
 \nonumber\\
 & -V^{\rho}V^{\sigma}C_{\rho}^{\phantom{\rho}\mu}C_{\sigma}^{\phantom{\sigma}\nu}
-V^{\tau}C_{\tau\sigma}(V^{\mu}C^{\nu\sigma}
+V^{\nu}C^{\mu\sigma}) \nonumber\\
 & +\frac{1}{2}(V^{\rho}V^{\sigma}C_{\rho \tau}C_{\phantom{\tau}\sigma}^{\tau}
-V^{\tau}V_{\tau}C^{\rho\sigma}C_{\rho\sigma}) ~.
\end{align}
In (\ref{ttorsionem1}) and (\ref{ttorsionem2}) we have used the expression
$C_{\mu}^{\phantom{\mu}\nu}
= C_{A}^{\phantom{A}B} e^{A}_{\phantom{A}\mu} e_{B}^{\phantom{B}\nu}$. The presence of
contributions of distinct irreducible tensors does not lead to interaction terms, apart
from the cases where the two classes of T-torsion are present.

In summary, an elementary torsion tensor
$T_{\mu\nu}^{\phantom{\mu\nu}\rho}=F_{\mu\nu}\Sigma^\rho$,
contributes to the
energy-momentum tensor with a symmetric tensor proportional to
\begin{eqnarray}
 {}^e{\bf t}_\mu^\nu =-2\Sigma^2F^{\nu\rho}F_{\mu\rho} +F^2\Sigma^\nu\Sigma_\mu
-\frac{1}{2}
F^2\Sigma^2\delta_\mu^\nu \,,
\label{t12bb}
\end{eqnarray}
where $\Sigma^2=\Sigma_\mu\Sigma^\mu$ and $F^2=F^{\mu\nu}F_{\mu\nu}$. Expression
(\ref{t12bb}),
through (\ref{t4}) and (\ref{t10}), is the final result involving also ordinary
perfect-fluid matter.

%%%%%%%%%%%%%%%%%%%%%%%%%%%%%%%%
\subsection{Torsion contributions to shear,
expansion, vorticity and acceleration}\label{svea}
%%%%%%%%%%%%%%%%%%%%%%%%%%%%%%%%%%%%

It has often been pointed out in the literature how torsion can
modify the behavior of fluids. In \cite{drls} it was shown that
the presence of torsion generated by a Weyssenhoff fluid
generalizes the Bernoulli theorem, through an extension of the
definition of the vorticity. In the same way such a modification
of the vorticity has led some authors to argue about the
possibility of having cosmological models with torsion which
could avert the initial singularity  \cite{Trautman:1973wy}. An extended
analysis of this problem has been made by re-writing the
Raychaudhuri equation in the presence of torsion for a
Weyssenhoff fluid \cite{Stewart:1973ux,Esposito:1990ub}.

In \cite{Demianski:1987hq} an inflationary Bianchi I
universe in the ECSK theory was considered. In this paper it was shown how
torsion could contribute to an isotropic expansion universe even
in anisotropic geometry, if the energy density of spin was
sufficiently large to counterbalance the anisotropic terms. As a
result, it followed that this model supplies a rapid isotropization
mechanism of the universe during the inflationary stage.

In  \cite{Ellis:1989jt,Ellis:1989ju,Ellis:1990gi} a gauge invariant and covariant
formalism for cosmological perturbations was formulated. In this derivation an
important role is attributed to the Raychaudhuri equation. Such
formulation has been extended recently in \cite{Palle:1998qf} for the
ECSK theory. It follows that one could construct tests for torsion in the
primordial universe through its effects on the spectrum of
perturbations. A complete study of perturbations for all the
irreducible torsion tensors can be useful to extend this program.

The previous considerations suggest to consider  how the
kinematical quantities are modified by each of the irreducible
components of torsion.

%%%%%%%%%%%%%%%%%%%%
\subsubsection{The kinematical quantities}
\begin{table*}[ht]
\begin{center}
 \begin{tabular}{|c|c|c|c|c|c|c|c|}\hline
&      &      \\
    & ${}^V T_{\mu\nu}^{\phantom{\mu\nu}\rho}$ & ${}^A
    T_{\mu\nu}^{\phantom{\mu\nu}\rho}$\\ \hline
    & & \\
  $\tilde\theta=$ &   $\theta-2T^{\rho}U_{\rho}$ & $\theta$ \\ \hline
   & &  \\
  $\tilde a_{\nu}=$ &  $a_{\nu}-T_{\nu}-T_{\mu}U^{\mu}U_{\nu}$ & $a_{\nu}$ \\ \hline
  & &  \\
  $\tilde\omega_{\mu\nu}=$ &   $\omega_{\mu\nu}$ &
$\omega_{\mu\nu}-\epsilon_{\mu\nu\rho\sigma}\sigma^{\sigma}U^{\rho}$
\\
\hline
  & & \\
  $\tilde\sigma_{\mu\nu}=$ &   $\sigma_{\mu\nu}$ & $\sigma_{\mu\nu}$ \\ \hline
\end{tabular}
\caption{Contributions of V-torsion and A-torsion to the
kinematical quantities.  }
\label{tableIold}
\end{center}
\end{table*}
\begin{table*}[ht]
\begin{center}
 \begin{tabular}
 {|c|c|c|c|c|c|c|c|}
 \hline
&      &
 \\
  & ${}^{T_1} T_{\mu\nu}^{\phantom{\mu\nu}\rho}$ &${}^{T_{2}}
T_{\mu\nu}^{\phantom{\mu\nu}\rho}$  \\ \hline
    & & \\
  $\tilde\theta=$ & $\theta$ & $\theta$     \\ \hline
   & &  \\
  $\tilde a_{\nu}=$ &$a_{\nu}+2V_{[\nu}C_{\mu]\rho}U^{\rho}U^{\mu} $&
$a_{\nu}-2\epsilon_{\lambda \tau \sigma\nu}V^{\lambda}C_{\lambda\rho}U^{\sigma}U^{\mu}
$       \\ \hline
  & &  \\
  $\tilde\omega_{\mu\nu}=$ & $\omega_{\mu\nu}+h_{\mu}^{\rho}h_{\nu}^{\sigma}V_{[\rho}
C_{\sigma]}^{\phantom{\sigma]}\lambda}U_{\lambda}$&
$\omega_{\mu\nu}+h_{\mu}^{\rho}h_{\nu}^{\sigma}\epsilon_{\lambda \tau
\rho \sigma}V^{\lambda}
C^{\tau \xi}U_{\xi}$  \\ \hline
  & & \\
  $\tilde\sigma_{\mu\nu}=$ &$\sigma_{\mu\nu}- 2
h_{\mu}^{\rho}h_{\nu}^{\sigma}(V_{\lambda}C_{(\rho\sigma)} - C_{\lambda(\rho}V_{\sigma)}
)U^{\lambda}$&
   $\sigma_{\mu\nu}- 2
h_{\mu}^{\rho}h_{\nu}^{\sigma}\epsilon_{\lambda
\tau \xi(\mu}V^{\lambda}C^{\tau}_{\phantom{\tau}\sigma)}U^{\xi}$
\\
\hline
\end{tabular}
\caption{ Contributions of the two T-torsions to the
kinematical quantities.  }
\label{tableIIold}
\end{center}
\end{table*}

One of the consequences of introducing torsion in a
space-time is that the definition of some quantities can be
modified. This is the case of the kinematical quantities,
defined from the following decomposition of  the covariant
derivative of the four velocity $U_\mu$  \cite{hawking}:
\begin{equation}\label{4vel}
 \tilde\nabla_{\mu}U_{\nu}= \tilde\sigma_{\mu\nu} + \frac{1}{3} h_{\mu\nu}
 \tilde\theta+\tilde\omega_{\mu\nu} -U_{\mu}\tilde a_{\nu},
\end{equation}
where $h_{\mu\nu}=g_{\mu\nu} +U_\mu U_\nu$, and
\begin{align}
\label{expansion}
 &\tilde\theta= \tilde\nabla_{\mu}U^{\mu}=\theta - 2T^{\rho}U_{\rho}, \\
\label{shear}
 &\tilde\sigma_{\mu\nu}
 =h_\mu^\rho h_\nu^\sigma\tilde\nabla_{(\rho}U_{\sigma)}
 =\sigma_{\mu\nu} + 2h_\mu^\rho h_\nu^\sigma K_{(\rho 
\sigma)}^{\phantom{(ab)}\lambda}\,U_{\lambda},
 \\
\label{vorticity}
 &\tilde\omega_{\mu\nu}
 =h_\mu^\rho h_\nu^\sigma\tilde\nabla_{[\rho}U_{\sigma]}
 =\omega_{\mu\nu} + 2h_\mu^\rho h_\nu^ \sigma 
K_{[\rho\sigma]}^{\phantom{[\rho\sigma]}\lambda}\,U_{\
lambda}, \\
\label{acceleration}
 &\tilde a_{\rho}
 = U^{\mu}\tilde\nabla_{\mu}U_{\rho}
 = a_{\rho}+ U^{\mu}K_{\mu\rho}^{\phantom{\mu\rho}\sigma}\,U_\sigma,
\end{align}
are respectively the expansion, shear, vorticity and acceleration. The quantities without
the tilde are those usually defined in GR. In Tables \ref{tableIold} and \ref{tableIIold}
we summarize the contributions to these objects given by the irreducible torsion tensors.

%%%%%%%%%%%%%%%%%%%%%%%%
\subsubsection{The Raychaudhuri
equation}
%%%%%%%%%%%%%%%%%%%%%%%%%%%%%

Given the four-velocity $U_\mu$ ($U_\mu U^\mu =-1$), having in mind
the identity
\begin{equation}\label{dd}
 U^{\nu}\tilde\nabla_{\rho}\tilde\nabla_{\nu}U_{\mu}=\tilde\nabla_{\rho}
 (U^{\nu}\tilde\nabla_{\nu}U_{\mu}) - \tilde\nabla_{\rho}U^{\nu}
 \tilde\nabla_{\nu}U_{\mu},
\end{equation}
and using Eq.  (\ref{Riemann}) to obtain
\begin{equation}
\label{ddd}
 U^{\nu}\tilde\nabla_{\rho}\tilde\nabla_{\nu}U_{\mu}=
 U^{\nu}\tilde\nabla_{\nu}\tilde\nabla_{\rho}U_{\mu} +
 R_{\rho \nu \mu}^{\phantom{\rho \nu\mu}\sigma}U_{\sigma}U^{\nu} - 2
 U^{\nu}T_{\mu\nu}^{\phantom{\mu\nu}\rho} \tilde\nabla_{\sigma}U_{\rho},
\end{equation}
we result to the equation
\begin{align}
\label{penultima}
 \frac{1}{3}h_{\rho\mu}\tilde\theta
 & +\tilde\sigma_{\rho\mu}+\tilde\omega_{\rho\mu}- U_{\rho}\tilde a_{\mu} = \nonumber\\
 & \tilde\nabla_{\rho}\tilde a_{\mu} -\Big( \frac{1}{9}h_{\rho\mu}\tilde\theta
 +\frac{2}{3}\tilde\theta\tilde\sigma_{\rho\mu} + 
\frac{2}{3}\tilde\theta\tilde\omega_{\rho\mu} \
nonumber\\
 & +2\tilde\sigma_{\rho}^{\nu} \tilde\omega_{\nu\mu} \tilde\sigma_{\rho}^{\nu} 
\tilde\sigma_{\nu\mu}
 + \tilde\omega_{\rho}^{\nu} \tilde\omega_{\nu\mu} - \frac{1}{3}U_{\rho}\tilde\theta\tilde 
a_{\mu} \
nonumber\\
 & -U_{\rho}\tilde a^{\nu} \tilde\sigma_{\nu\mu}-U_{\rho}\tilde a^{\nu} 
\tilde\omega_{\nu\mu}\Big) -
R_{\rho \nu \mu}^{\phantom{\rho \nu \mu}\sigma}\,U_{\sigma}U^{\nu} \nonumber\\
 & -2 U^{\nu}T_{\mu\nu}^{\phantom{\mu\nu}\sigma} 
\Big(\frac{1}{3}h_{\rho\sigma}\tilde\theta + \
tilde\rho_{\rho\sigma} +\tilde\omega_{\rho\sigma} - U_{\rho}\tilde a_{\sigma}\Big)\,.
\end{align}
Contracting the indices in (\ref{penultima}) one obtains immediately the most general
expression for the Raychaudhuri equation, namely
\begin{eqnarray}\label{raychau}
&&\dot{\tilde\theta}=\tilde\nabla_{\rho}\tilde a^{\rho}
-\frac{1}{3}\theta^{2} -\tilde\sigma^{\mu\nu}\tilde\sigma_{\mu\nu}
+ \tilde\omega^{\mu\nu}\tilde\omega_{\mu\nu} - R_{\mu\nu}U^{\mu}U^{\nu} \nonumber\\&&
\ \ \ \ \ \,
-2
U^{\nu}T_{\mu\nu}^{\phantom{\mu\nu}\sigma}
 \Big( \frac{1}{3}h_{\sigma}^{\mu}\tilde\theta +
\tilde\sigma_{\sigma}^{\mu}+\tilde\omega_{\sigma}^{\mu}- U_{\sigma}\tilde
a^{\mu} \Big) .
\end{eqnarray}
This is the most general form of the Raychaudhuri equation in presence of torsion. Simpler 
forms of 
this equation have been discussed in \cite{Stewart:1973ux, Esposito:1990ub, raych}.

%%%%%%%%%%%%%%%%%%%%%%%%%%%
%\subsection{Some remarks on torsion}
%%%%%%%%%%%%%%%%%%%%%%%%
As final remark, we have to say that a geometrical classification of torsion tensors is
possible. A decomposition of torsion into irreducible tensors is  given also in
\cite{McCrea:1992wa,Tsamparlis:1981xm}. For a systematic account, see \cite{Hehl:1994ue}.
Essentially, one has three classes of tensors: traceless, vector and totally
antisymmetric
ones. However, it is possible to add a classification scheme to this decomposition. The
proposal is based on the space-time properties of 4-vectors and bivectors, which can be
used to construct these torsion tensors. According to this classification,  it is
possible
to construct two tensors of the same irreducible class, with distinct properties, due to
the fact that one can use space-like, time-like, or null 4-vectors.

As a byproduct, one finds  also a second decomposition and
classification scheme based on elementary torsions. These
elementary tensors are given by the tensor product of simple
bivectors and vectors.  As a consequence, the classification of
these tensors is based on the space-time properties of the simple
bivectors (which we distinguished in {\it electric} and {\it
magnetic} bivectors), and on those of the vectors.

All the above classifications can help to distinguish
the physical situations associated to different torsion tensors.

%%%%%%%%%%%%%%%%%%%%%%%%%%%%%%%
%\pagebreak
%%%%%%%%%%%%%%%%%%%%%%%%%%%%%%%%%%%%%%%%%%
\section{ The case of Poincar\'e Gauge Gravity with torsion}
 \label{SecionPoincare}

A theory where torsion plays an important role is the so called Poincar\'e Gauge Gravity.
Following  the prescriptions of GR, the physical space-time is assumed to be a
four-dimensional differential manifold (see
\cite{Capozziello:2007ec,Capozziello:2009zza}
for a general discussion on gravity theories and their prescriptions). In Special
Relativity this manifold is the Minkwoski flat-space-time $M_{4}$, while in General
Relativity the underlying space-time is assumed to be curved in order to describe the
effects of gravitation.

Utiyama  \cite{Utiyama:1956sy} was the first to propose that General Relativity can be
seen as a gauge theory based on the local Lorentz group $SO(3$, $1)$, in much the same
way
that the Yang-Mills gauge theory \cite{Yang:1954ek} was developed on the basis of the
internal iso-spin gauge group $SU(2)$. In this formulation the Riemannian connection is
the gravitational counterpart of the Yang-Mills gauge fields. While $SU(2)$, in the
Yang-Mills theory, is an internal symmetry group, the Lorentz symmetry represents the
local nature of space-time rather than internal degrees of freedom. The Einstein
Equivalence Principle, asserted for GR, requires that the local space-time structure can
be identified with the Minkowski space-time possessing Lorentz symmetry. In order to
relate local Lorentz symmetry to the external space-time, we need to solder the local
space to the external space. The soldering tools are the tetrad fields. Utiyama regarded
the tetrads as objects given \textit{a priori}.

Soon after, Sciama \cite{Sciama} recognized that space-time should necessarily be endowed
with torsion in order to accommodate spinor fields. In other words, the gravitational
interaction of spinning particles requires the modification of the Riemann space-time of
GR, towards a (non-Riemannian) curved space-time with torsion. Although Sciama
used the tetrad formalism for his gauge-like handling of gravitation, his theory fell
shortcomings in treating tetrad fields as gauge fields.

Kibble \cite{Kibble:1961ba} made a comprehensive extension of the Utiyama gauge theory of
gravitation by showing that the local Poincar\'{e} symmetry $SO(3,1)\rtimes T\left(
3,1\right) $ ($\rtimes $ represents the semi-direct product) can generate a space-time
with torsion as well as curvature. The gauge fields introduced by Kibble include the
tetrads as well as the local affine connection. There have been a variety of gauge
theories of gravitation based on different local symmetries, which gave rise to several
interesting applications in theoretical physics  \cite{Grignani:1991nj,Hehl:1994ue,
Inomata:1979am,Ivanov:1981wn, Ivanov:1981wm,
Mansouri:1976df, Mansouri:1978pr, Yan:1982tq,Ivanenko:1984vf, Sardanashvily:2002vq,
Chang:1975fr} (for a review see  \cite{Hehlbook}).
Following the Kibble approach, it can be demonstrated how gravitation can be formulated
starting from a pure gauge viewpoint. In particular, the aim of this Section is to show,
in details, how a theory of gravitation is a gauge theory which can be obtained starting
from the local Poincar\'{e} symmetry.

In \cite{Ali:2007hu}, a gauge theory of gravity based on a nonlinear realization of the
local conformal-affine group of symmetry transformations was formulated. Moreover, the
coframe fields and gauge connections of the theory were obtained, and the tetrads and
Lorentz group metric were used to induce a space-time metric. In particular, the
inhomogenously transforming (under the Lorentz group) connection coefficients gave rise to
gravitational gauge potentials used to define covariant derivatives accommodating minimal
coupling of matter and gauge fields. On the other hand, the tensor valued connection
forms
were used as auxiliary dynamical fields associated with the dilation, special conformal
and deformation (shear) degrees of freedom inherent to the bundle manifold. This allowed
to define the bundle curvature of the theory. Then boundary topological invariants were
also constructed, serving as a prototype (source free) gravitational Lagrangian. Finally
the Bianchi identities, covariant field equations and gauge currents were extracted.

Starting from the Invariance Principle, we consider first the Global Poincar\'{e}
Invariance and then the Local Poincar\'{e} Invariance. This approach leads to formulate a
given theory of gravity as a gauge theory. This viewpoint, if considered in detail, can
avoid many shortcomings  and could be useful to formulate self-consistent schemes for
quantum gravity.

Before entering into the main discussion of this part, let us introduce the notations.
The metric in Minkowskian space $M_{4}$ is denoted by $\eta _{AB}~(A$,
$B=0$, $1$, $2$, $3)~$ with $\eta _{00}=-\eta _{11}=-\eta
_{22}=-\eta _{33}=1$ and $\eta _{AB}=0$ for $A\neq B$, whereas
the metric of curved space is given by $g_{\mu \nu }~(\mu $, $\nu =0$, $1$, $
2$, $3)$. The tetrads $e_{\mu }^{A}$ are defined by $g_{\mu \nu }=e_{\mu
}^Ae_{\nu }^B\eta _{AB}$. In particular, for holonomic coordinates $x^{\mu }$ the tetrads
are given by $e_{\mu }^{A}=\partial x^{A}/\partial x^{\mu }$.

%\section{Global Poincar\'e invariance}
%%%%%%%%%%%%%%%%%%%%%%%
\subsection{Invariance Principle}
%%%%%%%%%%%%%%%%%%%%%%%%%%%

As it is well-known, the field equations and conservation laws can be obtained from a
least action principle. Since this principle is the basis of any gauge theory, we start
from it to develop our considerations. In particular, we start from a least action
principle and the Noether theorem.

Let $\chi (x)$ be a multiplet field defined at a space-time point $x$ and $%
\mathcal{L}\{\chi (x)$, $\partial _{B}\chi (x)$; $x\}$ be the
Lagrangian density of the system. The action integral of the
system over a given space-time volume $\Omega $ is defined by
\begin{equation}
I(\Omega )=\int_{_{\Omega }}\mathcal{L}\{\chi (x)\text{, }\partial _{B}\chi
(x)\text{; }x\}\,d^{4}x\text{.}
\end{equation}%
Now let us consider the infinitesimal variations of the
coordinates
\begin{equation}
x^{A}\rightarrow x^{\prime }{}^{A}=x^{A}+\delta x^{A}\text{,}
\end{equation}%
and the field variables%
\begin{equation}
\chi (x)\rightarrow \chi ^{\prime }(x^{\prime })=\chi (x)+\delta \chi (x)%
\text{.}
\end{equation}%
Correspondingly, the variation of the action is given by
\begin{eqnarray}
&&\delta I=\int_{_{\Omega ^{\prime }}}\mathcal{L}^{\prime }(x^{\prime
})\,d^{4}x^{\prime }-\int_{_{\Omega }}\mathcal{L}(x)\,d^{4}x\nonumber\\
&&\ \ \ \,
=\int_{_{\Omega
}}\left[ \mathcal{L}^{\prime }(x^{\prime })||\partial _{B}x^{\prime
}{}^{B}||-\mathcal{L}(x)\right] \,d^{4}x\text{.}
\end{eqnarray}%
Since the Jacobian for the infinitesimal variation of coordinates becomes
$
||\partial _{B}x^{\prime }{}^{B}||=1+\partial _{B}(\delta x^{B})
$,
 the variation of the action takes the form
\begin{equation}
\delta I=\int_{_{\Omega }}\left[ \delta \mathcal{L}(x)+\mathcal{L}%
(x)\,\partial _{B}(\delta x^{B})\right] \,d^{4}x  \label{Action1}
\end{equation}%
where $
\delta \mathcal{L}(x)=\mathcal{L}^{\prime }(x^{\prime })-\mathcal{L}(x)$.

For any function $\Phi (x)$ of $x$, it is convenient to define the fixed
point variation $\delta _{0}$ by
\begin{equation}
\delta _{0}\Phi (x):=\Phi ^{\prime }(x)-\Phi (x)=\Phi ^{\prime }(x^{\prime
})-\Phi (x^{\prime })\text{.}
\end{equation}%
Expanding the function to first order in $\delta x^{B}$ as $
\Phi (x^{\prime })=\Phi (x)+\delta x^{B}\,\partial _{B}\Phi (x)$,
we obtain
\begin{eqnarray}
&&\!\!\!\!\!\!\!\!\!\!\!
\delta \Phi (x)=\Phi ^{\prime }(x^{\prime })-\Phi (x)=\Phi ^{\prime
}(x^{\prime })-\Phi (x^{\prime })+\Phi (x^{\prime })-\Phi (x)\nonumber\\
&&\ \ \
=\delta
_{0}\Phi (x)+\delta x^{B}\,\partial _{B}\Phi (x)\text{,}
\end{eqnarray}%
or
\begin{equation}
\delta _{0}\Phi (x)=\delta \Phi (x)-\delta x^{B}\partial _{B}\Phi (x)\text{.}
\end{equation}%
The advantage to have the fixed point variation is that $\delta
_{0}$ commutes with $\partial _{B}$, namely
$
\delta _{0}\partial _{B}\Phi (x)=\partial _{B}\delta _{0}\Phi (x)$.
Hence, for $\Phi (x)=\chi (x)$, we have
\begin{equation}
\delta \chi =\delta _{0}\chi +\delta x^{A}\partial _{A}\chi \text{,}
\end{equation}%
and%
\begin{equation}
\delta \partial _{A}\chi =\partial _{A}(\delta _{0}\chi )-\partial (\delta
x^{B})\partial _{A}\chi \text{.}
\end{equation}

Using the fixed point variation in the integrand of
(\ref{Action1}) gives
\begin{equation}
\delta I=\int_{_{\Omega }}\left[ \delta _{0}\mathcal{L}(x)+\partial
_{B}(\delta x^{B}\,\mathcal{L}(x))\right] \,d^{4}x\text{.}  \label{Action2}
\end{equation}%
If we require the action integral defined over any arbitrary region $\Omega $ to
be invariant, i.e. $\delta I=0$, then we must have
\begin{equation}
\delta \mathcal{L}+\mathcal{L}\partial _{B}(\delta x^{B})=\delta _{0}%
\mathcal{L}+\partial _{B}(\mathcal{L}\delta x^{B})=0\text{.}
\end{equation}%
If $\partial _{B}(\delta x^{B})=0$, then $\delta \mathcal{L}=0$, that is
the Lagrangian density $\mathcal{L}$ is invariant. In general, however, $
\partial _{B}(\delta x^{B})\neq 0$, and $\mathcal{L}$ transforms like a
scalar density. In other words, $\mathcal{L}$ is a Lagrangian density unless
$\partial _{B}(\delta x^{B})=0$.

For convenience, let us introduce a function $h(x)$ that behaves like a
scalar density, namely
\begin{equation}
\delta h+h\partial _{B}(\delta x^{B})=0\text{.}
\label{hxdefinition}
\end{equation}%
We further introduce the function $L$ through $\mathcal{L}(\chi ,\partial _{B}\chi
;x)=h(x)L(\chi ,\partial _{B}\chi ;x)$. Then we see that
\begin{equation}
\delta \mathcal{L}+\mathcal{L}\partial _{B}(\delta x^{B})=h\delta L,
\end{equation}
and hence the action integral remains invariant if
$\delta L=0$.
The newly introduced function $L(\chi ,\partial _{B}\chi ;x)$ is the scalar
Lagrangian of the system.

Let us calculate the integrand of (\ref{Action2}) explicitly. The fixed
point variation of $\mathcal{L}(x)$ is a consequence of a fixed point
variation of the field $\chi (x)$, which can be cast into the form
\begin{equation}
\delta _{0}\mathcal{L}=[\mathcal{L}]_{\chi }\delta _{0}\chi +\partial
_{B}\left[ \frac{\partial \mathcal{L}}{\partial (\partial _{B}\chi )}\delta
_{0}\chi \right],
\end{equation}
where
\begin{equation}
\lbrack \mathcal{L}]_{\chi }\equiv \frac{\partial \mathcal{L}}{\partial \chi
}-\partial _{B}\left[ \frac{\partial \mathcal{L}}{\partial (\partial
_{B}\chi )}\right] \text{.}
\end{equation}%
Consequently, we can write the action integral in the form
\begin{equation}
\delta I=\int_{_{\Omega }}\left\{ [\mathcal{L}]_{\chi }\delta _{0}\chi
+\partial _{B}\left[ \frac{\partial \mathcal{L}}{\partial (\partial _{B}\chi
)}\delta \chi -T_{C}^{B}\,\delta x^{C}\right] \right\} d^{4}x\text{,}
\end{equation}%
where
\begin{equation}
T^{B}\,_{C}:=\frac{\partial \mathcal{L}}{\partial (\partial _{B}\chi )}%
\partial _{C}\chi -\delta _{C}^{B}\,\mathcal{L}
\end{equation}%
is the canonical energy-momentum tensor density. If the variations are
chosen such that $\delta x^{B}=0$ over $\Omega $ and $\delta
_{0}\chi $ vanishes on the boundary of $\Omega $, then $\delta I=0$ gives us
the Euler-Lagrange equation
\begin{equation}
\lbrack \mathcal{L}]_{\chi }=\frac{\partial \mathcal{L}}{\partial \chi }%
-\partial _{B}\left[ \frac{\partial \mathcal{L}}{\partial (\partial _{B}\chi
)}\right] =0\text{.}
\end{equation}%
On the other hand, if the field variables obey the Euler-Lagrange equation $
[\mathcal{L}]_{\chi }=0$, then we have
\begin{equation}
\partial _{B}\left[ \frac{\partial \mathcal{L}}{\partial (\partial _{B}\chi )%
}\delta \chi -T^{B}\,_{C}\,\delta x^{C}\right] =0\text{,}
\end{equation}%
which gives rise, considering also the Noether theorem, to
conservation laws. These very straightforward considerations are
at the basis of the following discussion.

%%%%%%%%%%%%%%%%%%%%%%%
\subsection{Global Poincar\'{e} Invariance}
%%%%%%%%%%%%%%%%%%%%%%%%%%

As usual, we assert that our space-time in the absence of gravitation is a Minkowski
space
$M_{4}$. The isometry group of $M_{4}$ is the group of Poincar\'{e} transformations (PT)
which consists of the Lorentz group $SO(3$, $1)$ and the translation group $T(3$, $1)$.
The Poincar\'{e} transformations of coordinates are
\begin{equation}
x^{A}\overset{PT}{\rightarrow }x^{\prime }{}^{A}=a^{A}_{\ B}\,x^{B}+b^{A}\text{%
,}  \label{PoincareTransf}
\end{equation}%
where $a_{\ B}^{A}$ and $b^{A}$ are real constants, and $a_{\ B}^{A}$ satisfy
the orthogonality conditions $a_{\ C}^{A}a_{B}^{\ C}=\delta _{B}^{A}$. Thus, for
infinitesimal variations we have
\begin{equation}
\delta x^{\prime }{}^{A}=\chi ^{\prime }(x^{\prime })-\chi (x)=\varepsilon
^{A}\,_{B}\,x^{B}+\varepsilon ^{A},  \label{PoincareTrasf1}
\end{equation}%
where $\varepsilon _{AB}+\varepsilon _{BA}=0$ (the quantities  $\varepsilon
^{A}\,_{B}$ and $\varepsilon^{A}$ are the tensor and vector infinitesimal variations
respectively). While the Lorentz
transformation forms a six parameter group, the Poincar\'{e} group has ten
parameters. The Lie algebra for the ten generators of the Poincar\'{e} group
is
\begin{eqnarray}
&&\lbrack \Xi _{AB}\text{, }\Xi _{CD}] =\eta _{AC}\,\Xi _{BD}+\eta
_{BD}\,\Xi _{AC}\nonumber\\
&&
\ \ \ \ \ \ \ \ \ \ \ \ \ \ \ \ \ \ \ -\eta _{BC}\,\Xi _{AD}-\eta _{AD}\,\Xi _{BC},
\notag
\\
&&
\lbrack \Xi _{AB}\text{, }T_{C}] =\eta _{BC}T_{A}-\eta _{AC}T_{B}\notag
\\
&&
[T_{A}\text{, }T_{B}]=0,
\end{eqnarray}%
where $\Xi _{AB}$ are the generators of Lorentz transformations,
and $T_{A}$ are the generators of four-dimensional translations.
Obviously, $\partial
_{A}(\delta x^{A})=0$ for the Poincar\'{e} transformations (\ref{PoincareTransf}).
Therefore, our Lagrangian density $\mathcal{L}$, which is
the same as $L$ with $h(x)=1$, is invariant, namely $\delta
\mathcal{L}=\delta L=0$ for $\delta I=0$.

Suppose that the field $\chi (x)$ transforms under the infinitesimal Poincar%
\'{e} transformations as
\begin{equation}
\delta \chi =\frac{1}{2}\varepsilon ^{AB}S_{AB}\chi \text{,}
\end{equation}%
where the tensors $S_{AB}$ are the generators of the Lorentz group,
satisfying
\begin{eqnarray}
&&S_{AB}=-S_{BA}\nonumber\\
&&
[S_{AB}\text{, }S_{CD}]=\eta _{AC}\,S_{BD}+\eta
_{BD}\,S_{AC}\nonumber\\
&& \ \ \ \ \ \ \ \ \ \ \ \ \ \ \ \ \ \ \,
-\eta _{BC}\,S_{AD}-\eta _{AD}\,S_{BC}\text{.}
\label{LorentzGen}
\end{eqnarray}
Correspondingly, the derivative of $\chi $ transforms as
\begin{equation}
\delta (\partial _{C}\chi )=\frac{1}{2}\varepsilon ^{AB}S_{AB}\partial
_{C}\chi -\varepsilon ^{A}\,_{C}\partial _{A}\chi \text{.}
\end{equation}%
Since the choice of infinitesimal parameters $\varepsilon ^{A}$ and $%
\varepsilon ^{AB}$ is arbitrary, the vanishing variation of the
Lagrangian density  $\delta \mathcal{L}=0$ leads to the
identities
\begin{equation}
\frac{\partial \mathcal{L}}{\partial \chi }S_{AB}\chi +\frac{\partial
\mathcal{L}}{\partial (\partial _{C}\chi )}(S_{AB}\partial _{C}\chi +\eta
_{CA}\partial _{B}\chi -\eta _{CB}\partial _{A}\chi )=0\text{.}
\end{equation}%
Moreover, we obtain the conservation laws
\begin{equation}
\partial _{B}\,T_{C}^{B}=0,\ \partial _{C}\left(
S^{C}\,_{AB}-x_{A}T^{C}\,_{B}+x_{B}T^{C}\,_{A}\right) =0\text{,}
\end{equation}%
where
\begin{equation}
S^{C}\,_{AB}:=-\frac{\partial \mathcal{L}}{\partial (\partial _{C}\chi )}%
S_{AB}\chi \text{.}
\end{equation}%
These conservation laws imply that the energy-momentum and angular momentum,
respectively given by
\begin{eqnarray}
&&\!\!\!\!\!
P_{l}=\int T_{l}^{0}\,d^{3}x,\nonumber\\
&&\!\!\!\!\!
J_{AB}=\int \left[ S^{0}\,_{AB}\,-\left(
x_{A}T^{0}\,_{B}-x_{B}T^{0}\,_{A}\right) \right] d^{3}x\text{,}
\end{eqnarray}%
are conserved (the first term of
the angular momentum integral corresponds to the spin angular momentum, while the second
term gives the orbital angular momentum). This means that the system invariant under the
ten parameter symmetry group has ten conserved quantities. This is an example of Noether
symmetry.

The global Poincar\'{e} invariance of a system implies that, for the system, the
space-time is homogeneous (all space-time points are equivalent) as dictated by the
translational invariance, and is isotropic (all directions about a space-time point are
equivalent) as indicated by the Lorentz invariance. It is interesting to observe that the
fixed point variation of the field variables takes the form
\begin{equation}
\delta _{0}\chi =\frac{1}{2}\varepsilon ^{B}\,_{C}\Xi _{B}\,^{C}\,\chi
+\varepsilon ^{B}\,T_{B}\,\chi \text{,}
\end{equation}%
where
\begin{equation}
\Xi _{B}\,^{C}=S_{B}\,^{C}+\left( x^{B}\partial _{C}-x^{C}\partial
_{B}\right) \text{, \ }T_{B}=-\partial _{B}\text{.}
\end{equation}%
We mention that $\Xi _{B}\,^{C}$ are the generators of the Lorentz transformation and
$T_{B}$ are those of the translations.

As a next step, let us consider a modification of the infinitesimal Poincar\'{e}
transformations (\ref{PoincareTrasf1}) by assuming that the parameters $%
\varepsilon _{\ C}^{B}$ and $\varepsilon ^{B}$ are functions of the
coordinates and by writing them altogether as
\begin{equation}
\delta x^{\mu }=\varepsilon ^{\mu }\,_{\nu }(x)\,x^{\nu }+\varepsilon ^{\mu
}(x)=\xi ^{\mu }\text{,}
\end{equation}%
which we call the local Poincar\'{e} transformations (or the general coordinate
transformations). In order to make a distinction between the global
transformation and the local transformation, we use the Latin indices $(A$, $B=0$, $1$,
$2$, $3)$ for the former and the Greek indices $(\mu $, $\nu =0$, $1$, $2$, $3)$ for the
latter. The variation of the field variables $\chi (x) $ defined at a point $x$ is still
the same as that of the global Poincar\'{e} transformations, namely
\begin{equation}
\delta \chi =\frac{1}{2}\varepsilon _{AB}S^{AB}\,\chi \text{.}
\end{equation}%
The corresponding fixed point variation of $\chi $ takes the form,
\begin{equation}
\delta _{0}\chi =\frac{1}{2}\varepsilon _{AB}S^{AB}\chi -\xi ^{\nu }\partial
_{\nu }\chi .  \label{FixedPtVar}
\end{equation}%
Differentiating both sides of (\ref{FixedPtVar}) with respect to $x^{\mu }$,
we have
\begin{equation}
\delta _{0}\partial _{\mu }\chi =\frac{1}{2}\varepsilon ^{AB}S_{AB}\partial
_{\mu }\chi +\frac{1}{2}(\partial _{\mu }\varepsilon ^{AB})\,S^{AB}\chi
-\partial _{\mu }(\xi ^{\nu }\partial _{\nu }\chi )\text{.}
\end{equation}%
By using these variations, we obtain the variation of the
Lagrangian $L$, namely
\begin{eqnarray}
&&\!\!\!\!\!\!\!\!\!\!\!\!\!\!\!\!\!\!\!\!\!\!\!
\delta \mathcal{L}+\partial _{\mu }(\delta x^{\mu })\mathcal{L}=h\delta
L=\delta _{0}\mathcal{L}+\partial _{\nu }(\mathcal{L}\delta x^{\nu })
\nonumber\\
&&\ \ \ \ \ \ \ \ \
=-\frac{%
1}{2}(\partial _{\mu }\varepsilon ^{AB})\,S^{\mu }\,_{AB}-\partial _{\mu
}\xi ^{\nu }\,T_{\text{ }\nu }^{\mu \,}\text{,}
\end{eqnarray}
which is no longer zero unless the parameters $\varepsilon ^{AB}$ and $\xi
^{\nu }$ become constants. Accordingly, the action integral for the given
Lagrangian density $\mathcal{L}$ is not invariant under the local Poincar%
\'{e} transformations. We notice that while $\partial _{B}(\delta
x^{B})=0$ for the local Poincar\'{e} transformations, $\partial
_{\mu }\xi ^{\mu }$ does not vanish under local Poincar\'{e} transformations. Hence, as
expected $\mathcal{L}$ is not a Lagrangian scalar but a Lagrangian density. Therefore, as
mentioned earlier, in order to define the Lagrangian $L$ we have to choose an
appropriate non-trivial scalar function $h(x)$ satisfying
\begin{equation}
\delta h+h\partial _{\mu }\xi ^{\mu }=0\text{.}
\end{equation}

Proceeding forward we consider a minimal modification of the Lagrangian in order to make
the action integral invariant under the local Poincar\'{e} transformations. It is rather
obvious that if there is a covariant derivative $\nabla _{C}\chi $ which transforms as
\begin{equation}
\delta (\nabla _{C}\chi )=\frac{1}{2}\varepsilon ^{AB}S_{AB}\nabla _{C}\chi
-\varepsilon ^{A}\,_{C}\nabla _{A}\chi \text{,}
\end{equation}%
then a modified Lagrangian $L^{\prime }(\chi $, $\partial _{C}\chi $, $%
x)=L(\chi $, $\nabla _{C}\chi $, $x)$, obtained by replacing
$\partial _{C}\chi $ of $L(\chi $, $\partial _{C}\chi $, $x)$ by
$\nabla _{C}\chi $, remains invariant under the local Poincar\'{e}
transformations, that is
\begin{equation}
\delta L^{\prime }=\frac{\partial L^{\prime }}{\partial \chi }\delta \chi +%
\frac{\partial L^{\prime }}{\partial (\nabla _{C}\chi )}\delta (\nabla
_{C}\chi )=0\text{.}
\end{equation}%
To find such a $C$-covariant derivative we introduce the gauge fields $
V^{AB}\,_{\mu }=-V^{BA}\,_{\mu }$, and we define the $\mu $-covariant derivative
\begin{equation}
\nabla _{\mu }\chi :=\partial _{\mu }\chi +\frac{1}{2}V^{AB}\,_{\mu
}S_{AB}\chi  \label{mu-covariant}
\end{equation}%
in such a way that the covariant derivative transforms as
\begin{equation}
\delta _{0}\nabla _{\mu }\chi =\frac{1}{2}S_{AB}\nabla _{\mu }\chi -\partial
_{\mu }(\xi ^{\nu }\nabla _{\nu }\chi )\text{.}
\end{equation}%
The transformation properties of $V_{\;\;\;\mu }^{AB}$ are determined by $%
\nabla _{\mu }\chi $ and $\delta \nabla _{\mu }\chi $. Making use of
\begin{eqnarray}
&&\delta \nabla _{\mu }\chi =\frac{1}{2}\varepsilon ^{AB},_{\mu }S_{AB}\chi +%
\frac{1}{2}\varepsilon ^{AB}S_{AB}\partial _{\mu }\chi -\left( \partial
_{\mu }\xi ^{\nu }\right) \partial _{\nu }\psi
\nonumber\\
&&\ \ \ \ \ \ \ \ \  \ \,
+\frac{1}{2}\delta
V_{\;\;\;\mu }^{AB}S_{AB}\chi +\frac{1}{4}V_{\;\;\;\mu
}^{AB}S_{AB}\varepsilon ^{CD}S_{CD}\chi,
\end{eqnarray}%
and comparing with (\ref{mu-covariant}), we obtain
\begin{eqnarray}
&&\!\!\!\!\!\!\!\!\!\!\!\!\!\!\!\!
\delta V_{\;\;\;\mu }^{AB}S_{AB}\chi +%
\frac{1}{2}\left( V_{\;\;\;\mu }^{AB}\varepsilon ^{CD}-\varepsilon
^{AB}V_{\;\;\;\mu }^{CD}\right) S_{AB}S_{CD}\chi
\nonumber\\
&&\ \ \  \,
+\varepsilon ^{AB},_{\mu }S_{AB}\chi
+\left( \partial _{\mu }\xi
^{\nu }\right) V_{\;\;\;\nu }^{AB}S_{AB}\chi =0\text{.}  \label{inter}
\end{eqnarray}%
Using the antisymmetry in $AB$ and $CD$ to rewrite the term in
parentheses in (\ref{inter}) as $\left[
S_{AB},S_{CD}\right] V_{\;\;\;\mu }^{AB}\varepsilon ^{CD}\chi $,
we see the explicit appearance of the commutator $\left[
S_{AB},S_{CD}\right] $. Additionally, using the expression for the
commutator of Lie algebra generators
\begin{equation}
\left[ S_{AB}\text{, }S_{CD}\right] =\frac{1}{2}c_{\;\;\;\;\ \left[ AB\right]
\left[ CD\right] }^{\left[ EF\right] }S_{EF}\text{,}
\end{equation}%
where $c_{\;\;\;\;\left[ AB\right] \left[ CD\right] }^{\left[ EF\right] }$
(the square brackets denote anti-symmetrization) is the structure constants
of the Lorentz group (deduced below), we have%
\begin{equation}
\left[ S_{AB}\text{, }S_{CD}\right] V_{\;\;\;\mu }^{AB}\varepsilon ^{CD}=%
\frac{1}{2}\left( V_{\mu }^{AF}\varepsilon _{F}^{B}-V_{\mu }^{FB}\varepsilon
_{F}^{A}\right) S_{AB}\text{.}
\end{equation}%
The substitution of this relation and the consideration of the antisymmetry of $
\varepsilon _{A}^{\;B}=-\varepsilon _{\;A}^{B}$ enables us to write
\begin{eqnarray}
&&\!\!\!\!\!\!\!\!\!\!\!\!\!\!\!\!\!\!\!\!\!\!\!\!\!\!\!
\delta V^{AB}\,_{\mu }=\varepsilon ^{A}\,_{C}V^{CB}\,_{\mu }+\varepsilon
^{B}\,_{C}V^{AC}\,_{\mu }\nonumber\\
&&
-(\partial _{\mu }\xi ^{\nu })V^{AB}\,_{\nu
}-\partial _{\mu }\varepsilon ^{AB}.
\end{eqnarray}

We require the $C$-derivative and $\mu $-derivative of $\chi $ to be
linearly related as
\begin{equation}
\nabla _{C}\chi =e_{C}\,^{\mu }(x)\nabla _{\mu }\chi ,  \label{inter2}
\end{equation}%
where the coefficients $e_{C}\,^{\mu }(x)$ are position-dependent and behave
like a new set of field variables. From (\ref{inter2}) it is evident that $%
\nabla _{C}\chi $ varies as%
\begin{equation}
\delta \nabla _{C}\chi =\delta e_{C}^{\mu }\nabla _{\mu }\chi +e_{C}^{\mu
}\delta \nabla _{\mu }\chi \text{.}
\end{equation}%
Comparing with $\delta \nabla _{C}\chi =\frac{1}{2}\varepsilon
^{EF}S_{EF}\nabla _{C}\chi -\varepsilon _{\text{ }C}^{B}\nabla _{B}\chi $ we
obtain $
e_{\rho }^{C}\delta e_{C}^{\mu }\nabla _{\mu }\chi -\xi ^{\nu },_{\rho
}\nabla _{\nu }\chi +e_{\rho }^{C}\varepsilon _{\text{ }C}^{B}\nabla
_{B}\chi =0$.
Exploiting $\delta \left( e_{\rho }^{C}e_{C}^{\mu }\right) =0$ we find that the
quantity $e_{C}\,^{\mu }$ transforms according to
\begin{equation}
\delta e_{C}\,^{\mu }=e_{C}\,^{\nu }\partial _{\nu }\xi ^{\mu }-e_{A}\,^{\mu
}\varepsilon ^{A}\,_{C}.
\end{equation}%
It is also important to recognize that the inverse of $\det (e_{C}\,^{\mu })$
transforms like a scalar density as $h(x)$ does. For our minimal
modification of the Lagrangian density, we utilize this available quantity
for the scalar density $h$. In particular we let
\begin{equation}
h(x)=[\det (e_{C}\,^{\mu })]^{-1}\text{.}
\end{equation}%
In the limiting case, when we consider Poincar\'{e}
transformations that are not space-time dependent, $e_{C}\,^{\mu
}\rightarrow \delta _{C}^{\mu }$ so that $h(x)\rightarrow 1$. This
is a desirable property. Then we replace the Lagrangian density
$\mathcal{L}(\chi $, $\partial _{C}\chi $, $x)$, invariant under
the global Poincar\'{e} transformations, by a Lagrangian density
\begin{equation}
\mathcal{L}(\chi \text{, }\partial _{\mu }\chi \text{; }x)\rightarrow
h(x)L(\chi \text{, }\nabla _{C}\chi )\text{.}
\end{equation}
The action integral with the above modified Lagrangian density remains
invariant under the local Poincar\'{e} transformations. Since the
local Poincar\'{e} transformations $\delta x^{\mu }=\xi ^{\mu
}(x)$ are nothing else but  generalized coordinate
transformations, the newly introduced gauge fields $e_{A}^{\ \lambda
}$ and $V^{AB}\,_{\mu }$ can be interpreted, respectively, as the
tetrad ({\it vierbein}) fields which set the local coordinate
frame and as a local affine connection with respect to the tetrad
frame (see also  \cite{Basini:2005wq}).

Let us consider first  the case where the multiplet field $\chi $
is the Dirac field $\psi (x)$, which behaves like a four-component
spinor under the Lorentz transformations, and transforms as
\begin{equation}
\psi (x)\rightarrow \psi ^{\prime }(x^{\prime })=S(\Lambda )\psi (x)\text{,}
\end{equation}%
where $S(\Lambda )$ is an irreducible unitary representation of the Lorentz
group. Since the bilinear form $v^{C}=i\overline{\psi }\gamma ^{C}\psi $ is
a vector, it transforms according to%
\begin{equation}
v^{B}=\Lambda _{\text{ }C}^{B}v^{C}\text{,}
\end{equation}%
where $\Lambda _{\text{ }A}^{B}$ is a Lorentz transformation matrix
satisfying
$\Lambda _{AB}+\Lambda _{BA}=0$.
The invariance of $v^{A}$ (or the covariance of the Dirac
equation) under the
transformation $\psi (x)\rightarrow \psi ^{\prime }(x^{\prime })$ leads to%
\begin{equation}
S^{-1}(\Lambda )\gamma ^{\mu }S(\Lambda )=\Lambda _{\nu }^{\mu }\gamma ^{\nu
},  \label{Transf}
\end{equation}%
where the $\gamma ^{\prime }s$ are the Dirac $\gamma $-matrices satisfying
the anticommutator
$
\gamma _{A}\gamma _{B}+\gamma _{B}\gamma _{A}=\eta _{AB}\mathbf{1}$.
Furthermore, we notice that the $\gamma $-matrices have the following
properties:
{\small{
\begin{equation}
\left\{
\begin{array}{c}
\left( \gamma _{0}\right) ^{\dag }=-\gamma _{0}\text{, }\left( \gamma
^{0}\right) ^{2}=\left( \gamma _{0}\right) ^{2}=-1\text{, }\gamma
_{0}=-\gamma ^{0}\text{ and }\gamma _{0}\gamma ^{0}=1 \\
\\
\left( \gamma _{k}\right) ^{\dag }=\gamma _{k}\text{ ,}\left( \gamma
^{k}\right) ^{2}=\left( \gamma _{k}\right) ^{2}=1\text{; }(k=1,2,3)\text{
and }\gamma _{k}=\gamma ^{k} \\
\\
\left( \gamma _{5}\right) ^{\dag }=-\gamma _{5}\text{, }\left( \gamma
_{5}\right) ^{2}=-1\text{ and }\gamma ^{5}=\gamma _{5}\text{.}%
\end{array}%
\right.
\end{equation}
}}
We assume that the transformation $S(\Lambda )$ can be put into the form $%
S(\Lambda )=e^{\Lambda _{\mu \nu }\gamma ^{\mu \nu }}$. Expanding
$S(\Lambda)$ about the identity, retaining only terms up to first order in the
infinitesimals, and expanding $\Lambda _{\mu \nu }$ to first order in $
\varepsilon _{\mu \nu }$ as
\begin{equation}
\Lambda _{\mu \nu }=\delta _{\mu \nu }+\varepsilon _{\mu \nu },
\label{LorentzTransf}
\end{equation}%
we acquire
\begin{equation}
S(\Lambda )=1+\frac{1}{2}\varepsilon ^{AB}\gamma _{AB}\text{.}
\label{DiracTransf}
\end{equation}%
In order to determine the form of $\gamma _{AB}$, we substitute
(\ref{LorentzTransf})
and (\ref{DiracTransf}) into (\ref{Transf}) obtaining
\begin{equation}
\frac{1}{2}\varepsilon _{AB}\left[ \gamma ^{AB}\text{, }\gamma ^{C}\right]
=\eta ^{ki}\varepsilon _{BA}\gamma ^{B}\text{.}  \label{inter3}
\end{equation}%
Rewriting the r.h.s. of (\ref{inter3}) using the antisymmetry of $\varepsilon
_{AB}$ as
$
\eta ^{CA}\varepsilon _{BA}\gamma ^{B}=\frac{1}{2}\varepsilon _{AB}\left(
\eta ^{CA}\gamma ^{B}-\eta ^{CB}\gamma ^{A}\right)$
yields
\begin{equation}
\left[ \gamma ^{C}\text{, }\gamma ^{AB}\right] =\eta ^{CA}\gamma ^{B}-\eta
^{CB}\gamma ^{A}\text{.}
\end{equation}%
Assuming the solution to have the form of an antisymmetric product of two
matrices, we obtain the solution%
\begin{equation}
\gamma ^{AB}:=\frac{1}{2}\left[ \gamma ^{A}\text{, }\gamma ^{B}\right] \text{%
.}
\end{equation}

If $\chi =\psi $ then the group generator $S_{AB}$ appearing in (\ref{LorentzGen}) is
identified with
\begin{equation}
S_{AB}\equiv \gamma _{AB}=\frac{1}{2}(\gamma _{A}\gamma _{B}-\gamma
_{B}\gamma _{A}).
\end{equation}%
To be explicit, the Dirac field transforms under Lorentz transformations  (LT) as
\begin{equation}
\delta \psi (x)=\frac{1}{2}\varepsilon ^{AB}\gamma _{AB}\psi (x).
\end{equation}%
The Pauli conjugate of the Dirac field is denoted $\overline{\psi }$ and it is
defined by
\begin{equation}
\overline{\psi }(x):=i\psi ^{\dagger }(x)\,\gamma _{0}\text{, }i\in
\mathbb{C}
\text{.}
\end{equation}%
The conjugate field $\overline{\psi }$ transforms under LTs as,
\begin{equation}
\delta \overline{\psi }=-\overline{\psi }\frac{1}{2}\varepsilon ^{AB}%
\overline{\psi }\gamma _{AB}\text{.}
\end{equation}

Under local LTs, $\varepsilon _{AB}(x)$ becomes a function of space-time. Now, unlike
$\partial _{\mu }\psi (x)$, the derivative of $\psi ^{\prime
}(x^{\prime })$ is no longer homogenous due to the occurrence of the term $
\gamma ^{AB}\left[ \partial _{\mu }\varepsilon _{AB}(x)\right] \psi (x)$ in $
\partial _{\mu }\psi ^{\prime }(x^{\prime })$, which is non-vanishing unless
$\varepsilon _{AB}$ is constant. When going from locally flat to curved
space-time, we must generalize $\partial _{\mu }$ to the covariant derivative $%
\nabla _{\mu }$ to compensate for this extra term, allowing to
gauge the group of LTs. Thus, by using $\nabla _{\mu }$, we can
preserve the
invariance of the Lagrangian for arbitrary local LTs at each space-time point%
\begin{equation}
\nabla _{\mu }\psi ^{\prime }(x^{\prime })=S(\Lambda (x))\nabla _{\mu }\psi
(x)\text{.}
\end{equation}%
To determine the explicit form of the connection belonging to $\nabla _{\mu }
$, we study the derivative of $S(\Lambda (x))$. The transformation $%
S(\Lambda (x))$ is given by%
\begin{equation}
S(\Lambda (x))=1+\frac{1}{2}\varepsilon _{AB}(x)\gamma ^{AB}\text{.}
\end{equation}%
Since $\varepsilon _{AB}(x)$ is only a function of space-time for local
Lorentz coordinates, we can express this infinitesimal LT in terms of general
coordinates only by shifting all space-time dependence of the local
coordinates into the tetrad fields as
\begin{equation}
\varepsilon _{AB}(x)=e_{A}^{\text{ \ }\lambda }(x)e_{\text{ \ }B}^{\nu
}(x)\varepsilon _{\lambda \nu }\text{.}
\end{equation}%
Substituting this expression for $\varepsilon _{AB}(x)$, we obtain
$
\partial _{\mu }\varepsilon _{AB}(x)=\partial _{\mu }\left[ e_{A}^{\text{ \ }%
\lambda }(x)e_{\text{ \ }B}^{\nu }(x)\varepsilon _{\lambda \nu }\right]
$,
however since $\varepsilon _{\lambda \nu }$ has no space-time
dependence this reduces to
\begin{equation}
\partial _{\mu }\varepsilon _{AB}(x)=e_{A}^{\text{ \ }\lambda }(x)\partial
_{\mu }e_{B\lambda }(x)-e_{B}^{\text{ \ }\nu }(x)\partial _{\mu }e_{A\nu }(x)%
\text{.}  \label{inter4}
\end{equation}

Let us now introduce the quantity
\begin{equation}
\omega _{\mu BA}:=e_{B}^{\text{ \ }\nu }(x)\partial _{\mu }e_{A\nu }(x).
\label{inter9}
\end{equation}
Hence, the first and second terms in (\ref{inter4}) become
$e_{A}^{\lambda
}(x)\partial _{\mu }e_{B\lambda }(x)=\frac{1}{2}\omega _{\mu AB}$ and $%
e_{B}^{\nu }(x)\partial _{\mu }e_{A\nu }(x)=\frac{1}{2}\omega _{\mu BA}$
respectively. Using the identification
\begin{equation}
\partial _{\mu }\varepsilon _{AB}(x)=\omega _{\mu AB}\text{,}
\end{equation}%
we can write
\begin{equation}
\partial _{\mu }S(\Lambda (x))=-\frac{1}{2}\gamma ^{AB}\omega _{\mu AB}\text{%
.}
\end{equation}%
According to (\ref{mu-covariant}), the covariant derivative of the Dirac
spinor is
\begin{equation}
\nabla _{\mu }\psi =\partial _{\mu }\psi +\frac{1}{2}V^{AB}\,_{\mu }\gamma
_{AB}\psi \text{.}  \label{del-mu-psi}
\end{equation}%
Correspondingly, the covariant derivative of $\bar{\psi}$ is given by
\begin{equation}
\nabla _{\mu }\overline{\psi }=\partial _{\mu }\overline{\psi }-\frac{1}{2}%
V^{AB}\,_{\mu }\bar{\psi}\gamma _{AB}\text{.}
\end{equation}%
Using the covariant derivatives of $\psi $ and $\bar{\psi}$, we can show
that
\begin{equation}
\nabla _{\mu }v_{B}=\partial _{\mu }v_{B}-V^{A}\,_{B\mu }v_{A}\text{.}
\end{equation}%
The same covariant derivative should be used for any covariant vector $v_{C}$
under the Lorentz transformation. Since $\nabla _{\mu }(v_{A}v^{A})=\partial
_{\mu }(v_{A}v^{A})$, the covariant derivative for a contravariant vector $%
v^{A}$ must be
\begin{equation}
\nabla _{\mu }v^{A}=\partial _{\mu }v^{A}+V^{A}\,_{B\mu }v^{B}\text{.}
\end{equation}
Since the tetrad $e_{A}\,^{\mu }$ is a covariant vector under Lorentz
transformations, its covariant derivative must transform according to the
same rule. Using $\nabla _{A}=e_{A}^{\mu }(x)\nabla _{\mu }$, the covariant
derivatives of a tetrad in local Lorentz coordinates read%
\begin{eqnarray}
&&\nabla _{\nu }e_{A}\,^{\mu }=\partial _{\nu }e_{A}\,^{\mu }-V^{C}\,_{A\nu
}e_{C}\,^{\mu }\nonumber\\
&&
\nabla _{\nu }e^{A}\,_{\mu }=\partial _{\nu
}e^{A}\,_{\mu }+V^{A}\,_{C\nu }e^{C}\,_{\mu }\text{.}
\end{eqnarray}
Furthermore, the inverse of $e_{A}\,^{\mu }$ is denoted by $e^{A}\,_{\mu }$ and satisfies
\begin{equation}
e^{A}\,_{\mu }e_{A}\,^{\nu }=\delta _{\mu }\,^{\nu },~~~e^{A}\,_{\mu
}e_{B}\,^{\mu }=\delta ^{A}\,_{B}\text{.}
\end{equation}

In order to allow the transition to curved space-time, we take into account the general
coordinates of objects that are covariant under local Poincar\'{e} transformations. Here
we define the covariant derivative of a quantity $v^{\lambda }$ which behaves like a
contravariant vector under the local Poincar\'{e} transformation, namely
\begin{eqnarray}
&&
D_{\nu }v^{\lambda }\equiv e_{A}\,^{\lambda }\nabla _{\nu }v^{A}=\partial
_{\nu }v^{\lambda }+\Gamma ^{\lambda }\,_{\mu \nu }v^{\mu }\nonumber\\
&& D_{\nu
}v_{\mu }
\equiv e^{A}\,_{\mu }\nabla _{\nu }v_{A}=\partial _{\nu }v_{\mu
}-\Gamma ^{\lambda }\,_{\mu \nu }v_{\lambda }\text{,}
\end{eqnarray}%
where%
\begin{equation}
\Gamma ^{\lambda }\,_{\mu \nu }:=e_{A}\,^{\lambda }\nabla _{\nu
}e^{A}\,_{\mu }\equiv -e^{A}\,_{\mu }\nabla _{\nu }e_{A}\,^{\lambda }.
\end{equation}%
The definition of $\Gamma ^{\lambda }\,_{\mu \nu }$ implies
\begin{eqnarray}
&&\!\!\!\!\!\!\!\!\!\!\!\!\!
D_{\nu }e_{A}\,^{\lambda } =\nabla _{\nu }e_{A}\,^{\lambda }+\Gamma
^{\lambda }\,_{\mu \nu }e_{A}\,^{\mu }\nonumber\\
&&\ \ \ \
=\partial _{\nu }e_{A}\,^{\lambda
}-V^{C}\,_{A\nu }e_{C}\,^{\lambda }+\Gamma ^{\lambda }\,_{\mu \nu
}e_{A}\,^{\mu }=0,
\label{inter6}
\end{eqnarray}
\begin{eqnarray}
&&\!\!\!\!\!\!\!\!\!\!\!\!\!\!\!\!\!\!\!\!\!
D_{\nu }e^{A}\,_{\mu } =\nabla _{\nu }e^{A}\,_{\mu }-\Gamma ^{\lambda
}\,_{\mu \nu }e^{A}\,_{\lambda }\nonumber\\
&&
=\partial _{\nu }e^{A}\,_{\mu
}+V^{A}\,_{C\nu }e^{C}\,_{\mu }-\Gamma ^{\lambda }\,_{\mu \nu
}e^{A}\,_{\lambda }=0\text{.}
\label{inter6b}
\end{eqnarray}%
From (\ref{inter6}),(\ref{inter6b}) we find that
\begin{eqnarray}
&&\!\!\!\!\!\!\!\!\!\!\!\!\!\!
V^{A}\,_{C\nu }=e^{A}\,_{\lambda }\partial _{\nu }e_{C}\,^{\lambda }+\Gamma
^{\lambda }\,_{\mu \nu }e^{A}\,_{\lambda }e_{C}\,^{\mu }
\nonumber\\
&&\ \
=-e_{C}\,^{\lambda
}\partial _{\nu }e^{A}\,_{\lambda }+\Gamma ^{\lambda }\,_{\mu \nu
}e^{A}\,_{\lambda }e_{C}\,^{\mu },
\end{eqnarray}
or equivalently, in terms of $\omega $ defined in (\ref{inter9}):
\begin{eqnarray}
&&\!\!\!\!\!\!\!\!\!\!\!\!\!
V^{A}\,_{C\nu }=\omega _{\text{ }\nu C}^{A}+\Gamma ^{\lambda }\,_{\mu \nu
}e^{A}\,_{\lambda }e_{C}\,^{\mu }
\nonumber\\
&& \ \  \  =-\omega _{C\nu }^{\text{ \ \ \ }A}+\Gamma
^{\lambda }\,_{\mu \nu }e^{A}\,_{\lambda }e_{C}\,^{\mu }.
\end{eqnarray}%
Inserting into (\ref{del-mu-psi}) we may write
\begin{equation}
\nabla _{\mu }\psi =(\partial _{\mu }-\Gamma _{\mu })\psi \text{,}
\end{equation}%
where
\begin{equation}
\Gamma _{\mu }=\frac{1}{4}\left( \omega _{\text{ \ }B\mu }^{A}-\Gamma
^{\lambda }\,_{\mu \nu }e^{A}\,_{\lambda }e_{B}\,^{\nu }\right) \gamma
_{A}\,^{B},
\end{equation}%
is known as the Fock-Ivanenko connection.

We now study the transformation properties of $V_{\mu AB}$. Recall that $\omega
_{\mu AB}=e_{A}^{\text{ \ }\lambda }(x)\partial _{\mu }e_{B \lambda }(x)$
and since $\partial _{\mu }\eta _{AB}=0$, we write
\begin{equation}
\Lambda _{\text{ \ }\overline{A}}^{A}\eta _{AB}\partial _{\mu }\Lambda _{%
\overline{B}}^{\text{ \ }B}=\Lambda _{\text{ \ }\overline{A}}^{A}\partial
_{\mu }\Lambda _{A\overline{B}}\text{.}
\end{equation}%
Note that barred indices are equivalent to the primed indices used above.
Hence, the spin connection transforms as%
\begin{equation}
V_{\overline{A}\overline{B}\overline{C}}=\Lambda _{\overline{A}}^{\text{ \ }%
A}\Lambda _{\overline{B}}^{\text{ \ }B}\Lambda _{\overline{C}}^{\text{ \ }%
C}V_{ABC}+\Lambda _{\overline{A}}^{\text{ \ }A}\Lambda _{\overline{C}}^{%
\text{ \ }C}e_{\text{ \ }A}^{\mu }(x)\partial _{\mu }\Lambda _{\overline{B}%
C}.
\end{equation}%
In order to determine the transformation properties of
\begin{equation}
\Gamma _{ABC}=V_{ABC}-\left[ e_{\text{ \ }A}^{\mu }(x)\partial _{\mu }e_{%
\text{ \ }B}^{\nu }(x)\right] e_{\nu C}(x)\text{,}
\end{equation}
we consider the local LT of $\left[ e_{A}^{\text{ \ }\mu }(x)\partial _{\mu
}e_{\text{ \ }B}^{\nu }(x)\right] e_{\nu C}(x)$, which is
\begin{eqnarray}
&&\!\!\!\!\!\!\!\!\!\!\!\!\!\!\!\!\!\!\!
\left[ e_{\text{ \ }\overline{A}}^{\mu }(x)\partial _{\mu }e_{\text{ \ }%
\overline{B}}^{\nu }\right] e_{\nu \overline{C}}(x)=\Lambda _{\overline{A}}^{%
\text{ \ }A}\Lambda _{\overline{B}}^{\text{ \ }B}\Lambda _{\overline{C}}^{%
\text{ \ }C}\left[ V_{\text{ }AB}^{\nu }e_{\nu C}(x)\right]
\nonumber\\
&&\ \ \ \ \ \ \ \ \ \ \ \ \ \ \ \ \ \ \ \ \  \
+\Lambda _{%
\overline{A}}^{\text{ \ }A}\Lambda _{\overline{C}}^{\text{ \ }C}e_{\text{ \ }%
A}^{\mu }(x)\partial _{\mu }\Lambda _{C\overline{B}}\text{.}
\end{eqnarray}
From this result, we obtain the following transformation law:
\begin{equation}
\Gamma _{\overline{A}\overline{B}\overline{C}}=\Lambda _{\overline{A}}^{%
\text{ \ }A}\Lambda _{\overline{B}}^{\text{ \ }B}\Lambda _{\overline{C}}^{%
\text{ \ }C}\Gamma _{ABC}.
\end{equation}

We now explore the consequence of the antisymmetry of $\omega _{ABC}$ in $BC$. Recalling
the equation for $\Gamma _{ABC}$, exchanging $B$ and $C$ and
adding the two equations, we obtain%
\begin{eqnarray}
&&\!\!\!\!\!\!\!\!\!\!\!\!\!\!\!\!\!\!\!\!\!\!\!\!\!\!
\Gamma _{ABC}+\Gamma _{ACB}=-e_{\text{ \ }A}^{\mu }(x)
\left[ \left( \partial
_{\mu }e_{\text{ \ }B}^{\nu }(x)\right) e_{\nu C}(x)\right.\nonumber\\
&&\ \ \ \ \ \ \ \ \, \ \left.
+\left( \partial _{\mu
}e_{\text{ \ }C}^{\nu }(x)\right) e_{\nu B}(x)\right] \text{.}
\end{eqnarray}%
We know however, that
$\partial _{\mu }\left[ e_{\text{ \ }B}^{\nu }(x)e_{\nu C}(x)\right] =-e_{B}^{
\text{ \ }\lambda }(x)e_{\text{ \ }C}^{\nu }(x)\partial _{\mu }g_{\nu
\lambda }$
and thus we finally obtain
\begin{equation}
\Gamma _{ABC}+\Gamma _{ACB}=e_{A}^{\text{ \ }\mu }(x)e_{B}^{\text{ \ }%
\lambda }(x)e_{C}^{\text{ \ }\nu }(x)\partial _{\mu }g_{\nu \lambda },
\end{equation}
which  is equivalent to
\begin{equation}
\Gamma _{\overline{A}\overline{B}\overline{C}}+\Gamma _{\overline{A}\text{ }%
\overline{C}\overline{B}}=e_{\text{ \ }\overline{A}}^{\overline{\mu }}(x)e_{%
\text{ \ }\overline{B}}^{\overline{\lambda }}(x)e_{\text{ \ }\overline{C}}^{%
\overline{\nu }}(x)\partial _{\overline{\mu }}g_{\overline{\nu }\overline{%
\lambda }}\text{,}
\end{equation}%
and therefore
\begin{equation}
\Gamma _{\mu \lambda \nu }+\Gamma _{\mu \nu \lambda }=\partial _{\mu }g_{\nu
\lambda }\text{,}  \label{inter5}
\end{equation}%
which we recognize as the general coordinate connection.

It is known that
the covariant derivative for general coordinates is
\begin{equation}
\nabla _{\mu }V_{\nu }^{\text{ \ }\lambda }=\partial _{\mu }V_{\nu }^{\text{
\ }\lambda }+\Gamma _{\text{ \ }\mu \sigma }^{\lambda }V_{\nu }^{\text{ \ }%
\sigma }-\Gamma _{\text{ \ }\mu \nu }^{\sigma }V_{\sigma }^{\text{ \ }%
\lambda }\text{.}
\end{equation}%
In a Riemannian manifold, the connection is symmetric under the
exchange of $\mu
\nu $, i.e. $\Gamma _{\text{ \ }\mu \nu }^{\lambda }=\Gamma _{\text{ \ }%
\nu \mu }^{\lambda }$. Using the fact that the metric is a symmetric tensor
we can now determine the form of the Christoffel connection by cyclically
permuting the indices of the general coordinate connection equation (\ref{inter5}),
yielding
\begin{equation}
\Gamma _{\mu \nu \lambda }=\frac{1}{2}\left( \partial _{\mu }g_{\nu \lambda
}+\partial _{\nu }g_{\lambda \mu }-\partial _{\lambda }g_{\mu \nu }\right) .
\end{equation}%
Since $\Gamma _{\mu \nu \lambda }=\Gamma _{\nu \mu \lambda }$ is valid for
general coordinate systems, it follows that a similar constraint must hold
for local Lorentz transforming coordinates too, thus we expect $\Gamma
_{ABC}=\Gamma _{BAC}$. Recalling the equation for $\Gamma _{ABC}$ and
exchanging $A$ and $B$, we obtain
\begin{eqnarray}
&&\!\!\!\!\!\!\!\!\!\!\!\!\!\!\!\!\!\!
\omega _{ABC}-\omega _{BAC}=e_{\nu C}(x)\left[ e_{\text{ \ }A}^{\mu
}(x)\partial _{\mu }e_{\text{ \ }B}^{\nu }(x)\right.\nonumber\\
&&\left. \  \ \ \ \ \ \ \ \ \ \ \  \  \ \ \ \ \ \ \ \ \ \ \
-e_{\text{ \ }B}^{\mu
}(x)\partial _{\mu }e_{\text{ \ }A}^{\nu }(x)\right] \text{.}
\end{eqnarray}%
We now define the \textit{objects of anholonomicity} as
\begin{equation}
\Omega _{CAB}:=e_{\nu C}(x)\left[ e_{\text{ \ }A}^{\mu }(x)\partial _{\mu
}e_{\text{ \ }B}^{\nu }(x)-e_{\text{ \ }B}^{\mu }(x)\partial _{\mu }e_{\text{
\ }A}^{\nu }(x)\right] .
\end{equation}
Using $\Omega _{CAB}=-\Omega _{CBA}$, we permute indices in a similar manner
as was done for the derivation of the Christoffel connection above, yielding
\begin{equation}
\omega _{AB\mu }=\frac{1}{2}\left[ \Omega _{CAB}+\Omega _{BCA}-\Omega _{ABC}%
\right] e_{\text{ \ }\mu }^{C}\equiv \Delta _{AB\mu }.  \label{inter8}
\end{equation}%
For completeness, we determine the transformation law of the
Christoffel connection. Making use of $\Gamma _{\mu \nu }^{\lambda
}e_{\lambda
}=\partial _{\mu }e_{\nu }$ where%
\begin{equation}
\partial _{\overline{\mu }}\,e_{\overline{\nu }}=X_{\text{ \ }\overline{\mu }%
}^{\mu }X_{\text{ \ }\overline{\nu }}^{\nu }\partial _{\mu }e_{\nu }+X_{%
\text{ \ }\overline{\mu }}^{\mu }\left( \partial _{\mu }X_{\text{ \ }%
\overline{\nu }}^{\nu }\right) e_{\nu }\text{,}
\end{equation}
we can show that
\begin{equation}
\Gamma _{\text{ \ }\overline{\mu }\text{ }\overline{\nu }}^{\overline{%
\lambda }}=X_{\text{ \ }\overline{\mu }}^{\mu }X_{\text{ \ }\overline{\nu }%
}^{\nu }X_{\lambda }^{\text{ \ }\overline{\lambda }}\Gamma _{\text{ \ }\mu
\nu }^{\lambda }+X_{\text{ \ }\overline{\mu }}^{\mu }X_{\nu }^{\text{ \ }%
\overline{\lambda }}X_{\text{ \ }\mu \overline{\nu }}^{\nu },
\end{equation}%
where
$X_{\text{ \ }\mu \overline{\nu }}^{\nu }\equiv \partial _{\mu }\partial _{%
\overline{\nu }}x^{\nu }$.

In the light of the above considerations, we may regard infinitesimal local gauge
transformations as local rotations of basis vectors belonging to the tangent space  of
the
manifold \cite{Chang:1975fr,Mansouri:1977ej}. For this reason, given a local frame on a
tangent plane to the point $x$ of the base manifold, we can obtain all other frames on
the same tangent plane by means of local rotations of the original basis vectors.
Reversing this argument, we observe that by knowing all frames residing in the horizontal
tangent space to a point $x$ of the base manifold enables us to deduce the corresponding
gauge group of symmetry transformations.

%%%%%%%%%%%%%%%%%%%%%%%%
\subsection{Curvature, Torsion and Metric}
%%%%%%%%%%%%%%%%%%%%%%%%%%%%%%%

From the definition of the Fock-Ivanenko covariant derivative we
can find the second order covariant derivative as
\begin{eqnarray}
&&
\!\!\!\!\!\!\!\! \! \! \! \! \! \! \!
D_{\nu }D_{\mu }\psi  =\partial _{\nu }\partial _{\mu }\psi +\frac{1}{2}%
S_{CD}\left( \psi \partial _{\nu }V_{\mu }^{\text{ \ }CD}+V_{\mu }^{\text{ \
}CD}\partial _{\nu }\psi \right)
\notag \\
&& \ \ \ \ \ \,
+\Gamma _{\text{ \ }\mu \nu }^{\rho
}D_{\rho }\psi +\frac{1}{2}S_{EF}V_{\nu }^{\text{ \ }EF}\partial _{\mu
}\psi
\notag \\
&& \ \ \ \ \ \,
+\frac{1}{4}S_{EF}S_{CD}V_{\nu }^{\text{ \ }EF}V_{\mu }^{\text{ \ }CD}\psi
\text{.}
\end{eqnarray}%
Recalling that $D_{\nu }e^{C\mu }=0$ we can solve for the spin connection in
terms of the Christoffel connection
\begin{equation}
V_{\mu }^{\text{ \ }CD}=-e_{\text{ \ }\lambda }^{D}\partial _{\mu
}e^{C\lambda }-\Gamma _{\mu }^{\text{ \ }CD},
\end{equation}
and thus the derivative of the spin connection is then
\begin{equation}
\partial _{\mu }V_{\text{ \ \ }\nu }^{CD}=-e_{\text{ \ }\lambda
}^{D}\partial _{\mu }\partial _{{\nu }}e^{C\lambda }-\left( \partial _{\nu
}e^{C\lambda }\right) \partial _{\mu }e_{\lambda }^{\text{ \ }D}-\partial
_{\mu }\Gamma _{\text{ \ \ }\nu }^{CD}\text{.}
\end{equation}%
Noting that the Christoffel connection is symmetric and partial derivatives
commute, we find%
\begin{eqnarray}
&&\!\!\!\!\!\!\!\!\!\!\!\!\!\!\!\!
\left[ D_{\mu }\text{, }D_{\nu }\right] \psi =\frac{1}{2}S_{CD}\left[ \left(
\partial _{\nu }V_{\text{ \ \ }\mu }^{CD}-\partial _{\mu }V_{\text{ \ \ }\nu
}^{CD}\right) \psi \right]\nonumber\\
&&\
+\frac{1}{4}S_{EF}S_{CD}\left[ \left( V_{\text{ \
\ }\nu }^{EF}V_{\text{ \ \ }\mu }^{CD}-V_{\text{ \ \ }\mu }^{EF}V_{\text{ \
\ }\nu }^{CD}\right) \psi \right] ,
\end{eqnarray}%
where $
\partial _{\nu }V_{\text{ \ \ }\mu }^{CD}-\partial _{\mu }V_{\text{ \ \ }\nu
}^{CD}=\partial _{\mu }\Gamma _{\text{ \ \ }\nu }^{CD}-\partial _{\nu
}\Gamma _{\text{ \ \ }\mu }^{CD}$.
Relabeling running indices, we can write
{\small{
\begin{equation}
 S_{EF}S_{CD}\left( V_{\text{ \ \ }\nu }^{EF}V_{\text{ \ \ }\mu
}^{CD}-V_{\text{ \ \ }\mu }^{EF}V_{\text{ \ \ }\nu }^{CD}\right) \psi = \left[
S_{CD}\text{, }S_{EF}\right] V_{\text{ \ }\mu }^{EF}V_{\text{ \
\ }\nu }^{CD}\psi ,
\end{equation}
}}
and therefore, using $\left\{ \gamma _{A},\gamma _{B}\right\} =2\eta _{AB}$ to deduce
$
\left\{ \gamma _{A}\text{, }\gamma _{B}\right\} \gamma _{C}\gamma _{D}=2\eta
_{AB}\gamma _{C}\gamma _{D}$,
we find that the commutator of bi-spinors is given by
\begin{eqnarray}
&&\!\!\!\!\!\!\!\!\!\!\!\!\!\!\!\!\!\!\!\!\!\!\!\!\!\!\!
\left[ S_{CD}\text{, }S_{EF}\right] =\frac{1}{2}\left[
\eta _{CE}\delta
_{D}^{A}\delta _{F}^{B}-\eta _{DE}\delta _{C}^{A}\delta _{F}^{B}\right.\nonumber\\
&&\ \ \ \ \ \ \ \ \
\left.
+\eta
_{CF}\delta _{E}^{A}\delta _{D}^{B}-\eta _{DF}\delta _{E}^{A}\delta _{C}^{B}%
\right] S_{AB}.  \label{inter7}
\end{eqnarray}
Clearly the terms in brackets on the r.h.s. of (\ref{inter7}) are
antisymmetric in $CD$ and EF and  also antisymmetric under the
exchange of pairs of indices $CD$ and EF. Since the alternating
spinor is antisymmetric in $AB$, this must hold for the terms in
brackets too: this implies that the commutator does not vanish. Hence,
the term in brackets is totally antisymmetric under interchange of
indices $AB$, $CD$ and EF and exchange of these pairs of
indices. We identify this as the structure constant  of the
Lorentz group  \cite{DeWitt:1965jb}
{\small{
\begin{eqnarray}
&&\!\!\!\!\!\!\!\!\!\!\!\!\!\!\!\!\!\!\!\!
\left[ \eta _{CE}\delta _{D}^{A}\delta _{F}^{B}-\eta _{DE}\delta
_{C}^{A}\delta _{F}^{B}\right.\nonumber\\
&&\left.\ \ \ \
+\eta _{CF}\delta _{E}^{A}\delta _{D}^{B}-\eta
_{DF}\delta _{E}^{A}\delta _{C}^{B}\right]
={c_{[CD][EF]}}^{[AB]}\nonumber\\
&&\ \ \ \ \ \ \ \ \ \ \ \ \ \ \ \ \ \ \ \ \ \ \ \ \ \ \ \ \,\ \
\ \ \ \ \  \ \
={c^{[AB]}}%
_{[CD][EF]},
\end{eqnarray}
}}
with the aid of which we can write
{\small{
\begin{equation}
\frac{1}{4}\left[ S_{CD}\text{, }S_{EF}\right] V_{\text{ \ \ }\mu }^{EF}V_{
\text{ \ \ }\nu }^{CD}\psi =\frac{1}{2}S_{AB}\left[ V_{\text{ \ }E\nu
}^{A}V_{\text{ \ \ }\mu }^{EB}-V_{\text{ \ }E\nu }^{B}V_{\text{ \ \ }\mu
}^{AE}\right] \psi \text{,}
\end{equation}
}}
where
$
V_{\text{ \ }E\nu }^{A}V_{\text{ \ \ }\mu }^{EB}-V_{\text{ \ }E\nu }^{B}V_{%
\text{ \ \ }\mu }^{AE}=\Gamma _{\nu E}^{A}\Gamma _{\text{ \ \ }\mu
}^{EB}-\Gamma _{\nu E}^{B}\Gamma _{\text{ \ \ }\mu }^{EA}$.
Combining these results, the commutator of two $\mu $-covariant
differentiations gives
\begin{equation}
\lbrack \nabla _{\mu },\nabla _{\nu }]\chi =-\frac{1}{2}R^{AB}\,_{\mu \nu
}\,S_{AB}\chi ,
\end{equation}%
where
\begin{eqnarray}
&&
R^{A}\,_{B\mu \nu }=\partial _{\nu }V^{A}\,_{B\mu }-\partial _{\mu
}V^{A}\,_{B\nu }\nonumber\\
&& \ \ \ \ \ \ \ \ \ \ \ \ \
+V^{A}\,_{C\nu }V^{C}\,_{B\mu }-V^{A}\,_{C\mu }V^{C}\,_{B\nu
}\text{.}
\end{eqnarray}%
Using the Jacobi identities for the commutator of covariant
derivatives, it follows that the field strength  $R^{A}\,_{B\mu
\nu }$ satisfies the Bianchi identity
\begin{equation}
\nabla _{\lambda }R^{A}\,_{B\mu \nu }+\nabla _{\mu }R^{A}\,_{B\nu \lambda
}+\nabla _{\nu }R^{A}\,_{B\lambda \mu }=0\text{.}
\end{equation}%
Permuting indices, this can be put into the cyclic form
\begin{equation}
\varepsilon ^{\alpha \beta \rho \sigma }\nabla _{\beta }R_{\text{ \ }\rho
\sigma }^{AB}=0,
\end{equation}%
where $\varepsilon ^{\alpha \beta \rho \sigma }$ is the Levi-Civita
alternating symbol. Furthermore, $R^{AB}\,_{\mu \nu }=\eta
^{BC}R^{A}\,_{C\mu \nu }$ is antisymmetric with respect to both pairs of
indices, namely $
R^{AB}\,_{\mu \nu }=-R^{BA}\,_{\mu \nu }=R^{BA}\,_{\nu \mu }=-R^{AB}\,_{\nu
\mu }$. The above condition is known as the first curvature tensor identity.

In order to determine the analogue of $[\nabla _{\mu },\nabla _{\nu }]\chi
$ in local coordinates, we start from $\nabla _{C}\psi =e_{\text{
\ }C}^{\mu
}\nabla _{\mu }\psi $. From $\nabla _{C}\psi $ we obtain
$
\nabla _{D}\nabla _{C}\psi =e_{\text{ \ }D}^{\nu }\left( \nabla _{\nu }e_{%
\text{ \ }C}^{\mu }\right) \nabla _{\mu }\psi +e_{\text{ \ }D}^{\nu }e_{%
\text{ \ }C}^{\mu }\nabla _{\nu }\nabla _{\mu }\psi $, and permuting indices and
recognizing that $
e_{\mu }^{\text{ \ }A}\nabla _{\nu }e_{\text{ \ }C}^{\mu }=-e_{C}^{\text{ \ }%
\mu }\nabla _{\nu }e_{\text{ \ }\mu }^{A}$,
(which follows from $\nabla _{\nu }\left( e_{\mu }^{A}e_{C}^{\mu }\right) =0$), we arrive
at
\begin{eqnarray}
&&\!\!\!\!\!\!\!\!\!\!\!\!\!\!\!\!\!
e_{\text{ \ }D}^{\nu }\left( \nabla _{\nu }e_{\text{ \ }C}^{\mu }\right)
\nabla _{\mu }\psi -e_{\text{ \ }C}^{\mu }\left( \nabla _{\mu }e_{\text{ \ }%
D}^{\nu }\right) \nabla _{\nu }\psi \nonumber\\
&&
=\left( e_{\text{ \ }D}^{\mu }e_{\text{
\ }C}^{\nu }-e_{\text{ \ }C}^{\mu }e_{\text{ \ }D}^{\nu }\right) \left(
\nabla _{\nu }e_{\mu }^{\text{ \ }A}\right) \nabla _{A}\psi \text{.}
\end{eqnarray}%
Defining
\begin{equation}
C_{\;\;CD}^{A}:=\left( e_{\text{ \ }C}^{\mu }e_{\text{ \ }D}^{\nu }-e_{\text{
\ }D}^{\mu }e_{\text{ \ }C}^{\nu }\right) \nabla _{\nu }e_{\mu }^{\text{ \ }%
A},
\end{equation}
the commutator of the $C$-covariant differentiations takes the final form
 \cite{Kibble:1961ba}
\begin{equation}
\lbrack \nabla _{C},\nabla _{D}]\chi =-\frac{1}{2}R^{AB}\,_{CD}S_{AB}\,\chi
+C^{A}\,_{CD}\nabla _{A}\, \chi ,
\end{equation}%
where
$
R^{AB}\,_{CD}=e_{C}\,^{\mu }e_{D}\,^{\nu }R^{AB}\,_{\mu \nu }$.
As was done for $R^{A}\,_{B\mu \nu }$, using the Jacobi identities for
the commutator of covariant derivatives, we find  the Bianchi
identity in Einstein-Cartan space-time \cite{Hehl:1976kj, Blagojevic:2003cg}:
\begin{equation}
\varepsilon ^{\alpha \beta \rho \sigma }\nabla _{\beta }R_{\text{ }\rho
\sigma }^{AB}=\varepsilon ^{\alpha \beta \rho \sigma }C_{\beta \rho }^{\text{
\ \ \ }\lambda }R_{\text{ \ }\sigma \lambda }^{AB}.
\end{equation}%
The second curvature identity%
\begin{equation}
R_{\text{ \ }[\rho \sigma \lambda ]}^{C}=2\nabla _{\lbrack \rho }C_{\sigma
\lambda ]}^{\text{ \ \ \ \ }C}-4C_{[\rho \sigma }^{\text{ \ \ \ }%
B}C_{\lambda ]B}^{\text{ \ \ \ }C}
\end{equation}%
leads to
\begin{equation}
\varepsilon ^{\alpha \beta \rho \sigma }\nabla _{\beta }C_{\rho \sigma }^{%
\text{ \ \ \ }C}=\varepsilon ^{\alpha \beta \rho \sigma }R_{\text{ \ }B\rho
\sigma }^{C}e_{\text{ }\beta }^{B}.
\end{equation}%
Notice that if $
\Gamma _{\;\;\mu \nu }^{\lambda }=e_{A}^{\text{ \ }\lambda }\nabla _{\nu }e_{%
\text{ \ }\mu }^{A}=-e_{\mu }^{\text{ \ }A}\nabla _{\nu }e_{\text{ \ }%
A}^{\lambda }$
then
$
\Gamma _{\;\;\mu \nu }^{\lambda }-\ \Gamma _{\;\;\nu \mu }^{\lambda
}=e_{A}^{\lambda }\left( \nabla _{\nu }e_{\text{ \ }\mu }^{A}-\nabla _{\mu
}e_{\text{ \ }\nu }^{A}\right) $.
Contracting by $e_{C}^{\mu }e_{D}^{\nu }$, we obtain  \cite{Kibble:1961ba}:
\begin{equation}
C_{\;\text{\ }CD}^{A}=e_{C}^{\text{ \ }\mu }e_{D}^{\text{ \ }\nu }e_{\lambda
}^{\text{ \ }A}\left( \ \Gamma _{\;\;\mu \nu }^{\lambda }-\ \Gamma _{\;\;\nu
\mu }^{\lambda }\right) \text{.}
\end{equation}%
We therefore conclude that $C_{\;CD}^{A}$ is related to the antisymmetric
part of the affine connection
\begin{equation}
\Gamma _{\;\;\left[ \mu \nu \right] }^{\lambda }=e_{\mu }^{\text{ \ }%
C}e_{\nu }^{\text{ \ }D}e_{A}^{\text{ \ }\lambda }C_{\;\ CD}^{A}\equiv
T_{\;\;\mu \nu }^{\lambda },
\end{equation}%
which is usually interpreted as space-time {\it{torsion tensor}} $T_{\;\;\mu \nu
}^{\lambda }$.

Considering $\Delta _{AB\mu }$ defined in (\ref{inter8}), we see that the most
general connection in the Poincar\'{e} gauge approach to gravitation is%
\begin{equation}
V_{AB\mu }=\Delta _{AB\mu }-K_{AB\mu }+\Gamma ^{\lambda }\,_{\nu \mu
}e_{A\lambda }e_{B}\,^{\nu },
\end{equation}%
where
\begin{equation}
K_{abC}=-\left( T^{\lambda }\,_{\nu \mu }-T_{\nu \mu }^{\text{ \ \ }\lambda
}+T_{\mu \text{ \ }\nu }^{\text{ \ }\lambda }\right) e_{A\lambda
}e_{B}\,^{\nu }e_{C}^{\text{ \ }\mu }
\end{equation}%
is the contorsion tensor. Now, the quantity $R_{\sigma \mu \nu }^{\rho
}=e_{A}\,^{\rho }R^{A}\,_{\sigma \mu \nu }$ may be expressed as
\begin{equation}
R^{\rho }\,_{\sigma \mu \nu }=\partial _{\nu }\Gamma _{\text{ \ }\sigma \mu
}^{\rho }-\partial _{\mu }\Gamma _{\text{ \ }\sigma \nu }^{\rho }+\Gamma
^{\rho }\,_{\lambda \nu }\Gamma ^{\lambda }\,_{\sigma \mu }-\Gamma ^{\rho
}\,_{\lambda \mu }\Gamma ^{\lambda }\,_{\sigma \nu }\text{.}
\label{curvature}
\end{equation}%
Therefore, we can regard $R^{\rho }\,_{\sigma \mu \nu }$ as the
{\it{curvature tensor}} with respect to the affine connection $\Gamma
^{\lambda }\,_{\mu \nu }$. Additionally, by using the inverse of the tetrad we
can find the metric of the space-time manifold by
\begin{equation}
g_{\mu \nu }=e^{A}\,_{\mu }\,e^{B}\,_{\nu }\,\eta _{AB}.
\end{equation}%
From (\ref{inter6}),(\ref{inter6b}) and the fact that the Minkowski metric is constant, it
is obvious that the above metric is covariantly constant, namely
\begin{equation}
D_{\lambda }g_{\mu \nu }=0.
\end{equation}
Hence, the space-time thus specified by the local Poincar\'{e} transformation is
said to be metric space-time. It is not difficult to show that
\begin{equation}
\sqrt{-g}=[\det e^{A}\,_{\mu }]=[\det e_{A}\,^{\mu }]^{-1}\text{,}
\end{equation}%
where $g=\det g_{\mu \nu }$. Therefore, we may consider $\sqrt{-g}$ for the density
function $h(x)$ introduced in (\ref{hxdefinition}).

%%%%%%%%%%%%%%%%%%%%%%
\subsection{Field Equations for  Poincar\'{e}  Gravity}
%%%%%%%%%%%%%%%%%%%%%%%%%%%%%%

Finally, we are able to deduce the field equations for the
gravitational field. From the curvature tensor $R^{\rho }\,_{\sigma \mu \nu }$, given in
(\ref{curvature}), we can calculate the Ricci tensor as
\begin{equation}
R_{\sigma \nu }=R^{\mu }\,_{\sigma \mu \nu },
\end{equation}%
and the scalar curvature as
\begin{equation}
R=R^{\nu }\,_{\nu }=\overset{\text{L}}{R}+\partial _{A}K_{B}^{\text{ \ }%
AB}-T_{A}^{\text{ \ }BC}K_{BC}^{\text{ \ \ }A},
\end{equation}%
where $\overset{\text{L}}{R}$ denotes the usual\ Ricci scalar of
GR. Using this scalar curvature $R$ we construct the
Lagrangian density for free Einstein-Cartan gravity:
\begin{equation}
\mathcal{L}_{G}=\frac{1}{16 \pi G }\sqrt{-g}\left( \overset{\text{L}}{R}%
+\partial _{A}K_{B}^{\text{ \ }AB}-T_{A}^{\text{ \ }BC}K_{BC}^{\text{ \ \ }%
A}-2\Lambda \right) ,
\end{equation}%
where $G$ is a gravitational coupling constant, and $\Lambda
$ is the cosmological constant. These considerations can be easily
extended to any function of $\overset{\text{L}}{R}$ as in
 \cite{Capozziello:2008yx}. Observe that the second term is a divergence and
may be ignored.

The field equation can be obtained from the total action
\begin{equation}
S=\int \left\{ \mathcal{L}_{\text{field}}(\chi ,\partial _{\mu }\chi
,e_{A}\,^{\mu },V^{AB}\,_{\mu })+\mathcal{L}_{G}\right\} d^{4}x\text{,}
\end{equation}%
where the matter Lagrangian density is taken to be%
\begin{equation}
\mathcal{L}_{\text{field}}=\frac{1}{2}\left[ \overline{\psi }\gamma
^{A}D_{A}\psi -\left( D_{A}\overline{\psi }\right) \gamma ^{A}\psi \right]
\text{.}
\end{equation}%
Modifying the connection to include Christoffel, spin connection and contorsion
contributions, in order to be able to incorporate for generality spinoral arguments too,
we have
\begin{equation}
\Gamma _{\mu }=\frac{1}{4}g_{\lambda \sigma }\left( \Delta _{\text{ \ }\mu
\rho }^{\sigma }-\overset{\text{L}}{\Gamma }\text{ }_{\text{ \ }\rho \mu
}^{\sigma }-K_{\text{ \ }\rho \mu }^{\sigma }\right) \gamma ^{\lambda \rho }.
\end{equation}%
It is important to keep in mind that $\Delta _{\text{ \ }\mu \rho }^{\sigma }
$ act only on multi-component spinor fields, while $\overset{\text{L}}{%
\Gamma }$ $_{\text{ \ }\rho \mu }^{\sigma }$ act on vectors and arbitrary
tensors. The gauge covariant derivative for a spinor and adjoint spinor is
then given by%
\begin{equation}
D_{\mu }\psi =\left( \partial _{\mu }-\Gamma _{\mu }\right) \psi \text{, \ }%
D_{\mu }\overline{\psi }=\partial _{\mu }\overline{\psi }-\overline{\psi }%
\Gamma _{\mu }\text{.}
\end{equation}

The variation of the field Lagrangian is
\begin{equation}
\delta \mathcal{L}_{\text{field}}=\overline{\psi }\left( \delta \gamma ^{\mu
}D_{\mu }+\gamma ^{\mu }\delta \Gamma _{\mu }\right) \psi \text{.}
\end{equation}%
We know that the Dirac gamma matrices are covariantly vanishing, thus
$
D_{\kappa }\gamma _{\iota }=\partial _{\kappa }\gamma _{\iota }-\Gamma
_{\iota \kappa }^{\mu }\gamma _{\mu }+\left[ \gamma _{\iota }\text{, }%
\widehat{\Gamma }_{\kappa }\right] =0$.
The $4\times 4$ matrices $\widehat{\Gamma }_{\kappa }$\ are real matrices
used to induce similarity transformations on quantities with spinor
transformation properties \cite{Brill:1957fx} according to
$
\gamma _{A}^{\prime }=\widehat{\Gamma }^{-1}\gamma _{A}\widehat{\Gamma }%
$.
Solving for $\widehat{\Gamma }_{\kappa }$ leads to,%
\begin{equation}
\widehat{\Gamma }_{\kappa }=\frac{1}{8}\left[ \left( \partial _{\kappa
}\gamma _{\iota }\right) \gamma ^{\iota }-\Gamma _{\text{ \ }\iota \kappa
}^{\mu }\gamma _{\mu }\gamma ^{\iota }\right] ,
\end{equation}
and hence its variation reads
$
\delta \widehat{\Gamma }_{\kappa }
=\frac{1}{8}\left[ \left( \partial _{\kappa }\delta \gamma _{\iota
}\right) \gamma ^{\iota }-\left( \delta \Gamma _{\text{ \ }\iota \kappa
}^{\mu }\right) \gamma _{\mu }\gamma ^{\iota }\right] $.
Since we require the anticommutator condition on the gamma matrices $%
\{\gamma ^{\mu },\gamma ^{\nu }\}=2g^{\mu \nu }$ to hold, the variation of
the metric gives%
\begin{equation}
2\delta g^{\mu \nu }=\{\delta \gamma ^{\mu },\gamma ^{\nu }\}+\{\gamma ^{\mu
}\delta \gamma ^{\nu }\}\text{.}
\end{equation}%
One solution to this equation is,%
\begin{equation}
\delta \gamma ^{\nu }=\frac{1}{2}\gamma _{\sigma }\delta \gamma ^{\sigma \nu
}\text{.}
\end{equation}%
With the aid of this result, we can write
$
\left( \partial _{\kappa }\delta \gamma _{\iota }\right) \gamma ^{\iota }=%
\frac{1}{2}\partial _{\kappa }\left( \gamma ^{\nu }\delta g_{\nu \iota
}\right) \gamma ^{\iota }$.
Hence, exploiting also the anti-symmetry in $\gamma _{\mu \nu }$, we obtain%
\begin{equation}
\delta \widehat{\Gamma }_{\kappa }=\frac{1}{8}\left[ g_{\nu \sigma }\delta
\Gamma _{\mu \kappa }^{\text{ \ \ }\sigma }-g_{\mu \sigma }\delta \Gamma
_{\nu \kappa }^{\text{ \ \ }\sigma }\right] \gamma ^{\mu \nu }\text{.}
\end{equation}

In summary, the field Lagrangian defined in the Einstein-Cartan space-time can be
written explicitly in terms of its Lorentzian and contorsion components as
\cite{Carroll:1994dq, Hehl:1971qi, Hehl:1976kj, Shapiro:2001rz,Blagojevic:2003cg}:
\begin{eqnarray}
&&\!\!\!\!\!\!\!\!\!\!
\mathcal{L}_{\text{field}}=\frac{1}{2}\left[ \left( \overset{\text{L}}{D}%
_{\mu }\overline{\psi }\right) \gamma ^{\mu }\psi -\overline{\psi }\gamma
^{\mu }\overset{\text{L}}{D}_{\mu }\psi \right] \nonumber\\
&&\ \ \ \ \
-\frac{\hbar C}{8}K_{\mu
\alpha \beta }\overline{\psi }\left\{ \gamma ^{\mu }\text{, }\gamma ^{\alpha
\beta }\right\} \psi .
\end{eqnarray}%
Using the useful relations
\begin{eqnarray}
&&\!\!\!\!\!\!\!\!\!\!\!
- K_{\mu \alpha \beta }\overline{\psi }\left\{ \gamma ^{\mu }\text{%
, }\gamma ^{\alpha \beta }\right\} \psi = K_{\mu \alpha \beta }%
\overline{\psi }\gamma ^{\beta \alpha }\gamma ^{\mu }\psi
\nonumber
\\
&&\ \ \ \ \ \ \ \ \ \ \ \ \ \ \ \ \ \ \ \ \ \ \ \ \ \ \
 - K_{\mu
\alpha \beta }\overline{\psi }\gamma ^{\mu }\gamma ^{\alpha \beta }\psi
\text{,}\nonumber
\\
&&\!\!\!\!\!\!\!\!\!\!\!
\gamma ^{\mu }\gamma ^{\nu }\gamma ^{\lambda }\varepsilon _{\mu \nu \lambda
\sigma }=\left\{ \gamma ^{\mu }\text{, }\gamma ^{\nu \lambda }\right\}
\varepsilon _{\mu \nu \lambda \sigma }=3!\gamma _{\sigma }\gamma _{5}\text{,
\ }\nonumber \\
&&\!\!\!\!\!\!\!\!\!\!\!
\left\{ \gamma ^{\mu }\text{, }\gamma ^{\nu \lambda }\right\} =\gamma
^{\lbrack \mu }\gamma ^{\nu }\gamma ^{\lambda ]}\text{, \ }\nonumber
\end{eqnarray}%
we can obtain
\begin{equation}
K_{\mu \alpha \beta }\overline{\psi }\left\{ \gamma ^{\mu }\text{, }\gamma
^{\alpha \beta }\right\} \psi =\frac{1}{2i}K_{\mu \alpha \beta }\varepsilon
^{\alpha \beta \mu \nu }\left( \overline{\psi }\gamma _{5}\gamma _{\nu }\psi
\right) \text{.}
\end{equation}%
At this stage it is convenient to define the contorsion axial vector as
\begin{equation}
K_{\nu }:=\frac{1}{3!}\varepsilon ^{\alpha \beta \mu \nu }K_{\alpha \beta
\mu }.
\end{equation}%
Multiplying through by the axial current $j_{\nu }^{5}=\overline{\psi }%
\gamma _{5}\gamma _{\nu }\psi $, we acquire
$\left( \overline{\psi }\gamma _{5}\gamma _{\nu }\psi \right) \varepsilon
^{\alpha \beta \mu \nu }K_{\mu \alpha \beta }=-6ij_{\nu }^{5}K^{\nu }$.
Thus, the field Lagrangian density becomes
\begin{equation}
\mathcal{L}_{\text{field}}=\frac{1}{2}\left[ \left( \overset{\text{L}}{D}%
_{\mu }\overline{\psi }\right) \gamma ^{\mu }\psi -\overline{\psi }\gamma
^{\mu }\overset{\text{L}}{D}_{\mu }\psi \right] +\frac{3i\hbar C}{8}K_{\mu
}j_{5}^{\mu }\text{.}
\end{equation}

Assembling everything, the total action variation reads
\begin{eqnarray}
\delta I &=& \int \left( \delta \mathcal{L}_{G}+\delta \mathcal{L}_{\text{field}%
}\right) \sqrt{-g}d^{4}x\text{.}  \notag
\end{eqnarray}
Writing the metric in terms of the tetrads $g^{\mu \nu }=e_{\;A}^{\mu
}e^{\nu A}$, we observe that
$
\delta \sqrt{-g}=-\frac{1}{2}\sqrt{-g}\left( \delta e_{\;A}^{\mu }e_{\mu
}^{\;A}+e_{\nu A}\delta e^{\nu A}\right) $.
By using
$
\delta e^{\nu A}=\delta \left( \eta ^{AB}e_{\;B}^{\nu }\right) =\eta
^{AB}\delta e_{\;B}^{\nu }$,
we are able to deduce that
\begin{equation}
\delta \sqrt{-g}=-\sqrt{-g}e_{\mu }^{\;A}\delta e_{A}^{\;\mu }.
\end{equation}%
For the variation of the Ricci tensor $R_{A\nu }=e_{A}^{\;\mu }R_{\mu \nu }$
we have
$
\delta \overset{\text{L}}{R}_{A\nu }=\delta e_{A}^{\;\mu }\overset{\text{L}}{%
R}_{\mu \nu }+e_{A}^{\;\mu }\delta \overset{\text{L}}{R}_{\mu \nu }$
In an inertial frame, the Ricci tensor reduces to
$
\overset{\text{L}}{R}_{\mu \nu }=\partial _{\nu }\overset{\text{L}}{\Gamma }%
\text{ }_{\beta \mu }^{\beta }-\partial _{\beta }\overset{\text{L}}{\Gamma }%
\text{ }_{\nu \mu }^{\beta }$,
and thus
\begin{equation}
\delta \overset{\text{L}}{R}_{A\nu }=\delta e_{A}^{\;\mu }\overset{\text{L}}{%
R}_{\mu \nu }+e_{A}^{\;\mu }\left( \partial _{\nu }\delta \overset{\text{L}}{%
\Gamma }\text{ }_{\beta \mu }^{\beta }-\partial _{\beta }\delta \overset{%
\text{L}}{\Gamma }\text{ }_{\nu \mu }^{\beta }\right) \text{.}
\end{equation}%
The second term can be converted into a surface term and therefore it may be ignored.
Assembling our results, we have
\begin{eqnarray}
&&\delta g^{\mu \nu }=-g^{\mu \rho }g^{\nu \sigma }\delta g_{\rho \sigma },\nonumber\\
&&\delta \sqrt{-g}=-\frac{1}{2}\sqrt{-g}g_{\mu \nu }\delta g^{\mu \nu }=-\sqrt{
-g}e_{\mu }^{\;A}\delta e_{A}^{\;\mu }
,\nonumber\\
&&
\delta R_{\mu \nu }=g_{\rho \mu }\left( \nabla _{\lambda }\delta \Gamma _{
\text{ \ \ }\nu }^{\lambda \rho }-\nabla _{\nu }\delta \Gamma _{\text{ \ \ }%
\lambda }^{\lambda \rho }\right) +T_{\lambda \mu }^{\text{ \ \ }\rho }\delta
\Gamma _{\text{ \ \ }\rho \nu }^{\lambda }
,\nonumber\\
&&
\delta \overset{\text{L}%
}{R}_{A\nu }=\delta e_{A}^{\;\mu }\overset{\text{L}}{R}_{\mu \nu }
,\nonumber\\
&&
\delta R=\overset{\text{L}}{R}\text{ }^{\mu \nu }\delta g_{\mu \nu }+g^{\mu
\nu }\left( \nabla _{\lambda }\delta \overset{\text{L}}{\Gamma }\text{ }_{%
\text{ \ \ }\mu \nu }^{\lambda }-\nabla _{\nu }\delta \overset{\text{L}}{%
\Gamma }\text{ }_{\text{ \ }\mu \lambda }^{\lambda }\right)\nonumber\\
&&\ \ \ \ \ \ \ \,  -T_{A}^{\text{ \
}BC}\delta K_{BC}^{\text{ \ \ }A}\text{.}%
\end{eqnarray}%
From the above expressions, neglecting surface terms and using the four-current $v^{\mu
}$ introduced earlier, the action for the matter fields read  \cite{Brill:1957fx}:
{\small{
\begin{eqnarray}
&&\!\!\!\!\!\!\!
\delta I_{\text{field}} =\int \left[ \overline{\psi }\delta \gamma ^{\mu
}\nabla _{\mu }\psi +\overline{\psi }\gamma ^{\mu }\delta \widehat{\Gamma }%
_{\mu }\psi \right] \sqrt{-g}d^{4}x
\nonumber \\
&&=\int \left\{\left[ \frac{1}{2}g^{\mu \nu }\overline{\psi }\gamma _{A}\left( \nabla
_{\nu
}\psi \right) +T_{\text{ \ }\rho \sigma }^{\mu }T_{A}^{\text{ }\rho \sigma
}-\delta _{A}^{\mu }T_{\lambda \rho \sigma }T^{\lambda \rho \sigma }\right]
\delta e_{\text{ }\mu }^{A}\right.\nonumber\\
&&\left.
+\frac{1}{8}\left( g^{\rho \nu }v^{\mu }-g^{\rho \mu }v^{\nu }\right) \left(
g_{\mu \sigma }\delta \overset{\text{L}}{\Gamma }\text{ }_{\text{ \ }\nu
\rho }^{\sigma }-g_{\nu \sigma }\delta \overset{\text{L}}{\Gamma }\text{ }_{%
\text{ \ }\mu \rho }^{\sigma }\right)
\right\} \sqrt{-g}d^{4}x .\nonumber
 \end{eqnarray}}}
Removing the derivatives of variations of the metric appearing in $\delta
\Gamma _{\text{ \ }\nu \rho }^{\sigma }$ via partial integration, and
equating to zero the coefficients of $\delta g^{\mu \nu }$ and $\delta T_{%
\text{ \ }\nu \rho }^{\sigma }$\ in the variation of the action integral, we
obtain%
\begin{eqnarray}
&&\!\!\!\!\!\!\!\!\!\!\!
0 =\frac{1}{16\pi }\left( R_{\mu \nu }-\frac{1}{2}g_{\mu \nu }R-g_{\mu \nu
}\Lambda \right) \nonumber
\\
&&
+\left( \frac{1}{2}\overline{\psi }\gamma _{\nu }\nabla
_{\mu }\psi -\frac{1}{4}\nabla _{\mu }v_{\nu }\right)\nonumber
\\
&&+\nabla _{\sigma }T_{\mu \nu }^{\text{ \ \ }\sigma }+T_{\mu \rho \sigma
}T_{\nu }^{\text{ }\rho \sigma }-g_{\mu \nu }T_{\lambda \rho \sigma
}T^{\lambda \rho \sigma }  \notag
\label{Field-Eqn1}
\end{eqnarray}%
and%
\begin{equation}
T_{\rho \sigma \lambda }=8\pi \tau _{\rho \sigma \lambda }.
\label{Field-Eqn2}
\end{equation}%
Eqs.  (\ref{Field-Eqn1}) have the form of Einstein equations%
\begin{equation}
G_{\mu \nu }-g_{\mu \nu }\Lambda =8\pi \Sigma _{\mu \nu },
\end{equation}%
where the Einstein tensor and non-symmetric energy-momentum tensors are
\begin{equation}
G_{\mu \nu }=R_{\mu \nu }-\frac{1}{2}g_{\mu \nu }R\text{,}
\end{equation}%
\begin{equation}
\Sigma _{\mu \nu }=\Theta _{\mu \nu }+\mathfrak{T}_{\mu \nu }\text{,}
\end{equation}%
respectively. Here we identify $\Theta _{\mu \nu }$ as the canonical
energy-momentum%
\begin{equation}
\Theta _{\text{ \ }\nu }^{\mu }=\frac{\partial \mathcal{L}_{\text{field}}}{%
\partial (\nabla _{\mu }\chi )}\nabla _{\nu }\chi -\delta _{\text{ }\nu
}^{\mu }\mathcal{L}_{\text{field}}\text{,}
\end{equation}%
while $\mathfrak{T}_{\mu \nu }$ is the stress-tensor form of the non-Riemannian manifold.
For the case of spinor fields being considered here the explicit form of the
energy-momentum components are (after symmetrization of corresponding canonical source
terms in the Einstein equation) \cite{Schwinger:1963zz}
\begin{eqnarray}
&&\!\!\!\!\!\!\!\!\!\!\!\!\!\!\!\!\!\! \!\!\!\!\!\!\!\!
\Theta _{\mu \nu }=-\left[ \overline{\psi }\gamma _{\mu }\nabla _{\nu }\psi
-\left( \nabla _{\nu }\overline{\psi }\right) \gamma _{\mu }\psi\right.
\nonumber\\
&&\left.
+\overline{%
\psi }\gamma _{\nu }\nabla _{\mu }\psi -\left( \nabla _{\mu }\overline{\psi }%
\right) \gamma _{\nu }\psi \right],
\end{eqnarray}%
and by using the second field equation (\ref{Field-Eqn2}), we determine%
\begin{equation}
\mathfrak{T}_{\mu \nu }=\nabla _{\sigma }T_{\mu \nu }^{\text{ \ \ }\sigma
}+T_{\mu \rho \sigma }\tau _{\nu }^{\text{ }\rho \sigma }-g_{\mu \nu
}T_{\lambda \rho \sigma }\tau ^{\lambda \rho \sigma },
\end{equation}%
where $\tau _{\mu \nu }^{\text{ \ \ }\sigma }$ is the so-called
spin-energy
potential  \cite{Hehl:1971qi, Hehl:1976kj}%
\begin{equation}
\tau _{\mu \nu }^{\text{ \ \ }\sigma }:=\frac{\partial \mathcal{L}_{\text{%
field}}}{\partial (\nabla _{\sigma }\chi )}\gamma _{\mu \nu }\chi \text{.}
\end{equation}%
Explicitly, the spin energy potential reads $\tau ^{\mu \nu \sigma }=%
\overline{\psi }\gamma ^{\lbrack \mu }\gamma ^{\nu }\gamma ^{\sigma ]}\psi $. Finally,
the equation of motion obtained  from the variation of the
action with respect to $\overline{\psi }$ reads  \cite{Hehl:1971qi,
Hehl:1976kj}
\begin{equation}
\gamma ^{\mu }\nabla _{\mu }\psi +\frac{3}{8}T_{\mu \nu \sigma }\gamma
^{\lbrack \mu }\gamma ^{\nu }\gamma ^{\sigma ]}\psi =0\text{.}
\end{equation}%
It is interesting to observe that this generalized curved space-time Dirac
equation can be recast into the nonlinear equation of the Heisenberg-Pauli
type%
\begin{equation}
\gamma ^{\mu }\nabla _{\mu }\psi +\frac{3}{8}\varepsilon \left( \overline{%
\psi }\gamma ^{\mu }\gamma _{5}\psi \right) \gamma _{\mu }\gamma _{5}\psi =0.
\end{equation}

Although the gravitational field equation is similar in form to the Einstein field
equation, it differs from the original Einstein equation because the  curvature tensor,
containing space-time torsion, is non-Riemannian. Assuming that the Euler-Lagrange
equations for the matter fields are satisfied, we obtain the following conservation laws
for the angular momentum and energy  momentum respectively:
\begin{equation}
\left.
\begin{array}{c}
e_{\text{ \ }A}^{\mu }e_{\text{ \ }B}^{\nu }\Sigma _{\lbrack \mu \nu
]}=\nabla _{\nu }\mathcal{\tau }_{AB}^{\text{ \ }\nu }\text{,} \\
\\
e_{\mu }^{\text{ \ }C}\nabla _{\nu }\Sigma _{\text{ \ }\kappa }^{\nu
}=\Sigma _{\text{ \ }\kappa }^{\nu }T_{\text{ \ }\mu \nu }^{C}+\mathcal{\tau
}_{\text{ \ }AB}^{\nu }R_{\text{ \ }\mu \nu }^{AB}\text{.}%
\end{array}%
\right.
\end{equation}

In summary,  we have shown that  all the necessary ingredients for a theory of
gravitation
can be obtained from a gauge theory of local Poincar\'{e} symmetry. Gauge fields have
been
obtained by requiring the invariance of the Lagrangian density under local Poincar\'{e}
transformations.

The resulting Einstein-Cartan theory describes a space endowed with non-vanishing
curvature and torsion. The lowest order gravitational action is one that is linear in the
curvature scalar while being quadratic in torsion. However, the scheme can be immediately
extended to more general gravitational theories as in   \cite{Capozziello:2008yx}.

The Dirac spinors can be introduced as matter sources and it has been  found that they
couple to gravity via the torsion stress form $\mathfrak{T}_{\mu \nu }$ component of the
total energy-momentum $\Sigma _{\mu \nu }$. The field equations obtained from the action
by means of a standard variational principle describe a nonlinear equation of the
Heisenberg-Pauli type in the matter sector,  gravitational field equations similar to the
Einstein equations, and a constraint equation relating torsion to spin energy potential.
The generalized energy-momentum tensor is comprised of the usual canonical
energy-momentum
tensor of matter in addition to a torsion stress form. The stress form contains a torsion
divergence term as well as a term similar to an external non-spinor source to gravity. In
view of the structure of the generalized energy-momentum tensor, we remark that the
gravitational field equations here obtained are similar to the equations of motion found
in Einstein-Yang-Mills theory, with the torsion tensor playing the role of the Yang-Mills
field strength.

The Bianchi identities of Einstein-Cartan gravity differ from those of GR, since the
Riemann curvature tensor characterizing the non-Riemannian geometry does not exhibit the
usual symmetry properties. In the limit of vanishing torsion, the Bianchi identities
reduce to their usual form.

Additionally, having extracted the conservation laws for the angular momentum and the
energy-momentum, we deduced that the generalized energy-momentum tensor contains a
non-vanishing anti-symmetric component proportional to the divergence of the spin-energy
potential, while it is divergenceless only in the limit of vanishing torsion.

Clearly, in the Poincar\'{e} approach,  curvature and torsion  both contribute to the
overall dynamics where external degrees of freedom (space-time) and internal degrees of
freedom (spins) are dealt under the same standard.

%%%%%%%%%%%%%%%%%%%%%%%%%%%%
\section{Teleparallel gravity}
\label{SecionTeleparallelgrav}
%%%%%%%%%%%%%%%%%%%%%%%%%%%%%%%%
With the above considerations in mind, let us now discuss an alternative approach to the 
gravitation
al interaction named Teleparallel Gravity.
It corresponds to a gauge theory
for the translation group  \cite{moller, Mielke:1992te,deAndrade:1997qt, DeAndrade:2000sf, 
Hehl:1994ue}. It
was first introduced by Einstein himself  as an equivalent alternative to GR 
\cite{Unzicker:2005in}.
 Due to the features  of translations,  gauge theories including these transformations  
differ
from the standard internal gauge models in several  ways.  The most significant 
differences is the 
presence
of  tetrad fields. Besides,  tetrad fields can  be adopted in order  to
define a linear Weitzenb\"ock connection, which is a connection related to torsion
without  curvature. A tetrad field can also  be used to define a
Riemannian metric, in terms of which the Levi-Civita connection can be
derived. In such a case,  the connection gives rise to curvature and
 not to  torsion.

As we saw in detail in the previous Section, it is crucial to mention that torsion and
curvature are properties of the  connection \cite{Aldrovandi:1996ke}, and several  
different
connections can be defined on the same space-time \cite{koba}. In other words,  the
presence of  nontrivial tetrad field in a gauge theory induces both  teleparallel and
Riemannian structures on a given   space-time. The first is related to the Weitzenb\"ock 
connection
\cite{weitz}, and the second to the Levi-Civita connection. Due to the universality of the
gravitational interaction,  it is possible to link both these geometrical
structures to the gravitational interaction.

In the context of teleparallel gravity, both curvature and torsion  provide
equivalent descriptions of gravity. However, conceptual differences have to be taken into 
account. 
According to GR, curvature is used to give a geometric picture  space-time. It  
successfully 
describes the gravitational interaction. On the contrary, teleparallelism  relates  
gravity  to 
torsion, and then  torsion accounts for
gravitation interaction
not by a geometric picture of  the interaction, but by acting as a {\it force}. 
This implies
that, in the teleparallel equivalent of GR, geodesic equation is  analogous to the Lorentz
force equation of electrodynamics \cite{deAndrade:1997qt}. However, the geodesic structure
is preserved in the teleparallel version of GR.
Thus, we can say that the gravitational interaction can be
described {\em alternatively} in terms of curvature, as it is usually done in GR, or in
terms of torsion, in  the so called teleparallel gravity. Therefore, at this level,
requiring for gravitational interaction   a ``curved'' or a ``twisted'' space-time turns 
out to be a
matter of convention.

%%%%%%%%%%%%%%%%%%%%%%%%%%%
\subsection{Teleparallel equivalent of General Relativity}
%%%%%%%%%%%%%%%%%%%%%%%%%%%%
Let us now describe the teleparallel equivalent of GR starting from the tetrad fields and 
the 
Weitzenb\"ock connection. The Greek alphabet $(\mu, \nu, \rho, \dots = 0,1,2,3)$ will be 
used to 
denote
indices related to space-time (base space), and the capital Latin alphabet $(A,B,C,
\dots = 0,1,2,3)$ to denote indices related to the tangent space (fiber),
assumed to be a Minkowski space with  metric
$\eta_{AB}=\mbox{diag}(+1,-1,-1,-1)$.
A gauge transformation can be defined as a local translation on the
tangent-space  as
\begin{equation}
\delta x^{A} = \delta\alpha^BP_Bx^{A} \; ,
\end{equation}
where $P_A = \partial /\partial x^A$  are the translation generators, and
$\delta \alpha^A$ the corresponding infinitesimal parameters. Defining the
gauge potentials as $A^B{}_{\mu}$, the gauge covariant derivative of a
given  matter field $\Psi$ is  \cite{deAndrade:1997qt}
\be
{\mathcal D}_\mu \Psi = e^B{}_{\mu} \; \partial_B \Psi \; ,
\ee
where
\be
e^B{}_{\mu} = \partial_{\mu}x^B + A^B{}_{\mu}\,.
\label{2.17}
\ee
This is a nontrivial tetrad field.  We assume  the speed of  light as $c=1$.
From the covariance of $D_{\mu} \Psi$,  one get  the gauge potential transformation
\begin{equation}
A^{B}{}_{\mu}^{\prime} = A^B{}_{\mu} -
 \partial_{\mu}\delta\alpha^B \; .
\end{equation}
As it is standard in  the Abelian gauge theories, the field strength is given by
\begin{equation}
F^B{}_{\mu \nu} = \partial_{\mu} A^B{}_{\nu}
- \partial_{\nu}A^B{}_{\mu} \; ,
\label{core}
\end{equation}
which satisfies the relation
\begin{equation}
[{\mathcal D}_{\mu}, {\mathcal D}_{\nu}] \Psi =
  F^B{}_{\mu \nu} P_B \Psi \; .
\end{equation}
It is worth stressing  that, whereas the tangent space indices are
raised and lowered by the metric $\eta_{AB}$, the space-time indices are
raised and lowered by the Riemann metric
\be
g_{\mu \nu} = \eta_{AB} e^A{}_{\mu} \, e^B{}_{\nu} \; .
\label{gmn}
\ee
A nontrivial tetrad field induces on space-time a teleparallel structure
directly related to  the gravitational field. In other words,
given a nontrivial tetrad, it is possible to define the so called
Weitzenb\"ock connection
\be
\hat{\Gamma}^{\rho}{}_{\mu \nu} = e_A{}^{\rho}\partial_{\nu}e^A{}_{\mu} \;  ,
\label{carco}
\ee
which is a connection without curvature and  presenting only torsion 
\cite{Aldrovandi:1996ke}. As
a direct consequence of this definition, the Weitzenb\"ock covariant derivative of
the tetrad field vanishes identically:
\be
\nabla_{\nu}e^A{}_{\mu} \equiv \partial_{\nu}e^A{}_{\mu} -
\hat{\Gamma}^{\rho}{}_{\mu \nu} \, e^A{}_{\rho} = 0 \; .
\label{cacd}
\ee
This is the so called absolute parallelism condition. Besides, the torsion related to  the 
Weitzenb\
"ock
connection is
\be
T^{\rho}{}_{\mu \nu} = \hat{\Gamma}^{\rho}{}_{\nu \mu} -
\hat{\Gamma}^{\rho}{}_{\mu \nu} \; ,
\label{tor}
\ee
from which  the gravitational "force" results from  the torsion
written in the tetrad basis, that is
\be
F^A{}_{\mu \nu} =  e^A{}_{\rho}T^{\rho}{}_{\mu \nu} \; .
\ee
A nontrivial tetrad field can  be adopted also to define a torsionless linear
connection which is  the Levi-Civita connection of the metric (\ref{gmn}), that is
\be
{\stackrel{\circ}{\Gamma}}{}^{\sigma}{}_{\mu \nu} = \frac{1}{2}
g^{\sigma \rho} \left[ \partial_{\mu} g_{\rho \nu} + \partial_{\nu}
g_{\rho \mu} - \partial_{\rho} g_{\mu \nu} \right] \; .
\label{lci}
\ee
The Weitzenb\"ock and the Levi--Civita connections are then related by the formula
\be
\hat{\Gamma}^{\rho}{}_{\mu \nu} =
{\stackrel{\circ}{\Gamma}}{}^{\rho} {}_{\mu \nu} +
K^{\rho}{}_{\mu \nu} \; ,
\label{rela}
\ee
where
\be
K^{\rho}{}_{\mu \nu} = {\textstyle \frac{1}{2}} \left( T_{\mu}{}^{\rho}{}_{\nu}
+ T_{\nu}{}^{\rho}{}_{\mu} - T^{\rho}{}_{\mu \nu} \right)
\label{contorsiontensor}
\ee
is the so  contorsion tensor defined above.
As already remarked, the curvature of the Weitzenb\"ock connection vanishes identically 
being
\be
{R}^{\rho}{}_{\lambda \mu \nu} = \partial_\mu
\hat{\Gamma}^{\rho}{}_{\lambda \nu} + \hat{\Gamma}^{\rho}{}_{\sigma
\mu} \; \hat{\Gamma}^{\sigma}{}_{\lambda \nu} - (\mu  \leftrightarrow \nu)
\equiv 0 \; .
\label{r}
\ee
Substituting $\hat{\Gamma}^{\rho}{}_{\mu \nu}$ as given in (\ref{rela}), we get
\be
{R}^{\rho}{}_{\lambda \mu \nu} =
{\stackrel{\circ}{R}}{}^{\rho}{}_{\lambda \mu \nu} +
Q^{\rho}{}_{\lambda \mu \nu} \equiv 0 \; ,
\label{relar}
\ee
where ${\stackrel{\circ}{R}}{}^{\rho}{}_{\lambda \mu \nu}$ is the
curvature of the Levi--Civita connection, and
\be
Q^{\rho} {}_{\lambda \mu \nu} = {D}_{\mu}{}{K}^{\rho}{}_{\lambda \nu} -
{D}_{\nu}{}{K}^{\rho}{}_{\lambda \mu} + {K}^{\sigma}{}_{\lambda \nu}
\; {K}^{\rho}{}_{\sigma \mu} - {K}^{\sigma}{}_{\lambda \mu} \;
{K}^{\rho}{}_{\sigma \nu}
\label{qdk}
\ee
is a tensor written in terms of the Weitzenb\"ock connection only.
$D_\mu$ is the teleparallel covariant derivative, which is  the
Levi-Civita covariant derivative of GR rephrased in terms
of the Weitzenb\"ock connection  \cite{deAndrade:1997cj}.
Operating on a space-time vector
$V^\mu$,  its explicit form is
\be
D_\rho \, V^\mu \equiv \partial_\rho V^\mu +
\left( \hat{\Gamma}^\mu{}_{\lambda \rho} - K^\mu{}_{\lambda \rho} \right) V^\lambda \; .
\label{tcd}
\ee
Eq. (\ref{relar}) can be straightforwardly  interpreted:  the contribution
${\stackrel{\circ}{R}}{}^{\rho}{}_{\lambda \mu \nu}$ coming from the
Levi--Civita connection compensates  the contribution
$Q^{\rho}{}_{\lambda \mu \nu}$ coming from the Weitzenb\"ock connection: this fact yields
an  curvature tensor ${R}^{\rho}{}_{\lambda \mu \nu}$ identically zero. This is a
constraint satisfied by the Levi--Civita and Weitzenb\"ock connections, and it is the
key  of the equivalence between the Riemannian and the teleparallel descriptions
of gravitational interaction.

Adopting the above results, the gauge gravitational field Lagrangian can be written as
 \be
{\cal L}_G =
\frac{e  }{16 \pi G} \; S^{\rho \mu \nu} \; T_{\rho \mu \nu} \; ,
\label{gala}
\ee
where $e = {\rm det}(e^{a}{}_{\mu})$, and
\be
S^{\rho \mu \nu} = - S^{\rho \nu \mu} \equiv {\textstyle \frac{1}{2}}
\left[ K^{\mu \nu \rho} - g^{\rho \nu} \; T^{\lambda \mu}{}_{\lambda} + g^{\rho \mu}
\; T^{\lambda \nu}{}_{\lambda} \right]
\label{Ssuperpotdef}
\ee
is a tensor written in terms of the Weitzenb\"ock connection. As standard in the
gauge theories, it is quadratic in the field strength. This approach has been
first presented in \cite{Maluf:1994ji}  and later on in \cite{deAndrade:1997qt}. In that
case, a tensor $\Sigma^{abc}$, equivalent to $S^{\rho\mu\nu}$, has been introduced
allowing for a substantial simplification of the field equations.

%%%%%%%%%%%%%%%%%%%%%%
\subsubsection{The field equations}
%%%%%%%%%%%%%%%%%%%%%%

Starting from  relation (\ref{rela}), the Lagrangian (\ref{gala}) can be reformulated  in 
terms of
the Levi-Civita connection only. A part a  divergence, the result is exactly the
Hilbert--Einstein Lagrangian of GR
\be
{\cal L} = - \frac{1}{16 \pi G} \;  \sqrt{-g} \, {\stackrel{\circ}{R}} \; ,
\ee
with  the identification $e = \sqrt{-g}$.
By  varying  the gauge Lagrangian ${\cal L}_G$  with respect to the gauge
field $A_B{}^\rho$, one obtains the teleparallel version of the gravitational field
equations, that is
\be
\partial_\sigma(e S_B{}^{\sigma \rho}) -
 4 \pi G  \, (e j_B{}^{\rho}) = 0 \; ,
\label{tfe1}
\ee
where $S_B{}^{\sigma \rho} \equiv e_B{}^{\lambda}
S_{\lambda}{}^{\sigma \rho}$. As for the Yang-Mills theories, the quantity
\be
e j_B{}^{\rho} \equiv \frac{\partial {\cal L}_G}{\partial e^B{}_{\rho}} = -
\frac{1}{4 \pi G} \, e e_B{}^{\lambda} S_{\mu}{}^{\nu \rho}
T^\mu{}_{\nu \lambda} + e_B{}^{\rho} {\cal L}_G
\label{ptem1}
\ee
is the gauge current that now is  the energy and
momentum of the gravitational field  \cite{deAndrade:2000kr}. The quantity  $(e 
S_B{}^{\sigma
\rho})$ is the  {\it superpotential} in the sense that its derivative yields the gauge
current $(e j_B{}^{\rho})$. Due to the anti-symmetry of $S_B{}^{\sigma \rho}$
in the last two indices, $e j_B{}^{\rho}$ is conserved as a consequence of the
field equations, i.e.
\be
\partial_\rho (e j_B{}^\rho) = 0 \; .
\label{conser1}
\ee
By   the identity
\be
\partial_\rho e \equiv e {\stackrel{\circ}{\Gamma}}{}^{\nu}{}_{\nu \rho} =
e \left( {\hat\Gamma}^{\nu}{}_{\rho \nu} - K^{\nu}{}_{\rho \nu} \right) \; ,
\label{id1}
\ee
the above conservation law can  be also  written in the form
\be
D_\rho \, j_B{}^\rho \equiv \partial_\rho j_B{}^\rho +
\left( {\hat\Gamma}^\rho{}_{\lambda \rho} - K^\rho{}_{\lambda \rho} \right)
j_B{}^\lambda = 0 \; ,
\label{conser2}
\ee
with $D_\rho$ the teleparallel  covariant derivative defined in (\ref{qdk}) following
\cite{deAndrade:1997cj}.

%%%%%%%%%%%%%%%%%%%%%%%%%%%%%%%%%%%
\subsubsection{Gravitational energy-momentum current}
%%%%%%%%%%%%%%%%%%%%%%%%%%%%%%%%
An important issue has to be discussed at this point. As can be easily proved, the current
$j_A{}^\rho$ transforms  under a general coordinate covariant  transformation;
it is invariant under local gauge translation of the tangent-space coordinates, and
transforms  under a covariant global Lorentz transformation in tangent--space. This 
property implies
that $j_A{}^\rho$, despite not being covariant under a local Lorentz transformation, is a
 space-time and gauge tensor \cite{deAndrade:2000kr}.
The relation between the above gauge
approach and GR can be easily found out. By using Eq.~(\ref{carco}) to express
$\partial_\rho e_A{}^\lambda$,  field Eqs. (\ref{tfe1}) can be rewritten in
a space-time form as
\be
\partial_\sigma(e S_\lambda{}^{\sigma \rho}) -
 4 \pi G  \, (e t_{\lambda}{}^{\rho}) = 0 \; ,
\label{tfe2}
\ee
where
\be
e t_{\lambda}{}^{\rho} =
\frac{  e}{4 \pi G} \, \Gamma^{\mu}{}_{\nu \lambda} S_{\mu}{}^{\nu \rho}
+ \delta_\lambda{}^{\rho} {\cal L}_G
\label{ptem2}
\ee
stands for the canonical energy-momentum pseudotensor of the gravitational field
\cite{Virbhadra:1990vs,Virbhadra:1990zr,Aguirregabiria:1995qz,Shirafuji:1996im}.  
Eq.~(\ref{tfe2}) is symmetric in
$(\lambda \rho)$. Besides, by  Eq.  ~(\ref{rela}), it can be rewritten in
terms of the Levi-Civita connection. As expected, due to the equivalence
between the corresponding Lagrangians, the Einstein field  equations are reproduced:
\be
\frac{e}{2} \left[{\stackrel{\circ}{R}}_{\mu \nu} -
\frac{1}{2} \, g_{\mu \nu}
{\stackrel{\circ}{R}} \right] = 0 \; .
\ee
Hence, since the field equations coincide completely with those of GR: due to this 
property, we can 
deal with  the  ``Teleparallel Equivalent of General Relativity'' (TEGR)
\cite{Unzicker:2005in}.

It is worth  noticing that the canonical energy-momentum pseudotensor
$t_{\lambda}{}^{\rho}$ is not simply the gauge current $j_A{}^\rho$ with the
algebraic index ``$A$'' changed to the space-time index ``$\lambda$''. It
incorporates also an extra term coming from the derivative term of
Eq.~(\ref{tfe1}), that is
\be
t_\lambda{}^\rho = e^A{}_\lambda \, j_A{}^\rho +
\frac{1}{4 \pi G} \, \Gamma^{\mu}{}_{\lambda \nu} S_{\mu}{}^{\nu \rho} \; .
\label{ptem3}
\ee
As a consequence, one can  see the origin of the connection-term which transforms the 
gauge
current $j_A{}^\rho$ into the energy-momentum pseudotensor $t_\lambda{}^\rho$.
By the same mechanism, it is possible to  exchange  terms
between the derivative and the current terms into the  field Eqs. (\ref{tfe2}). This fact
gives rise to different definitions of the energy-momentum pseudotensor, each
one connected to a different {\it superpotential} $(e S_\lambda{}^{\rho \sigma})$.

Like in the case of  the gauge current $(e j_A{}^\rho)$, the pseudotensor $(e 
t_\lambda{}^\rho)$
is conserved as a consequence of the field equations. That is
\be
\partial_\rho (e t_\lambda{}^\rho) = 0 \; .
\label{conser3}
\ee
However, despite
 what occurs with $j_A{}^\rho$, due to the pseudotensor properties of
$t_\lambda{}^\rho$, this conservation law cannot be rewritten as a covariant
derivative.

Thanks to these features, the TEGR approach to gravitation seems to
be  more appropriate than GR to deal with the energy problem of  gravity. In fact,  as 
pointed out 
by M{\o}ller,  a satisfactory solution to
the problem of gravitational  energy  could be obtained in the
framework of tetrad gravity. In our notation, the  M{\o}ller expression of the 
gravitational
energy-momentum density is  \cite{moller}
\be
e t_\lambda{}^\rho = \frac{\partial {\cal L}}{\partial \partial_\rho e^A{}_\mu} \;
\partial_\lambda e^A{}_\mu + \delta_\lambda{}^\rho \, {\cal L} \; ,
\ee
which is nothing else but the standard canonical energy-mo\-mentum density given  by the
Noether  theorem. Using  the gauge Lagrangian (\ref{gala}), it is
 easy  to verify that M{\o}ller's expression coincides exactly with the
teleparallel energy-momentum density appearing in the field equation
(\ref{tfe2})-(\ref{ptem2}). Since $j_A{}^\rho$ is a  space-time tensor, whereas
$t_\lambda{}^\rho$ is not, the gauge current $j_a{}^\rho$ is a version of the M{\o}ller 
energy-
momentum density $t_\lambda{}^\rho$.
Mathematically, they can be obtained from each other by using the relation
(\ref{ptem3}). It should be remarked, however, that both of them transform
covariantly  under  global  tangent-space Lorentz transformations. The lack
of a  local Lorentz covariance in the  tetrad teleparallel gravity  can be
considered as the teleparallel manifestation of the pseudotensor character of the
gravitational energy-momentum density of GR
\cite{deAndrade:2000kr,Krssak:2015rqa,Krssak:2015lba,Krssak:2015oua}.

For the sake of completeness, some remarks are in order at this point. The problem
of the localization of the gravitational energy-momentum tensor has been   discussed in
detail in \cite{Maluf1}. The first investigation of the  gravitational energy localization
in the context of  TEGR was developed in \cite{Maluf2}  and the gravitational
energy-momentum four-vector $P^a$ was  first introduced in \cite{Maluf3}, and further
developed in \cite{Maluf4}. The integral of the gauge current $e\,j_{B}{}^{\rho}$, given 
in
\eqref{ptem1}, over a finite volume of the 3-dimensional space, is related to $P^a$.
Furthermore, a real energy-momentum tensor (not a gauge current) was firstly
presented  in \cite{Maluf5}  where consistent conservation laws and expressions for the
gravitational energy-momentum  flux were derived and discussed. In all these studies, the
physical interpretation of the tetrad fields,  where a frame is adapted to moving
observers, is considered. Specifically,  for a given metric, a  tetrad field set,
adapted to the  observer, is chosen. A detailed discussion of this mathematical structure
 is  reported in Ref. \cite{Maluf1}. Finally, as reported above, by rewriting the field
equations in a purely space-time form, the gauge current $ j_A{}^{\rho}$ reduces to the
canonical energy-momentum pseudo-tensor of the gravitational field. Alternatively,
 as reported  in Ref. \cite{Maluf5}, the true energy-momentum tensor exactly yields the
gravitational energy-momentum four-vector $P^a$. As a final remark, we have to say that
the correct definitions of the gravitational energy-momentum tensor and current are
crucial issues of the teleparallel picture of GR. These topics deserve further
investigations.

%%%%%%%%%%%%%%%%%%%
\subsubsection{Bianchi Identities in teleparallel gravity}
%%%%%%%%%%%%%%%%%%%%%%%%%%
Let us consider the second Bianchi identity of GR
\be
{\stackrel{\circ}{\nabla}}_\sigma {\stackrel{\circ}{R}}_{\lambda \rho \mu \nu} +
{\stackrel{\circ}{\nabla}}_\nu {\stackrel{\circ}{R}}_{\lambda \rho \sigma \mu} +
{\stackrel{\circ}{\nabla}}_\mu {\stackrel{\circ}{R}}_{\lambda \rho \nu \sigma} = 0
\; ,
\label{birg2}
\ee
where ${\stackrel{\circ}{\nabla}}_\mu$ is the  Levi-Civita covariant
derivative. Its contracted form is
\be
{\stackrel{\circ}{\nabla}}_\mu \left[ {\stackrel{\circ}{R}}{}^\mu{}_\nu -
\textstyle{\frac{1}{2}} \delta^\mu{}_\nu {\stackrel{\circ}{R}} \right] = 0 \; .
\ee
From Eq.  (\ref{relar}),  it is
possible to rewrite it in terms of the Weitzenb\"ock connection. The result is
\be
D_\rho \left[ \partial_\sigma(e S_\lambda{}^{\sigma \rho}) -
 4 \pi G  \, (e t_{\lambda}{}^{\rho}) \right] = 0 \; ,
\label{bi3}
\ee
where $D_\rho$ is the teleparallel covariant derivative, defined in Eq.  (\ref{conser2}).
This is the TERG second Bianchi identity. It expresses
the fact that the teleparallel covariant derivative of the sourceless field Eqs.
(\ref{tfe2})  identically vanishes.

In the presence of a  generic matter field, the TEGR  field Eqs.
(\ref{tfe2}) becomes
\be
\partial_\sigma(e S_\lambda{}^{\sigma \rho}) -
 4 \pi G  \, (e t_{\lambda}{}^{\rho}) =
 4 \pi G  \, [e\, {T^{(m)}}_{\lambda}{}^{\rho}] \; ,
\label{tfe3}
\ee
with ${T^{(m)}}_{\lambda}{}^{\rho}$ the matter energy-momentum tensor. From the Bianchi 
identity (\
ref{bi3}), and  Eq. (\ref{id1}), we obtain
\be
D_\rho\, {T^{(m)}}_\lambda{}^\rho = 0 \; .
\label{telecon}
\ee
This is the conservation law of matter energy-momentum tensor.  It is worth noticing
that, in teleparallel gravity, it is not the Weitzenb\"ock covariant derivative
$\nabla_\mu$, but the teleparallel covariant derivative (\ref{conser2}) that
yields the conservation law of  energy-momentum tensor of matter
fields. It should be remarked that (\ref{telecon}) is the only conservation law compatible
with the corresponding conservation law of GR, that is
\be
{\stackrel{\circ}{\nabla}}{}_\mu \,{T^{(m)}}^\mu{}_\rho \equiv
\partial_\mu \, {T^{(m)}}^\mu{}_\rho +
{\stackrel{\circ}{\Gamma}}{}^{\mu}{}_{\lambda \mu} \,{T^{(m)}}^\lambda{}_\rho -
{\stackrel{\circ}{\Gamma}}{}^{\lambda}{}_{\rho \mu} \,{T^{(m)}}^\mu{}_\lambda = 0 \; ,
\label{grcon}
\ee
as can be   verified by the  relation (\ref{rela}).

%%%%%concl
In GR,  the gravitational field is expressed by the torsionless Levi--Civita
metric--connection, whose curvature determines the strength of the gravitational field.
On the other hand, in TEGR, the presence of a
gravitational field is expressed by the flat Weitzenb\"ock connection, whose torsion is
 responsible for determining the intensity of gravitational field. Therefore, gravity can 
be 
described  either in terms of curvature or
in terms of torsion. In other words, whether  gravity  requires a curved or a twisted 
space-time is 
a matter of convention.

An important feature of the teleparallel equivalent of GR is that it allows to
define  an energy-momentum gauge current $j_A{}^\rho$ for the gravitational field,
which is covariant under a space-time general coordinate transformation, and transforms
covariantly under a global tangent-space Lorentz transformation. This means essentially
that $j_A{}^\rho$ is a space-time tensor, but not a tangent--space tensor. Then, by
rewriting the gauge field equation in a  space-time form, it becomes the Einstein field
equations: in this sense, the gauge current $j_A{}^\rho$ reduces to the canonical 
energy-momentum
pseudotensor of the gravitational field. In other words, TEGR  seems to
provide a more appropriate framework to deal with the energy problem, since in the
standard context of GR, the energy-momentum density for the gravitational field is
always  represented by a pseudotensor.

In GR, the conservation law of the energy-momentum tensor of matter  fields
can be obtained from the Bianchi identities. In the case of TEGR, the
energy-momentum tensor turns out to be conserved by the teleparallel covariant
derivative, which is the Levi-Civita covariant derivative of GR rephrased in terms of the
Weitzenb\"ock connection \cite{deAndrade:1997cj}.

As final remark, we obtained the gravitational analog of the Lorentz force
equation. This  is an equation written in the  Weitzenb\"ock space-time.
According to this approach, the trajectory of a particle is described  as the Lorentz 
force 
describes the trajectory of a charged particle in the presence
of  electromagnetic fields.  Here, the  torsion plays the role of the force. When 
rewritten in
terms of   metric structure, such an equation becomes the geodesic equation of
GR, which is an equation written in the  Riemann space-time. Since both
equations are deduced from the same equation, they are just equivalent ways of describing
the same physical trajectory induced by the presence of the gravitational field.

%%%%%%%%
\section{Modified gravity in terms of torsion: The $f(T)$ extension}
\label{SecionfT}
%%%%%%%%%%%%%%%%%%%%%%%%%%%%%%%

In the previous Section we saw that one can alternatively formulate a gravitational
theory in terms of torsion. On the other hand, the last two decades there is a huge
effort in the literature to modify gravity \cite{Capozziello:2011et} in order to be able
to describe the observed universe evolution, as well as alleviate the
non-renormalizability issues of GR \cite{Stelle:1976gc,Biswas:2011ar}. Hence, even if one
decides to take the serious step of modifying gravity, there is still the question of
what
formulation of gravity to modify. Most of the works in the literature start from the
usual, curvature-based formulation, and modify/extend the Einstein-Hilbert action.
However, having in mind the discussion of the previous Section, one could reasonably
think
to start from the teleparallel equivalent of general relativity (TEGR), and use it as a
basis to build a gravitational modification. The simplest such modification, namely
$f(T)$
gravity, will be presented in detail in this Section.

The need of constructing a modified gravitational theory, that possesses GR as a
particular limit, arose from the detailed observational data. In fact, along with the
rapid development of observational cosmology since the late 1990s, cosmologists
believe that the universe may have undergone two phases of cosmic acceleration. The first
phase is called inflation, which was proposed in 1980s to solve several conceptual
puzzles
of standard hot Big Bang cosmology   \cite{Guth:1980zm, Linde:1981mu, Albrecht:1982wi}
(see also \cite{Starobinsky:1980te,Fang:1980wi, Sato:1980yn}). This cosmic acceleration
is suggested to occur about $10^{-35}$ second after the Big Bang. Based on this paradigm,
the corresponding perturbation theory predicted a nearly scale invariant power spectrum
of
primordial density fluctuations \cite{Press:1980zz,Mukhanov:1981xt} (see
\cite{Mukhanov:1990me} for a comprehensive review) and this significant prediction was
later verified to high precision by a series of cosmological observations
\cite{Ade:2013zuv, Smoot:1992td, Bennett:2003bz, Abazajian:2003jy}. The second stage of
the cosmic acceleration is happening right now, of which the underlying mystery is
addressed as an existence of an unknown dark energy component
\cite{Weinberg:1988cp,Carroll:1991mt,Peebles:2002gy,
Padmanabhan:2002ji,Capozziello:2003tk,Copeland:2006wr,
Albrecht:2006um,Linder:2008pp,Frieman:2008sn,
Caldwell:2009ix,Silvestri:2009hh,Cai:2009zp,Li:2011sd}. The speeding up of our
universe at present was discovered by two independent observational signals on distant
Type Ia supernovae (SNIa) in 1998 \cite{Riess:1998cb, Perlmutter:1998np}, and later it
was confirmed by a number of observations, such as those concerning the large scale
structure (LSS) \cite{Tegmark:2003ud, Tegmark:2006az}, baryon acoustic oscillations
(BAO) \cite{Eisenstein:2005su, Percival:2007yw}, cosmic microwave background (CMB)
radiation  \cite{Ade:2013zuv,Komatsu:2008hk,Komatsu:2010fb,Hinshaw:2012aka}, and their
cross-correlations.

Although phenomenological studies of the above two cosmic accelerations have been
successful, the fundamental theory of nature that could explain the microscopic physics
of
these phenomena remains unknown at present. One possible approach is to introduce novel,
exotic, forms of matter, the simplest example of which are scalar fields such as the
inflaton and quintessence ones. It is phenomenologically viable to construct a
particular profile of the potential for the scalar field such that the dynamical
evolution can fit to the observations.

There exists, however, a second approach to explain these accelerating phases, that is to
modify the gravity sector itself
\cite{Capozziello:2002rd,Capozziello:2003tk,Carroll:2003wy,
Vollick:2003aw, Dolgov:2003px, Nojiri:2003ft, Chiba:2003ir, Nojiri:2003ni,Carroll:2004de,
Capozziello:2005ku, Woodard:2006nt, Amendola:2006kh, Capozziello:2006dj,
Starobinsky:2007hu}. As we mentioned above,  most of the works in the literature start
from the usual, curvature-based formulation, and modify/extend the Einstein-Hilbert
action, with the simplest example being the $f(R)$ paradigm in which the Lagrangian is
considered to be a non-linear function of the curvature scalar (see
\cite{Capozziello:2007ec,Sotiriou:2008rp, DeFelice:2010aj,
Nojiri:2010wj,Clifton:2011jh,Capozziello:2011et} for reviews).

Nevertheless, it is also reasonable to think to start from the Teleparallel Equivalent of
General Relativity (TEGR), namely from the torsional formulation of gravity, and try to
construct various extensions. As one could straightforwardly be inspired by the $f(R)$
extensions of curvature-based gravity, the simplest class of these torsion-based
modifications is the paradigm of $f(T)$ gravity \cite{Ferraro:2006jd, Ferraro:2008ey,
Bengochea:2008gz}, in which the Lagrangian is taken to be a non-linear function of the
TEGR Lagrangian $T$. The crucial issue is that although TEGR coincides completely with
general relativity at the level of equations, $f(T)$ is different from $f(R)$ gravity,
with novel features (amongst others note the significant advantage that the field
equations of $f(T)$ gravity are of second order while those of $f(R)$ are of fourth
order). The $f(T)$ gravity has interesting cosmological solutions, which provide
alternative interpretations for the accelerating phases of the universe
\cite{Ferraro:2006jd,Ferraro:2008ey,
Bengochea:2008gz,Linder:2010py,Myrzakulov:2010vz,Wu:2010mn,Wu:2010xk,Tsyba:2010ji,
Chen:2010va,Wu:2010av,Bamba:2010iw,Myrzakulov:2010tc,Karami:2011np,
Karami:2013rda, Bamba:2010wb,
Geng:2011aj,Liu:2012fk,Banijamali:2012nx,Dong:2012pp,Behboodi:2012ak,Daouda:2012wt,
Ferraro:2012wp,Tamanini:2012hg,Dong:2012en,Liu:2012kk,Setare:2012vs,Karami:2012if,
Yang:2012hu,Jamil:2011mc,Farajollahi:2011af,Wu:2011xa,Wei:2011mq,Meng:2011ne,Wei:2011jw,
Khatua:2011gv,Chattopadhyay:2011fp,Karami:2010bu,Yang:2010hw,Jamil:2012ju,Karami:2012fs,
Jamil:2012yz,Rahaman:2012qk,Daouda:2011yf,Yerzhanov:2010vu,Karami:2010xy,Ferraro:2011zb,
Nashed:2015pda,Fayaz:2015yka,Darabi:2014dla,Wu:2015naa,Aghamohammadi:2014vsa,Qi:2014yxa,
Bamba:2014zra,Farooq:2013ava,Li:2013xea,Setare:2013xh,Setare:2012ry,Darabi:2012zh,
Chattopadhyay:2012eu,Ghosh:2012dy,Nashed:2014baa,Abedi:2015cya}.

In the present section we give a review of the basic setup of $f(T)$ gravity and the
corresponding background dynamics, as well a brief survey of linear perturbations.

%%%%%%%%%%%%%%%%%%
\subsection{Equations of motion}
\label{eomsbasicfT}
%%%%%%%%%%%%%%%%%%%%%%%%%%%%%%%%%%%

As we described in detail in the previous sections, the dynamical variable of the
teleparallel gravity, as well as of its $f(T)$ extension, is the vierbein field ${\bf
e}_A(x^\mu)$. As usual we use Greek indices to run over all space-time coordinates and
capital Latin indices to denote the coordinates of the tangent space-time. The vierbein
field forms an orthonormal basis for the tangent space at each point $x^\mu$ of the
manifold, i.e., ${\bf e}_A\cdot{\bf e}_B = \eta_{AB}$ where $\eta_{AB} = {\rm
diag}(+1,-1,-1,-1)$ is the Minkowski metric for the tangent space. As we mentioned above,
we analyze the vierbein vector with the use of its component form ${\bf
e}_A = e_
A^\mu\partial_\mu$, and thus the metric tensor can be expressed as in (\ref{gmn}), namely
\begin{eqnarray}
 g_{\mu\nu}(x) = \eta_{AB} e^A_\mu(x)e^B_\nu(x) ~.
\end{eqnarray}
Moreover, the vierbein components follow the usual relations
$ e^\mu_A e^A_\nu = \delta^\mu_\nu$ and $e^\mu_A e^B_\mu = \delta^B_A $.

In $f(T)$ gravity, similarly to TEGR, one uses the Weitzenb\"ock connection \cite{weitz},
defined in (\ref{carco}) as
\begin{eqnarray}
 \hat\Gamma^\lambda_{\mu\nu} \equiv e^\lambda_A \partial_\nu e^A_\mu
 = -e_\mu^A \partial_\nu e_A^\lambda ~.
\end{eqnarray}
This definition obviously leads to zero curvature, but nonzero torsion, and accordingly
one can write down the torsion tensor as in (\ref{tor}), namely
\begin{eqnarray}
\label{torsiontensor}
 T^\lambda_{\mu\nu} \equiv \hat\Gamma^\lambda_{\mu\nu} -
\hat\Gamma^\lambda_{\nu\mu}
 = e^\lambda_A ~ ( \partial_\mu e^A_\nu - \partial_\nu e^A_\mu) ~.
\end{eqnarray}
As we mentioned in detail in the previous Section the difference between the Levi-Civita
and Weitzenb\"ock connections is the contortion tensor $
 K^{\mu\nu}_{\:\:\:\:\rho} = - \frac{1}{2} \Big( T^{\mu\nu}_{\:\:\:\:\rho} - {
T}^{\nu\mu}_{\:\:\:\:\rho}
 - T_{\rho}^{\:\:\:\:\mu\nu} \Big)$, while it proves useful to define the
superpotential
\begin{equation}  \label{Stensor}
 S_\rho^{\:\:\:\mu\nu} = \frac{1}{2}\Big(K^{\mu\nu}_{\:\:\:\:\rho} +\delta^\mu_\rho
{T}^{\alpha\nu}_{\:\:\:\:\alpha}
 -\delta^\nu_\rho T^{\alpha\mu}_{\:\:\:\:\alpha}\Big) ~.
\end{equation}
Using these quantities one can writing down the teleparallel Lagrangian density, which is
nothing
other than the torsion scalar  \cite{Hayashi:1979qx, Maluf:1994ji, Arcos:2005ec}, as
\begin{equation}  \label{telelag}
 T \equiv S_\rho^{\:\:\:\mu\nu} ~ T^\rho_{\:\:\:\mu\nu}.
\end{equation}
Then in the present formalism all the information concerning the gravitational field is
included in the torsion tensor $T^\lambda_{\:\mu\nu} $, and the torsion scalar $T$
arises from it in a similar way as the curvature scalar arises from the Riemann curvature
tensor. Finally, the torsion scalar gives rise to the dynamical equations for the
vierbein, which imply the Einstein equations for the metric.

While in teleparallel gravity the action is constructed by the teleparallel Lagrangian
density $T$, the idea of $f(T)$ gravity is to generalize $T$ to an arbitrary function
$f(T)$, which is similar in spirit to the generalization of the Ricci scalar $R$ in the
Einstein-Hilbert action to a function $f(R)$. In particular, the action in a universe
governed by $f(T)$ gravity reads:
\begin{eqnarray}  \label{action_fT}
 {\cal S} = \int d^4x ~|e|~ \left[ \frac{f(T)}{16\pi G} + L^{(m)} \right],
\end{eqnarray}
where $|e| = \text{det}(e_{\mu}^A) = \sqrt{-g}$ and $L^{(m)}$ stands for the matter
Lagrangian. For convenience, we would like to rewrite $f(T)=T+F(T)$ and mention that
since the Ricci scalar $R$ and the torsion scalar $T$ differ only by a total derivative
\cite{Weinberg:2008zzc}, the action (\ref{action_fT}) is equivalent to GR
in the case of a vanishing $F(T)$ term, i.e. in the case where $f(T)$ gravity becomes
TEGR.

Variation of the action \eqref{action_fT} with respect to the tetrad $e^A_\nu$ leads to
the field equations as
\begin{eqnarray}\label{eom_fT_general}
&&\!\!\!\!\!\!\!
 e^{-1}\partial_{\mu}(e e^\rho_A
S_{\rho}{}^{\mu\nu}) [1+F_{T}] - e_{A}^{\lambda}{\cal
T}^{\rho}{}_
{\mu\lambda}S_{\rho}{}^{\nu\mu}[1+F_{T}] \nonumber\\
&&\!\!\!\!\!\!\!
 + e^\rho_A S_{\rho}{}^{\mu\nu}(\partial_{\mu}{T})F_{TT} +\frac{1}{4}e_{A}^{\nu}
[T+F({T})]
 = 4\pi G e_{A}^{\rho} {T^{(m)}}_{\rho}{}^{\nu}\,,\nonumber\\
\end{eqnarray}
where $F_{T}$ and $F_{TT}$ denote the first and second order derivatives of $F(T)$ with
respect to the torsion scalar $T$, and ${T^{(m)}}_{\rho}{}^{\nu}$ is the energy-momentum
tensor constructed by the matter field Lagrangian.

If we assume the background manifold to be a spatially flat Friedmann-Robertson-Walker
(FRW) universe then the vierbein takes the form
\begin{equation}  \label{weproudlyuse}
 e_{\mu}^A=\mathrm{diag}(1,a,a,a) ~,
\end{equation}
with $a$ the scale factor of the universe (hence the dual vierbein is $
e^{\mu}_A=\mathrm{diag}(1,
a^{-1},a^{-1},a^{-1})$ and the determinant $e=a^3$). One can immediately see that this
choice gives rise to the well-known FRW metric
\begin{equation}
 ds^{2}=dt^{2}-a^{2}(t)\delta_{ij}dx^idx^j\,.
\label{metriccosmo}
\end{equation}
Moreover, assuming a perfect fluid for matter, its energy momentum tensor takes the form
\begin{eqnarray}
 {T^{(m)}}_{\mu\nu} = -p_mg_{\mu\nu} - (\rho_{m} + p_m )u_\mu u_\nu ~,
\end{eqnarray}
where $p_m$, $\rho_{m}$ and $u^\mu$ are the pressure, energy density and four
velocity of the matter fluid. For the spatially flat FRW background, equations
\eqref{eom_fT_general} lead to the following effective Friedmann equations:
\begin{align}
\label{background11}
 H^{2} &= \frac{8\pi G}{3}\rho_{m} - \frac{F({T})}{6} - 2f_{T}H^2 ~, \\
\label{background22}
 \dot{H} &= -\frac{4\pi G(\rho_{m} + p_m)}{1+ F_{T} -12H^2 F_{TT} } ~.
\end{align}
In the above expressions we have introduced the Hubble parameter $H\equiv \dot{a}/a$ to
describe the expansion rate of the universe, where a dot denotes a derivative with
respect to the cosmic time. Additionally, we have a very useful relation for the torsion
scalar, namely
\begin{equation} \label{TH2}
 T = -6H^{2} ~,
\end{equation}
which can be derived straightforwardly from the definition \eqref{telelag} in the case of
the FRW vierbein (\ref{weproudlyuse}).

Usually, one can introduce an equation-of-state parameter $w_m = p_m /
\rho_{m}$ to characterize the dynamics of the matter fluid, where the fluid satisfies
the continuity equation
\begin{eqnarray}\label{eom_rho_m}
 \dot\rho_{m} + 3H(1+w_m) \rho_{m} = 0 ~.
\end{eqnarray}
With the above background equations at hand, one can investigate cosmological dynamics
governed by
any $f(T)$ gravity model.

\subsection{Lorentz violation}
\label{LorViolsusection}

In general, the theory of $f(T)$ gravity is different from $f(R)$ gravity in several
perspectives. Firstly, the $f(T)$ gravity cannot be reformulated as a teleparallel action
plus a scalar field through the conformal transformation due to the appearance of
additional scalar-torsion coupling terms  \cite{Yang:2010ji} (see also 
\cite{Bamba:2013jqa}). Secondly, as we 
have
already mentioned, the equations of motion of $f(T)$ gravity remain second-order rather
than the fourth-order equations derived in $f(R)$. Thirdly, in $f(T)$ gravity one can
obtain more degrees of freedoms compared with $f(R)$ theories, which correspond to one
massive vector field  \cite{Li:2011rn, Li:2011wu}.

One crucial difference, that has triggered a big discussion in the literature concerns
Lorentz invariance. In particular, in the standard formulation of $f(T)$ gravity local
Lorentz invariance is either completely absent or strongly restricted
\cite{Li:2010cg,Sotiriou:2010mv,Ferraro:2014owa}, due to the strong imposition made in
\cite{Ferraro:2006jd,Bengochea:2008gz,Ferraro:2008ey,Linder:2010py, Li:2010cg} that the
spin
connection vanishes. This assumption has the  good motivation to make the
theory simpler in order to derive solutions: in fact, the spin connection has in
general 24 degrees of freedom, however under the teleparallel condition and the fact that
the 20 independent components of the Riemann tensor are zero, the spin connection
components reduce and hence they can be set to zero through a local Lorentz
transformation. However, in the case of $f(T)$ gravity, in general this
assumption makes the theory frame-dependent
since a solution of the field equations depends on the choice of the frame
\cite{Li:2010cg}.

A simple way of examining the local Lorentz violation in the usual formulation of $f(T)$
gravity is to study the corresponding conformal rescaling. It is well know that, for the
action of $f(R)$ gravity, there exists an equivalent theory that is described by Einstein
gravity minimally coupled with a scalar field. Similarly, one can perform the following
conformal transformation:
\begin{eqnarray}
 \tilde{e}_\nu^A = e^{\frac{\tilde\phi}{2\sqrt{3}}} e_\nu^A~,
\end{eqnarray}
by introducing a dimensionless scalar field $\tilde{\phi}$. Then the action
\eqref{action_fT} can
be rewritten as
\begin{eqnarray}
\label{action_fT_conformal}
&&\!\!\!\!\!\!\!\!\!\!
S = \int d^4x \sqrt{-\tilde{g}} \frac{1}{16\pi G}
 \left[ \tilde{R} +
\frac{1}{2}\partial_\mu\tilde{\phi}\partial^\mu\tilde{\phi} - U(\tilde{\phi})
\right.\nonumber\\
&&\left.\  \ \ \ \ \ \ \  \ \  \ \ \ \ \ \ \  \  \ \  \ \ \ \ \ \
 - \frac{2}{\sqrt{3}} \tilde T^{\lambda\nu}_\lambda \partial_\nu\tilde{\phi}
\right]
~,
\end{eqnarray}
with $U(\tilde{\phi}) = \frac{TF_{T}-F}{(1+F_{T})^2}$. In the expression of the action
we have applied the relations $|e|=\sqrt{-g}$ and $T = -R -2\nabla^\nu {\cal
T}^\lambda_{\nu\lambda}$.

From \eqref{action_fT_conformal} one can immediately observe the following interesting
properties: Firstly, the sign in front of the kinetic term of the scalar field is
negative
and thus it indicates a potential instability for perturbations, unless there is some
symmetry or dynamical mechanism to stabilize it. Secondly, the last term appearing in
\eqref{action_fT_conformal} explicitly shows a mixing between the kinetic term of the
scalar field and the torsion fields. This term is not a local Lorentz scalar, and this
feature clearly reflects the violation of local Lorentz invariance in the usual
formulation of $f(T)$ gravity \cite{Li:2010cg}.

The Lorentz violation problem is potentially a severe one. One can neglect this issue and
investigate solutions in particular frames (this is in analogy with the investigation of
electromagnetism in the particular class of inertial frames), however strictly speaking
the problem is there and will become obvious when questions about frame transformations
and Lorentz invariance are raised, which is usual for instance in the case of spherically
symmetric solutions.

Hence, the only way to solve this severe problem is to construct a consistent,
covariant, formulation of $f(T)$ gravity. This is achieved if instead of the
pure-tetrad  teleparallel gravity  we start from the covariant teleparallel gravity 
\cite{Obukhov:2002tm,Obukhov:2006sk,Lucas:2009nq, Pereirabook,Krssak:2015rqa,
Krssak:2015lba},
using both the vierbein and the spin connection in a way that for every vierbein choice a
suitably constructed connection makes the whole theory covariant, as it was recently
done in \cite{Krssak:2015oua}. This is performed in subsection \ref{restoringLI} below.

%%%%%%%%%%%%%%%%%%%%%%%%%
\subsection{Degrees of freedom in $f(T)$ gravity}
\label{DOFinFT}
%Based on  \cite{Li:2011rn,Li:2011wu}
%%%%%%%%%%%%%%%%%%%%%%%%%%%%%%%

In order to examine the number of degrees of freedom in $f(T)$ gravity, one can
straightforwardly perform the Hamiltonian analysis and study the corresponding constraint
structure \cite{Li:2011rn}. This approach was also developed in \cite{Maluf:2000ag,
daRochaNeto:2011ir} to investigate the constraint structure of teleparallel gravity.
To begin with, we would like to slightly reformulate the
Lagrangian density of $f(T)$
(recall that $f(T) = T + F(T)$ as introduced in   subsection \ref{eomsbasicfT})
in the form of the Brans-Dick theory as follows:
\begin{eqnarray} \label{L_fT_BD}
 {\cal L}_{BD} = \frac{|e|}{16\pi G} \left[ \phi T - W(\phi) \right] ~,
\end{eqnarray}
with
\begin{eqnarray} \label{phi_W_BD}
 \phi \equiv 1+F_{T}~,~~W(\phi) = TF_{T}-F ~,
\end{eqnarray}
being introduced respectively as one auxiliary field and the corresponding potential,
which arise from the regular approach of Lagrange multipliers.
Note that, one can write $T$ as a function of $\phi$ by solving the first equation of
\eqref{phi_W_BD} inversely.

Ignoring the term $16\pi G$ for simplicity, one can define the conjugate momenta for the
vierbein $e_{A\mu}$ and the scalar field $\phi$ as
\begin{align}
\label{Pi_momenta_BD}
 \Pi^{A\mu} &\equiv \frac{\partial {\cal L}_{BD}}{\partial (\partial_0 e_{A\mu})}
 = 4\phi |e| S^{A0\mu} ~, \\
\label{pi_constraint_BD}
 \pi  &\equiv \frac{\partial {\cal L}_{BD}}{\partial (\partial_0 \phi)} = 0 ~,
\end{align}
respectively. It is obvious that \eqref{pi_constraint_BD} itself automatically
yields a constraint in the dynamical system, since there is no time derivative of $\phi$.
Moreover, one can easily observe that
\begin{eqnarray} \label{Pi^A0_constraint_BD}
 \Pi^{A0} = 0~,
\end{eqnarray}
due to the vanishing time derivatives of $e_{A0}$. In addition, there exist another
class of primary constraint equations which are expressed as:
\begin{eqnarray}
&&\!\!\!\!\!\!\!\!\!\!\!\!\!\!\!\!\!\!
\Gamma^{AB} = \Pi^{AB} - \Pi^{BA}
\nonumber\\
&&\!\!\!\!\!
+
2\phi e \big[ e^{A l}e^{Bk}T_{lk}^0
 \!-\!(e^{Al}e^{B0}-e^{Bl}e^{A0}) T^k_{lk} \big] \approx 0 .
\end{eqnarray}
Then, one can write down the total Hamiltonian density as
\begin{align} \label{H_total_BD}
 H = H_0 + \lambda_{AB}\Gamma^{AB} + \lambda\pi ~.
\end{align}
The first term of \eqref{H_total_BD} is given by
{\small{
\begin{eqnarray} \label{H_0_BD}
 &&\!\!\!\!\!\!\!\!\!\!
 H_0 = \Pi^{A\mu} \dot e_{A\mu} + \pi \dot\phi -{\cal L}_{BD} \nonumber\\
 &&= e W(\phi) -e_{A0}\partial_k \Pi^{Ak} -{\frac{e}{4g^{00}}} \phi
 \left( g_{ik}g_{jl}P^{ij}P^{kl}- \frac{1}{2}P^2 \right)\nonumber\\
 &&\ \ \
 +\phi e \left( \frac{1}{4}g^{il}g^{kj}T^A\,_{lk}T_{Aij}
 + \frac{1}{2}g^{lj}T^a\,_{kl}T^k\,_{aj}\right.\nonumber\\
 &&\ \ \  \ \  \ \  \ \ \ \
 \left.
 -g^{il}T^a\,_{ai}{\cal
T}^b\,_{bl} \right) ~,
\end{eqnarray}
}}
where
\begin{align}
 P^{ij} &\equiv \frac{1}{\phi e}\Pi^{(ij)}-\Delta^{ij}~,~~P\equiv P^{ij}g_{ij} ~, \\
 \Delta^{ij} &\equiv -g^{0l} \big( g^{jk}T^i\,_{lk}+g^{ik}{\cal
T}^j\,_{lk}-2g^{ij}T
^k\,_{lk} \big)\nonumber\\
&\ \ \ \,
 - \big( g^{jl}g^{0i}+g^{il}g^{0j} \big) T^k\,_{lk} ~,
\end{align}
in which the bracket in the subscript appearing in the expression of $P^{ij}$ represents
 anti-symmetrization. Moreover, $\lambda_{AB}$ and $ \lambda$ introduced in
\eqref{H_total_BD} are Lagrange multipliers that yield the primary constraints derived
above.  Finally, the Poisson brackets of the canonical variables are expressed as
\begin{align}
 &
 \{ e_{A\mu}(t,\vec{x}), \Pi^{B\nu}(t,\vec{x}') \}
 = \delta_A^B \delta_\mu^\nu \delta^{(3)}(\vec{x} -\vec{x}') ~,\nonumber\\
 &
 \{ \phi(t,\vec{x}), \phi(t,\vec{x}') \} = \delta^{(3)}(\vec{x}-\vec{x}') ~.
\end{align}

One can derive secondary constraints in the above Hamiltonian system. In particular,
by applying the relation $\{ \Pi^{A0}, H \} \approx 0$ one obtains four secondary
constraints, given by
\begin{align}\label{C^A_fT_DoF}
 {\cal C}^A =& \phi e~ e^{A0} \bigg[ \frac{1}{4}g^{ij}g^{kl}T^B\,_{jk}{\cal
T}_{Bil}
 + \frac{1}{2}g^{ij}T^b\,_{ki}T^k\,_{bj}\nonumber\\
 &\ \ \   -g^{ij}{\cal
T}^a\,_{ai}{\cal
T}^b\,_{bj}
 -\frac{1}{4g^{00}} \left( g_{ij}g_{kl}P^{ik}P^{jl} -\frac{P^2}{2} \right) \bigg]
\nonumber\\
 & -\phi e~ e^{Ai} \bigg( g^{0j}g^{kl}T^B\,_{il}T_{Bjk}
 +g^{jk}T^0\,_{lj}T^l\,_{ik}
\nonumber\\
 &\ \ \
 +g^{0j}T^k\,_{lj}T^l\,_{ki}
 -2g^{0j}T^k\,_{kj}T^l\,_{li} -2g^{jk}T^0\,_{ij}T^l\,_{lk}
\bigg) \nonumber\\
 & -\frac{\phi e}{2g^{00}} \big( g_{ij}g_{kl}\gamma^{Aik}P^{jl}
-\frac{1}{2}g_{ij}\gamma^{Aij}P \big) \nonumber\\
&
 -\partial_i \Pi^{Ai} + e e^{A0} W(\phi) ~,
\end{align}
with
\begin{align}
\!\!\!\!\!\!\!\!\!\!\!
 \gamma^{Aij} \equiv
 & -e^{Ak}\bigg[ g^{00} \big( g^{jl}T^i\,_{kl}+g^{il}{\cal
T}^j\,_{kl}
 +2g^{ij}T^l\,_{lk} \big)\nonumber\\
 &\ \ \ \ \ \ \ \ \
 + g^{0l} \big( g^{0j}{\cal
T}^i\,_{lk}+g^{0i}{\cal
T}^j\,_{lk} \big)
 -2g^{0i}g^{0j}T^l\,_{lk} \nonumber\\
 &\ \ \ \ \ \ \ \ \ +(g^{jl}g^{0i}+g^{il}g^{0j}-2g^{ij}g^{0l})T^0\,_{lk} \bigg]
\nonumber\\
 &
 - \frac{1}{2ke} \left(e^{Ai}e^{B0} e^{Cj} +e^{Aj}e^{B0}e^{Ci}\right) \Gamma_{BC} ~.
\end{align}

One may notice that there is no further secondary constraint in teleparallel gravity and
the above constraints $\Gamma^{AB}$, $H_0$, $H_i$, and $\Pi^{A0}$ are all first class
 \cite{Maluf:2000ag}. Particularly, $\Gamma^{AB}$ components are the generators of six
local Lorentz transformations, while $H_0$ and $H_i$ are those of four general
coordinates transformations. The constraints $\Pi^{A0}$ can be applied to determine the
tetrad fields $e_{A0}$, and therefore this is in agreement with the fact that $e_{A0}$
are not dynamical. Hence, one can extract the following formula for the number of degrees
of freedom (D.o.F):
\begin{eqnarray}\label{DoF_fT_DoF}
 {\rm D.o.F.} = \frac{1}{2} \Big( 2n^{f} -2n^{(1)}- n^{(2)} \Big) ~,
\end{eqnarray}
in which $n^{f}$ denotes the number of fields, $n^{(1)}$ represents the number of first
class constraints, and $n^{(2)}$ the number of second class constraints. Note that one
can
identify that the number of degrees of freedom for TEGR equals $2$, which is exactly the
same as in the GR case, as expected.

However, the constraint structure of $f(T)$ gravity is very different from that of
teleparallel gravity. For instance, Poisson brackets between $\Gamma^{AB}$,  $H_0$, and
$\pi$ are non-vanishing since the auxiliary field $\phi$ is a function of space-time
coordinates. In the following we review the analysis of the degrees of freedom for $f(T)$
gravity in even dimension and odd dimension,
respectively.

\subsubsection{Degrees of freedom in even dimensions}

We first consider the case of even dimension. For simplicity, we investigate the degrees
of freedom for $f(T)$ gravity in a $4D$ space-time. Recall that $\Pi^{A0}$ and $H_i$ are
independent of $\phi$ and $\pi$. Therefore, Poisson brackets between $\Pi^{A0}$, as well
as $H_i$, and the associated constraints are vanishing. Then one can list the
non-vanishing
Poisson brackets as below:
\begin{eqnarray}
\label{constraints1_fT_DoF}
 &&\!\!\!\!\!\!\!\!\!\!\!\!\!\!\!\!\!
 \{\Gamma^{AB}(\vec{x}), \Gamma^{CD}(\vec{y})\} \approx  \Big[ \eta^{BC}{\cal G}^{AD}
 +\eta^{AD}{\cal G}^{BC} -\eta^{AC}{\cal G}^{BD}\nonumber\\
 &&\ \ \ \ \ \ \  \ \ \ \ \ \  \ \ \ \ \ \  \ \
 -\eta^{BD}{\cal G}^{AC} +\Big]
\delta^{(3)}(\vec{x}
-\vec{y}) ~,
\end{eqnarray}
\begin{eqnarray}
\label{constraints2_fT_DoF}
 &&\!\!\!\!\!\!\!\!\!\!\!\!\!\!\!\!\!\!\!\!\!
 \{\Gamma^{AB}(\vec{x}), \pi(\vec{y})\} \approx  2 e \Big[ \big(
e^{Bi}e^{A0}-e^{Ai}e^{B0} \big) T^j\,_{ij}
\nonumber\\
 &&\ \ \ \ \ \ \  \ \ \ \ \ \  \ \ \ \ \,  +
e^{Ai}e^{Bj}{\cal
T}^0\,_{ij}
 \Big]
\delta^{(3)}(\vec{x}-\vec{y}) ~,
\end{eqnarray}
\begin{eqnarray}
\label{constraints3_fT_DoF}
 &&\!\!\!\!\!\!\!\!\!\!\!\!\!\!\!\!
 \{H_0(\vec{x}), \Gamma^{AB}(\vec{y})\}
 \approx
 e
 \Big\{ \frac{ P^{kl} }{g^{00}}
 \big( g_{ik}g_{jl} -\frac{1}{2}g_{ij}g_{kl} \big)  \nonumber\\
 &&\ \ \ \ \ \ \ \ \ \ \ \ \ \ \ \ \ \ \ \ \ \cdot \left[ \big(
e^{Aj}e^{Bm}-e^{Am}e^{Bj}
\big) g^{0i}\right.
\nonumber\\
&&\ \ \ \ \ \ \ \  \ \ \ \ \ \ \ \ \ \ \ \ \ \ \ \,
 +\big( e^{Am}e^{B0}-e^{Bm}e^{A0} \big) g^{ij}\nonumber\\
&&\ \ \ \ \ \ \ \  \ \ \ \ \ \ \ \ \ \ \ \ \ \ \ \, \left. +(e^{A0}e^{Bj}
-e^{Aj}e^{B0})g^{im}
\right]
 \nonumber \\
 &&\ \ \ \ \ \ \ \ \ \ \
+2 \left[ \big( e^{Ak}e^{Bm}-e^{Bk}e^{Am} \big) T^j\,_{kj}\right.\nonumber\\
&&\ \ \ \ \ \ \ \  \ \ \   \left.
 -e^{Ak}e^{Bj}T^m\,_{kj} \right] \Big\} (\partial_m \phi)
\delta^{(3)}(\vec{x}-\vec{y}) ,
\end{eqnarray}
\begin{eqnarray}
\label{constraints4_fT_DoF}
 &&\!\!\!\!\!\!\!\!\!\!\!\!\!\!
 \{\pi(\vec{x}), H_0(\vec{y})\} \approx  \Big[
\frac{e}{4g^{00}} \big(
g_{ik}g_{jl}P^{ij}P^{kl}
 -\frac{P^2}{2} \big)
 - e W_{,\phi}
 \nonumber \\
 &&\ \ \ \ \ \ \ \ \ \ \ \ \ \  \,  -\frac{1}{2\phi g^{00}} \big(
g_{ik}g_{jl}-\frac{1}{2}g_{ij}g_{kl}
\big) P^{
kl}\Pi^{ij}
 \nonumber \\
 &&\ \ \ \ \ \ \ \ \ \ \ \ \ \  \, -e \big( \frac{1}{4} g^{ik}g^{lj}{\cal
T}^A\,_{kl}T_{Aij} -\frac{1}{2}g^{kj}{\cal
T}^i\,_{lk}T^l\,_{ij}\nonumber \\
 && \ \ \ \ \ \ \ \ \ \ \ \ \ \  \,
 -g^{ik}T^j\,_{ji}T^l\,_{lk} \big) \Big] \delta^{(3)}(\vec{x}-\vec{y}) ~,
\end{eqnarray}
where ${\cal G}^{AB} \equiv 2e(e^{Aj}e^{B0}-e^{Bj}e^{A0}) (\partial_j \phi)$ has been
introduced. Note that, the symbol ``$\approx$'' appeared in the above equations denotes
the Dirac's weak equality. The nonzero relations governed by
(\ref{constraints1_fT_DoF})-(\ref{constraints3_fT_DoF}) indicate that the local Lorentz
invariance is broken as well.

One can further derive the rest secondary constraints. The consistency of constraints
$H_0$, $\Gamma^{AB}$ and $\pi$ yields the following relations:
{\small{
\begin{align}\label{equ_fT_DoF}
 \{ H_0, H \} =& \{ H_0, H_0 \} +\{ H_0, \Gamma^{CD} \} \lambda_{CD} +\{ H_0, \pi \}
\lambda
 \approx 0, \nonumber\\
 \{ \Gamma^{AB}, H \} =& \{ \Gamma^{AB}, H_0 \} +\{ \Gamma^{AB}, \Gamma^{CD} \}
\lambda_{CD}
 +\{ \Gamma^{AB}, \pi \} \lambda \approx 0,\nonumber\\
 \{ \pi, H \} =& \{ \pi, H_0 \} +\{ \pi, \Gamma^{CD} \}\lambda_{CD} \approx 0 ~,
\end{align}
}}
with $\lambda$ and $\lambda_{CD}$ being introduced in the total Hamiltonian density
\eqref{H_total_BD}. There are $8$ equations but only $7$ unknown quantities for $\lambda$
and $\lambda_{CD}$. Accordingly, it is obvious to see an existence of one secondary
constraint from the above relations. For simplicity, one further introduces
$\Gamma^i=e_A^{\ 0}e_B^{\ i}\Gamma^{AB}$ and $\Gamma^{ij}=e_
A^{\ i}e_B^{\ j}\Gamma^{AB}$. Correspondingly, one derives $\{ \Gamma^i, \Gamma^j \}
\approx 0$ and $\{ H_0, \Gamma^i \} \approx \Pi^{(ik)} \partial_k \ln\phi$. In the
following, we can define
\begin{eqnarray}
\label{y_fT_DoF}
 &&y_i=\{ H_0, \Gamma^i \} ~,~~~ y_4=\{ H_0, \Gamma^{12} \} ~,\nonumber\\
 &&
 y_5=\{H_0, \Gamma^{13} \} ~,~~~ y_6=\{ H_0, \Gamma^{23} \} ~,
\end{eqnarray}
\begin{eqnarray}
\label{x_fT_DoF}
&& x_0=\{ H_0, \pi \} ~,~~~ x_i=\{\Gamma^i, \pi \} ~,~~~ x_4=\{\Gamma^{12}, \pi \}
~,\nonumber\\
&&
 x_5=\{\Gamma^{13}, \pi \} ~,~~~ x_6=\{\Gamma^{23}, \pi \} ~,
\end{eqnarray}
and
\begin{eqnarray}
\label{A_B_fT_DoF}
&& \!\!\!\!\!\!\!\!\!\!\!\!\!\!\!\!
{\cal A}_{i1} = \{\Gamma^i, \Gamma^{12} \}
 \approx 2e \left[ g^{0i}(g^{01}g^{2k} -g^{02}g^{1k})\right.
\nonumber\\
&&\ \ \ \ \ \ \  \ \ \ \ \ \  \ \ \ \  \ \ \  \left. +g^{1i}(g^{0k}g^{02}
-g^{2k}g^{00})\right.
\nonumber\\
&&\ \ \ \ \ \ \  \ \ \ \ \ \  \ \ \ \  \ \ \  \left.
-g^{i2}(g^{0k}g^{01} -g^{1k}g^{00}) \right] \partial_k\phi ~,
\nonumber\\
&& \!\!\!\!\!\!\!\!\!\!\!\!\!\!\!\!
{\cal A}_{i2} = \{\Gamma^i, \Gamma^{13} \}
 \approx 2e \left[ g^{0i}(g^{01}g^{3k} -g^{03}g^{1k}) \right.
\nonumber\\
&&\ \ \ \ \ \ \  \ \ \ \ \ \  \ \ \ \  \ \ \ \left.
+g^{1i}(g^{0k}g^{03} -g^{3k}g^{00})
\right.
\nonumber\\
&&\ \ \ \ \ \ \  \ \ \ \ \ \  \ \ \ \  \ \ \ \left.
 -g^{i3}(g^{0k}g^{01} -g^{1k}g^{00}) \right] \partial_k\phi ~, \nonumber\\
 && \!\!\!\!\!\!\!\!\!\!\!\!\!\!\!\!
 {\cal A}_{i3} = \{\Gamma^i, \Gamma^{23} \}
 \approx 2e \left[ g^{0i}(g^{02}g^{3k} -g^{03}g^{2k})
 \right.
\nonumber\\
&&\ \ \ \ \ \ \  \ \ \ \ \ \  \ \ \ \  \ \ \ \left.
+g^{2i}(g^{0k}g^{03} -g^{3k}g^{00})
\right.
\nonumber\\
&&\ \ \ \ \ \ \  \ \ \ \ \ \  \ \ \ \  \ \ \ \left.
 -g^{i3}(g^{0k}g^{02} -g^{2k}g^{00}) \right] \partial_k\phi ,
 \end{eqnarray}
 \begin{eqnarray}
&& \!\!\!\!\!\!\!\!\!\!\!\!\!\!\!\!
{\cal B}_{12} = \{\Gamma^{12}, \Gamma^{13} \}
 \approx 2e \left[ g^{12}(g^{1k}g^{03} -g^{3k}g^{01})\right.
\nonumber\\
&&\ \ \ \ \ \ \  \ \ \ \ \ \  \ \ \ \  \ \ \ \left.
-g^{11}(g^{2k}g^{03} -g^{3k}g^{02})
\right.
\nonumber\\
&&\ \ \ \ \ \ \  \ \ \ \ \ \  \ \ \ \  \ \ \ \left.
 +g^{13}(g^{2k}g^{01} -g^{1k}g^{02}) \right] \partial_k\phi ~,
 \nonumber\\
&& \!\!\!\!\!\!\!\!\!\!\!\!\!\!\!\!
{\cal B}_{13} = \{\Gamma^{12}, \Gamma^{23} \}
 \approx 2e \left[ g^{22}(g^{1k}g^{03} -g^{3k}g^{01})\right.
\nonumber\\
&&\ \ \ \ \ \ \  \ \ \ \ \ \  \ \ \ \  \ \ \ \left.
-g^{12}(g^{2k}g^{03} -g^{3k}g^{02})
\right.
\nonumber\\
&&\ \ \ \ \ \ \  \ \ \ \ \ \  \ \ \ \  \ \ \ \left.
 -g^{23}(g^{1k}g^{02} -g^{2k}g^{01}) \right] \partial_k\phi ~, \nonumber\\
&& \!\!\!\!\!\!\!\!\!\!\!\!\!\!\!\!
{\cal B}_{23} = \{\Gamma^{13}, \Gamma^{23} \}
 \approx 2e \left[ g^{23}(g^{1k}g^{03} -g^{3k}g^{01})
 \right.
\nonumber\\
&&\ \ \ \ \ \ \  \ \ \ \ \ \  \ \ \ \  \ \ \ \left.
-g^{33}(g^{1k}g^{02} -g^{2k}g^{01})
\right.
\nonumber\\
&&\ \ \ \ \ \ \  \ \ \ \ \ \  \ \ \ \  \ \ \ \left.
 +g^{13}(g^{3k}g^{02} -g^{2k}g^{03}) \right] \partial_k\phi .
\end{eqnarray}

Then the constraint equations in \eqref{equ_fT_DoF} can be combined as a compact matrix
equation
\begin{equation}\label{MTheta_fT_DoF}
 {\cal M} ~ \Theta =0,
\end{equation}
where we define the vector $\Theta=(1, \theta_1, \theta_2, \theta_3, \theta_4, \theta_5,
\theta_6, \theta_7)^T$ with $\theta_i = e_{A0}e_{Bi}\Gamma^{AB}$, $\theta_4 =
e_{A1}e_{B2}\Gamma^{AB}$, $\theta_5 = e_{A1}e_{B3}\Gamma^{AB}$, $\theta_6 =
e_{A2}e_{B3}\Gamma^{AB}$ and $\theta_7=\lambda$.
Moreover, the matrix ${\cal M}$ is expressed as
{\small{
\begin{equation}
 {\cal M}=\left(
  \begin{array}{cccccccc}
   0 & y_1 & y_2 & y_3 & y_4 & y_5 & y_6 & x_0 \\
   -y_1 & 0 & 0 & 0 & {\cal A}_{11} & {\cal A}_{12} & {\cal A}_{13} & x_1 \\
   -y_2 & 0 & 0 & 0 & {\cal A}_{21} & {\cal A}_{22} & {\cal A}_{23} & x_2 \\
   -y_3 & 0 & 0 & 0 & {\cal A}_{31} & {\cal A}_{32} & {\cal A}_{33} & x_3 \\
   -y_4 & -{\cal A}_{11} & -{\cal A}_{21} & -{\cal A}_{32} & 0 & {\cal B}_{12} & {\cal
B}_{13} & x_
4 \\
   -y_5 & -{\cal A}_{12} & -{\cal A}_{22} & -{\cal A}_{32} & -{\cal B}_{12} & 0 & {\cal
B}_{23} & x_
5 \\
   -y_6 & -{\cal A}_{13} & -{\cal A}_{23} & -{\cal A}_{33} & -{\cal B}_{13} & -{\cal
B}_{23}  & 0  &
 x_6 \\
   -x_0 & -x_1 & -x_2 & -x_3 & -x_4 & -x_5 & -x_6 & 0
\end{array}
\right).
\end{equation}}}
Since the matrix equation \eqref{MTheta_fT_DoF} allows for a nontrivial solution of
$\Theta$, the determinant of ${\cal M}$ ought to vanish. As a result, one arrives an
additional constraint $| {\cal M} | \approx 0$. Substituting the expressions
 \eqref{y_fT_DoF}-\eqref{A_B_fT_DoF} into the determinant of ${\cal
M}$, one can further find that $x_0$ does not contribute to $| {\cal M} |$. Thus, one can
simplify the above constraint by requiring
\begin{eqnarray}
 m_q \equiv \sqrt{|{\cal M}|}_{x_0=0} = 0~.
\end{eqnarray}
After performing a very lengthy calculation, one eventually obtains the following
constraint equation
\begin{eqnarray}\label{mq_fT_DoF}
&&\!\!\!\!\!\!\!\!\!\!\!\!\!\!\!\!\!\!\!\!\!\!\!\!\!\!\!\!\!\!
\epsilon_{ijk} \partial_l\phi \partial_m\phi \partial_n\phi \Big( \Pi^{(mn)} g^{il}
g^{jh} g^{kg} {
\cal T}^0\,_{hg} \nonumber\\
&&\ \ \ \ \ \ \
 -\Pi^{0m} g^{in} g^{jh} g^{kg} T^l\,_{hg} \Big) \approx 0 ~.
\end{eqnarray}

We recall that ${\cal M}$ is an $8\times8$ antisymmetric matrix with a vanishing
determinant and hence, the rank of ${\cal M}$ is $6$. Thus, after imposing $m_q=0$, there
are only $6$ independent equations in \eqref{MTheta_fT_DoF} for $7$ Lagrange multipliers.
However, the consistency condition of the constraint $m_q$, which yields
\begin{equation}
\label{Cmq_fT_DoF}
 \{ m_q, H \}=\{  m_q, H_0 \}+\{  m_q, \Gamma^{cd} \}\lambda_{cd} +\{  m_q, \pi
\}\lambda
 \approx 0 ~,
\end{equation}
leads to another equation for Lagrange multipliers. Hence, the combination of
\eqref{Cmq_fT_DoF} and \eqref{MTheta_fT_DoF} presents $7$ independent equations for
Lagrange multipliers. As a result, all Lagrange multipliers can be determined and there
are no further secondary constraints.

In the last step we turn to analyze the structure of constraints. Recall that the Poisson
brackets between ($\Pi^{A0}, H_i, H$) and ($\Gamma^{AB}, \pi $) are all zero.
Consequently, the Poisson brackets among the constraints ($m_q, \pi, \Gamma^{AB},
\Pi^{A0}, H_i, H$) take the form
\begin{widetext}
$$
\begin{array}{c@{\hspace{-5pt}}l}
{\cal N}=\left(\begin{array}{ccc|ccc}
 0&\{m_q, \pi \} & \{m_q, \Gamma^{AB} \}&\{m_q, \Pi^{A0} \} &\{m_q, H_i \} & 0 \\
 \{\pi , m_q \} & 0 & \{\pi, \Gamma^{AB} \} & 0 & 0 &   0 \\
 \{\Gamma^{cd}, m_q\} & \{\Gamma^{CD}, \pi \} & \{\Gamma^{CD}, \Gamma^{AB}\} & 0 & 0 & 0
\\
 \hline
 \{\Pi^{A0}, m_q\} & 0 & 0 & 0 &0 &0 \\
 \{H_i, m_q\} & 0 & 0 & 0 & 0 & 0 \\
 0 & 0 & 0 & 0 & 0 & 0
\end{array}\right)
 &\begin{array}{l}\left.\rule{0mm}{6mm}\right\}8 \\ \\
 \left. \rule{0mm}{6mm}\right\}8
 \end{array}\\[-5pt]
\begin{array}{cc} ~~~~ \underbrace{\rule{55mm}{0mm}}_8 &~
\underbrace{\rule{40mm}{0mm}}_8\end{array}&
\end{array}
$$
\end{widetext}
which is a $16 \times 16$ antisymmetric matrix. This matrix can be separated to four
parts as indicated above. The left top part is an $8 \times 8$ non-singular matrix based
on the argument that all the Lagrange multipliers are solvable. The left bottom part is
another $8 \times 8$ matrix with nonzero components, of which the rank is at most unity.
Since the whole matrix is antisymmetric, it is easy to make the same conclusion for the
right top part. The rest part, shown in the right bottom regime, is merely a zero matrix.
Consequently, one can easily find that the nonzero part of the above matrix can only
become a $9\times9$ antisymmetric matrix of which the rank is $8$.

Eventually, the analysis performed above implies that there exist $8$ second class
constraints as well as $8$ first class constraints. Applying the relation
\eqref{DoF_fT_DoF} with $n^{f}=17$ (the extra one is the scalar field $\phi$),
$n^{(1)}=n^{(2)}=8$, one can conclude that there exist $5$ dynamical degrees of freedom
in
the theory of $f(T)$ gravity \cite{Li:2011rn}.

\subsubsection{Degrees of freedom in odd dimensions}

We start the investigation of the number of degrees of freedom of $f(T)$ gravity in 3D,
and then we shall extend the discussion in a background space-time with arbitrary
dimensions, deriving a generic formula for counting the number of degrees of freedom.

Repeating the same analysis performed in the previous paragraph, one can arrive at $10$
constraints in 3D, which are ($H_0$, $H_i$, $\Pi^{A0}$, $\Gamma^{1}$, $\Gamma^{2}$,
$\Gamma^{12}$, $\pi$),
where ``$A=$'' runs from 0 to 2 and $i=1,2$. The structure of the constraints is very
similar to the case of 4D. Particularly, one can define the following quantities:
\begin{align}
\label{3Dy_fT_DoF}
 &\!\!\!\!
 y_i = \{ H_0, \Gamma^i \} ~,~~~ y_3 = \{ H_0, \Gamma^{12} \} ~, \\
\label{3Dx_fT_DoF}
 &\!\!\!\! x_0 = \{ H_0, \pi \} ~,~~~ x_i = \{\Gamma^i, \pi \} ~,~~~ x_3 =
\{\Gamma^{12},\pi \}
~,
\\
\label{3DA_fT_DoF}
 &\!\!\!\! A_i = \{\Gamma^i, \Gamma^{12} \} \approx 2e \big[ g^{0i}(g^{01}g^{2j} -
g^{02}g^{1j})
\nonumber\\
&
\ \ \ \  + g^{1i} (g^{0j}g^{02} - g^{2j}g^{00}) - g^{i2}(g^{0j}g^{01} -
g^{1j}g^{00}) \big]
\partial_j \phi
 ~,
\end{align}
and rewrite the self-consistent equations (\ref{equ_fT_DoF}) as a matrix
equation:
${\cal M}
 \Theta=0$. Note that, in 3D $\Theta=(1, \theta_1, \theta_2, \theta_3, \theta_4)^T$ and
the matrix
${\cal M}$ is given by,
\begin{equation}
{\cal M}=\left(
\begin{array}{ccccc}
 0 & y_1 & y_2 & y_3 & x_0 \\
 -y_1 & 0 & 0 & A_1 & x_1 \\
 -y_2 & 0 & 0 & A_2 & x_2 \\
 -y_3 & -A_1 & -A_2 & 0 & x_3 \\
 -x_0 & -x_1 & -x_2 & -x_3 & 0
\end{array}
\right).
\end{equation}

Again, since the matrix equation has a nontrivial solution for $\Theta$, the determinant
of ${\cal M}$ is vanishing. However, this requirement is automatically satisfied in 3D
due to the fact that ${\cal M}$ is a $5 \times 5$ antisymmetric matrix. Therefore,
differently from $f(T)$ gravity in 4D, there is no further constraint in the case of
3D. Hence, one can easily read that the rank of ${\cal M}$ in 3D is $4$, which also
implies that there are $4$ independent equations for those $4$ Lagrange multipliers. To
be specific, one can calculate all Lagrangian multipliers as follows:
\begin{align}\label{multipliers_fT_DoF}
 & \theta_1 = \frac{A_2 x_0+x_3 y_2-x_2 y_3}{A_1 x_2-A_2 x_1} ~,~~~
 \theta_2 = \frac{-A_1 x_0-x_3 y_1+x_1 y_3}{A_1 x_2-A_2 x_1} ~, \nonumber\\
 & \theta_3 = \frac{y_1 x_2-y_2 x_1}{A_1 x_2-A_2 x_1} ~,~~~
 \theta_4 = \frac{A_1y_2-A_2 y_1}{A_1 x_2-A_2 x_1} ~.
\end{align}
To keep a generic discussion, one may assume $A_1 x_2-A_2 x_1\neq 0$. Accordingly, one
can easily figure out that there are $6$ first class constraints, which are governed by
$H$, $H_i$, $\Pi^{A0}$. Moreover, there are $4$ second class constraints, which
correspond
to $\Gamma^{1}$, $\Gamma^{2}$, $\Gamma^{12}$ and $\pi$. Then one can again apply the
formula \eqref{DoF_fT_DoF} and conclude that there are only $2$ dynamical degrees of
freedom for $f(T)$ gravity in 3D space-time.

We close this paragraph by generalizing the conclusion to a space-time of arbitrary
dimensionality, based on the analyses in the above two examples. For a general
space-time of $d$ dimensions, one can have $\frac{1}{2}d(d-3)+d-1$ degrees of freedom in
$f(T)$ gravity \cite{Li:2011rn}. This result can be understood as follows. Firstly, we
notice that the rank of a $\frac{1}{2}d(d-1) \times \frac{1}{2}d(d-1)$ matrix in
\eqref{constraints1_fT_DoF} is $2(d-2)$. Thus, applying the second equation of
\eqref{equ_fT_DoF}, one can determine the Lagrangian multiplier $\lambda$ and also obtain
$\frac{1}{2}d(d-1)-2(d-2) -1$ secondary constraints. Secondly, one can substitute
$\lambda$ into \eqref{equ_fT_DoF} and make use of those secondary constraints, and
thus he obtains $2(d-2)+1$ independent equations for $\frac{1}{2}d(d-1)$ Lagrangian
multipliers $\lambda_{AB}$.
Thirdly, the consistency check of the secondary constraints can yield another
$\frac{1}{2}d(d-1)-2(d-2)-1$ equations for $\lambda_{AB}$. As a consequence, one can fix
all the Lagrangian multipliers with none of secondary constraints being left. Therefore,
one can find that there are $2d$ first class constraints and $d(d-1)-2(d-2)$
second class constraints, and accordingly the number of degrees of freedom in $f(T)$
gravity is $\frac{1}{2}d(d-3)+d-1$. As a side remark, following the above analysis it is
interesting to observe that $f(T)$ gravity is non-dynamical in a 2D space-time, a special
property first observed in \cite{Ferraro:2011us}.

%%%here manos%%%%%%

\subsection{Restoring local Lorentz invariance in $f(T)$ gravity}
\label{restoringLI}

As we described in detail in subsection \ref{LorViolsusection}, in the usual formulation
of $f(T)$ gravity local Lorentz invariance is either completely absent or strongly
restricted  \cite{Li:2010cg,Sotiriou:2010mv,Ferraro:2014owa}. The reason behind this is
the strong imposition made in
\cite{Ferraro:2006jd,Bengochea:2008gz,Ferraro:2008ey,Linder:2010py} that the spin
connection vanishes, which makes the theory in general frame-dependent.
%In particular,
%since the spin connection is not a tensor and under the local Lorentz transformation
%transforms non-covariantly, it is always possible to transform to a frame that the spin
%connection is transformed to zero.
Such a consideration  helps to make the
theory simpler, and it can lead to the extraction of many solutions, as long as one does
not ask questions about frame transformations and Lorentz invariance. Obviously, the
Lorentz violation problem is potentially a severe one, and it becomes obvious when
questions about frame transformations and Lorentz invariance are raised, which is usual
for instance in the case of spherically symmetric solutions.

The only way to solve this severe problem is to construct a consistent, covariant,
formulation of $f(T)$ gravity. This is achieved if instead of the pure-tetrad
teleparallel gravity we start from the covariant teleparallel gravity, using both the
vierbein and the spin connection in a way that for every vierbein choice a
suitably constructed connection makes the whole theory covariant, as it was recently
done in \cite{Krssak:2015oua}. In this case the theory is frame-independent and Lorentz
invariant, since every frame transformation will be accompanied by a suitable connection
transformation.

Unfortunately, the above covariant re-formulation of $f(T)$ gravity was absent till
recently \cite{Krssak:2015oua}, and thus many works on the subject were instead
devoted in choosing/constructing suitable non-diagonal tetrads in order to incorporate
the Lorentz invariance issues. In particular, a large number of papers appeared,
discussing the relation between diagonal and non-diagonal vierbeins, the relation between
``good'' and ``bad'' tetrads, and the whole discussion was mainly focused on spherically
symmetric solutions where frame transformation is apparent. Such an approach is not
wrong and it can indeed lead to interesting solutions (see Sec.
\ref{sectionBlackHoles}), but clearly it faces the problem only in a skin-deep way,
without trying to restore Lorentz invariance fully and correctly. The latter is possible
only if one re-formulates $f(T)$ gravity keeping an arbitrary spin connection, and this
is the goal of this subsection following \cite{Krssak:2015oua}.

In a general geometrical formulation, the fundamental variable is the tetrad
$e^A_{\ \mu}$, while the spin connection $\omega^A_{\ B\mu}$ determines the rule of
parallel transportation. In the case where we keep a
non-vanishing $\omega^A_{\ B\mu}$, the torsion tensor (\ref{torsiontensor}) is extended to
\begin{equation}
T^A_{\ \mu\nu}(e^A_{\ \mu},\omega^A_{\ B\mu})=
\partial_\mu e^A_{\ \nu} -\partial_\nu e^A_{\ \mu}+\omega^A_{\ B\mu}e^B_{\ \nu}
-\omega^A_{\ B\nu}e^B_{\ \mu}.
\label{tordefkraas}
\end{equation}
We stress that in principle the above relation is the correct one, while
(\ref{torsiontensor}) is the simplification made in
\cite{Ferraro:2006jd,Bengochea:2008gz,Ferraro:2008ey,Linder:2010py} under
the imposition of a zero $\omega^A_{\ B\mu}$. The expression for the torsion scalar
(\ref{telelag}), namely $T= T^A_{\ \mu\nu}S_A^{\ \mu\nu}$, as well as the expressions for
the contortion tensor $K^{\mu\nu}_{\  \ A}=\frac{1}{2}
\left(T^{\ \mu\nu}_{A}
+T^{\nu\mu}_{\ \ A}
-T^{\mu\nu}_{\  \ A}
\right)$ and the superpotential\footnote{In this paragraph we follow the conventions of
\cite{Krssak:2015oua}, used in TEGR literature, that slightly differ from those in $f(T)$
gravity works by a factor of 2. Therefore, the superpotential here is two times the one
defined in (\ref{Stensor}).}
$S_A^{\ \mu\nu}= e_A^{\ \rho}S_\rho^{\
\mu\nu}=K^{\mu\nu}_{\ \ A}
-e^A_{\ \nu}T^{ \mu}
+e^A_{\ \mu}T^{ \nu}$, are the same as in the usual formulation, however since the
torsion tensor now includes the non-trivial teleparallel spin connection, the final 
results are
different.

The action of this covariant $f(T)$ gravity still writes as the one in
(\ref{action_fT}), namely
\begin{eqnarray}
\label{action_fTkrass}
 {\cal S} = \int d^4x ~|e|~ \left[ \frac{f(T)}{16\pi G} + L^{(m)} \right],
\end{eqnarray}
with $|e| = \text{det}(e_{\mu}^A) = \sqrt{-g}$ and $L^{(m)}$  the matter
Lagrangian, {\bf{however in the present case $T$ includes $\omega^A_{\ b\mu}$ which is 
determined from the reference tetrad by the method proposed in  \cite{Krssak:2015oua}, and 
which is briefly discussed later on. Then, the equations of motion are obtained by 
variation with respect to the tetrad only, and they read as  \cite{Krssak:2015oua}}}:
\begin{eqnarray}
 &&\!\!\!\!\!\!\!\!\!\!\!\!\!\!\!\!\!\!
 e^{-1} f_T \partial_{\nu}\left( e S_A^{\ \mu\nu} \right)
+
f_{TT} S_A^{\ \mu\nu} \partial_{\nu} T
-
f_T T^B_{\ \nu A }S_B^{\ \nu\mu}
\nonumber\\
&&+
f_T \,\omega^B_{\ A\nu}S_B^{\ \nu\mu}+
\frac{1}{4}f(T) e_A^{\ \mu}=4\pi G e_{A}^{\rho} {T^{(m)}}_{\rho}{}^{\nu},\
\label{ftequationkrass}
\end{eqnarray}
where ${T^{(m)}}_{\rho}{}^{\nu}$ is the matter energy-momentum tensor. These equations
obviously coincide with (\ref{eom_fT_general}) if we set $\omega^A_{\ b\mu}=0$.

One can easily see that the action (\ref{action_fTkrass}) is invariant with respect to
local Lorentz transformations, while the field equations (\ref{ftequationkrass}) are
covariant, which was actually the main goal of this subsection. In order to verify this,
let us consider a local Lorentz transformation represented by the matrix
$\Lambda^A_{\
B}=\Lambda^A_{\ B}(x)
$  that obeys
\begin{equation}
\eta_{AB}=\eta_{CD}\Lambda^C_{\ A}\Lambda^D_{\ B},
\end{equation}
under which the tetrad and spin connection respectively transform as
\begin{eqnarray}
\label{lortranskrass1}
&&
e'{}^A_{\ \mu}=\Lambda^A_{\ B}e^B_{\ \mu},
\\
&&
\omega'{}^A_{\ B\mu}=\Lambda^A_{\ C}\omega^C_{\ D\mu}\Lambda_B^{\ D}+\Lambda^A_{\ C}
\partial_\mu \Lambda_B^{\ C},
\label{lortranskrass2}
\end{eqnarray}
with primes denoting the transformed quantities. As one can very easily see, under the
above transformation the field equations (\ref{ftequationkrass}) are covariant, since the
transformation of the connection can compensate the transformation of the vierbeins. On
the other hand, if one had set the spin connection to zero, as it is done in the usual
formulation of $f(T)$ gravity, the above capability is lost, and that is why the theory
does not have local Lorentz invariance.

In summary, in this subsection we have indeed managed to consistently formulate $f(T)$
gravity with local Lorentz invariance. In particular, we can always construct a
non-vanishing spin connection that restores local Lorentz invariance in $f(T)$ gravity
for
{\it{every}} vierbein. This is a crucial result that one must have in mind when
discussing Lorentz invariance issues in $f(T)$ gravity.

We would like to mention here that indeed in such a consistent $f(T)$ gravity
formulation, i.e. with a non-vanishing spin connection, it is harder to obtain solutions.
The reason is the following: In the usual $f(T)$ gravity the spin connection is absent,
and thus the field equations are equations for the vierbeins only. On the other hand, in
the general case the field equations do depend on  both the vierbeins and the spin
connection, which now appears explicitly as a
dynamical variable, and thus additional methods are needed in order to match each to
other. Note the interesting feature that in simple TEGR this is not the case, even if one
keeps a non-zero spin connection. In order to see this, one starts  by explicitly
writing the torsion scalar as \cite{Krssak:2015rqa,Krssak:2015lba,Krssak:2015oua}
\begin{equation}
T (e^A_{\ \mu},\omega^A_{\ B\mu})=
T (e^A_{\ \mu},0) +  \frac{4}{e}\partial_\mu \left(
e\, \omega^{AB}_{\ \ \nu}e_A^{\ \nu}e_{B}^{\ \mu}
\right),
\label{holtkrass}
\end{equation}
where $T (e^A_{\ \mu},0)$ is the usual torsion scalar corresponding to zero spin
connection. Therefore, although in the case of TEGR Lagrangian the spin connection
terms form a total derivative and hence one can find solutions easily, first   solving for
the tetrad and then proceeding to finding the spin connection, in the case of
$f(T)$ gravity this is not possible since  the solution of the field equation depends on
the choice of the spin connection. Clearly, one needs a method of finding the
spin connection to the given tetrad, which does not rely on the solution of the field
equations. It turns out that this is possible, if we make some reasonable
assumption about the reference tetrad, given by the symmetry of
the problem being investigated  \cite{Krssak:2015oua,Krssak:2015rqa}.

We close this subsection referring very briefly on how the above procedure would work in
an FRW geometry, i.e formulating a consistent and Lorentz invariant $f(T)$ cosmology, as
well as in spherically symmetric cases.
\begin{itemize}

\item FRW geometry

We
start with the Cartesian coordinate system and the diagonal tetrad that
represents the FRW metric, namely
\begin{equation}
e^A_{\ \mu}=\text{diag} (1,a(t),a(t),a(t)), \label{tetradfrwc}
\end{equation}
which leads to the Friedmann equations (\ref{background11}),(\ref{background22}). In
the spherical coordinate system the above vierbein writes as
\begin{equation}
e^A_{\ \mu}=\text{diag} \left(1,  a, r a,  r a \sin\theta
\right),
\label{tetradfrwl}
\end{equation}
however note that in this case the field equations (\ref{ftequationkrass}) for this
tetrad and vanishing spin connection lead to the condition $f_{TT}=0$,
which is satisfied only in the TEGR case. Nevertheless, we can easily find the
non-vanishing, purely inertial, spin connection corresponding to (\ref{tetradfrwl}),
which writes as  \cite{Krssak:2015oua}
\begin{eqnarray}
&&\omega^{{1}}_{\  {2}\theta}=-1,
\nonumber\\
&&
\omega^{ {1}}_{\  {3}\phi}=- \sin\theta, \nonumber\\
&&
\omega^{ {2}}_{\  {3}\phi}=-\cos\theta,
\label{spfrwl}
\end{eqnarray}
with all its other components being zero. Using this spin connection, it is easy to check
that the vierbein (\ref{tetradfrwl}) leads to the same field equations
(\ref{background11}),(\ref{background22}) as the ones obtained  from the tetrad
(\ref{tetradfrwc}).

\item Spherically symmetric geometry

The spherically symmetric space-time is characterized by the metric
\begin{equation}
\ \ \ \ \ \ ds^2=A(r)^2 dt^2-B(r)^2 dr^2-r^2d\theta^2-r^2\sin^2\theta d\phi^2,
\label{metricspherical}
\end{equation}
where $A(r)$ and $B(r)$ are arbitrary functions of $r$. The most natural
choice of the  tetrad that corresponds to this metric has the simple  diagonal form
\begin{equation}
e^A_{\ \mu}=\text{diag}\left(A(r),B(r),r,r\sin\theta\right)
\label{Reviewschwtet}.
\end{equation}
It is straightforward to check that if we assume the trivial spin connection $\omega^A_{\
B\mu}=0$, then the $2-\theta$ field equation from  (\ref{ftequationkrass}), namely
$-\frac{8 f_{TT}  }{r^5}\cot\theta=0$, gives us necessarily  the condition
$f_{TT}=0$, which restricts the theory to TEGR. In the literature this feature is
wrongly interpreted as ``the diagonal tetrad is not a good tetrad for
sperically-symmetric solutions in $f(T)$ gravity''
\cite{Ferraro:2011us,Ferraro:2011ks,Tamanini:2012hg,Wang:2011xf,HamaniDaouda:2011iy,
Nashed:2012ms,Atazadeh:2012am,Nashed:2014sea,
Aftergood:2014wla,Bhadra:2014jea,delaCruz-Dombriz:2014zaa,Abbas:2015xia,Junior:2015dga,
Das:2015gwa,Zubair:2015cpa,Ruggiero:2015oka,Nashed:2015pga}.

Following \cite{Krssak:2015oua} we will show that the above issue is an artifact of the
non-covariant formulation of $f(T)$ gravity. In particular, using the covariant
formulation presented above we can calculate the appropriate spin connection which will
allow to use any tetrad giving the metric (\ref{metricspherical}), without restricting
the
functional dependence of the Lagrangian. This reads as \cite{Krssak:2015oua}
\begin{equation}
\ \ \ \ \ \ \ \ \ \ \omega^{ {1}}_{\  {2}\theta}=-1, \quad
\omega^{ {1}}_{\  {3}\phi}=-\sin\theta, \quad
\omega^{ {2}}_{\  {3}\phi}=-\cos\theta.
\label{Reviewspschw1}
\end{equation}
 Hence, the field equations for every $f(T)$-form can
be satisfied by all tetrads related through Lorentz transformation and corresponding to
the spherically-symmetric metric (\ref{metricspherical}), and not only by specifically
constructed ones.

We stress that the field equations of covariant $f(T)$ gravity \cite{Krssak:2015oua},
generated from the diagonal tetrad (\ref{Reviewschwtet}) and the non-zero spin connection
(\ref{Reviewspschw1}), coincide with those obtained in the usual, non-covariant,
formulation of $f(T)$ gravity, for zero spin connection but for the specific and peculiar
non-diagonal tetrad \cite{Ferraro:2011us,Ferraro:2011ks,Tamanini:2012hg}
{\small{
\begin{equation}
\ \ \ \ \ \ \ \tilde{e}^A_{\ \mu}=
\left( \begin{array}{cccc}
A & 0 & 0 & 0\\
0 & B \cos\phi\sin\theta & r\cos\phi\cos\theta & -r \sin\phi\sin\theta \\
0 & -B\cos\theta & r \sin\theta & 0 \\
0 & B\sin\phi\sin\theta & r\sin\phi\cos\theta & r\cos\phi\sin\theta
\end{array}
\right).
\label{offd}
\end{equation} }}
This coincidence of the field equations can be easily explained. The off-diagonal tetrad
(\ref{offd}) is related to the diagonal tetrad (\ref{Reviewschwtet}) through a local
Lorentz transformation of the form
$\tilde{e}^A_{\ \mu}=\Lambda^A_{\ B} e^B_{\ \mu}$,
where the Lorentz matrix is given explicitly by
\begin{equation}
\ \ \ \ \ \ \ \ \ \ \Lambda^A_{\ B}=
\left( \begin{array}{cccc}
1 & 0 & 0 & 0\\
0 & \cos\phi\sin\theta & \cos\phi\cos\theta & -\sin\phi \\
0 & -\cos\theta & \sin\theta & 0 \\
0 & \sin\phi\sin\theta & \sin\phi\cos\theta & \cos\phi
\end{array}
\right).
\label{specLorentz}
\end{equation}
We should now recall that a local Lorentz transformation simultaneously
transforms both the tetrad and spin connection through
(\ref{lortranskrass1}),(\ref{lortranskrass2}), and thus the
spin connection (\ref{Reviewspschw1}) gets transformed as well. Interestingly enough,
the
transformed spin connection through (\ref{specLorentz}) is identically zero, namely
$\tilde{\omega}^A_{\ B\mu}=0$.
Hence, we can see that the off-diagonal tetrad (\ref{offd}) is a proper tetrad, i.e.
a tetrad in which the inertial spin connection vanishes, and that is why the obtained
field equations coincide with the ones of the covariant formulation.

In other words, in
the usual, non-covariant, formulation of $f(T)$ gravity, one considers specific
peculiar non-diagonal tetrads, and thus making the theory frame-dependent, as a naive way
to be consistent with a vanishing spin connection. However, as we showed, the correct and
general way to acquire consistency is to use the covariant formulation of $f(T)$ gravity,
in which case frame-dependence is absent. In particular, one is allowed to use any
form of the tetrad provided that he calculates the corresponding spin connection. The
off-diagonal tetrad (\ref{offd}) has no privileged position anymore, and it is just a
specific tetrad in which the corresponding spin connection happens to be zero.
Hence, there are not ``good'' and ``bad'' tetrads in $f(T)$ gravity, there is
no-frame dependence, as long as one abandons the strong imposition of zero spin
connection

\end{itemize}

In summary, covariant $f(T)$ gravity is the correct and consistent way to modify TEGR.
There is an additional difficulty to find the spin connection, however one can follow the
above method in order to achieve it and extract solutions. In particular, one is allowed
to use an arbitrary tetrad in an arbitrary coordinate system along with the corresponding
spin connection, resulting always to the same physically relevant field equations
\cite{Krssak:2015oua}.

%%%%%%%%%%%%%%%%%%%%%%%%%%%%%%%
\subsection{Perturbations in $f(T)$ gravity}
\label{sec:pert_fT_gravity}
%%%%%%%%%%%%%%%%%%%%%%%%%%%%%%%%%%%%%%%

In this subsection, we review in detail the linear-order dynamics of perturbations of
$f(T)$ gravity within the cosmological frame, following \cite{Chen:2010va} (see also
 \cite{Wu:2012hs}). For simplicity, we investigate scalar perturbations in the Newtonian
gauge and provide the full set of gravitational and energy-momentum tensor up to linear
order. Then we examine the stability of the theory from the aspect of scalar
perturbations.

%%%%%%%%%%%%%%%%%%%%%%%%%%%%%%%
\subsubsection{Scalar perturbations in Newtonian gauge}
%%%%%%%%%%%%%%%%%%%%%%%%%%%%%%%%%%
Let us focus on the spatially flat cosmological scenarios governed by $f(T)$ gravity.
Using the symbol ${e}_{\mu}^A$ for the perturbed vierbein and $\bar{e}_{\mu}^A$ for the
unperturbed one, scalar perturbations can be included by writing
\begin{eqnarray}
 e_{\mu}^A = \bar{e}_{\mu}^A + {\mathbb{E}}_{\mu}^A ~,
 \label{basicveirpertChen}
\end{eqnarray}
where
\begin{eqnarray}
 \label{Chen_pert1}
 \bar{e}_{\mu}^0 = \delta_{\mu}^0, \,\,\,\,\, \bar{e}_{\mu}^a = \delta_{\mu}^aa,
\,\,\,\,\,
 \bar{e}^{\mu}_0 = \delta^{\mu}_0, \,\,\,\,\, \bar{e}^{\mu}_a = \frac{\delta^{\mu}_a}{a}
~,
\end{eqnarray}
and
\begin{eqnarray}
 \label{Chen_pert2}
 \! \! \!\!{\mathbb{E}}_{\mu}^0 = \delta_{\mu}^0\psi, \,\,\, {\mathbb{E}}_{\mu}^a
=-\delta_{\mu}^a a\phi, \,
\,\,
 {\mathbb{E}}^{\mu}_0 = -\delta_{0}^{\mu}\psi, \,\,\, {\mathbb{E}}^{\mu}_a =
\frac{\delta^{\mu}_
a}{a}\phi .\
\end{eqnarray}
In the above expansion we have introduced the scalar modes of perturbations $\phi$ and
$\psi$, which are functions of space ${\bf x}$ and time $t$. Note that these symbols, as
well as their associated coefficients, were conveniently chosen such that the vierbein
perturbation can match with a metric perturbation of the known form in Newtonian gauge,
namely
\begin{eqnarray}\label{pertmetric}
 ds^2 = (1 + 2\psi)dt^2 -a^2(1-2\phi)\delta_{ij}dx^idx^j ~.
\end{eqnarray}
Accordingly, the determinant is given by $e = \textrm{det}(e_{\mu}^A) = a^3(1+\psi -
3\phi)$.

To proceed, one can calculate the expressions of $T^{\lambda}{}_{\mu\nu}$ from
(\ref{torsiontensor}) and of $S_{\lambda}{}^{\mu\nu}$ from (\ref{Stensor}), at linear
order under the perturbations \eqref{Chen_pert1} and \eqref{Chen_pert2}. To be explicit,
the torsion expression can be written as
\begin{eqnarray}
\label{Tpert}
 T^{\lambda}{}_{\mu\nu} = (\bar{e}^{\lambda}_A +
{\mathbb{E}}^{\lambda}_A)[\partial_{\mu}(\bar{e}_{\nu}^A + {\mathbb{E}}_{\nu}^A) -
\partial_{\nu}(\bar{e}_{\mu}^A + {\mathbb{E}}_{\mu}^A)] ~.
\end{eqnarray}
After some algebra one can derive the component expressions to be:
\begin{align}
 & T^{0}{}_{\mu\nu} = \partial_{\mu}\psi\delta_{\nu}^0 - \partial_{\nu}\psi\delta_{\mu}^0
~,~~ T^{i}
{}_{0i} = H - \dot{\phi} ~, \nonumber\\
 &S_{0}{}^{0i} = \frac{\partial_i \phi}{a^2} ~,~~ S_{i}{}^{0i} = -H + \dot{\phi} + 2H\psi
~, \nonumber\\
 & T^{i}{}_{ij} = \partial_j\phi ~,~~ S_{i}{}^{ij} = \frac{1}{2a^2}\partial_j (\phi -
\psi) ~.
\end{align}
Additionally, the perturbed torsion scalar from (\ref{telelag}) can be expressed as
\begin{eqnarray}
 {T}\equiv S_{\rho}{}^{\mu\nu}T^{\rho}{}_{\mu\nu} =T_0+T_1,
\end{eqnarray}
with $ T_0=-6H^2 $ being the background part and
\begin{eqnarray}
\label{Chen_T1}
 T_1=12H(H\psi + \dot{\phi}) ~,
\end{eqnarray}
the first-order part. Further, one can expand the form of $F(T)$ (recall that $F(T)\equiv
f(T)-T$)
and its derivatives up to first order to be:
\begin{align}
\label{Chen_FTexpansion}
 &F(T)=F(T_0)+T_1\frac{dF(T)}{dT}\Big|_{T=T_0}\equiv F_0+F_1 \nonumber\\
 &F_T(T)=\frac{dF(T)}{dT}\Big|_{T=T_0}+T_1\frac{d^2F(T)}{dT^2}\Big|_{T=T_0}\equiv F'_0
+F'_1 \nonumber\\
 &F_{TT}(T)=\frac{d^2F(T)}{dT^2}\Big|_{T=T_0}+T_1\frac{d^3F(T)}{dT^3}\Big|_{T=T_0}\equiv
F''_0 +F''_1
~.
\end{align}

Then, we calculate the perturbed energy momentum tensor. Note that, the perturbed
components of the energy moment tensor are expressed as follows:
\begin{align}
\label{Chen_T00pert}
 \delta {T^{(m)}}_0{}^0 &= -\delta\rho_m \\
 \delta {T^{(m)}}_0{}^i &= a^{-2}(\rho_m + p_m)(-\partial_i \delta u) \\
\label{Chen_Ta0pert}
 \delta {T^{(m)}}_i{}^0 &= (\rho_m + p_m)(\partial_i\delta u) \\
\label{Chen_Tabpert}
 \delta {T^{(m)}}_i{}^j &= \delta_{ij}\delta p_m+\partial_i\partial_j\pi^{S} ~,
\end{align}
where $\pi^S$ is the scalar part of the anisotropic stress.

Combining all the above expressions we eventually obtain the perturbed
gravitational equations of motion up to first order \cite{Chen:2010va}:
\begin{align}
\label{Chen_E00}
 E_{0}^0 &= -4\pi G \delta\rho_m \nonumber\\
 &= (1+F'_0)(\nabla^2\phi) + 6(1+F'_0)H\dot{\phi} + 6(1+F'_0)H^2\psi \nonumber\\
 &\ \ \
 -3F'_1H^2
-\frac{T_1
+F_1}{4}
 ~, \\
\label{Chen_E0a}
 E_{0}^i &= -4\pi G(\rho_m + p_m)\partial_i \delta u \nonumber\\
 &= (1+F'_0)\partial_i \dot{\phi} + (1+F'_0)H\partial_i \psi -12H\dot{H}F''_0\partial_i
\phi
  ~, \\
\label{Chen_Ea0}
 E_{a}^0 &= 4\pi G (\rho_m + p_m) \partial_i\delta_a^i\delta u \nonumber\\
 &=
12H^2\partial_i\delta_a^i(\dot{\phi}+H\psi)F''_0-(1+F'_0)\partial_i\delta_a^i(\dot{\phi}
+
H\psi)
 ~,
\end{align}
\begin{align}
\label{Chen_Eaa}
 E_{a}^i\delta_i^a &= \frac{4\pi G}{a}(p_m\phi + \delta p_m) \nonumber\\
 &= \frac{F'_1}{a}\left(-3H^2-\dot{H}\right) + \frac{F''_1}{a}\left(12H^2\dot{H}\right)
-\frac{T_1 +F_1}{4a}
 \nonumber\\
 & \ \ \
 -\frac{(1+F'_0)}{2a}\sum_{b\neq
a}\partial^j\delta_j^b\partial_i\delta^i_b(\psi-\phi)
-\frac{\phi (
T_0 +F_0)}{4a}
\nonumber \\
 &~~ + \frac{(1+F'_0)}{a}[6H\dot{\phi} + 6H^2\psi-3H^2\phi \nonumber\\
 &\ \ \ \ \ \ \ \ \ \ \  \ \,  \ \ \ \ \ +\ddot{\phi}
+\dot{H}(2\psi
-\phi) + H\dot{\psi}]
\nonumber\\
 &~~ + \frac{F_0''}{a}(-24H\dot{H}\dot{\phi} -48\psi H^2\dot{H}
\nonumber\\
 &\ \ \ \ \ \ \ \ \ \ \
-12H^2\ddot{\phi}
-12H^3\dot{\psi}
+ 12H^2\dot{H}\phi)~, \\
\label{Chen_Eab2}
 E_{b}^i\delta_i^a &= 4\pi G a^2\partial_j\delta_b^j\partial^i\delta_i^a\pi^{S}
\nonumber\\
 &= \frac{(1+F'_0)}{2}\partial_j\delta_b^j\partial^i\delta_i^a(\phi-\psi) ~, ~~ ({\rm
when} ~b\neq
a)
\end{align}
where the indices are not summed up unless the $\Sigma$ symbol is shown explicitly.

So far the equations of motion derived above are general to most cosmological models
governed by $f(T)$ gravity. In many situations, one can further simplify the above
equations by assuming a vanishing scalar anisotropic stress $\pi^S$. This assumption is
often in agreement with the observational fact that the contribution of anisotropic
expansion of the Universe is sub-dominant. Under the zero-anisotropic-stress assumption,
Eq. \eqref{Chen_Eab2} implies $\phi=\psi$, and along with $\delta p_m=0$ may lead the
system of perturbation equations to be overdetermined. Once the vanishing of the
anisotropic stress is implemented, one has four equations determining the three
remaining perturbation variables $\delta\rho_m$, $\phi$ and $\delta u$. However, in
the limit $F_{TT}(T)\simeq0$, equations \eqref{Chen_E0a} and \eqref{Chen_Ea0} become
identical and then remove the over-determination. Therefore, one can conclude that the
requirement of no anisotropic stress imposes another constraint on $f(T)$ models, namely
that $F_{TT}(T)\simeq0$. Moreover, one may notice that this requirement can be relaxed for
more general vierbien choices than \eqref{weproudlyuse}.

\subsubsection{Stability analysis}

After having derived the perturbed gravitational equations at linear order, one can
examine the stability of $f(T)$ gravity. In the following we focus on the stability issue
of the gravity sector by ignoring the matter field contributions in the set of equations
(\ref{Chen_E00})-(\ref{Chen_Eab2}). One notices that, if the pure gravitational sector is
unstable, the theory would be ill-defined even after introducing the matter components.
Working in Fourier space, we introduce the mode-
expansion of $\phi$ to be
\begin{eqnarray}
\label{Chen_phiexpansion}
 \phi(t,\bx)=\int \frac{d^3k}{(2\pi)^\frac{3}{2}} ~ \tilde{\phi}_k(t)e^{i\bk\cdot\bx} ~.
\end{eqnarray}
Then, inserting this expansion into Eq. (\ref{Chen_T1}), one obtains the Fourier mode of
$T_1$ as follows:
\begin{eqnarray}
\label{Chen_T1k}
 \tilde{T}_{1k}=12H^2\tpk+12H\dot{\tilde{\phi}}_k ~.
\end{eqnarray}
Moreover, Eq. (\ref{Chen_FTexpansion}) yields the following Fourier modes:
\begin{align}
\label{Chen_f1k}
 \tilde{F}_{1k} &= \tilde{T}_{1k} \frac{dF(T)}{dT} \Big|_{T=T_0} = \tilde{T}_{1k} F'_0 ~,
\\
\label{Chen_f1kprime}
 \tilde{F}'_{1k} &= \tilde{T}_{1k} \frac{d^2F(T)}{dT^2}\Big|_{T=T_0} = \tilde{T}_{1k}
F''_0 ~,\\
\label{Chen_f1kdoubleprime}
 \tilde{F}''_{1k} &= \tilde{T}_{1k} \frac{d^3F(T)}{dT^3}\Big|_{T=T_0} = \tilde{T}_{1k}
F'''_0 ~,
\end{align}
which are the Fourier modes of the functions of $F_1$, $F'_1$, and $F''_1$, with $F_0$,
$F'_0$ and $F''_0$ being independent of $\phi$.

Substituting equations (\ref{Chen_phiexpansion})-(\ref{Chen_f1kdoubleprime}) into
(\ref{Chen_Eaa}), one obtains
\begin{eqnarray}
\label{Chen_phiddk}
 \ddot{\tilde{\phi}}_k +\Gamma \dot{\tilde{\phi}}_k +\omega^2 \tilde{\phi}_k=0 ~,
\end{eqnarray}
where $\omega^2$ and $\Gamma$ are given by
\begin{align}
\label{Chen_omega2}
 \omega^2 &= \big( {1+F'_0-12H^2F''_0} \big)^{-1}
 \left[ \frac{3H^2}{2} +\dot{H}
-\frac{F_0}{4} +\dot{H}F'_0
\right.
\nonumber\\
&\left.\ \ \ \ \  \
-(36H^4-48H^2\dot{H} ) F''_
0
+144H^4\dot{H} F'''_0 \right] ~,
\\
\label{Chen_Gamma}
 \Gamma &= \big( {1+F'_
0-12H^2F''_0} \big)^{-1}4H
\Big[1 +F'_0
\nonumber\\
&\ \ \ \ \ \ \ \ \ \  \ \ \
-(12H^2 +9\dot{H} )F''_0 +36H^2
\dot{H}F'''_0 \Big]    ~,
\end{align}
respectively. Eq.   (\ref{Chen_phiddk}) allows one to study the stability of any given
model. For instance, a model for which $\omega^2$ is negative is obviously unstable
against gravitational perturbations. Additionally, as mentioned previously, it is
sufficient to consider a scenario without any matter content. In such a pure
gravitational system, the background equation (\ref{background22}) (that is with
$\rho_m=0$, $p_m=0$) leads to a constant value of $H$. Therefore, Eq. (\ref{Chen_omega2})
can be simplified as
\begin{eqnarray}
\label{Chen_simple}
 \omega^2=\frac{\frac{3H^2}{2}-\frac{F_0}{4}-36H^4 F''_0}{1+F'_0-12H^2F''_0} ~.
\end{eqnarray}

Accordingly, one can make use of the relation (\ref{Chen_simple}) to justify if a
specific model of $f(T)$ gravity is free of instabilities. For example, we insert the
power-law ansatz $F(T)=\alpha (-
T)^n$, with $\alpha=(6H_0^2)^{1-n}/(2n-1)$ in the absence of matter ($H_0$ is the present
Hubble parameter), and the exponential ansatz $F(T)=-\alpha T\,(1-e^{pT_0/T})$, with
$\alpha=1/[1-(1-2p)e^p]$ in the absence of matter, considered in  \cite{Linder:2010py},
and
then we can straightforwardly determine the allowed values for the ansatz-parameters
numerically. It is easy to check that in both these cases the stability condition is
satisfied for the phenomenologically relevant ranges of parameters, that is $0<n<1$ for
the power-law model and $0<p<1$ for the exponential model.

\subsubsection{Growth of matter perturbations}

Having examined the stability of $f(T)$ gravity, we can now switch on the matter
sector, and examine the fluctuations about the FRW background in the presence of matter.
This is a crucial subject in any cosmological scenario.  In order to study the growth of
perturbations, we assume for simplicity a matter-only universe, i.e. we impose
$p_m=0=\delta p_m$. As usual, we define the matter overdensity $\delta$ as $ \delta
\equiv
\frac{\delta\rho_m}{\rho_m}$.

Now we can rewrite Eq.  (\ref{Chen_E00}) as
\begin{eqnarray}
\label{Chen_poisson}
&& 3H (1+F'_0-12H^2F''_0 )\dot{\tilde{\phi}}_k
\nonumber\\
&&+
[ (3H^2+k^2/a^2 ) (1+F'_0 )-36H^4 F''_0 ]
\tilde{\phi}_k \nonumber\\
&&+4\pi G\rho_m\tilde{\delta}_k=0 ~.
\end{eqnarray}
Note that this is the relativistic version of the Poisson equation in $f(T)$ gravity. As
a result, equations (\ref{Chen_phiddk}) and (\ref{Chen_poisson}) can be applied to evolve
$\delta$ for a given $f(T)$ model, which will allow for a comparison with observational
data.

Before proceeding to the investigation of equation (\ref{Chen_poisson}), let us make a
comment concerning the relation to GR. We notice an interesting feature
that emerges from the above analysis, both of the pure gravitational sector of the
previous paragraph as well as of the matter inclusive sector of the present paragraph,
namely that the linear perturbations in $f(T)$ gravity can locally reduce to those of
General Relativity in the limit where $F(T)$ is constant. For example, if $F(T)={\rm
const.}=-2\Lambda$, where $\Lambda$ is the cosmological constant, then one can
immediately
see that equations (\ref{Chen_E00})-(\ref{Chen_Eab2}) reduce to the well-known
first-order equations of GR  \cite{Mukhanov:1990me,Weinberg:2008zzc}.
Furthermore, equations (\ref{Chen_phiddk}) and (\ref{Chen_poisson}) reduce to the
well-known equations for the growth of perturbations in $\Lambda$CDM cosmology (see e.g.
 \cite{Ma:1995ey}). However, one should keep in mind that new degrees of freedom might
arise at non-perturbative level  \cite{Li:2011rn}.

We now proceed to the investigation of the physical implications of a non-trivial
$f(T)$-ansatz, studying the growth of the overdensity for a specific model. We choose the
power-law model suggested in  \cite{Linder:2010py}:
\begin{eqnarray}
\label{Chen_powerlaw}
 F({T}) = \alpha(-T)^n,
\end{eqnarray}
where $\alpha= (6H_0^2 )^{(1-n)} (1-\Omega_{m0} )/(2n-1)$ and $H_0$ and $\Omega_{m0}$
refer to the Hubble parameter and the matter density parameter at present.
\begin{figure}[ht]
\begin{center}
\epsfig{file=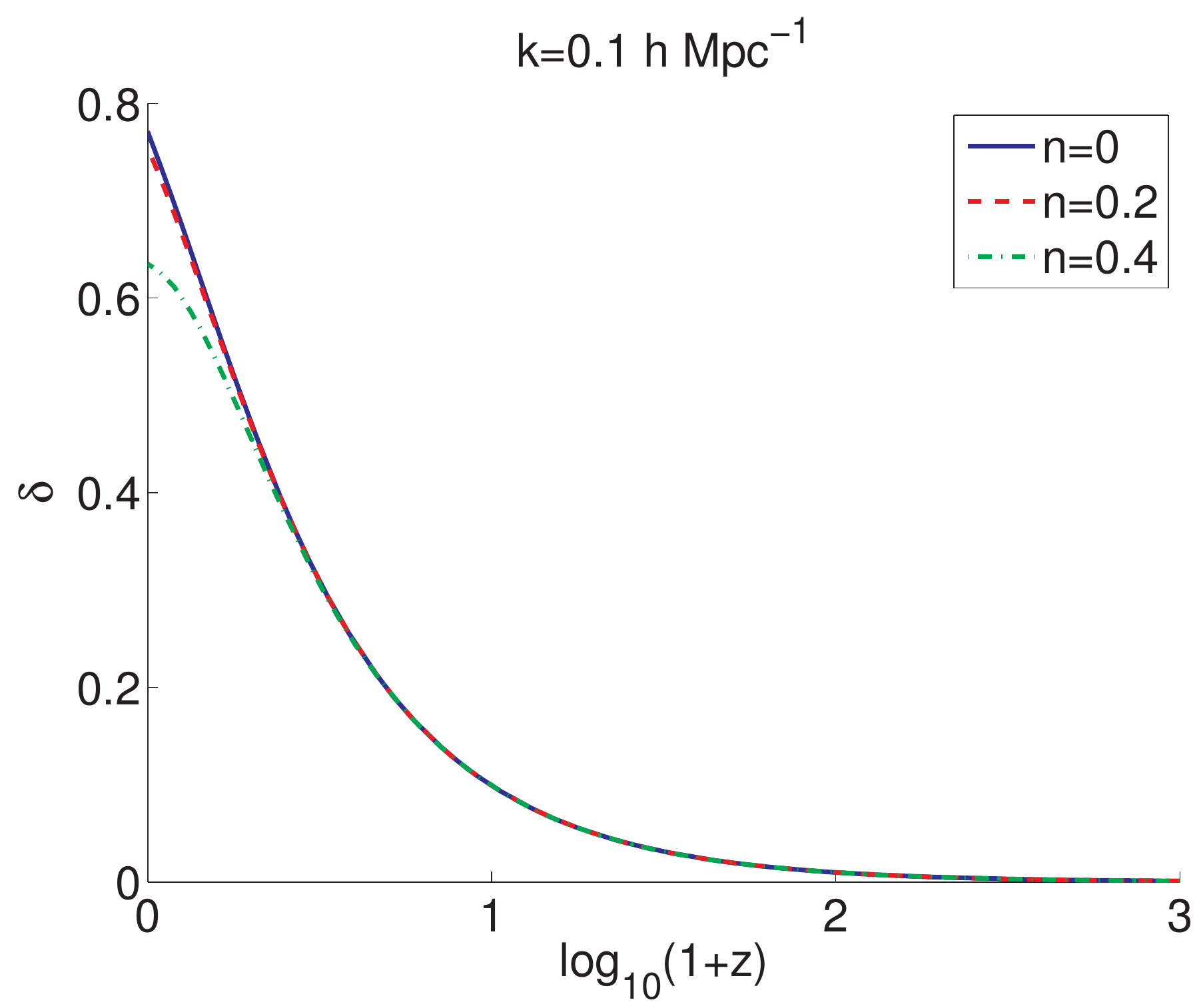,height=60mm}
\caption{
\label{Chen_k0p1}
\textit{The evolution of the matter overdensity $\delta$ as a function of the redshift
$z$, on a
scale of $k=0.1h$ Mpc$^{-1}$, for three choices of $n$, for the power-law model given by
(\ref{Chen_powerlaw}). From  \cite{Chen:2010va}. } }
\end{center}
\end{figure}
\begin{figure}[!]
\begin{center}
\epsfig{file=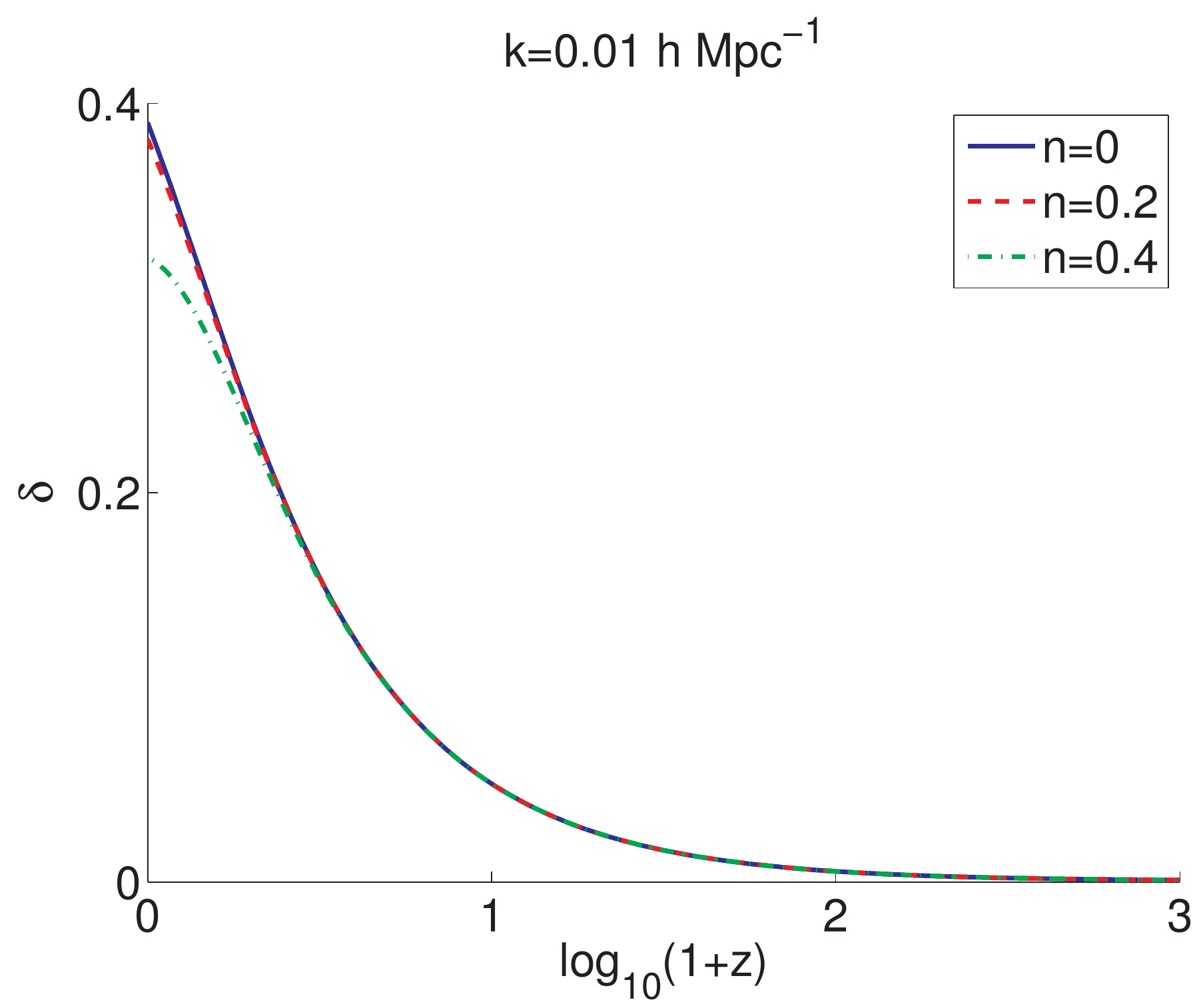,height=60mm}
\caption{
\label{Chen_k0p01}
\textit{The evolution of the matter overdensity $\delta$ as a function of the redshift
$z$, on a
scale of $k=0.01h$ Mpc$^{-1}$, for three choices of $n$, for the power-law model given by
(\ref{Chen_powerlaw}). From  \cite{Chen:2010va}. } }
\end{center}
\end{figure}
\begin{figure}[!]
\begin{center}
\epsfig{file=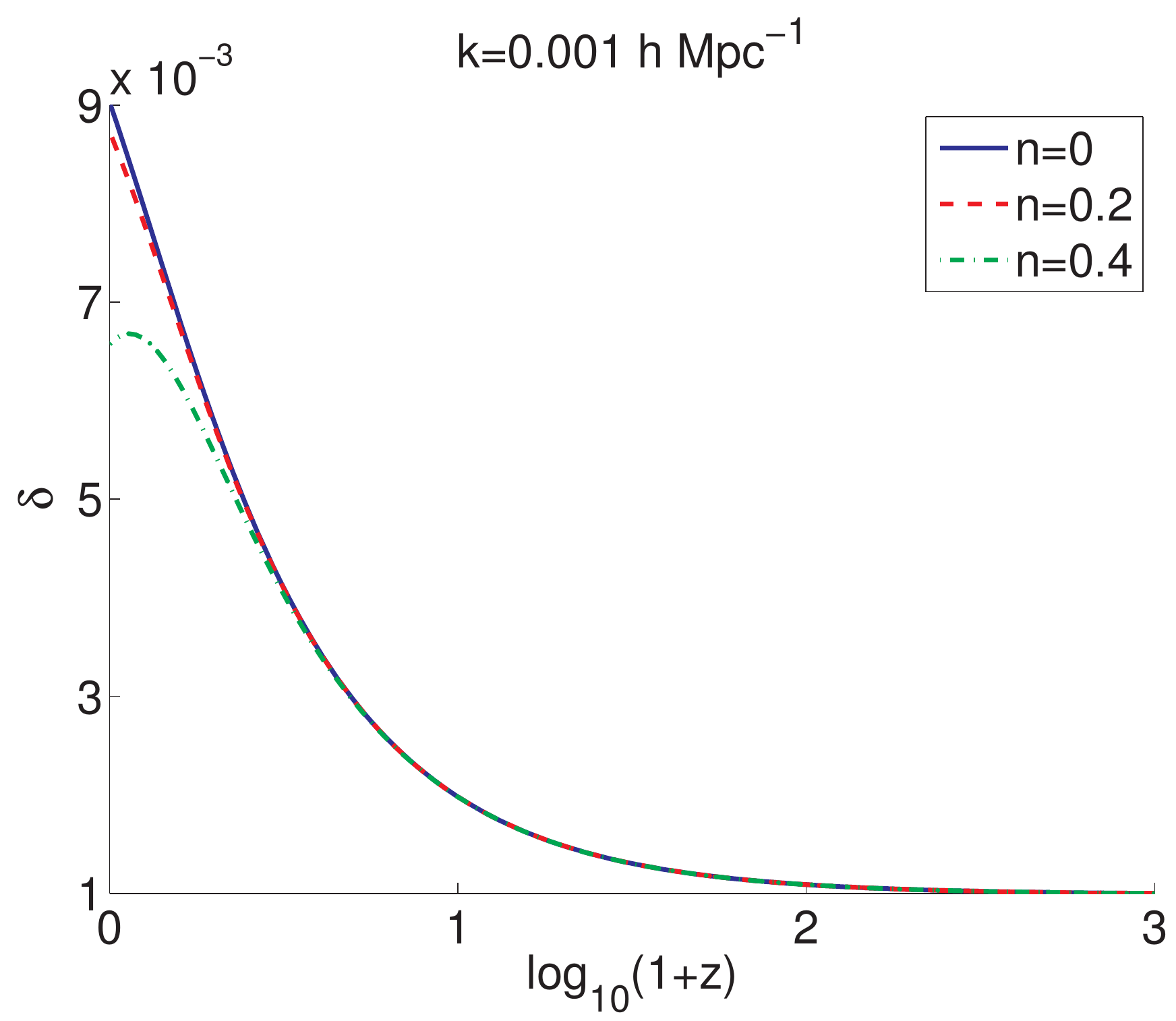,height=60mm}
\caption{
\label{Chen_k0p001}
\textit{The evolution of the matter overdensity $\delta$ as a function of the redshift
$z$, on a
scale of $k=0.001h$ Mpc$^{-1}$, for three choices of $n$, for the power-law model given
by
 (\ref{Chen_powerlaw}). From  \cite{Chen:2010va}. } }
\end{center}
\end{figure}
\begin{figure}[!]
\begin{center}
\epsfig{file=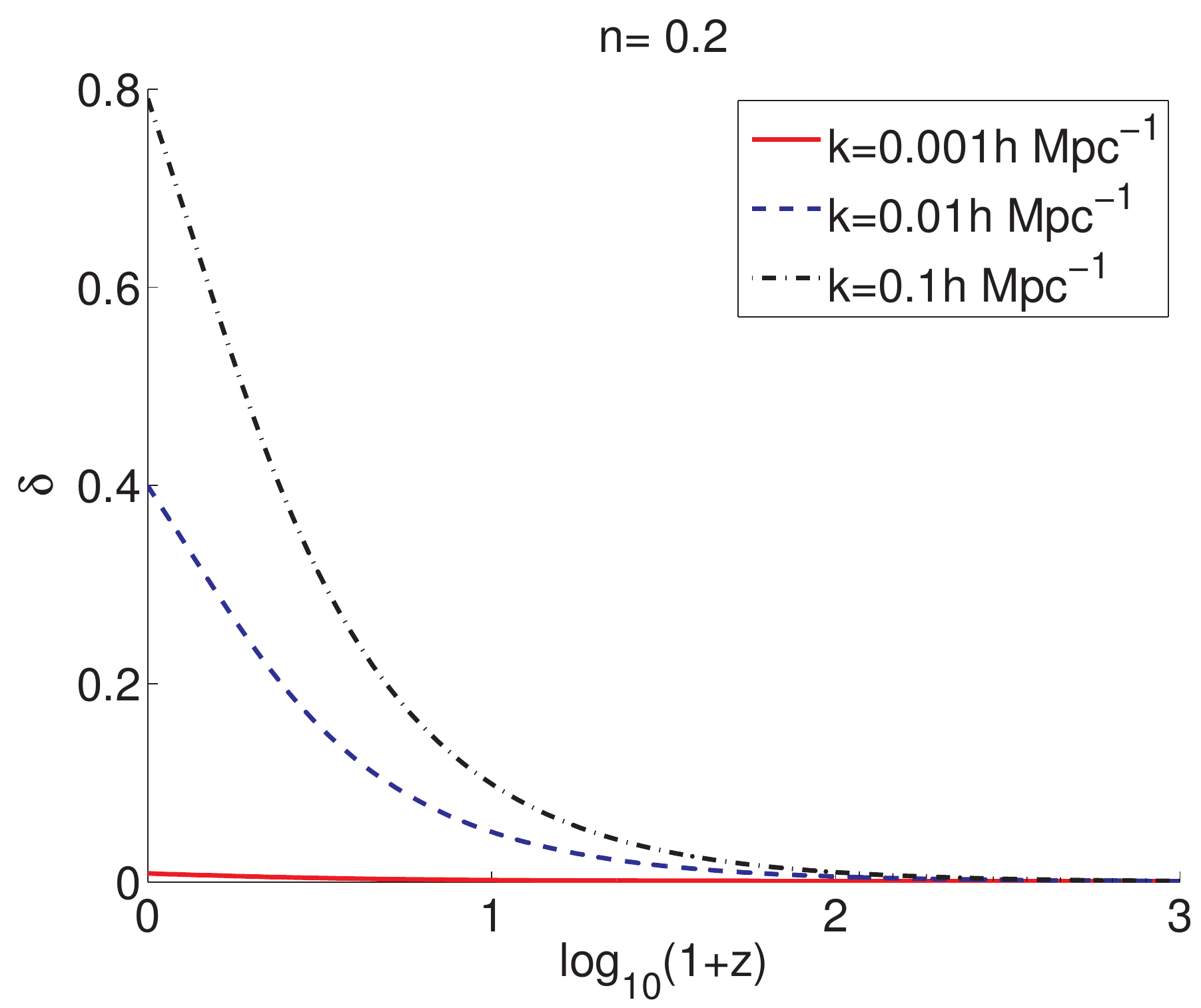,height=60mm}
\caption{
\label{Chen_vark}
\textit{The evolution of the matter overdensity $\delta$ as a function of the redshift
$z$, for a
fixed $n$ for the power-law model given by (\ref{Chen_powerlaw}), for three different
scales. From  \cite{Chen:2010va}. } }
    \end{center}
\end{figure}

Our results for the growth of perturbations, arising from a numerical elaboration, are
presented in Figures \ref{Chen_k0p1}-\ref{Chen_k0p001}. In these figures we follow the
growth of the matter overdensity $\delta$, from the time of last scattering to the
present one, for different choices of $n$, for three different $k$-scales. In Fig.
\ref{Chen_vark} we depict the evolution of $\delta$ for fixed $n=0.2$,
 but for three different scales. As usual, we use the redshift $z$ as the independent
variable, defined as  $1 + z =a_0/a$ with $a_0$ the present scale-factor value.

As expected, the $n=0$ case is identical to $\Lambda$CDM scenario  \cite{Dent:2008ia}.
However, as $n$ increases we find that there is a suppression of growth at smaller
redshifts, which can act as a clear distinguishing feature of these models. We also
notice that larger scales are more strongly affected than smaller ones. Therefore,
observations which target sub-horizon scales close to the horizon could, in principle,
constrain these models.

In addition, as we mentioned earlier, in order for a theory of $f(T)$ gravity to be free
of over-determination (since we have imposed the zero-anisotropic-stress assumption which
reduces the degrees of freedom) we need to require $F''(T)\simeq0$. This condition in the
case of the power-law ansatz requires $n\ll 1$, which is what we have used for the
numerical analysis. Interestingly enough, it is exactly the same condition that is needed
in order to acquire an observationally compatible dark-energy and Newton constant
phenomenology at the background level  \cite{Linder:2010py}.

Finally, we mention that, as has been studied in  \cite{Li:2011wu}, the extra degrees of
freedom of $f(T)$ gravity that are designed to explain the observed late time
acceleration, generally decay at small scales. Therefore, their effects on scales such as
galaxies and clusters are difficult to be probed. However, on large scales the new
perturbation modes could produce large deviations from the standard $\Lambda$CDM
cosmology, leading to stringent constraints on theories of $f(T)$ gravity.

\subsubsection{Vector and Tensor  Perturbations}
\label{vectortensorpert}

In the previous paragraphs we focused on the scalar perturbations of $f(T)$ gravity,
since
they are sufficient to reveal the basic features of the theory, allowing for a discussion
of the growth of matter overdensities. For completeness, following \cite{Chen:2010va} we
now extend our analysis in order to include the vector and tensor sectors of the theory,
in the absence of matter.

The general perturbed vierbein at linear order reads as
\begin{eqnarray}
\label{fullpert}
e_{\mu }^{0} &=&\delta _{\mu }^{0}(1+\psi )+a\left( G_{i}+\partial
_{i}F\right) \delta _{\mu }^{i} \nonumber\\
e_{\mu }^{A} &=&a\delta _{\mu }^{A}(1-\phi )+a\left(
h_{i}{}^{A}+\partial _{i}\partial ^{A}B+\partial
_{i}C^{A}+\partial ^{A}C_{i}\right) \delta _{\mu
}^{i}\nonumber \\
e_{0}^{\mu } &=&\delta _{0}^{\mu }(1-\psi )\nonumber \\
e_{A}^{\mu } &=&\frac{1}{a}\left[\delta _{A}^{\mu }(1+\phi
)+\left( h^{i}{}_{A}+\partial ^{i}\partial _{A}B+\partial
^{i}C_{A}+\partial _{A}C^{i}\right) \delta _{i}^{\mu
}\right]\nonumber\\
&\ &-\left( G_{i}+\partial _{i}F\right) \delta _{A}^{i}\delta
_{0}^{\mu }.
\end{eqnarray}%
In these expressions, apart from the scalar modes $\phi$ and $\psi$ used in
(\ref{basicveirpertChen})-(\ref{Chen_pert2}), we have introduced the transverse vector
modes $G_i$ and $C_i$, the transverse traceless tensor mode $h_i^A$, and the scalar
modes $F$ and $B$, the divergence of which will also contribute to
the vector sector. Similarly to the simple scalar case, the
coefficients on the above expressions have been chosen in order
for this vierbein perturbation to give rise to a perturbed metric
of the familiar form:
\begin{align}
&g_{00} =1+2\psi\nonumber \\
&g_{i0} =a\left[ \partial _{i}F+G_{i}\right] \nonumber\\
&g_{ij} =-a^{2}\left[ (1-2\phi )\delta _{ij}+h_{ij}+\partial
_{i}\partial _{j}B+\partial _{j}C_{i}+\partial _{i}C_{j}\right].
\end{align}

Let us make a comment here concerning the number of degrees of
freedom of the perturbed theory. As may be deduced straightaway,
$T^{\lambda}_{\mu\nu}$, $K^{\rho}_{\mu\nu}$ and $S^{\rho\mu\nu}$
are spacetime tensors under an infinitesimal coordinate
transformation of the form
 \be \label{gct} x^{\mu}\rightarrow
x^{\mu}+\epsilon^{\mu}.
 \ee
  This implies that the torsion scalar $T$ is generally a covariant scalar, and thus
actions of the form of (\ref{action_fT}) are generally covariant as well as invariant
under (\ref{gct}). As a result, for our choice of vierbien \eqref{weproudlyuse}, the
number of degrees of freedom (DOF) is identical to General Relativity. In particular, in
$3+1$ spacetime dimensions, the metric, being symmetric, has 10
independent DOF. This is reflected in (\ref{fullpert}), which
comprises
\begin{itemize}
\item 4 scalar DOF $\psi$, $\phi$, $F$ and $B$ \item 4 vector DOF,
2 associated with each of the divergenceless vectors $G_i$ and
$C_i$ \item 2 tensor DOF associated with the transverse, traceless
and symmetric tensor $h_{ij}$.
\end{itemize}
However, not all of these DOF are independent as there exist $3+1$
DOF associated with the coordinate transformation (\ref{gct}) (the
temporal part of $\epsilon^{\mu}$ is the scalar $\epsilon_{0}$,
and its spatial part can be decomposed into the gradient of a
scalar plus a divergenceless vector: $\partial_i \epsilon^S +
\epsilon_i^V$, leading to a total of 2 scalar and 2 vector DOF).
Subtracting these, we are left with a total of 6 DOF: 2 scalar, 2
vector, and 2 tensor, just as in the case of GR.

We can therefore work in the Newtonian gauge, setting $F$ and $B$ to zero. This is easily
understood since under (\ref{gct}), the gauge transformation of $B$ is
$-(2\epsilon^S)/a^2$, while that of $F$ is $(1/a)(-\epsilon_0 -\dot{\epsilon}^S +
2H\epsilon^S)$. Therefore, $\epsilon^S$ can be chosen in order to give rise to $B=0$, and
similarly an accompanying choice of $\epsilon_0$ will lead to $F=0$. Additionally, we
choose a gauge where the vector mode $C_i$ vanishes (through an appropriate choice of
$\epsilon_i^{V}$). As usual, the vector modes are transverse, while the tensor mode
is transverse and traceless, namely
\begin{eqnarray}
\partial_i C^i =
\partial_i G^i = 0\,;\ \ \ \
\partial_ih^{ij} =
\delta^{ij} h_{ij} = 0.
\end{eqnarray}
Finally, we easily deduce the following relation between the
tensor perturbations in the vierbein and inverse vierbein:
$h_{\rho=1}{}^{A=1} = -h_{A=1}{}^{\rho=1}$.

Using the above relations, the perturbed torsion tensor
(\ref{Tpert}) becomes
\begin{eqnarray}
T^0{}_{\mu\nu} &=& \partial_{\mu}\psi\delta^0_{\nu} - \partial_{\nu}\psi\delta_{\mu}^0 +
a(\partial_
{\mu}G_{\nu} -\partial_{\nu} G_{\mu})\nonumber\\
T^{i}{}_{0i} &=& H-\dot{\phi} + {\dot{h}_i{}^c}\delta_c^i\nonumber\\
T^{i}{}_{ij} &=& \partial_j\phi +
\partial_{i}h_{j}{}^c\delta_c^i-\partial_{j}h_{i}{}^c\delta_c^i,
\end{eqnarray}
with small Latin letters spanning the tangent-space spatial coordinates, and where for
notational compactness we have introduced a $G_0$ part of
$G_i$, which is zero. This torsion tensor inserted into
(\ref{Stensor}) leads to
\begin{eqnarray}
S_0{}^{0i} &=& \frac{1}{a^2}\partial_i\phi +
\frac{1}{2a^2}\sum_j\partial_jh_{i}{}^a\delta_a^j + \frac{H}{a}G_i\nonumber\\
S_0{}^{ij} &=&\frac{1}{4a^3}\left(\partial_iG_j - \partial_jG_i\right)\nonumber\\
S_i{}^{0i} &=& -H + \dot{\phi} + 2H\psi + \frac{1}{2}{\dot{h}}_i{}^a\delta_a^i\nonumber\\
S_i{}^{j0} &=&
\frac{1}{4a}\left(\partial_iG_j-\partial_jG_i\right)\nonumber\\
\nonumber S_i{}^{ij} &=& \frac{1}{2a^2}\partial_j(\phi-\psi)
+\frac{1}{2a^2}\left(\partial_kh_j{}^a -
\partial_jh_k{}^a\right)\delta_a^k\\ &\ &\ + \frac{H}{a}G_j
+\frac{1}{2a}\dot{G}_j.
\end{eqnarray}
However, the torsion scalar is unaffected by the vector and tensor
modes, and thus it is still given by (\ref{Chen_T1}), namely
\begin{eqnarray}
T\equiv T_0+T_1 = -6H^2 + 12H^2\psi + 12H\dot{\phi},
\end{eqnarray}
and likewise the determinant $e$ is still given by
$e = a^3(1+\psi-3\phi)$.

We now have all the necessary  machinery in order to extract the
equations of motion for the vector and tensor sectors. Following
the steps of the previous paragraph, we can similarly decompose the
energy-momentum tensor into its vector and tensor components and
ignore the vector and tensor anisotropic stresses. We finally
obtain
\begin{eqnarray}
\label{vectoreqn0}
 [1+F'(T)]\nabla^2 G_j =0
\end{eqnarray}
for the vector mode. Since the quantity in square brackets is zero
only for the unphysical model $f(T)=T+F(T)=0$ (for which the action
(\ref{action_fT}) does not describe the gravitational sector
anymore), we can eliminate it, resulting in
\begin{eqnarray}
\label{vectoreqn} \nabla^2 G_j =0.
\end{eqnarray}
Therefore, the vector modes in $f(T)$ gravity decay as $1/a^2$, which is similar to the
GR
case.

For the tensor mode, we obtain
 {\small{
\begin{equation}
\label{tensoreqn}
\!\!\bigg\{[1+F'(T)]\bigg(\frac{{\ddot{h}}_i{}^a}{2a}
-\frac{\nabla^2h_i{}^a}{2a}+\frac{3H{\dot{h}}_i{}^a}{2a}\bigg)
 - \frac{6H\dot{H}f''(T){\dot{h}}_i{}^a}{a}\bigg\}\delta_a^j = 0.
\end{equation}}}
Similarly to the scalar case, one can Taylor-expand the derivatives of $F(T)$ using
(\ref{Chen_FTexpansion}), and Fourier-expand the vector and tensor modes, in order to
obtain the corresponding dispersion relations.  Moreover, one can split the tensor sector
into left-handed and right-handed polarizations. This detailed analysis is performed in
Section \ref{SecionGravwaves} below.

Concerning the tensor equation (\ref{tensoreqn}), although there is a new friction term,
there are no new mass terms, which is a behavior similar to the scalar case of the
previous paragraph. Therefore, we can safely conclude that, in general, $f(T)$ theories
do
not introduce massive gravitons. Thus, when $f(T)$ tends to a constant we do not obtain
the typical problems of massive gravity, which is a significant advantage of $f(T)$
gravity. Additionally, note that similar to the scalar case, in the limit where $f(T)$
tends to a constant we do recover the behavior of GR at linear order,
which is a self-consistency test of the construction. Lastly, from the equations of
motion
for scalar, vector and tensor perturbations presented above, it is clear that these three
classes of perturbations decouple from one another in $f(T)$ gravity, just as they do in
GR.

%%%%%%%%%%%%%%%%%%%%%%%%%%%
\subsection{Implications for thermodynamics }
%%%%%%%%%%%%%%%%%%%%%%%%%%%

In this subsection we turn to discuss thermodynamics in the frame of $f(T)$ gravity in
order to further examine the viability of a given $f(T)$ model. In particular, we follow
the procedure developed in  \cite{Miao:2011ki, Bamba:2011pz, Sharif:2014kda,
Bamba:2012vg,Sharif:2013zga,Sharif:2014caa,Zubair:2015yma} and examine
whether the first and second laws of thermodynamics can be verified during the late-time
acceleration realized by $f(T)$ gravity. We recall that the fundamental connection
between
gravitation and thermodynamics was found in black hole physics  \cite{Bardeen:1973gs,
Bekenstein:1973ur, Hawking:1974sw, Gibbons:1977mu} (and see  \cite{Padmanabhan:2009vy} for
a comprehensive review). Especially, the standard Einstein equation in GR
was derived from the Clausius relation in thermodynamics  \cite{Jacobson:1995ab}. This
formalism was extensively applied to generalized gravity theories in the
literature  \cite{Eling:2006aw, Elizalde:2008pv, Bamba:2009gq, Brustein:2009hy}.

\subsubsection{First law}

In a flat FRW geometry, the  apparent horizon radius is   $ \tilde{r}_A = 1/H$. 
Moreover, the 
dynamical apparent horizon is determined by the expression
$h^{\alpha \beta} \partial_{\alpha} \tilde{r} \partial_{\beta} \tilde{r} =0$. Taking the 
time
derivative of $\tilde{r}_A$ one obtains $ -d\tilde{r}_A/\tilde{r}_A^3 = \dot{H}H dt $, 
and thus inserting it into the effective Friedmann
equation \eqref{background22}, one can find that $ \left[
1/\left(4\pi G\right)\right] d\tilde{r}_A 
=\tilde{r}_A^3 H \left(
\rho_{\mathrm{t}}+P_{\mathrm{t}} \right) dt $, with  $\rho_{\mathrm{t}}$ and 
$P_{\mathrm{t}}$ denoting respectively the total energy density
and pressure of the whole universe, namely corresponding to both matter and 
torsion-induced dark energy. In GR, the
Bekenstein-Hawking horizon entropy is given by $ S=\mathcal{A}/\left( 4G \right) $,
where $\mathcal{A}=4\pi \tilde{r}_A^2$ is  the apparent horizon area. Hence, using this 
relation along with the horizon entropy, one can derive
\begin{eqnarray}
\label{Bamba_eq:VA-4.8}
 \frac{1}{2\pi \tilde{r}_A} dS=4\pi \tilde{r}_A^3 H \left(
\rho_{\mathrm{t}}+P_{\mathrm{t}} \right)
dt ~.
\end{eqnarray}

It is well known that the temperature of the apparent horizon is associated with the
Hawking temperature, which is expressed as $ T_{\mathrm{H}} =
|\kappa_{\mathrm{sg}}|/\left(2\pi\right)$.
Moreover, the surface gravity $\kappa_{\mathrm{sg}}$ is given by  \cite{Cai:2005ra}
\begin{align}
\label{Bamba_eq:VA-3.12}
 \kappa_{\mathrm{sg}} &= \frac{1}{2\sqrt{-h}}
 \partial_\alpha \left(
\sqrt{-h}h^{\alpha\beta} \partial_\beta \tilde{r} \right) \nonumber\\
 &= -\frac{1}{\tilde{r}_A} \left( 
 1-\frac{\dot{\tilde{r}}_A}{2H\tilde{r}_A} \right)
=-\frac{\tilde{
r}_A}{2} \left( 2H^2+\dot{H} \right)
\nonumber\\
 &= -\frac{2\pi G}{3F} 
 \tilde{r}_A \left( \rho_{\mathrm{t}}-3P_{\mathrm{t}} \right)
 = -\frac{2\pi G}{3F}
 \tilde{r}_A \left(1 -3w_{\mathrm{t}}
 \right) \rho_{\mathrm{t}} ~,
\end{align}
with $h$   the determinant of the metric $h_{\alpha\beta}$ and where $w_{\mathrm{t}} 
\equiv
P_{\mathrm{
t}}/\rho_{\mathrm{t}}$ is the total equation-of-state parameter. As one can see 
  from  (\ref{Bamba_eq:VA-3.12}), for  $w_{\mathrm{t}} \le 1/3$ we obtain
$\kappa_{\mathrm{sg}} \le 0$. As
a result,  inserting (\ref{Bamba_eq:VA-3.12}) into $T_{\mathrm{H}} =
|\kappa_{\mathrm{sg}}|/\left(2\pi\right)$ gives rise to
\begin{equation}
\label{Bamba_eq:VA-3.14}
 T_{\mathrm{H}}=\frac{1}{2\pi \tilde{r}_A} \left(
1-\frac{\dot{\tilde{r}}_A}{2H\tilde{r}_A} \right)
~.
\end{equation}
Finally, combining  (\ref{Bamba_eq:VA-4.8}) with  (\ref{Bamba_eq:VA-3.14}) one
derives
\begin{equation}
\label{Bamba_eq:VA-4.9}
 T_{\mathrm{H}} dS = 4\pi \tilde{r}_A^3 H \left(\rho_{\mathrm{t}}+P_{\mathrm{t}} \right)
dt
 -2\pi  \tilde{r}_A^2 \left(\rho_{\mathrm{t}}+P_{\mathrm{t}} \right) d\tilde{r}_A ~.
\end{equation}

We remind that the Misner-Sharp energy  \cite{Misner:1964je} is defined through $ E 
=
\tilde{r}_A/\left(2G\right) = V \rho_{\mathrm{t}} $, where  $V=4\pi \tilde{r}_A^3/3$ is
the volume enclosed inside the apparent
horizon  \cite{Bak:1999hd}. Note that from the above definition we deduce that 
  $E$ is 
equivalent to the total
intrinsic
energy. Assembling we obtain:
\begin{eqnarray}
\label{Bamba_eq:VA-4.11}
 dE = -4\pi \tilde{r}_A^3 H \left(\rho_{\mathrm{t}}+P_{\mathrm{t}} \right) dt
 +4\pi \tilde{r}_A^2 \rho_{\mathrm{t}} d\tilde{r}_A ~.
\end{eqnarray}
Now, the work density is defined as \cite{Hayward:1997jp, Hayward:1998ee, Cai:2006rs}
$ W \equiv -\left (1/2\right) \left( T^{(\mathrm{m})
\alpha\beta} h_{\alpha\beta} +
T^{(\mathrm{DE})\alpha\beta}
h_{\alpha\beta} \right) = \left (1/2\right)  \left(
\rho_{\mathrm{t}}-P_{\mathrm{t}} \right) $, with 
$T^{(\mathrm{m})\alpha\beta}$ and
$T^{(\mathrm{DE})\alpha\beta}$ respectively the energy-momentum tensors of matter and
of effective dark energy (induced by the $f(T)$ terms) sectors.

In summary, using (\ref{Bamba_eq:VA-4.11}) and the work density $W$, we can express the 
first law of (equilibrium) thermodynamics as
\begin{equation}
\label{Bamba_eq:VA-4.12}
 T_{\mathrm{H}} dS=-dE+W dV ~.
\end{equation}
Hence, we have implemented an equilibrium description of thermodynamics.
Additionally, one can easily find the useful relation:
$ \dot{S} = 8\pi^2 H \tilde{r}_A^4 \left(\rho_{\mathrm{t}}+P_{\mathrm{t}}\right) = \left(
6\pi/G \right) \left( \dot{T}/T^2 \right) = -\left( 2\pi/G \right) \left[ \dot{H}/\left(
3H^3 \right) \right] $. 
This relation implies that in an expanding universe  the horizon
entropy $S$ is always increasing in the case where the null energy condition
$\rho_{\mathrm{t}}+P_{\mathrm{t}} \ge 0$ is satisfied (in which case the universe is not 
super-accelerating, i.e $\dot{H} \leq 0$).

%%%%%%%%%%%%%%%%%%%%%%%%%%%%%%
\subsubsection{Second law}
%%%%%%%%%%%%%%%%%%%%%%%%%%%%%%%

Let us now study the second law of thermodynamics through the usual equilibrium 
description (see
also  \cite{Bamba:2011pz, Karami:2012fu,Bamba:2012rv,Ghosh:2012pg,Bamba:2012vg} and
references therein for related investigations). In terms of all fluids that are enclosed 
by the horizon, the Gibbs equation reads as
$
T_{\mathrm{H}} dS_{\mathrm{t}} 
= d\left( \rho_{\mathrm{t}} V 
\right) +P_{\mathrm{t}} dV =
V d\rho_{\mathrm{t}} + \left( \rho_{\mathrm{t}} +P_{\mathrm{t}}
\right) dV $. Hence, the second law of thermodynamics writes as $ 
dS_{\mathrm{sum}}/dt
\equiv dS/dt + dS_{\mathrm{t}}/dt \geq 0 $, where $ S_{\mathrm{sum}} \equiv S +
S_{\mathrm{t}} $, with $S_{\mathrm{t}}$ the entropy corresponding to the total energy 
inside the horizon and $S$ the entropy of the horizon itself. Thus, using $V=4\pi 
\tilde{r}_A^3/3$,
the second Friedmann equation \eqref{background22}, expression (\ref{Bamba_eq:VA-3.14}), 
and the
relation $ \dot{S} = 8\pi^2 H \tilde{r}_A^4 \left(\rho_{\mathrm{t}}+P_{\mathrm{t}}\right)
= \left( 6\pi/G \right) \left( \dot{T}/T^2 \right) $, one can obtain
\begin{equation}
\label{Bamba_eq:VB-4.A003}
 \frac{dS_{\mathrm{sum}}}{dt} = -\frac{6\pi}{G}\left( \frac{\dot{T}}{T} \right)^2
\frac{1}{4HT + \dot{T}} ~.
\end{equation}

In summary, in order for the second law of thermodynamics to be valid, from  
~(\ref{Bamba_eq:VB-4.A003}) we extract the requirement  \cite{Bamba:2011pz} $ Y \equiv 
-\left( 4HT + \dot{T} \right) =
12H \left( 2H^2 + \dot{H} \right) \geq 0 $. We remind that in an expanding accelerating 
universe in $f(T)$ cosmology the relation $2H^2 + \dot{H} \geq 0$ is always
satisfied \cite{Bamba:2012vg}, and therefore the generalized second law of thermodynamics
is valid.

%---------------------------------------
\section{ $f(T)$ cosmology}
\label{SecionfTcosmology}

As we mentioned in the beginning of the previous Section, one important goal is to
construct viable models of $f(T)$ or modified teleparallel gravity that can accommodate
the expanding history of our background universe. In particular, cosmological
observations such the SNIa, the CMB, the LSS, and the BAO
data, have revealed that the universe is currently accelerating. This profound mystery
leads us to the prospect that, either about $70\%$ of the universe is made up of a
substance known as dark energy, about which we have almost no knowledge at all, or that
GR is modified at cosmological scales. It is manifest that $f(T)$ gravity can
provide a mechanism of phenomenologically realizing the present cosmic
acceleration, alternative to dark energy or curvature-based modified gravity.
Based on the considerations of the previous Section, in this Section we investigate in
detail the cosmological applications of $f(T)$ gravity.

\subsection{Late time acceleration}

Let us consider the flat FRW background space-time, of which the metric is given by
\begin{eqnarray}
 ds^2 = dt^2 - a^2(t) d\vec{x}^2 ~,
\end{eqnarray}
and hence the form of the vierbein is expressed as in \eqref{weproudlyuse}, namely
$e_{\mu}^A=\mathrm{diag}(1,a,a,a)$. In this case the background Friedmann equations were
given in \eqref{background11} and
\eqref{background22}. It is interesting to note that one can reformulate the Friedmann
equations to be the same as the ordinary ones in GR, namely
\begin{align}
\label{FRW1_fT_DE}
 H^2 &= \frac{8\pi G}{3} (\rho_r+\rho_b+\rho_{dm} + \rho_{DE}) ~,\\
\label{FRW2_fT_DE}
 \dot{H} &= -4\pi G (\rho_r+\rho_b+\rho_{dm} + \rho_{DE} + P_r + P_{DE}) ~,
\end{align}
by defining the energy density and pressure of an ``effective'' dark energy component as
\begin{align}
 \rho_{DE} &= \frac{2TF_{T}-F}{16\pi G} ~,\\
 P_{DE} &= \frac{F-TF_{T}+2T^2F_{TT}}{16\pi G (1+F_{T}+2TF_{TT})} ~.
\end{align}
(note that $F=f-T$ as was introduced below Eq.  \eqref{action_fT}). The above cosmic
system involves the radiation $\rho_r$, the baryon matter $\rho_{b}$, the cold dark
matter
$\rho_{dm}$ and the ``effective'' dark energy $\rho_{DE}$ arisen from $F(T)$'s
contribution. Furthermore, the continuity equations for the various matter components are
written as
\begin{eqnarray}\label{continuity_fT_DE}
 \dot\rho_i+3H(1+w_i)\rho_i=0~,
\end{eqnarray}
where the equation-of-state parameters $w_i\equiv \frac{P_i}{\rho_i}$ are defined as the
ratio of pressure to energy density. In particular, they read as $w_r=\frac{1}{3}$ for
radiation, $w_b=0$ for baryon matter, $w_{dm}=0$ for cold dark matter, respectively. For
the contribution of $f(T)$ gravity, one can effectively define:
\begin{eqnarray}\label{wde_fT_DE}
 w_{DE} \equiv \frac{P_{DE}}{\rho_{DE}} = -1
 + \frac{(F-2TF_{T}-T)(F_{T}+2TF_{TT})}{(F-2TF_{T})(1+F_{T}+2TF_{TT})}
\,.\nonumber\\
\end{eqnarray}
In particular, when $F_{T}=0$, one gets $w_{DE}=-1$ and hence this case corresponds to
the concordant model, i.e., the $\Lambda$CDM. One can further generalize the
equation-of-state parameter of the $i$-th component as a function of the redshift
$w_i(z)$, with the redshift given as $1+z=\frac{a_0}{a}$ (where we can set the present
scale factor $a_0$ to 1). Therefore, the evolutions of various energy densities are
described by
\begin{eqnarray}
 \rho_i = \rho_{i0} ~ \exp\left\{{3\int_0^z[1+w_i(\tilde z)]d\ln(1+\tilde z)}\right\} ~.
\end{eqnarray}

According to today's observations one learns that $\rho_{DE}$ is of the order of
$10^{-47}~{\rm GeV}^4$ and the corresponding equation of state parameter is approximately
$w=-1$. To our knowledge, the energy scale of mysterious dark component is far below any
cut-off or symmetry-breaking scales in quantum field theory. Therefore, if dark energy
originates from the nature of particle physics, namely, the energy density of quantum
vacuum states, one has to search for a delicate mechanism of canceling over large vacuum
energy density, but leaving unaffected the rest contributions that slightly deviate from
zero. This theoretical challenge is known as the famous cosmological constant problem in
the concordant cosmological model.

The idea of constructing dark energy models from the space-time torsion has already
appeared in the context of the TEGR  \cite{Bengochea:2008gz, Shie:2008ms, Ao:2010mg}.
Later, the $f(T)$ realization of dark energy models were extensively studied in the
literature, for example see  \cite{Wu:2010mn, Myrzakulov:2010vz, Bengochea:2010sg}. In
these models, people have noticed that the background field equations are always second
order differential equations. This profound advantage makes the cosmologies in $f(T)$
theory much simpler than those in $f(R)$ gravity, at the background level.
%The torsion
%tensor is formed solely from products of first derivatives of the tetrad and the
%space-time torsion
%manifests itself as a generator of gravitational repulsion.
%This enables the theory to
%produce universe acceleration.

\subsubsection{Background evolutions}
\label{subsubsec:DE_background}

In subsection \ref{eomsbasicfT} we presented the background equations of $f(T)$
cosmology. Hence, in this subsection we proceed to the investigation of a specific
example, which will allow as to perform a quantitative analysis. We start with the
study of exponential $f(T)$ theory suggested in  \cite{Linder:2010py}, of which the form
reads
\begin{eqnarray}\label{DEmodel_fT_background}
 f(T) = T+ \xi T \left[ 1- \exp \left( \frac{\alpha T_0}{T} \right) \right] ~,
\end{eqnarray}
where $\xi$ satisfies
\begin{eqnarray}
 \xi = -\frac{1-\Omega_m^{(0)}}{1-(1-2\alpha)e^\alpha} ~.
\end{eqnarray}
In the above expression, $\Omega_m^{(0)} \equiv 8\pi G\rho_m^{(0)}/3H_0^2$ is the
dimensionless density parameter of dust matter and $T_0 \equiv -6H_0^2$ denotes the value
of the torsion scalar in the current universe, which can be fixed by observations.
Thus, this model only involves one single dimensionless parameter $\alpha$.
\begin{figure}[!]
\centering
\includegraphics[width=6cm]{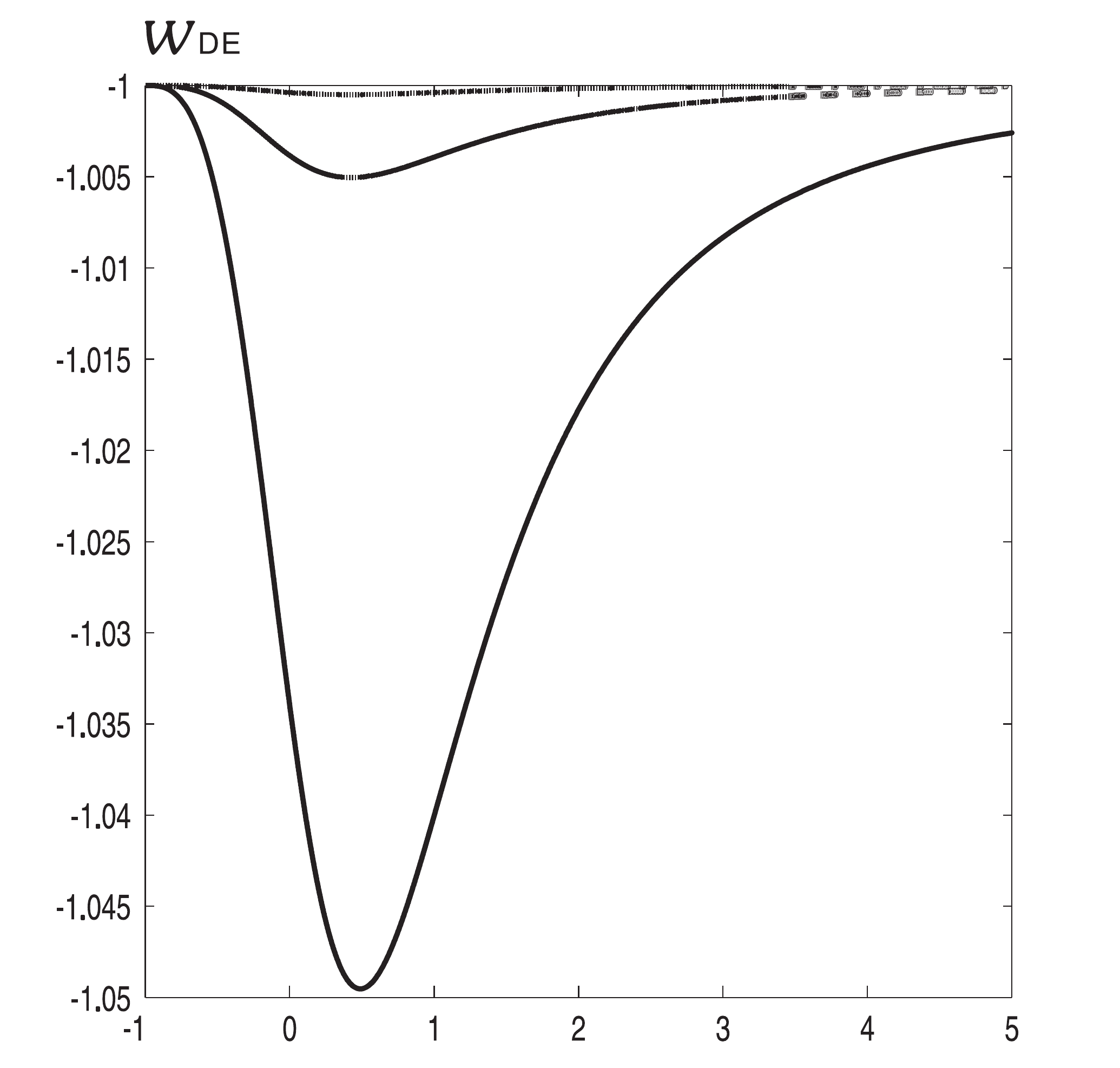}
\includegraphics[width=6cm]{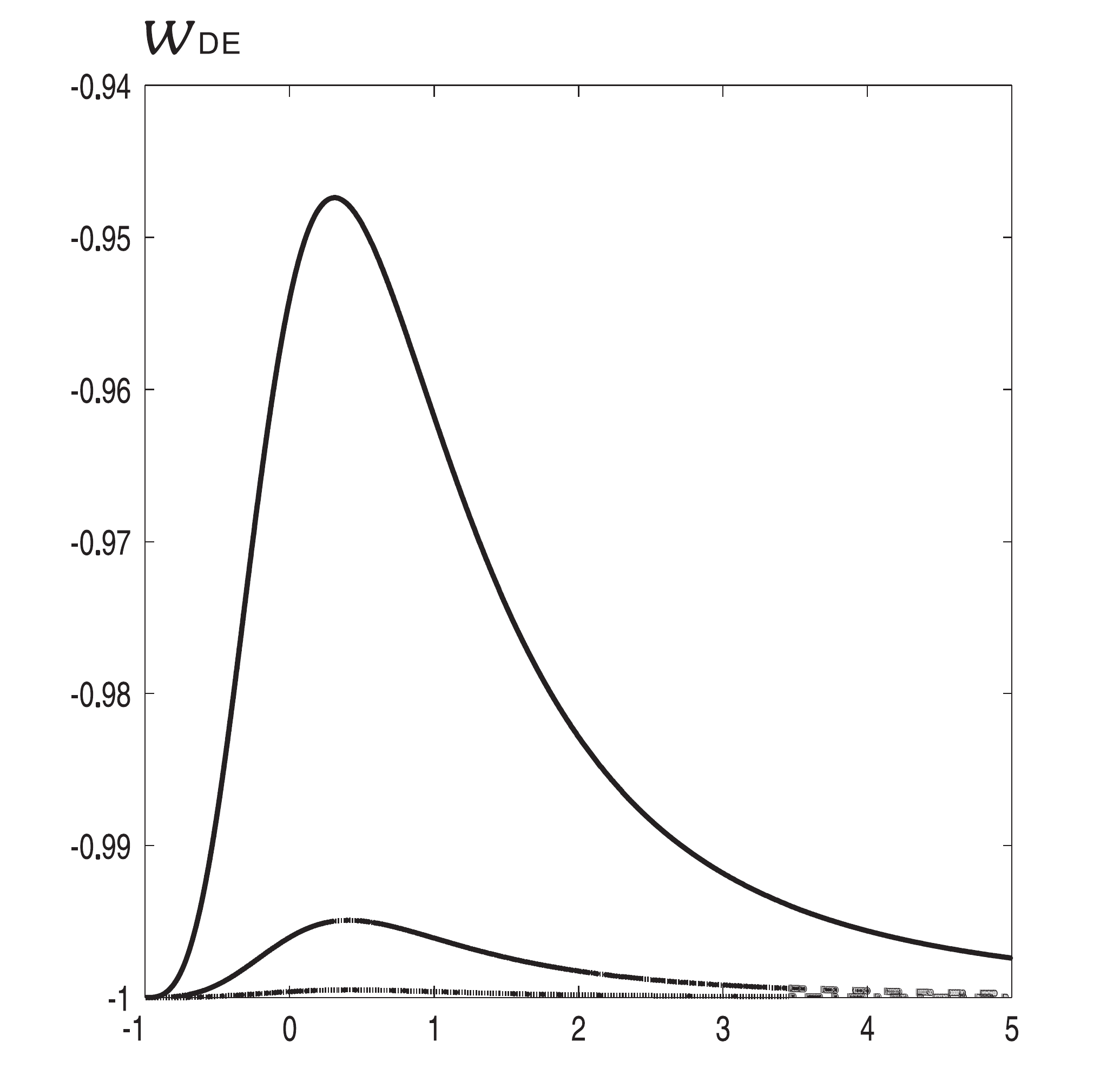}
\caption{\it Numerical evolutions of the equation of state parameter $w_{DE}$ of the
effective dark energy component arisen from the exponential $f(T)$ gravity as a function
of the redshift $z$. In the numerical calculation one sets the present value
$\Omega_m^{(0)}=0.26$. In the upper panel the solid line depicts the evolution of
$w_{DE}$ with $\alpha = 0.1$, dashed line is for $\alpha=0.01$, and the dash-dotted curve
is for $\alpha=0.001$, respectively. The lower panel shows the same cases except that
$\alpha$ takes negative values. From \cite{Bamba:2010wb}}.
\label{Fig_wde_fT_background}
\end{figure}

Utilizing the background Friedmann equation \eqref{FRW1_fT_DE}, the continuity equation
\eqref{continuity_fT_DE} and the expression \eqref{wde_fT_DE}, one can numerically
extract
the evolution of the equation-of-state parameter $w_{DE}$ of the effective
dark energy arising from the contributions of the $f(T)$ terms. In particular, we
consider the cases of $\alpha=\pm 0.1$, $\pm 0.01$, and $\pm 0.001$, respectively, and
adopt
$\Omega_m^{(0)}=0.26$ in the detailed numerical analyses.
The results are listed in Fig. \ref{Fig_wde_fT_background}. One can read from the figure
that $w_{DE}$ can be either phantom-like ($w<-1$) or quintessence-like ($w>-1$),
depending on the positivity of the model parameter $\alpha$. In particular, the values of
$w_{DE}$ in today's universe at the redshift $z=0$ are numerically determined as $-1.03$,
$-1.003$, and $-1.0003$ for $p=0.1$, $0.01$,
and $0.001$, respectively; while, they are found to be $-0.954$, $-0.996$, and $0.999$
for $p=-0.1$, $-0.01$, and $-0.001$, respectively. However, in all cases the solutions
approach to a quasi-de Sitter phase with $w_{DE}=-1$ in the far future, which indicates
that a final de Sitter stage might be a stable solution. This will be verified in the
following subsection, where we perform a detailed dynamical analysis.
\begin{figure}[!]
\centering
\includegraphics[width=6cm]{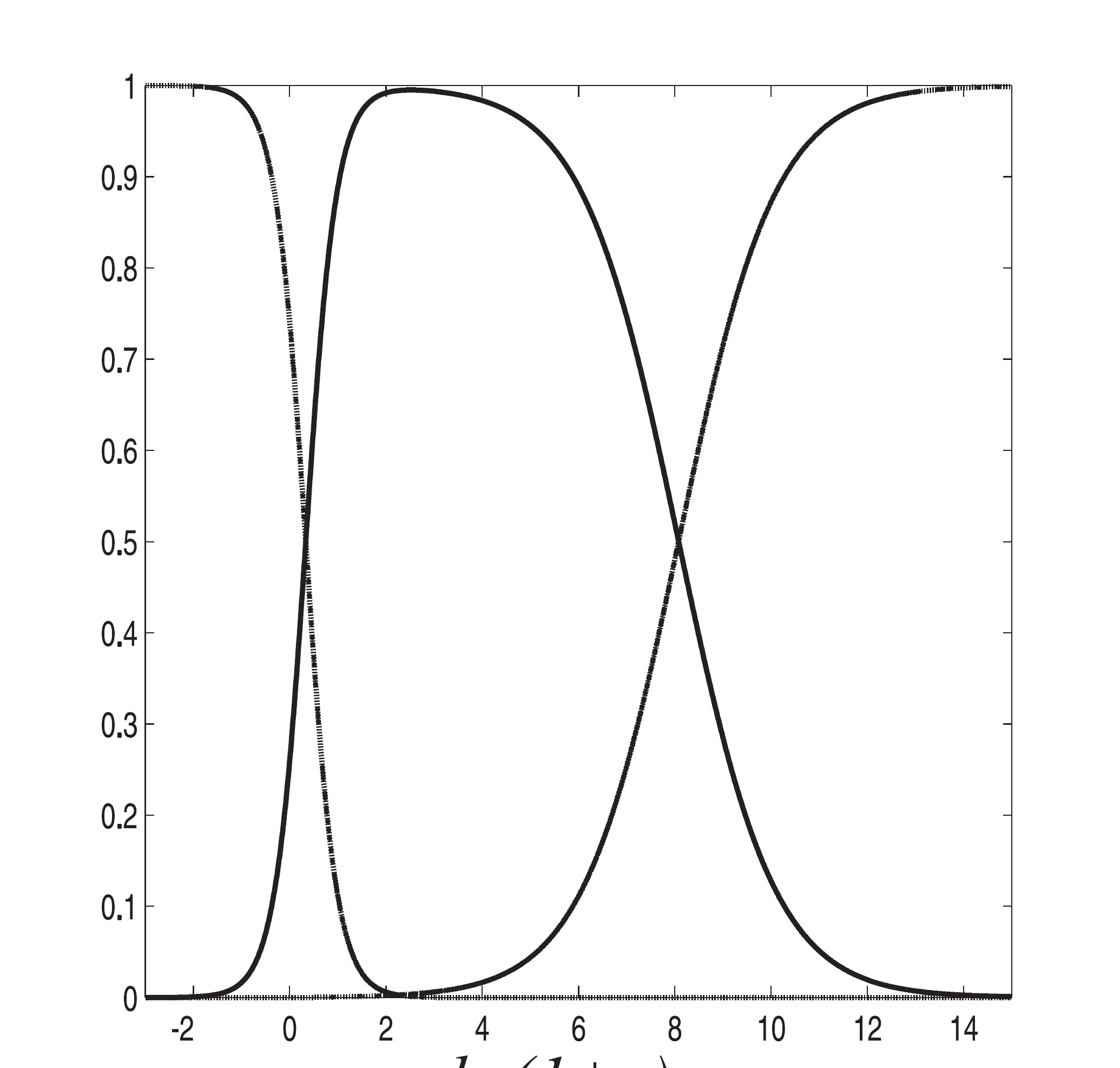}
\includegraphics[width=6cm]{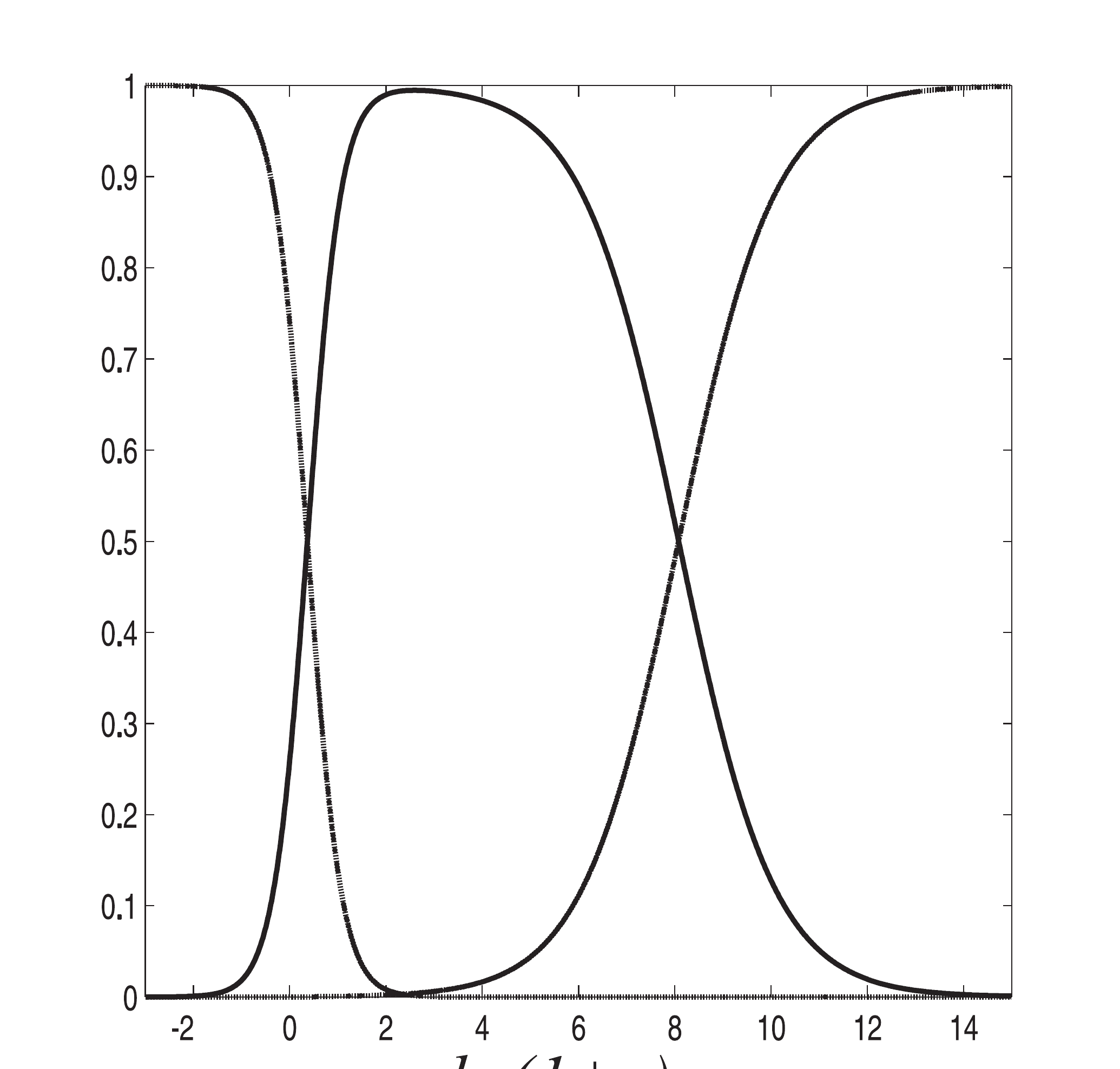}
\caption{\it Numerical evolutions of the dimensionless density parameter $\Omega_{DE}$ of
the effective dark energy component arisen from the exponential $f(T)$ gravity as a
function of $\ln(1+z)$. The model parameter $p$ takes the value of $p=0.1$ in the upper
panel and $p=-0.1$ in the lower panel, while the other parameters are the same with Fig.
\ref{Fig_wde_fT_background}. From \cite{Bamba:2010wb}. }
\label{Fig_omegade_fT_background}
\end{figure}

Fig. \ref{Fig_omegade_fT_background} depicts the evolutions of dimensionless density
parameter $\Omega_i\equiv 8\pi G\rho_i/3H^2$ for radiation (the dash-dotted curve), dust
matter (the solid curve), and the effective dark energy (the dashed line), respectively.
In the numerical calculation, one can apply the initial condition at today by taking the
realistic values for $\Omega_r$, $\Omega_m$ and $\Omega_{DE}$, and also set $\alpha=\pm
0.1$. From this figure one can read that the universe is dominated by radiation at the
high redshift regime, then it becomes matter dominated in the intermediate regime, and
eventually it enters the accelerating phase in the low-redshift area. Therefore, these
numerical results nicely demonstrate that the $f(T)$ gravity can explain the present
cosmic acceleration and in the meanwhile accommodate with the thermal expanding
history as observed in our universe.

\subsubsection{Dynamical analysis}
\label{subsubsec:DE_dynamical}
%Based on  \cite{Wu:2010xk, Zhang:2011qp, Jamil:2012nm}

The interesting cosmological behavior of $f(T)$ gravity makes it necessary to perform a
phase-space and stability analysis, examining in a systematic way the possible
cosmological behaviors, focusing on the late-time stable solutions. This procedure allows
to bypass the high non-linearities of the cosmological equations, which prevent any
complete analytical treatment, obtaining a (qualitative) description of the global
dynamics of these models, that is independent of the initial conditions and the specific
evolution of the universe.

In order to perform the phase-space and stability analysis of the scenario at hand, one
has to transform the aforementioned dynamical system into its autonomous form
$\label{eomscol}
\textbf{X}'
=\textbf{f(X)}$  \cite{Copeland:1997et, Ferreira:1997au, Chen:2008ft,Leon:2013qh}, where
$\textbf{X}$
is the
column vector constituted by suitable auxiliary variables, \textbf{f(X)} the
corresponding  column vector of the autonomous equations, and prime denotes derivative
with respect to $M=\ln a$. Then one can extract its critical points $\bf{X_c}$
satisfying
$\bf{X}'=0$, and in order to determine the stability properties of these critical points,
one can expand around $\bf{X_c}$, setting $\bf{X}=\bf{X_c}+\bf{U}$, with $\textbf{U}$ the
perturbations of the variables considered as a column vector. Thus, for each critical
point one can expand the equations for the perturbations up to first order as:
$\label{perturbation} \textbf{U}'={\bf{Q}} \cdot \textbf{U}$, where the
matrix ${\bf {Q}}$ contains the coefficients of the perturbation equations. Finally, for
each critical point, the eigenvalues of ${\bf {Q}}$ determine its type and stability.

Let us now investigate the dynamical properties of a cosmic system governed by  $f(T)$
gravity. Concrete analyses on this topic in various $f(T)$ models were performed in Refs.
 \cite{Wu:2010xk, Zhang:2011qp, Jamil:2012nm,Feng:2014fsa}. The main goal of this analysis
is to demonstrate that there indeed exists a stable accelerating solution at late times,
such that the present cosmic speeding up can be explained naturally.

For convenience, we consider that the energy components in the universe are only dust
matter (which includes both dark matter and baryons) $\rho_m$ and radiation $\rho_r$.
Then one can rewrite the equations of motion as a dynamical system by introducing the
following dimensionless variables:
\begin{eqnarray}\label{xyz_fT_dynamical}
 x \equiv -\frac{f}{6H^2} ~,~~~ y \equiv \frac{T f_{T}}{3H^2} ~,~~~
 z \equiv \Omega_r\equiv\frac{8\pi G \rho_r}{3H^2} ~,
\end{eqnarray}
where $\Omega_r$ is the dimensionless energy density parameter of radiation. Applying the
background Friedmann equations and the continuity equations for all matter components,
one
can deduce
\begin{align}
\label{x_fT_dynamical}
 x' &= -(2x+y)\frac{z+3-3x-3y}{2my-2+y} ~, \\
\label{y_fT_dynamical}
 y' &= 2my\frac{z+3-3x-3y}{2my-2+y} ~, \\
\label{z_fT_dynamical}
 z' &= -4z-2z\frac{z+3-3x-3y}{2my-2+y} ~,
\end{align}
where the prime denotes the derivative with respect to $\ln a$, and where we have defined
\begin{eqnarray}\label{dm_fT_dynamical}
 m \equiv \frac{T F_{TT}}{F_{T}} ~.
\end{eqnarray}

Note that, for a large family of $f(T)$ models, one can express $T$ as a function of
$y/x$, based on their definitions given in \eqref{xyz_fT_dynamical}. Then $m$ can  be
expressed in terms of $y/x$, too. To be specific, consider $f(T)= T+ \alpha
[(-T)^p-\beta]^q$. In this type of models, one can easily calculate that $m= p-1 +
\frac{1-q}{2q} (y/x)$. As a result, for a given form of $f(T)$, the
dynamical system governed by equations \eqref{x_fT_dynamical}, \eqref{y_fT_dynamical},
\eqref{z_fT_dynamical} becomes autonomous. Accordingly, the main Friedmann equation
\eqref{FRW1_fT_DE}
yields
\begin{eqnarray}\label{om_fT_dynamical}
 \Omega_m \equiv \frac{8\pi G \rho_m}{3H^2} = 1-x-y-z ~,
\end{eqnarray}
where $\Omega_m$ is the dimensionless density parameter of matter. Also the effective
equation of
state parameter for ``dark energy'' can be expressed as
\begin{eqnarray}
 w_{DE} = -\frac{x+y/2-my}{(1-y/2-my)(x+y)} ~.
\end{eqnarray}

In order to extract the dynamical properties of the above autonomous system, one solves
the combined equations: $x'=0$, $y'=0$, and $z'=0$. As a consequence, one can
obtain two isolated critical points and a continuous critical line, which list as
\begin{align}
& {\rm Point ~ A}: \quad
   x_c=0,\; y_c=0, \; z_c=1\;, \nonumber\\
& {\rm Point ~ B}: \quad
   x_c=0,\; y_c=0, \; z_c=0\;, \nonumber\\
& {\rm Line ~ C}: \quad ~
   x_c=1-y_c, \; z_c=0\;. \nonumber
\end{align}
In the following, we discuss these solutions and we examine their stability.

$\bullet$ Point A corresponds to a radiation dominated solution, since at this point one
can easily derive that $\Omega_{r}=1$. Then one can study the stability of this solution
by calculating the eigenvalues of the above linearized system. They are found to be
\begin{equation}
\!\left( 1 ~,~~~  2 \Big(1 - m \pm \sqrt{1 + 2 m + m^2 - 2 \frac{dm}{d(y/x)} }\Big)
\right).
\end{equation}
Hence, this critical point is unstable due to the presence of a positive eigenvalue,
which means that the universe evolves away if there exists any small classical
perturbation on the background trajectory.

$\bullet$ Point B corresponds to a matter dominated solution, since it
exhibits $\Omega_{m}=1$. One can further calculate the eigenvalues at
this point as
\begin{eqnarray}
\! \left(-1 ~,~~~  2\Big(1 - m \pm \sqrt{1 + 2 m + m^2 - 2 \frac{dm}{d(y/x)}
}\Big)\right).
\end{eqnarray}
Since the cosmological features of this point are not favored by observations, we
deduce that for a realistic cosmology one expects that the universe should deviate from
it along the background expansion. This can be satisfied if the real part
of one of the above eigenvalues is positive definite.

$\bullet$ Line C corresponds to the solution dominated by the $f(T)$ contribution. In
this solution, $x_c+y_c=1$ indicates that both $\Omega_m$ and $\Omega_r$ are vanishing.
Moreover, it is easy to read that $w_{DE}=-1$ and that $H$ becomes constant on this
critical line. Therefore, the background evolution on this critical line corresponds to a
de Sitter expanding phase. Accordingly, the eigenvalues of the linearized system in this
solution are given by $(-4, 0,-3)$, which are either negative or vanishing. This implies
that the background evolution described by the last solution is always (marginally)
stable. Therefore, the solution of Line C is of cosmological interest in explaining the
present acceleration of our universe.

Having the above general analysis in mind, one can proceed and study the detailed
dynamics of the cosmological evolution governed by a specific $f(T)$ model. To be
explicit, one may consider a power-law model of the following form
\cite{Bengochea:2008gz, Linder:2010py}:
\begin{eqnarray}\label{PL_fT_dynamical}
 f(T)= T+ f_0 M_p^2 \Big( \frac{-T}{M_p^2} \Big)^p ~,
\end{eqnarray}
where $f_0$ and $p$ are two dimensionless parameters. Depending on the choice of the
model parameters, this type of $f(T)$ gravity can be connected with several
representative cosmological models that are already familiar in the literature. For
instance, the above model can reduce to the $\Lambda$CDM if $p=0$. Additionally, it
can mimic the Dvali-Gabadadze-Porrati (DGP) model for $p=1/2$, while for $p=1$ the
background equation can be expressed as $H^2=\frac{8\pi
G}{3(1-f_0)}(\rho_m+\rho_r)$, which then recovers a cold dark matter (CDM) cosmology
after rescaling the gravitational constant via $G\rightarrow G/(1-f_0)$. Furthermore, if
one desires this specific $f(T)$ gravity to be phenomenologically reasonable, then the
value of $p$ has to be between $0$ and unity.

Specifically, one can substitute the form of $f(T)$ \eqref{PL_fT_dynamical} into the
background Friedmann equation \eqref{FRW1_fT_DE}, and then it is easy to observe that the
Hubble parameter can be a non-zero constant, by setting the energy densities of matter and
radiation to be vanishing and considering that $p\neq 1$. This corresponds to the
solution of a de Sitter expansion represented by Line C. Additionally, one can
also examine the case of Point B in the specific model under consideration. Then the
eigenvalues of the linearized system at Point B are given by $-1$, $2$, $-2p$,
respectively. The appearance of a positive value manifestly implies that this critical
point is unstable along the cosmic expansion.

Moreover, combining all the above critical points, it is interesting to observe that the
universe can evolve from a radiation dominated phase (described by Point A) to a matter
dominated era (described by Point B) and eventually enter a quasi de Sitter expanding
phase (described by Line C). Thus, the present model can provide a cosmological solution
that exactly describes the expanding history as observed in our universe.
\begin{figure}[ht]
\centering
\includegraphics[width=6cm]{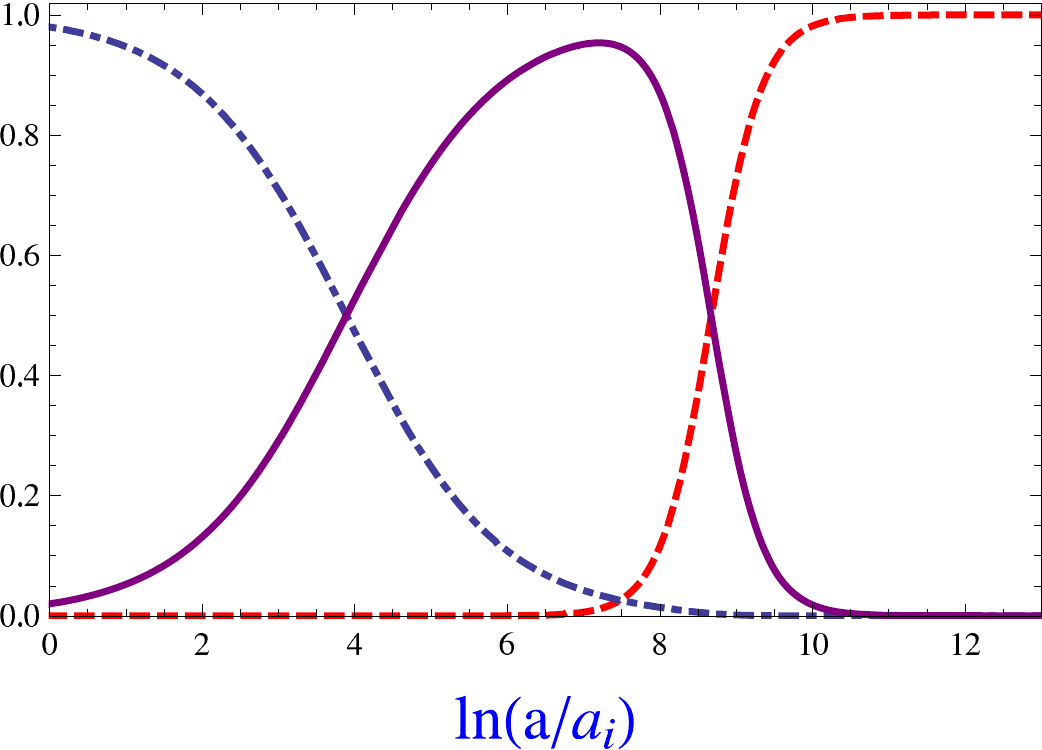}
\includegraphics[width=6cm]{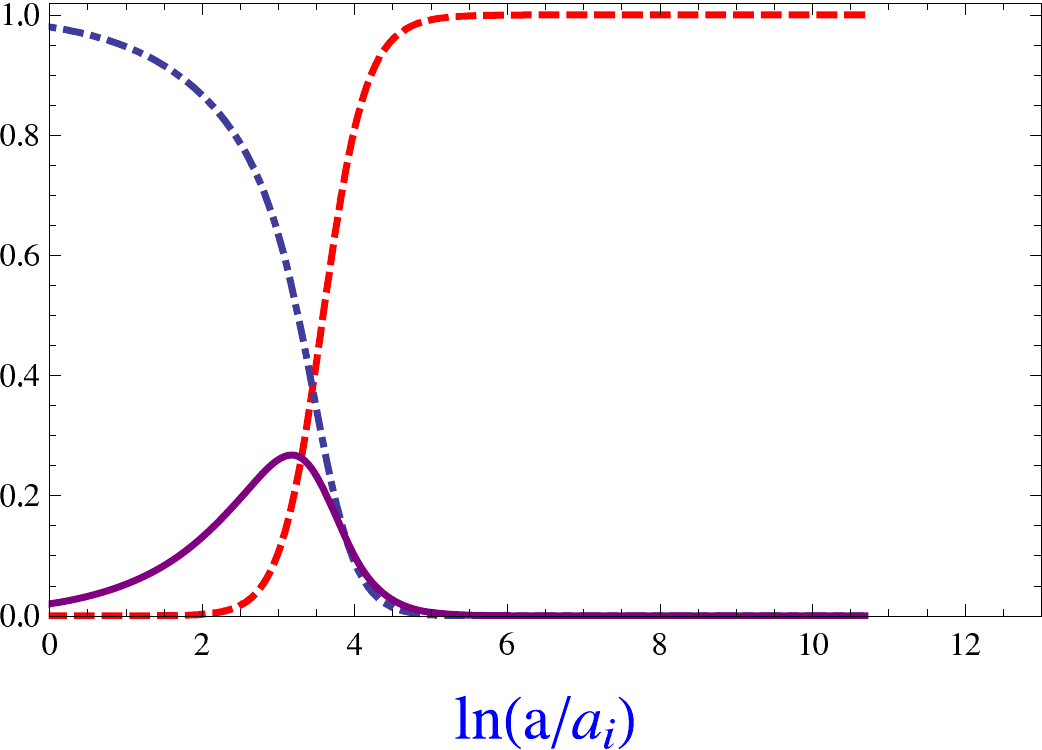}
\caption{\it Numerical results of the background evolutions for the power-law $f(T)$
gravity \eqref{PL_fT_dynamical} .
The model parameter $p$ is chosen to be $0.2$. $a_i$ is the initial value of the scalar
factor. The dot-dashed, solid, and dashed curves correspond to the cosmic evolutions of
dimensionless density parameters $\Omega_r$, $\Omega_m$, and $\Omega_{DE}$, respectively.
In the upper panel the initial conditions are taken as: $x_i = y_i = 10^{-13}$ and $z_i =
0.98$, while in lower panel they are given by $x_i = y_i = 10^{-5}$ and $z_i=0.98$. From
 \cite{Wu:2010xk}.}
\label{Fig_fT_dynamical}
\end{figure}

In order to demonstrate such a possibility, one can numerical solve the background
evolutions governed by the specific power-law $f(T)$ gravity \eqref{PL_fT_dynamical}.
Fig. \ref{Fig_fT_dynamical} depicts numerical evolutions under different initial
conditions. In the numerical computation one specifically considers $p=0.2$. It is
manifest to see that the initial values of $x$ and $y$ have to be suitably small, in
order to obtain a sufficiently long era of radiation domination, in agreement with
observations. From the upper panel of the figure we indeed observe that the universe
was initially in the radiation dominated phase, then it became matter dominated in the
intermediate regime, and finally it entered the cosmic acceleration at late
times. As a result, for $f(T)$ gravity of the power-law form,  the conditions for the
universe to evolve from an initial phase of radiation domination to a cosmic acceleration
at late times are simply $x_i \ll 1$, $y_i \ll 1$ and $p\neq 1$. Moreover, in order to
satisfy the current observational constraints, $|p|\ll1$ is also required
\cite{Bengochea:2008gz, Linder:2010py, Wu:2010mn}.

Uo to now we have only considered a simple example of $f(T)$ gravity with a power-law
function, that is minimally coupled to regular matter. This example is already sufficient
to demonstrate the occurrence of the late-time acceleration. Moreover, it is important to
note that in the above model the famous coincidence problem is not completely
overcome since one needs to fine tune the values of initial conditions such that the
universe can start the cosmic acceleration in the recent cosmological past. To address
the issue of cosmological coincidence, one can phenomenologically introduce interaction
terms between the torsion scalar and the matter sector. Based on this scenario, one can
perform a phase space analysis and search for attractor solutions where the matter and
dark energy density parameters are of the same order \cite{Jamil:2012nm}, since this
could alleviate the coincidence problem.

\subsubsection{Phantom crossing/Quintom scenario}
\label{subsubsec:DE_quintom}

One interesting property from the $f(T)$ gravity is the realization of the so-called
quintom scenario, which is also dubbed as the phantom-divide crossing in the literature.
This phenomenological scenario is motivated by the  mildly favored observational signal
that the equation-of-state parameter of dark energy might have crossed the cosmological
constant boundary from above to below in the recent cosmological past \cite{Feng:2004ad,
Wang:2006ts, Zhao:2006qg, Xia:2007km, Zhao:2012aw,Xia:2013dea}. However, it was noticed
that a consistent scalar field model that realizes
the quintom scenario is extremely difficult to be achieved  \cite{Vikman:2004dc,
Hu:2004kh,Caldwell:2005ai}. For instance, for a regular dark energy model constructed by a
quintessence field, the corresponding equation of state parameter is limited in the
regime $-1 \leq w \leq 1$; while for a phantom field it is always $w<-1$. The difficulty
for the model buildings of realizing the phantom crossing is based on the proof of a
``No-Go'' theorem for dynamical dark energy models  \cite{Xia:2007km}. The study upon
this
topic was reviewed comprehensively in \cite{Cai:2009zp}, which states that for dark
energy models described by a single perfect fluid or a single scalar field with a
Lagrangian of K-essence form \cite{ArmendarizPicon:1999rj,
Garriga:1999vw, ArmendarizPicon:2000dh}, their cosmological perturbations could
encounter a severe divergence when the background equation-of-state parameter is forced
to cross $-1$. Additionally, it was pointed out in \cite{Carroll:2003st, Cline:2003gs}
that such a process might violate the null energy condition and hence would bring
potential quantum instabilities unless one considers non-conventional dark energy models
such as Galileon/Horndeski scalar field constructions \cite{Horndeski:1974wa,
Nicolis:2008in,Deffayet:2011gz}, bi-scalar constructions \cite{Padilla:2013jza}, or
spinor field scenarios \cite{Cai:2008gk, Wang:2009ae}.

In the literature many dark energy models were put forward to realize the phantom
crossing behavior, but it is interesting to note that such a possible phenomenon can be
effectively realized in theories of modified gravity. The realization of the scenario of
crossing the cosmological constant boundary within $f(R)$ gravity were extensively
studied in \cite{Abdalla:2004sw, Nojiri:2006ri, Amendola:2007nt, Hu:2007nk,
Bamba:2008hq,Nojiri:2013ru}.
As an analogue, it is expected that the same phenomenon can be realized in the context of
$f(T)$ gravity. In the following, we consider a specific model, which was proposed in
\cite{Bamba:2010wb} (see also \cite{Farajollahi:2011af}) to demonstrate that the
phantom crossing behavior is achievable in this theory.

The explicit form of $f(T)$ gravity is constructed as  \cite{Bamba:2010wb}:
\begin{equation}
\label{DEmodel_fT_quintom}
 f(T) = T+ \tilde\xi T \left[ 1- \exp \left( \frac{\alpha T_0}{T} \right)
 - \sqrt{\frac{T_0}{\alpha T}}\, \ln \left(\frac{\alpha T_0}{T}\right) \right] ~,
 \end{equation}
where the first two terms of the r.h.s. are the same as the exponential $f(T)$ model in
\eqref{DEmodel_fT_background}, except that
\begin{eqnarray}
 \tilde\xi = -\frac{1-\Omega_m^{(0)}}{1-(1-2\alpha)e^\alpha+2/\sqrt{\alpha}} ~.
\end{eqnarray}
However, the last term appearing in \eqref{DEmodel_fT_quintom} takes a logarithmic form.
From the above expression one can immediately find that $\alpha$ has to be positive in
order for the model to be well defined. Recall that in paragraph
\ref{subsubsec:DE_background} above we have pointed out that
$w_{DE}<-1$ when $\alpha>0$. But, the last term of \eqref{DEmodel_fT_quintom} can yield a
quintessence-like component with an effective $w_{DE}>-1$. As a result, the combined
effect of these two parts naturally provides a possible dynamical realization of the
behavior of crossing over $w_{DE} = -1$.

\begin{figure}[htbp]
\centering
\includegraphics[width=6cm]{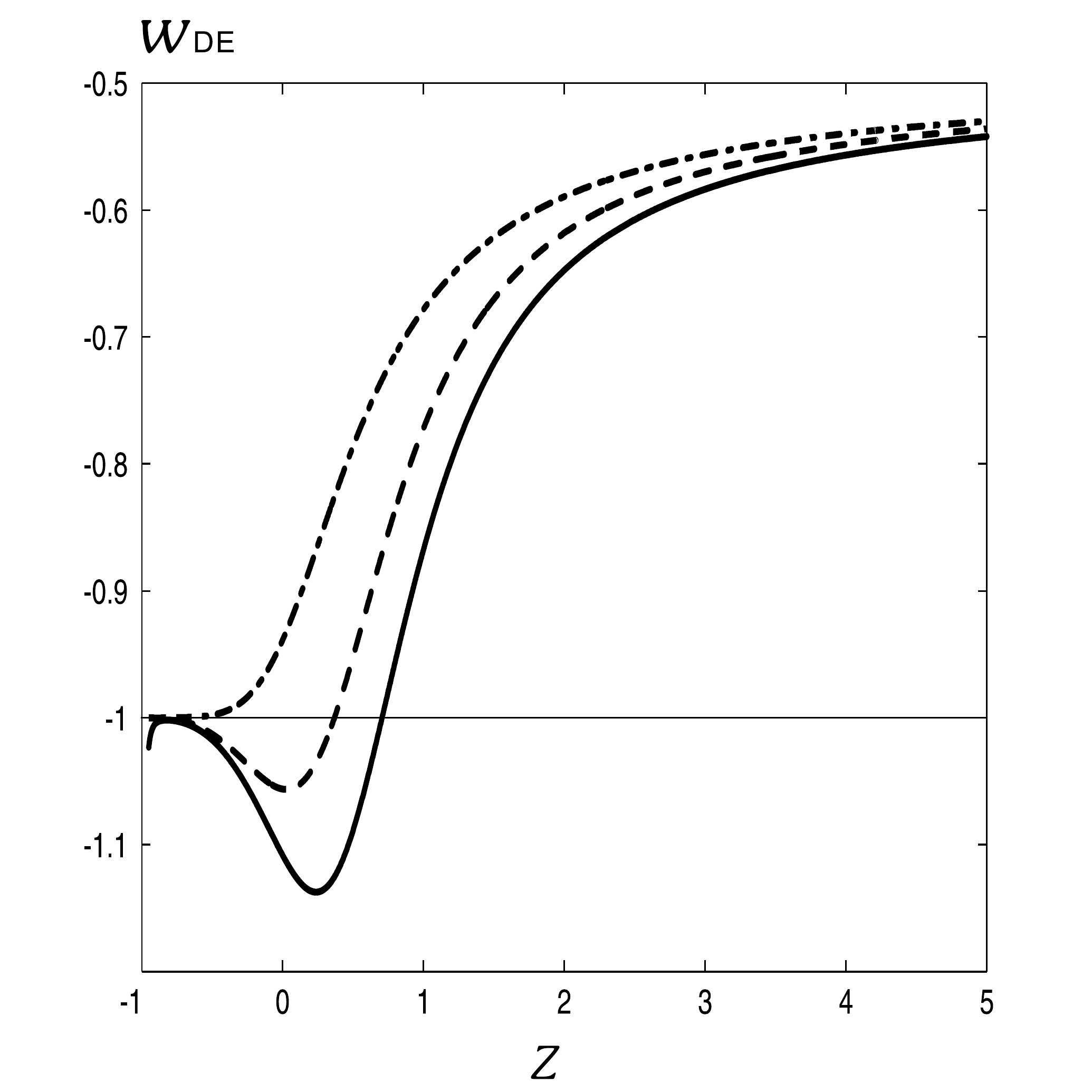}
\includegraphics[width=6cm]{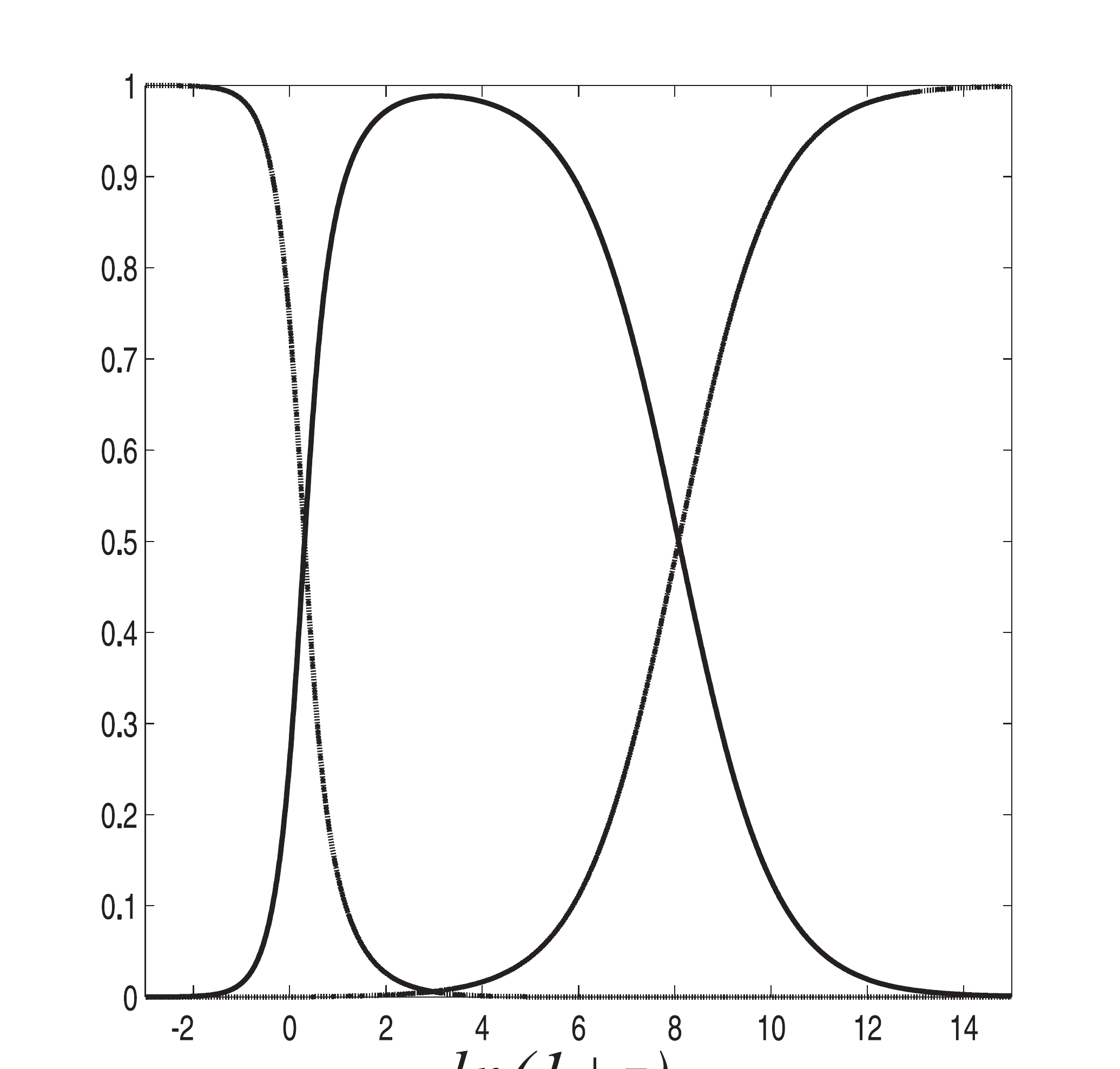}
\caption{\it Numerical evolutions of the effective dark energy equation-of-state parameter
$w_{DE}$ and the dimensionless density parameters in the model of quintom-like $f(T)$
gravity \eqref{DEmodel_fT_quintom} as functions of the redshift $z$ or its log scale
$\ln(1+z)$. In the numerical calculation one considers $\Omega_m^{(0)}=0.26$. In the
upper panel we use the solid line to depict the evolution of $w_{DE}$ with $\alpha=1$,
dashed line for $\alpha=0.8$, and the dash-dotted curve for $\alpha=0.5$, respectively.
The lower panel shows the evolution curves of $\Omega_{DE}$ (dashed line), $\Omega_m$
(solid line) and $\Omega_r$ (dash-dotted line) with a fixed model parameter
$\alpha=1$. From \cite{Bamba:2010wb}.}
\label{Fig_fT_quintom}
\end{figure}

Similarly to the case considered in paragraph \ref{subsubsec:DE_background}, the
present model also involves only one single parameter $\alpha$. Thus, it is not difficult
to make use of the background Friedmann equation \eqref{FRW1_fT_DE} and the continuity
equation \eqref{continuity_fT_DE} to numerically extract the cosmological solutions. In
the upper panel of Fig. \ref{Fig_fT_quintom} we straightforwardly depict the evolutions of
the effective dark energy equation-of-state parameter $w_{DE}$ along with the redshift
$z$, for different choices of the model parameter $\alpha$. From the solid (where
$\alpha=1$) and dashed (where $\alpha=0.8$) curves, one can manifestly observe that
$w_{DE}$ is able to cross over the cosmological constant boundary. In both cases, the
universe evolves from a quintessence-like phase ($w_{DE}> -1$) into a phantom-like phase
($w_{DE}<-1$) with a smooth crossing behavior, and eventually approaches to the
exponential expanding phase. Moreover, the values of the redshift for the occurrence of
quintom scenario are determined as $z=0.70$ and $z=0.36$ for the solid and dashed curves,
respectively. It is interesting to notice that if $\alpha\leq0.5$ then the universe would
asymptotically approach to the $w_{DE}=-1$ expanding phase, without crossing behavior,
since in this case the exponential term in the expression \eqref{DEmodel_fT_quintom}
would have become secondary in $f(T)$ gravity.

Moreover, the evolutions of dimensionless density parameter $\Omega_i$ for radiation (the
dash-dotted curve), dust matter (the solid curve), and the effective dark energy (the
dashed line) are presented in the lower panel of Fig. \ref{Fig_fT_quintom}, respectively.
In the numerical calculation one sets the model parameter  $\alpha = 1$, and chooses the
initial conditions by fixing the present values of $\Omega_r^{(0)}$, $\Omega_m^{(0)}$ and
$\Omega_{DE}^{(0)}$. The numerical result explicitly shows that the model can
easily accommodate with the regular thermal expanding history, and hence it can very
efficiently be consistent with cosmological observations.

In the above we have analyzed the background dynamics of the $f(T)$ gravity described by
\eqref{DEmodel_fT_quintom} and we have showed explicitly that this type of model can
realize an effective dark energy equation-of-state parameter crossing $w_{DE}=-1$. By
numerically evolving the dimensionless density parameters we could roughly deduce that
the model could be consistent with cosmological observations.
However, it is important to quantitatively justify how well this $f(T)$ model can fit the
data. Thus, we would like to investigate the observational constraints on the model
parameter $\alpha$ and the present fractional dust matter density $\Omega_m^{(0)}$ based
on the $\chi^2$ method utilizing the experimental data. In particular, the SNIa data from 
the 
Supernova Cosmology Project (SCP) Union2 compilation
 \cite{Amanullah:2010vv}, BAO data from the Two-Degree
Field Galaxy Redshift Survey (2dFGRS), Sloan Digital Sky Survey (SDSS) data
release 7 \cite{Percival:2009xn}, as well as CMB data from WMAP7
\cite{Komatsu:2010fb}, are adopted.

\begin{figure}[tbph]
\includegraphics[scale=0.4]{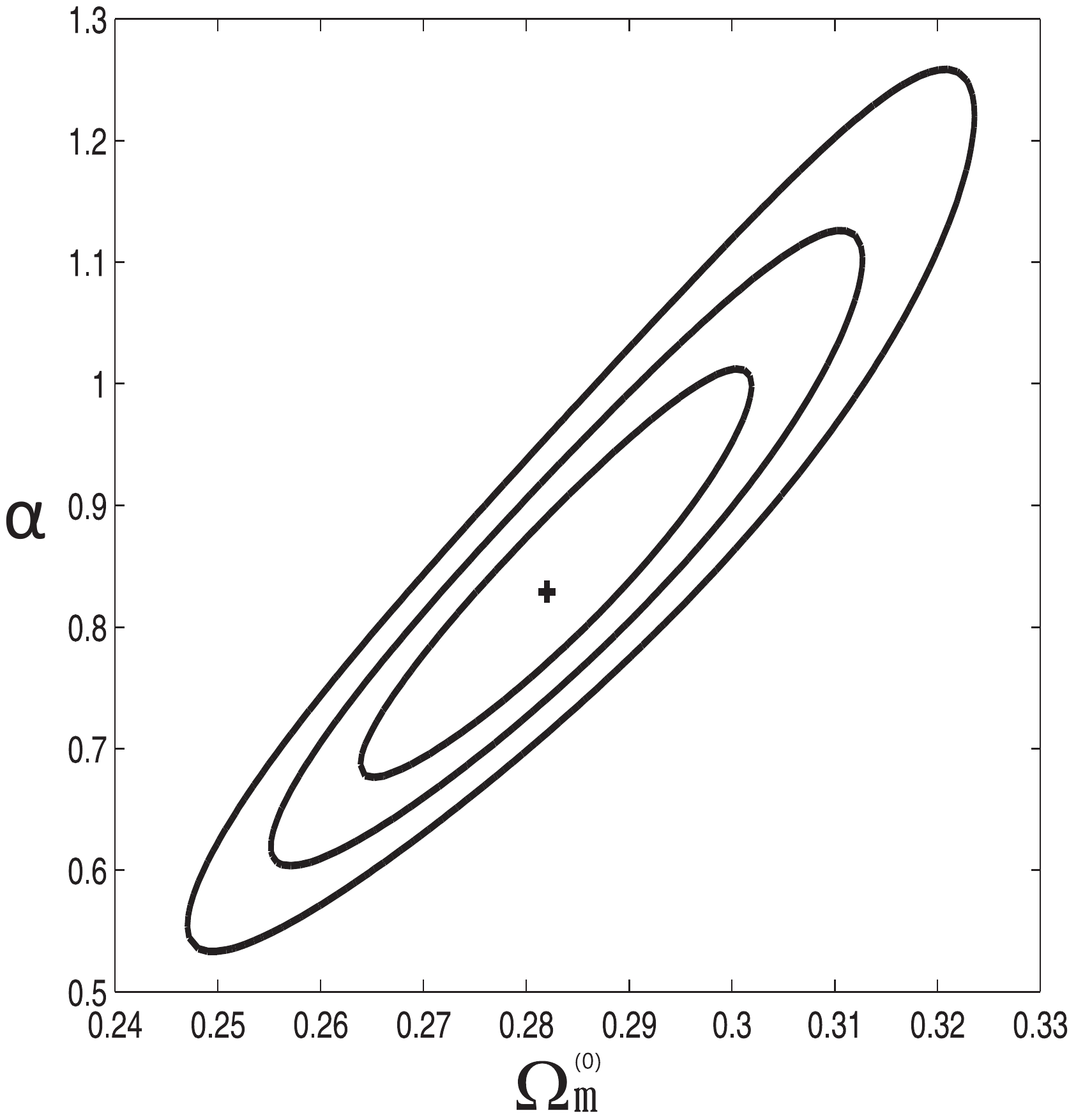}
\caption{\it Observational constraints on the model parameters in the ($\alpha$ -
$\Omega_{m}^{(0)}$) plane at $1\sigma$, $2\sigma$, and $3\sigma$ C.L.(from inside to
outside), respectively, based on the combined SNIa, BAO and CMB data, for the $f(T)$
gravity model described by \eqref{DEmodel_fT_quintom}. The best-fit point is denoted
by the plus symbol in the center region of the
plot. From  \cite{Bamba:2010wb}. }
\label{Fig_fitting_fT_quintom}
\end{figure}

The contours of observational bounds from the SNIa, BAO and CMB data at $1\sigma$,
$2\sigma$ and $3\sigma$ confidence level (C.L.) in the ($\alpha$ - $\Omega_{m}^{(0)}$)
plane upon the $f(T)$ model described by \eqref{DEmodel_fT_quintom} are shown in Fig.
\ref{Fig_fitting_fT_quintom}. According to the numerics, one concludes that at $2\sigma$
C.L., the model parameters are limited to be: $0.6<\alpha<1.13$ and
$0.255<\Omega_m^{(0)}<0.312$. This result is obviously in a good agreement with that
derived in the $\Lambda$CDM. Moreover, the best-fit values of the cosmological
parameters in the $f(T)$ model, which corresponds to the minimum of $\chi^2$, are given
by $\alpha=0.829$, $\Omega_m^{(0)} = 0.282$, and $h\equiv H_0/100/[\mathrm{km} ~
\mathrm{sec}^{-1} ~ \mathrm{Mpc}^{-1}] = 0.691$, with $\chi_{\rm min}^{2} = 544.56$.

In addition, one can compare this specific $f(T)$ model with the standard $\Lambda$CDM
one, by listing the best-fit values of all relevant model parameters as shown in Table
III.
\begin{table}[ht]
\begin{center}
\begin{tabular}
{ccccc}
\hline
Model  &  $u$  &  $\Omega_{\mathrm{m}}^{(0)}$  &  $h$  &  $\chi_{\mathrm{min}}^{2}$
\\[0mm]
\hline
$f(T)$  &  $0.829$  &   $0.282$  &  $0.691$  &  $544.56$  \\[1mm]
$\Lambda$CDM  &  &  $0.275$  &  $0.707$  &  $545.23$  \\[1mm]
\hline
\end{tabular}
\caption{The best-fit values of $u$, $\Omega_{\mathrm{m}}^{(0)}$, $h$ and
$\chi_{\mathrm{min}}^{2}$
for the $f(T)$  model \eqref{DEmodel_fT_quintom} and $\Lambda$CDM paradigm. }
\end{center}
\label{table_fT_quintom}
\end{table}
From this table, one can see that the minimum value of $\chi^2$ in $f(T)$ gravity is
slightly smaller than that of the $\Lambda$CDM, and hence in this regard the specific
$f(T)$ gravity model \eqref{DEmodel_fT_quintom} can explain cosmological observations
better than the $\Lambda$CDM paradigm. However, one ought to keep in mind that this is
mostly due to the presence of one extra model parameter $\alpha$.

We close this subsection by summarizing the analyses and the obtained results. We
started with a general discussion of cosmological applications of $f(T)$ gravity, and
then we devoted to a study of a specific model that nicely demonstrates the possibility of
realizing the cosmic acceleration, alternatively to a cosmological constant or
scalar-field models. After that, we performed a dynamical analysis on the phase space and
we proved that a quasi de Sitter expansion is an attractor solution at late times,
which explains why $f(T)$ can provide a good dynamical mechanism for obtaining the
present accelerating phase. Motivated by observational signals, we continued by
investigating an interesting realization of the phantom-divide crossing behavior in a
specific model of $f(T)$ gravity. Finally, we presented the recent observational
constraints on the model parameters.

\subsection{Inflation}

After having reviewed the cosmological implications of $f(T)$ gravity at late times, in
this subsection we devote ourselves to the study of early universe physics. It was first
obtained in  \cite{Ferraro:2006jd} that a so-called Born-Infeld modified teleparallel
gravity can solve the particle horizon problem in a spatially flat FRW universe, by
realizing an early time acceleration, without introducing any inflaton field. Thus, in the
present subsection we review such a remarkable inflationary solution derived in modified
teleparallel gravity, and then we comment on its generalization into $f(T)$ version.
%%%%%%%%%%%%%%%%%%%%%%%%%%%%%%%%%

\subsubsection{Inflation from modified teleparallel gravity}
%%%%%%%%%%%%%%%%%%%%%%%%%%%%%%%%%%%

The idea of Born-Infeld Lagrangian appeared initially from a modified field theory that
aimed to smooth possible singularities \cite{Born:1934gh}. Nowadays this type of
Lagrangian has been widely applied under the great developments of low-energy effective
descriptions of string theory \cite{Aharony:1999ti, Myers:1999ps}. In recent years,
models of gravitational Born-Infeld analogue were extensively studied in the literature
(for instance see \cite{Deser:1998rj, Ferraro:2006jd,
Banados:2010ix,Gullu:2010pc,Gullu:2010st,Delsate:2012ky,
Avelino:2012ue}). Briefly speaking, for a regular
Lagrangian density ${\cal L} = \sqrt{-g}L$, the Born-Infeld modification suggests:
$$
 {\cal L} \rightarrow {\cal L}_{\rm BI} = \sqrt{-g} \lambda \left[
\sqrt{1+\frac{2L}{\lambda}} -1 \right] ~,
$$
where $\lambda$ is associated with some unspecified UV physics. For example, for a
D-brane moving towards a warped throat in string theory, the parameter $\lambda$ is
related to the warp factor of the AdS throat.

Inspired by the above scenario, an interesting modification of teleparallel gravity was
proposed in \cite{Ferraro:2006jd}, in which the model is of the Born-Infeld type
\begin{equation}\label{Ferraro_gravedadmodificada}
 \mathcal{L}_{\textbf{{BI}}}=\frac{\lambda}{16\pi G} ~e~
 \left[ \sqrt{ 1 +\frac{2 S_\mu^{\ \ \nu\rho} T^\mu_{\ \ \nu\rho}}{\lambda}} -1 \right] ~.
\end{equation}
Note that, if one performs a local Lorentz transformation, a boundary term arises from
$S_\rho^{\ \ \mu\nu}\, T^\rho_{\ \ \mu\nu}$ and is trapped inside the square root.
Therefore, it renders the above model sensitive to local Lorentz transformations.
Moreover, similar to the regular Born-Infeld field Lagrangian, the dynamics governed by
this model would approach to those of standard Einstein equations when $S_\rho^{\ \
\mu\nu}\, T^\rho_{\ \
\mu\nu} \ll \lambda$.

In order to study the cosmological dynamics induced by this model, one can take the
ansatz for the vierbein as
\begin{equation}\label{Ferraro_tetrada}
 e^A_\mu =\emph{diag}(N(t), a(t), a(t), a(t)) ~,
\end{equation}
where $N(t)$ is the lapse function. Under this ansatz the metric of the background
space-time takes the form of
\begin{equation}\label{Ferraro_metrica friedmann}
 g_{\mu\nu} = \emph{diag}( N^2(t), - a(t)^2, - a(t)^2, - a(t)^2) ~,
\end{equation}
which coincides with the metric of a spatially flat FRW universe. Note that one can
absorb the lapse function by rescaling the time scale. At the moment we keep this
function undetermined and study the background equations of motion.

We consider the background matter component to be a homogeneous and isotropic fluid,
and hence one acquires $T^{\rho}
_{~~ \sigma} = \emph{diag}( \rho, -p, -p, -p)$ in the comoving frame. Then the highly
symmetric background ansatz eventually leads to only two independent background equations
of motion, which are a first order equation:
\begin{equation}
\label{Ferraro_n(t)=1}
 \Big(1-\frac{12\, H^2}{N^2\lambda}\Big)^{-\frac{1}{2}} -1
 = \frac{16 \pi G}{\lambda}N^2 \rho ~,
\end{equation}
which results from varying with respect to $e^0_0$ ($H(t)=\dot{a}(t)/a(t)$ is the Hubble
parameter), and a second order one:
\begin{equation}
\label{Ferraro_ecuacion para a(t)}
 \Big(\frac{16 H^2}{N^2\lambda} + \frac{4 H^2}{N^2\lambda} q-1 \Big)
 \Big(1-\frac{12 H^2}{N^2\lambda}\Big)^{-\frac{3}{2}}+1
 = \frac{16 \pi G}{\lambda} p ~,
\end{equation}
which results from varying with respect to $e^A_\sigma$   ($q =
-\ddot{a}\, a/\dot{a}^2$ is the deceleration parameter). The above two equations can also
be derived by varying the Lagrangian \eqref{Ferraro_gravedadmodificada} with respect to
the lapse function $N(t)$ and the scale factor $a(t)$. Note that $S_\mu^{\ \ \nu\rho}\,
T^\mu_{\ \ \nu\rho}=-6\, H(t)^2/N(t)^2$, and thus $\lambda$ in
(\ref{Ferraro_gravedadmodificada}) will prevent the Hubble parameter from becoming
infinite. Note also that Eq. (\ref{Ferraro_n(t)=1}) is not a dynamical equation for
$N(t)$, but a constraint for $a(t)$, and therefore one has the freedom to fix $N(t)$,
namely to set $N(t)=1$.

The continuity equation for the matter component can be guaranteed by differentiating Eq.
(\ref{Ferraro_n(t)=1}) with respect to $t$ and combining it with Eq.
(\ref{Ferraro_ecuacion para a(t)}),
which then yields
\begin{equation}
 \frac{d}{dt}\, (\rho\, a^3)\ =\ -p\, \frac{d}{dt} a^3.
\end{equation}
For example, for a perfect fluid with the background equation of state being $\omega =
p/\rho$, one obtains
\begin{equation}\label{Ferraro_conservacion}
 a^{3(1+\omega)}\, \rho\ =\ constant \ =\ a_0^{3(1+\omega)}\, \rho_0 ~,
\end{equation}
where $a_0$ and $\rho_0$ indicate the present-day values.

Combining Eqs.  (\ref{Ferraro_n(t)=1}) and (\ref{Ferraro_ecuacion para a(t)}), one can
derive
\begin{equation}\label{Ferraro_desaceleracion}
 1 + q
 = \frac{3}{2}\frac{\left( 1+\omega\right)}{\left(1+\frac{16\pi G}{\lambda}\rho \right)
 \left(1+\frac{8\pi G}{\lambda}\rho\right)} ~.
\end{equation}
In the limit of GR (i.e., $\lambda\rightarrow\infty$) an accelerated expansion ($q < 0$)
is only possible if $\omega < -1/3$ (negative pressure). However, it is interesting to
observe that in Born-Infeld modified teleparallelism an accelerated expansion can be
realized without resorting to negative pressure, since a large energy density $\rho$ is
sufficient:
\begin{equation}\label{Ferraro_rhoinflat}
 \frac{32\pi G}{\lambda}\rho\ >\ -3+\sqrt{13+12\, \omega}  ~,
\end{equation}
which can be achieved in early universe. Note that for $\rho\rightarrow\infty$ in
(\ref{Ferraro_desaceleracion}) one gets $q\rightarrow -1$, and the expansion becomes
exponential.

If one takes into account the spatial curvature term, Eq.  (\ref{Ferraro_n(t)=1}) would
define an effective critical density $\rho_c$ making the universe spatially flat. Thus,
it is important to measure the fractional energy density contributed from each component
by introducing $\Omega_i\ =\
\rho_i/\rho_c$. To combine Eqs.  (\ref{Ferraro_n(t)=1}) and Eq.
(\ref{Ferraro_conservacion}), one
obtains
\begin{equation}\label{Ferraro_valores iniciales}
 \Bigg(1-\frac{12\dot{a}^{2}}{\lambda a^{2}}\Bigg)^{-\frac{1}{2}}-1
 = \frac{16\, \pi\, G}{\lambda} \sum_i{\rho_0}_i
\left(\frac{a}{a_0}\right)^{-3(1+\omega_i)} ~.
\end{equation}
This equation can be also reformulated as
\begin{equation}\label{Ferraro_ecdiferencial}
 \dot{\texttt{x}}^2 + \mathcal{V}(\texttt{x})
 =0 ~,~~~ \texttt{x}=\frac{a}{a_0} ~,
\end{equation}
with $\mathcal{V}(\texttt{x})$ being an effective potential given by
\begin{equation}\label{Ferraro_potencial}
 \mathcal{V}(\texttt{x}) = \frac{\lambda}{12} \texttt{x}^2
 \Big[\big(1 + \beta_0 \sum_i{\Omega_0}_i \texttt{x}^{-3(1+\omega_i)}\big)^{-2}-1\Big] ~,
\end{equation}
where $\beta_0\equiv (1-12 H_0^2/\lambda)^{-1/2}-1$ is a constant. The potential is
always negative and vanishes with null derivative when $a\rightarrow 0$, for any value of
$\omega$. For example, if $\omega > -1/3$, the potential asymptotically approaches zero
when \texttt{x} goes to infinity. Instead, if $\omega < -1/3$, then $\mathcal{V}$ becomes
a decreasing function. Moreover, if $\omega > -1$ then the initial behavior is in general
$a(t)\propto \exp[(\lambda/12)^{1/2} t]$. Therefore, the Hubble parameter is equal to the
maximum value $H_{\emph{max}}=(\lambda/12)^{1/2}$
at the early stage, and Eq. (\ref{Ferraro_ecdiferencial}) gives rise to the following
solution:
\begin{equation}\label{Ferraro_hubble}
 H(z)^2 = H_{\emph{max}}^2 [1
 -(1 + \beta_0 \sum_i{\Omega_0}_i (1+z)^{3(1+\omega_i)})^{-2}] ~,
\end{equation}
where $z=a_0/a(t)-1$ is the redshift.

As an example we consider a single component with $\Omega=1$. Then, it is easy to find
that Eq. (\ref{Ferraro_ecdiferencial}) leads to the following compact relation:
\begin{equation}\label{Ferraro_resultado}
 \ln \left[2 (1+ v) +2 \sqrt{ v (2+ v)}\right] -\sqrt{v^{-1} (2+v)} = \mathcal{T} ~,
\end{equation}
where we have defined $v \equiv \beta_0\, (a/a_0)^{-3 (1+\omega)}$ and $\mathcal{T}
\equiv
-3 (1+\omega)\ H_{\emph{max}}\ t$.

In Fig. 10
%\ref{Ferraro_factor22}
we depict the dimensionless scale factor $a(t)/a_{0}$ as a
function of $H_{0}t$ for several values of $\alpha=H_{max}/H_0$, as indicated by Eq.
(\ref{Ferraro_resultado}) with $\omega=1/3$. The standard ($a/a_{0}=(2H_{0}t)^{1/2}$)
behavior is plotted as a reference (dashed) curve. It is obvious to see that modified
teleparallelism naturally gives rise to an exponential expansion at early times, and then
the universe can smoothly exit into a radiation dominated one. Such a cosmological
solution can easily satisfy the observational constraints, such as the BBN bound.
\begin{figure}[ht]
\centering
\includegraphics[scale=.29]{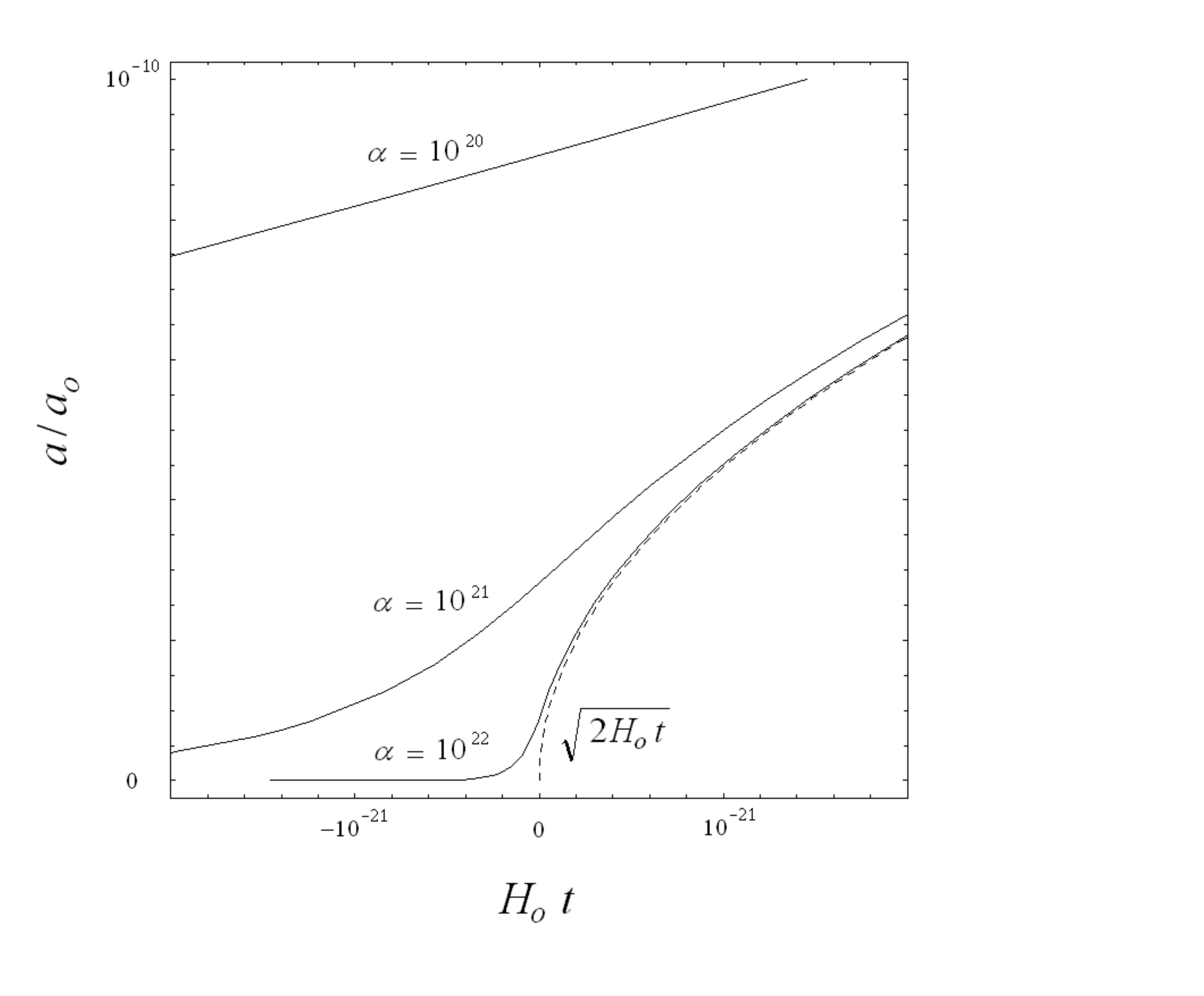}
\label{Ferraro_factor22}
\caption[]{\it Scale factor as a function of the cosmological time, for $\omega=1/3$ and
different values of $\alpha=H_{max}/H_0$. The dashed line represents the solution of
General Relativity. From
 \cite{Ferraro:2006jd}. }
\end{figure}
The main feature of the scale factor evolution is its asymptotic exponential behavior for
any value of $\omega$. This means that $H(z)$ becomes a constant when $z$ goes to
infinity. This feature implies that the particle horizon radius
$\sigma=a_{0}\int_{0}^{a_{0}}(a\dot{a})^{-1} da$ diverges. Hence, the whole space-time
ends up being causally connected, in agreement with the isotropy of the CMB background.
This fact appears as an essential property of modified teleparallelism, which does not
require any special assumption for the sources of the gravitational field, as for example
an inflaton field.

Now, the Standard Big Bang cosmology successfully explains the relative abundances of
light elements. Therefore, a modified gravity theory cannot noticeably change the standard
evolution of the universe from the epoch of nucleosynthesis. This means that $H(z)$ at
$z_{nuc}\sim 10^9-10^{10}$ should not appreciably differ from its standard value. Fig. 11
%\ref{Ferraro_corrimiento}
shows how the Hubble parameter move away from the case of General
Relativity, represented by the dashed line, to approach the value $H_{max}$ as the
redshift increases. The redshift $z_t$ characterizing the transition between both
behaviors can be defined as the value of $z$ at which the asymptotic lines
intersect. Since in GR with only one component there is
$\log(H/H_0)=(3/2)(1+\omega)\log(1+z)$, one obtains
%\begin{equation}
 $$ (1 + z_t)^{3(1+\omega)/2} = \frac{H_{max}}{H_0}~. $$
%\end{equation}
The condition $z_t >> z_{nuc}$ implies a lower bound for $H_{max}$. For a radiation
dominated universe ($\omega=1/3$) one obtains that $H_{max}/H_0 >> 10^{18}$.
\begin{figure}[ht]
\centering
\includegraphics[scale=.27]{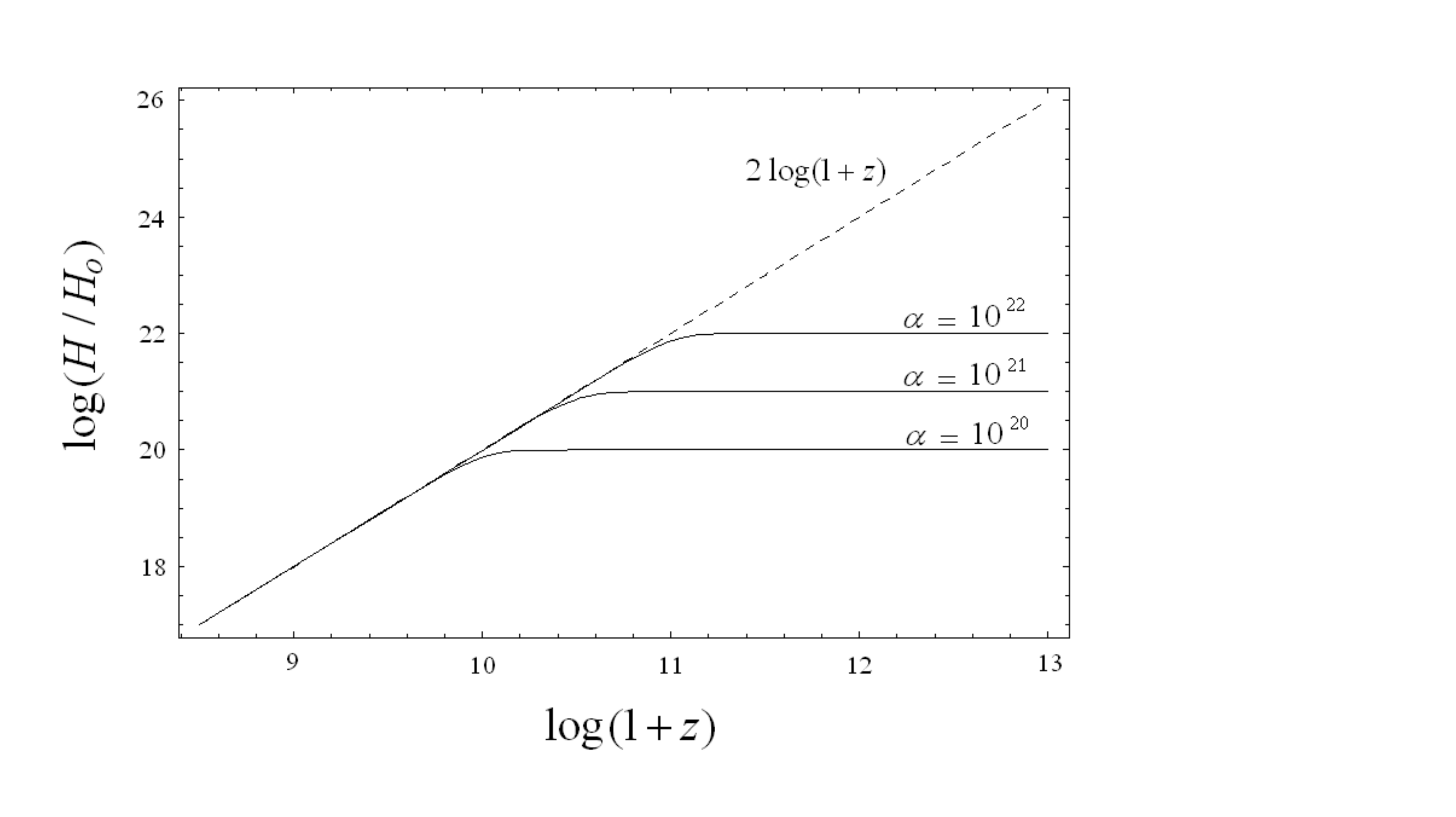}
\label{Ferraro_corrimiento}
\caption{\it Hubble parameter as a function of the redshift for $\omega=1/3$ and different
values of $\alpha=H_{max}/H_0$. The dashed line represents the solution of General
Relativity. From  \cite{Ferraro:2006jd}. }
\end{figure}

We mention that the realization of inflationary cosmology without any inflaton field does
not only exist in modified teleparallel gravity, but it also appears in other gravity
theories such as the Einstein-Cartan framework \cite{Gasperini:1986mv, Demianski:1987hq}.
However, one should be aware that the generation of primordial perturbations in these
theories are unclear. This is an important topic that deserves a close study.

\subsubsection{Inflation from $f(T)$ gravity and primordial perturbations}

In the previous paragraph we investigated the realization of inflationary cosmology by
virtue of modified teleparallel gravity. In the present paragraph we desire to study
inflation realization in the context of $f(T)$ cosmology. We point out that, for the case
of $f(T)$ gravity, since the torsion scalar can be explicitly expressed by the Hubble
parameter via $T = -6 H^2$, it is completely possible to reconstruct a viable model of
$f(T)$ gravity that can nicely coincide with the cosmological evolution as observed.
Therefore, a period of inflation at early times can also be allowed in this method. The
reconstruction of a suitable $f(T)$ theory, that can describe the early universe
inflation as well as late-time acceleration, was studied in detail in \cite{Bamba:2012vg}
(see also
\cite{Nashed:2014vsa,Hanafy:2014bsa,Hanafy:2014ica,Nashed:2014lva,Gudekli:2015uma,
Rezazadeh:2015dza,Hanafy:2015lda,Bamba:2016wjm}).
However, the authors of that paper did not address the cosmic perturbations and hence it
remains unclear how these type of gravity models can explain the CMB and LSS observations
in a consistent way.

Let us briefly review inflationary perturbations obtained in $f(T)$ gravity, as they were
studied in \cite{Wu:2011kh}. We recall that our starting point is the action $ S = \int
d^4x |e|
f(T)/(16\pi G) $. For convenience in studying the quadratic perturbed action of this
theory, in this paragraph we temporarily impose the convention $8\pi G=1$. One
usually takes the
conformal
transformation through the following formulation
\begin{align}
 f(T) = {\cal F} T - 2V ~,~~ {\cal F} \equiv \frac{df}{dT} ~,~~ V = \frac{{\cal
F}T-f}{2}~.
\end{align}
Accordingly, one can define
\begin{align}
 \hat{e}_\mu^A = \sqrt{F} e_\mu^A \equiv \Omega e_\mu^A~,
\end{align}
and the torsion scalar then is transformed as
\begin{align}
 T = \Omega^2 [ \hat{T} -4\hat{\partial}^\mu\omega \hat{T}^\rho_{~\rho\mu}
+6(\hat{\nabla}\omega)^2
]~,
\end{align}
with $\hat{\partial}^\mu\omega\equiv \hat{\partial}^\mu\Omega/\Omega$ being introduced.
Note that we use the hat $\hat{}$ in order to denote the Jordan frame. As a result,
the original $f(T)$ action can be rewritten as
\begin{align}
 S = \int d^4x \frac{|e|}{2} \Big[ \hat{T}
-\frac{4}{\sqrt{6}}\hat{\partial}^\mu\varphi\hat{T}^\rho_
{~\rho\mu} +(\hat{\partial}\varphi)^2 - 2U(\varphi) \Big] ~,
\end{align}
where we have introduced a canonical scalar field $d\varphi = \sqrt{6} d\omega = \sqrt{6}
dF/2F$ and its potential $U=V/F^2$.

From the above reformulated action, and as we discussed after \eqref{action_fT_conformal}
too, it is obvious to observe that there exists an ``additional''
scalar-torsion coupling term, $\hat{\partial}^\mu\varphi\hat{T}^\rho_{~\rho\mu}$, which
manifestly breaks the local Lorentz invariance and leads to extra degrees of freedom. The
corresponding field equation can be obtained by varying the action with respect to the
vierbein $\hat{e}_\mu^A$, where one can formulate the covariant representation as
\begin{align}\label{GTH_sec:inflation}
 G_{\mu\nu} = T^{(\varphi)}_{\mu\nu} + H_{\mu\nu} ~,
\end{align}
where
\begin{align}
 T^{(\varphi)}_{\mu\nu} = \hat{\partial}_\mu\varphi\hat{\partial}_\nu\varphi -
\frac{\hat{g}_{\mu\nu}}{2}(\hat{\nabla}\varphi)^2 +\hat{g}_{\mu\nu} U~
\end{align}
is the energy-momentum tensor of the scalar field $\varphi$, while
\begin{align}
 H_{\mu\nu} =& \frac{2}{\sqrt{6}} \Big( \hat{g}_{\mu\nu} \hat{\partial}^\lambda\varphi
\hat{T}^\rho_
{~\rho\lambda} - \hat{\partial}^\lambda\varphi \hat{T}_{\nu\mu\lambda} -2
\hat{\partial}_\nu\varphi
\hat{T}^\rho_{~\rho\mu} \Big) \nonumber\\
 &- \frac{2}{\sqrt{6}} \hat{e}^{-1} \hat{g}_{\nu\lambda} \hat{e}^A_\mu
\hat{\partial}_\alpha \big[ \hat{e}(\hat{\partial}^\lambda\varphi e^\alpha_A
-\hat{\partial}^\alpha\varphi e^\lambda_A) \big] ~
\end{align}
is derived from the variation of the scalar-torsion coupling. Additionally, the variation
with respect to the scalar field $\varphi$ yields the equation of motion as
\begin{eqnarray}
 \Box\varphi + \frac{\partial U}{\partial\varphi} +\frac{2}{\sqrt{6}} \hat{\nabla}^\mu
\hat{T}^\rho_{~\rho\mu} =0~.
\end{eqnarray}

Moreover it is interesting to notice that $G_{\mu\nu}$ and $T^{(\varphi)}_{\mu\nu}$ are
symmetric tensors, but $H_{\mu\nu}$ in general is not necessarily a symmetric tensor.
Thus, the field equation \eqref{GTH_sec:inflation} can be further split into
\begin{align}
G_{\mu\nu} = T^{\varphi}_{\mu\nu} + H_{(\mu\nu)} ~,~~
H_{[\mu\nu]} = 0 ~, \label{extra_sec:inflation}
\end{align}
where $H_{\mu\nu}=H_{(\mu\nu)}+H_{[\mu\nu]}$. The second equation of
(\ref{extra_sec:inflation})
determines that the field equation \eqref{GTH_sec:inflation} contains $16$ component
equations rather than $10$, and hence, it indicates that the $f(T)$ theory may involve
more degrees of freedom than the linear version of teleparallel gravity. The analysis of
degrees of freedom in this theory was presented in detail in subsection \ref{DOFinFT}.

The computation of the quadratic action is commonly used in ADM formalism for the
perturbation
theory of inflationary cosmology. The ADM decomposition for torsion gravity theories can
be found
in Refs.  \cite{Blagojevic:2002du, Mielke:1992te, Maluf:1994ji}. It was found in
 \cite{Wu:2011kh}
that the extra degrees of freedom do not affect the main part of the quadratic action for
scalar
perturbations during inflation. Here we skip the lengthy calculation but
straightforwardly
write
down the quadratic actions for scalar perturbations as follows:
\begin{align}
 S_\zeta = \frac{1}{2} \int dt d^3x a^3 \left[ {\cal P}_\zeta \dot\zeta^2
-\frac{c_\zeta^2}{a^2}(\partial_i\zeta)^2 \right]~,
\end{align}
where
\begin{align}
 {\cal P}_\zeta &= 3 - \frac{U}{(H-\dot\varphi/\sqrt{6})^2}~,\\
 c_\zeta^2 &= \frac{1}{a}\frac{d}{dt}\left( \frac{a}{H-\dot\varphi/\sqrt{6}}\right) -1~.
\end{align}
It is easy to see that the coefficient ${\cal P}_\zeta$ characterizes the positivity of
the kinetic energy of scalar modes, while $c_\zeta^2$ is the square of the speed of sound
parameter which describes the gradient propagations of scalar perturbations. In order to
ensure that the model is stable against ghost and gradient instability, one needs to
impose the conditions: ${\cal
P}_\zeta
> 0$ and $c_\zeta^2 > 0$.

\subsection{Cosmic bounce}

While inflation is often considered as a crucial part of the cosmic evolution, this
``standard model'' of the universe is known to suffer from several conceptual
challenges. For instance, inflationary cosmology does not resolve the problem of the
initial singularity inherited from the hot Big Bang  \cite{Borde:1993xh}. Moreover, it is
known that the Planck-mass suppressed corrections to the inflaton potential generally
lead to masses of the order of the Hubble scale, and then spoil the slow roll
conditions rendering a sustained inflationary stage impossible  \cite{Copeland:1994vg}.
This issue could be even worse if the field variation of the inflaton is super-Planckian
\cite{Lyth:1996im}. From the perspective of perturbation theory, if we trace
backwards the cosmological perturbations observed today, their length scales could
go beyond the Planck length at the onset of inflation  \cite{Brandenberger:1999sw,
Martin:2000xs}. Additionally, in order to study quantum field theory during inflation, it
is inevitably necessary to systematically study the nonlinear corrections of field
fluctuations that are on one side not ultraviolet (UV) complete, and on the other side
yield observably large infrared (IR) effects that were not detected in experiments
\cite{Tsamis:1992sx, Mukhanov:1996ak}. Therefore, it is worth searching for possibly
extended or alternative paradigms that might not only be as successful as inflation in
phenomenologically explaining the CMB and LSS of our universe, but that can also resolve
or at least circumvent some conceptual issues mentioned above.

Non-singular bouncing cosmologies can resolve the initial singularity problem of the
inflationary $\Lambda$CDM model and hence have attracted a lot of attention in the
literature \cite{Mukhanov:1991zn, Brandenberger:1993ef}. Such scenarios have been
constructed through various approaches to modified gravity, such as the Pre-Big Bang
 \cite{Veneziano:1991ek, Gasperini:1992em} and Ekpyrotic  \cite{Khoury:2001wf,
Khoury:2001bz} models, gravitational actions with higher order corrections
 \cite{Brustein:1997cv}, the gravitational Lagrangian modified as in Ho\u{r}ava
gravity  \cite{Calcagni:2009ar, Kiritsis:2009sh, Brandenberger:2009yt}, non-relativistic
gravitational action \cite{Cai:2009in}, Lagrange-multiplier gravity  \cite{Cai:2010zma,
Cai:2011bs}, nonlinear massive gravity  \cite{Cai:2012ag}, non-local gravity
 \cite{Biswas:2005qr, Biswas:2006bs} and the loop quantum cosmology
 \cite{Bojowald:2008zzb,
Ashtekar:2008zu, Singh:2006im, Cai:2014zga}. A non-singular bounce solution can also be
achieved by making use of matter fields with the NEC violation, such as in the quintom
bounce  \cite{Cai:2006dm,Cai:2007qw,Cai:2007zv},
 the Lee-Wick bounce  \cite{Cai:2008qw,Bhattacharya:2013ut}, the ghost condensate bounce
 \cite{Buchbinder:2007ad,Creminelli:2007aq, Lin:2010pf}, the braneworld bounce
 \cite{Kehagias:1999vr,
Shtanov:2002mb, Saridakis:2007cf}, the Galileon bounce  \cite{Qiu:2011cy, Easson:2011zy},
and the bounce models with Horndeski operators  \cite{Cai:2012va, Cai:2013vm, Cai:2013kja,
Koehn:2013upa, Battarra:2014tga}. A non-singular bounce may also be achieved in a
universe with non-flat spatial geometry  \cite{Solomons:2001ef, Martin:2003sf}. It was
found that a cosmology of non-singular bounce could explain the combined constraints of
CMB
observations better than that done by pure inflation models \cite{Cai:2008qb, Liu:2010fm,
Xia:2014tda}. We refer to \cite{Novello:2008ra, Lehners:2008vx, Cai:2014bea,
Battefeld:2014uga} for recent reviews of various bouncing cosmologies.

One interesting feature of $f(T)$ gravity is that the null energy condition could be
effectively violated. Accompanied with this feature, it is not surprising to look for a
series of nontrivial cosmological solutions that may resolve the initial singularity
problem \cite{Cai:2011tc}. The avoidance of the Big Bang singularity by using torsion
arisen from a cosmic spinor field can be found in
\cite{ArmendarizPicon:2003qk,Alexander:2008vt, Alexander:2014eva,Alexander:2014uaa}.
Moreover, it has been known for many years that the coupling between the torsion tensor
and the cosmic spinor field can lead to interesting gravitational repulsion and thus
avoid curvature singularities by violating the energy condition
\cite{Kopczynski:1972,Tafel:1973,
Hehl:1974cn,Kuchowicz:1978,Gasperini:1998eb,Poplawski:2011jz,Magueijo:2012ug}. In this
subsection, we are interested in searching for a non-singular bouncing solution in the
early universe within the frame of $f(T)$ gravity.

%%%%%%%%%%%%%%%%%%%%%%%%
\subsubsection{Background solutions}
%%%%%%%%%%%%%%%%%%%%%%%%%%%%

We are interested in examining how cosmological scenarios governed by $f(T)$ gravity can
give rise to a non-singular bounce following  \cite{Cai:2011tc} (see also 
\cite{Bamba:2012ka,deHaro:2012zt,Amoros:2013nxa,Odintsov:2015uca,Haro:2014wha,
Astashenok:2013kka}).
There are two distinct points in such an investigation. The first is to examine whether
the background evolution allows for bouncing solutions. If this is indeed possible, then
the second point is to examine the evolution of perturbations through the bounce. The
first task is the subject of this paragraph, while the second one will be investigated in
the next paragraph. Finally, we mention that in order to be closer to the convention of
the literature on this field, we use  $G$ as the gravitational constant, and the
connection to other works of the literature is obtained through the known relations
$8\pi G=\frac{1}{M_P^2}=\kappa^2$.

Whether a universe is expanding or contracting merely depends on the positivity of the
Hubble parameter. In the contracting phase that exists prior to the bounce, the Hubble
parameter $H$ is negative, while in the expanding one that exists after it we have $H>0$.
By making use of the continuity equations it follows that at the bounce point $H=0$.
Finally, it is easy to see that throughout this transition $\dot H> 0$. On the other
hand,
for the transition from expansion to contraction, that is for the cosmological
turnaround,
we have $H>0$ before and $H<0$ after, while exactly at the turnaround point we have
$H=0$.
Throughout this transition $\dot{H} < 0 $.

Having in mind the above general requirements for a cosmic bounce, and observing the
background Friedmann equations \eqref{background11}, \eqref{background22}, we can find
that such a behavior can be easily obtained in principle in the context of $f(T)$
cosmology. In particular, one can start with a specific, desirable form of the bouncing
scale factor $a(t)$, and thus can derive explicitly $H(t)$. Concerning the matter fluid
content of the universe, with an equation-of-state parameter $w_{m}$, its evolution
equation \eqref{eom_rho_m} straightforwardly gives the solution $\rho_{m}(t)$ since
$a(t)$ is already known. Then, we can insert these relations into \eqref{background11},
and determine the form of $f(T)$, which generates a non-singular bouncing solution. In
principle the above procedure can always be done numerically and exactly. However, in
order to better understand the cosmological implications, we would like to present a
specific example of a non-singular bounce that allows the analytical calculation.

To be explicit, we consider a bouncing solution with the scale factor of the universe
evolving as
\begin{equation} \label{at_fT_bounce}
 a(t)=a_{B}\left( 1 +\frac{3}{2}\sigma t^{2} \right)^{1/3} ~,
\end{equation}%
where $a_{B}$ is the scale factor at the bouncing point, and $\sigma$ is a positive
parameter which describes how fast the bounce takes place. Such an ansatz presents the
non-singular bouncing behavior, corresponding to matter-dominated contraction and
expansion. In addition, it has the advantage of allowing for semi-analytic solutions. In
this ansatz the cosmic time $t$ varies between $-\infty $ and $+\infty $, with $t=0$ the
bounce point. In the present subsection we normalize the bounce scale factor $a_{B}$ to
unity.

Using the definition of the Hubble parameter, one can directly derive
\begin{equation} \label{sol_HT_fT_bounce}
 H(t)=\frac{\sigma t}{(1+3\sigma t^{2}/2)} ~,~~
 T(t)=-\frac{6\sigma^{2}t^{2}}{ \left( 1+\frac{3}{2}\sigma t^{2}\right) ^{2}} ~.
\end{equation}
Therefore, provided $-\sqrt{{2}/{3\sigma }}\leqslant t\leqslant \sqrt{{2}/{3\sigma}}$,
the
inversion of the above expression yields the following relation
\begin{equation}\label{tinvT_fT_bounce}
 t(T) = \pm \left( -\frac{4}{3T}-\frac{2}{3\sigma } +\frac{4\sqrt{T\sigma^{3}
 +\sigma^{4}}}{3T \sigma^{2}} \right)^{1/2} ~,
\end{equation}%
where we have kept the solution pair that gives the correct ($t=0$ at $T=0$) behavior.
Notice that when $t>\sqrt{{2}/{3\sigma}}$ and $t < -\sqrt{{2}/{3\sigma}}$, we have
assumed that the usual Einstein gravity, i.e. the TEGR, is the prevailing framework, thus
negating the need to pursue an $f\left( T\right)$ action in that region. Furthermore, we
assume the matter content of the universe to be dust, with an equation-of-state
parameter $w_{m}\approx 0$. Inserting this matter fluid into the continuity equation, one
can easily arrive at the usual dust evolution, namely $\rho_{m} = \rho_{mB}
a_{B}^{3}/a^{3}$, with $\rho_{mB}$ its value at the bouncing point.

Inserting the above expressions into (\ref{background11}) we obtain a differential
equation for the
reduced form $F(t)$, which can be easily solved analytically as
\begin{eqnarray}\label{ft_fT_bounce}
 &&\!\!\!\!\!\!\!\!\!\!\!\!\!\!\!\!\!\!\!
 F(t) = \frac{4t}{(2+3\sigma t^{2})M_{P}^2}\times \bigg[\frac{\rho _{mB}}{t}
 +\frac{ 6tM_{P}^2\sigma^{2}}{2+3t^{2}\sigma }\nonumber\\
 &&\ \ \ \ \ \ \ \  \ \ \  \ \ \
 +\sqrt{6\sigma }\rho _{mB} \,
 \text{ArcTan} \left( \sqrt{\frac{3\sigma }{2}} t\right) \bigg] ~.
\end{eqnarray}
We mention that in the calculation we have set the integration constant to be zero, in
order for the solution to be consistent with the Friedmann equation. Thus, the
corresponding $f(T)$ expression that generates a bouncing scale factor of the form
\eqref{at_fT_bounce} arises from expression \eqref{ft_fT_bounce} with the insertion of
the
$t(T)$ relation from \eqref{tinvT_fT_bounce}. Note that the solution \eqref{ft_fT_bounce}
is an even function of $t$, and thus the $\pm$ solutions of \eqref{tinvT_fT_bounce}
correspond to the contraction and expansion phase respectively. Obviously, they give the
same form of $f(T)$.

\begin{figure}[tbph]
\includegraphics[scale=0.3]{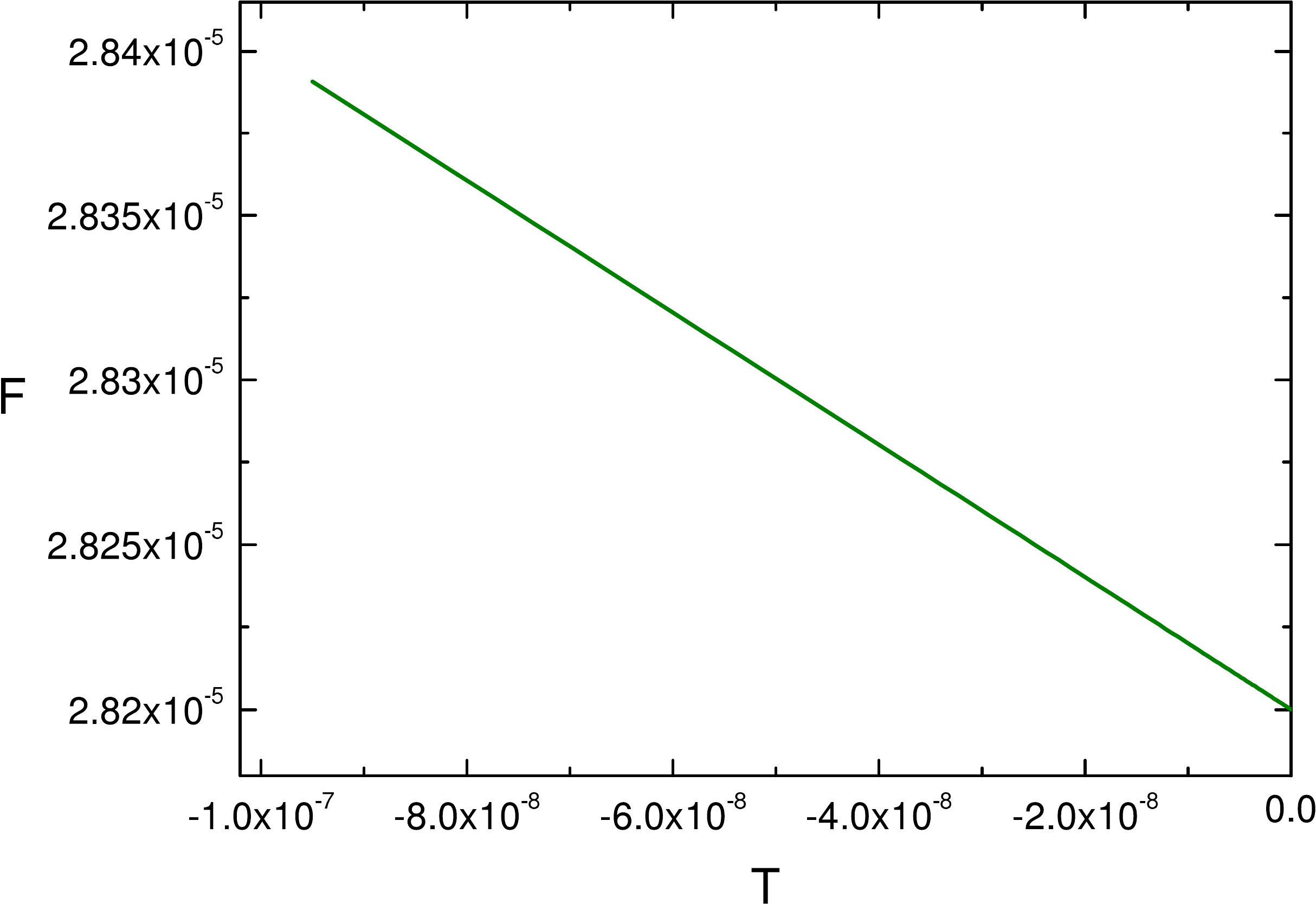}
\caption{\it Numerical result on the reduced form of $F$ as a function of the torsion
scalar $T$ in the dust-dominated non-singular bounce cosmology. The model parameters are
chosen as: $\sigma =
7 \times
10^{-6} M_{P}^2$ and $\rho _{mB} = 1.41 \times 10^{-5}M_{P}^{4}$, respectively. The graph
is plotted in units of $M_{P}$. From  \cite{Cai:2011tc}. }
\label{Fig-fofT_fT_bounce}
\end{figure}

In order to present the above process more clearly, in Fig. \ref{Fig-fofT_fT_bounce} we
numerically depict the reduced form of $F(T)$ that generates the dust-dominated bouncing
solution as desired. We particularly choose the parameters as follows:
$a_B=1$, $\sigma = 7 \times 10^{-6} M_{P}^2$, and $\rho_{mB} = 1.41 \times
10^{-5}M_{P}^{4}$. We note that the value of $\sigma$ mainly relies on the amplitude of
the CMB spectrum, and that of $\rho_{mB}$ depends on how fast the standard Einstein
gravity is recovered in $f(T)$ gravity.

\begin{figure}[tbph]
\includegraphics[scale=0.3]{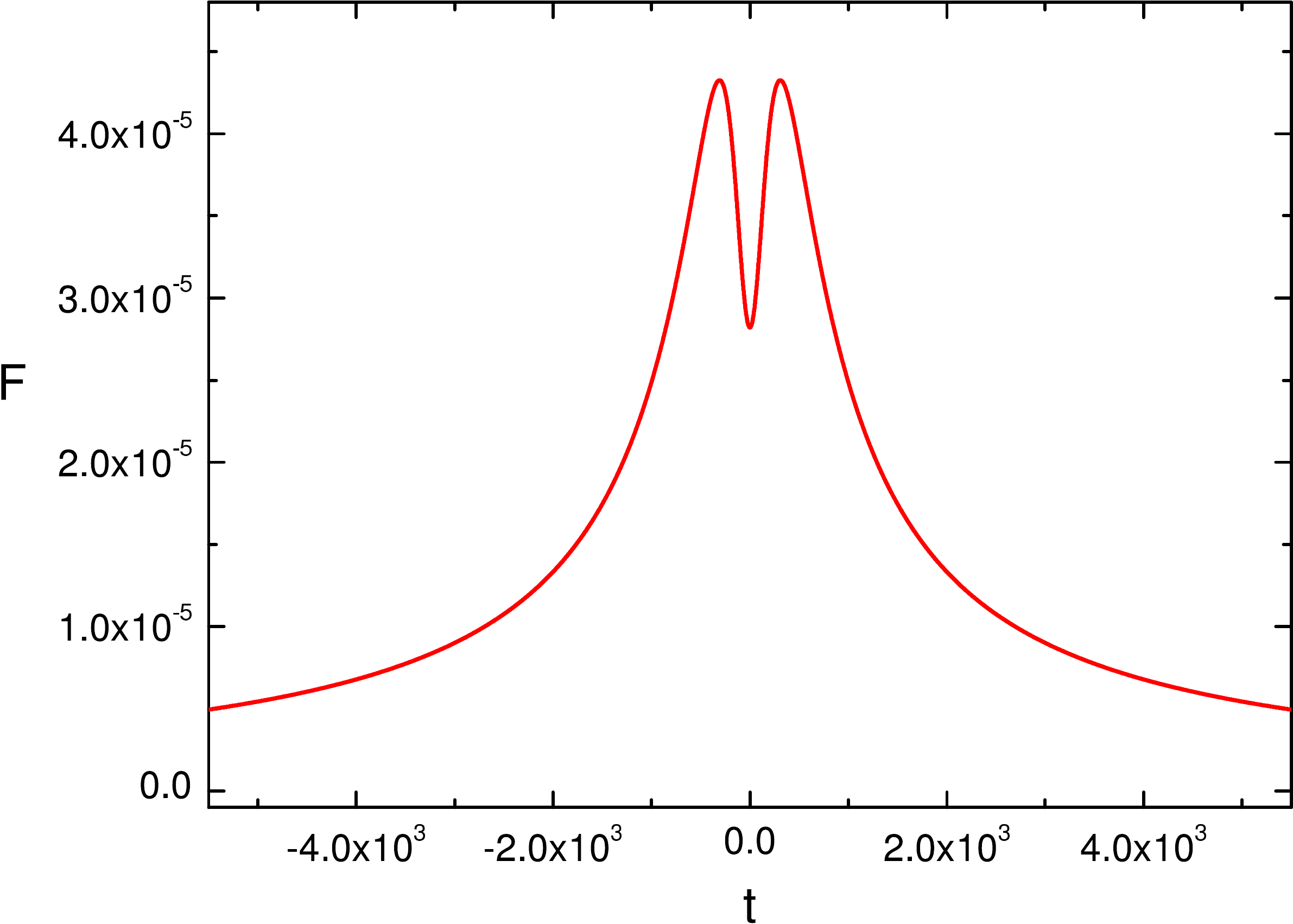}
\caption{\it Numerical result on the evolution of the reduced form $F$ in terms of the
cosmic time $t$, in the dust-dominated non-singular bounce cosmology. The model
parameters are the same as those provided in Fig. \ref{Fig-fofT_fT_bounce}. The graph is
plotted in units of $M_{P}$. From \cite{Cai:2011tc}. }
\label{Fig-ft_fT_bounce}
\end{figure}
\begin{figure}[tbph]
\includegraphics[scale=0.3]{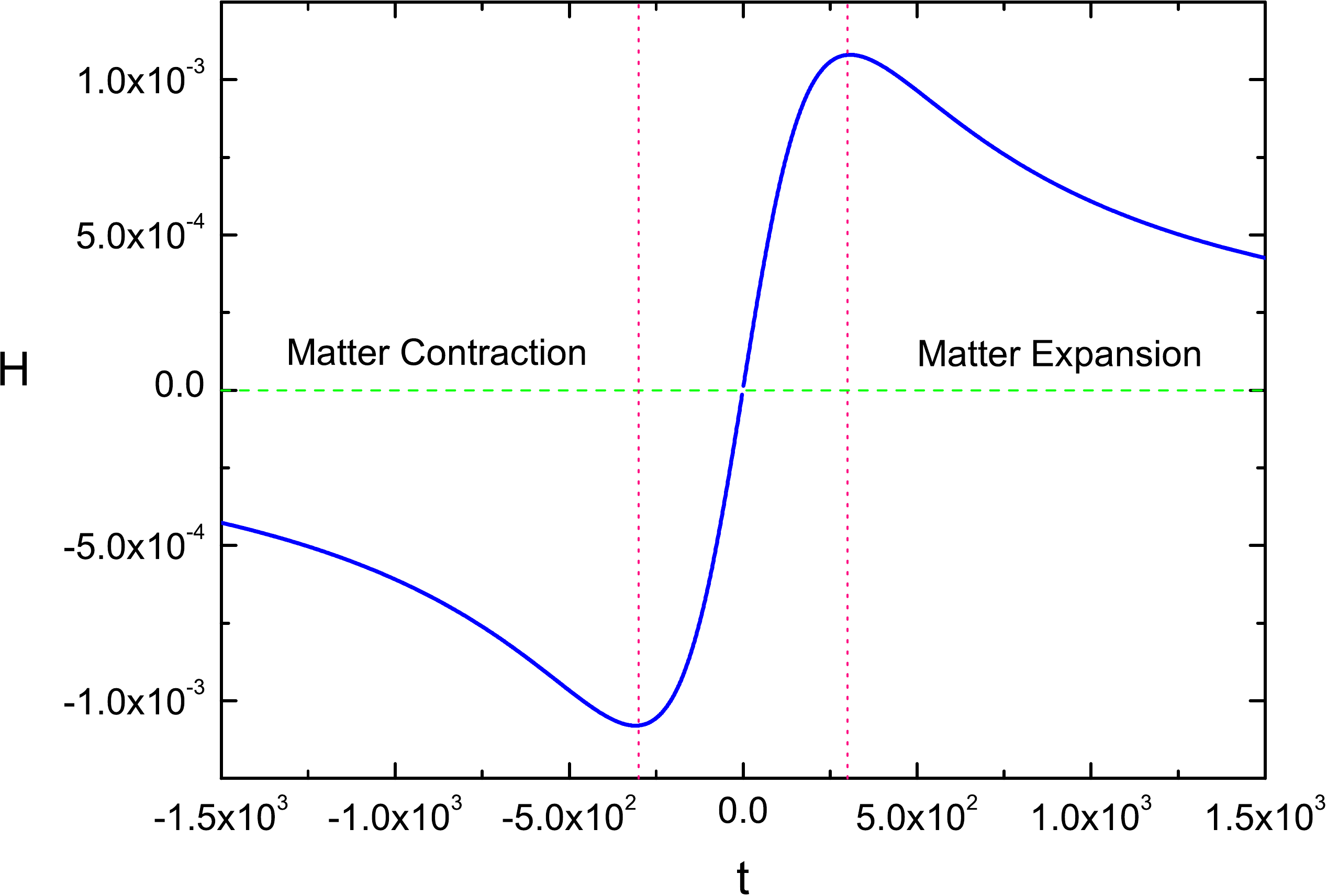}
\caption{\it Numerical result on the evolution of the Hubble parameter $H$ in terms of the
cosmic time $t$, in the dust-dominated non-singular bounce cosmology. The model parameters
are the same as those provided in Fig. \ref{Fig-fofT_fT_bounce}. The graph is plotted in
units of $M_{P}$. From  \cite{Cai:2011tc}. }
\label{Fig-hubble_fT_bounce}
\end{figure}

Furthermore, we numerically evolve the $F(T)$ and the Hubble parameter $H$ as functions
of the cosmic time in Figs. \ref{Fig-ft_fT_bounce} and \ref{Fig-hubble_fT_bounce}
respectively. In particular, Fig. \ref{Fig-ft_fT_bounce} shows that the evolution of
$F(T)$ is symmetric with respect to the bouncing point $t_B=0$. At the bouncing point
$F(T)$ arrives at a minimal value $\rho_{mB}/M_{P}^2$, which happens to cancel the
contribution of normal matter fields, and thus leads to the non-singular bounce. From Fig.
\ref{Fig-hubble_fT_bounce}, one can deduce that the background evolution of the universe
follows the usual Einstein gravity away from the bouncing phase, but it is dominated by
$F(T)$ in the middle period. These feature are completely consistent with the
analytical designs of the scenario as discussed previously.

\subsubsection{Perturbation analysis}

In order to examine the stability issue and to explore an interpretation of the CMB
observations alternative to inflation, in the following we perform a perturbation
analysis within the non-singular bouncing background. In our scenario, cosmological
perturbations arise from quantum fluctuations of space-time in the contracting phase.
Along with the dust-dominated contraction, these quantum fluctuations can exit the Hubble
radius, since the Hubble radius decrease faster than the physical wavelengths of
perturbation modes. When passing through the bouncing point, the background evolution
could affect the scale dependence of the perturbations at ultra-violet scales. However,
the observable primordial perturbations, responsible for the large scale structure of our
universe, are mainly originated in the infrared regime, where the modified gravity effect
becomes very limited (see for example  \cite{Wands:1998yp,Cai:2008ed, Cai:2008qw,
Cai:2009hc, Xue:2013bva}). In the following, we study the perturbation theory in $f(T)$
cosmology in
detail, and we verify these statements in a specific model of matter bounce cosmology.

To begin with, and following the analysis of subsection \ref{sec:pert_fT_gravity}, we
shall work in the longitudinal gauge which only involves scalar-type
metric perturbations as
\begin{eqnarray}
 ds^2=(1+2\Phi)dt^2-a^2(t)(1-2\Psi)d\vec{x}^2~.
\end{eqnarray}
As usual, the scalar metric perturbations are characterized by two functions $\Phi$ and
$\Psi$. Correspondingly, the perturbation of the torsion scalar at leading order is given
by
\begin{eqnarray}
 \delta{T} = 12 H (\dot\Phi+H\Psi)~.
\end{eqnarray}
By expanding the gravitational equations of motion to linear order, we obtain the $(00)$,
$(0i)$, $(
ij)$ and $(ii)$ component perturbation equations, which take the forms of
\begin{align}
\label{delta_00_fT_bounce}
 &\left( 1+F_{T} \right) \frac{\nabla^2}{a^2}\Psi -3(1+F_{T})H\dot\Psi
 -3(1+F_{T})H^2\Phi\nonumber\\
 &\ \ \ \ \ \
 +36F_{TT}H^3(\dot\Psi+H\Phi) = 4{\pi}G ~ \delta\rho_m ~,
\end{align}
\begin{align}
\label{delta_0i_fT_bounce}
 (1+F_{T}-12H^2F_{TT}) (\dot\Psi+H\Phi) = 4{\pi}G ~ \delta q ~,
\end{align}
\begin{align}
\label{delta_ij_fT_bounce}
 (1+F_{T})(\Psi-\Phi) = 8{\pi}G ~ \delta s ~,
\end{align}
and
\begin{align}
\label{delta_ii_fT_bounce}
 & (1+F_{T}-12H^2F_{TT})\ddot\Psi +H(1+F_{T}-12H^2F_{TT})\dot\Phi  \nonumber\\
 & +3H(1+F_{T} -12H^2F_{TT} -12\dot{H}F_{TT}+48H^2\dot{H}F_{TTT})\dot\Psi \nonumber\\
 & +[3H^2(1+F_{T}-12H^2F_{TT})
\nonumber\\
&\ \ \ \
 +2\dot{H}(1+F_{T}-30H^2F_{TT} +72H^4F_{TTT})]\Phi
\nonumber\\
 & +\frac{1+F_{T}}{2a^2}\nabla^2(\Psi-\Phi) = 4{\pi}G ~\delta{p_m} ~,
\end{align}
respectively. The functions $\delta\rho_m $, $\delta{p}_m$, $\delta{q}$, and $\delta{s}$
are the fluctuations of energy density, pressure, fluid velocity, and anisotropic stress,
respectively. If we consider the matter sector as a canonical scalar field $\phi$ with a
Lagrangian of the form of $\mathcal{L} = \frac{1}{2}\partial_{\mu}\phi\partial^{\mu}\phi
-V(\phi)$, then we can write down
\begin{align}
\label{delta_rho_fT_bounce}
 \delta \rho_m &= \dot{\phi}(\delta \dot{\phi}-\dot{\phi}\Phi) +V_{,\phi} \delta\phi ~, \\
\label{delta_q2_fT_bounce}
 \delta q &= \dot{\phi}\delta \phi ~, \\
\label{delta_s1_fT_bounce}
 \delta s &= 0~, \\
\label{delta_p_fT_bounce}
 \delta p_m &= \dot{\phi}(\delta \dot{\phi}-\dot{\phi}\Phi) -V_{,\phi} \delta\phi ~,
\end{align}
respectively. The above equations can be simplified as follows. We first insert the
relation \eqref{delta_s1_fT_bounce} into Eq. \eqref{delta_ij_fT_bounce} and obtain
$\Psi=\Phi$ due to a vanishing anisotropic stress. Moreover, we combine
\eqref{delta_0i_fT_bounce} and \eqref{delta_q2_fT_bounce} and then find that $\Phi$ is
completely determined by the scalar field fluctuation $\delta\phi$ in our case. We notice
this nice property applies only to the present case but is not true in general situation
 \cite{Li:2011rn}. Accordingly, for our choice of the tetrad given in
\eqref{weproudlyuse}, there exists only a single scalar degree of freedom. Note that
there
is another equation of motion to describe the dynamics of cosmological perturbations,
namely the perturbed equation for $\delta \phi$. However, it can be shown that this
equation is consistent with \eqref{delta_00_fT_bounce} and \eqref{delta_ii_fT_bounce} by
applying \eqref{delta_0i_fT_bounce}.

Combining \eqref{delta_00_fT_bounce}, \eqref{delta_0i_fT_bounce} and
\eqref{delta_ii_fT_bounce}, one can derive the key equation of motion for one Fourier
mode of the gravitational potential $\Phi_k$ as
\begin{eqnarray}  \label{eom_Phi_com_fT_bounce}
 \ddot\Phi_k + \alpha \dot\Phi_k + \mu^2 \Phi_k + c_s^2 \frac{k^2}{a^2}\Phi_k = 0 ~,
\end{eqnarray}
with
\begin{align}
 \alpha &= 7H +\frac{2V_{,\phi}}{\dot\phi} -\frac{36H\dot{H} (F_{TT}
 -4H^2F_{TTT})}{1+F_{T} -12H^2F_{TT}} ~, \\
 \mu^2 &= 6H^2 +2\dot{H} +\frac{2HV_{,\phi}}{\dot\phi}
\nonumber\\
&\ \ \
-\frac{36H^2\dot{H}(F_{TT}
 -4H^2F_{TTT})}{1+F_{T} -12H^2F_{TT}} ~, \\
 c_s^2 &= \frac{1+F_{T}}{1+F_{T} -12H^2F_{TT}} ~.
\end{align}
The functions $\alpha$, $\mu^2$ and $c_s^2$ stand for the frictional term, the
effective mass, and the sound speed parameter for the gravitational potential $\Phi$,
respectively. Moreover, we recall that the background equation of motion for the matter
field takes the form of:
$\ddot\phi +
3H\dot\phi + V_{,\phi} = 0$; and the second Friedmann equation in our case reads: $( 1
+F_{T} -
12H^2F_{TT}) \dot{H} = -4\pi G\dot\phi^2$. Consequently, making use of these two
background equations, Eq. (\ref{eom_Phi_com_fT_bounce}) can be further simplified to be
\begin{equation}  \label{eom_Phi_sim_fT_bounce}
 \ddot\Phi_k + \left( H-\frac{\ddot{H}}{\dot{H}} \right) \dot\Phi_k
 + \left( 2\dot{H}-\frac{H\ddot{H}}{\dot{H}} \right) \Phi_k
 + \frac{c_s^2 k^2}{a^2}\Phi_k = 0 ~.
\end{equation}
As a result, we find that the perturbation equation for the gravitational potential in
the present scenario is the same as in the case of standard Einstein gravity
\cite{Mukhanov:1990me}, apart from the newly introduced sound speed square parameter
$c_s^2$. This important feature is a key for us to explore potential implications of
$f(T)$ gravity in cosmological surveys.

%\item{Perturbation variables}

One often uses a gauge-invariant variable $\zeta$, the curvature fluctuation in comoving
coordinates, to characterize the cosmological inhomogeneities. In the scenario under
consideration, we take the form of $\zeta$ the same as that defined in standard
perturbation theory, namely
\begin{eqnarray}  \label{zeta_fT_bounce}
 \zeta \equiv \Phi - \frac{H}{\dot H} \big( \dot\Phi +H\Phi \big) ~,
\end{eqnarray}
however note that strictly speaking, even using the same variables and equations, one
should now talk about ``torsion fluctuation''.
A useful relation for the time derivative of $\zeta$ can be obtained by applying Eq.
\eqref{eom_Phi_sim_fT_bounce}, that is
\begin{eqnarray}  \label{dot_zeta_fT_bounce}
 \dot\zeta_k = \frac{H}{\dot{H}} \frac{c_s^2k^2}{a^2} \Phi_k ~.
\end{eqnarray}
In a generic expanding universe $\dot\zeta_k$ approaches zero at large length scales,
namely $k \rightarrow 0$, since the dominant mode of $\Phi_k$ is then nearly constant.
However, in bounce cosmology the metric perturbation $\Phi_k$ in the contracting phase is
dominated by a growing mode, and therefore $\zeta$ keeps increasing until the bounce
point \cite{Cai:2008ed}.

We mention that the variable $\zeta$ could become ill-defined when $\dot H$ changes the
sign. This issue was extensively studied in many aspects of cosmological perturbation
theory. At present, our understanding on a well-defined cosmological perturbation theory
is to require the metric perturbation and the corresponding extrinsic curvature to behave
smoothly throughout the background evolution. The discussion on this topic appeared in
\cite{Wands:1998yp}, and we refer to \cite{Xia:2007km, Cai:2008ed,
Wands:2008tv,Cai:2009hc,
Cai:2008qw, Xue:2013bva} for detailed analyses from various perspectives. In the
following calculations we still use $\zeta$ merely since it is convenient to perform the
analytic calculation away from the bouncing phase.

In order to further simplify the calculation, we introduce a canonical variable of the
cosmological perturbation through
\begin{eqnarray}  \label{v_var_fT_bounce}
 v = z\zeta ~,
\end{eqnarray}
with $z \equiv a\sqrt{2\epsilon}$ and $\epsilon\equiv-\frac{\dot{H}}{H^2}$. Afterwards,
one can rewrite the key perturbation equation as
\begin{eqnarray}  \label{v_eom_g_fT_bounce}
 v_k^{\prime \prime } +(c_s^2k^2 -\frac{ z^{\prime \prime} }{z})v_k=0 ~,
\end{eqnarray}
where the prime denotes derivative with respect to the comoving time $\tau \equiv
\int dt/a$. In order to perform a specific analysis we recall that in the matter-like
contracting phase (as obtained in the previous paragraph of the background solution) the
scalar factor evolves as $a \sim t^{2/3} \sim \tau^2$, and that $z\propto a$. By solving
the background equations one obtains the following approximate relations
\begin{eqnarray}\label{fTmatter_fT_bounce}
 H\simeq\frac{2}{3t}~,~
 F(T) \simeq \left( -1+\frac{\rho_{mB}}{2M_{P}^2\sigma} \right) T ~,
\end{eqnarray}
that hold far before the beginning of the bouncing era $t_m = -\sqrt{\frac{2}{3\sigma}}$.
As we mentioned above, $\rho_{mB}$ is the energy density of the matter field at the
bouncing point and $\sigma$ describes how fast the bounce takes place. Therefore, the
sound speed of the curvature perturbation reverts to $c_s^2\simeq1$ in the matter-like
contracting phase.

Inserting the $F(T)$ form of \eqref{fTmatter_fT_bounce} into the action
\eqref{action_fT}, we find that the standard Einstein gravity will automatically be
recovered when $\rho_{mB} \simeq 2M_{P}^2\sigma$. Particularly, when $\rho_{mB}$ exactly
equals to $2M_{P}^2 \sigma$, we get $F(T)\sim O(T^2)$, which will dilute faster than the
Ricci scalar during late-time evolution. Even when $\rho_{mB}$ is not equal to
$2M_{P}^2\sigma$, it is clear that the system satisfies GR
with a rescaled gravitational constant. Thus, the combination of $\rho_{mB}$ and $\sigma$
could, in principle, be constrained by measurements of the gravitational constant. In our
computation we choose $\rho_{mB}$ to be slightly different from $2M_{P}^2\sigma$, thus
our model is able to approach the standard Einstein theory far away from the bounce. As a
consequence, the perturbation equation in the contracting phase becomes
\begin{eqnarray} \label{v_eom_m_fT_bounce}
 v_{k}^{\prime \prime }+\left( k^{2}-\frac{2}{\tau ^{2}}\right) v_{k}\simeq 0 ~.
\end{eqnarray}
 Initially the $k^{2}$-term dominates in
\eqref{v_eom_m_fT_bounce} and thus we can neglect the gravitational term. This implies
that the fluctuation corresponds to a free scalar propagating in a flat space-time, and
thus the initial condition naturally takes the form of
the Bunch-Davies vacuum: $v_{k}\simeq \frac{e^{-ik\tau }}{\sqrt{2k}}$. Making use of the
vacuum initial condition we can solve the perturbation equation exactly and yield
\begin{eqnarray}
 v_{k}=\frac{e^{-ik\tau }}{\sqrt{2k}} \big(1-\frac{i}{k\tau}\big) ~.
\end{eqnarray}
From this result we can find that the quantum fluctuations could become classical
perturbations, after exiting the Hubble radius, due to the gravitational term in Eq.
\eqref{v_eom_m_fT_bounce}. Moreover, the amplitude of the metric perturbations keeps
increasing until the universe arrives at the bouncing phase at the moment $t_{m}$.

From the definition of the power spectrum we see that $\zeta\sim k^{3/2}|v_k|$ is
scale-invariant in our model, which can also be achieved in inflationary cosmology.
However, the coefficient $\epsilon$ takes the value $\frac{3}{2}$ in the matter-like
contraction and thus it is unable to amplify the power spectrum of metric perturbation as
in inflation. A detailed calculation provides the expression of the primordial power
spectrum for the $f(T)$ matter bounce as
\begin{eqnarray}
 P_{\zeta} \equiv \frac{k^3}{2\pi^2}\left|\frac{v_k}{z}\right|^2
 = \frac{H_m^2}{48\pi^2M_{P}^2} ~,
\end{eqnarray}
where $H_m=\sqrt{{\sigma}/{6}}$ is the absolute value of the Hubble parameter at the
beginning moment of the bouncing phase.

It is important to notice that the bouncing cosmology analyzed above yields the amplitude
of primordial scalar perturbation to be of the same order of the tensor spectrum. This
implies that the scenario of $f(T)$ matter bounce suffers from the issue of explaining
the small value of tensor-to-scalar ratio, defined as $r\equiv P_T/P_\zeta$, as observed
in the CMB experiments. Consequently, the specific model discussed above is enough to
demonstrate the theoretical fact that the initial Big Bang singularity can be avoided in
the $f(T)$ gravity, as well as to explain part of the CMB observations. However, in order
to make this model consistent with cosmological observations on the value of $r$, one
needs to design certain mechanisms to magnify the amplitude of scalar-type metric
perturbations. This issue can be resolved by introducing additional light scalar fields,
as in the bounce curvaton scenario  \cite{Cai:2011zx, Cai:2011ci, Cai:2014xxa}. These
scalars are able to seed isocurvature fluctuations, and then transfer to a
scale-invariant
spectrum of the adiabatic fluctuations during the non-singular bouncing phase, through
the
so-called kinetic amplification. Thus, we obtain a mechanism for enhancing the primordial
adiabatic fluctuations and suppressing the tensor-to-scalar ratio. The study on more
general mechanisms is an important direction for future projects in the field of $f(T)$
early universe.

%%%%%%%%%%%%%%%%%%%%%%%%%%%%%%%%%%%
\subsection{Cosmological solutions by Noether Symmetry Approach}
\label{FRWNetherr}
 %%%%%%%%%%%%%%%%%%%%%%%%%%%%%%%%%%%%%%%%%

In this subsection we will present a different but very helpful way to extract
cosmological solutions in $f(T)$ gravity, based on Noether symmetries,
following \cite{Wei:2011aa,Atazadeh:2011aa}. We start by constructing the   point-like
Lagrangian of $f(T)$ gravity. Starting from the usual action
$ {\cal S}=\int d^4 x\,|e|\,f(T)+{\cal S}_m$, and following
 \cite{Capozziello:2008ch,Vakili:2008ea,Terzis:2014cra,Terzis:2015mua,
Paliathanasis:2015mxa}, one
 can define a canonical Lagrangian
 ${\cal L}={\cal L}(a,\dot{a},T,\dot{T})$, with
 ${\cal Q}=\{a,T\}$ the configuration space and
 ${\cal TQ}=\{a,\dot{a},T,\dot{T}\}$ the related tangent
 bundle on which $\cal L$ is defined.
 The scale factor $a(t)$
 and the torsion scalar
 $T(t)$ are taken as independent
 dynamical variables, and hence one can use the  Lagrange
 mutipliers method in order to set $T$ as a constraint of the dynamics (recall that
$T=-6H^2$), namely
 \cite{Wei:2011aa,Atazadeh:2011aa}
 \begin{equation}
\label{eq12wei}
 {\cal S}= \int dt\,a^3\left[f(T)-\lambda\left(T+
 6\frac{\dot{a}^2}{a^2}\right)-\frac{\rho_{m0}}{a^3}\right],
 \end{equation}
 where $\lambda$ is a Lagrange multiplier, and $\rho_{m0}$ is the value of the matter
energy density at the present scale factor $a_0=1$. Since variation with
 respect to $T$  gives
$ \lambda=f_T$, the action (\ref{eq12wei}) gives  the point-like Lagrangian
 \begin{equation}
\label{eq15wei}
 {\cal L}(a,\dot{a},T,\dot{T})=a^3\left(f-f_T T\right)-
 6f_T a\dot{a}^2-\rho_{m0}\,.
\end{equation}

The above point-like Lagrangian can give rise to the Friedmann equations of $f(T)$
gravity, namely (\ref{background11}),(\ref{background22}). Firstly, substituting
(\ref{eq15wei}) into the Euler-Lagrange
equation $\frac{d}{dt}\left(\frac{\partial \cal L}{\partial\dot{q}_i}
\right)-\frac{\partial \cal L}{\partial q_i}=0$, where in the present case the
generalized coordinates $q_i$ of the configuration space $\cal Q$ are just
$a$ and $T$, we obtain
 \begin{eqnarray}
 &&a^3 f_{TT}\left(T+
 6\frac{\dot{a}^2}{a^2}\right)=0\,,\label{eq17wei}\\
 &&f-f_T T+2f_T H^2+
 4\left(f_T\frac{\ddot{a}}{a}+Hf_{TT}\dot{T}\right)=0\,.\ \ \ \ \label{eq18wei}
 \end{eqnarray}
 Note that if $f_{TT}\not=0$ then (\ref{eq17wei}) recovers the relation
 $T=-6H^2$ as the Euler constraint of the dynamics. Substituting it into (\ref{eq18wei})
 one additionally recovers the Raychaudhuri equation (\ref{background22}) (with
$p_m=0$), namely
\begin{equation}
 48H^2 f_{TT}\dot{H}-4f_T\left(3H^2+\dot{H}\right)-f=0\,.
\end{equation}
Secondly, the Hamiltonian ${\cal H}=\sum\limits_i\frac{\partial \cal L}
 {\partial\dot{q}_i}\dot{q}_i-{\cal L}\ $ corresponding  to Lagrangian $\cal L$ becomes
 \begin{equation}
 \label{eq22wei}
 {\cal H}(a,\dot{a},T,\dot{T})=a^3\left(-6f_T
 \frac{\dot{a}^2}{a^2}-f+f_T T+\frac{\rho_{m0}}{a^3}\right),
\end{equation}
and hence the Hamiltonian constraint ${\cal H}=0$ eventually gives rise to the first
Friedmann equation (\ref{background11})(with $\rho_{m}=\rho_{m0}/a^3$), namely
 \begin{equation}
 12H^2 f_T+f=\frac{\rho_{m0}}{a^3}\,.
\end{equation}

As we know, Noether symmetry can be a useful tool to obtain specific solutions motivated
at a fundamental level
 \cite{deRitis:1990ba,Capozziello:2008ch,Vakili:2008ea,Capozziello:2009te}.
In the case of $f(T)$ cosmology, the generator of Noether symmetry is the vector
 \cite{Wei:2011aa,
Atazadeh:2011aa}
\begin{equation}
 \label{eq24wei}
 {\bf X}=\alpha\frac{\partial}{\partial a}+
 \beta\frac{\partial}{\partial T}+
 \dot{\alpha}\frac{\partial}{\partial\dot{a}}+
 \dot{\beta}\frac{\partial}{\partial\dot{T}}\,,
\end{equation}
where $\alpha=\alpha(a,T)$ and $\beta=\beta(a,T)$ are functions of the generalized
coordinates $a$ and $T$. Hence, a Noether symmetry exists if the equation
\begin{equation}
\label{eq25wei}
 L_{\bf X}{\cal L}={\bf X}{\cal L}=
 \alpha\frac{\partial \cal L}{\partial a}+
 \beta\frac{\partial \cal L}{\partial T}+
 \dot{\alpha}\frac{\partial \cal L}{\partial\dot{a}}+
 \dot{\beta}\frac{\partial \cal L}{\partial\dot{T}}=0,
\end{equation}
where $L_{\bf X}{\cal L}$ is the Lie derivative of the Lagrangian $\cal L$ with respect
to
the
vector $\bf X$,  has a solution, and the corresponding Noether charge will be
 \cite{Wei:2011aa,Atazadeh:2011aa}
\begin{equation}
 \label{eq26wei}
 Q_0=\sum\limits_i\alpha_i\frac{\partial \cal L}{\partial\dot{q}_i}
 =\alpha\frac{\partial \cal L}{\partial\dot{a}}+
 \beta\frac{\partial \cal L}{\partial\dot{T}}=const.
\end{equation}
Substituting  (\ref{eq15wei}) into (\ref{eq25wei}), and using that $\dot{\alpha} =
(\partial\alpha/\partial a)\,\dot{a} +(\partial\alpha/\partial T)\,\dot{T}$, $\dot{\beta}
= (\partial\beta/\partial
a)\,\dot{a} +(\partial\beta/\partial T) \, \dot{T}$, we find
\begin{eqnarray}
\label{eq27wei}
 &&
 3\alpha a^2\left(f-f_T T\right)-\beta a^3 f_{TT}T-
 12a\dot{a}\dot{T}\frac{\partial\alpha}{\partial T}\nonumber\\
 &&
 -6\dot{a}^2\left(\alpha f_T+\beta a f_{TT}+
 2a f_T\frac{\partial\alpha}{\partial a}\right)=0\,.
\end{eqnarray}
Thus, requiring the coefficients of $\dot{a}^2$, $\dot{T}^2$ and $\dot{a}\dot{T}$ to be
zero
separately  \cite{Wei:2011aa,Atazadeh:2011aa}, we acquire
\begin{eqnarray}
 &&a\frac{\partial\alpha}{\partial T}=0\,,\label{eq28wei}\\
 &&\alpha f_T+\beta a f_{TT}+
 2a f_T\frac{\partial\alpha}{\partial a}=0\,,\label{eq29wei}\\
 &&3\alpha a^2\left(f-f_T T\right)-
 \beta a^3 f_{TT}T=0\,.\label{eq30wei}
\end{eqnarray}
The constraint (\ref{eq30wei}) is the Noether condition and the corresponding Noether
charge (\ref{eq26wei}) reads
\begin{eqnarray}
 \label{eq31wei}
 Q_0=-12\alpha f_T a\dot{a}=const.
\end{eqnarray}

If one finds explicit forms $\alpha$ and $\beta$, with at least one of them different
from zero, that can satisfy (\ref{eq28wei}), (\ref{eq29wei}) and
(\ref{eq30wei}), then Noether symmetry exists. After some simple algebra, one can find
 \cite{Wei:2011aa,Atazadeh:2011aa}
 \begin{eqnarray}
 &&f(T)=\mu T^n\,,\label{eq37wei}
  \label{eq36wei}\\
 &&\alpha(a)=\alpha_0\,a^{1-3/(2n)}
  \label{eq37wei}\\
 &&\beta(a,T)
 =-\frac{3\alpha_0}{n}\,a^{-3/(2n)}\,T\,,
 \label{eq38wei}
 \end{eqnarray}
 where $\mu$ and $\alpha_0$ are integration constants.
 Finally, one can easily find the solution for the scale factor that corresponds to the
above $f(T)$ form, namely  \cite{Wei:2011aa,Atazadeh:2011aa}
  \begin{eqnarray}
\label{eq42wei}
 a(t)
 =(-1)^{1+2n/3}\frac{3}{2n}\,(c_2 t-c_3)^{2n/3}\,,
 \end{eqnarray}
 where
  \begin{eqnarray}
  c_2=\left[\frac{Q_0}{-12\alpha_0\mu n(-6)^{n-1}}
 \right]^{1/(2n-1)},
  \end{eqnarray}
 and
 $c_3$ is an integration constant. Hence, at late times ($|c_2 t|\gg |c_3|$)   the
universe expands in a power-law form (for $n>0$).

Lastly, one can calculate the physical quantities corresponding to the above exact
solution. In particular, the effective  dark energy density and pressure
straightforwardly become
  \begin{eqnarray}
 &&\rho_{DE}=6H^2-f-12H^2 f_T\,,\label{eq46wei}\\
 &&p_{DE}=-\rho_{DE}-4\left(12H^2 f_{TT}-f_T+
 1\right)\dot{H}\,.\label{eq47wei}
  \end{eqnarray}
 Hence, the dark energy equation-of-state parameter becomes
  {\small{\begin{eqnarray}
  \label{eq48}
 w_{DE}
 =-\frac{8n(n-1)\,3^n}{8n^2\cdot 3^n-3\mu(8^n
 -n\cdot 2^{1+3n})(-n^2)^n\,t^{2(1-n)}},\
 \end{eqnarray}}}
 while the total one reads
 \begin{eqnarray}
 w_{tot}\equiv\frac{p_{tot}}{\rho_{tot}}=
 \frac{1}{3}\left(2q-1\right)=\frac{1}{n}-1\,.
 \end{eqnarray}
 Therefore, for $n>3/2$ we have $w_{tot}<-1/3$, that is we acquire acceleration.

We close this subsection by mentioning that with the above Noether symmetry approach we
can find which $f(T)$ forms can have a theoretical justification, instead of considering
them arbitrarily by hand. For more details the reader could see
\cite{Atazadeh:2011aa,Wei:2011aa,Jamil:2012fs,Sadjadi:2012xa,Dong:2013rea,
Basilakos:2013rua}.

\subsection{Anisotropic cosmology}
\label{anisotrcosm}

So far we have investigated the cosmological solutions provided by $f(T)$ gravity in the
frame of FRW universe. In a more generic case, the background manifold could be
anisotropic. In the standard cosmological paradigm described by GR, the
evolution from an anisotropic universe into an FRW one can be achieved by a period of
inflationary expansion. However, it is interesting to examine whether $f(T)$ gravity can
accommodate with an anisotropic universe. Especially, $f(T)$ gravity itself is
constructed upon the vierbein, that naturally gives rise to a non-vanishing anisotropic
stress at the perturbative level, as it was shown in Sec. \ref{sec:pert_fT_gravity}. The
solutions of an anisotropic universe governed by $f(T)$ gravity have been studied in
detail in the literature (for instance see \cite{Sharif:2011bi, Daouda:2011rt,
Rodrigues:2012qua,
Daouda:2012nj,Rodrigues:2013iua,Aslam:2013coa,Rodrigues:2014xam,Fayaz:2014swa,
Amir:2015wja}).

In the present subsection we study the realization of anisotropic evolution in $f(T)$
cosmology. In particular, we assume more general cosmological metrics than the FRW
one, namely the Bianchi type-I, the Kantowski-Sachs (KS) and the Bianchi type-III
ones, in order to provide a more general description of the cosmological evolution. Note
that anisotropic cosmological metrics have been already studied in the context of GR with
the presence of isotropic and anisotropic fluids \cite{Barrow:2005qv, Chen:2001fh,
Leon:2010pu}. Moreover, we desire to obtain several representative results in this
scenario, such as power-law and de Sitter (dS) expansions. Since power-law and dS
solutions can provide a good description for some specific phases of the universe
evolution, their reconstruction in $f(T)$ gravity becomes a crucial point in order to
consider this class of theories as a serious candidate for explaining the cosmological
history.

Let us consider the energy momentum tensor of the matter sector to be: $T^\mu_\nu =
{diag} (1, -\omega_x, -\omega_y, -\omega_z) \rho_m$, where the $\omega_i$ are the
effective equation-of-state parameters corresponding to the pressures $p_i$ along spatial
directions with $i=x, y, z$. The general Bianchi type-III case  metric reads
\begin{equation} 
\label{Rodrigues_metrictype3}
 dS^2=dt^2-A^2(t)dx^2-e^{-2\alpha x}B^2(t)dy^2-C^2(t)dz^2 ~,
\end{equation}
with a constant $\alpha$. Hence, Bianchi type-I can be obtained for
$\alpha=0$, while Kantowski-Sachs is acquired  for $\alpha=0$ and $B(t)=C(t)$. 
Accordingly,
we take the following set of diagonal tetrads producing the metric
(\ref{Rodrigues_metrictype3}):
\begin{eqnarray}
\label{Rodrigues_matrixtype3}
  e^{A}_{\;\;\mu} = \mathrm{diag} \left( 1, A(t), e^{-\alpha x} B(t), C(t) \right) ~,
\end{eqnarray}
with determinant $e=e^{-\alpha x}ABC$.
Correspondingly, inserting the above vierbein into the torsion tensor
(\ref{tor}) and into the contorsion tensor (\ref{contorsiontensor}) we obtain their
components as
\begin{align}
\label{Rodrigues_torsiontype3}
 & T^{1}_{\;\;01} = \frac{\dot{A}}{A} ~,~~
 T^{2}_{\;\;02} = \frac{\dot{B}}{B} ~,~~
 T^{2}_{\;\;21} = \alpha ~,~~
 T^{3}_{\;\;03} = \frac{\dot{C}}{C} ~, \\
\label{Rodrigues_contorsiontype3}
 & K^{01}_{\;\;\;\;1} = \frac{\dot{A}}{A} ~,~~
 K^{02}_{\;\;\;\;2} = \frac{\dot{B}}{B} ~,~~
 K^{12}_{\;\;\;\;2} = \frac{\alpha}{A^2} ~,~~
 K^{03}_{\;\;\;\;3} = \frac{\dot{C}}{C} ~,
\end{align}
respectively. Additionally, the components of the superpotential tensor
$S_{\alpha}^{\;\;\mu\nu}$ defined in (\ref{Ssuperpotdef}) become
\begin{eqnarray}\label{Rodrigues_tensortype3}
 &&S_{0}^{\;\;01} = S_{3}^{\;\;31}=\frac{\alpha}{2A^2} ~,~~
 S_{1}^{\;\;10}=\frac{1}{2}\left(\frac{\dot{B}}{B}+\frac{\dot{C}}{C}\right) ~,\nonumber\\
 &&
 S_{2}^{\;\;20} = \frac{1}{2}\left(\frac{\dot{A}}{A}+\frac{\dot{C}}{C}\right) ~,~~
 S_{3}^{\;\;30}=\frac{1}{2}\left(\frac{\dot{A}}{A}+\frac{\dot{B}}{B}\right) .\ \ \ \
\end{eqnarray}
Inserting these into
the torsion scalar (\ref{telelag}) we acquire
\begin{eqnarray}
\label{Rodrigues_torsionScalar1}
 T = -2 \left( \frac{\dot{A}
 \dot{B}}{AB} +\frac{\dot{A}\dot{C}}{AC}
+\frac{\dot{B}\dot{C}}{BC}
\right) ~.
\end{eqnarray}
Finally, the equations of motion (\ref{eom_fT_general}) then become
\cite{Rodrigues:2012qua}
\begin{equation}
\label{Rodrigues_densitytype3}
 16\pi \rho_m = f +4f_{T}
 \Big[\frac{\dot{A}\dot{B}}{AB} +\frac{\dot{A}\dot{C}}{AC}
 +\frac{\dot{B}\dot{C}}
 {BC} -\frac{\alpha^2}{2A^2}\Big] ~,
\end{equation}
\begin{eqnarray}
\label{Rodrigues_radialpressuretype3}
 &&-16\pi p_x = f +2f_{T}
 \left[\frac{\ddot{B}}{B} +
 \frac{\ddot{C}}{C}
+\frac{\dot{A}\dot{B}}{AB}
 +\frac{\dot{A}\dot{C}}{AC}
+2\frac{\dot{B}\dot{C}}{BC}\right]\nonumber\\
&&\ \ \ \ \ \ \ \ \ \ \ \ \   \,
+2\left(\frac{\dot{B}}{B}
 +\frac{\dot{C}}{C}\right)\dot{T}f_{TT} ~,
\end{eqnarray}
\begin{eqnarray}
\label{Rodrigues_tangentialpressure1type3}
 &&-16\pi p_y = f +2f_{T} \left[
 \frac{\ddot{A}}{A} +\frac{\ddot{C}}{C}
+\frac{\dot{A}\dot{B}}{AB}
 +2\frac{\dot{A}\dot{C}}{AC} +
 \frac{\dot{B}\dot{C}}{BC}\right]\nonumber\\
 &&\ \ \ \ \ \ \ \ \ \ \ \ \   \,
+2\left(\frac{\dot{A}}{A}
 +\frac{\dot{C}}{C} \right)
 \dot{T}f_{TT} ~,
 \end{eqnarray}
 \begin{eqnarray}
\label{Rodrigues_tangentialpressure2type3}
 &&-16\pi p_z = 2f_{T} 
 \left[\frac{\ddot{A}}{A} 
 +\frac{\ddot{B}}{B}
+2\frac{\dot{A}\dot{B}}{AB}
 +\frac{\dot{A}\dot{C}}{AC}
 +\frac{\dot{B}\dot{C}}{BC} -\frac{\alpha^2}{A^2} \right]
\nonumber\\
 &&\ \ \ \ \ \ \ \ \ \ \ \ \   \,
+f  +2\left(\frac{\dot{A}}{A} 
+\frac{\dot{B}}{B}\right) \dot{T}f_{TT} ~,
  \end{eqnarray}
   \begin{eqnarray}
\label{Rodrigues_constraint1}
 0 = \frac{\alpha}{2A^2} \left[
 \left( \frac{\dot{A}}{A} -\frac{\dot{B}}{B}\right)f_{T}
-\dot{T}f_
{TT} \right] ~,
\end{eqnarray}
\begin{eqnarray}
\label{Rodrigues_constraint2}
 0 = \alpha \left( 
 \frac{\dot{A}}{A}
 -\frac{\dot{B}}{B} \right) f_{T}  ~.
\end{eqnarray}
Note that, the constraint equation (\ref{Rodrigues_constraint2}) appears in $f(R)$
gravity too \cite{Sharif:2011fa}. However, one can acquire a second constraint
equation, namely (\ref{Rodrigues_constraint1}), which appears as a generalization of
\eqref{Rodrigues_constraint2} due to a second derivative of the function $f(T)$ with
respect to $T$. As mentioned above, by setting $\alpha=0$ one can recover the Bianchi
type-I case, while the equations of motion corresponding to KS model are obtained by
setting $\alpha=0$ and $B=C$.

Let us now try to reconstruct the $f(T)$ action for some particular metric ansatzen 
analyzed previously. In particular, in separate paragraphs we
consider solutions of de Sitter and power-law expansions. We mention that these kind of 
evolutions  have been studied in detail in $f(R)$ and Gauss-Bonnet modified gravity 
\cite{Goheer:2009ss, Nojiri:2009kx, Elizalde:2010jx,
Myrzakulov:2010gt}.

\subsubsection{De Sitter solutions}

The de Sitter solution is well known since it is a perfect approximation of early universe
inflation,  corresponding to  exponential expansion. We start by considering for 
simplicity Bianchi type-I and
Kantowski-Sachs metrics, thus taking $\alpha=0$ in (\ref{Rodrigues_metrictype3}). We
consider exponential expansion in each
spatial direction, namely
\begin{eqnarray} \label{Rodrigues_D2}
 A=A_0 e^{a t} ~,~~ B=B_0 e^{b t} ~,~~ C=C_0 e^{c t} ~,
\end{eqnarray}
and thus the corresponding expansion rates are given by
\begin{equation}
\label{Rodrigues_D3}
 H_x = \frac{\dot{A}}{A}=H_{x0} ~,~~ H_y 
 = \frac{\dot{B}}{B}=H_{y0} ~,~~ H_z =
\frac{\dot{C}}{C}=H_{
z0} ~,
\end{equation}
where $H_{x0}=a,H_{y0}=b,H_{z0}=c$ are constants. Then, the conservation equation for
the energy-momentum tensor can be easily obtained, namely
\begin{equation}
\label{Rodrigues_D1}
 \dot{\rho}_m +\left(H_x+H_y+H_z\right)
 \rho_m +H_x p_x+H_y p_y +H_z p_z=0 .
\end{equation}
 Additionally, the torsion scalar from (\ref{Rodrigues_torsionScalar1}) becomes
\begin{eqnarray}
\label{Rodrigues_D4}
 T_0 = -2\left(H_{x0}H_{y0}
 +H_{x0}H_{z0} +H_{y0}H_{z0}\right) ~.
\end{eqnarray}
Imposing $p_x=p_y=p_z=p_m$ and an equation-of-state parameter $w=p_m/\rho_m$,      
Eq. (\ref{Rodrigues_D1}) for the choice  (\ref{Rodrigues_D2}) can be solved 
as
\begin{eqnarray}
\label{Rodrigues_D5}
 \rho_m=\rho_{m0}
 e^{-(H_{x0}+H_{y0}+
 H_{z0})(1+w)t} ~.
\end{eqnarray}
Thus, the equations of motion (\ref{Rodrigues_densitytype3})-(\ref{Rodrigues_constraint2})
become
\begin{align}
\label{Rodrigues_D6}
 &16\pi \rho_{m0} e^{-(H_{x0} +H_{y0} 
 +H_{z0})(1+w)t} = f(T_0)\nonumber\\
 &\ \ \ \ \ \ \ \ \ \ \ \ \ \
+4\left[H_{x0}H_{y0}
+H_{z0}(H_{x0} +H_{
y0})\right]f_{T}(T_0)\ ,
\\
\label{Rodrigues_D7}
& -16\pi w\rho_{m0} e^{-(H_{x0} 
+H_{y0} +H_{z0})(1+w)t} = f(T_0)\nonumber\\
 &\ \ \ \ \ \ \ \ \ \ \ \ \  \ 
 +2(H_{y0} +H_{z0})(H_{x0}
+H_{y0} +H_{
z0})f_{T}(T_0)\ , \\
\label{Rodrigues_D8}
& -16\pi w\rho_{m0} e^{-(H_{x0}
+H_{y0} +H_{z0})(1+w)t} = f(T_0) \nonumber\\
 &\ \ \ \ \ \ \ \ \ \ \ \ \ 
 \ +2(H_{x0} +H_{z0})(H_{x0}
+H_{y0} +H_{
z0})f_{T}(T_0)\ , \\
\label{Rodrigues_D9}
& -16\pi w\rho_{m0} e^{-(H_{x0}
+H_{y0} +H_{z0})(1+w)t} = f(T_0)\nonumber\\
 &\ \ \ \ \ \ \ \ \ \ \ \ \ \ 
 +2(H_{x0} +H_{y0})(H_{x0}
+H_{y0} +H_{
z0})f_{T}(T_0)\ .
\end{align}
We mention that in the presence of a perfect fluid the only possible solution is the one 
arising from $w=-1$ (unless $H_{x0}+H_{y0}+H_{z0}=0$, but such a  case would correspond to
decelerating expansion
at least in one direction). Therefore,  the only possible solution reads
\begin{equation}
 A(t) =B(t) =C(t) ,
 \label{Rodrigues_D10}
\end{equation}
i.e. $H_{x0} =H_{y0} =H_{z0} 
=H_0 $. Hence, in this case  the metric 
(\ref{Rodrigues_metrictype3}) becomes the usual FRW one with
exponential expansion, namely $A(t)=A_0
e^{H_0 t}$. Summarizing, the only solution  when the
pressures are equal, i.e. $p_x=p_y=p_z$,  for a pure de Sitter expansion in  Bianchi-I
and Kantowski-Sachs metrics leads to an FRW geometry. Finally, in this case the equation 
system (\ref{Rodrigues_D4})-(\ref{Rodrigues_D9}) reduces to the sole independent
equation
\begin{eqnarray}
 16 \pi \rho_{m0} = f(T_0)
 +12H_{0}^2f_{T}(T_0).
 \label{Rodrigues_D11}
\end{eqnarray}

Considering \cite{Rodrigues:2012qua}
\begin{eqnarray}
 f(T) = \left(-T\right)^n\ ,
\label{Rodrigues_D12}
\end{eqnarray}
where $n$ is a   constant,  equation (\ref{Rodrigues_D11}) leads to the solution
\begin{eqnarray}
 H_{0}^2 = \frac{1}{6}
 \left(\frac{16\pi\rho_{m0}}{1-2n}\right)^{1/n}\ ,
\label{Rodrigues_D14}
\end{eqnarray}
and thus  $n\leq1/2$. In this case the de Sitter solution is triggered by $\rho_{m0}$,
which is a cosmological-constant term due to the fact that $w=-1$.

Considering \cite{Rodrigues:2012qua}
\begin{eqnarray}
 f(T) = C_1 T 
 +C_2 \left(-T\right)^n\ ,
\label{Rodrigues_D15}
\end{eqnarray}
with  $C_1,C_2,n$   constants, equation (\ref{Rodrigues_D11}) in vacuum can be
explicitly solved for specific $n$.  For example, for $n=2$ it leads to the solution
\begin{eqnarray}
 H_{0}=\sqrt{\frac{C_1}{18C_2}}\ ,
\label{Rodrigues_D17}
\end{eqnarray}
and thus the de Sitter phase is triggered by the $f(T)$ sector itself.

%%%%%%%%%%%%%%%%%%%%
\subsubsection{Power-law solutions}
%%%%%%%%%%%%%%%%%%%%%%

We now study the cosmological evolution corresponding to power-law expansions along 
spatial
coordinates. In this situation, the scale functions for the Bianchi-I and
Kantowski-Sachs metric (\ref{Rodrigues_metrictype3}) (i.e. for $(\alpha=0)$) become
\begin{equation}
 A(t)=A_0t^a ~,~~ 
 B(t)=B_0t^b ~,
 ~~ C(t)=C_0t^c ~,
\label{Rodrigues_D18}
\end{equation}
where $a,b,c$ and $A_0,
B_0,C_0$ are parameters. The various
expansion rates become
\begin{equation}
 H_x = \frac{a}{t} ~,~~
 H_y=\frac{b}{t} ~,~~ H_z=\frac{c}{t} ~,
\label{Rodrigues_D19}
\end{equation}
and moreover the torsion scalar (\ref{Rodrigues_torsionScalar1})
reads
\begin{equation}
 T = -2\left(\frac{ab}{t^2}+
 \frac{ac}{t^2}+
 \frac{bc}{t^2}\right)\ .
\label{Rodrigues_D20}
\end{equation}
Substituting the above expressions into the field equations
(\ref{Rodrigues_densitytype3})-(\ref{Rodrigues_tangentialpressure2type3}) we obtain
\cite{Rodrigues:2012qua}:
\begin{align}
&\!16\pi \rho_m(T) = f(T) -2Tf_{T}(T)\ , \label{Rodrigues_D21} \\
&\!\!\!\!
-16\pi w\rho_m(T)
= f(T) +\frac{(b+c)(1-a-b-c)}{bc+a(b+c)}Tf_{T}(T)\nonumber\\
 &\ \ \ \ \ \ \ \ \ \ \ \ \ \ \ \  \ \ \
+2\frac{(b+c)}{bc+a(b+c)}T^2f_{,
TT}(T)\ ,
\label{Rodrigues_D22}\\
&\!\!\!\!-16\pi w\rho_m(T) = f(T) +\frac{(a+c)(1-a-b-c)}{bc+a(b+c)}Tf_{T}(T) \nonumber\\
 &\ \ \ \ \ \ \ \ \ \ \ \ \ \ \ \  \ \ \
+2\frac{(a+c)}{bc+a(b+c)}T^2
f_{TT}(T)\ ,
\label{Rodrigues_D23} \\
&\!\!\!\!-16\pi w\rho_m(T) = 
f(T) +\frac{(a+b)(1-a-b-c)}{bc+a(b+c)}Tf_{T}(T) \nonumber\\
 &\ \ \ \ \ \ \ \ \ \ \ \ \ \ \ \  \ \ \
+2\frac{(a+b)}{bc+a(b+c)}
T^2f_{TT}(T)\ , \label{Rodrigues_D24}
\end{align}
where we have made the simplified assumption $p_x=p_y=p_z=p_m$ as well as $w=p_m/\rho_m$.
The continuity equation (\ref{Rodrigues_D1}) leads to the solution
\begin{eqnarray}
&&\!\!\!\!\!\!\!\!\! \rho_m=\rho_{m0}\, t^{-(a+b+c)(1+w)}\nonumber\\
 &&=\rho_{m0}
\left(-\frac{T}{2(ab+ac+bc)}\right)^{\frac{(a+b+c)(1+w)}{2}}\
,
\label{Rodrigues_D25}
\end{eqnarray}
while equations  (\ref{Rodrigues_D21})-(\ref{Rodrigues_D24}) give
\begin{eqnarray}
 f(T)=c_1 \sqrt{-T}+c_2\ T^{\frac{(1+w)(a+b+c)}{2}}\ ,
\label{Rodrigues_D26}
\end{eqnarray}
with $c_1$ an integration constant and  $c_2$ a constant given by
 {\small{
\begin{equation}
c_2=\frac{2^{4-(1+w)(a+b+c)/2}\pi \rho_{m0}}{\left[-1+(1+w)(a+b+c)\right]
\left[-bc-a(b+c)\right]^{\frac{(1+w)(a+b+c)}{2}}}\ .
\label{Rodrigues_D28}
\end{equation}}}
Note that in order to have real solutions, $(1+w)(a+b+c)=$ has to be an even number.
Finally, in order to obtain consistency, the following relations
have to be imposed:
\begin{enumerate}
\item $ c=\frac{1-w(a+b)}{w} $, with $w\neq 0$. This case corresponds to an anisotropic
solution, with $A(t)$, $B(t)$ and $C(t)$ being different than
(\ref{Rodrigues_D18}).

\item $a=b=c$. In this case the cosmological evolution expressed by
\eqref{Rodrigues_D18} reduces to the
FRW case.
\end{enumerate}
Lastly, we mention that in vacuum, only an FRW solution is possible.

%%%%%%%%%%%%%%%%%%%%%%%
\subsubsection{Bianchi type-III solutions}
%%%%%%%%%%%%%%%%%%%%

Let us now examine the more complicated  Bianchi-III case. The complication arises from
the constraint equations (\ref{Rodrigues_constraint1}) and (\ref{Rodrigues_constraint2}).
In particular, (\ref{Rodrigues_constraint2}), gives ${\dot{A}}/{A} =
{\dot{B}}/{B}$, which combining with (\ref{Rodrigues_constraint1}) yields
\begin{eqnarray} \label{Rodrigues_constraint3}
 \dot{T}f_{TT}=0 ~.
\end{eqnarray}
This implies that either $\dot{T}=0$ or $f_{TT}=0$. Since the second case corresponds to
TEGR, we restrict our analysis to the first one.

Relation $\dot{T}=0$ implies a constant torsion scalar, namely
(\ref{Rodrigues_torsionScalar1}) gives
\begin{eqnarray} \label{Rodrigues_steph111}
\frac{\dot{A}^2}{A^2}+2\frac{\dot{A}\dot{C}}{AC}=K ~,
\end{eqnarray}
with $K$ a positive constant. Considering $A=C^n$, with $n>0$ or $n<-2$,
(\ref{Rodrigues_steph111}) yields
\begin{eqnarray}
 &&\!\!\!\!\!\!\!\!\!\!\!\!\!\!\!\!\!\!\!
 C(t) = c_3 e^{\left( \pm \sqrt{\frac{K}{n(n+2)}}
t \right) } \\
\label{Rodrigues_vincent18}
 &&\!\!\!\!\!\!\!\!\!\!\!\!\!\!\!\!\!\!\!
 A(t) = B(t) = \left\{\begin{array}{ll}
 c_3^n e^{ \left( -n \sqrt{\frac{K}{n(n+2)}} t \right) } \,, \quad
\mbox{for}\,\,\,
n<-2 \\
 c_3^n e^{ \left( n \sqrt{\frac{K}{n(n+2)}} t \right) } \,, \;\;\;\quad
\mbox{for}\,\,\, n>0
\,\,\,, \end{array} \right.
\end{eqnarray}
where $c_3$ is a positive constant. Hence, the expansion rate for all three
spatial directions is constant, i.e. we obtain 
a de Sitter universe.

The remaining equations, namely
(\ref{Rodrigues_densitytype3})-(\ref{Rodrigues_tangentialpressure2type3}), can now be
straightforwardly solved, leading to \cite{Rodrigues:2012qua}
\begin{eqnarray}
 f(T) = c_4\exp{\left[R(n)T\right]}~,
\end{eqnarray}
with  $c_4$ is an integration constant and
\begin{eqnarray}
 R(n) = \frac{n(n+2)(\omega_z+1)}{2K\left[2n(n-1)\omega_x
-(\omega_z+1)(2n^2+3n+1)\right]}
~. \nonumber
\end{eqnarray}
We mention that for $n=1$ and $\omega_x=\omega_z$ we re-obtain
(\ref{Rodrigues_D10}) as expected.

\subsection{Observational constraints}

As we have analyzed in detail, the paradigm of $f(T)$ gravity can give rise to a
late-time accelerated expansion in a universe filled with regular matter. In this
mechanism, the cosmic speed up is driven by the space-time torsion, that can be
effectively treated as a dark fluid whose equation-of-state parameter can be determined
by the form of $f(T)$. Hence, given a specific $f(T)$ scenario, one can compare the
dynamical evolution predicted by this model with the
observational data and then examine how it is constrained or whether it is ruled out
\cite{Wu:2010mn,
Bengochea:2010sg, Fu:2011zze,Cardone:2012xq, Iorio:2012cm,Xie:2013vua,
Nesseris:2013jea,Iorio:2015rla,Wu:2014wra}.

In this subsection we follow the work of \cite{Cardone:2012xq} and consider two
particular examples for $f(T)$, which possess the nice property to emulate the
phantom-divide crossing. Using these $f(T)$ forms, one can examine their viability
confronting the predicted background evolution with the observed one through the SNIa and 
Gamma Ray 
Bursts (GRBs), the measurement of the rate expansion $H(z)$, the BAOs at different 
redshifts, and 
the CMB distance priors. One may further distinguish them
relying on the different growth factors. We would like to mention that all data sets
are not up-to-date, since the study in the present subsection is merely to demonstrate
the data analysis method of constraining $f(T)$ models rather than to report on the
latest observational data.

We are interested in models that are able to describe the phantom-divide crossing. Two
such models are \cite{Wu:2010av}
\begin{align}
\label{Cardone_eq:ftmods}
 F(T) \equiv f(T) - T &= \alpha (-T)^n \tanh \Big( \frac{T_0}{T} \Big), \\
\label{Cardone_eq:ftmods1}
 F(T) \equiv f(T) - T &= \alpha (-T)^n \Big[ 1 - \exp{\Big( - p \frac{T_0}{T} \Big) } 
\Big],
\end{align}
where the subscript $0$ marks the present value. In the first model (called ``the tanh
model'')  one must set $n > 3/2$ in order to obtain an effective dark-energy fluid
with a positive-defined energy density, whereas for the second model (called ``the exp
model'') the same requirement leads to $n > 1/2$. The parameter $\alpha$ can
be expressed in terms of the present-day quantities as
\begin{align}
\label{Cardone_eq: alpha}
 \alpha_{tanh} &= 
 - \frac{\left 
 ( 6 H_0^2
 \right )
 ^{1 - n}
 (1 - \Omega_{m0} -
\Omega_{r0})}{2
{\rm sech}^2
{(1)} + (1 - 2n) \tanh{( 1)}}, \\
\label{Cardone_eq: alpha1}
 \alpha_{exp} &= 
 - \frac{\left ( 6 H_0^2 
 \right )^{1 - n}
 (1 - \Omega_{m0} -
\Omega_{r0})}{1 
- 2n - (1 - 2n
+ 2p) 
{\rm e}^p},
\end{align}
for the tanh and exp model respectively.
\begin{figure}[ht]
\centering
\includegraphics[width = 7cm]{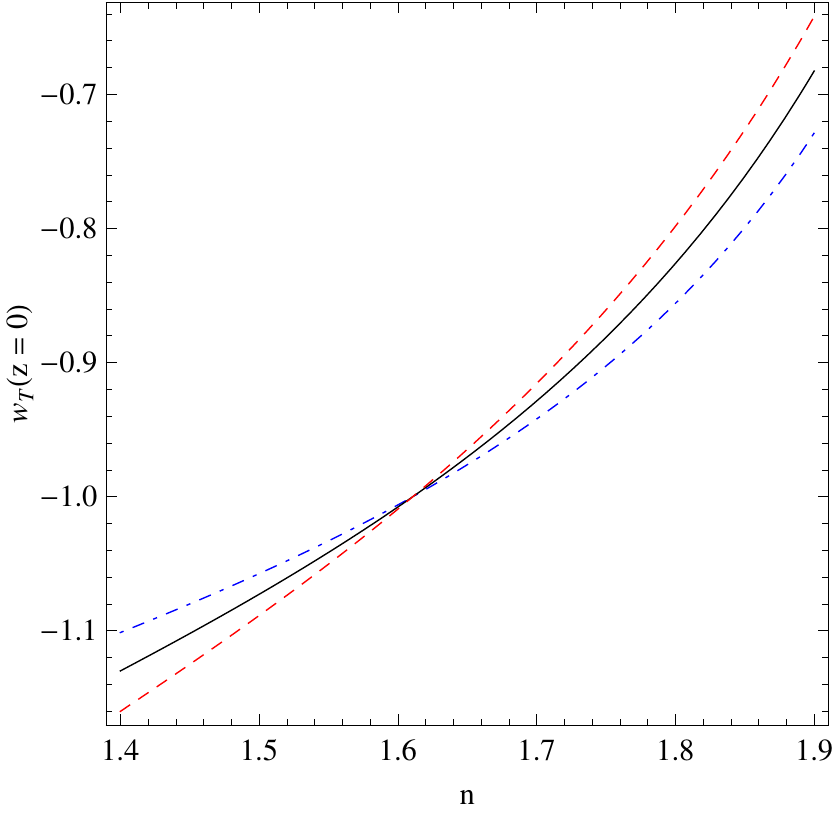}
\caption{\it Today's value of the torsion-induced, effective dark-energy
equation-of-state parameter, for the tanh model in (\ref{Cardone_eq:ftmods}), for
$\Omega_{m0} = 0.20$ ( dashed-dotted blue), $\Omega_{m0} = 0.25$ (solid-black), and 
$\Omega_{m0} = 0.30$ (dashed-red).
From \cite{Cardone:2012xq}. }
\label{Cardone_fig: wzefftanh}
\end{figure}
\begin{figure}[ht]
\centering
\includegraphics[width = 7cm]{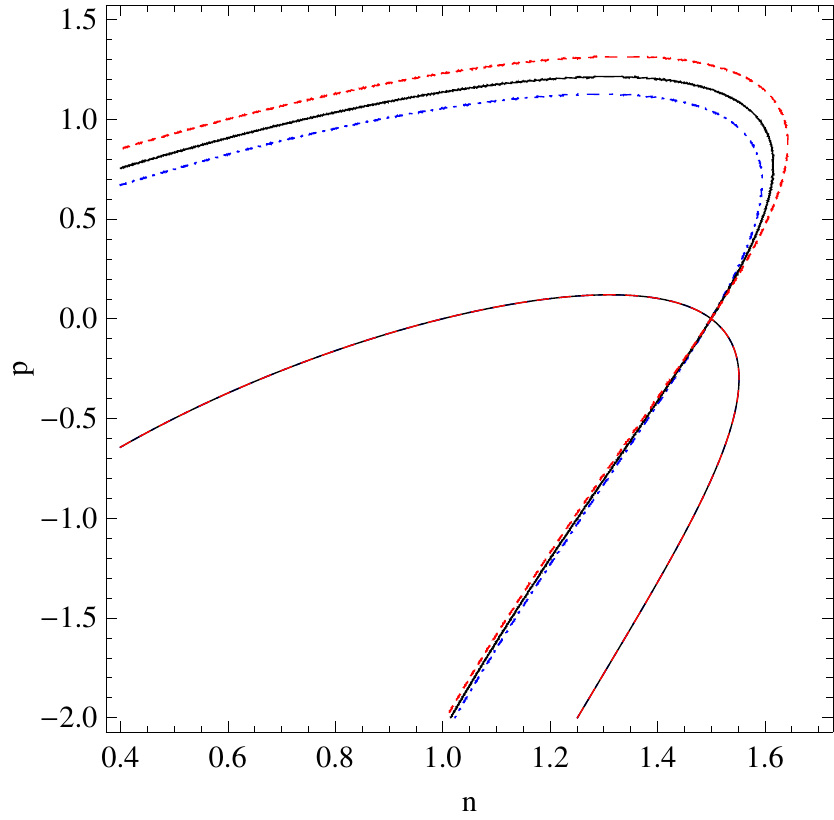}
\caption{\it Likelihood contours in the $(n, p)$ plane  for 
the
exp model in (\ref{Cardone_eq:ftmods1}), having set $w_{T}(z = 0) = -1$, and 
  $\Omega_{m0} = 0.20$ (dashed-dotted blue),
$\Omega_{m0} = 0.25$ (solid black), and $\Omega_{m0} = 0.30$ (dashed red). From 
\cite{Cardone:2012xq}. }
\label{Cardone_fig: wzeffexp}
\end{figure}

We recall that the effective equation-of-state parameter of the dark fluid arising from
$f(T)$ sector has been defined in (\ref{wde_fT_DE}). One can insert
(\ref{Cardone_eq:ftmods}),(\ref{Cardone_eq:ftmods1}) into the definition 
(\ref{wde_fT_DE}) and derive the explicit
expressions for the two models. In Figs.~\ref{Cardone_fig: wzefftanh} and
\ref{Cardone_fig: wzeffexp} we can see the dependence of its present day value
$w_{T}(z = 0)$ on the $f(T)$ parameters, respectively for the tanh and exp model. Since
$\Lambda$CDM paradigm fits the data very efficiently, we expect that
in order to be able to fit the data the parameters of both models would be constrained
in a region corresponding to $w_{T}( 0) \approx -1$.
Fig.~\ref{Cardone_fig: wzefftanh} reveals that this favors
values of $n \simeq 1.6$ for the tanh model,  and that the larger 
the $n$ is the smaller $\Omega_{m0}$ must
be in order to obtain an effective equation-of-state parameter close to the $\Lambda$CDM 
one
at present. For the exp model Fig.~\ref{Cardone_fig: wzeffexp} shows that for given
$(\Omega_{m0}, n)$ values there can be models with different $p$  but with the same
$w_{T}(0) = -1$. The $(n, p)$ parameters are thus degenerate and hence the
condition $w_{T}( 0) = -1$ cannot distinguish amongst different parameter
sets. Hence, taking into account the evolution of $w_{T}$, and not only its present
value, will break the degeneracy among $(n, \Omega_{m0})$ for the tanh model and among
$(n, p)$ for the exp one.

\subsubsection{$f(T)$ gravity versus cosmological data}
\label{subsubsec: Cardone_data}

In order to examine whether $f(T)$ gravity can be consistent with the observed Universe, 
we study 
the parameter space of a fixed model by analyzing the likelihood function as follows
\begin{equation} 
\label{Cardone_eq: totlike}
 {\cal{L}}({\bf p})
 = {\cal{L}}_{\mu}
 ({\bf p})
 \times {\cal{L}}_{H}
 ({\bf p})
 \times
{\cal{L}}_{\rm
BAO}({\bf p}) 
\times 
{\cal{L}}_{\rm CMB}({\bf p}) ~,
\end{equation}
where   model parameters are
\begin{displaymath}
 {\bf p} = \left \{
 \begin{array}{l}
 (\Omega_m, 
 h, n) \\
%~ \\
 (\Omega
 _m, h,
 n, p) \\
 \end{array}
 \right .
\end{displaymath}
respectively for the tanh and exp models. Note that $h$ is the Hubble constant $H_0$ in
units of $100 \ {\rm km s^{-1}Mpc^{-1}}$, and today's radiation density parameter is set 
to be
\begin{displaymath}
 \Omega_{r} = \omega_{\gamma} h^{-2} (1 + 0.2271 N_{eff}) ~,
\end{displaymath}
with $(\omega_{\gamma}, N_{eff}) = (2.469 \times 10^{-5}, 3.04)$ in agreement with WMAP7
data \cite{Komatsu:2010fb} (note that the data are not up-to-date since here we are
interested in demonstrating the procedure of data analysis rather than to investigate the
latest observational constraints).

All terms in (\ref{Cardone_eq: totlike}) are expressed as:
\begin{displaymath}
{\cal{L}}_i({\bf p}) =
\frac{\exp{\left 
[ - \chi^2_i({\bf p})/2 \right ]}}{(2
\pi)^{{\cal{N}}_i/2} 
\Gamma_i^{1/2}} ~,
\end{displaymath}
where $({\cal{N}}_i, \Gamma_i, \chi^2_i)$ are determined by the detailed dataset. The 
first refers to the 
Hubble diagram.
This gives the distance modulus $\mu$ as a function of the redshift $z$ via the following 
relation:
\begin{equation}\label{Cardone_eq: defmu}
 \mu(z) = 25 + 5 \log{d_L(z)} = 25 + 5 \log{\left [ (1 + z) r(z) \right ]} ~,
\end{equation}
where $d_L(z) = (1 + z) r(z)$ is the luminosity distance and
\begin{equation}\label{Cardone_eq: defrz}
 r(z) = \frac{c}{H_0} \int_{0}^{z}{\frac{dz^{\prime}}{E(z^{\prime}, {\bf p})}}
\end{equation}
is the comoving distance. For instance, one can apply both the Union2 SNIa dataset 
\cite{Amanullah:2010vv}, comprising ${\cal{N}}_{\rm SNeIa} = 557$ objects with $0.015 \le 
z \le 1.40$, and ${\cal{N}
}_{\rm GRB} = 64$ GRBs to probe the redshift range $1.48 \le z \le 5.60$ with the 
distance 
modulus estimated in \cite{Cardone:2011ga}. For both data 
sets we have
\begin{eqnarray}
 &&\Gamma_i = 
 \prod_{j = 1}^{{\cal{N}}_{i}}
 {\left 
 ( \sigma_{j}^2 
 + \sigma_{int}^2 \right
)}
~,\nonumber\\
&& \chi^2_{i} 
= \sum_{j = 1}^{{\cal{N}}_{j}}
{ \Big[ 
\frac{\mu_{obs}(z_j) -
\mu_{th}(z_j,
{\bf p})}{\sqrt{\sigma_{j}^2
+ \sigma_{int}^2}} \Big]^2} ~,
\end{eqnarray}
where $\sigma_j$ is the observational error for the $j$\,-\,th object, and $\sigma_{int}$
depicts the intrinsic scatter of the tracer around the relations incorporated to
estimate the value of $\mu$. For SNIa, one can set $\sigma_{int} = 0$, and for GRBs
this is a nuisance parameter that can be marginalized over. As the Hubble diagram probes 
the
integrated expansion rate, the second data set refers to the Hubble parameter
$H(z)$ which can be determined through the differential age method \cite{Jimenez:2001gg}. 
Applying
red envelope galaxies as cosmic chronometers \cite{Stern:2009er}, $H(z)$ for the
redshift interval $0.10 
\le z \le  
1.75$ was estimated in \cite{Stern:2009ep}. In this
sample one adds the $H_0$ determination from local distance ladders acquired from 
SHOES \cite{Riess:2009pu} and extract a total $\chi^2$ similarly to SNIa.

Except for SNIa and GRBs, the BAOs can also work as standardizable candles to probe the 
distance-redshift relation. Thus, one can add ${\cal{L}}_{BAO}$ to the full likelihood 
following the method 
of \cite{Blake:2011en}.
Firstly, one can apply the 6dFGRS \cite{Beutler:2011hx} and the SDSS 
\cite{Percival:2009xn} to 
find the scaled volume distance parameter through
\begin{equation}
\label{Cardone_eq: defdz}
 d_z = \frac{r_s(z_d)}
 {d_V(z)} = r_s(z_d)
 \times
 \left [
 \frac{c z r^2(z)}
 {H_0 E(z)}
\right 
]^{-\frac{1}
{3}} ~,
\end{equation}
where the sound horizon to distance $z$ reads
\begin{equation}
\label{Cardone_eq: defrs}
 r_s(z) 
 = \frac{c}{\sqrt{3} H_0}
 \int_{z}^{\infty}
 {\frac{E^{-1}(z) 
 dz^{\prime}}
 {\sqrt{1 +
(3 \omega_b)
/(4 \omega_r) (1 +
z^{\prime})^{-1}}}} ,
\end{equation}
with
$z_d$ the drag redshift. The value of $d_z$ at $z = 0.106$ and the corresponding error 
is from \cite{Beutler:2011hx}, and  \cite{Percival:2009xn} provides $d_z$ for $z = 0.20$ 
and $z = 0.35$ along with the corresponding covariance matrix.
Secondly, BAO constraints from the WiggleZ survey \cite{Blake:2011en} quantify the 
acoustic 
parameter \cite{Eisenstein:2005su}:
\begin{equation}\label{Cardone_eq: defaz}
 {\cal{A}}(z)
 = 
 {\sqrt
 {\Omega_m H_0^2}
 d_V(z)}/(c z) .
\end{equation}
Then one can apply the observed values and their covariance matrix for ${\cal{A}}(z)$ 
estimations at $z = (0.44, 0.60, 0.70)$ given in \cite{Blake:2011en}, and define the 
BAO likelihood as 
follows:
\begin{displaymath}
 {\cal{L}}_{\rm BAO}({\bf p})  
 = {
 \cal{L}}_{\rm 6dFGRS}({\bf p})
 \times {\cal{L}}_{\rm 
SDSS}({\bf p})
 \times {\cal{L}}_{\rm
 WiggleZ}({\bf p}) ,
\end{displaymath}
where:
\begin{align}
 {\cal{L}}_{\rm 6dFGRS} 
 &= \frac{1}{\sqrt{2 
 \pi \sigma_{0.106}^2}} \times
 \exp{\left \{ 
 - \frac{1}{2} \Big[ 
 \frac{d_{0.106}^{obs} 
 - d_{0.106}^{th}({\bf
p})}{
\sigma_{0.106}}
 \Big]^2 \right \}}
 ~, \nonumber\\
 {\cal{L}}_{\rm SDSS}
 &= \frac{1}{(2 \pi)^{{\cal{N}}_{SDSS}}
 | {\bf C}_{SDSS} |^{1/2}}
\nonumber\\
&\ \ \
\times
 \exp{ \Big[ - 
 \frac{{\bf D}_{SDSS}^{T}({\bf p})
 {\bf C}_{SDSS}^{-1} {\bf D}_{SDSS}({\bf
p})}{2} \Big]}
~, \nonumber\\
 {\cal{L}}_{\rm WiggleZ}
 & = \frac{1}{(2 \pi)^{{\cal{N}}_{WiggleZ}}
 | {\bf C}_{WiggleZ}
|^{1/2}}
\nonumber\\
&\ \ \
\times
 \exp{ \Big[ 
 -\frac{{\bf
 D}_{WiggleZ}^{T}({\bf p})
 {\bf C}_{WiggleZ}^{-1} {\bf
D}_{WiggleZ}({\bf p}
)}{2} \Big]}  ~, \nonumber
\end{align}
with ${\bf D}$ a ${\cal{N}}_i$ dimensional vector accounting for the difference between 
predicted and observed values and ${\bf C}_i$ the accompanying covariance matrix.

The last term in the likelihood \eqref{Cardone_eq: totlike} refers to the CMB distance 
priors, 
which are viewed as a very efficient approach to include the CMB constraints without 
calculating the 
complete anisotropic spectrum. Based on \cite{Komatsu:2010fb}, the CMB likelihood can be 
determined 
similarly to the SDSS and WiggleZ ones. However, the observable quantities become the 
redshift $z_{\star}$ at the last scattering surface, calculated by the expressions of 
\cite{Hu:1995en}, 
the shift parameter 
${\cal{R}}$ \cite{Bond:1997wr}
\begin{equation}\label{Cardone_eq: defshiftpar}
 {\cal{R}} = {\sqrt{\Omega_m} r(z_{\star})H_0}/{c} ~,
\end{equation}
and
the 
acoustic scale $\ell_A = \pi r(z_{\star})/r_s(z_{\star})$. 
Additionally, to use the distance priors one needs to assume that the early universe is 
dominated by matter and that the dark energy is negligible. 
Despite the fact that the $f(T)$ contribution does 
not vanish at high $z$, one can easily check that the effective energy density of the 
torsion-induced fluid is smaller than the matter-fluid  energy density  by many 
orders of magnitude  at $z_{\star}$, so 
that the CMB determination of the distance priors is applicable.

Combining the above, one can use a Markov Chain Monte Carlo (MCMC) code, apply multiple 
chains and use  the Gelman-Rubin criterium in order to check the convergence, 
and then efficiently examine the 
parameter spaces of models under consideration \cite{Gelman:1992zz}. The best-fit 
parameters are 
the ones that maximize the total likelihood, however the most reliable constraints on 
a parameter 
$p_i$ are acquired through marginalization over all parameters apart from the $i$-th one. 
Applying this, one can 
determine the mean and median as a 
reference value, and apply the $68\%$ and $95\%$ 
confidence ranges 
as the $1\sigma$ and the $2 \sigma$ errors.

\begin{table}[ht]
\begin{center}
\begin{tabular}{cccccc}
\hline
~ & $x_{BF}$ & $\langle x \rangle$ & $\tilde{x}$ & $68\% \ {\rm CL}$ & $95\% \ {\rm CL}$
\\
\hline \hline
~ & ~ & ~ & ~ & ~ & ~ \\
$\Omega_m$ & 0.286 & 0.286 & 0.287 & (0.274, 0.299) & (0.264, 0.311) \\
~ & ~ & ~ & ~ & ~ & ~ \\
$h$ & 0.719 & 0.722 & 0.722 & (0.712, 0.734) & (0.702, 0.745) \\
~ & ~ & ~ & ~ & ~ & ~ \\
$n$ & 1.616 & 1.610 & 1.615 & (1.581, 1.636) & (1.547, 1.667) \\
~ & ~ & ~ & ~ & ~ & ~ \\
\hline
\end{tabular}
\caption{Constraints on   the tanh-model parameters of (\ref{Cardone_eq:ftmods}).
The columns correspond to: 1.) Parameter name, 2.) Best-fit value, 3.) Mean value, 
4.) Median value, 5.),
6.) $68\%$ and $95\%$ confidence
levels respectively. From \cite{Cardone:2012xq}. }
\label{Cardone_tab: tanhfit}
\end{center}
\end{table}

\begin{table}[ht]
\begin{center}
\begin{tabular}{cccccc}
\hline
~ & $x_{BF}$ & $\langle x \rangle$ & $\tilde{x}$ & $68\% \ {\rm CL}$ & $95\% \ {\rm CL}$
\\
\hline \hline
~ & ~ & ~ & ~ & ~ & ~ \\
$\Omega_m$ & 0.284 & 0.286 & 0.287 & (0.276, 0.297) & (0.265, 0.308) \\
~ & ~ & ~ & ~ & ~ & ~ \\
$h$ & 0.724 & 0.731 & 0.731 & (0.723, 0.740) & (0.713, 0.749) \\
~ & ~ & ~ & ~ & ~ & ~ \\
$n$ & 1.152 & 0.757 & 0.736 & (0.577, 0.939) & (0.514, 1.103) \\
~ & ~ & ~ & ~ & ~ & ~ \\
$p$ & 0.814 & -0.110 & -0.100 & (-0.263, 0.046) & (-0.395, 0.131) \\
~ & ~ & ~ & ~ & ~ & ~ \\
\hline
\end{tabular}
\caption{The same as Table \ref{Cardone_tab: tanhfit} but for the exp model of 
(\ref{Cardone_eq:ftmods1}). From \cite{Cardone:2012xq}. }
\label{Cardone_tab: expfit}
\end{center}
\end{table}

\begin{figure}[!]
\centering
\includegraphics[width=7.8cm]{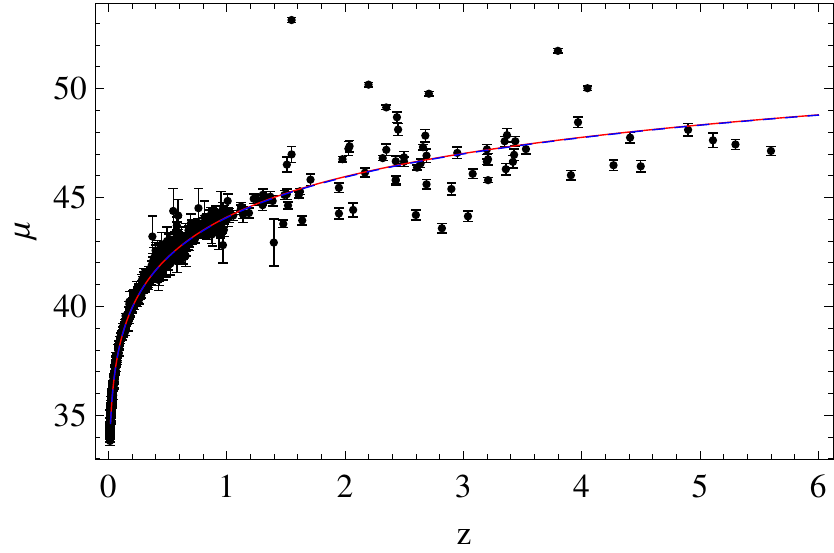}
\includegraphics[width=7.8cm]{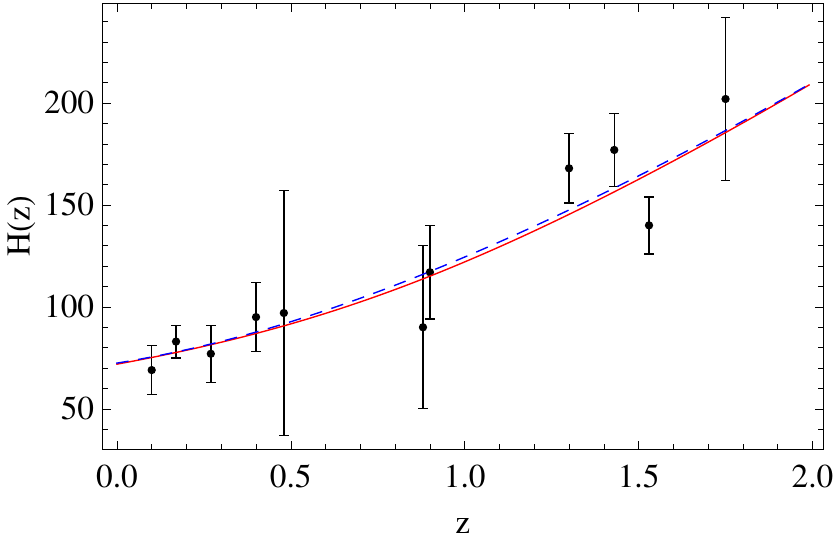}
\caption{\it Best-fit curves on top of the SNIa\,+\,GRB Hubble diagram (upper) and
$H(z)$ data (lower), for the tanh model  of 
(\ref{Cardone_eq:ftmods}) (solid red) and for the exp model  of 
(\ref{Cardone_eq:ftmods1}) (dashed blue). We mention that in the Figure resolution scale  
the $\mu(z)$ curves are almost  superimposed and hence  no significant difference is 
visible. From
 \cite{Cardone:2012xq}. }
\label{Cardone_fig: bfplot}
\end{figure}

Based on the analysis of \cite{Cardone:2012xq}, the best-fit values and marginalized 
constraints of 
parameters for the tanh and exp models are listed in Tables \ref{Cardone_tab: tanhfit} and 
\ref{Cardone_tab: expfit} respectively. From Fig.~\ref{Cardone_fig: bfplot} we can see 
the agreement of the best-fit model 
predictions, with the SNIa\,+\,GRB Hubble diagram and $H(z)$ data. The total efficiency 
 of the 
fitting procedure could be also become obvious through comparison of the model 
predictions for the BAO and CMB quantities 
with the ones arise from  observations. In particular, for the tanh model, one 
obtains \cite{Cardone:2012xq}
\begin{displaymath}
\left \{
 \begin{array}{lll}
 d_{0.106}^{bf} = 0.3418 & 
 {\rm vs} & d_{0.106}^{obs} = 0.336 \pm 0.015 \\
 %~ & ~ & ~ \\
 d_{0.200}^{bf} = 0.1864 
 & {\rm vs} & d_{0.200}^{obs} = 0.1905 \pm 0.0061 \\
 %~ & ~ & ~ \\
 d_{0.350}^{bf} = 0.1173 
 & {\rm vs} & d_{0.350}^{obs} = 
 0.1097 \pm 0.0036 \\
 \end{array}
 \right . \ ,
\end{displaymath}
\begin{displaymath}
 \left \{
 \begin{array}{lll}
 {\cal{A}}^{bf}(0.44) = 0.467
 & {\rm vs} & {\cal{A}}^{obs}
 (0.44) = 0.474 \pm 0.034 \\
 %~ & ~ & ~ \\
 {\cal{A}}^{bf}(0.60) = 0.442 &
 {\rm vs} & {\cal{A}}^{obs}(0.60) 
 = 0.442 \pm 0.020 \\
 %~ & ~ & ~ \\
 {\cal{A}}^{bf}(0.73) = 0.422 
 & {\rm vs} & 
 {\cal{A}}^{obs}(0.73) = 0.424 \pm 0.021 \\
 \end{array}
 \right . \ ,
\end{displaymath}
\begin{displaymath}
 \left \{
 \begin{array}{lll}
 \ell_A^{bf} = 302.66 & {\rm vs}
 & \ell_A^{obs} = 302.09 \pm 0.76 \\
 %~ & ~ & ~ \\
 {\cal{R}}^{bf} = 1.733 & 
 {\rm vs} & {\cal{R}}^{obs} = 1.725 \pm 0.018 \\
 %~ & ~ & ~ \\
 z_{\star}^{bf} = 1092.04 &
 {\rm vs} & z_{\star}^{obs} = 1091.30 \pm 0.91 \\
 \end{array}
 \right . \ ,
\end{displaymath}
and thus the best-fit tanh-model predictions lie far inside the $1 \sigma$ region
comparing to  the observed values. 
Moreover, for the exp model, one obtains \cite{Cardone:2012xq}
\begin{displaymath}
 \left \{
 \begin{array}{lll}
 d_{0.106}^{bf} = 0.3428 & 
 {\rm vs} & d_{0.106}^{obs} = 0.336 \pm 0.015 \\
 %~ & ~ & ~ \\
 d_{0.200}^{bf} = 0.1865 &
 {\rm vs} & d_{0.200}^{obs} =
 0.1905 \pm 0.0061 \\
 %~ & ~ & ~ \\
 d_{0.350}^{bf} = 0.1121 & 
 {\rm vs} & d_{0.350}^{obs} = 0.1097 \pm 0.0036 \\
 \end{array}
 \right . \ ,
\end{displaymath}
\begin{displaymath}
 \left \{
 \begin{array}{lll}
 {\cal{A}}^{bf}(0.44) =
 0.465 & {\rm vs} & {\cal{A}}^{obs}(0.44)
 = 0.474 \pm 0.034 \\
 %~ & ~ & ~ \\
 {\cal{A}}^{bf}(0.60) = 0.439 & {\rm vs} 
 & {\cal{A}}^{obs}(0.60) = 0.442 \pm 0.020 \\
 %~ & ~ & ~ \\
 {\cal{A}}^{bf}(0.73) = 0.418 & {\rm vs} &
 {\cal{A}}^{obs}(0.73) = 0.424 \pm 0.021 \\
 \end{array}
 \right . \ ,
\end{displaymath}
\begin{displaymath}
 \left \{
 \begin{array}{lll}
 \ell_A^{bf} = 302.54 &
 {\rm vs} & \ell_A^{obs} = 302.09 \pm 0.76 \\
 %~ & ~ & ~ \\
 {\cal{R}}^{bf} = 1.735
 & {\rm vs} & {\cal{R}}^{obs} = 1.725 \pm 0.018 \\
 %~ & ~ & ~ \\
 z_{\star}^{bf} = 1092.12 & 
 {\rm vs} & z_{\star}^{obs} = 
 1091.30 \pm 0.91 \\
 \end{array}
 \right . \ ,
\end{displaymath}
which are also in agreement with the BAO and CMB observations.

The upper graph of Fig.~\ref{Cardone_fig: bfplot} shows that two best-fit models cannot 
be 
discriminated by the Hubble diagram data. This can be read from the lower graph too, 
where $H(z)$ is similar within these two models, with the largest deviations appearing in 
the redshift interval 
$(0.5, 1.5)$. Since it is only a few percent, it is relatively obvious  
that the required integration in order to calculate the distance modulus, 
 dilutes out the differences between the above two 
models. Therefore, as shown in Fig.~\ref{Cardone_fig: bfplot}, this yields the 
very efficient fitting.
\begin{figure}[ht]
\centering
\subfigure{\includegraphics[width=8cm]{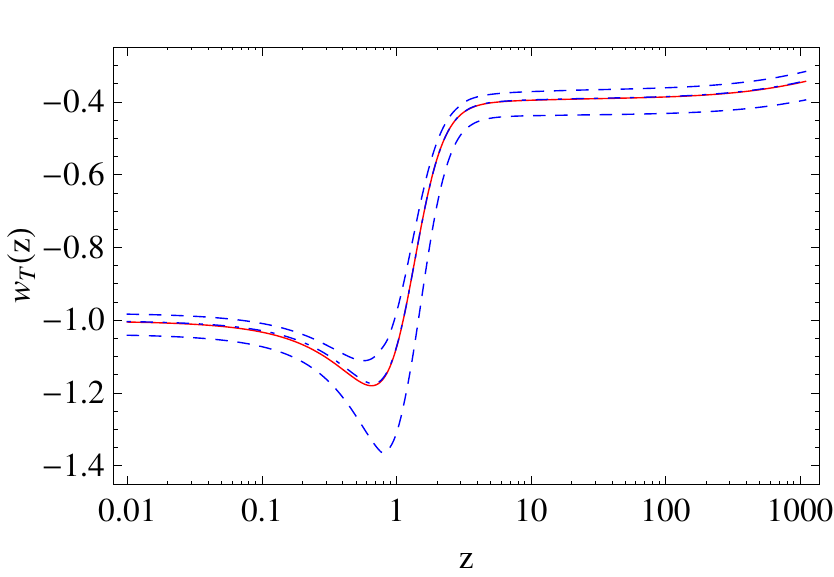}}
\subfigure{\includegraphics[width=8cm]{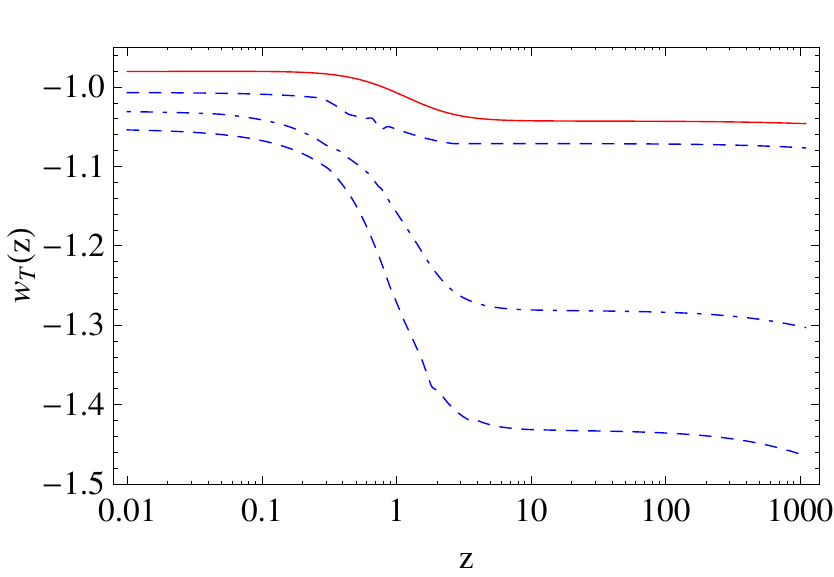}}
\caption{\it The constrained torsion-induced effective equation-of-state parameter 
$w_T(z)$, for the tanh model of (\ref{Cardone_eq:ftmods})
(upper) and for the exp model of (\ref{Cardone_eq:ftmods1}) (lower). From
\cite{Cardone:2012xq}. }
\label{Cardone_fig: eosplot}
\end{figure}

An alternative way to understand why two models lead to similar predictions is to analyze 
the 
reconstructed effective equation-of-state parameter induced by torsional gravity. In 
Fig.~\ref{Cardone_fig: eosplot} we present $w_T(z)$, constrained by the likelihood 
analysis. In this Figure 
the red solid curve corresponds to the best-fit model, and the blue dot-dashed (dashed) 
one provides  the median value (the $68\%$ confidence range) 
for each $z$, as they are extracted by calculating 
$w_T(z, {\bf p}
)$ for the  parameters of the model in the MCMC chain. For two best-fit models (the red 
curves), $w_T$ is 
in the vicinity of the cosmological constant $w = -1$ up to $z \sim 1$, and thus the 
dynamics is 
roughly the same with the $\Lambda$CDM scenario. On the contrary, $w_T$ exhibits a 
significantly different behavior at high redshifts in these two models. Particularly, for 
the tanh model we obtain an 
increasing $w_T$; whereas for the exp model it yields an approximately constant $w_T$ to 
a value 
slightly smaller than the present one and hence it can mimic the $\Lambda$CDM paradigm 
better than 
the tanh model. However, both models lead to similar predictions for the CMB 
distance priors 
due to the very limited contribution of the torsion induced dark energy up to the last 
scattering 
surface.

\subsubsection{Discriminations from the growth factor}
\label{subsubsec: Cardone_growth}

The aforementioned analysis showed that both tanh and exp $f(T)$ models are in agreement 
with observations, however it revealed that looking only at the background evolution 
cannot fully distinguish between them and the $\Lambda$CDM 
paradigm. Hence, in order to extract better constraints one should study  the evolution of 
the perturbations. To be explicit, one can consider
the growth factor $g = d\ln{\delta_m}/d\ln{a}$, with the matter overdensity reading 
as $\delta_m \equiv \delta \rho_m/\rho_m$. The dynamics of $\delta_m$ arise from the
equation \cite{Zheng:2010am,Nesseris:2013jea}
\begin{equation}
 \ddot{\delta}_m + 2 H \dot{\delta}_m 
 - 4 \pi G_{eff} \rho_m \delta_m = 0 ~,
\label{Cardone_eq: growthfac}
\end{equation}
which coincides to the $\Lambda$CDM one as long as one replaces the gravitational 
constant $G$ by an effective one, namely $G_{eff} = G/(1 + f_T)$. We would like to 
mention that
(\ref{Cardone_eq: growthfac}) is valid only in the sub-horizon limit, that is for 
 $k >> {\cal{H}}, {\cal{H}}^{\prime}/{\cal{H}}, {\cal{H}}^2$, where  
${\cal{H}}$ is the 
comoving Hubble parameter. In the general case, one obtains a more complicated equation
\cite{Li:2011wu}, which is the one that should be used in order to calculate 
 large-scale probes, for example the matter power spectrum or weak-lensing observables. 
For a preliminary investigation however, it is adequate to use
Eq. (\ref{Cardone_eq: growthfac}), which is safely applicable at galactic and
galaxy-cluster scales.

\begin{figure}[ht]
\centering
\includegraphics[width=8.2cm]{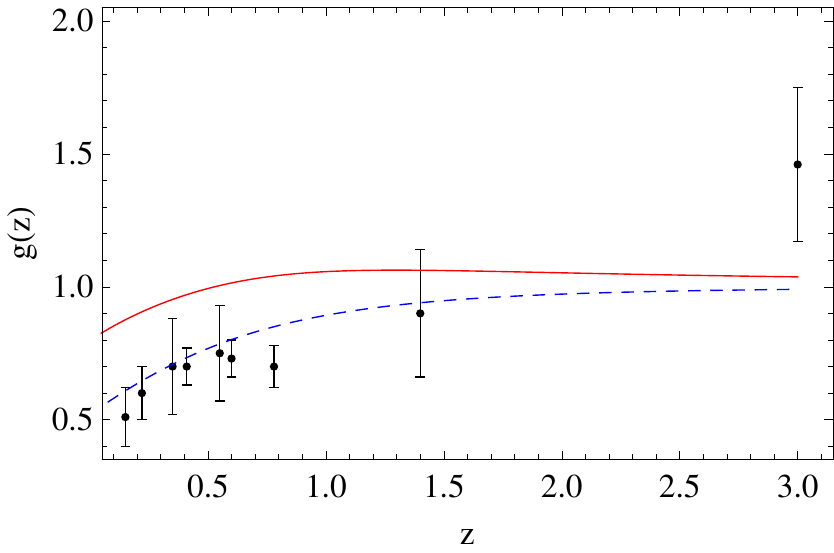}
\caption{\it The evolution of the growth factor $g(z)$ for the tanh model of 
(\ref{Cardone_eq:ftmods}) (solid red) and of the exp model of 
(\ref{Cardone_eq:ftmods1}) 
(dashed blue). From  \cite{Cardone:2012xq}. }
\label{Cardone_fig: gzplot}
\end{figure}

In Fig.~\ref{Cardone_fig: gzplot} we depict the growth factor evolution for the best-fit
tanh and the median exp models, along with a $g(z)$-measurement compilation given in  
\cite{Nesseris:2007pa, Blake:2011rj}. Apparently, these two models become easily
distinguishable from the different $g(z)$ behavior, although their induced background 
evolution is
almost identical, as it becomes obvious from the very good agreement of the 
corresponding $\mu(z)$
and $H(z)$ curves depicted in Fig.~\ref{Cardone_fig: bfplot}. These $g(z)$
values imply   that the  structure growth is slower in the exp model than in the tanh 
tone, and hence 
observationally determining $g(z)$ allows to distinguish  between these 
two
$f(T)$ models.

Using the data depicted in the Figures, for
the tanh model one can calculate $\tilde{\chi}^2 = 15.3$, 
which implies a significant disagreement. Nevertheless, for the exp model one acquires 
better results, namely $\tilde{\chi}^2 = 2.1$. We stress that these values have
to be used with   caution, 
and should not be taken as an  evidence that
the exp model is more efficient than the tanh one, for the following reasons: Firstly, 
most  measurements for $g(z)$ used in \cite{Nesseris:2007pa}
refer in practice to
$g(z)/b(z)$, where $b(z)$ stands for the galaxy population    bias  that traces 
the growth
factor. 
However, this bias is related to the specific underlying gravitational theory, and 
therefore a modified Lagrangian is expected to induce
  deviations from the usual collapse scenario, and hence
  to lead to a different bias, that might also be scale dependent. 
Secondly, $g(z)$ can be also acquired by  modeling   the matter power spectrum 
preliminarily \cite{Blake:2011rj}, but this procedure is partially based 
on the consideration of a $\Lambda$CDM-like model in order to  convert   the
redshift to   real space clustering. Hence, in summary, although the background evolution 
is almost the same
as the one in $\Lambda$CDM cosmology up to intermediate redshifts, the power-spectrum 
modeling    and the description  of its
distortions may be different.

\subsection{Cosmography in $f(T)$-gravity}

In the previous subsection we reviewed the regular method of data analysis that can
constrain different specific models of $f(T)$ gravity responsible for the present cosmic
acceleration. However, it is interesting to note that, if the universe is always
homogeneous and isotropic on large scales, there exists an almost model-independent
method to investigate $f(T)$ cosmology, namely the method of {\it cosmography}. In this
subsection we follow \cite{Capozziello:2011hj, Bamba:2012cp} and
\cite{Capozziello:2015rda} and we present how one could use cosmography in order to
extract the confidence ranges for $f(T)$ derivatives up to the fifth order, and then we
investigate the viability of given models.

\subsubsection{Cosmographic parameters and $f(T)$ derivatives}

Standard candles (such as SNIa and, to a limited extent, gamma ray bursts) are ideal
tools in modern cosmology, since they make it possible to reconstruct the Hubble diagram,
i.e. the redshift - distance relation, up to high redshift values. It is then customary to
assume a parameterized model (such as the concordance $\Lambda$CDM one or any other dark
energy scenario) and confront it against the data in order to check its viability and
constrain its characterizing parameters. As it is clear, such an approach is model
dependent and thus some doubts always remain on the validity of the constraints on
derived quantities,such as the present day values of the deceleration parameter and the
age of the universe.

In order to overcome this problem, one may resort to cosmography, i.e. expand the scale
factor in Taylor series with respect to the cosmic time. Such expansions lead to a
distance-redshift relation which only relies on the assumption of the FRW metric, and
thus it is fully model-independent since it does not depend on the particular form of the
solution of cosmic equations. To this aim, it is convenient to introduce the following
functions:
\begin{eqnarray}\label{CCFR_par}
 &&
 H = \frac{1}{a}\frac{da}{dt} ~,~~
 q = -\frac{1}{a}\frac{d^2a}{dt^2}H^{-2} ~,\nonumber\\
 &&
 j = \frac{1}{a}\frac{d^3a}{dt^3}H^{-3} ~,~~
 s = \frac{1}{a}\frac{d^4a}{dt^4}H^{-4} ~,\nonumber\\
 &&
 l = \frac{1}{a}\frac{d^5a}{dt^5}H^{-5} ~,
\end{eqnarray}
which are usually referred to as the Hubble, deceleration, jerk, snap and lerk
parameters, respectively. Their present day values (denoted with the subscript ``$0$'')
may be used to characterize the evolutionary status of the universe. For instance, $q_0 <
0$ denotes an accelerated expansion, while $j_0$ allows to distinguish amongst different
accelerating models. It is then a matter of algebra to demonstrate the following useful
relations:
\begin{align}\label{CCFR_dotH}
 & \dot H = -H^2 (1+q) ~, \nonumber\\
 & \ddot H = H^3 (j+3q+2) ~, \nonumber\\
 & \dddot H = H^4 [s-4j-3q(q+4)-6] ~, \nonumber\\
 & H^{(iv)} = H^5 [l-5s+10(q+2)j+30(q+2)q+24] ~,
\end{align}
where a dot denotes derivative with respect to the cosmic time $t$ and
$H^{(iv)}=d^4H/dt^4$. Eqs. (\ref{CCFR_dotH}) make it possible to relate the derivative of
the Hubble parameter to the other cosmographic parameters.

Rather than choosing a parameterized expression for $f(T)$ and then numerically solving
the modified Friedmann equations for given values of the boundary conditions (as we did
in
the previous subsection), one can relate the present values of its derivatives to the
cosmographic parameters $(q_0, j_0, s_0, l_0)$, and hence constraining them in a
model-independent way will provide a hint for what kind of $f(T)$ model could be
able to fit the observed Hubble diagram. Differentiating the torsion scalar $T$ with
respect to $t$, one can easily get the following relations:
\begin{align}
\label{CCFR_dif1T}
 & \dot T = -12H\dot H ~,  \\
\label{CCFR_dif2T}
 & \ddot T = -12[\dot H^2+H\ddot H] ~, \\
\label{CCFR_dif3T}
 & \dddot T = -12[3\dot H\ddot H+H\dddot H] ~, \\
\label{CCFR_dif4T}
 & T^{(iv)} = -12[3\ddot H^2+4\dot H\dddot H+HH^{(iv)}] ~.
\end{align}
The modified Friedmann Eqs.  (\ref{background11}) and (\ref{background22}) can then be
rewritten as
\begin{align}
\label{CCFR_friedfried}
 H^2 &=\frac{-1}{12f_T(T)}[T\Omega_{m}+f(T)] ~,\\
\label{CCFR_acceacce}
 \dot H&=\frac{1}{4f_T(T)}[T\Omega_{m}-4H\dot{T}f_{TT}(T)] ~,
\end{align}
where $\Omega_m$ represents the dimensionless matter density parameter. However, in order
to calculate all cosmographic parameters one needs to differentiate (\ref{CCFR_acceacce})
three more times. As a result, one derives \cite{Capozziello:2011hj}:
\begin{align}
\label{CCFR_expddH}
 \ddot H &= \frac{\Omega_m}{4Hf_T(T)}[H\dot T -T(3H^2+2\dot H)]\nonumber\\
 &\ \ \
 -\frac{1}{f_T(T)}[(2\dot H\dot T+H\ddot T)f_{TT}(T)+H\dot T^2f_{TTT}(T)] ~, \\
\label{expdddH}
 \dddot H &=
 \frac{\Omega_m}{4H^2f_T(T)} \Big[T(9H^4 +6H^2\dot H +4\dot H^2)\nonumber\\
 &\ \ \ \ \ \ \ \ \ \ \ \ \ \ \ \ \ \
 -H\dot T(3\dot H +6H^2) +H(H\ddot T-2\ddot H T)\Big]
 \nonumber \\
 & ~ -\frac{1}{Hf_T(T)}
 \Big[(2\dot H^2\dot T +3H\ddot H\dot T
 +4H\dot H\ddot T +H^2\dddot T)f_{TT}(T)\nonumber \\
 &\ \ \ \ \ \ \ \ \ \ \ \ \ \ \ \,    +
 \dot H\ddot Hf_T(T) +
  +H^2\dot T^3f^{(iv)}(T) \nonumber \\
 &\ \ \ \ \ \ \ \ \ \ \ \ \ \ \ \,  +H\dot T(4\dot H\dot T+3H\ddot T)f_{TTT}(T)\Big] ~,
\end{align}
and
\begin{align}
\label{CCFR_expddddH}
 H^{(iv)} &= \frac{\Omega_m}{4H^3f_T(T)}
 \Big[T(10H\dot H\ddot H +12H^3\ddot H
 -27H^6 \nonumber\\
 & \ \ \ \ \ \ \ \ \ \ \ \ \ \ \ \  \ \ \    \ \ \,
 -12H^2\dot H^2 -8\dot H^3 -2H^2\dddot H)
 \nonumber \\
 & \ \ \ \ \ \ \ \ \ \ \ \ \ \ \ \  \ \ \,
 +H^3\dddot T  +H^2\dot T(9H\dot H+27H^3-5\ddot
H)\nonumber\\
 & \ \ \ \ \ \ \ \ \ \ \ \ \ \ \ \  \ \ \,
 -3H^2\ddot T(3H^2+\dot H)
 +7H\dot H^2\dot T\Big]
 \nonumber \\
 & ~ -\frac{1}{H^2f_T(T)} \Big[(3H\dot H\dddot H +\dot H^2\ddot H +H\ddot
H^2)f_T(T)\nonumber\\
 & \ \ \ \ \ \ \ \ \ \ \ \ \ \ \ \  \
 +H^2\dot T^2(7\dot H\dot T+6H\ddot T)f^{(iv)}(T) \nonumber \\
 & \ \ \ \ \ \ \ \ \ \ \ \ \ \ \   \ \  +\Big(4H^2\dddot H\dot T +2\dot H^3\dot T
+7H^2\dot H\dddot T
 \nonumber\\
 & \ \ \ \ \ \ \ \ \ \ \ \ \ \ \   \ \ \ \ \ \ \,
 +10H\dot H^2\ddot T
 +7H^2\ddot H\ddot T \nonumber\\
 & \ \ \ \ \ \ \ \ \ \ \ \ \ \ \   \ \ \ \ \ \ \,
 +11H\dot H\ddot H\dot T +H^3T^{(iv)}\Big)f_{TT}(T)
\nonumber \\
 & \ \ \ \ \ \ \ \ \ \ \ \ \ \ \ \  \ \ \, +H\Big(10\dot H^2\dot T^2 +7H\ddot H\dot T^2
+21H\dot H\dot T\ddot T
 \nonumber\\
 & \ \ \ \ \ \ \ \ \ \ \ \ \ \ \   \ \ \ \ \ \ \,  \ \, \ \, \ \,
 +3H^2\ddot T^2 +4H^2\dot T\dddot T\Big) f_{TTT}(T) \nonumber \\
 & \ \ \ \ \ \ \ \ \ \ \ \ \ \ \ \    \ \ \,
 +H^3\dot T^4f^{(v)}(T)\Big] ~,
\end{align}
with $ H^{(iv)}\equiv d^4 H(t)/dt^4 $,  $f^{(iv)}(T) \equiv d^4f(T)/dT^4$ and $f^{(v)}(T)
\equiv d^5f(T)/dT^5$. Since Eqs.
\eqref{CCFR_friedfried}-\eqref{CCFR_expddddH} have to hold along the full evolutionary
history of the universe, one can evaluate them at present day $t_0$ obtaining:
\begin{align}
\label{CCFR_friedcosmo}
 & H_0^2 = \frac{-1}{12f_T(T_0)}[T_0\Omega_{m0} +f(T_0)] ~, \\
\label{CCFR_accecosmo}
 & \dot{H_0} = \frac{1}{4f_T(T_0)}[T_0\Omega_{m0} -4H_0\dot{T_0}f_{TT}(T_0)] ~,
\end{align}
and similarly for the next three derivatives.

In summary, one obtains five equations, namely (\ref{CCFR_friedcosmo}),
(\ref{CCFR_accecosmo}) and (\ref{CCFR_expddH})-(\ref{CCFR_expddddH}) evaluated at
present time. We call these, ``final equations'', which will turn out to be useful in the
following. But, one another relation is expected in order to close the system and
determine the six unknown quantities: $f(T_0)$, $f_T(T_0)$,
 $f_{TT}(T_0)$, $f_{TTT}(T_0)$, $f^{(iv)}(T_0)$ and $f^{(v)}(T_0)$. This can be easily
obtained by noticing that, inserting back the physical units, Eq.  (\ref{background11})
reads:
\begin{equation}
\label{CCFR_backfried}
 H^2 = \frac{8\pi G}{6f_T(T)} \left[ \rho_m -\frac{f(T)}{16\pi G} \right] ~,
\end{equation}
which clearly shows that, in $f(T)$ gravity, the Newtonian gravitational constant $G$
has to be replaced by an effective (time varying) coupling $G_{eff}$. However, the
present value of Newtonian gravitational constant has to be recovered, and then:
\begin{equation}\label{CCFR_constraint}
 G_{eff}(z=0) = G \Rightarrow f_T(T_0)=1 ~,
\end{equation}
which implies the recovery of TEGR.

Let us now assume that $f(T)$ may be well approximated by its fifth order Taylor
expansion in $T-T_0$, i.e. we set:
\begin{align}
\label{CCFR_taylor}
 f(T) & \approx f(T_0) +f_T(T_0)(T-T_0) +\frac{1}{2}f_{TT}(T_0)(T-T_0)^2\nonumber \\
 & \ \ \
+\frac{1}{6}f_{TTT}(T_0)(T-T_0)^3 +\frac{1}{24}f^{(iv)}(T_0)(T-T_0)^4\nonumber \\
 & \ \ \  +\frac{1}{120}f^{(v)}(T_0)(T-T_0)^5 ~.
\end{align}
%In such an approximation, $f^{(n)}(T) =d^nf/dT^n =0$ for $n\geq6$.
Evaluating Eqs. (\ref{CCFR_dif1T})-(\ref{CCFR_dif4T}) at present time and using Eqs.
(\ref{CCFR_dotH}), one obtains:
\begin{align}
\label{CCFR_T}
 & T_0 = -6H_0^2 ~, \\
\label{CCFR_dotT}
 & \dot{T_0} = 12H_0^3 (1 +q_0) ~, \\
\label{CCFR_ddotT}
 & \ddot{T_0} = -12H_0^4 [q_0(q_0+5) +j_0 +3] ~, \\
\label{CCFR_dddotT}
 & \dddot{T_0} = -12H_0^5 [s_0 -j_0(3q_0+7) -3q_0(4q_0+9) -12] ~, \\
\label{CCFR_ddddotT}
 & T^{(iv)}_0 = -12H_0^6 \Big[l_0 -s_0(4q_0+9) +j_0(3j_0 +44q_0 +48)\nonumber\\
 &\ \ \ \ \ \ \ \ \ \
 +3q_0(4q_0^2
+39q_0 +56)
+60\Big] ~.
\end{align}
After inserting all these expressions into the ``final equations'', one can solve them
under the constraint (\ref{CCFR_constraint}) with respect to the present day values of
$f(T)$ and its derivatives up to the fifth order. After some algebra, one ends up with
the
desired result, namely \cite{Capozziello:2011hj}:
\begin{align}
\label{CCFR_fT}
 & \frac{f(T_0)}{6H_0^2} = \Omega_{m0} -2 ~, \\
\label{CCFR_f1T}
 & f_T(T_0) = 1 ~, \\
\label{CCFR_f2T}
 & \frac{f_{TT}(T_0)}{(6H_0^2)^{-1}} = \frac{-3\Omega_{m0}}{4(1 +q_0)} +\frac{1}{2} ~, \\
\label{CCFR_f3T}
 & \frac{f_{TTT}(T_0)}{(6H_0^2)^{-2}} = \frac{-3\Omega_{m0}(3q_0^2 +6q_0 +j_0 +2)}{8(1
+q_0)^3} +\frac{
3}{4} ~, \end{align}
\begin{align}
\label{CCFR_f4T}
 & \frac{f^{(iv)}(T_0)}{(6H_0^2)^{-3}} = \frac{-3\Omega_{m0}}{16(1 +q_0)^5}
 \Big[9+j_0(6q_0^2 +17q_0 +3j_0 +5) \nonumber \\
 &\ \ \ \ \ \ \ \ \ \ \ \ \ \ \ \ \ \ \ \ \ \ \ \ \ \ \ \ \ \
  +3q_0(5q_0^3 +20q_0^2 +29q_0 +16)\nonumber \\
 &\ \ \ \ \ \ \ \ \ \ \ \ \ \ \ \ \ \ \ \ \ \ \ \ \ \ \ \ \ \   +s_0(1 +q_0)
 \Big] +\frac{15}{8} ~,
  \end{align}
\begin{align}
\label{CCFR_f5T}
 & \frac{f^{(v)}(T_0)}{(6H_0^2)^{-4}} = \frac{-3\Omega_{m0}}{32(1 +q_0)^7}
 \Big[l_0(1 +q_0)^2  +10j_0s_0(1 +q_0) \nonumber \\
 &\ \ \ \ \ \ \ \ \ \ \ \ \ \ \
 +s_0(10q_0^3 +43q_0^2 +46q_0 +13)\nonumber \\
  &\ \ \ \ \ \ \ \ \ \ \ \ \ \ \
 +5j_0^2 (6q_0^2 +22q_0 +3j_0 +7)\nonumber \\
 &\ \ \ \ \ \ \ \ \ \ \ \ \ \  \  +j_0(45q_0^4 +225q_0^3
+412q_0^2 +219q_0 +32)
\nonumber\\
 &\ \ \ \ \ \ \ \ \ \ \ \ \ \ \  +3q_0 (35q_0^5 +210q_0^4 +518q_0^3 +666q_0^2\nonumber
\\
  &\ \ \ \ \ \ \ \ \ \ \ \ \ \ \   +448q_0 +150) +60\Big] +\frac{105}{16} ~.
\end{align}
Eqs.  (\ref{CCFR_fT})-(\ref{CCFR_f5T}) make it possible to estimate the present day values
of $f(T)$ and its first five derivatives as functions of the Hubble constant $H_0$ and
the cosmographic parameters $(q_0, j_0, s_0, l_0)$, provided a value for the matter
density parameter $\Omega_{m0}$ is fixed.

In order to acquire a first hint on the possible values of $f(T)$ and its derivatives, one
can reproduce the cosmographic parameters for the $\Lambda$CDM model as a standard case.
This is a minimal approach but it is useful to probe the self-consistency of the model.
The cosmographic parameters for the $\Lambda$CDM model read \cite{Capozziello:2011hj}
\begin{align}
\label{CCFR_qlambda}
 q &= -\left(\frac{H_0}{H}\right)^2 \left(1 -\Omega_{m0} -\frac{\Omega_{m0}}{2a^3}\right)
~, \\
\label{jlambda}
 j &= \left(\frac{H_0}{H}\right)^3 \left(1 -\Omega_{m0}
+\frac{\Omega_{m0}}{a^3}\right)^{3/2} ~, \\
\label{CCFR_slambda}
 s &= \left(\frac{H_0}{H}\right)^4 \left(1 -2\Omega_{m0} -\frac{5\Omega_{m0}}{2a^3}
\right. \nonumber\\
& \ \ \ \ \ \ \ \ \  \ \ \ \ \ \ \ \left.
+\Omega_{m0}^2
 +\frac{5\Omega_{m0}^2}{2a^3} -\frac{7\Omega_{m0}^2}{2a^6}\right) ~,\\
\label{CCFR_llambda}
 l &= \left(\frac{H_0}{H}\right)^5 \sqrt{1-\Omega_{m0}
+\frac{\Omega_{m0}}{a^3}} \left(1-2\Omega_{m0} +\frac{5\Omega_{m0}}{a^3}\right.
\nonumber\\
& \ \ \ \ \ \ \ \ \  \ \ \ \ \ \ \ \left.
 +\Omega_{m0}^2 -\frac{5\Omega_{m0}^2}{a^3}
 +\frac{35\Omega_{m0}^2}{2a^6}\right)   ~,
\end{align}
which, evaluated at the present time, yield:
    $q_0=-1+\frac{3}{2}\Omega_{m0}$,
    $j_0=1$,
    $s_0=1-\frac{9}{2}\Omega_{m0}$,
    $l_0=1+3\Omega_{m0}+\frac{27}{2}\Omega_{m0}^2$.
Inserting the previous equations into Eqs. (\ref{CCFR_f2T})-(\ref{CCFR_f5T}) we obtain
\begin{equation}
\label{CCFR_constraint2}
    f_{TT}(T_0)=f_{TTT}(T_0)=f^{(iv)}(T_0)=f^{(v)}(T_0)=0,
\end{equation}
and in the absence of these terms $f(T)$ reduces to $f(T)\sim T-2\Lambda$. This is
consistent with what we expected for the $\Lambda$CDM model and can be used as a
consistency check.

\subsubsection{Results of observational constraints}

In order to outline the form of $f(T)$ through its own value and those of its
derivatives at present, one needs to constrain observationally the cosmographic
parameters by using appropriate distance indicators. Moreover, one has to take care that
the expansion of the distance related quantities in terms of $(q_0, j_0, s_0, l_0)$
closely follows the exact expressions over the range probed by the data used. Taking SNIa 
and a 
fiducial $\Lambda$CDM model as a test case, one needs to check that the
approximated luminosity distance (see \cite{Capozziello:2008qc} for the analytical
expression) deviates from the $\Lambda$CDM case less than the measurement uncertainties
up
to $z \simeq  1.5$, to avoid introducing any systematic bias. Since the goal is to
constrain $(q_0, j_0, s_0, l_0)$, one needs to expand the luminosity distance $D_L$ up to
the fifth order in $z$, which can track the $\Lambda$CDM expression with an error less
than $1\%$ over the full redshift range. This also applies for the angular diameter
distance $D_A = D_L(z)/(1 + z)^2$ and the Hubble parameter $H(z)$ which, however, is only
expanded up to the fourth order to avoid introducing any further cosmographic parameter.

To constrain the parameters $(h, q_0, j_0, s_0, l_0)$, one may make use of the Union2
SNIa dataset \cite{Amanullah:2010vv} and the BAO data from the analysis of the SDSS
seventh release \cite{Percival:2009xn} (or even more recent data) adding a prior on $h$
from the recent determination of the Hubble constant by the SHOES team
 \cite{Riess:2009pu}. To update, one can also add the measurement of $H(z)$ obtained in
\cite{Stern:2009er} from the age of passively evolving galaxies and in
\cite{Gaztanaga:2008xz} from the radial BAO. By exploring such a five
dimensional parameter space with a Markov Chain Monte Carlo method, one then obtains the
constraints  \cite{Capozziello:2011hj} summarized in Table~\ref{CCFR_tab: cosmofit}. It
is
easy to check that these results are consistent with other analyses in the literature
\cite{Vitagliano:2009et,Xu:2010hq, Capozziello:2011tj, Ester}.
\begin{table}[ht]
\begin{center}
\begin{tabular}{cccccc}
\hline
$x$ & $x_{BF}$ & $\langle x \rangle$ & $x_{med}$ & $68\%$ CL & $95\%$ CL \\
\hline \hline
~ & ~ & ~ & ~ & ~ & ~ \\
$h$ & 0.718 & 0.706 & 0.706 & (0.693, 0.719) & (0.679, 0.731) \\
~ & ~ & ~ & ~ & ~ & ~ \\
$q_{0}$ & -0.64 & -0.44 & -0.43 & (-0.60, -0.30) & (-0.71, -0.26) \\
~ & ~ & ~ & ~ & ~ & ~ \\
$j_{0}$ & 1.02 & -0.04 & -0.15 & (-0.88, -0.90) & (-1.07, 1.40) \\
~ & ~ & ~ & ~ & ~ & ~ \\
$s_{0}$ & -0.39 & 0.18 & 0.02 & (-0.57, 1.07) & (-1.04, 1.78) \\
~ & ~ & ~ & ~ & ~ & ~ \\
$l_{0}$ & 4.05 & 4.64 & 4.54 & (2.99, 6.48) & (1.78, 8.69) \\
~ & ~ & ~ & ~ & ~ & ~ \\
\hline
\end{tabular}
\end{center}
\caption{Constraints on the cosmographic parameters. Columns are: 1. Parameter name, 2.
Best-fit, 3. Mean from the marginalized likelihood, 4. Median from the marginalized
likelihood; 5. $68\%$ confidence range, 6. $95\%$ confidence range. From
 \cite{Capozziello:2011hj}. }
\label{CCFR_tab: cosmofit}
\end{table}

Note that due to the degeneracies, among the five cosmographic parameters the best-fit
values can also be different from the median ones, which is indeed the case. This is,
however, not a shortcoming of the fitting analysis, but a consequence of the Bayesian
approach giving more importance to sampling the marginalized parameters distributions
rather than to looking for the best-fit accordance within a given model and the available
dataset. Qualitatively, one can say that the best-fit value of, e.g. $q_0$, is less
important than the median one since the best-fit $q_0$ is the correct one only if the
other parameters also take their best-fit values, while the median one is more reliable
since it describes the full distribution whatever are the values of the other parameters.
In particular, here, the best-fit values are quite close to those predicted for the
$\Lambda$CDM model (for instance, $j_0 = 1$ for a $\Lambda$ dominated universe), while
the
median ones allow for significant deviations (with the $\Lambda$CDM values being,
however,
within the $95\%$ confidence ranges).

In order to translate the constraints on the cosmographic parameters on similar
constraints on the present day values of $f(T)$ and its derivatives, one can use Eqs.
({\ref{CCFR_fT})-(\ref{CCFR_f5T}) evaluating them along with cosmographic parameters and
then looking at the corresponding histograms. To this end, one additionally needs to set
the value of $\Omega_{m0}$, which is not constrained by the fitting analysis described
above. In order to overcome this difficulty, it is useful to take into account the CMB
determination of the physical matter density (namely $\omega_m = \Omega_{m0} h^2 = 0.1329$
in WMAP7) and, for each value of $h$, one fixes $\Omega_{m0} = \omega_m/h^2$ having
neglected the error on $\omega_m$ since it is subdominant with respect to the one on $h$.
Note that the adopted estimate of $\omega_m$ comes from the fit to the CMB anisotropy
spectrum and it mainly depends on the early universe physics. Since it is reasonable to
expect that GR is recovered in this limit, one can safely assume the
validity of this result whichever $f(T)$ model is considered.
Defining for shortness
$$f_n = f^{(n)}(T_0)/(6 H_0^2)^{-(n-1)} ~,$$
one then obtains the constraints summarized in Table~\ref{CCFR_tab: ftfit} and presented
in Fig.~\ref{CCFR_fig: clplots}, where the degeneracy between some couples of parameters
is shown as an example.
\begin{table}[ht]
\begin{center}
\begin{tabular}{cccccc}
\hline
$x$ & $x_{BF}$ & $\langle x \rangle$ & $x_{med}$ & $68\%$ CL & $95\%$ CL \\
\hline \hline
~ & ~ & ~ & ~ & ~ & ~ \\
$f_{0}$ & -1.742 & -1.733 & -1.733 & (-1.743, -1.723) & (-1.751, -1.712) \\
~ & ~ & ~ & ~ & ~ & ~ \\
$f_{2}$ & -0.033 & 0.113 & 0.147 & (0.007, 0.208) & (-0.153, 0.226) \\
~ & ~ & ~ & ~ & ~ & ~ \\
$f_{3}$ & -0.092 & 0.530 & 0.815 & (0.172, 0.921) & (-1.483, 1.033) \\
~ & ~ & ~ & ~ & ~ & ~ \\
$f_{4}$ & 0.294 & -0.955 & 1.061 & (0.193, 2.306) & (-18.307, 3.603) \\
~ & ~ & ~ & ~ & ~ & ~ \\
$f_{5}$ & 8.690 & -68.893 & 6.371 & (2.956, 11.014) & (-370.966, 31.004) \\
~ & ~ & ~ & ~ & ~ & ~ \\
\hline
\end{tabular}
\end{center}
\caption{Constraints on the $f_i$ values from the Markov Chain for the cosmographic
parameters.
Columns order is the same as in Table~\ref{CCFR_tab: cosmofit}. From
 \cite{Capozziello:2011hj}. }
\label{CCFR_tab: ftfit}
\end{table}
\begin{figure*}
\centering
\subfigure{\includegraphics[width=5.2cm]{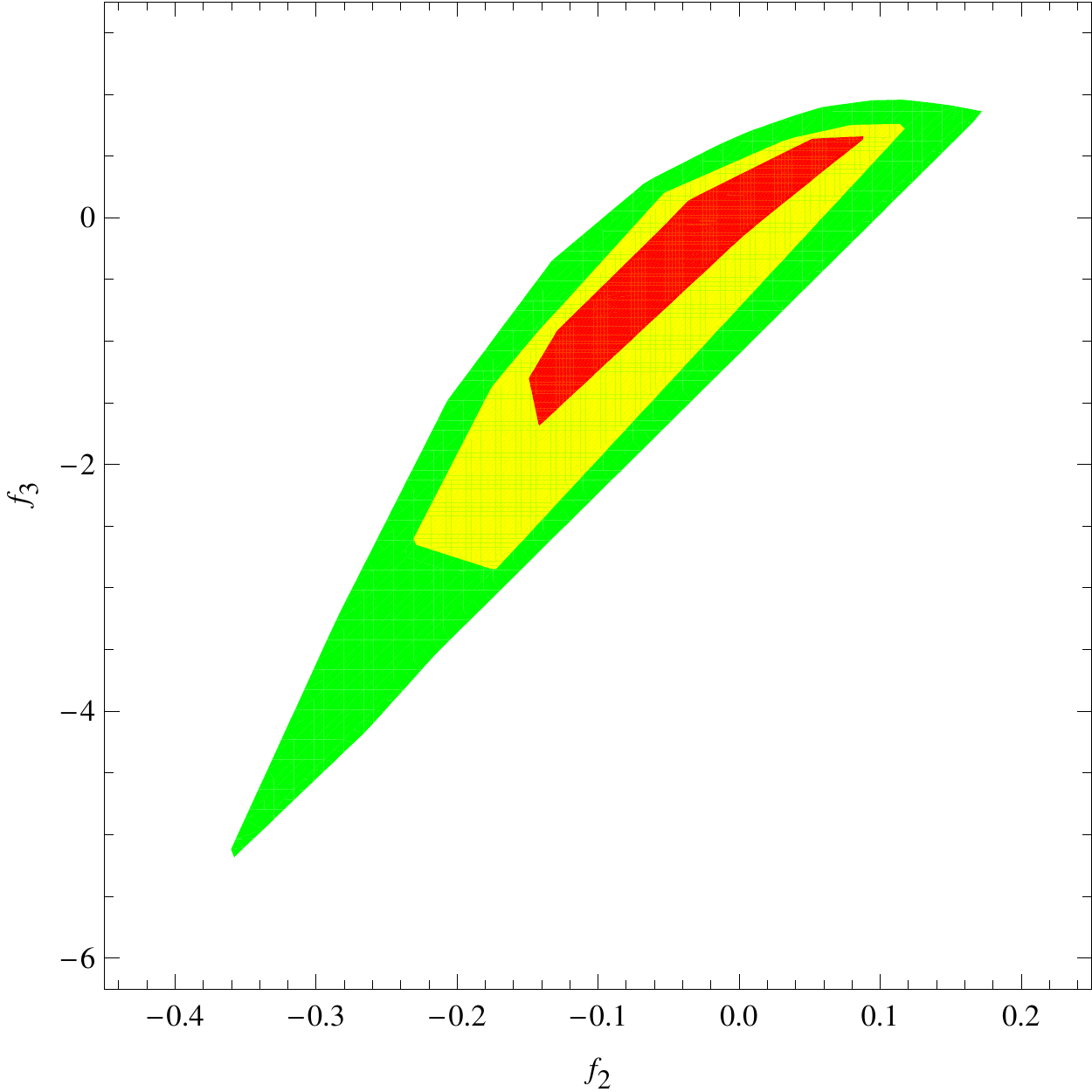}}
\subfigure{\includegraphics[width=5.2cm]{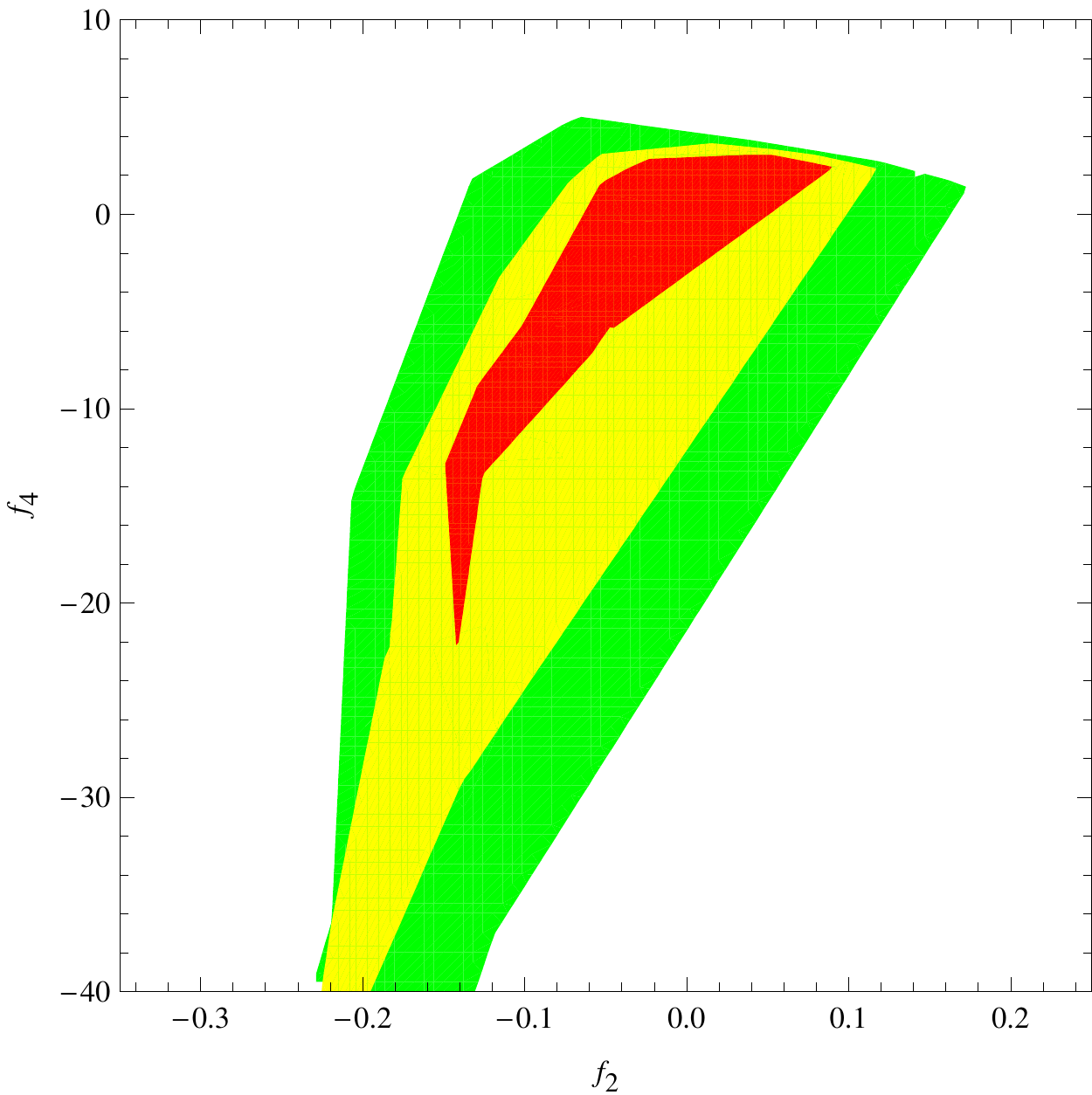}}
\subfigure{\includegraphics[width=5.2cm]{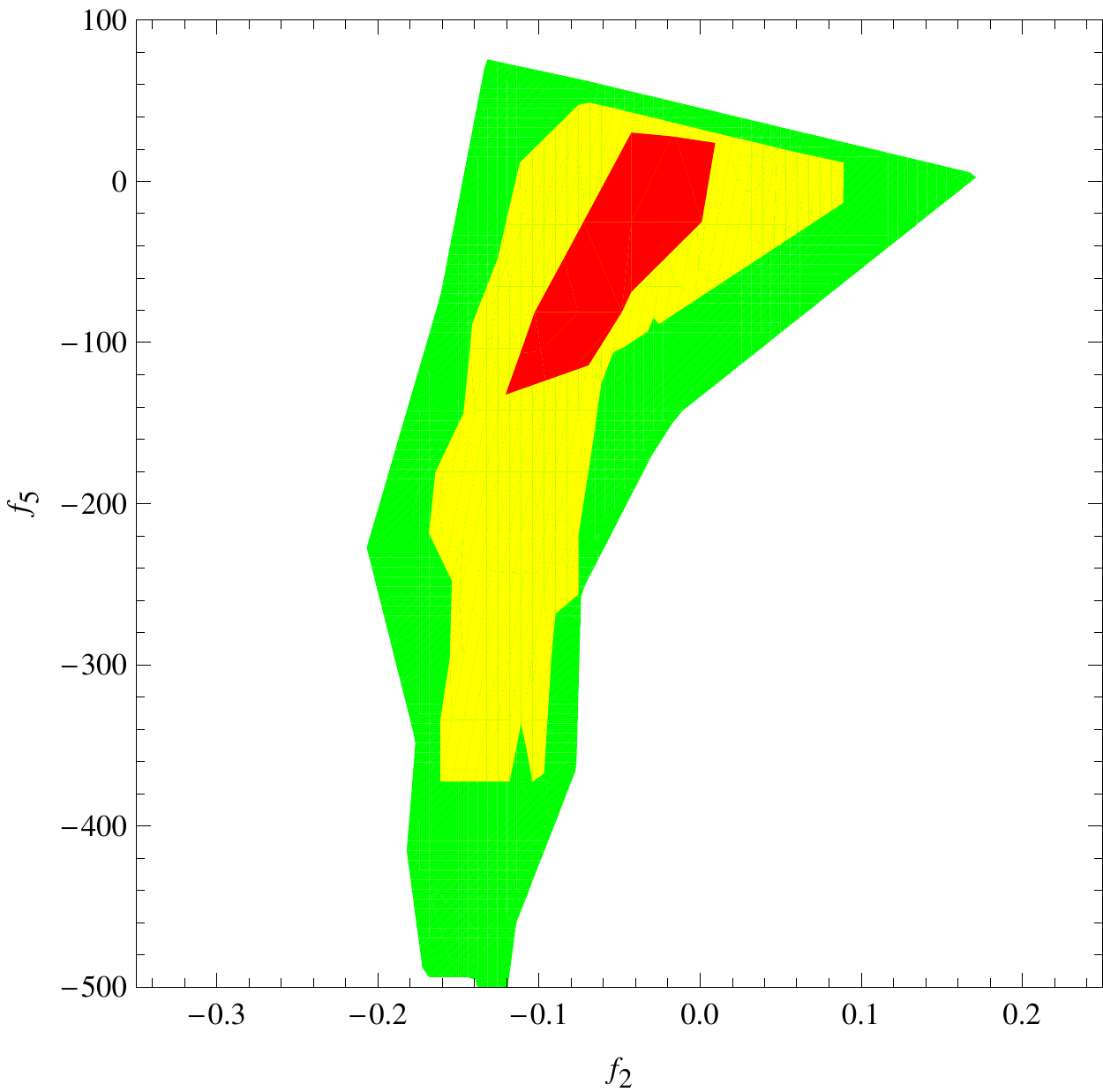}}
\caption{\it Iso-likelihood ($68\%$, $95\%$ and $99\%$ CL) contours for the $f_i$
quantities. The fuzziness is due to numerical artifacts. From  \cite{Capozziello:2011hj}.}
\label{CCFR_fig: clplots}
\end{figure*}

We mention that as best-fit value here we refer to the one obtained by fixing the
cosmographic parameters to the best-fit values. However, because of the degeneracies
among $(q_0, j_0, s_0, l_0)$ and the nonlinear behavior of the relations with $f_n$, it is
possible that the best-fit $f_n$ are quite different from their median values which is
indeed the case (in particular, for $f_5$). Note also that the confidence ranges become
larger as the order $n$ of the derivative increases. This is indeed an expected result,
since the higher is $n$ the larger is the number of cosmographic parameters involved, and
hence the weakness of the constraints on the higher order cosmographic parameters
and the degeneracies among them makes the constraints on $f_n$ weaker and weaker as $n$
gets larger. From a different point of view, such a behavior simply reflects the naive
expectation that one has to go to deeper redshifts to probe the exact functional
shape of $f(T)$ and hence put severe constraints on the value of its high order
derivatives. As a further remark, we note that the constraints on $(f_3, f_4, f_5)$ are
strongly asymmetric, with a long tail extending towards negative values causing a large
offset between the mean and the median. This is actually a consequence of the terms $(1 +
q_0)^{-\alpha}$, with $\alpha = (3, 5, 7)$ for $(f_3, f_4, f_5)$
respectively, which appear as common factors in Eqs. (\ref{CCFR_f3T})-(\ref{CCFR_f5T}).
As $q_0$ comes close to -1, these terms become increasingly large and thus make $f_{n}$
explode.

Although we have checked that the fifth order expansion closely matches the exact
luminosity and angular diameter distances and the Hubble parameter within less than
$1\%$, it is worth noting that a decent approximation is also obtained if one stops the
expansion to the third or fourth order. Cutting the expansion to order three (four) means
that one can only constrain cosmographic parameters up to the jerk $j_0$ (the snap $s_0$)
and hence estimate confidence ranges for the $f(T)$ derivatives up to the third (fourth)
order. It is nevertheless worth exploring how the constraints depend on the order of the
expansion. In order to perform this, one can fit the same dataset as above
with both the third and fourth order expansion of the involved quantities and then use
the
corresponding Markov Chains to estimate confidence limits on $(f_0, f_2, f_3)$. For
instance, from the third order fit one obtains (median and $68\%$ and $95\%$ CL)
\cite{Capozziello:2011hj}:
\begin{eqnarray}
 &&f_0 = -1.741_{-0.008 \ -0.016}^{+0.009 \ +0.017} ~,\nonumber\\
 &&f_2 = 0.005_{-0.069 \ -0.154}^{+0.054 \ +0.098} ~,\nonumber\\
 &&f_3 = 0.097_{-0.515 \ -1.552}^{+0.303 \ +0.475} ~,
\end{eqnarray}
while the fourth-order fit yields\,:
\begin{eqnarray}
 &&f_0 = -1.733_{-0.009 \ -0.019}^{+0.011 \ +0.021} ~,\nonumber\\
 &&f_2 = 0.043_{-0.076 \ -0.195}^{+0.061 \ +0.113} ~,\nonumber\\
 &&f_3 = 0.439_{-0.441 \ -1.536}^{+0.266 \ +0.439} ~.
\end{eqnarray}
Comparing the value of $f_i$ for different fits (including the fifth-order one in
Table~\ref{CCFR_tab: ftfit}) allows us to draw some interesting lessons. Firstly, although
the median values are different, the confidence ranges are well overlapped, thus
indicating that the expansion order should not have any statistically meaningful impact
on the constraints. However, more accurate studies have to be performed in order to
confirm this statement. Secondly, increasing the expansion order shifts away from
the $\Lambda$CDM one (i.e. $f_i = 0$ for $i > 1$). This is actually a subtle effect of
the degeneracy among the cosmographic parameters. Indeed, increasing the order $n$ of the
expansion adds further parameters to the fit, therefore allowing for much more
combinations of the cosmographic parameters able to fit well the same data. As a
consequence, the constraints on $q_0$ will become weaker, allowing for models with $q_0$
closer to $-1$ and hence $(f_2, f_3)$ values far away from the fiducial $\Lambda$CDM
ones.
Nevertheless, we recommend the reader to refer to the results in Table
\ref{CCFR_tab: ftfit}, since the fifth order expansion provides a better approximation to
the underlying expansion history and hence the fit is less affected by any bias due to
any error in the approximation.

The constraints discussed above have been obtained under two basic underlying
assumptions. Firstly, we have set $f_T(T_0) = 1$ in order to recover an effective
gravitational constant which matches the Newton one today. Nevertheless, although it is a
reasonable assumption, there are no compelling arguments why the Newton constant which is
measured in laboratory experiments is the same as the cosmological one. As such, it is
worth wondering how the above results would change should we allow for deviations from
the
$G = G_{cosmo}$ assumption. Secondly, we have used the CMB constraints on the physical
matter density $\omega_m$ in order to infer the present day matter density parameter and
then use Eqs. (\ref{CCFR_fT})-(\ref{CCFR_f5T}) to constrain $(f_0, f_2, f_3, f_4, f_5)$
from the cosmographic parameters. It is worth noting that, if one takes into account the
perturbation analysis such as reviewed in previous subsections, it is more likely to find
out remarkable differences of $f(T)$ models with respect to the standard GR. As a
consequence, one cannot exclude the possibility to recover a correct growth of structure
even if $f(T))$ does not reduce to GR in the early universe. Should this be the case, the
use of the WMAP7 $\omega_m$ value is inconsistent. Taking care of these possible effects
is actually quite easy. Indeed, some algebra shows that Eqs.
(\ref{CCFR_fT})-(\ref{CCFR_f5T}) can all be recast as
\begin{displaymath}
 f_n = \frac{{\cal{P}}_n(q_0, j_0, l_0, s_0)}{(1 + q_0)^{\alpha_n}} \Omega_m + \kappa_n
(1
+ \varepsilon),
\end{displaymath}
with ${\cal{P}}_n(q_0, j_0, l_0, s_0)$ a polynomial function of its arguments, $\alpha_n
=(0, 1, 3,
 5, 7)$ for $n = (0, 2, 3, 4, 5)$, $\kappa_n$ a constant depending on $n$, and where we
have set $f_T(T_0) = 1 + \varepsilon$. Using this simple formula allows to
immediately scale our constraints to different values of $\omega_m$ and $\varepsilon$,
provided one has a theoretical or observational estimate of these quantities.

\subsubsection{Cosmography versus $f(T)$ models}

Up to now we have not yet assumed any fixed form of $f(T)$, and the constraints in
Table~\ref{CCFR_tab: cosmofit} indeed hold for the full class of such theories, provided
one can approximate $f(T)$ by its fifth order Taylor series over the redshift range
probed
by the data. Such a result can also be read in a different way. Given a specific $f(T)$
model, its characterizing parameters must be chosen in such a way that the constraints in
Table~\ref{CCFR_tab: cosmofit} are satisfied. This consideration offers an interesting
route to check the viability of a given $f(T)$ model without the need of explicitly
solving the field equations and fitting the data.

In order to demonstrate the use of cosmography in constraining $f(T)$ gravity, let us
consider the following model  \cite{Myrzakulov:2010vz} as an example:
\begin{equation}
\label{CCFR_eq: ftmyrza}
 f(T) = \alpha T + \beta T^{\delta} \ln{T} ~.
\end{equation}
Imposing Eq.  (\ref{CCFR_fT}) and $f_T(T_0) = 1$ yields:
\begin{align}
\label{CCFR_eq: am}
 \alpha &= \frac{2 -\Omega_{m0} - [1 +(\Omega_{m0} - 2) \delta] \ln{T_0}}{1 + (\delta -1)
\ln{T_0}}
~, \\
\label{CCFR_eq: bm}
 \beta &= \frac{( \Omega_{m0} -1) T_0^{1 - \delta}}{1 +(\delta -1) \ln{T_0}} ~,
\end{align}
and thus one can express explicitly $f_i$ for $i = (2, 3, 4, 5)$ as functions of
$\delta$ only. For each $f_2$ value of the sample obtained above from the cosmographic
parameters analysis, one gets $\hat{f}_2(\delta) = f_2$. Since this equation has two
roots, one can store them and then compute $(f_3, f_4, f_5)$ for both values thus
obtaining a histogram for the model prediction of these quantities. The median and $68\%$
and $95\%$ confidence ranges read:
\begin{eqnarray}
 &&f_3 = -0.296_{-0.115 \ -0.149}^{+0.272 \ +0.599} ~,\nonumber\\
&& f_4 = 0.891_{-0.799 \ -1.797}^{+0.330 \ +0.424}  ~,\nonumber\\
 &&f_5 = -3.568_{-1.274 \ -1.633}^{+3.143 \ +7.176} ~,
\end{eqnarray}
for the smaller $\delta$ solution and
\begin{eqnarray}
&& f_3 = 8.779_{-0.088 \ -0.131}^{+0.193 \ +0.415} ~,\nonumber\\
 &&f_4 = -3.120_{-0.024 \ -0.050}^{+0.018 \ +0.032}  ~,\nonumber\\
 &&f_5 = -31.033_{-0.811 \ -1.810}^{+0.371 \ +0.525} ~,
\end{eqnarray}
for the larger one. Since the $95\%$ CL in Table~\ref{CCFR_tab: ftfit} are quite large
due to the impact of $q_0$, it is simpler to use only the $68\%$ confidence ranges in
order to compare them with the above constraints. For the lower $\delta$ solution,
both $f_3$and $f_5$ are smaller than the $68\%$ CL from cosmographic parameters, while
the range for $f_4$ has a marginal overlap. On the other hand, choosing the larger
$\delta$ solution leads to $(f_3, f_4, f_5)$ values that fully disagree with the
model-independent constraints. In this regard, one could conclude that the model
(\ref{CCFR_eq: ftmyrza}) is disfavored by the observational data.

\subsection{A brief summary of $f(T)$ cosmology}

We close this section with a brief summary of the presented results. In particular, we
have reviewed various cosmological solutions obtained in $f(T)$ gravity. These solutions
present various phenomenological behaviors that are of significant cosmological interest.
In particular, we first showed that a large class of $f(T)$ models are able to give rise
to a late-time acceleration in an FRW universe, and hence the observation of the present
cosmic speed up can be explained without introducing any dark energy component. Upon the
late-time acceleration, we performed the procedure of dynamical analysis to study the
classical stability of those solutions. Moreover, within the frame of $f(T)$ gravity, the
interesting scenario of phantom-divide crossing (or quintom scenario) becomes very easy
to be realized and thus the theory may explain the cosmological observations in a better
way. Let us mention here that one can extend the whole discussion in braneworld scenarios
and higher dimensions as in
\cite{Nozari:2012qi,Bamba:2013fta,Menezes:2014bta,Behboodi:2014tda,Atazadeh:2014joa,
Geng:2014nfa,Behboodi:2015uqa,Nashed:2014iua}.

After having studied the late-time accelerating solutions, we proceeded to the
implications of $f(T)$ gravity at early universe, namely we investigated inflationary and
bouncing cosmologies. Nevertheless, the realization of inflationary paradigm implies that
the horizon problem existing in the Big Bang cosmology described by GR can be
successfully
resolved in $f(T)$ gravity even without any inflaton field. Moreover, the appearance of
the bouncing solution shows that the universe governed by $f(T)$ gravity could be smooth
and non-singular throughout the whole cosmic evolution. This result demonstrates that the
problem of initial singularity that puzzles cosmologists for decades might be solved in
the frame of modified gravity theories. Furthermore, we commented on more general
situations in $f(T)$ cosmology, by considering the background universe to be an
anisotropic one.

Afterwards, we devoted ourselves to the observational constraints on $f(T)$ gravity, by
making use of various cosmological data. Choosing a specific $f(T)$ model, we combined
observational data from SNIa, GRB, BAO as well as CMB, to constrain the relevant model
parameters. At the background level, we have shown that the present datasets are
difficult
to differentiate $f(T)$ models from the $\Lambda$CDM cosmology. After taking into account
the perturbation analysis, however, it becomes possible to discriminate them by examining
the growth factor of the matter density fluctuation. Finally, we introduced the nearly
model-independent method of cosmography (note however that it still relies on certain
assumptions), which is a method that allows us to investigate the confidence ranges for
$f(T)$ derivatives up to the fifth order and then examine backwards the viability of a
given model.

%%%%%%%%%%%%%%%%%%%%%%%%%%%%%%%%%%%%%%%%
\section{Gravitational waves in $f(T)$ gravity}
\label{SecionGravwaves}

Gravitational waves offer a remarkable opportunity to see the universe from a new
perspective, providing an access to astrophysical insights that are not available in any
other way. Theoretical and experimental studies have been developed to understand the
mechanisms for the production of gravitational waves, both in astrophysics and in
cosmology \cite{Sathyaprakash:2009xs, Maggiore}. But even today, many conceptual
problems and technical issues related to the production of gravitational waves from
self-gravitating systems, have not fully been resolved. In particular, extended theories
of gravity seem to be a viable scheme to overcome shortcomings related to gravitational
waves. In this Section we investigate gravitational waves in $F(T)$ gravity.

\subsection{Gravitational waves in teleparallelism}
%%%%%%%%%%%%%%%%%%%%%%%%%%%%%%%%%%%%%%%%%%%%

%It is explicitly shown that the gravitational wave modes in $F(T)$ gravity are equivalent
%to those in GR.

We start our discussion by recalling that, as we showed in detail in subsection
\ref{DOFinFT} above, for $F(T)$ gravity in four dimensional space-time there exist three
extra degrees of freedom, namely a massive vector field or a massless
vector field with a scalar \cite{Li:2011rn}. This implies that gravitational wave modes
in
$F(T)$ gravity are equivalent to those in GR since such further modes do not contribute
to the gravitational radiation in the post-Minkowskian limit. We will verify this
consequence by using the preservative analysis in the weak-field limit around the flat
background in the scalar-tensor representation of $F(T)$ gravity. Thus, our main result
is
consistent with that of Ref. \cite{Li:2011rn} and the analysis of perturbations in
$F(T)$ gravity performed in Ref. \cite{Izumi:2012qj}. As usual we apply units where
$k_\mathrm{B} = c = \hbar = 1$ and we use the gravitational constant $G^{-1/2} = 1.2
\times 10^{19}$\,\,GeV.

Following the discussions of Refs.
\cite{DeLaurentis:2011tp,DeLaurentis:2013zv,DeLaurentis:2013fra,DeLaurentis:2012dq},
in order to obtain gravitational waves the most natural starting point is to use
linearized gravity. This implies that we adopt the weak-field limit approximation
\cite{Maggiore, DeLaurentis:2010bv}. The weak-field limit is achieved by
assuming that the space-time metric $g_{\mu\nu}$ is represented by the sum of the
Minkowski space-time plus a small perturbation, namely
\begin{eqnarray}
g_{\mu\nu}\,=\,\eta_{\mu\nu}+h_{\mu\nu}\,,
\label{metricgw}
\end{eqnarray}
with $h_{\mu\nu}$ being small and of first order ($ {\cal O}(h^2) \ll 1$). This implies
that the gravitational field is required to be weak, and furthermore
that the coordinate system is constrained to be approximately the Cartesian one.
It is straightforward to demonstrate that the relation (\ref{metricgw}) is equivalent to
the usual one connecting the metric with the vierbeins, namely $g_{\mu\nu}
= \eta_{AB}e^A_\mu e^B_\nu$, if we impose the identifications
\begin{eqnarray}
\label{eq:metric}
&&\!\!\!\!\!\!\!\!\!\!\!
g_{\mu\nu} = \eta_{AB}e^A_\mu e^B_\nu\,=\,
\eta_{AB}\left(\delta^A_\mu+{\mathbb{E}}^A_\mu\right)\left(\delta^B_\nu+{\mathbb{E}}
^B_\nu\right)
\nonumber
\\&&
=  \eta_{AB}\delta^A_\mu \delta^B_\nu+\eta_{AB}\delta^A_\mu
{\mathbb{E}}^B_\nu+\eta_{AB}{\mathbb{E}}^A_\mu\delta^B_\nu+\eta_{AB}{\mathbb{E}}^A_\mu
{\mathbb{E}}^B_\nu
\nonumber
\\&&
=\eta_{\mu\nu}+\underbrace{\eta_{\mu B}{\mathbb{E}}^B_\nu+
\eta_{A\nu}{\mathbb{E}}^A_\mu}_{\simeq\,
h_{\mu\nu}}+{\cal O}(
{\mathbb{E}}^2)\,,
\label{veribeinexpansionGW}
\end{eqnarray}
where ${\mathbb{E}}^A_\mu$ is the small vierbein perturbation. In expression
(\ref{veribeinexpansionGW}) we only keep terms linear in $h_{\mu\nu}$, while higher
order terms are discarded since we need to maintain the smallness of the perturbation.
Hence, the perturbation on the vierbeins (tangent space) is connected to the metric
perturbation on the manifold. The metric perturbation, as it is well known, encapsulates
gravitational waves, but it contains additional non-radiative degrees of freedom as well.
We mention that the metric perturbation $h_{\mu\nu}$ transforms as a tensor under the
Lorentz transformations, but not under general coordinate transformations.

Let us now compute all quantities which are needed in order to describe linearized
gravity. In particular, the Ricci tensor at first order of approximation in term of the
perturbation is given by
\begin{eqnarray}
R_{\mu\nu}^{(1)}\,
=\, \frac{1}{2}\left(\partial_\rho\partial_\nu {h^\rho}_\mu + \partial^\rho
\partial_\mu h_{\nu\rho} - \Box h_{\mu\nu} - \partial_\mu\partial_\nu
h\right)\,,\nonumber\\
\label{eq:ricci}
\end{eqnarray}
where $h = {h^\mu}_\mu$ is the trace of the metric perturbation, and $\displaystyle{\Box
= \partial_\rho\partial^\rho =  \partial_t^2}-\nabla^2 $ is the wave
operator. Here $R_{\mu\nu}^{(n)}$ denotes the term in $R_{\mu\nu}$ that is of $n$-th
order in $h_{\mu\nu}$.
 Contracting one more time we obtain the scalar curvature as
\begin{eqnarray}
R^{(1)} = {R^\mu}_\mu = \partial_\rho\partial^\mu {h^\rho}_\mu - \Box h\,,
\label{eq:scalar}
\end{eqnarray}
and finally we are able to construct the Einstein tensor as
\begin{eqnarray}
G_{\mu\nu} &=& R_{\mu\nu} - \frac{1}{2}\eta_{\mu\nu} R
\nonumber\\
&=& \frac{1}{2}\left(\partial_\rho\partial_\nu {h^\rho}_\mu + \partial^\rho
\partial_\mu h_{\nu\rho} - \Box h_{\mu\nu} - \partial_\mu\partial_\nu h
\right.\nonumber\\&& \left.
-\eta_{\mu\nu}\partial_\rho\partial^\sigma {h^\rho}_\sigma + \eta_{\mu\nu} \Box
h\right)\,.
\label{eq:einstein_h}
\end{eqnarray}
Now, we have all the necessary ingredients in order to write the field equations %Eqs.
(\ref{eom_fT_general}) in terms of the perturbation as
\begin{equation}
f_T \left( R_{\mu\nu}^{(1)} - \frac{1}{2} g_{\mu\nu} R^{(1)} \right)
+ \frac{1}{2}g_{\mu\nu} \left[ f(T) - f_T T \right]
  = {8\pi G}\, T^{(m)}_{\mu\nu},
\label{eq:2pert}
\end{equation}
where we have discarded the terms higher than first order in metric quantities
(we recall that we use the action $
 {\cal S} = \int d^4x |e| \left[ \frac{f(T)}{16\pi G} + L^{(m)} \right],
$ with $f(T)=T+F(T)$).
Concerning $f(T)$ we assume that it has an analytic form, which implies that
\begin{eqnarray}\label{sertay}
f(T)&=&\sum_{n}\frac{f^n(T_0)}{n!}(T-T_0)^n\nonumber\\&\simeq&
f_0+f'_0T+\frac{1}{2}f''_0T^2+...\,,\nonumber\\
\end{eqnarray}
with $f_0\equiv f(T_0)$, $f'_0\equiv f_T(T_0)$ and $f''_0\equiv f_{TT}(T_0)$. Hence,
the gravitational field equations (\ref{eq:2pert}) can be expressed as
\begin{eqnarray}
&&\left(f_0^{\prime}+T f_0^{\prime\prime} \right) \left( R_{\mu\nu}^{(1)} - \frac{1}{2}
g_{\mu\nu} R^{
(1)} \right)
\nonumber\\&&+ \frac{1}{2}g_{\mu\nu} \left[
\left(T\,f_0^{\prime}+\frac{T}{2}f_0^{\prime\prime}\right)
-\left( f_
0^{\prime} T+Tf_0^{\prime\prime}\right)T \right]\nonumber\\
 &&\ \ \ \ \ \ \ \  \ \   \  = 8\pi G\,
T^{(0)(m)}_{\mu\nu}\,.
\label{eq:2pert1}
\end{eqnarray}

Here, $T^{(0)(m)}_{\mu\nu}$ is fixed at the zeroth-order in Eq. (\ref{eq:2pert1}), since
in this perturbation scheme the first order on the Minkowski space has to be connected
with the zeroth order of the standard energy-momentum tensor of matter. Finally, after
some simplifications, we obtain
\begin{eqnarray}
&&\!\!\!\!\!\!\!\!\!\!\!\!
f_0^{\prime}\left[ R_{\mu\nu}^{(1)} - \frac{1}{2} g_{\mu\nu} R^{(1)} \right]
+  Tf_0^{\prime\prime} \left[ R_{\mu\nu}^{(1)} - \frac{1}{2} g_{\mu\nu} R^{(1)} \right]
\nonumber
\\
&&\ \ \ \ \ \,
-\frac{1}{4}
T^2f_0^{\prime\prime}
  = {8\pi G}\,
T^{(0)(m)}_{\mu\nu}\,.
\label{eq:2pert1s}
\end{eqnarray}
At this point, we can make a further assumption on $T$, namely that in our case
the usual relation $R = -T - 2\nabla^{\mu} \left( T^{\nu}_{\verb| |\mu\nu} \right)$
becomes $R\approx-T$ since the second term is of higher order.

%%%%%
The reason why we have truncated the expansion of $f(T)$ at the second order in terms of
$T$ is that we here study the weak-field region. Since the absolute value of the torsion
scalar $T$ is actually related to that of curvature shown above, when we examine the
weak-field region it is considered that the higher order terms in $T$ can be neglected.
In
other words, even if we include the higher order terms in $T$, for instance those in
proportional to $T^3$, the qualitative consequences would not be changed.
%%%%%
In this way, the gravitational field equation is assumed to be
\begin{eqnarray}
&&f_0^{\prime}\left[ R_{\mu\nu}^{(1)} - \frac{1}{2} g_{\mu\nu} R^{(1)} \right]
+  R^{(1)}f_0^{\prime\prime} \left[ R_{\mu\nu}^{(1)} - \frac{1}{2} g_{\mu\nu} R^{(1)}
\right] \nonumber\\&&-\frac{1}{4}(R^{(1)})^2f_0^{\prime\prime}
  = {8\pi G}
T^{(0)(m)}_{\mu\nu}\,,
\label{eq:2pert1s}
\end{eqnarray}
that in terms of perturbation becomes
\begin{eqnarray}
&& \frac{1}{2}f_0^{\prime}\left(\partial_\rho\partial_\nu {h^\rho}_\mu + \partial^\rho
\partial_\mu h_{\nu\rho} - \Box h_{\mu\nu} - \partial_\mu\partial_\nu h
-\right.\nonumber\\ && \left.\eta_{\mu\nu}\partial_\rho\partial^\sigma {h^\rho}_\sigma +
\eta_{\mu\nu} \Box
h\right){}+f_0^{\prime\prime}\left(\partial_\rho\partial^\mu {h^\rho}_\mu - \Box
h\right) \nonumber \\
&&
\times \left(\partial_\rho\partial_\nu {h^\rho}_\mu + \partial^\rho
\partial_\mu h_{\nu\rho} - \Box h_{\mu\nu} - \partial_\mu\partial_\nu h
\right.\nonumber\\&& \left.-\eta_{\mu\nu}\partial_\rho\partial^\sigma {h^\rho}_\sigma +
\eta_{\mu\nu} \Box
h\right) \nonumber \\
&&
{}-\frac{1}{4}f_0^{\prime\prime} \left(\partial_\rho\partial^\mu {h^\rho}_\mu - \Box
h\right)^2\,=\,
{8\pi G}
T^{(0)(m)}_{\mu\nu}\,.
\label{FEh}
\end{eqnarray}
This expression can be simplified significantly using the {\it trace-reversed}
perturbation $\displaystyle{\bar h_{\mu\nu} = h_{\mu\nu} - \frac{1}{2}\eta_{\mu\nu} h}$,
where $\displaystyle{\bar {h}^\mu_\mu = -h}$.
Replacing $h_{\mu\nu}$ with
$\displaystyle{\bar h_{\mu\nu} + \frac{1}{2}\eta_{\mu\nu} h}$ in
Eq. ~(\ref{FEh}) and expanding the equation,
we find that all the terms with the trace $h$ are canceled.
As a result, what remains is
\begin{eqnarray}
&& \frac{1}{2}f_0^{\prime}\left(\partial_\sigma\partial_\nu {{\bar h}^\rho}_{\ \mu} +
\partial^\rho \partial_\mu \bar h_{\mu\nu} - \Box \bar h_{\mu\nu} -\eta_{\mu\nu}
\partial_\rho\partial^\sigma {{\bar h}^\rho}_{\
\sigma}\right)\,\nonumber\\&&
\ \ \ \ \ \ \ \
=\,{8\pi G}
T^{(0)(m)}_{\mu\nu}\,.\label{FEhs}
\end{eqnarray}
Applying the Lorentz gauge condition $\partial^\mu \bar h_{\mu\nu} = 0$ to the above
expression, we see that all but one term vanishes, namely:
\begin{eqnarray}
 -\frac{f_0^{\prime}}{2}\Box \bar h_{\mu\nu}\,=\,{8\pi G}
T^{(m)}_{\mu\nu}\,.\label{GW}
\end{eqnarray}
Thus, in the Lorentz gauge, the gravitational field equation for $f(T)$ gravity is simply
reduced to the wave operator acting on the trace-reversed
metric perturbation (up to a factor $-\displaystyle{\frac{f'_0}{2 }}$)
as in GR.
Therefore, the linearized field equation reads
\begin{equation}
\Box \bar h_{\mu\nu} = - \frac{16 \pi}{f^{\prime}_0} T^{(m)}_{\mu\nu}\,.
\label{eq:elin}
\end{equation}
In vacuum, this equation reduces to
\begin{equation}
\Box \bar h_{\mu\nu} = 0\,.
\label{eq:elin1}
\end{equation}
Similarly to GR, Eq. ~(\ref{eq:elin}) admits a class of homogeneous solutions which are
superpositions of plane
waves, that is
\begin{displaymath}
{\bar h}_{\mu\nu}({\bf x},t) = {\rm Re} \int d^3 k \ A_{\mu\nu}({\bf k}) e^{i
  ({\bf k} \cdot {\bf x} - \omega t)}\,,
\label{eq:planewaves}
\end{displaymath}
with $\omega = |{\bf k}|$. The complex coefficients $A_{\mu\nu}({\bf
k})$ depend on the wavevector ${\bf k}$, but are independent of ${\bf
x}$ and $t$. They are subject to the constraint $k^\mu A_{\mu\nu} = 0$
(which follows from the Lorentz gauge condition) with $k^\mu =
(\omega,{\bf k})$, but are otherwise arbitrary. These solutions are the
gravitational waves.

From this result it becomes clear that $f(T)$ cannot be a ``signature'' to discriminate
further gravitational wave modes or polarizations at the first order in linearized
theory. It is important to note that this result is completely different from that in
$f(R)$ gravity, where it is evident that there exist further degrees of freedom of the
gravitational field \cite{Capozziello:2008rq,Bogdanos:2009tn}. In particular, it can be
found that besides a massless spin-2 field (the standard graviton), $f(R)$ gravity
theories contain also spin-0 and spin-2 massive modes, with the latter being, in general,
ghost modes. As shown in Ref.~ \cite{Li:2011rn} and discussed in detail in subsection
\ref{DOFinFT} above, there are 3 extra degrees of freedom for $f(T)$ gravity in 4
space-time dimensions. These modes do not contribute to the gravitational radiation if it
considered, as standard, at first-order perturbation theory.

%%%%%%%%%%%%%%%%%%%%%
\subsection{Scalar-tensor representation and gravitational waves}
%\pagebreak

Let us now consider, by analogy to $f(R)$ gravity, the scalar-tensor representation of
$f(T)$ gravity, where it is straightforward to check that the scalar mode does not
propagate, unlike the case of $f(R)$ gravity.

We start by re-writing the action (\ref{action_fT}) for $f(T)$ gravity as
\be
S=\int d^4x\ |h| \left(\phi T-V(\phi)\right)\ .
\label{ST1}
\ee
By varying the action with respect to $\phi$, the correspondence between the action
(\ref{ST1}) and (\ref{action_fT}) is obtained if
\be
V'(\phi)=T  \Rightarrow  \phi=\phi(T) \,.
\label{ST2}
\ee
Hence, in principle, we can solve the first equation in (\ref{ST2}) in terms of $\phi$
as the second equation.
This yields
\be
f(T)= \phi(T) T- V(\phi(T))\ .
\label{ST3}
\ee
Moreover, the scalar field and its potential can be written in terms of the function
$f(T)$ as
\be
\phi=f_T(T)\ , \quad V(\phi)=f_T(T)\ T-f(T)\ .
\label{ST3a}
\ee
On the other hand,
the gravitational field equations can be easily obtained by variation of
the action (\ref{ST1}) with respect to the vierbein $e^{A}_{\;\;\mu}$ as:
\begin{eqnarray}
&&\!\!\!\!\!\!\!\!\!\!\!\!\!\!\!\!\!\!\!
\left[
\partial_\mu \left( e e_A^\nu S_\nu^{\verb| |\lambda\rho} \right)
-e e_A^\rho S^{\mu\nu\lambda} T_{\mu\nu\rho} \right] \phi
+e e_A^\nu S_\nu^{\verb| |\lambda\rho} \partial_\rho \phi
\nonumber
\\
&&\ \ \ \ \ \ \ \ \ \
+ \frac{1}{2}  e e_A^\lambda \left[\phi T-V(\phi)\right] =
{8\pi G}
{T^{(m)}}_A^\lambda\ .
\label{ST4}
\end{eqnarray}
In a covariant form, the gravitational field equation (\ref{ST4}) reads
\be
\left( R_{\mu\nu} - \frac{1}{2} g_{\mu\nu} R \right) \phi
- \frac{1}{2}g_{\mu\nu} V(\phi)
+  S_{\nu\mu\rho} \nabla^{\rho} \phi  = {8\pi G}
{T^{(m)}}_{\mu\nu}\,.
\label{ST5}
\ee
Thus, as showed above, we may explore the weak-field limit by providing that
$g_{\mu\nu}\,=\,\eta_{\mu\nu}+h_{\mu\nu}$ and that the scalar field can be described by
a constant background value $\phi_0$ plus a small perturbation $\delta\phi$ around it as
\be
\phi=\phi_0+\delta\phi\ .
\label{ST6}
\ee
Furthermore, the scalar potential can be expanded in powers of the perturbations as
\be
V(\phi)=V_0+V_0^{\prime}\delta\phi+{\cal O}(\delta\phi^2)\ .
\label{ST7}
\ee
Accordingly, the gravitational field equation (\ref{ST5}) at first order of the
perturbations becomes
\be
\!\!\!
\left(\! R_{\mu\nu}^{(1)} - \frac{1}{2} \eta_{\mu\nu} R^{(1)}\! \right) \phi_0
- \frac{1}{2}\left(h_{\mu\nu} V_0+\eta_{\mu\nu}V_0^{\prime}\delta\phi\right)  =
{8\pi G}
{T^{(m)}}_{\mu\nu}.
\label{ST8}
\ee
We mention that the tensor $S_\rho^{\verb| |\mu\nu}$ consists of only first derivatives
of the tetrads and therefore it becomes null at the zeroth order. Additionally, note
that the last term in the left-hand side of Eq. (\ref{ST5}) is null at the
first order in perturbations. Moreover, the scalar torsion $T$ also becomes null at
the zeroth order, and by the scalar field equation (\ref{ST2}) the derivative of the
scalar potential evaluated at $\phi = \phi_0$ is given by
\be
V_0^{\prime}=T_0=0\ ,
\label{ST9}
\ee
where $T_0$ is the value of $T$ at $\phi = \phi_0$. Summarizing, Eq. ~(\ref{ST8})
is approximated as
\be
R_{\mu\nu}^{(1)} - \frac{1}{2} \eta_{\mu\nu} R^{(1)}
+h_{\mu\nu} \Lambda  = \frac{8\pi G}{\phi_0}
{T^{(m)}}_{\mu\nu}\ .
\label{ST10}
\ee
This coincides with the Einstein equation at the first order of perturbations in the
presence of a cosmological constant $\Lambda=-\frac{1}{2}V_0=\frac{1}{2}f(T = 0)$, and
hence the well known result of GR for gravitational waves is recovered. Consequently, in
$f(T)$ gravity, unlike in $f(R)$ gravity, there are no propagating scalar modes in the
gravitational waves, at least when a flat background is assumed.

In summary, in this section we have investigated gravitational waves in $f(T)$ gravity.
In particular, in the Minkowskian limit and for a class of analytic $f(T)$ functions in
the Lagrangian, we have explicitly shown that the gravitational wave modes in $f(T)$
gravity are the same as those in GR \cite{Bamba:2013ooa}. By using this representation,
it
has been shown that the scalar field does not propagate at the first order of
perturbations, because the only remaining terms of the scalar field in the perturbed
equations are the zeroth order, and hence the Einstein equation of GR with a cosmological
constant proportional to $f(T = 0)$ can be recovered. It should be emphasized that the
cosmological constant can exist only if $f(T = 0) \neq 0$.

%---------------------------------------

%%%%%%%%%%%%%%%%%%%%%%%%%%%
\section{Black holes and other solutions}
\label{sectionBlackHoles}
%%%%%%%%%%%%%%%%%%%%%%%%%%%%%%%%%%%%%%%%

In this section we will investigate solutions of $f(T)$ gravity for non-cosmological
geometries, such as the spherically symmetric and black hole solutions, cylindrical
solutions and wormhole solutions. Let us make a crucial comment on this issue. As we
discussed earlier in subsection \ref{restoringLI}, the current formulation of $f(T)$
gravity is based on the strong imposition that the spin connection vanishes, which makes
the theory in general frame-dependent. This consideration is allowed, it helps to make
the
theory simpler, and it can lead to the extraction of many solutions, as long as one does
not ask questions about frame transformations and Lorentz invariance (this is in analogy
with the investigation of electromagnetism in the particular class of the inertial
frames). If one desires to examine frame transformation and Lorentz invariance then he
must go back, start from the covariant teleparallel gravity, instead of the pure-tetrad
one, and re-formulate the theory with a general spin connection in a covariant way
\cite{Krssak:2015oua}, as we did in subsection \ref{restoringLI}. In this case the theory
is frame-independent and Lorentz invariant, since every frame transformation will be
accompanied by a suitable connection transformation. However, covariant $f(T)$ gravity
is a bit more complicated, and as we discussed one must follow specific methods
in order to extract solutions \cite{Krssak:2015oua}.

Hence, in this section we will keep the standard formulation of $f(T)$ gravity, namely
imposing a zero spin connection, having in mind however that this would lead us to be
very careful with the frame choice. In particular, and following the existing literature,
we will seek for solutions in the simplest case of diagonal vierbeins, as well as in the
case of non-diagonal vierbeins, separately. In this case, different choices of frames
lead to different classes of solutions, and vice versa a given metric solution is
realized by a specific choice of vierbeins. Nevertheless, once again we stress that these
differences disappear in the consistent, covariant $f(T)$ gravity.

\subsection{Black hole solutions}

\subsubsection{Neutral spherically symmetric solutions}

\paragraph{Diagonal and non-diagonal vierbein choice}
\label{diagandnondiagBHSw}

Let us extract spherically symmetric solutions in $f(T)$ gravity. In order to clarify the
discussion on the diagonal and non-diagonal vierbein choice in the standard formulation
of $f(T)$ gravity, i.e with zero spin connection imposition, we first give a simple
example following  \cite{Ferraro:2011ks}. In particular, we consider a general $f(T)$
ansatz and we want to examine which vierbein choice would lead to the extraction of
exactly the Schwarzschild solution (note that this is a different question from the
one to start with a given veirbein and try to find how a given $f(T)$ gravity adds
corrections to the Schwarzschild solution).

For completeness and convenience for the reader we remind that we consider the
action (\ref{action_fT}), namely
\begin{equation}
 {\cal S}=\frac{1}{16\pi G}\int d^{4}x\;e\;f(T)+\int d^{4}x\;e\;
L_m\, ,  \label{actionBHSw}
\end{equation}
where $L_m$ refers to the standard matter Lagrangian, which leads to the
general equations of motion (\ref{eom_fT_general}), namely:
\begin{eqnarray}
&&\!\!\!\!\!\!\!\!\!\!\!\!\!\!\!\!\!\!\!
e^{-1}\partial _{\mu }(e\,e_{A}^{\ \ \rho }S_{\rho}^{\ \ \mu \nu })\ \,f_{T }+e_{A}^{\
\ \lambda }S_{\rho }^{\ \ \nu \mu }\ T_{\ \ \mu \lambda
}^{\rho }\,\ f_{T}
\notag \\
&&\!\!\!\!\!\!\!\!\!\!\!\!
+S_{A}^{\ \ \mu \nu }\,\partial _{\mu }T\ \,f_{TT }+ \frac{1}{4}\
e_{A}^{\ \ \nu }\,f=4\pi G\,
e_{A}^{\ \, \mu}\,
{T^{(m)}}_{\mu}{}^{\nu},
\label{BHsectioneoms}
\end{eqnarray}
with
\begin{equation}
S_{\ \mu \nu }^{\rho }=\frac{1}{4}\,(T_{\ \mu \nu }^{\rho }-T_{\mu
\nu }^{\ \ \ \rho }+T_{\nu \mu }^{\ \ \ \rho })+\frac{1}{2}\ \delta
_{\mu }^{\rho }\ T_{\sigma \nu }^{\ \ \ \sigma }-\frac{1}{2}\ \delta
_{\nu }^{\rho }\ T_{\sigma \mu }^{\ \ \,\sigma },  \label{tensor}
\end{equation}
and
where ${T^{(m)}}_{\mu}{}^{\nu}$ stands for the matter energy-momentum tensor.

Let us now consider the Schwarzschild metric
\begin{equation}
ds^2=\left(1-\frac{2M}{r}\right)\,
dt^2-\frac{dr^2}{1-\frac{2M}{r}}-\, r^2\, d\Omega^2\, .
\label{metusual}
\end{equation}
According to the usual relation $g^{\mu\nu}\,e^A_\mu\,e^B_\nu\ =\ \eta^{AB}$, it is clear
that the above diagonal metric can be generated either by the diagonal vierbein
\begin{eqnarray}
e^0&=&\left(1-\frac{2M}{r}\right)^{1/2}\ dt\, , \notag \\
e^1&=&\left(1-\frac{2M}{r}\right)^{-1/2}\ dr\, , \notag\\
e^2&=&r\ d\theta\, , \notag\\
e^3&=&r \ sin\,\theta\ d\varphi\, , \label{naiveviebrein}
\end{eqnarray}
or by any other Lorentz transformed, non-diagonal, choice. However, substituting the
diagonal vierbein choice (\ref{naiveviebrein}) into the $r$-$\theta$ equation of
(\ref{BHsectioneoms}), we obtain the equation
\begin{equation}
f_{TT}\,(16 M^3-8M^2r-2Mr^2+r^3)=0\, ,
\end{equation}
which clearly cannot be satisfied unless we are in the simple TEGR, i.e when
$f(T)=T-2\Lambda$ and thus $f_{TT}=0$. Hence, we deduce that in order to obtain a
realization of the Schwarzschild metric in $f(T)$ gravity (in its standard formulation
with zero spin connection), we need to consider a suitably designed non-diagonal vierbein.
In the following we will briefly describe the construction of such a vierbein, following
 \cite{Ferraro:2011ks}.

Firstly, it proves more convenient to transform the spherically symmetric Schwarzschild
metric in isotropic coordinates as
\begin{equation}
ds^2=A(\rho)^2\, dt^2-B(\rho)^2\, \left(dx^2+dy^2+dz^2\right)\, ,
\label{Swain isotropicc}
\end{equation}%
where  $A$ and $B$ are functions of the radial coordinate
$\rho=\sqrt{x^2+y^2+z^2}$, reading as
\begin{equation}
A(\rho)=\frac{2\rho-M}{2\rho+M},\,\,\,\,\,\,\,B(\rho)=\Big(1+
\frac{M}{2\rho}\Big)^2\, .
\label{ABformSw}
\end{equation}
Note that the isotropic interval covers the exterior region of the
Schwarzschild space-time only, as it becomes clear from the relation between the radial
coordinate $r$ and the coordinate $\rho$, namely
$ \sqrt{r^2-2M r}+r-M=2\rho$. Now, let us introduce the ``asymptotic'' vierbein,
by taking the ``squared root'' of the metric
(\ref{Swain isotropicc}) as
\begin{eqnarray}
e^0&=&A(\rho)\,dt\, , \notag \\
e^1&=&B(\rho)\,dx\, , \notag\\
e^2&=&B(\rho)\,dy\, , \notag\\
e^3&=&B(\rho)\,dz\, .
\label{transformedframeBHSw}
\end{eqnarray}
This non-diagonal vierbein, unlike (\ref{naiveviebrein}), partially has the features we
are looking for, since it captures the asymptotic geometrical meaning of the
parallelization process, reflected in the fact that at spatial infinity  one obtains the
Minkowskian frame $e^a_{\mu}(\infty)=\delta^a_{\mu}$, which yields a vanishing
torsion tensor. Nevertheless, the  asymptotic frame  (\ref{transformedframeBHSw}) is also
not adequate to globally describe the Schwarzschild geometry, as can
be checked by replacing it in the equations of motion (\ref{BHsectioneoms}). Hence, one
needs to seek for a Lorentz transformation in such a way that acting on
(\ref{transformedframeBHSw}) to obtain a non-diagonal vierbein  able to achieve a null
torsion scalar $T$. This can be achieved by using  a radial boost depending solely
on the radial coordinate $\rho$. The most general boosted (radial) frame coming from
(\ref{transformedframeBHSw}) writes as  \cite{Ferraro:2011ks}
\begin{eqnarray}
\bar{e}^0&=&A(\rho)\gamma(\rho)\,dt\nonumber\\
&&
-\frac{B(\rho)}{\rho}
\sqrt{\gamma^{2}(\rho)-1}\,[x\,dx+y\,dy+z\,dz]\, , \notag \\
\bar{e}^1&=&-\frac{A(\rho)}{\rho}\,\sqrt{\gamma^{2}(\rho)-1}\,x
\,dt
\nonumber\\
&&
+B(\rho)\Big[(1+\frac{\gamma(\rho)-1}{\rho^2}x^2)\,dx+
\frac{\gamma(\rho)-1}{\rho^2}\,x
\,y\,dy
\nonumber\\
&& \ \ \ \ \ \ \ \ \ \ \,
+\frac{\gamma(\rho)-1}{\rho^2} \,x \,z\,dz\Big]\, ,\notag \\
\bar{e}^2&=&-\frac{A(\rho)}{\rho}\,\sqrt{\gamma^{2}(\rho)-1}\,y
\,dt
\nonumber\\
&&
+B(\rho)\Big[\frac{\gamma(\rho)-1}{\rho^2}\,x
\,y\,dx+(1+\frac{\gamma(\rho)-1}{\rho^2}y^2)\,dy
\nonumber\\
&& \ \ \ \ \ \ \ \ \ \ \,
+
\frac{\gamma(\rho)-1}{\rho^2}\,y
\,z\,dz\Big]\, ,\notag \\
\bar{e}^3&=&-\frac{A(\rho)}{\rho}\,\sqrt{\gamma^{2}(\rho)-1}\,z
\,dt
\nonumber\\
&&
+B(\rho)\Big[\frac{\gamma(\rho)-1}{\rho^2}\,x\,
z\,dx+\frac{\gamma(\rho)-1}{\rho^2}\,y\,
z\,dy
\nonumber\\
&& \ \ \ \ \ \ \ \ \ \ \,
+(1+\frac{\gamma(\rho)-1}{\rho^2}z^2)\,dz\Big]\, ,\notag \\
\label{nondiagvierbBHSw}
\end{eqnarray}
where
\begin{equation}
\gamma(\rho)=\Big[1-\beta^{2}(\rho)\Big]^{-\frac{1}{2}},\,\,\,\,\,
\beta(\rho)=v(\rho)/c\,  .
\label{gammabetaBHSw}
\end{equation}
One can verify that the torsion scalar $T$ vanishes  if and only if
the Lorentz factor $\gamma(\rho)$ is chosen as
\begin{equation}
\gamma(\rho)=\frac{4\rho^2+M^2}{4\rho^2-M^2}\, .
\end{equation}
Hence, in summary, one can see that the Schwarzschild metric (\ref{Swain isotropicc}) can
be realized in $f(T)$ gravity (as long as $f(T)$ includes the usual TEGR case, i.e if
it can be written as $T+ \mathcal{O}(T^2)$  \cite{Ferraro:2011ks}) only for the class of
non-diagonal vierbeins (\ref{nondiagvierbBHSw}).

From the above discussion we deduce that in usual, ``pure tetrad'' $f(T)$ gravity,
different subclasses of solutions would be revealed using different vierbein choices,
although the metric is the same. Definitely, as we discussed above, and in detail in
subsection \ref{restoringLI}, this will not be the case anymore if one re-formulates
$f(T)$ gravity in a covariant way,  using a non-zero spin connection. In this case any
metric solution is able to be realized by any vierbein choice (along with a suitable spin
connection), there are not ``good'' and ``bad'' tetrads, and hence all classes of
solutions will be extracted at once. However, in covariant $f(T)$ gravity specific
methods
have to be followed in order to be able to by-pass the complexity of the equations
and extract spherically symmetric solutions \cite{Krssak:2015oua}, as we did for simple
examples in subsection \ref{restoringLI}. Hence, in the following paragraphs we remain in
the  usual $f(T)$ gravity and we discuss some simple spherically symmetric solutions for
diagonal and non-diagonal vierbein choices separately.

\paragraph{Spherically symmetric solutions with diagonal vierbeins}

We look for spherically symmetric solutions of the field equations (\ref{BHsectioneoms})
of the form
\begin{align}
  ds^2 = e^{a(r)} dt^2 - e^{b(r)} dr^2 -R(r)^2 d\Omega^2\,,
  \label{metricdiagBHSw}
\end{align}
where $d\Omega^2 = d\theta^2 + \sin^2\negmedspace\theta\, d\varphi^2$ and where $a$, $b$
and $R$
are three unknown functions. Concerning the vierbein choice that realizes the above
metric we choose the diagonal one, namely:
 \begin{align}
  e^A{}_{\mu} = \text{diag}(e^{a(r)/2},e^{b(r)/2},R(r),R(r)\sin \theta)\,.
  \label{tetradddiagBHSw}
\end{align}
Thus, its determinant $e$ is simply  $e=e^{(a+b)/2}R^2\sin\theta$. Using the above
diagonal vierbein, the torsion scalar $T$ from (\ref{telelag}) becomes
\begin{eqnarray}
  && T(r) = 2e^{-b}\frac{R'}{R} \left(a'+\frac{R'}{R}\right)\,,
  \label{TscalardiagBHSw}
\end{eqnarray}
while the field equations (\ref{BHsectioneoms}) become:
{\small{
\begin{eqnarray}
 && \!\!\!\!\!\!\!\!\!\!\!
 4 \pi \rho_m=\left\{T-\frac{
1}{R^2}-e^{-b}\left[(a'+b')\frac{R'}{R}-2\frac{R''}{R}\right]\right\}\frac{f_T}{2}-\frac{f
} {4}\nonumber\\
&&\ \ \ \ \ \ \ \ \ \ +
e^{-b}\frac{R'}{R}T' f_{TT}
\,,
  \label{field:t}\\
&& \!\!\!\!\!\!\!\!\!\!\! 4\pi p_m= \left(\frac{1}{R^2}-T\right)\frac{f_T}{2}+\frac{f}{4}
\,,
  \label{field:r}\\
&& \!\!\!\!\!\!\!\!\!\!\!4 \pi p_m=
-\frac{e^{-b}}{2}\left(\frac{a'}{2}+\frac{R'}{R}\right)T'f_{TT}+\frac{f}{4}
  \nonumber\\
 && \!\!\!\!\! \!\!\!\!\!\!\!-\left\{ \frac{T}{2}+e^{-b}\left[\frac{R''}{
R}+\frac{a''}{2}+\left(\frac{a'}{4}+\frac{R'}{2
R}\right)(a'-b')\right]\right\}\frac{f_T}{2} \,,
  \label{field:theta}
 \\
&& \!\!\!\!\! \!\!\! \!\!\!\frac{e^{-b/2}\cot\theta}{2R^2}T'f_{TT}=0\,,
  \label{field:rtheta}
\end{eqnarray}}}
with primes denoting derivative with respect to $r$, and where $\rho_m$ and $p_m$ are
respectively the matter energy density and pressure assuming that the matter energy
momentum tensor corresponds to an isotropic perfect fluid. Note that the last equation is
the $(r,\theta)$ one, which is absent in the case of simple TEGR, but it is non-trivial
in
the general $f(T)$ gravity. Hence, in the case of general $f(T)$ gravity, i.e. where
$f_{TT} \neq0$, we deduce that the $r$-derivative of the torsion scalar $T$
must vanish.

The  field equations (\ref{field:t})-(\ref{field:rtheta}) are four independent
equations for six unknown functions, namely $a(r)$, $b(r)$, $R(r)$, $\rho_m(r)$, $p_m(r)$
and $f(T)$, i.e they form an  under-determined system. In order to solve it we have to
determine two of the above functions, and the usual assumption is to impose the matter
equation of state $\rho_m=\rho_m(p_m)$, as well as the function $f(T)$ which implies to
choose a specific $f(T)$ theory. However, since such a procedure would in general lead to
mathematical complications which forbid analytical treatment, we follow
\cite{Boehmer:2011gw} and we either make assumptions on the metric functions, or
we impose conditions on useful combinations of terms which simplify the field equations.

\begin{itemize}

\item {Solutions with $T=0$}

We start by extracting the simplest solution subclass, namely the one that corresponds to
$T=0$. Inserting this condition into (\ref{TscalardiagBHSw}) gives
\begin{equation}
e^{a(r)} = \frac{c_1}{R(r)},
 \label{solT0boemher1}
\end{equation}
where $c_1$ is a constant of integration. Additionally, note that due to
$T=0$, $f$ and its derivatives are constants, equal to $f(0)$ and $f_T(0)$
respectively. Inserting the above into the field equations
(\ref{field:t})-(\ref{field:rtheta}) gives  \cite{Boehmer:2011gw}
 \begin{align}
 e^{b(r)} = \frac{R(r) R'(r)^2}{c_2-4R(r)},
  \label{solT0boemher2}
\end{align}
with $c_2$ an integration constant, and moreover setting $f_T(T=0)=0$ we obtain
\begin{equation}
  \rho_{m0}=-p_{m0}=\frac{f_0}{16 \pi}\,.
\end{equation}
In summary, the metric becomes
\begin{align}
ds^2= \frac{c_1}{R} dt^2 -\frac{R R'^2}{c_2-4R} dr^2 - R^2d\Omega^2\,,
\label{metricT0BHBoemer}
\end{align}
and it has a singularity when $R \rightarrow 0$.

We mention that in the above solution the $f(T)$ form is arbitrary, as long as its value
(or its derivative) at the origin remains finite. Finally, note that the above solution
yields a constant energy density $\rho_{m0}$ and pressure $p_{m0}$, with equation of
state
equal to $-1$.

\item {Solutions with $T'=0$}

Let us now investigate the solution subclass with $T'=0$, that is
\begin{equation}
T(r)=T_0=const.\,
\end{equation}
Imposing additionally the gauge $R(r)=r$  the field equation~(\ref{field:r}) becomes
\begin{align}
  4\pi p_m=\frac{T_0}{2}f_T(T_0)-\frac{f(T_0)}{4}+\frac{f_T(T_0)}{2r^2} \,,
\end{align}
with $f(T_0)$ and $f_T(T_0)$ constants, which at $r\rightarrow\infty$ leads to
\begin{align}
  f(T_0) = c_3 \sqrt{T_0} - 16\pi p_{m\infty} \,,
\end{align}
with $p_{m\infty}$ the pressure value at infinity and $c_3$ a constant of integration
($p_{m\infty}$ plays the role of a cosmological constant). Hence, unfortunately, note
that
this solution subclass does not exist for the simple TEGR, but it corresponds to a
special $f(T)$ form.

Concerning the functions $a(r)$ and $b(r)$ these are hard to be
extracted analytically, however one can find approximate expressions in the case where
$r^2T_0\ll 1$, namely  \cite{Boehmer:2011gw}
\begin{align}
  e^{a(r)}=\frac{k_1}{r}\,, \quad e^{b(r)} = \frac{r}{k_2-4}\,,
 \label{solTptim0boemher}
\end{align}
with $k_1$,$k_2$ integration constants,
which coincides with solution (\ref{solT0boemher1}),(\ref{solT0boemher2}) for $T=0$.
Therefore, at zeroth order in $r^2T_0$ the general solution matches the $T=0$ solution,
as
expected. On the other hand for $r^2T_0\gg 1$
one finds  \cite{Boehmer:2011gw}
 \begin{eqnarray}
&&  e^{a(r)}=  \left(\frac{r^8T_0-2k_3}{k_4}\right)^{-\frac{1}{8}} \,\\
&& e^{b(r)}=\frac{k_3}{r^6}-\frac{T_0}{2}r^2 \,,
\end{eqnarray}
with $k_3$,$k_4$ integration constants. In summary, in the regime $r^2T_0\gg 1$ the
metric reads
{\small{
\begin{equation}
  ds^2 = \left(\frac{k_4}{T_0r^8-2k_3}\right)^{1/8} dt^2 -
\left(\frac{2k_3-T_0r^8}{2r^6}\right)
\,dr^2 -r^2 d\Omega^2\,,
\end{equation}}}
which is  singular at $r\rightarrow \infty$ and presents a horizon at
$r=(2k_1/T_0)^{1/8}$.

 \end{itemize}

\paragraph{Spherically symmetric solutions with non-diagonal vierbeins}

In this paragraph we will examine the realization of the same spherically symmetric
metric (\ref{metricdiagBHSw}) of the previous paragraph but with a non-diagonal vierbein
choice. Without loss of generality, we can consider  \cite{Pereirabook}
\begin{align}
  e^A{}_{\mu} = \left(
\begin{array}{cccc}
 e^{a/2} & 0 & 0 & 0 \\
 0 & e^{b/2} \sin\theta\cos\phi  & R\cos \theta\cos\phi & -R\sin \theta\sin \phi \\
 0 & e^{b/2} \sin\theta\sin\phi & R\cos \theta\sin\phi& R\sin\theta \cos\phi \\
 0 & e^{b/2} \cos \theta  & -R \sin\theta & 0
\end{array}
\right)\,,
  \label{tetradoffworm}
\end{align}
 with determinant $e=e^{(a+b)/2}R(r)^2\sin\theta$. In such a case the torsion scalar
(\ref{telelag}) reads as
\begin{align}
  T(r) = \frac{2 e^{-b} \left(e^{b/2}-R'\right)\left(e^{b/2}-R'-R a '\right)}{R^2}\,.
\end{align}
Hence, substitution into the general  field equations (\ref{BHsectioneoms}) yields
\begin{eqnarray}
 && 4\pi\rho_m = \frac{e^{-b/2}}{R}(R'e^{-b/2}-1)
T'f_{TT}+\left(\frac{T}{4}-\frac{1}{2R^2}\right)f_T \nonumber\\
 &&\ \ \ \ \ \ \ \  +\frac{e^{-b}}{2 R^2} \left(2 RR''-R R'b
'+R'^2\right)f_T-\frac{f}{4}\,,
  \label{field:toff}
\end{eqnarray}
\begin{eqnarray}
  4\pi p_m =\left[ \frac{1}{2R^2}-\frac{T}{4}-\frac{e^{-b}}{2R^2}
R'(R'+Ra')\right]f_T+\frac{f}{4}\,,
  \label{field:roff}
\end{eqnarray}
{\small{
\begin{eqnarray}
&&\!\!\!\!\!\!\!\!\!\!\!\!\!\!\!\!\!\!4\pi p_m = \frac{f}{4}
-\frac{e^{-b}}{2}\left(\frac{a'}{2}+\frac{R'}{R}-\frac{e^{b/2}}{R}\right)T'
f_{TT} \nonumber
\\
&&\!\!\!\!\!\!\!\!\!\!\!\!\!\!\!\!\!\!\!\!\!\! -f_T\left\{ \frac{T}{4}+\frac{e^{-b}}{2 R}
\left[R''+\left(\frac{R'}{2}+\frac{Ra'}{4}\right) \left(
a'-b'\right)+\frac{Ra''}{2}\right]\right\}\!.
  \label{field:thetaoff}
\end{eqnarray}}}
Note that there is no equation enforcing the constancy of the torsion scalar, contrary to
the case of diagonal vierbeins of the previous paragraph.

\begin{itemize}

\item {Solutions with $b=0$}

Let us first look for solutions in the subclass $b(r)=0$. Imposing additionally the gauge
$R(r)=r$ we find that for all values of $a$ and $r$ the torsion scalar
(\ref{telelag}) becomes
\begin{align}
  T(r)=0\,,
\end{align}
and thus  $f$ and its derivative are constants, denoted by  $f_0$ and $f_1$ respectively.
In this case, the field equations (\ref{field:toff})-(\ref{field:thetaoff}) give
 \cite{Boehmer:2011gw}
\begin{align}
 e^{a(r)} =  c_2 (r^2 + c_1)^2\,,
\end{align}
with $c_1$,$c_2$ integration constants, and moreover
\begin{align}
  \rho_{m0} &= -\frac{f_0}{16\pi} \,,\\
  p_m& =-\frac{1}{2\pi(r^2+c_1)}f_1+\frac{f_0}{16\pi} \,,
\end{align}
where the pressure is regular everywhere provided that $c_1 > 0$. Hence, in summary, in
this case the metric  (\ref{metricdiagBHSw}) becomes
\begin{align}
  ds^2= c_2 (r^2 + c_1)^2 dt^2- dr^2- r^2d\Omega^2\,.
  \label{metricb0}
\end{align}

\item {Solutions with  $f_T(T=0)=0$}

 Let us now keep the gauge  $R(r)=r$ but impose the function $b(r)$ to be related
with $a(r)$ through
  \begin{align}
  b(r)=2\ln\left[1+r\,a'(r)\right]\,.
  \label{barnondiagonBoehmer}
\end{align}
Such a choice leads to $T=0$ for all $a(r)$. Assuming additionally that $f_T(T=0)=0$
we finally obtain \cite{Boehmer:2011gw}:
\begin{align}
  p_m=-\rho_m =\frac{f(0)}{16\pi} \,.
  \label{rhoparnondiagonBoehmer}
\end{align}
Note that the imposition $f_T(T=0)=0$  is satisfied by a large class of $f(T)$
forms, including the standard TEGR. Hence, if $f(T)=T^n$ then
(\ref{rhoparnondiagonBoehmer}) leads to  $p_m=\rho_m=0$, while if for instance
$f(T)=\cos(kT)$
then $p_m=-\rho_m=1/(16\pi)$. For all these models we acquire an infinite class of
solutions, given by choosing an arbitrary form for $a(r)$ and extracting $b(r)$ through
the condition (\ref{barnondiagonBoehmer}). For example,
considering a Schwarzschild-like form for $a(r)$, namely
\begin{align}
 e^{ a(r)}=\left(1-\frac{2M}{r}\right)\,,
\end{align}
we find
\begin{align}
 e^{ b(r)}=\left(1-\frac{2M}{r}\right)^{-2}\,,
\end{align}
and the metric  (\ref{metricdiagBHSw}) becomes
\begin{equation}
 \ \ \ \ \ \
 ds^2= \left(1-2M/r\right)dt^2-\left(1-2M/r\right)^{-2}dr^2-r^2d\Omega^2\,,
\end{equation}
which is similar, but not exactly the Schwarzschild solution.

\item {Solutions with $T'=0$}

In the previous case we finally found that $T=0$ for all $r$. However, one can extract
solutions with constant $T$ for some specific $f(T)$ models. We fix the gauge  $R(r)=r$
and we consider  \cite{Boehmer:2011gw}
\begin{align}
  a(r)=\ln(q\,r)-T_0\,r^2\,, \qquad b(r)=\ln 4 \,,
  \label{barnondiagonBoehmer13}
\end{align}
with $q$ and $T_0$ constants, corresponding to the metric
\begin{align}
  ds^2= q\,r\,e^{-T_0r^2} dt^2 - 4 dr^2 - r^2d\Omega^2\,.
\label{barnondiagonBoehmer14}
\end{align}
Note that solution (\ref{barnondiagonBoehmer13}) immediately implies $T=T_0$. Thus,
in this case the field equations (\ref{field:toff})-(\ref{field:thetaoff}) lead to
\begin{align}
p_m=-\rho_m=\frac{f(T_0)}{4} \,,
\end{align}
along with the requirement $f_T(T_0)=0$. Note that this last constraint implies that not
all $f(T)$ forms allow for the solution  (\ref{barnondiagonBoehmer14}). One $f(T)$
example that accepts the above solution is the $f(T)=T-\frac{T^2}{2T_0}$, in which case
one obtains standard TEGR for $T\ll T_0$. Note that the energy density and pressure
become constant and equal to $p_m=-\rho_m=T_0/8$ (hence behaving like a cosmological
constant), and thus energy positivity requires $T_0<0$.

\item {Power-law solutions}

 Let us consider the ansatzen    \cite{Boehmer:2011gw}
 \begin{align}
 \ \ \ \ \  a(r)=m\ln(p\,r)\,, \qquad\mbox{and}\qquad b(r)=2\ln n \,,
\end{align}
 which correspond to the metric
\begin{align}
  ds^2= (p\,r)^mdt^2-\frac{1}{n^2}dr^2-r^2d\Omega^2\,,
  \label{barnondiagonBoehmer16}
\end{align}
with $n$ and $m$ constants and  $p$ a parameter with  dimension 1/length. In this case
the torsion scalar (\ref{telelag}) is calculated as
\begin{align}
  T=\frac{2 (n-1) (-m+n-1)}{n^2 r^2} \,,
  \label{015}
\end{align}
(with $n\neq 0$, $n\neq 1$ and $n\neq m+1$ in order to obtain a non-zero and regular
$T$).
In this case, and for $m\neq 2n+2$, the field equations
(\ref{field:toff})-(\ref{field:thetaoff}) lead to
\begin{align}
&  f(T)=\gamma+\frac{\xi}{N}T^N \,, \end{align}
with
\begin{align}
  \qquad N =
\frac{m^2+4(n-1)^2}{4(2+m-2n)} \,,
  \label{017}
\end{align}
and $\gamma$, $\xi$ constants. Hence, for all these power-law $f(T)$ models one obtains
the spherically symmetric solution (\ref{barnondiagonBoehmer16}). We mention that TEGR is
recovered when $N=1$, which can happen for a wide choice of the parameters $n$ and $m$.
Finally, the field equations yield also
\begin{align}
  16 \pi\rho_m &= \xi\,h_1(n,m) \left[\frac{2(n-1) (-m+n-1)}{n^2
r^2}\right]^N
 \nonumber\\
  & \ \ \ -\gamma
\,,\\
  16 \pi p_m &=\xi\, h_2(n,m) \left[\frac{2(n-1) (-m+n-1)}{n^2 r^2}\right]^N\nonumber\\
   & \ \ \ + \gamma
\,,
\end{align}
where $h_1$ and $h_2$ are two constants that depend on $n$ and $m$ in a complicated
way. Notice that if $N>0$ when $r\rightarrow\infty$ we acquire $p_m=-\rho_m=\gamma$,
while if
$N<0$ this happens at $r=0$. The explicit forms of $\rho_m$ and $p_m$ depend upon the
choice of $n$ and $m$. In summary, we were able to extract a solutions with a
non-constant $T$, for the wide class of power-law $f(T)$ forms.

 \end{itemize}

We close this paragraph by mentioning that one can find many other subclasses of
neutral spherically symmetric solutions, imposing various considerations. The reader is
refereed to  \cite{Ferraro:2011ks, Wang:2011xf,
HamaniDaouda:2011iy,Boehmer:2011gw,Nashed:2011fz,
Deliduman:2011ga,
Nashed:2012ms,Atazadeh:2012am,Nashed:2014sea,Aftergood:2014wla,
Bhadra:2014jea, delaCruz-Dombriz:2014zaa, Kofinas:2015hla, Kofinas:2015zaa, Abbas:2015xia,
Junior:2015fya,Junior:2015dga,Das:2015gwa,Debnath:2014yya,Zubair:2015cpa,Ruggiero:2015oka,
Abbas:2015yma,Nashed:2015qza,Hanafy:2015yya,Nashed:2015pga,Nashed:2015wia}
for more details.

\subsubsection{Spherically Symmetric solutions by Noether symmetry approach}

In this paragraph we will present a different but very helpful way to extract spherically
symmetric solutions in $f(T)$ gravity, based on Noether symmetries, following
 \cite{Paliathanasis:2014iva}. We start by generalizing the $f(T)$ gravity Lagrangian
(\ref{actionBHSw}) (neglecting matter for the moment)
through the use of a Lagrange multiplier. In particular, we can write it as
\begin{equation}
L\left( x^{k},x^{\prime k},T\right) =2f_{T}\bar{\gamma}_{ij}\left(
x^{k}\right) x^{\prime i}x^{\prime j}+M\left( x^{k}\right) \left(
f-Tf_{T}\right) ,  \label{L1Paliath}
\end{equation}%
where $x^{\prime }=\frac{dx}{d\tau }$, $M(x^{k})$ is the Lagrange multiplier
and $\bar{\gamma}_{ij}$ is a second rank tensor which is related to the
frame [one can use $eT(x^{k},x^{\prime k})$] of the background space-time. In
the same lines, the Hamiltonian of the system is written as
\begin{equation}
H\left( x^{k},x^{\prime k},T\right) =2f_{T}\bar{\gamma}_{ij}\left(
x^{k}\right) x^{\prime i}x^{\prime j}-M\left( x^{k}\right) \left(
f-Tf_{T}\right) \;.  \label{L2Paliath}
\end{equation}
Hence,  the system is autonomous and because of that $\partial _{\tau
} $ is a Noether symmetry with corresponding Noether integral the
Hamiltonian $H$. Additionally, since the coupling function $M$ is a function
of $x^{k}$, it is implied that the Hamiltonian (\ref{L2Paliath}) vanishes
 \cite{Tsamparlis:2013aza}. In this framework, considering $\{x^{k},T\}$ as the canonical
variables of
the configuration space, we can derive, after some algebra, the general
field equations of $f(T)$ gravity. Indeed, starting from the Lagrangian (\ref%
{L1Paliath}), the Euler-Lagrange equations
\begin{equation}
\frac{\partial L}{\partial T}=0,\;\;\;\;\frac{d}{d\tau }\left( \frac{%
\partial L}{\partial x^{\prime k}}\right) -\frac{\partial L}{\partial x^{k}}%
=0\,,  \label{Lf.01}
\end{equation}%
give rise to
\begin{equation}
f_{TT}\left( 2\bar{\gamma}_{ij}x^{\prime i}x^{\prime j}-MT\right) =0,
\label{Lf.04}
\end{equation}%
\begin{equation}
x^{i\prime \prime }+{\bar \Gamma}_{jk}^{i}x^{j\prime }x^{k\prime }+\frac{%
f_{TT}}{f_{T}}x^{i\prime }T^{\prime }-M^{,i}\frac{\left( f-Tf_{T}\right) }{%
4f_{T}}=0\;.  \label{Lf.06}
\end{equation}%
We mention here that, for convenience, the functions ${\bar \Gamma}_{jk}^{i}$
are considered: they are exactly the Christoffel symbols for the metric $%
\bar{\gamma}_{ij}$. Therefore, the system is determined by the two
independent differential equations (\ref{Lf.04}),(\ref{Lf.06}), and the
Hamiltonian constrain $H=0$ where $H$ is given by (\ref{L2Paliath}).

The point-like Lagrangian (\ref{L1Paliath}) determines completely the related
dynamical system in the minisuperspace $\{x^{k},T \}$, implying that one can
easily recover some well known cases. In brief, these are the flat FRW space-time
analyzed
in subsection \ref{FRWNetherr}, the Bianchi type I, and the static spherically symmetric
space-time that we are interested in the current paragraph, written more conveniently as:
\begin{eqnarray}
\label{SSPaliath}
&&ds^{2}=a^{2}\left( \tau \right) dt^{2}-\frac{1}{N^{2}\left( a\left( \tau
\right) ,b\left( \tau \right) \right) }d\tau ^{2}  \notag \\
&&\ \ \ \ \ \ \ \ \, -b^{2}\left( \tau \right) \left( d\theta ^{2}+\sin
^{2}\theta d\phi ^{2}\right),
\end{eqnarray}
arising from the diagonal vierbein (for the shake of simplicity we do not examine here
the non-diagonal case)
\begin{equation}
\label{SSAPaliath}
e_{i}^{A}=\left( a\left( \tau \right) ,\frac{1}{N\left( a\left( \tau \right)
,b\left( \tau \right) \right) },b\left( \tau \right) ,b\left( \tau \right)
\sin \theta \right) \;,
\end{equation}
where $a(\tau)$ and $b(\tau)$ are functions which need to be determined.
Therefore, the line element of $\bar{\gamma}_{ij}$ and $M\left( x^{k}\right)$
are respectively given by
\begin{eqnarray}  \label{SSA1}
&&ds_{\bar{\gamma}}^{2}=N\left( 2b~da~db+a~db^{2}\right),  \notag \\
&&M(a,b)=\frac{ab^{2}}{N}.
\end{eqnarray}

Let us now apply the Noether Symmetry Approach \cite{Tsamparlis:2011wf} (see also
\cite{Christodoulakis:2014wba}) to a general
class of $f(T)$ gravity models where the corresponding Lagrangian of the field
equations is given by (\ref{L1Paliath}). We start by performing the analysis
for arbitrary space-times, and then we focus on static spherically-symmetric
geometries. The Noether symmetry condition for the Lagrangian (\ref{L1Paliath}) is given
by  \cite{Paliathanasis:2014iva}
\begin{equation}
X^{\left[ 1\right] }L+L\xi ^{\prime }=g^{\prime },
\label{LL2Paliath}
\end{equation}%
where the generator $X^{\left[ 1\right] }$ is written as
\begin{eqnarray}
&&X^{\left[ 1\right] }=\xi \left( \tau ,x^{k},T\right) \partial _{\tau
}+\eta ^{k}\left( \tau ,x^{k},T\right) \partial _{i}  \notag \\
&&\ \ \ \ \ \ \ \ \ +\mu \left( \tau ,x^{k},T\right) \partial _{T}+\left(
\eta ^{\prime i}-\xi ^{\prime }x^{\prime i}\right) \partial _{x^{\prime i}}.
\end{eqnarray}%
For each term of the Noether condition (\ref{LL2Paliath}) for the Lagrangian (\ref%
{L1Paliath}) we respectively obtain
\begin{eqnarray*}
X^{\left[ 1\right] }L &=&2f_{T}\bar{g}_{ij,k}\eta ^{k}x^{\prime i}x^{\prime
j}+M_{,k}\eta ^{k}\left( f-Tf_{T}\right) \\
&&+2f_{TT}\mu \bar{g}_{ij}x^{\prime i}x^{\prime j}-Mf_{TT}\mu \\
&&+4f_{T}\bar{g}_{ij}x^{\prime i}\left( \eta _{,\tau }^{j}+\eta
_{,k}^{j}x^{\prime k}+\eta _{,T}^{j}T^{\prime }\right. \\
&&\left. -\xi _{,\tau }x^{\prime j}-\xi _{,k}x^{\prime j}x^{\prime k}-\xi
_{,T}x^{\prime j}T^{\prime }\right) ,
\end{eqnarray*}%
\begin{eqnarray*}
&&L\xi ^{\prime }=\left[ 2f_{T}\bar{g}_{ij}x^{\prime i}x^{\prime j}+M\left(
x^{i}\right) \left( f-Tf_{T}\right) \right] \\
&&\ \ \ \ \ \ \ \ \ \cdot \left( \xi _{,\tau }+\xi _{,k}x^{\prime k}+\xi
_{,T}T^{\prime }\right) ,
\end{eqnarray*}%
\begin{equation*}
g^{\prime }=g_{,\tau }+g_{,k}x^{\prime k}+g_{,T}T^{\prime }\;.
\end{equation*}%
Inserting these expressions into (\ref{LL2Paliath}) we find the Noether symmetry
conditions
\begin{equation}
\xi _{,k}=0~,~\xi _{,T}=0~,~g_{,T}=0~,~\eta _{,T}=0,
\end{equation}%
\begin{equation}
4f_{T}\bar{\gamma}_{ij}\eta _{,\tau }^{k}=g_{,k},  \label{LL3}
\end{equation}%
{\small {\
\begin{equation}
M_{,k}\eta ^{k}\left( f-Tf_{T}\right) -MTf_{TT}\mu +\xi _{,\tau }M\left(
f-Tf_{T}\right) -g_{,\tau }=0,  \label{LL4}
\end{equation}%
}}
\begin{equation}
L_{\eta }\bar{\gamma}_{ij}=\left( \xi _{,\tau }-\frac{f_{TT}}{f_{T}}\mu
\right) \bar{\gamma}_{ij},  \label{LL6Paliath}
\end{equation}%
where $L_{\eta }\bar{\gamma}_{ij}$ is the Lie derivative with respect to the
vector field $\eta ^{i}(x^{k})$. Furthermore, from (\ref{LL6Paliath}) we deduce
that $\eta ^{i}$ is a Conformal Killing Vector of the metric $\bar{\gamma}%
_{ij}$, and the corresponding conformal factor is
\begin{equation}
2\bar{\psi}\left( x^{k}\right) =\xi _{,\tau }-\frac{f_{TT}}{f_{T}}\mu =\xi
_{,\tau }-S(\tau ,x^{k})\;.  \label{LL7}
\end{equation}%
Finally, utilizing simultaneously Eqs. (\ref{LL4}), (\ref{LL6Paliath}), (\ref{LL7})
and the condition $g_{,\tau }=0$, we rewrite (\ref{LL4}) as
\begin{equation}
M_{,k}\eta ^{k}+\left[ 2\bar{\psi}+\left( 1-\frac{Tf_{T}}{f-Tf_{T}}\right) S%
\right] M=0\;.  \label{LL9}
\end{equation}%
Considering that $S=S(x^{k})$ and using the condition $g_{,\tau }=0$, we
acquire $\xi _{,\tau }=2\bar{\psi}_{0},\bar{\psi}_{0}\in \mathbb{R}$ with $%
S=2(\bar{\psi}_{0}-\bar{\psi})$. At this point, we have to deal with the
following two situations:

Case 1. In the case of $S=0$, the symmetry conditions are
\begin{eqnarray}
&&L_{\eta }\bar{\gamma}_{ij}=2\bar{\psi}_{0}\bar{\gamma}_{ij},  \notag \\
&&M_{,k}\eta ^{k}+2\bar{\psi}_{0}M=0,
\end{eqnarray}%
implying that the vector $\eta ^{i}(x^{k})$ is a Homothetic Vector of the
metric $~\bar{\gamma}_{ij}$. The latter implies that for arbitrary $f\left(
T\right) \neq T^{n}$ functional forms, our dynamical system could possibly
admit extra (time independent) Noether symmetries.

Case 2. If $S\neq 0$ then Eq.  (\ref{LL9}) immediately leads to the following
differential equation
\begin{equation}
\frac{Tf_{T}}{ f-Tf_{T}}=C,
\end{equation}
which has the solution
\begin{equation}
f(T)=T^{n}, \;\;\;\;C\equiv \frac{n}{1-n} \;.
\end{equation}

Let us now apply the above  in the specific case of static spherically symmetric
geometry (\ref{SSPaliath}). Using the general expressions provided above, we
can deduce the Noether algebra of the metric (\ref{SSA1}). In particular, the Lagrangian
(\ref{L1Paliath})
and the Hamiltonian (\ref{L2Paliath}) now become  \cite{Paliathanasis:2014iva}
\begin{equation}
L=2f_{T}N\left( 2ba^{\prime }b^{\prime }+ab^{\prime 2}\right) +M(a,b)\left(
f-f_{T}T\right),
\end{equation}
\begin{equation}
H=2f_{T}N\left( 2ba^{\prime }b^{\prime }+ab^{\prime 2}\right)-M(a,b)\left(
f-f_{T}T\right)\equiv 0 \;,  \label{HLf.06}
\end{equation}
where $M(a,b)$ is given by (\ref{SSA1}) (one can immediately deduce that TEGR
and thus GR is restored as soon as $f(T)=T$, while if $N=1$,
$\tau=r$ and $ab=1$ we fully recover the standard Schwarzschild solution).

Using this method we can
now determine all the functional forms
of $f(T) $ for which the above dynamical system admits Noether point
symmetries beyond the trivial one $\partial _{\tau}$ related to the energy,
and we summarize the results in Tables \ref{tablepaliath1} and  \ref{tablepaliath2} taken
from  \cite{Paliathanasis:2014iva}.
Thus, we can use the
obtained Noether integrals in order to classify the analytical solutions.

\begin{table}[ht]
\tabcolsep %
6pt
\begin{tabular}{ccc}
\hline
$N(a,b)$ & Symmetry & Integral \\ \hline\hline
$\frac{1}{a^{3}}N_{1}\left( a^{2}b\right) $ & $-\frac{a}{2b^{3}}\partial
_{a}+\frac{1}{b^{2}}\partial _{b}$ & $\frac{N_{1}\left( a^{2}b\right) }{%
2a^{3}b^{2}}\left( 2ba^{\prime }+ab^{\prime }\right) f_{T}$ \\
$N_{2}\left( b\sqrt{a}\right) $ & $-2a\partial _{a}+b\partial _{b}$ & $%
N_{2}\left( b\sqrt{a}\right) \left( b^{2}a^{\prime }-abb^{\prime }\right)
f_{T}$ \\
$aN_{3}\left( b\right) $ & $\frac{1}{ab}\partial _{a}~$ & $N_{3}\left(
b\right) b^{\prime }f_{T}$ \\ \hline
\end{tabular}%
\caption{Noether symmetries and integrals for arbitrary $f(T)$ forms. From
 \cite{Paliathanasis:2014iva}.}
\label{tablepaliath1}
\end{table}

\begin{table*}[ht]
\tabcolsep 6pt
\begin{tabular}{ccc}
\hline
$N(a,b)$ & Symmetry & Integral \\ \hline\hline
arbitrary & $2\bar{\psi}_{0}\tau\partial _{\tau} +\frac{2\bar{\psi}%
_{0}\left( C-1\right) }{2C+1}a\partial _{a}+\frac{2\bar{\psi}_{0}-2\bar{\psi}%
_{4}}{C}T\partial _{T}$ & $2\psi _{0}n\frac{C-1}{1+2C}abN\left( a,b\right)
T^{n-1}b^{\prime }~$ \\
& $-2a\partial _{a}+b\partial _{b}-\frac{2\bar{\psi}_{5}}{C}T\partial _{T}$
& $nN\left( a,b\right) T^{n-1}\left( b^{2}a^{\prime }-abb^{\prime }\right) $
\\
& $-\frac{a}{2}b^{-\frac{3\left( 1+2C\right) }{4C}}\partial _{a}+b^{-\frac{%
3+2C}{4C}}\partial _{b}-\frac{2\bar{\psi}_{6}}{C}T\partial _{T}$ & $\frac{n}{%
2}N\left( a,b\right) T^{n-1}\left( 2b^{\frac{2C-3}{4C}}a^{\prime }+ab^{-%
\frac{3+2C}{4C}}b^{\prime }\right) $ \\
& $a^{-\frac{1}{2C}}b^{-\frac{1+2C}{4C}}\partial _{a}-\frac{2\bar{\psi}_{7}}{%
C}T\partial _{T}~$ & $N\left( a,b\right) na^{-\frac{1}{2C}}b^{-\frac{1+2C}{4C%
}}T^{n-1}b^{\prime }$ \\
&  &  \\
arbitrary & $2\bar{\psi}_{0}\tau\partial _{\tau} +\frac{3\bar{\psi}_{0}}{2}%
a\ln \left( a^{2}b\right) \partial _{a}+\frac{2\bar{\psi}_{0}-2\bar{\psi}%
_{4}^{\prime }}{C}T\partial _{T}$ & $\frac{3}{2}\psi _{0}N\left( a,b\right)
T^{-\frac{1}{2}}ab\ln \left( a^{2}b\right) b^{\prime }$ \\
& $b\partial _{b}-\frac{2\bar{\psi}_{5}}{C}T\partial _{T}$ & $\frac{1}{2}%
N\left( a,b\right) T^{-\frac{1}{2}}\left( b^{2}a^{\prime }+abb^{\prime
}\right) $ \\
& $-a\ln \left( ab\right) \partial _{a}+b\ln b\partial _{b}-\frac{2\bar{\psi}%
_{6}}{C}T\partial _{T}$ & $\frac{1}{2}N\left( a,b\right) T^{-\frac{1}{2}%
}b\left( b\ln b~a^{\prime }-a\ln a~b^{\prime }\right) $ \\
& $a\partial _{a}-\frac{2\bar{\psi}_{7}}{C}T\partial _{T}~$ & $\frac{1}{2}%
N\left( a,b\right) T^{-\frac{1}{2}}ab~b^{\prime }$ \\ \hline
\end{tabular}
\caption{Extra Noether symmetries and integrals for $f(T)=T^{n}$ with $C=%
\frac{n}{1-n}$. The last four lines correspond to the special case where $%
n=1/2$. Notice, that ${\bar \protect\psi}_{5-7}$ are the conformal factors
defined as ${\bar \protect\psi}=\frac{1}{\mathrm{dim}{\bar \protect\gamma}%
_{ij}}\protect\eta^{k}_{;k}$. We notify that the power-law case also admits
the Noether symmetries of Table \ref{tablepaliath1}. From  \cite{Paliathanasis:2014iva}.}
\label{tablepaliath2}
\end{table*}

Using the Noether symmetries and the corresponding integral of motions
obtained in the previous section, we can extract all the static
spherically-symmetric solutions of $f(T)$ gravity. We stress that, in this
way, we obtain new solutions, that could not be obtained by the standard
methods applied in  \cite{Wang:2011xf,Boehmer:2011gw,HamaniDaouda:2011iy,Daouda:2011rt,
Gonzalez:2011dr}.

Without loss of generality, we choose the conformal factor $N(a,b)$ such as $%
N(a,b)=ab^{2}$ (or equivalently  $M(a,b)=1$). In order to simplify the current
dynamical problem  we consider the coordinate transformation
\begin{equation}
b=\left( 3y\right) ^{\frac{1}{3}},\;\;\;\;\;a=\sqrt{\frac{2x}{\left(
3y\right) ^{\frac{1}{3}}}}\;.  \label{L4S001}
\end{equation}%
Substituting the above variables into the field equations (\ref{Lf.04}), (%
\ref{Lf.06}), (\ref{HLf.06}) we immediately obtain
\begin{eqnarray}
&&x^{\prime \prime }+\frac{f_{TT}}{f_{T}}x^{\prime }T^{\prime }=0,
\label{L4S03} \\
&&y^{\prime \prime }+\frac{f_{TT}}{f_{T}}y^{\prime }T^{\prime }=0,
\label{L4S04} \\
&&H=4f_{T}x^{\prime }y^{\prime }-\left( f-Tf_{T}\right) ,  \label{L4S02}
\end{eqnarray}%
while the torsion scalar   is given by
\begin{equation}
T=4x^{\prime }y^{\prime }\;.  \label{L4S01}
\end{equation}%
Finally, the generalized Lagrangian (\ref{L1Paliath}) acquires the simple form
$
L=4f_{T}x^{\prime }y^{\prime }+\left( f-Tf_{T}\right).
$

Since the previous analysis revealed two classes of
Noether symmetries, namely for arbitrary $f(T)$, and $f(T)=T^{n}$, in the
following we investigate them separately. We would like
to mention that the solutions provided below have been extracted under the
assumption $f_{TT}\neq 0$, that is when $f(T)$ is not a linear function of $T
$ and therefore these solutions cannot be extrapolated back to the GR solutions
where $f(T)=T$ (additionally note that these two cases exhibit different
phase space and different Noether symmetries, and thus the obtained
solutions do not always have a $f_{TT}\rightarrow0$ limit  \cite{Paliathanasis:2014iva}).

\begin{itemize}
\item  {Arbitrary $f(T)$}

In the case where $f(T)$ is arbitrary, a ``special solution'' of the system
(\ref{L4S03})-(\ref{L4S01})  can be found, leading to the metric
 \cite{Paliathanasis:2014iva}
\begin{equation}
ds^{2}=A\left( r\right) dt^{2}-\frac{1}{c_{3}^{2}}\frac{1}{A\left( r\right)
}dr^{2}-r^{2}\left( d\theta ^{2}+\sin ^{2}\theta d\phi ^{2}\right) ,
\label{yLLa}
\end{equation}%
with
\begin{equation}
A\left( r\right) =\frac{2c_{1}}{3c_{3}}r^{2}-\frac{2c_{\mu }}{c_{3}r}%
=\lambda _{A}\left(1-\frac{r_{\star }}{r}\right)R(r),  \label{ALLa}
\end{equation}%
and
\begin{equation}
R(r)=\left( \frac{r}{r_{\star }}\right) ^{2}+\frac{r}{r_{\star }}+1.
\label{RLLa}
\end{equation}%
In these expressions, we have defined $c_{\mu }=c_{1}c_{4}-c_{2}c_{3}$, $%
\lambda _{A}=\left( \frac{8c_{1}c_{\mu }^{2}}{3c_{3}^{3}}\right) ^{1/3}$,
and $r_{\star }=(\frac{3c_{\mu }}{c_{1}})^{1/3}=(\frac{3c_{3}\lambda _{A}}{%
2c_{1}})^{1/2}$ is a characteristic radius with the restriction $c_{\mu
}c_{1}>0$. As we can observe, if we desire to obtain a Schwarzschild-de Sitter-like metric
we need to select the constant $c_{3}$ such as $c_{3}\equiv 1$. On the other
hand, the function $R(r)$ can be viewed as a distortion factor which
quantifies the deviation from the pure Schwarzschild solution. Thus, the $%
f(T)$ gravity on small spherical scales ($r \rightarrow r_{\star}^{+}$)
tends to create a Schwarzschild solution.

\item { The case $f(T)=T^{n}$}

In the $f(T)=T^{n}$ case, the field Eqs.  (\ref{Lf.04}), (\ref%
{Lf.06}), (\ref{HLf.06}) and the torsion scalar (\ref{L4S01}) give rise to
the following dynamical system:
\begin{eqnarray}
&&T=4x^{\prime }y^{\prime},  \label{L4S1} \\
&& 4nT^{n-1}x^{\prime }y^{\prime }-\left( 1-n\right) T^{n}=0,  \label{L4S2}
\\
&& x^{\prime \prime }+\left( n-1\right) x^{\prime }T^{-1}T^{\prime }=0,
\label{L4S3} \\
&& y^{\prime \prime }+\left( n-1\right) y^{\prime }T^{-1}T^{\prime }=0\;.
\label{L4S4}
\end{eqnarray}
It is easy to show that combining Eq. (\ref{L4S1}) with the Hamiltonian (\ref%
{L4S2}), we can impose constraints on the value of $n$, namely $n=1/2$.
Under this condition, we can solve the system of equations (\ref{L4S3}) and (\ref{L4S4}),
and then we can find the solution for the metric as  \cite{Paliathanasis:2014iva}
\begin{equation}
\label{MM44Paliath}\ \ \ \ \ \ \
ds^{2}=A(r)dt^{2}-B(r)dr^{2}-r^{2} \left( d\theta +\sin ^{2}\theta
d\phi^{2}\right),
\end{equation}
where
\begin{equation}  \label{AA44}
A(r)=\frac{2}{3}r^{2}+\frac{2c_\sigma}{r} =\lambda_{A}(1-\frac{r_{\star}}{r}%
)R(r),
\end{equation}
with $\lambda_{A}=\left( \frac{8c^{2}_{\sigma}}{3}\right)^{1/3}$ %with
and
$
B(r)=\frac{F_{,r}^{2}}{A(r)r^{4}}$.
The functional form of the distortion parameter $R(r)$ is given by relation (%
\ref{RLLa}), in which the characteristic distance becomes $%
r_{\star}=(-3c_{\sigma})^{1/3}=\left( \frac{3\lambda_{A}}{2}\right)^{1/2}$,
implying $c_{\sigma}<0$.

\end{itemize}

Note that the above novel class of spherically symmetric solutions can be useful
in order to deal with anisotropic deformations of neutron star  instead of searching for
exotic form of matter  \cite{Nelmes:2012uf}, since the solutions are written in terms of
the well known Schwarzschild space-time, modified by a distortion function that depends on
a characteristic radius. Once again we stress that the obtained solution classes are
more general and cannot be obtained by the usual solutions methods.

\subsubsection{Charged spherically symmetric solutions}

In this paragraph we investigate the charged spherically symmetric solutions in $f(T)$
gravity following  \cite{Gonzalez:2011dr,Capozziello:2012zj}. For this shake, alongside
the $f(T)$ gravity Lagrangian (\ref{actionBHSw}), we add the
electromagnetic
Lagrangian
\begin{equation}
\mathcal{L}_{F}=-\frac{1}{2}F\wedge ^{\star }F,
  \label{EMlagrangian}
\end{equation}
where $F=dA$, with $A\equiv A_{\mu
}dx^{\mu }$ the electromagnetic potential 1-form, and where
$\star $ stands for the Hodge dual operator and $\wedge $ for the
usual wedge product. In order to be as general as possible, we will work in arbitrary
space-time dimensions $D$, having in mind that the most interesting case is $D=4$.
Proceeding to extract the charged spherically symmetric solutions in the $D$-dimensional
Maxwell-$f(T)$ theory, we consider the metric form
\begin{equation}
\label{metricchargedvasquez}
ds^{2}=F\left( r\right) ^{2}dt^{2}-\frac{1}{G\left( r\right) ^{2}}
dr^{2}-r^{2}\sum_{i=1}^{i=D-2}dx_{i}^{2}~,
\end{equation}
which arises from the diagonal  vierbein
ansatz
\begin{eqnarray}
&&e^{0}=F\left( r\right) dt~,\text{ \ }e^{1}=\frac{1}{G\left( r\right) }dr~,
\text{ \ }\nonumber\\
&&
e^{2}=rdx_{1}~,\text{ \ \ }e^{3}=rdx_{2}~, \text{ \  }\ldots.
\label{diagonalchargedvasquez}
\end{eqnarray}
For the shake of simplicity  we will study the diagonal vierbein case only, since the
incorporation of non-diagonal ansatzen makes the extraction of analytical
solutions very difficult, which is also the case if one applies the
covariant $f(T)$ gravity presented in subsection
\ref{restoringLI}, with both vierbein and spin connection as dynamical variables. Note
however that the diagonal choice is
adequate to reveal the novel features of the theory and the solution structure. Finally,
concerning the electric sector of the electromagnetic 2-form we assume
\begin{equation}
F=dA=E_{r}\left( r\right) e^{1}\wedge e^{0}+\sum_{i=1}^{i=D-2}E_{i}( r
)e^{0}\wedge e^{i+1}~,
\end{equation}
where $E_{r}$ is the radial electric field, neglecting for the moment the
magnetic part.

The field equations of Maxwell-$f(T)$ theory, for the above ansatzen, write as
 \cite{Capozziello:2012zj}
\begin{equation}
f_T T-\left[ f\left( T\right) -Tf_T
\right] +2\Lambda
+\frac{1}{2}E_{r}^{2}-\frac{1}{2}\sum_{i=1}^{i=D-2}E_{i}^{2}=0~,
\label{fieldequation1vasquez}
\end{equation}
\begin{eqnarray}
&&f_T \left[ -\frac{G\left( r\right) G^{\prime
}\left( r\right) }{r}+\frac{F^{\prime }\left( r\right) G\left( r\right)
^{2}%
}{rF\left( r\right) }\right] \nonumber\\
&&\ \ \
-f_{TT}T^{\prime }\left(
r\right) \frac{G\left( r\right)
^{2}}{r}-\frac{1}{2}\sum_{i=1}^{i=D-2}E_{i}^{2}=0~,
\label{fieldequation2vasquez}
\end{eqnarray}
{\small{
\begin{eqnarray}
&&f_T\left[ -\frac{F^{\prime \prime
}\left(
r\right) G\left( r\right) ^{2}}{F\left( r\right) }-\frac{F^{\prime }\left(
r\right) G^{\prime }\left( r\right) G\left( r\right) }{F\left( r\right) }+
\frac{F^{\prime }\left( r\right) G\left( r\right) ^{2}}{rF\left(
r\right)}\right]\nonumber
\\
&&- f_T \left[\left( D-3\right)
\frac{G^{\prime }\left( r\right) G\left( r\right) }{r}
-\left( D-3\right) \frac{G\left( r\right) ^{2}}{r^{2}}\right]\nonumber\\
&&
- f_{TT}T^{\prime }\left( r\right) \left[ \frac{F^{\prime
}\left( r\right) G\left( r\right) ^{2}}{F\left( r\right) }+\left(
D-3\right)
\frac{G\left( r\right) ^{2}}{r}\right]\nonumber\\
&&
+\frac{1}{2}E_{r}^{2}-\frac{1}{2}%
E_{1}^{2}=0~,\label{fieldequation3vasquez}
\end{eqnarray}}}
\begin{equation}
E_{r}E_{j}=0\text{ \ \ }j=1,\ldots ,D-2~,
\label{electric1vasquez}
\end{equation}
\begin{equation}
E_{i}E_{j}=0\text{ \ \ }i,j=1,\ldots ,D-2~(i\ne j)~,
\label{electric2vsquez}
\end{equation}
where the torsion scalar becomes
\begin{equation}
T\left( r\right) =2\left( D-2\right) \frac{F^{\prime }\left( r\right)
G\left( r\right) ^{2}}{rF\left( r\right) }+\left( D-2\right) \left(
D-3\right) \frac{G\left( r\right) ^{2}}{r^{2}}\text{ }.
\label{scalartorsion1vasquez}
\end{equation}
The remaining field equations are equivalent to equation
(\ref{fieldequation3vasquez}), that is the $a=j$ equation is similar to
(\ref{fieldequation3vasquez}), but with $-\frac{1}{2}E_{1}^{2}$ replaced by
$-\frac{1}{2}E_{j-1}^{2}$ .

A first observation is that from (\ref{fieldequation1vasquez}) we deduce that $T$
has, in general, an $r$-dependence, which disappears for a zero
electric charge. Such a behavior reveals the new features that are brought
in by the richer structure of the addition of the electromagnetic sector.
Moreover, from (\ref{electric1vasquez}) and (\ref{electric2vsquez}), we deduce that we
cannot have simultaneously two non-zero electric field components. This
result is similar to the known no-go theorem of 3D GR-like gravity
 \cite{Cataldo:2002fh,Blagojevic:2008xz}, which states that configurations with
two non-vanishing components of the Maxwell field are dynamically not
allowed. However, it is not valid anymore if we add the magnetic sector, as
we will see later on (it holds only for D=3).
Therefore, in the following we investigate the cases of radial electric
field, of non-radial electric field, and of magnetic and radial
electric field, separately.

 \begin{itemize}
 \item  {Radial electric field}

We first consider the case where there exists only radial electric field.
Thus, the Maxwell equations give
\begin{equation}
E_{r}=\frac{Q}{r^{D-2}}~,
\end{equation}
where $Q$ is an integration constant which, as usual, coincides with the
electric charge of the black hole. Now, integrating equation
(\ref{fieldequation2vasquez}) we find the very simple and helpful result
\begin{equation}
F\left( r\right) =G\left( r\right)f_T~.
\label{expressionvasquez}
\end{equation}
Then, using equations (\ref{expressionvasquez}) and (\ref{scalartorsion1vasquez}) we
obtain
{\small{
\begin{equation}\label{G}
\ \ \ \ G\left( r\right) ^{2}=\frac{1}{f_T^2\,r^{D-3}}
\left[ \frac{1}{\left( D-2\right) }\int f_T^2\,r^{D-2}\,T\left( r\right) dr+Cnst\right],
\end{equation}
\begin{equation}
\label{Fsolvasquez}
F\left( r\right) ^{2}=\frac{1}{r^{D-3}}\left[ \frac{1}{\left( D-2\right) }
\int f_T^2\,r^{D-2}\,T\left( r\right)
dr+Cnst\right],
\end{equation}}}
where $Cnst$ is an integration constant related to the mass of the
spherical object.

In order to proceed, and similar to  \cite{Gonzalez:2011dr}, we will
consider Ultraviolet (UV) corrections to $f(T)$ gravity. In particular, we examine
the modifications on the solutions caused by UV
modifications of  D-dimensional gravity and we consider a representative
ansatz
of the form $f(T)=T+\alpha T^{2}$. This is the first order correction in
every realistic $f(T)$ gravity, in which we expect $\alpha T^{2}\ll T$
 \cite{Wu:2010mn,Iorio:2012cm}, since $T$ (like $R$) is small in
$8\pi G$-units. Thus, for $\alpha \neq 0$, equation
(\ref{fieldequation1vasquez}) leads to
\begin{equation}
\label{TUVvasquez}
T\left( r\right) =\frac{-1\pm \sqrt{1-24\alpha \Lambda -6\alpha
Q^{2}r^{4-2D}
}}{6\alpha}~,
\end{equation}
with the upper and lower signs corresponding to the positive and negative
branch solutions respectively
(note that if $\alpha=0$ then (\ref{fieldequation1vasquez}) becomes linear having
only one solution, which is given by the $\alpha\rightarrow0$ limit of the
positive branch of (\ref{TUVvasquez}), namely $T(r)=-Q^2 r^{4-2D}/2-2\Lambda$, in
which case teleparallel gravity is restored). In this case, the obtained solution reads
 \cite{Capozziello:2012zj}
\begin{widetext}
\begin{eqnarray}\label{G2vasquez}
&&G\left( r\right) ^{2}=\frac{1}{\left( \frac{2}{3}\pm
\frac{1}{3}\sqrt{1-24\alpha \Lambda -6\alpha
Q^{2}r^{4-2D}}\right) ^{2}r^{D-3}}
\Big\{ \frac{1}{\left( D-2\right) }\nonumber\\
&&\Big\{ \frac{1}{54\alpha}
\Big[-\frac{18\alpha Q^{2}r^{3-D}}{3-D}-\frac{\left( 1+72\alpha \Lambda
\right) r^{D-1}}{D-1}\Big]\nonumber\\
 &&  \pm \frac{\sqrt{r^{4D}\left( 1-24\alpha \Lambda
-6\alpha
Q^{2}r^{4-2D}\right) }}{54\alpha}\Big[ \frac{6\alpha
Q^{2}r^{3-3D}}{2D-5}-\frac{\left(
-1+24\alpha \Lambda \right) r^{-1-D}}{D-1}\Big]\nonumber\\
 && \mp \frac{\left( D-2\right) ^{2}\left( -1+24\alpha \Lambda
\right) Q^{2}r^{3+D}
\sqrt{1+\frac{6\alpha Q^{2}r^{4-2D}}{-1+24\alpha \Lambda }}\ _2F_1\left(
\frac{D-3}{2\left( D-2\right) },\frac{1}{2},\frac{3D-7}{2\left( D-2\right)
};\frac{6\alpha Q^{2}r^{4-2D}}{1-24\alpha \Lambda }\right)}{ 3\left(
D-3\right) \left( 2D-5\right) \left( D-1\right) \sqrt{r^{4D}\left(
1-24\alpha \Lambda -6\alpha Q^{2}r^{4-2D}\right)}}\Big\}\nonumber\\
&&+Const\Big\} \ \ \
\ \  \
\end{eqnarray}
and
\begin{eqnarray}
\label{Fsol2vasquez}
&&F\left( r\right) ^{2}=\frac{1}{r^{D-3}}
\Big\{ \frac{1}{\left( D-2\right) }\nonumber\\
&&\Big\{ \frac{1}{54\alpha}
\Big[-\frac{18\alpha Q^{2}r^{3-D}}{3-D}-\frac{\left( 1+72\alpha \Lambda
\right) r^{D-1}}{D-1}\Big]\nonumber\\
 &&  \pm \frac{\sqrt{r^{4D}\left( 1-24\alpha \Lambda
-6\alpha
Q^{2}r^{4-2D}\right) }}{54\alpha}\Big[\frac{6\alpha
Q^{2}r^{3-3D}}{2D-5}-\frac{\left(
-1+24\alpha \Lambda \right) r^{-1-D}}{D-1}\Big]\nonumber\\
 && \mp \frac{\left( D-2\right) ^{2}\left( -1+24\alpha \Lambda
\right) Q^{2}r^{3+D}
\sqrt{1+\frac{6\alpha Q^{2}r^{4-2D}}{-1+24\alpha \Lambda }}\ _2F_1\left(
\frac{D-3}{2\left( D-2\right) },\frac{1}{2},\frac{3D-7}{2\left( D-2\right)
};\frac{6\alpha Q^{2}r^{4-2D}}{1-24\alpha \Lambda }\right)}{ 3\left(
D-3\right) \left( 2D-5\right) \left( D-1\right) \sqrt{r^{4D}\left(
1-24\alpha \Lambda -6\alpha Q^{2}r^{4-2D}\right)}}\Big\}\nonumber\\
 &&+Const\Big\},\ \ \
\ \  \
\end{eqnarray}
\end{widetext}
where
$ \
_2F_1(a,b,c;x)$ is the hypergeometric function. We mention that the last
argument of this function, namely $\left( 6\alpha Q^{2}r^{4-2D}\right)
/\left(1-24\alpha \Lambda \right)$, must be negative, while from
(\ref{TUVvasquez}) it is required that  $1-24\alpha \Lambda
-6\alpha Q^{2}r^{4-2D}$ must be positive, therefore we deduce that
$\alpha$ should be negative.
 Finally,
one can straightforwardly check that in the limit $\alpha\rightarrow0$ (of
the positive branch since in this case the negative branch disappears) one
re-obtains the usual charged GR solutions. Lastly,
for completeness we mention that
the special point  in the parameter space $\Lambda =1/(24\alpha )$, as well as the case
$D=3$, have to be analyzed separately  \cite{Gonzalez:2011dr,Capozziello:2012zj}.

\item {Zero radial field}

Let us for the moment assume that we have zero radial field. In this case
equation (\ref{electric2vsquez}) implies that we can have at most one non-zero
component of the electric field along the non-radial (transversal)
directions. However,
as we mentioned below equation (\ref{scalartorsion1vasquez}), for $D>3$ the
remaining field equations are similar to equation (\ref{fieldequation3vasquez})
but with $-\frac{1}{2}E_{1}^{2}$ replaced by $-\frac{1}{2}
E_{j-1}^{2}$, therefore subtracting these equations we acquire the
conditions $E_{i}^{2}=E_{j}^{2}$, with $i$ and $j$ running from $1$ to
$D-2$. These conditions, along with equation (\ref{electric2vsquez}), yield
$E_{i}=0$ ($i=1,...,D-2$) for $D>3$, that is the electric field
is completely zero. The only cases where zero radial electric field does
not lead to a disappearance of the total electric field is for $D=3$
(where a non-zero azimuthal electric field is possible) which was analyzed
in detail in  \cite{Gonzalez:2011dr}, or if we consider simultaneously
non-zero non-radial electric field with magnetic field, case which lies
beyond the scope of the present investigation.

\item{Magnetic field and radial electric field}

For completeness we also examine the case where magnetic field is present.
While in $D=3$ we deduce that electric field must be absent
 \cite{Gonzalez:2011dr}, for $D>3$ one can simultaneously have non-zero
magnetic and electric fields. As an explicit example we consider an
electromagnetic strength 2-form in four dimensions given by
\begin{equation}
F=E_{r}\left( r\right) e^{1}\wedge e^{0}+B_{23}\left( r\right) e^{2}\wedge
e^{3}~,
\end{equation}
that is we consider a radial electric field $E_{r}$ and a magnetic
field $B_{23}$ both depending on the radial coordinate $r$ only. From the
Maxwell equations in four dimensions for the electric
field we immediately obtain
$E_{r}\left( r\right) =Q/r^{2},$
while incorporating the equations of motion analogous to
(\ref{fieldequation1vasquez})-(\ref{electric2vsquez}) we can see that a
solution is obtained by
$B_{23}\left( r\right) =P/r^{2},$
leading to the metric coefficients (\ref{Fsol2vasquez})  with
$Q^{2}+P^{2}$ in place of $Q^{2}$ (and for $D=4$).

\item {Schwarzschild and Reissner-Nordstr{\"{o}}m solutions}

Let us make some comments on the black-hole solution structure of $f(T)$ gravity. By
definition, every solution of TEGR is also
a solution of $f(T)$ gravity, for the special case $f(T)=T$. In principle,
a non-trivial $f(T)$ ansatz, that is a non-trivial correction to TEGR,
will give rise to non-trivial corrections to the solutions of TEGR, that
is to the solutions of GR. This was analyzed in detail in
the above paragraphs.
However, one could follow a different approach and
ask what should be the specific $f(T)$ ansatz in order to obtain exactly
the same solutions with TEGR. Thus, in the following we extract the  $f(T)$
form in  order for the theory to accept as exact solutions the usual
Reissner-Nordstr{\"{o}}m or Schwarzschild black holes. In this case we
impose the diagonal vierbeins corresponding to
the metric (for completeness we allow also for a general transverse
curvature and for a cosmological constant)
\begin{equation}\label{metricbb}
ds^{2}=G\left( r\right) ^{2}dt^{2}-\frac{1}{G\left( r\right) ^{2}}
dr^{2}-r^{2}d\Omega_k~,
\end{equation}
with
\begin{equation}\label{Gbb}
G(r)^2=k-\frac{2M}{r}+\frac{Q^2}{r^2}-\frac{\Lambda}{3}r^2~,
\end{equation}
where  $k= 1, -1, 0$ represents the curvature of the transverse section,
corresponding to a spherical, hyperbolic or plane section respectively.
Concerning the electric sector of the electromagnetic 2-form we assume
$
F=dA=E_{r}\left( r\right) e^{1}\wedge e^{0}~,
$
where $E_{r}$ is the radial electric field. Inserting the above
ansatzen
in the field equations amongst others we obtain    \cite{Capozziello:2012zj}
the equation
\begin{equation}
f_{TT}T^{\prime }\left(
r\right)=0~,
\label{fieldequation2vasquezbb}
\end{equation}
which is the familiar equation (\ref{field:rtheta}). In this case, as usual, we deduce
that either
$f_{TT}=0$, in which case we re-obtain TEGR, or $T^{\prime
}\left(r\right)=0$, in which case we acquire $k=0$, $Q=0$
and $f(T)=T+\alpha\sqrt{T}$, with $\alpha$ an integration constant. Therefore, we
conclude that in usual formulation of $f(T)$ gravity there is no $f(T)$ ansatz that can
lead to the usual charged black-hole solution (Reissner-Nordstr{\"{o}}m-de Sitter black
hole) of TEGR, in the case of diagonal vierbeins, as was proved in paragraph
\ref{diagandnondiagBHSw}, and this is the reason that one should use non-diagonal
vierbeins. Note however, that there is indeed a specific $f(T)$ ansatz, namely
$f(T)=T+\alpha\sqrt{T}$, that leads to the Schwarzschild-de Sitter-like
solution with flat transverse section, even for the case of diagonal vierbeins. This is
similar to the corresponding
analysis of $f(R)$ gravity, where one finds that although a general $f(R)$
form leads to corrections in the black-hole solutions of General
Relativity   \cite{Pun:2008ae,delaCruzDombriz:2009et}, there is a specific
form, namely $f(R)=\alpha\sqrt{R+\beta}$, which leads exactly to the
Schwarzschild solution  \cite{PerezBergliaffa:2011gj,Mazharimousavi:2012cb} (however note
that in $f(R)$ case the solution is exactly Schwarzschild, and not
Schwarzschild-like with flat transverse section). However, we should have in mind that,
as we have already discussed in subsection \ref{restoringLI}, if we use the covariant
formulation of $f(T)$ gravity \cite{Krssak:2015oua} then there is no frame-dependence
anymore, and obviously an arbitrary vierbein in an arbitrary coordinate system along with
the corresponding spin connection results always to the same physically relevant field
equations.

\item {Singularities and horizons}

We close the discussion on the charged spherically symmetric solutions by examining the
singularities and the horizons of the obtained solutions. The first step is to find at
which $r$ do the functions $G\left(
r\right) ^{2}$ and $F\left( r\right) ^{2}$ become zero or infinity.
However, since these singularities may correspond to  coordinate
singularities, the usual procedure is to investigate various invariants,
since if these invariants diverge at one point they will do that
independently of the specific coordinate basis, and thus the corresponding
point is a physical singularity (note that the opposite is not true, that
is the finiteness of an invariant is not a proof that there is not a physical singularity
there). In standard black-hole literature of curvature-formulated gravity (either GR or
its modifications), one usually studies the Ricci scalar, the Kretschmann scalar, or
other
invariants constructed by the Riemann tensor and its contractions.

In teleparallel description of gravity, one has, in principle, two approaches of finding
invariants. The first is to use the solution for the vierbein and the Weitzenb{\"{o}}ck
connection in order to calculate torsion invariants such as the torsion scalar $T$, or
the teleparallel equivalent of the Gauss Bonnet term $T_G$ (see subsection
\ref{TTGgravity}). The second is to use the solution for the corresponding metric in
order to construct the Levi-Civita connection, and then use it to calculate curvature
invariants such are the Ricci and Kretschmann scalars (a calculation of curvature scalars
using straightaway the Weitzenb{\"{o}}ck connection leads to zero by construction).
Obviously, the physical information must be independent on the formalism and the
intermediate quantities.

We mention however that in order to see the coincidence of the torsion and curvature
analysis, one needs to go beyond the simple invariants, such as $T$ and $R$, and examine
higher-order invariants such as $T_G$ and  Kretschmann scalar. The reason is that  $T$
and $R$ are not capable in revealing all the singularities and horizon structure,
similarly to the standard curvature gravity where $R$ is not adequate and one should
study
the Kretschmann scalar too. In other words, similarly to the case where $R$ is finite at a
point where the Kretschmann scalar is not, $T$ can be finite at a point where $R$ is not,
however $T_G$ will not be finite and hence equivalence between torsion and curvature
formulation is obtained.

Concerning the physical results, one can show that the black-hole solutions of charged
$f(T)$ gravity may possess a horizon at $r_H$ that shields the physical
singularities  \cite{Gonzalez:2011dr,Capozziello:2012zj} . However, firstly it is
not guaranteed that $r_H$ exists, since there could be parameter choices for which $G(r)$
has no roots, that is the physical singularity at $r=0$ becomes naked.
Secondly, even if $r_H$ exists it is not guaranteed that it will shield the second
physical singularity of the charged negative branch at $r_s$, since this will
depend on the specific parameter choice. This is not the case for $f(T)\rightarrow0$ or
$Q\rightarrow0$, in which, as we mentioned, $r_s$ disappears. Therefore, we conclude that
the cosmic censorship theorem, namely that there are always horizons that shield the
physical singularities, does not always hold for charged $f(T)$ gravity, independently of
the space-time dimensionality \cite{Gonzalez:2011dr,Capozziello:2012zj}.

 \end{itemize}

\subsubsection{Cylindrical solutions}

In this paragraph we will investigate cylindrical solutions in $f(T)$ gravity, following
 \cite{Houndjo:2012sz}. We start by imposing the cylindrically symmetric
metric in the Weyl static gauge, namely
 \cite{Stephani:2003tm}
\begin{eqnarray}
&&ds^2=e^{2u(r)}\,
dt^2-e^{2[k(r)-u(r)]}   \,dr^2- w(r)^2e^{-2u(r)}\,
d\varphi^2\nonumber\\
&&\ \ \ \ \ \ \ \
\,-e^{2[k(r)-u(r)]}\,dz^2\,.
\label{cylindricalmetric}
\end{eqnarray}
For simplicity we will focus on the diagonal tetrad
\begin{equation}
e^{A}_{\mu}=diag\Big[ e^u(r),\, e^{k(r)-u(r)},\, we^{-u(r)},\, e^{k(r)-u(r)} \big]\,.
\end{equation}
Hence, the torsion scalar (\ref{telelag}) becomes
\begin{eqnarray}
T=\frac{e^{2[u(r)-k(r)]}}{w(r)}\Big[2w(r)u'(r)^2-2k'(r)w'(r)\Big]\,.
\label{cylsolTscalar}
\end{eqnarray}
Finally, as usual, we assume a perfect fluid with energy-momentum tensor of the form
\begin{eqnarray}
 T_{\mu\nu}=diag(\rho,-p_r,-p_\varphi,-p_z)\,.
\end{eqnarray}
 Under the above considerations, the field equations read   \cite{Houndjo:2012sz}
 \begin{widetext}
 \begin{eqnarray}\label{cylsoleq1}
4\pi\rho=\frac{f}{4}+\frac{e^{2(u-k)}}{2w}\Big(2wu''+2w'u'-w''-k'w'-wk''\Big)f_T-\frac{e^{
4(u-k)}}{
w^3}f_{TT}\times\\ \nonumber
\Big[
4w^3u'^2u''-2w^2w'u'u''-2w^3k'u'u''+4w^3u'^4-2w'w^2u'^3-6w^3k'u'^3+2w^3k'^2u'^2-
2k'w'w^2u'^2\\
\nonumber -2k'w''w^2u'+4wk'u'w'^2+6w'u'k'^2w^2-2w'w^2k''u'+wk'w'w''+w^2k'^2w''\\ \nonumber
-k'w'^3+ww'^2k''-3wk'^2w'^2+w^2w'k'k''-2w^2w'k'^3
\Big]\,\,,
\end{eqnarray}
\begin{eqnarray}
-4\pi p_r=\frac{f}{4}+\frac{e^{2(u-k)}}{w}(wu'^2-k'w')f_{T}\,\,,\label{cylsoleq2}\\
-4\pi
p_{\varphi}=\frac{f}{4}+\frac{e^{2(u-k)}}{2w}(k'w'+wk'')f_{T}+\frac{e^{4(u-k)}}{w^2}f_{TT}
\times\\ \nonumber
\Big[2k'w^2u'u''+2k'w^2u'^3-2k'^2w^2u'^2-2k'^2ww'u'-k'^2ww''+w'^2k'^2+2k'^3ww'-wk'
w'k''\Big]\,\,,\label{cylsoleq3}
\end{eqnarray}
\begin{eqnarray}\label{cylsoleq4}
-4\pi p_z=\frac{f}{4}-\frac{w''}{2w}e^{2(u-k)}f_{T}+\frac{e^{4(u-k)}}{w^3}f_{TT}\times\\
\nonumber
\Big[
2w^2w'u'u''+2w^2w'u'^3-2k'w^2w'u'^2-2wk'w'^2u'-wk'w'w''+k'w'^3+2k'^2ww'^2-ww'^2k''\Big
]\,\,.
\end{eqnarray}
\end{widetext}
As we observe, and as expected, the system (\ref{cylsoleq1})-(\ref{cylsoleq4}) in the
case of the TEGR, i.e. for $f(T)=T$, reduces to the standard field equations
\cite{Stephani:2003tm}.

In the following, for simplicity we will examine  vacuum solutions, namely we will impose
$\rho=p_r=p_{\varphi}=p_z=0$. In particular, we will consider specific forms of $f(T)$
and for each case we will do the analysis separately.

\begin{itemize}

\item  {Solutions for $f(T)=T+\beta T^2$}

Let us start by considering an $f(T)$ corresponding to a quadratic correction in the
TEGR, namely  $f(T)=T+\beta T^2$, which as discussed above is always a good
approximation in cases where $T\ll1$. In this case, one can find the solution of
(\ref{cylsoleq1})-(\ref{cylsoleq4}) as  \cite{Houndjo:2012sz}
   \begin{eqnarray}
&&u(r) =c_1, \nonumber\\
&&k(r)=c_2+\frac{1}{2}\log\left[\frac{5\beta}{-c_2-\frac{3r^2}{2}e^{
-2c_1 }
}\right]+ \frac{c_1^2}{2}r^2,\nonumber\\
&&w(r)=c_0 r \,\,,
\label{cylsolsimple}
\end{eqnarray}
while the torsion scalar (\ref{cylsolTscalar}) becomes
   \begin{eqnarray}
T=-\,{\frac {3{{\rm e}^{-2\,c_{{2}}-{c_{{1}}}^{2}{r}^{2}}}}{5\beta}}\,\,.
\end{eqnarray}
Note that in the limit $r\rightarrow\infty$ we obtain $T=0$. Hence, this solution is
asymptotically torsionless, similarly to the asymptotically flat solutions
of GR. For a more transparent presentation of the above solution, we rewrite the
cylindrical metric (\ref{cylindricalmetric}) as
$g_{\mu\nu}dx^\mu dx^\nu=d\tilde{t}^2-\gamma_{ij}dx^idx^j$, with  the following
coordinate redefinitions:
\begin{eqnarray}
\tilde{z}=z\,\,,\\
\tilde{t}=e^{c1}t\,\,,\\
\tilde{\varphi}=c_0e^{-c_1}\varphi\,\,,\\
\tilde{r}=e^{c_2-c_1}r\,\,,
\end{eqnarray}
where
 \begin{equation}
\gamma_{ij}dx^idx^j=-\frac{5\beta
e^{\left(\frac{c_1}{c_2-c_1}\right)\tilde{r}^2}}{c_2+\frac{3}{2}e^{-2c_2}\tilde{r}^2}
\left(d\tilde { r } ^2+dz^2\right)+\tilde{r}^2d\tilde{\varphi}^2\label{gamma2}.
\end{equation}

\item{Solution for $f(T)=T+\beta T^2+\Lambda$}

Let us try to extend the previous results, by additionally considering an explicit
cosmological constant in the action, namely choosing $f(T)=T+\beta T^2+\Lambda$. The
source of such solution may be a rod, therefore it proves convenient to introduce the
logarithmic Newtonian potential as $u(r)=\frac{2}{3}\log [\cos(\lambda r)]$. Moreover, we
impose the gauge  \cite{Tian:1986zz}
$w(r)=\lambda ^{
-1} \sin(\lambda r)\cos^{1/3}(\lambda r)$, with $\lambda=\frac{1}{2}\sqrt{3\Lambda}$.
In this case, one can extract the remaining component of the metric as
 \begin{eqnarray}
k(r)\approx\frac{1}{2}\log\left[\frac{-2\beta}{c_1+\int \frac{r}{\cos^{4/3}(\lambda
r)}dr}\right],
\end{eqnarray}
which is a valid approximation in the limit
$\beta\leq \frac{9}{20\Lambda}\ll1$. As expected, in the limit $\Lambda\rightarrow0$ the
above expressions give (\ref{cylsolsimple}). Furthermore, the torsion scalar
(\ref{cylsolTscalar}) acquires a complicated $r$-dependence of the form
\begin{eqnarray}
T&=&-\frac{2\lambda}{3}\frac{\Sigma}{\left[ \sin \left( \lambda\,r \right)
\right]},
\end{eqnarray}
with
\begin{eqnarray}
&&\Sigma =   r \left[ \cos \left(
\lambda\,r \right)  \right]^{2}-\frac{1}{4}\,r+\frac{2{\it
c_1}}{3}\,\lambda\,\sqrt
[3]{\cos
 \left( \lambda\,r \right) }\,\sin \left( \lambda\,r
 \right)
 \nonumber\\
&&\ \ \ \ \ \,
-\frac{2{\it c_1}}{3}\,\lambda\, \left[ \cos \left( \lambda\,r \right)
 \right] ^{7/3}\,\sin \left( \lambda\,r \right)  \,\,.
\nonumber\\
&&\ \ \ \ \ \,
+\frac{2}{3}\,\lambda\,\sin \left( \lambda\,r \right)  \left\{
\sqrt [3]{\cos \left( \lambda\,r \right) }- \left[ \cos \left( \lambda
\,r \right)  \right] ^{7/3} \right\} \nonumber\\
&&\ \ \ \ \ \, \ \
\cdot \int
\!{\frac {r}{ \left[ \cos
 \left( \lambda\,r \right)  \right] ^{4/3}}}{dr}.
\end{eqnarray}
Hence, the above metric can be seen as the torsion-based correction to the standard
Linet and Tian  solution of GR  \cite{Linet:1986ba,Tian:1986zz}.
Finally, we mention here that the above solutions present a dual family which
can be obtained  imposing the transformations
$t\rightarrow iz,\ \ z\rightarrow it,\ \ 2u\rightarrow u$, where the first two
transformations are just the Wick rotation of the coordinate $t$ or $z$, while the
last one can be seen as a scaling invariance transformation of the function $u$.
Therefore, applying these transformations one can obtain new solution sub-classes.
Lastly, applying the Ehler's transformation one can acquire the dual stationary
solution sub-class, similarly to \cite{Momeni:2005uc}.

\item{General solution with finite values of $u(r)$ on the axis $r=0$}

In Weyl coordinates one introduces the following ansatz for metric function
$u(r)$:
\begin{eqnarray}
u(r)=\frac{1}{\pi}\int_{0}^{\pi}h(i r \cos \Theta)d\Theta\label{uj},
\end{eqnarray}
adopting also the gauge $w(r)=r$, compatible with the standard GR vacuum solutions.
Concerning the arbitrary function $h$, one could start by examining the case
$h(\phi)=e^{\phi}$, while for the $f(T)$ form we consider as before the quadratic
correction to TEGR, namely   $f(T)
=T+\beta T^2$. Inserting the above into the field equations
(\ref{cylsoleq1})-(\ref{cylsoleq4}) in the
vacuum case, we acquire the following solution:
\begin{eqnarray}
k(r)=c_2-\frac{1}{2}\log \left[c_1+\int re^{-2J_0(r)}dr\right]\,,
\end{eqnarray}
which is regular on the axis $r=0$. Additionally, note that this solution leads to a
constant torsion scalar, since in this case  (\ref{cylsolTscalar}) gives  $T=e^{-2c_1}$.
Hence, in this case we obtain an effective cosmological constant
$\Lambda_{eff}=\frac{e^{-2c_1}}{2}$.

\end{itemize}

\subsection{Wormhole solutions}

In this subsection we will investigate the wormhole solutions in $f(T)$ gravity,
following  \cite{Bohmer:2011si} (see also \cite{Sharif:2014bsa}). Wormholes are
hypothetical tunnels in space-time, through which observers may freely
traverse~ \cite{Morris:1988cz,Hawking:1988ae,Lobo:2007zb}. In usual GR, wormhole
space-times are supported by ``exotic'' forms of matter, which involves an
energy-momentum tensor violating the null energy condition (NEC), namely
$T_{\mu\nu}k^{\mu}k^{\nu}<0$, where $k^{\mu}$ is any null vector. Although in principle a
NEC violating matter is problematic, we stress
that in the context of modified theories of gravity one can have an effective
energy-momentum tensor violating NEC while normal matter still satisfies all energy
conditions  \cite{Lobo:2008zu}.

We start by considering the static spherically symmetric metric
\begin{equation}
  ds^2 = e^{a(r)} dt^2 - e^{b(r)} dr^2 -r^2 \left(d\theta^2 + \sin^2\theta\,
  d\varphi^2\right)\,,
  \label{metric1offworm}
\end{equation}
where $a(r)$ and $b(r)$ are functions of the coordinate $r$. A static and spherically
symmetric
wormhole is produced when
\begin{equation}
  e ^{-b(r)} = 1-\frac{\beta(r)}{r} \,,
  \label{metricwormhole}
\end{equation}
where $a(r)$ denotes the ``redshift function'', since it is related to the gravitational
redshift, while $\beta(r)$ denotes the ``shape function'', since it determines the shape
of the wormhole  \cite{Morris:1988cz,Hawking:1988ae,Lobo:2007zb}. The coordinate $r$ is
non-monotonic and it decreases from $+\infty$ to a minimum value $r_0$, denoting the
location of the wormhole throat where $b(r_0)=r_0$, and then it increases from $r_0$ to
$+\infty$. Hence, in oder to have a wormhole solution one must impose the flaring out of
the throat, which is provided by the condition $(\beta-\beta'
r)/2\beta^2>0$, while at the
throat we
have that $\beta '(r_0)<1$   \cite{Morris:1988cz,Hawking:1988ae,Lobo:2007zb}.

As usual in the teleparallel and $f(T)$ formulation of gravity, one should specify the
vierbein choice that correspond to the above metric. As discussed in subsection
\ref{restoringLI}, if we re-formulate $f(T)$ gravity keeping a general spin connection,
that is if we use covariant $f(T)$ gravity \cite{Krssak:2015oua}, then all vierbein
choices corresponding to the same metric are ``equally good'' and they lead to the same
physical solution.  However, in the usual simplified $f(T)$ formulation, where zero spin
connection is imposed, different choices of frames lead to different classes of
solutions,
and vice versa a given metric solution is in general realized by a specific choice of
vierbeins. Thus, similarly to the case of the Schwarzschild solution discussed in
\ref{diagandnondiagBHSw}, it proves that in order to obtain a static and spherically
symmetric wormhole solution in the standard simple formulation of $f(T)$ gravity one must
use a non-diaginal vierbein choice  \cite{Boehmer:2011gw}, since the
diagonal vierbein does not allow for the extraction of a solution with the necessary
form (\ref{metricwormhole}), as shown in  \cite{Bohmer:2011si}. Hence, in the
following we consider  that the metric (\ref{metric1offworm}) is realized by the vierbein
\begin{eqnarray}
  e^A{}_{\mu} = \left(
\begin{array}{cccc}
  e^{a/2} & 0 & 0 & 0 \\
  0 & e^{b/2} \sin\theta\cos\phi  & r\cos \theta\cos\phi & -r\sin \theta\sin \phi \\
  0 & e^{b/2} \sin\theta\sin\phi & r\cos \theta\sin\phi& r\sin\theta \cos\phi \\
  0 & e^{b/2} \cos \theta  & -r \sin\theta & 0
\end{array}
\right)
  \label{tetradoffworm},\nonumber
\end{eqnarray}
 where for simplicity from now on we omit the $r$-dependence in the metric functions.

Under the above considerations, the torsion scalar (\ref{telelag}) becomes
\begin{eqnarray}
  T(r) &=& \frac{2 e^{-b} \left(e^{b/2}-1\right)\left(e^{b/2}-1-r a '\right)}{r^2}\,,
  \label{defToffworm}
\end{eqnarray}
while the general field equations (\ref{BHsectioneoms}) read  \cite{Bohmer:2011si}
{\small{
\begin{eqnarray}
 \!\!\!\!\!\!\!\!\!\!\! 4\pi\rho_m(r) &=&
\frac{e^{-b/2}}{r}(1-e^{-b/2}) T'f_{TT}-\left(\frac{T}{4}-\frac{1}
  {2r^2}\right)f_T
  \nonumber \\
 \!\!\!\!\!\!\!\!\!\!\! &&+\frac{e^{-b}}{2 r^2} \left(rb '-1\right)f_T-\frac{f}{4}\,,
  \label{field:toffworm}\\
 \!\!\!\!\!\!\!\! 4\pi p_{mr}(r) &=&\left[- \frac{1}{2r^2}+\frac{T}{4}+\frac{e^{-b}}{2r^2}
  (1+ra')\right]f_T-\frac{f}{4}\,,
  \label{field:roffworm}\\
\!\!\!\!\!\!\!\!\!\!\!  4\pi p_{mt}(r) &=&
\frac{e^{-b}}{2}\left(\frac{a'}{2}+\frac{1}{r}-\frac{e^{b/2}}{r}\right)T'
  f_{TT}
  \nonumber \\
 \!\!\!\!\!\!\!\!&&\hspace{-2.0cm}\ \, + f_T\left\{ \frac{T}{4}+\frac{e^{-b}}{2 r}
\left[\left(\frac{1}
 {2}+\frac{ra'}{4}\right) \left(a'-b'\right)+\frac{ra''}{2}\right]\right\}-
 \frac{f}{4}\;,
  \label{field:thetaoffworm}
\end{eqnarray}}}
where $\rho_m(r)$ is the matter energy density, $p_{mr}(r)$ is the matter radial
pressure, and $p_{mt}(r)$ stands for the matter
pressure of the tangential directions, orthogonal to the radial direction. Note
that with the above non-diagonal vierbein choice, and contrary to the diagonal vierbein
choice which led to equation (\ref{field:rtheta}), there is no equation enforcing the
constancy of the torsion scalar \cite{Bohmer:2011si,Boehmer:2011gw}. The above equations
of motion provide three independent equations for six unknown functions, namely
$\rho_m(r)$, $p_{mr}(r)$, $p_{mt}(r)$, $a(r)$, $b(r)$ and $f(T)$, i.e they form an
under-determined system. In order to solve it we have to make some assumptions for some
of
them, and this is performed separately in the following paragraphs. Lastly, note the
self-consistency test that for $f(T)=T$, that is for the TEGR case, all the above
equations coincide with those of standard GR \cite{Oliveira:2011vu} as expected.

 \begin{itemize}
 \item  {Solutions with $T(r)= 0$}

Let us start with the simple solution sub-class corresponding to $T(r)= 0$. In this
case the energy-momentum tensor components become  \cite{Bohmer:2011si}
{\small{
\begin{eqnarray}
  4\pi \rho_m(r)&=&\frac{\beta'}{2r^2}f_T(0)+\frac{f(0)}{4} \,,
  \label{T0rho1worm}\\
  4\pi p_{mr}(r)&=&-\frac{1}{2r^2}\left[1-\left(1-\frac{\beta}{r}\right)\left(1+ra'
    \right)\right]f_T(0)\nonumber\\
  &&-\frac{f(0)}{4} \,,
  \label{T0pr1worm}\\
  4\pi p_{mt}(r)&=& \frac{1}{4r^2} \left( 1-\frac{\beta}{r} \right)\Bigg[r^2a''+
    \left(1+\frac{a'r}{2}\right)\times
    \nonumber \\
    && \left(ra'-\frac{\beta' r-\beta}{r(1-\beta/r)}\right)  \Bigg]f_T(0)-\frac{f(0)}{4}.
\label{T0pt1worm}
\end{eqnarray}}}
The weak energy condition (WEC), namely $\rho_m(r)\geq 0$ and
$\rho_m(r)+p_{mr}(r)\geq 0$,
imposes the positivity of the r.h.s of (\ref{T0rho1worm}), as well as of the expression
\begin{eqnarray}
  &&\!\!\!\!
  4\pi [\rho_m(r)+p_{mr}(r)] =
\frac{1}{2r}\left[\left(1-\frac{\beta}{r}\right)a'\right.\nonumber\\
&&\left.\ \ \ \ \ \ \ \ \ \ \ \ \ \ \ \ \ \ \ \ \ \ \ \ \ \   \ \ \ \ \ \
  -\frac{\beta-\beta'r}{r^2} \right] f_T(0)  .
  \label{WEC2worm}
\end{eqnarray}
Similarly, the null energy condition (NEC), in addition to imposing the positivity of
the l.h.s. of (\ref{WEC2worm}), imposes the positivity along
the tangential direction of the following expression:
{\small{
\begin{eqnarray}
&&4\pi [\rho_m(r)+p_{mt}(r)] = \frac{1}{2r^2}f_T -
\frac{(\beta-\beta'r)}{4r^3}\left(1-\frac{ra'}
{2}\right)f_T  \nonumber  \\
&&  -\frac{1}{2r^2} \left(1-\frac{\beta}{r}\right)\left[1-\frac{1}{2}\left(1+\frac{ra'}{2}
\right)r a' -\frac{r^2a''}{2} \right] f_T.
\label{NEC2worm}
\end{eqnarray}
}}
Thus, using the above expressions it is straightforward to deduce specific constraints at
the throat. In particular, evaluating (\ref{T0rho1worm}) and (\ref{WEC2worm}) at the
throat respectively gives
\begin{eqnarray}
  4\pi \rho_m|_{r_0}&=& \frac{\beta'_0}{2r_0^2}f_T(0) +\frac{f(0)}{4}\,,
 \label{pos_rhoworm} \\
 4\pi \left(\rho_m+p_{mr} \right)|_{r_0}&=&-\frac{\beta(r)-r\beta'(r)}{2r^3}\Big|_{r_0}
f_T(0) \,.
\end{eqnarray}
Hence, from the flaring out condition at the throat, namely $(\beta-r
\beta')/(2\beta^2)|_{r_
0}>0$, in order to acquire $4\pi \left(\rho_m+p_{mr} \right)|_{r_0}>0$ the condition
$f_T(0)<0$ has to be imposed. Nevertheless, in order to obtain an asymptotically flat
space-time, with vanishing energy-momentum components
at infinity, one can verify from the field
equations (\ref{T0rho1worm})-(\ref{T0pt1worm}) that $f(0)
=0$. This constraint implies the
positivity of the r.h.s. of  (\ref{pos_rhoworm}), and therefore taking into
account $f_T(0)<0$ leads the form
function to obey $\beta'_0<0$. Furthermore, one can extract more constraints evaluating
(\ref{NEC2worm}) at the throat,
namely
\begin{eqnarray}
 && \!\!\!\!\!\!
 4\pi \left(\rho_m+p_{mt} \right)|_{r_0}=\frac{1}{2r_0^2}f_T(0)\nonumber\\
  && \ \ \ \ \ \ \ \ \ \ \  \ \ \ \ \   \
  -
  \frac{(1-\beta'_0)}{4r_0^2}\left(1- \frac{r_0 a'_0}{2} \right) f_T(0)\,,
  \label{NECtangworm}
\end{eqnarray}
and thus for a zero redshift function at the throat, namely $a_0'=0$, and considering
$4\pi
\left(\rho_m+p_{mt} \right)|_{r_0} \geq 0$, we acquire the constraint $\beta_0'\leq -1$.
For the general case $a_0'\neq 0$, we obtain the restriction $r_0 a_0' \leq
[1-2/(1-\beta_0')]$.

Let us now try to solve (\ref{defToffworm}) under  $T(r)= 0$. In this case, and
imposing the condition $e^{b/2}-1\neq 0$ (see  \cite{Boehmer:2011gw} for the detailed
discussion of the  $e^{b/
2}-1=0$ case), we obtain the
differential equation
\begin{equation}
  \sqrt{1-\frac{\beta(r)}{r}}-1-ra'=0 \,,
  \label{odeToffworm}
\end{equation}
which can be solved considering specific choices for the shape function $\beta(r)$.
For example, if one considers the Ellis
wormhole  \cite{Ellis:1973yv}
\begin{equation}
  \beta(r)=\frac{r_0^2}{r} \,,
\end{equation}
in which case $\beta '_0=-1$,  then
 (\ref{odeToffworm}) leads to the redshift function
\begin{equation}
  e^{a(r)}=\frac{1}{2}\left(1+\sqrt{1-\frac{r_0^2}{r^2}}  \right)\,,
\end{equation}
and therefore $e^{a(r)}\rightarrow 1$ as $r \rightarrow \infty$. Hence, the
energy-momentum components for this specific case write as
{\small{
\begin{eqnarray}
&&\!\!\!\!  4\pi \rho_m(r)=\frac {r_0^2}{2r^4}|f_T(0)| \,,\label{T0rho2worm}\\
&&\!\!\!\! 4\pi p_{mr}(r)=\frac{1}{2r^2} \left[\left(\sqrt{1-\frac{r_0^2}{r^2}}-1\right) +
\frac{r_0^2}{r^2}\right]|f_T(0)|
  \,, \label{T0pr2worm} \\
 &&\!\!\!\! 4\pi p_{mt}(r)= \frac{1}{4r^2}
\left[\left(\sqrt{1-\frac{r_0^2}{r^2}}-1\right)
-
\frac{r_0^2}{2r^2}
\right]|f_T(0)|.\
  \label{T0prnewworm}
\end{eqnarray}
}}
Note that the energy density is positive throughout the space-time due to the condition
$f_T(0) < 0$. Additionally, it is easy to show that the NEC is always satisfied
too.

\item  {Solutions for $f(T)=T+T_0T^2$ with $a(r)=0$, $b(r)=r_0^2/r$ }

Let us consider the solution subclass obtained for the case of a quadratic correction to
TEGR, i.e for $f(T)=T+T_0T^2$, imposing additionally that the redshift and shape
functions respectively write as
\begin{equation}
  a(r)=0, \qquad b(r)=\frac{r_0^2}{r}\,.
\end{equation}
  Substituting the above into the energy-momentum tensor components
 (\ref{field:toffworm})-(\ref{field:thetaoffworm}), leads to
 \cite{Bohmer:2011si}:
{\small{
\begin{eqnarray}
  4\pi \rho_m(r)&=&-\frac{r_0^2}{2r^4}\Bigg[1+16\frac{T_0}{r_0^2}\sqrt{1-\frac{r_0^2}
      {r^2}}\left(3-5\frac{r_0^2}{r^2} \right)
    \nonumber  \\
    &&-2\frac{T_0}{r_0^2}\left(24-52\frac{r_0^2}{r^2}+17\frac{r_0^4}{r^4}\right)
\Bigg]\,,\\
  4\pi p_{mr}(r)&=&-\frac{r_0^2}{2r^4}\Bigg[1+16\frac{T_0}{r_0^2}\sqrt{1-\frac{r_0^2}
      {r^2}}\left(1-\frac{r_0^2}{r^2} \right)
    \nonumber  \\
    &&-2\frac{T_0}{r_0^2}\left(8-12\frac{r_0^2}{r^2}+3\frac{r_0^4}{r^4}\right) \Bigg]\,,\\
  4\pi p_{mt}(r)&=&\frac{r_0^2}{2r^4}\Bigg[1+8\frac{T_0}{r_0^2}\sqrt{1-\frac{r_0^2}
      {r^2}}\left(2-5\frac{r_0^2}{r^2} \right)
    \nonumber  \\
    &&-2\frac{T_0}{r_0^2}\left(8+24\frac{r_0^2}{r^2}-9\frac{r_0^4}{r^4}\right) \Bigg]\,.
\end{eqnarray}}}
One can show that the energy density is positive at the throat, although it changes sign
for a specific value of the radial coordinate. Finally, we mention the interesting
feature that the NEC and WEC are satisfied both at the throat and at its neighborhood,
which is not the case for their GR counterparts.

 \end{itemize}

%\pagebreak

%%%%%%%%%%%%%%%%%%%%%
\section{Extensions of  $f(T)$ gravity}
\label{SecionextensionsfT}
%%%%%%%%%%%%%%%%%%%%%%%%%%%%%%%%%%%%%

In this section we present some extensions of $f(T)$ gravity, inspired as usual by the
corresponding curvature extensions of $f(R)$ gravity. The important and interesting
feature, is that these extensions are different from their corresponding curvature
constructions, and hence they deserve investigation, both at the theoretical as well as
at the cosmological level.

\subsection{Non-minimally coupled scalar-torsion gravity}

In GR, apart from the $f(R)$ extension, one can be based on the
quintessence scenario, and introduce a non-minimal coupling between the scalar
field and gravity
 \cite{Sahni:1998at,Uzan:1999ch,Bartolo:1999sq,Bertolami:1999dp,Faraoni:2000wk}, or more
generally extend it to the scalar-tensor paradigm  \cite{Maedabook}. One can also use a
phantom instead of a canonical field  \cite{Elizalde:2004mq,Setare:2008mb,Gupta:2009kk},
or the combination of both these fields in the unified quintom scenario
 \cite{Setare:2008pc}. In particular, in this GR extension one starts from the action
\begin{equation}
S=\int\ud^{4}x\sqrt{-g}\Bigg[\frac{R}{16\pi G}
+ \frac{1}{2} \Big(\partial_{\mu}\phi\partial^{\mu}\phi+\xi
R\phi^{2}\Big) - V(\phi)+\mathcal{L}_m\Bigg] \label{actionnonminimalquintess},
\end{equation}
with $G$ the gravitational constant, $c=1$, $\phi$ the scalar field with $V(\phi)$ its
potential,
and $\xi$ the non-minimal coupling parameter \footnote{Note
the difference in the metric signature that exists amongst the various
works in the literature and the corresponding sign changes in the action,
since a change in the metric signature leads $g_{\mu\nu}$,
$\Box$ and $R_{\mu\nu}$ to change sign, while $R$ and the energy-momentum
tensor remain
unaffected
 \cite{Sahni:1998at,Uzan:1999ch,Bartolo:1999sq,Bertolami:1999dp,Faraoni:2000wk}.
Thus, under the convention of (\ref{actionnonminimalquintess}), the conformal value of
$\xi$
is $-1/6$.}. Additionally, concerning the cosmological application one includes the
matter
Lagrangian $\mathcal{L}_m$ too. We mention that one can extend the above non-minimal
coupling $\xi R\phi^{2}$  to the general form $\xi B(\phi) R$, that is
allowing for an arbitrary scalar-field coupling function.

In the case of a flat Friedmann-Robertson-Walker (FRW) background metric
the Friedmann equations write as
\begin{eqnarray}
\label{FR1nonminimalquint}
&&
H^{2}=\frac{8\pi G}{3}\Big(\rho_{\phi}+\rho_{m}\Big),\
\\
\label{FR2nonminimalquint}
&&
\dot{H}=-4\pi G\Big(\rho_{\phi}+p_{\phi}+\rho_{m}+p_{m}
\Big),~~~~
\end{eqnarray}
where the energy density and pressure of the non-minimally coupled scalar field write as
\begin{eqnarray}
\label{rhophinonminimalquint}
 &&\!\!\!\!\!\!\!\!\!\!
 \rho_{\phi}=  \frac{1}{2}\dot{\phi}^{2} + V(\phi)
-  6 \xi H\phi\dot{\phi} -  3\xi H^{2}\phi^{2},\\
 &&\!\!\!\!\!\!\!\!\!\!
 p_{\phi}=  \frac{1}{2}(1+4\xi)\dot{\phi}^{2} - V(\phi)  + 2 \xi
(1+6\xi)\dot{H}\phi^{2}  \nonumber\\
 \label{pphi}
  &&\!\!\!\!\!\!\!  \ \ \ \ \  -   2 \xi
H \phi\dot{\phi}
+  3\xi(1+8\xi)H^{2}\phi^{2} - 2\xi\phi V'(\phi),
\label{pphinonminimalquint}
\end{eqnarray}
where a prime denotes derivative with respect to $\phi$. We mention that we have
simplified the above relations by using the expression $R=6(\dot{H}+2H^2)$ that holds
in FRW geometry. In this scenario, the dark energy sector is attributed to
the non-minimal scalar field, and thus its equation-of-state  parameter reads $
w_{DE}\equiv w_\phi=p_\phi/\rho_\phi$.
 Finally, the equations close by considering the
evolution equation for the scalar field, namely
 \cite{Sahni:1998at,Uzan:1999ch,Bartolo:1999sq,Bertolami:1999dp,Faraoni:2000wk}
\begin{equation}
\ddot{\phi}+3H\dot{\phi}-6\xi(\dot{H}+2H^2)\phi+   V'(\phi)=0,
\end{equation}
which can alternatively be written in the standard form
$\dot{\rho}_\phi+3H(1+w_\phi)\rho_\phi=0$.

Let us now follow the same recipe in torsional gravity following  \cite{Geng:2011aj}. In
particular, we start from the Teleparallel Equivalent of GR, and we add a non-minimally
coupled
scalar field. Since in TEGR the scalar that incorporates the
gravitational field, at the lowest order, is the torsion scalar $T$, this will be the
gravitational scalar that will be coupled to $\phi$.
Thus, the action will  read:
\begin{equation}
S=\int\ud^{4}x e\Bigg[\frac{T}{16\pi G}
+ \frac{1}{2} \Big(\partial_{\mu}\phi\partial^{\mu}\phi+\xi
T\phi^{2}\Big) - V(\phi)+\mathcal{L}_m\Bigg]. \label{action2teleDE}
\end{equation}
Variation  with respect to the vierbein fields yields the equations of motion,
namely  \cite{Geng:2011aj}
%\begin{widetext}
{\small{
\begin{eqnarray}\label{eom2}
\left(\frac{1}{4\pi G}+2 \xi
\phi^2 \right)\left[e^{-1}\partial_{\mu}(ee_A^{\rho}S_{\rho}{}^{\mu\nu} )
-e_{A}^{\lambda}T^{\rho}{}_{\mu\lambda}S_{\rho}{}^{\nu\mu}
-\frac{1}{4}e_{A}^{\nu
}T\right]\nonumber\\
-
e_{A}^{\nu}\left[\frac{1}{2}
\partial_\mu\phi\partial^\mu\phi-V(\phi)\right]+
  e_A^\mu \partial^\nu\phi\partial_\mu\phi\ \ \nonumber\\
+ 4\xi e_A^{\rho}S_{\rho}{}^{\mu\nu}\phi
\left(\partial_\mu\phi\right)
=e_{A}^{\rho} T_{\ \ \ \ \rho}^{(m)\ \nu},~~~~~~~~~~~~
\label{eom2teleDE}
\end{eqnarray}}}
%\end{widetext}
with the superpotential $S_{\rho}{}^{\mu\nu}$ defined in (\ref{Stensor}) and $T_{\ \ \ \
\rho}^{(m)\ \nu}$ the matter energy-momentum tensor.
Therefore, applying these equations in the FRW geometry (i.e on the vierbein choice
(\ref{weproudlyuse})) we obtain the Friedmann equations
(\ref{FR1nonminimalquint}),(\ref{FR2nonminimalquint}),
however now the scalar field energy density and pressure become:
\begin{eqnarray}
\label{teleDErho}
 &&\!\!\!\!\rho_{\phi}=  \frac{1}{2}\dot{\phi}^{2} + V(\phi)
-  3\xi H^{2}\phi^{2},\\
 &&\!\!\!\!p_{\phi}=  \frac{1}{2}\dot{\phi}^{2} - V(\phi) +   4 \xi
H \phi\dot{\phi}
 + \xi\left(3H^2+2\dot{H}\right)\phi^2,\ \ \ \ \ \
 \label{teleDEp}
\end{eqnarray}
where as usual we have used the useful relation
$T=-6H^2$. Furthermore, variation of the action with respect to the scalar field
provides its evolution equation
\begin{equation}
\ddot{\phi}+3H\dot{\phi}+6\xi H^2\phi+   V'(\phi)=0.
\label{fieldevol2teleDE}
\end{equation}
In this scenario, similar to standard quintessence, dark
energy is attributed to the scalar field, and thus its equation-of-state
parameter   is defined as $w_{DE}\equiv w_\phi=p_\phi/\rho_\phi$,
but $\rho_\phi$ and $p_\phi$ are now given by (\ref{teleDErho}) and
(\ref{teleDEp}), respectively. Moreover,  the scalar field evolution
(\ref{fieldevol2teleDE}) can be again re-written as the standard relation
$\dot{\rho}_\phi+3H(1+w_\phi)\rho_\phi=0$.  Thus, this scenario is named ``teleparallel
dark energy''  \cite{Geng:2011aj}.

Interestingly enough, although GR coincides with TEGR, and although minimal quintessence
coincides with minimal teleparallel quintessence (as can be seen by comparing
(\ref{rhophinonminimalquint}),(\ref{pphinonminimalquint}) with
(\ref{teleDErho}),(\ref{teleDEp}) for $\xi=0$,  and one
can verify that at the level of perturbations too),  when the non-minimal coupling is
switched on the resulting theories exhibit different behavior.
This is expected since,
concerning the gravitational sector, TEGR is identical with GR, and in the
minimal case one just adds a distinct scalar
sector, thus making no difference whether it is added in either of the two
theories. However, things are different if we switch on the non-minimal
coupling, since while in GR one couples the
scalar field with the only suitable gravitational scalar of lowest order, namely the Ricci
scalar $R$, in the latter one couples the scalar field with the only
suitable gravitational scalar of lowest order, namely the torsion scalar $T$. Clearly,
teleparallel dark energy, under the non-minimal coupling, is a different theory.

The richness of the resulting theory comparing to GR quintessence is additionally
manifested in the fact that, although in the latter one can perform a
conformal transformation and transit to an ``equivalent'',
minimally-coupled, theory with transformed field and potential
 \cite{Sahni:1998at,Uzan:1999ch,Bartolo:1999sq,Bertolami:1999dp,Faraoni:2000wk}, in the
former such a transformation does not exist
since one obtains extra terms depending on the torsion tensor itself, as
can be easily verified transforming the vierbeins as
$e^\mu_A\rightarrow \Omega\tilde{e}^\mu_A$ (one
applies in our case the similar analysis of  \cite{Yang:2010ji} of the
case of $f(T)$ scenarios). Thus, teleparallel dark
energy cannot be transformed to an
``equivalent'' minimally coupled form, and this indicates its richer
structure. Such an absence of conformal transformation exists in other
cosmological scenarios too, for example in scalar-field models with
non-minimal derivative couplings, where it is also known that the resulting
theories possess a richer structure
\cite{Amendola:1993uh,Daniel:2007kk,Saridakis:2010mf}.

Similarly, to the case of simple $f(T)$ gravity, in the present extension of
scalar-torsion theory  a Lorentz-violating term appears (the last term in the left hand
side of (\ref{eom2teleDE})), despite the fact that the theory is linear in $T$.
However, no new degree of freedom will appear at the background level on which we focus
on
this work. Clearly, going beyond background evolution and examine whether the Lorenz
violations do indeed appear under cosmological geometries and scales (we have checked
that
at the low-energy limit, the theory's basic Parametrized Post Newtonian parameters are
consistent with Solar System observations), and if they can be detected, is an
interesting
and open subject, as it is in usual $f(T)$ gravity too  \cite{Li:2010cg,Li:2011wu}.
However, we stress that the above Lorentz violation problem will be solved if one
formulates the covariant version of the theory, as we did in subsection \ref{restoringLI}
for simple $f(T)$ theory.

Finally, note that one can extend the above non-minimal scalar-torsion coupling $\xi
T\phi^{2}$  to the general form $\xi B(\phi) T$
 \cite{Otalora:2013tba,Otalora:2013dsa,Skugoreva:2014ena,Jarv:2015odu}, that is
allowing for an arbitrary scalar-field coupling function\footnote{Along these lines one
can construct further extensions through the non-minimal coupling of the four-derivative
of the scalar field with the ``vector torsion'' \cite{Otalora:2014aoa}, or of the scalar
field with both the torsion scalar and the divergence of the torsion vector
\cite{Bahamonde:2015hza}, or even couple the torsion scalar to a fermionic instead of a
scalar field \cite{Kucukakca:2014vja}.},
similarly to the general curvature non-minimal quintessence. In this case, equations
(\ref{eom2teleDE}), (\ref{teleDErho}), (\ref{teleDEp}) and (\ref{fieldevol2teleDE})
respectively become \cite{Otalora:2013tba,Otalora:2013dsa,Skugoreva:2014ena,Jarv:2015odu}:
\begin{eqnarray}
&&\!\!\!\!\!\!\!\!\!\!\! \left[\frac{1}{4\pi G}+2 \xi\,B(\phi)\right]\!\left[
e^{-1}\partial_{\mu}(ee_A^{\rho}S_{\rho}{}^{\mu\nu})
-e_{A}^{\lambda}T^{\rho}{}_{\mu\lambda}S_{\rho}{}^{\nu\mu}+\frac{1}{
4} e_{A}^{\nu}T
\right]\nonumber
\\
&&\ \ \ \ \ \ \ \ \ \  \,
-
e_{A}^{\nu}\left[\frac{1}{2}
\partial_\mu\phi\partial^\mu\phi-V(\phi)\right]+
  e_A^\mu \partial^\nu\phi\partial_\mu\phi,
\nonumber
\\
&&\ \ \ \ \ \ \ \ \ \  \, + 2\xi
e_A^{\rho}S_{\rho}{}^{\mu\nu}B'(\phi)
\left(\partial_\mu\phi\right)= e_{A}^{\rho} T_{\ \ \ \ \rho}^{(m)\ \nu},
 \label{eomsteleDEgeneral}
\end{eqnarray}
 \begin{eqnarray}
\label{teleDErhogeneral}
 &&\rho_{\phi}=  \frac{1}{2}\dot{\phi}^{2} + V(\phi)
-  3\xi H^{2} B(\phi),\\
 &&p_{\phi}=  \frac{1}{2}\dot{\phi}^{2} - V(\phi) +   2 \xi
H   B'(\phi)   \dot{\phi}\nonumber\\
&&\ \ \ \ \ \ \
 + \xi\left(3H^2+2\dot{H}\right) B(\phi),
 \label{teleDEpgeneral}
\end{eqnarray}
  and
\begin{equation}
\label{phieomteleDEgeneral}
\ddot{\phi}+3 H\dot{\phi}+3\xi H^2 B'(\phi)+V'(\phi)=0.
\end{equation}

\subsubsection{Late-time cosmological behavior}

Let us first investigate the late-time cosmological behavior of the above scenario of
teleparallel dark energy. First of all, one deduces the straightforward result that
using (\ref{teleDErho}),(\ref{teleDEp}) he can obtain
a  dark-energy sector  possessing dynamical nature, as well as that it can
drive the universe acceleration. However, the most interesting and direct
consequence of the teleparallel dark energy density and pressure,  is that the
corresponding dark energy
equation-of-state parameter can lie in the quintessence regime
($w_{DE}>-1$), in the phantom regime ($w_{DE}<-1$), or exhibit the
phantom-divide crossing during cosmological evolution. This is a
radical difference with the non-minimal quintessence scenario, in which dark energy lies
always above the phantom divide, and thus this feature reveals the
capabilities of the construction.

In order to present the above features in a more transparent way,
one evolves numerically the cosmological system for dust matter
($w_m\approx0$), using the redshift $z=a_0/a-1$ as the independent
variable, imposing the present scale factor $a_0$ to be equal to
1, the  dark energy density
$\Omega_{DE}\equiv 8\pi G\rho_{\phi}/(3H^2)$ at present to be
$\approx0.72$ and its initial value to be $\approx0$   \cite{Geng:2011aj}. Finally,
concerning the scalar field potential we use the exponential
ansatz of the form $V=V_0e^{\lambda\phi}$.
\begin{figure}[ht]
\mbox{\epsfig{figure=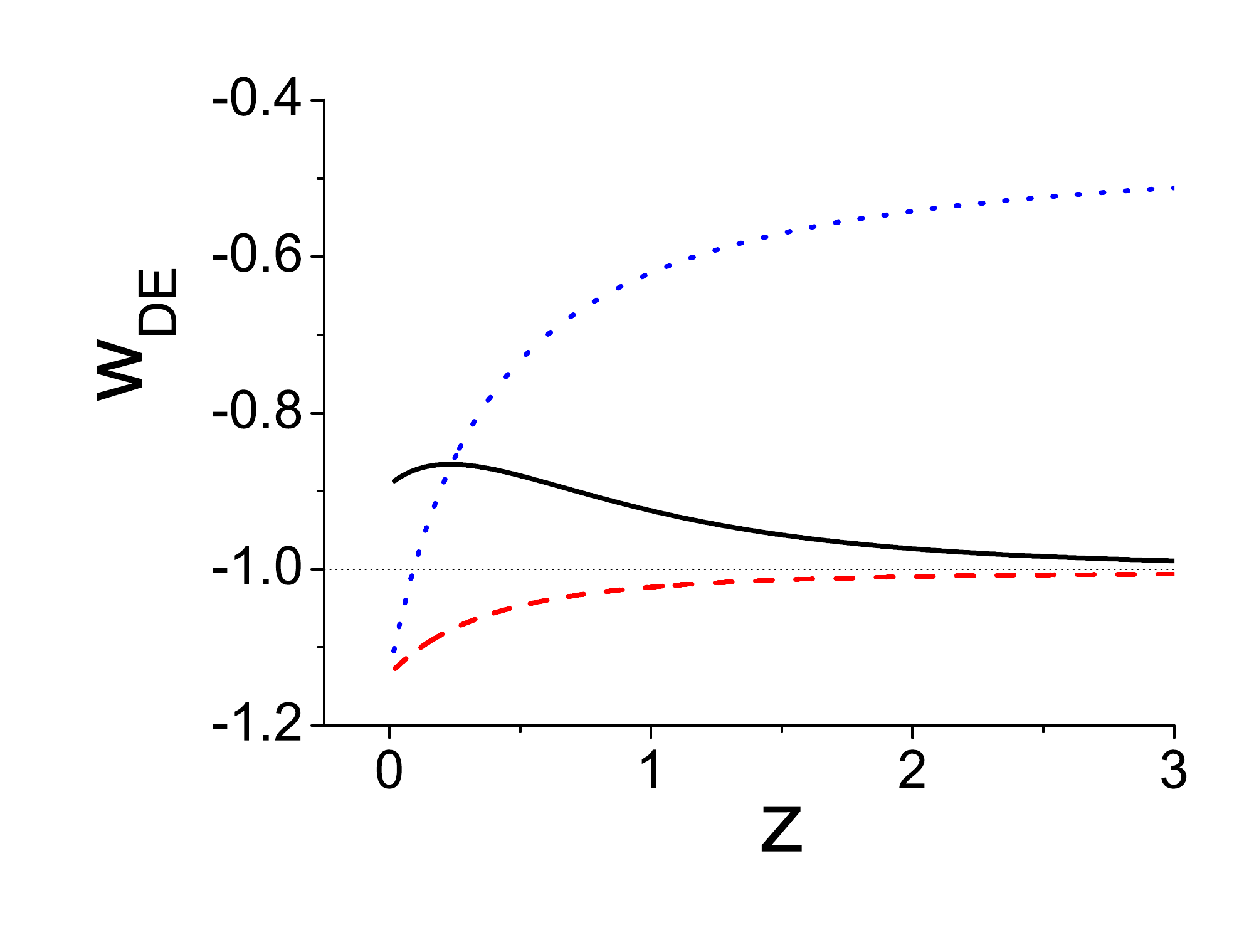,width=8.9cm,angle=0}} \caption{{\it
Evolution of the dark energy equation-of-state parameter $w_{DE}$
as a function of the redshift $z$, for three cases of the
teleparallel dark energy scenario, in the exponential scalar-field
potential ansatz of the form $V=V_0e^{\lambda\phi}$. The
black-solid curve presents quintessence-like behavior and
corresponds to $\xi=-0.4$, $\lambda=1.5$ and $V_0\approx2\times
10^{-13}$, the red-dashed curve presents phantom-like behavior and
corresponds to $\xi=-0.8$, $\lambda=0.05$ and $V_0\approx
10^{-13}$, and the blue-dotted curve presents the phantom-divide
crossing and corresponds to $\xi=-0.25$, $\lambda=40$ and
$V_0\approx 10^{-12}$. $\lambda$ and $V_0$ are measured in
$8\pi G$-units and the -1-line is depicted for convenience. The Figure is from
 \cite{Geng:2011aj}.}}
\label{plot}
\end{figure}

In Fig.~\ref{plot} we depict the $w_{DE}$-evolution for three realizations of the
scenario
at hand. In the case of the black-solid curve the teleparallel dark energy behaves like
quintessence, in the red-dashed curve it behaves like a phantom, while in blue-dotted
curve the dark energy exhibits the phantom-divide crossing during the evolution.
Note that the crossing behavior in Fig.~\ref{plot} is the one favored by the
observational
data, in contrast with viable $f(R)$-gravity models where it is the opposite one
\cite{Bamba:2010iy}. We remark that in the above figure one focuses on the qualitative
features, and in particular we maintain the same potential just to stress that in
principle one can obtain the various behaviors with the same potential. Clearly, one
could
be quantitatively more accurate and impose the observational $w_{DE}(z)$ as an input,
reconstructing the corresponding potential.

We stress that we can obtain the above quintessence-like, phantom-like, or quintom-like
behaviors, although the scalar field is canonical. Clearly this behavior is much richer
comparing to GR with a scalar field. Moreover, the fact that the
phantom regime can be described without the need of phantom fields, which have ambiguous
quantum behavior  \cite{Cline:2003gs}, is a significant advantage. In summary, the rich
behavior of teleparallel dark energy makes it a promising cosmological scenario.

\subsubsection{Dynamical analysis}

The interesting cosmological behavior of teleparallel dark energy makes it necessary to
perform a phase-space and stability analysis, examining in a systematic way the possible
cosmological behaviors, focusing on the late-time stable solutions.  In order to
transform the system to its autonomous form we introduce the auxiliary variables
\cite{Xu:2012jf}:\footnote{For the dynamical analysis of more general scalar-torsion
models see  \cite{Otalora:2013tba,Otalora:2013dsa,Jamil:2012vb}, while for the
investigation of the
pure gravitational sector, i.e. in the absence of matter, see  \cite{Skugoreva:2014ena}.}
\begin{eqnarray}
&&x=\frac{\sqrt{8\pi G}\dot{\phi}}{\sqrt{6}H}\nonumber\\
&&y=\frac{\sqrt{8\pi G}\sqrt{V(\phi)}}{\sqrt{3}H}\nonumber\\
&&z=\sqrt{|\xi|}\sqrt{8\pi G}\phi. \label{auxiliaryteleDE}
\end{eqnarray}
Using these variables the first Friedmann equation, with the teleparallel dark energy
density \eqref{teleDErho}, becomes
\begin{equation}
\label{Fr1normteleDE}
x^2+y^2-z^2 \text{sgn}(\xi )+\frac{8\pi G \rho_m}{3 H^2}=1.
\end{equation}
This constraint allows for expressing $\rho_m$ as a function of the
auxiliary variables (\ref{auxiliaryteleDE}). Therefore, using (\ref{Fr1normteleDE})
and (\ref{teleDErho}) we can write the density parameters as:
\begin{eqnarray}
&&\Omega_m\equiv\frac{8\pi G\rho_m}{3H^{2}}=1-x^2-y^2+z^2 \text{sgn}(\xi )\nonumber\\
 &&\Omega_{DE}\equiv\frac{8\pi G\rho_{\phi}}{3H^{2}}=x^2+y^2-z^2 \text{sgn}(\xi ),
 \label{OmegasteleDE}
\end{eqnarray}
while for the dark-energy equation-of-state parameter $w_{DE}\equiv
w_\phi=p_\phi/\rho_\phi$, using (\ref{teleDErho}),(\ref{teleDEp}) we
obtain:
\begin{eqnarray}\label{wdephaseteleDE}
&&\!\!\!\!\!\!\!\!\!\!\!\!\!\!\!\!\!\!\!\!
w_{DE}= \left[1+z^2\text{
sgn}(\xi)
\right]^{-1}\left[x^2+y^2-z^2\text{sgn}(\xi)\right]^{-1}\nonumber\\
&&  \cdot
\Big\{x^2-y^2+4\sqrt{\frac{2}{3}}zx\sqrt{|\xi|}\text{sgn}
(\xi)\nonumber\\
&&\ \ \
-z^2w_m
\text{sgn}(\xi)\left[1-x^2-y^2+z^2\text{sgn}(\xi)\right]\Big\}.
\end{eqnarray}
 Without loss of generality, in the following
we restrict the analysis in the dust matter case, that is we assume that
$w_m=0$.

It is convenient to introduce two additional quantities with great physical
significance, namely the ``total'' equation-of-state parameter:
{\small{
\begin{equation}\label{wtotteleDE}
w_{tot}\equiv\frac{p_\phi}{\rho_\phi+\rho_m}=w_{DE}\Omega_{DE}
=\frac{x^2-y^2+4\sqrt{\frac{2}{ 3}}zx\sqrt{|\xi|}\text{sgn}
(\xi)}{1+z^2\text{
sgn}(\xi)},
\end{equation}}}
and the deceleration parameter $q$:
\begin{eqnarray}
&&\!\!\!\!\!\!\!\!\!\!\!\!\!\!\!
q\equiv-1-\frac{\dot{H}}{H^2}=\frac{1}{2}+\frac{3}{2}w_{tot}\nonumber\\
&&
\!\!\!\!\!\!\!\!\!\!=
\frac{1+3(x^2-y^2)+\left[z+4\sqrt{6}x\sqrt{|\xi|}\right]z\text {
sgn}(\xi)}{2\left[
1+z^2\text {
sgn}(\xi)
\right]}.
\label{deccteleDE}
\end{eqnarray}

Finally, concerning the scalar potential $V(\phi)$ the usual assumption in
the literature is to assume an exponential potential of the form
\begin{equation}
\label{exppotteleDE}
V=V_0\exp(-\sqrt{8\pi G}\lambda\phi),
\end{equation}
since exponential potentials are known to be significant in various cosmological models
 \cite{Copeland:1997et,Ferreira:1997au,Chen:2008ft}. Moreover, note that the exponential
potential was used as an example in the initial work on teleparallel dark energy
 \cite{Geng:2011aj}.

In summary, using the auxiliary variables (\ref{auxiliaryteleDE}) and considering the
exponential potential (\ref{exppotteleDE}), the equations of motion  in
the case of dust matter can be transformed into the following autonomous system
\cite{Xu:2012jf}:
\begin{align}
 &x'=\frac{3 x^3}{2  z^2 \text{sgn}(\xi )+2}+\frac{2 \sqrt{6} z
   \sqrt{|\xi |} x^2\text{sgn}(\xi )}{  z^2\text{sgn}(\xi )+2}\nonumber\\
&\ \ \ \ \ \ \
   -\frac{3
x}{2}
+\frac{1}{2} y^2
   \left[\sqrt{6} \lambda -\frac{3 x}{\text{sgn}(\xi )
   z^2+1}\right]\nonumber\\
&\ \ \ \ \ \ \
-\sqrt{6} z \sqrt{|\xi|} \text{sgn}(\xi)\nonumber\\
&y'=\frac{3 y x^2}{2
   \text{sgn}(\xi ) z^2+2}+\frac{3}{2} y
\left[1-\frac{y^2}{ z^2\text{sgn}(\xi )
  +1}\right]\nonumber\\
&\ \ \ \ \ \ \
-\frac{\sqrt{\frac{3}{2}} yx
\left\{\left[\lambda  z-4   \sqrt{|\xi |}\right] z\text{sgn}(\xi
)+\lambda\right\} }{1+
   z^2\text{sgn}(\xi )}\nonumber\\
&z'=\sqrt{6} \sqrt{|\xi |} x,
\label{autonomousteleDE}
\end{align}
where we have used that
for every
quantity $F$ we acquire $\dot{F}=HF'$.
Since $\rho_m$ is
nonnegative, from (\ref{Fr1normteleDE}) we obtain that  $x^2+y^2-z^2
\text{sgn}(\xi )\leq 1$.
Thus, we deduce that for $\xi<0$ the system \eqref{autonomousteleDE}
defines a flow on the compact phase space
$\Psi=\left\{ x^2+y^2-z^2 \text{sgn}(\xi )\leq 1, y\geq 0\right\},$ for
$\xi=0$ the phase space is compact and it is reduced to the circle
$\Psi=\left\{ x^2+y^2\leq 1, y\geq 0\right\},$  while for
$\xi>0$ the phase space $\Psi$ is unbounded.

Before proceeding we make two comments on the degrees of freedom and the choice of
auxiliary variables. Firstly, as we have already mentioned, in the minimal coupling case,
that is when $\xi=0$, the model at hand coincides with standard quintessence, whose
phase-space analysis is well known using two degrees of freedom, namely the variables $x$
and $y$ defined above  \cite{Copeland:1997et}. On the other hand, when $\xi\neq0$ we have
three degrees of freedom and all $x$, $y$, $z$ are necessary. Therefore, in order to
perform the analysis in a unified way, we use the three variables defined above, having
in
mind that for $\xi=0$ the variable $z$ becomes zero and thus irrelevant. Secondly,
apart from the standard choices of the variables $x$ and $y$, one must be careful in
suitably choosing the variable $z$ in order not to lose dynamical information. For
example, although the choice for the variable $z=\frac{\sqrt{8\pi G}\rho_m}{\sqrt{3}H}$ is
another reasonable choice, however it still loses the critical points that lie at
``infinity''. Therefore, in order to completely cover the phase-space behavior one should
additionally use the Poincar\'e central projection method to investigate the dynamics at
``infinity'' \cite{Xu:2012jf}. The negligence of this point was the reason of the
incomplete phase-space analysis of teleparallel dark energy performed in
\cite{Wei:2011yr}.

Let us now proceed to the phase-space analysis. The real and physically meaningful (that
is corresponding to an expanding universe, i.e possessing $H>0$) critical points $(x_c,
y_c, z_c)$ of the autonomous system (\ref{autonomousteleDE}), obtained by setting the
left hand sides of the equations to zero, are presented in Table \ref{critteleDE}. In the
same table we provide their existence conditions. For each critical point of Table
\ref{critteleDE} we examine the sign of the real part of the eigenvalues of the
$3\times3$ matrix ${\bf{Q}}$ of the linearized perturbation equations, in order to
determine the type and stability of the point. The various eigenvalues are presented in
Table \ref{eigenteleDE}. Hence, in  Table \ref{crit1teleDE} we summarize the results, and
additionally for each critical point we calculate the values of $\Omega_{DE}$,
$w_{DE}$, $w_{tot}$ and $q$  given by (\ref{OmegasteleDE}), (\ref{wdephaseteleDE}),
(\ref{wtotteleDE}) and (\ref{deccteleDE}) respectively.
 \begin{table*}[!]
\begin{center}
\begin{tabular}{|c|c|c|c|c|}
\hline
&&&&   \\
 Cr. P.& $x_c$ & $y_c$ & $z_c$ & Exists for \\
\hline \hline
$A$& 0 & 0 & 0 &  all $\xi$,$\lambda$
\\
\hline
 $B$& 1& 0 & 0 &  $\xi=0$, all $\lambda$ \\
\hline
 $C$& -1& 0 & 0 &  $\xi=0$, all $\lambda$ \\
\hline
$D$& $\frac{\lambda}{\sqrt{6}}$ & $\sqrt{1-\frac{\lambda^2}{6}}$ &
0
&  $\xi=0$,   $ \lambda^2\leq6$ \\
\hline
 $E$& $\sqrt{\frac{3}{2}}\frac{1}{\lambda}$ &
$\sqrt{\frac{3}{2}}\frac{1}{\lambda}$ &
0 &  $\xi=0$,    $\lambda^2\geq3$ \\
\hline
 $F$& 0 & $\sqrt{\frac{2\xi -2\sqrt{\xi  \left(\xi -\lambda
^2\right)}}{\lambda ^2}}$ &
$\frac{\left[\xi -\sqrt{\xi  \left(\xi -\lambda ^2\right)}\right]
\sqrt{|\xi |}}{\lambda  \xi }$& $0<\lambda^2\leq\xi$ \\
\hline
 $G$&0 & $\sqrt{\frac{2\xi +2\sqrt{\xi  \left(\xi -\lambda
^2\right)}}{\lambda ^2}}$ &
$\frac{\left[\xi +\sqrt{\xi  \left(\xi -\lambda ^2\right)}\right]
\sqrt{|\xi |}}{\lambda  \xi }$ & $0<\lambda^2\leq\xi$ or
$\xi<0$ \\
\hline
$J$&0 & 1 & 0& $\xi\neq0$ and $\lambda=0$ \\
\hline
\end{tabular}
\end{center}
\caption[critteleDE]{\label{critteleDE} The real and physically meaningful
critical points of the autonomous system (\ref{autonomousteleDE}), along with their
existence
conditions. From  \cite{Xu:2012jf}.}
\end{table*}
\begin{table*}[htp]
\begin{center}
\begin{tabular}{|c|c|c|c|c|}
\hline
 Cr. Point & Exists for & $\nu_1$ & $\nu_2$ & $\nu_3$
\\
\hline \hline
 A & all $\xi$,$\lambda$  & $\frac{3}{2}$ & $ \frac{1}{4}
\left(-3-\sqrt{9-96 \xi }\right)$ & $\frac{1}{4} \left(-3+\sqrt{9-96 \xi
}\right)$ \\
\hline
 B & $\xi=0$, all $\lambda$ & $3$ & $3-\sqrt{\frac{3}{2}}\lambda$ & 0 \\
\hline
C & $\xi=0$, all $\lambda$ & $3$ & $3+\sqrt{\frac{3}{2}}\lambda$ & 0 \\
\hline
D& $\xi=0$,  $\lambda^2\leq6$ & $\lambda^2-3$ & $\frac{1}{2} \left(\lambda
^2-6\right)$ &
0   \\
\hline
 E   & $\xi=0$,  $\lambda^2\geq3$ & $\alpha^+(\lambda)$ &
$\alpha^-(\lambda)$  &0 \\
\hline
F & $\lambda^2\leq\xi$  & $-3$ &
$\beta^+(\lambda,\xi)$ &
$\beta^-(\lambda,\xi)$ \\
\hline
G  & $ \lambda^2\leq\xi$
or $\xi<0$ & $-3$ & $\gamma^+(\lambda,\xi)$ &
$\gamma^-(\lambda,\xi)$\\
\hline
J & $\xi\neq0$
and $\lambda=0$ & $-3$ &
$\frac{-3+\sqrt{9-24 \xi }}{2}$ &
$ \frac{-3-\sqrt{9-24\xi}}{2}$\\
\hline
\end{tabular}
\end{center}
\caption[eigenteleDE]{\label{eigenteleDE} The eigenvalues of the matrix
${\bf {Q}}$ of the perturbation equations of the autonomous system
(\ref{autonomousteleDE}). Points $B$-$E$ exist only for $\xi=0$, and for
these points the variable $z$ is zero and thus irrelevant. Although for
 these points the eigenvalue associated to the z-direction is zero,
the stability conditions are obtained by analyzing the eigenvalues of the
non-trivial $2\times2$ submatrix of  ${\bf {Q}}$. We introduce the notations
$\alpha^\pm(\lambda)=\frac{3}{4} \left(-1\pm\frac{\sqrt{24
\lambda ^2-7 \lambda ^4}}{\lambda ^2}\right),$
$\beta^\pm(\lambda,\xi)=\frac{-3\pm\sqrt{9-24\sqrt{\xi^2-\lambda^2\xi}}}{2}$ and
$\gamma^\pm(\lambda,\xi)=\frac{-3\pm\sqrt{9+24\sqrt{\xi^2-\lambda^2\xi}
}}{2}.$ From  \cite{Xu:2012jf}.}
\end{table*}
\begin{table*}[!]
\begin{center}
\begin{tabular}{|c|c|c|c|c|c|}
\hline
&&&&&   \\
 Cr. P.&
Stability & $\Omega_{DE}$ &  $w_{DE}$ & $w_{tot}$ & $q$\\
\hline \hline
$A$&  saddle point &   0 & arbitrary & 0 &$\frac{1}{2}$ \\
\hline
 $B$&   unstable for
$\lambda<\sqrt{6}$ &
   &     &   & \\
  &    saddle point otherwise &
1  &   1 & 1 & 2\\
\hline
 $C$&   unstable for
$\lambda>-\sqrt{6}$ &
   &     &   & \\
  &   saddle point otherwise &
1  &   1 & 1 & 2\\
\hline
$D$& stable node for $\lambda^2<3$ &   1 &
$-1+\frac{\lambda^2}{3}$ & $-1+\frac{\lambda^2}{3}$ &
$-1+\frac{\lambda^2}{2}$\\
& saddle point for $3<\lambda^2<6$&    &  & &\\
\hline
 $E$& stable
node for
 $3<\lambda^2<\frac{24}{7}$ & $\frac{3}{\lambda^2}$ & 0 & 0 &
$\frac{1}{2}$\\
& stable spiral for
 $\lambda^2>\frac{24}{7}$&    &  & &\\
\hline
 $F$&
stable node for  $\lambda^2<\xi$ &  1 & $-1$ & $-1$ & $-1$\\
\hline
 $G$& saddle point &  1 & $-1$ & $-1$ & $-1$\\
\hline
$J$& stable spiral for
$\frac{3}{8} <\xi$ & 1 & $-1$ & $-1$ & $-1$\\
&    stable node for
$0<\xi<\frac{3}{8}$
&    &  & &\\
&  saddle point for
 $\xi<0$
&    &  & &\\
\hline
\end{tabular}
\end{center}
\caption[crit1teleDE]{\label{crit1teleDE} The real and physically meaningful critical
points of the autonomous system (\ref{autonomousteleDE}) along with their stability
conditions, and the
corresponding values of the dark-energy
density parameter $\Omega_{DE}$, of the  dark-energy equation-of-state
parameter $w_{DE}$, of the total equation-of-state parameter
  $w_{tot}$ and of the deceleration parameter $q$. From  \cite{Xu:2012jf}.}
\end{table*}

Apart from the above finite critical points, the scenario at hand possesses critical
points at infinity too, due to the fact that the dynamical system
\eqref{autonomousteleDE} is non-compact for the choice $\xi>0$. In order to extract and
analyze them, one needs to apply the Poincar\'e central projection method
 \cite{PoincareProj, Leon:2014rra}. However, since the obtained four critical points at
infinity are non-stable, we do not describe the whole procedure in detail, and we refer
the reader to the original work  \cite{Xu:2012jf}.

Having performed the complete phase-space analysis of teleparallel dark
energy, we can now discuss the corresponding cosmological behavior. A
first remark is that in the minimal case (that is $\xi=0$) we do verify
that the scenario at hand coincides completely with standard quintessence.
Therefore, we will make a brief review on the subject and then focus
on the non-minimal case.

The points $B$ to $E$ exist only for the minimal case, that is only for
$\xi=0$. Points $B$ and $C$ are not stable, corresponding to a non-accelerating,
dark-energy dominated universe, with a stiff dark-energy equation-of-state parameter
equal
to 1. Both of them exist in standard quintessence  \cite{Copeland:1997et}.

Point $D$ is a stable for $0<\lambda^2<3$, and thus it can attract the universe at late
times. It corresponds to a dark-energy dominated universe, with a dark-energy
equation-of-state parameter lying in the quintessence regime, which can be
accelerating or not according to the $\lambda$-value. This point
exists in standard quintessence  \cite{Copeland:1997et}. It is  the most
important one in that scenario, since it is both stable and possesses a
$w_{DE}$ compatible with observations.

Point $E$ is a stable one for $3<\lambda^2$. It has the advantage that the
dark-energy density parameter lies in the interval $0<\Omega_{DE}<1$, that
is it can alleviate the coincidence problem, since dark matter and dark
energy density parameters can be of the same order (in order to treat the
coincidence problem one must explain why the present dark energy and matter
density parameters are of the same order of magnitude although they follow
different evolution behaviors). However, it has the disadvantage that
$w_{DE}$ is 0 and the expansion is not accelerating, which are not favored
by observations. This point exists in standard quintessence too \cite{Copeland:1997et}.

Let us now analyze the case $\xi\neq0$. In this case we obtain the
critical points $A$, $F$, $G$ and $J$. Point $A$ is saddle point, and
thus it cannot be late-time solution of the universe. It corresponds to a
non-accelerating, dark-matter dominated universe, with arbitrary
dark-energy equation-of-state parameter. Note that this trivial point
exists in the standard quintessence model too  \cite{Copeland:1997et}, since it is
independent of $\xi$.

The present scenario of teleparallel dark energy, possesses two
additional, non-trivial critical points that do not exist in standard
quintessence. Thus, they account for the new information of this richer
scenario, and as expected they depend on the non-minimal coupling $\xi$. In
particular, point $F$ is stable if $\lambda^2<\xi$, and thus it can attract
the universe at late times. It corresponds to an accelerating
universe with complete dark energy domination, with $w_{DE}=-1$, that is
dark energy behaves like a cosmological constant. We stress that this
$w_{DE}$ value is independent of $\lambda$ and $\xi$, which is an important
and novel result. Thus, while point $D$ (the important point of standard
quintessence) needs to have  a very flat, that is quite tuned, potential in
order to possess a $w_{DE}$ near the observed value $-1$, point $F$
exhibits this behavior for every $\lambda$-value, provided that
$\lambda^2<\xi$. This feature is a significant advantage of the scenario at
hand, amplifying its generality, and offers a mechanism for the
stabilization of $w_{DE}$ close to the cosmological-constant value.
Similarly, we have the   point $G$, which has the same cosmological
properties with $F$, however it is a saddle point and thus it cannot be a
late-time solution of the universe, but the universe can spend a
large period of time near this solution.

Finally, when $\xi\neq0$ and for the limiting case $\lambda=0$, that is for a constant
potential, the present scenario exhibits the critical point $J$, which is stable for
$\xi>0$. It corresponds to a dark-energy dominated, de Sitter universe, in which dark
energy behaves like a cosmological constant.

Before closing this section, let us make a comment on another crucial
difference of teleparallel dark energy, comparing with standard
quintessence, that is of the $\xi\neq0$ case comparing to the $\xi=0$ one.
In particular, when $\xi=0$, in which teleparallel dark energy coincides
with standard quintessence, $w_{DE}$ is always larger that $-1$, not only
at the critical points, but also throughout the cosmological evolution as
well. However, for $\xi\neq0$, during the cosmological evolution $w_{DE}$
can be either above or below $-1$, and only at the stable critical point
it becomes equal to $-1$. Such a cosmological behavior is much
richer, and very interesting, both from the theoretical as well as from
the observational point of view, since it can explain the dynamical
behavior of $w_{DE}$ either above or below the phantom divide, and
moreover its stabilization to the cosmological-constant value, without
the need for parameter-tuning.
 \begin{figure}[ht]
\begin{center}
\includegraphics[scale=0.4]{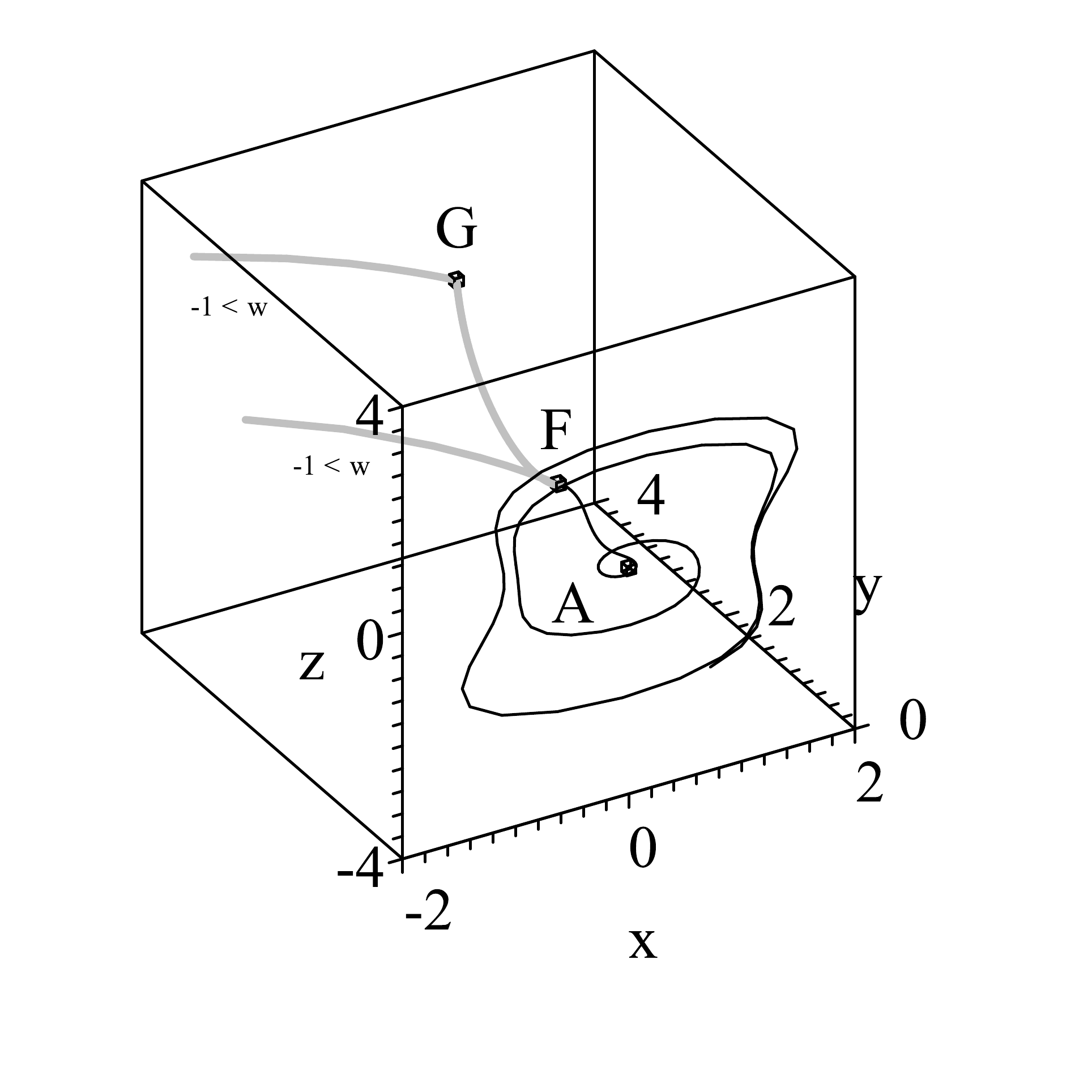}
\caption{{\it{The phase-space evolution for the teleparallel dark
energy scenario with $\lambda=0.7$ and $\xi=1$. The trajectories
are attracted by the cosmological-constant-like stable point $F$,
however the evolution towards it possesses a $w_{DE}$ being
quintessence-like, phantom-like, or experiencing the
phantom-divide crossing, depending on the specific initial
conditions. The orbits with a thick gray curve on the top left are those
with $w_{DE}>-1$ initially, while the thin black curve corresponds to
$w_{DE}<-1$ initially.}} } \label{fig4teleDEphasespace}
\end{center}
\end{figure}

In order to present the aforementioned behavior more transparently, one evolves the
autonomous system (\ref{autonomousteleDE}) numerically, and in
Fig. \ref{fig4teleDEphasespace} we observe the corresponding phase-space behavior.
In this case the universe at late times is attracted  by the
cosmological-constant-like stable point $F$. However, during the evolution
towards this point the dark-energy equation-of-state parameter $w_{DE}$
presents a very interesting behavior, and in particular, depending on the
specific initial conditions, it can be quintessence-like, phantom-like, or
experience the phantom-divide crossing. Such a behavior is much richer
than standard quintessence, and reveals the capabilities of teleparallel
dark energy scenario. Finally, one should also investigate in detail
whether the scenario at hand is stable  at the perturbation level,
especially in the region $w_{DE}<-1$. Such an analysis can be heavily based
on  \cite{Chen:2010va} and one can find that  $f(T)$ cosmology and
teleparallel dark energy are stable even in the region $w_{DE}<-1$.

\subsubsection{Observational constraints}

Let us now  use observations in order to constrain the parameters of teleparallel dark
energy, following \cite{Geng:2011ka}. In particular, one uses data from SNIa from the 
Supernova 
Cosmology Project (SCP) Union2 compilation \cite{Amanullah:2010vv}, BAO data from the
Two-Degree Field Galaxy Redshift Survey (2dFGRS) and the Sloan Digital Sky Survey Data
Release 7 (SDSS DR7)~ \cite{Percival:2009xn}, and the Cosmic Microwave Background (CMB)
radiation data from Seven-Year Wilkinson Microwave Anisotropy Probe (WMAP)
observations~ \cite{Komatsu:2010fb}, in order to plot likelihood-contours for the present
values of the dark-energy equation of state $w_{DE_0}$, for the matter density parameter
$\Omega_{m0}$ and for the non-minimal coupling parameter $\xi$. Since the model includes
the scalar-field potential, one performs the analysis separately for power-law (quartic)
$V(\phi)=V_0 \phi^4$ and exponential $V(\phi)=V_0 e^{-\sqrt{8\pi G}\lambda\phi}$
potentials (see  \cite{Geng:2011ka} for the incorporation of the inverse  hyperbolic
cosine  potential).

In Fig. \ref{phi41} we present the likelihood contours for $w_{DE_0}$  and
$\Omega_{m0}$ with the teleparallel dark energy scenario under the quartic
potential.
\begin{figure}[ht]
\begin{center}
\includegraphics[width=6.5cm]{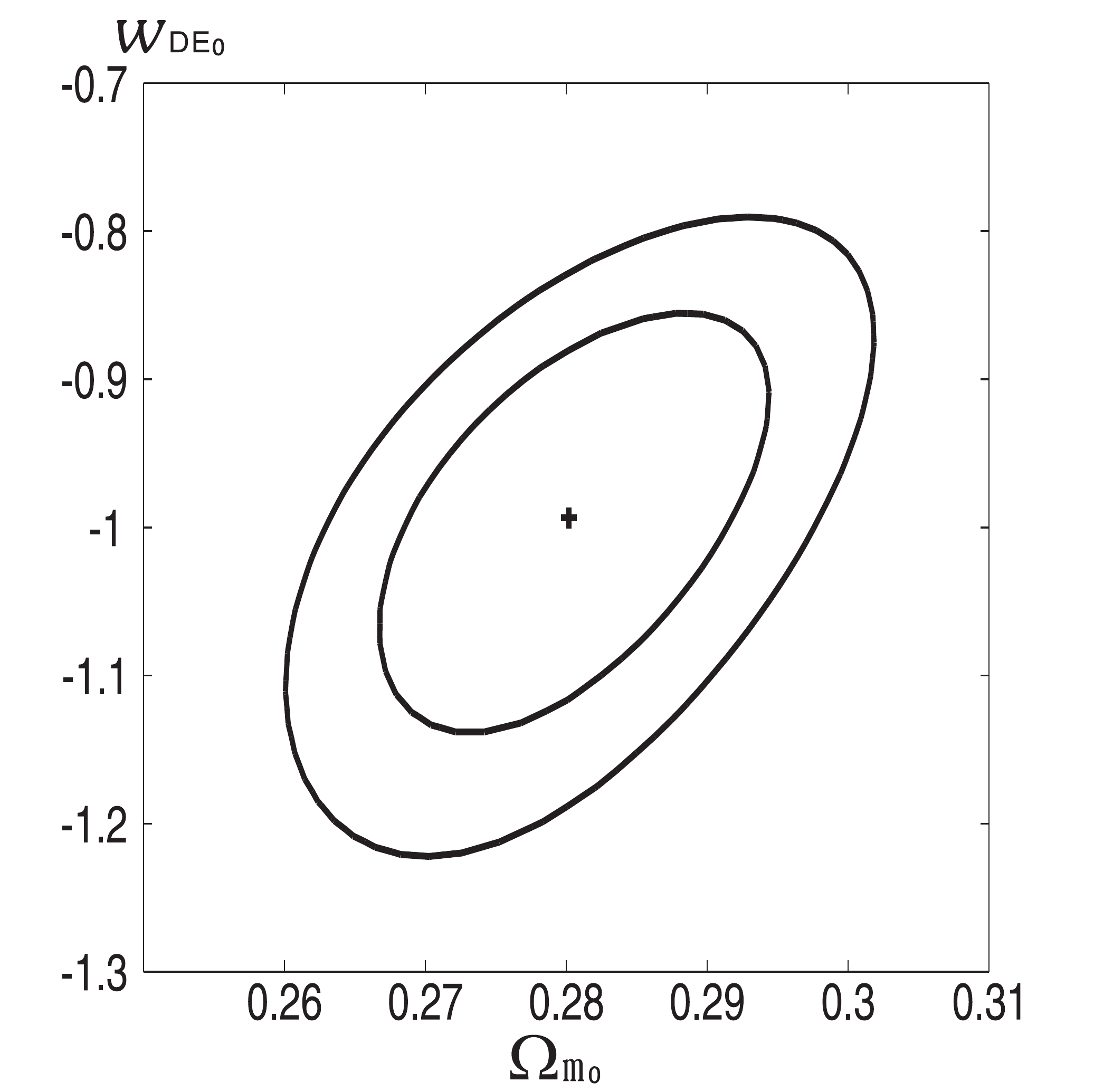}
\caption{{\it{ Contour plots of the present dark-energy
equation-of-state parameter $w_{DE_0}$ versus the present matter
density parameter $\Omega_{m0}$ under SNIa, BAO and CMB
observational data in the teleparallel dark energy scenario with the
quartic potential $V(\phi)=V_0 \phi^4$. The curves correspond
to 1$\sigma$ and 2$\sigma$ confidence levels, respectively, and the cross
marks the best-fit point. From  \cite{Geng:2011ka}.}} }
\label{phi41}
\end{center}
\end{figure}
As we observe, the scenario at hand is in agreement with observations, and
as expected, it can describe both the quintessence and phantom regimes.
Since the scalar field is canonical, it is a great advantage of
the present model.

In Fig. \ref{phi42} we present the likelihood contours for $w_{DE_0}$  and
the non-minimal coupling parameter $\xi$, for the quartic potential.
\begin{figure}[ht]
\begin{center}
\includegraphics[width=6.5cm]{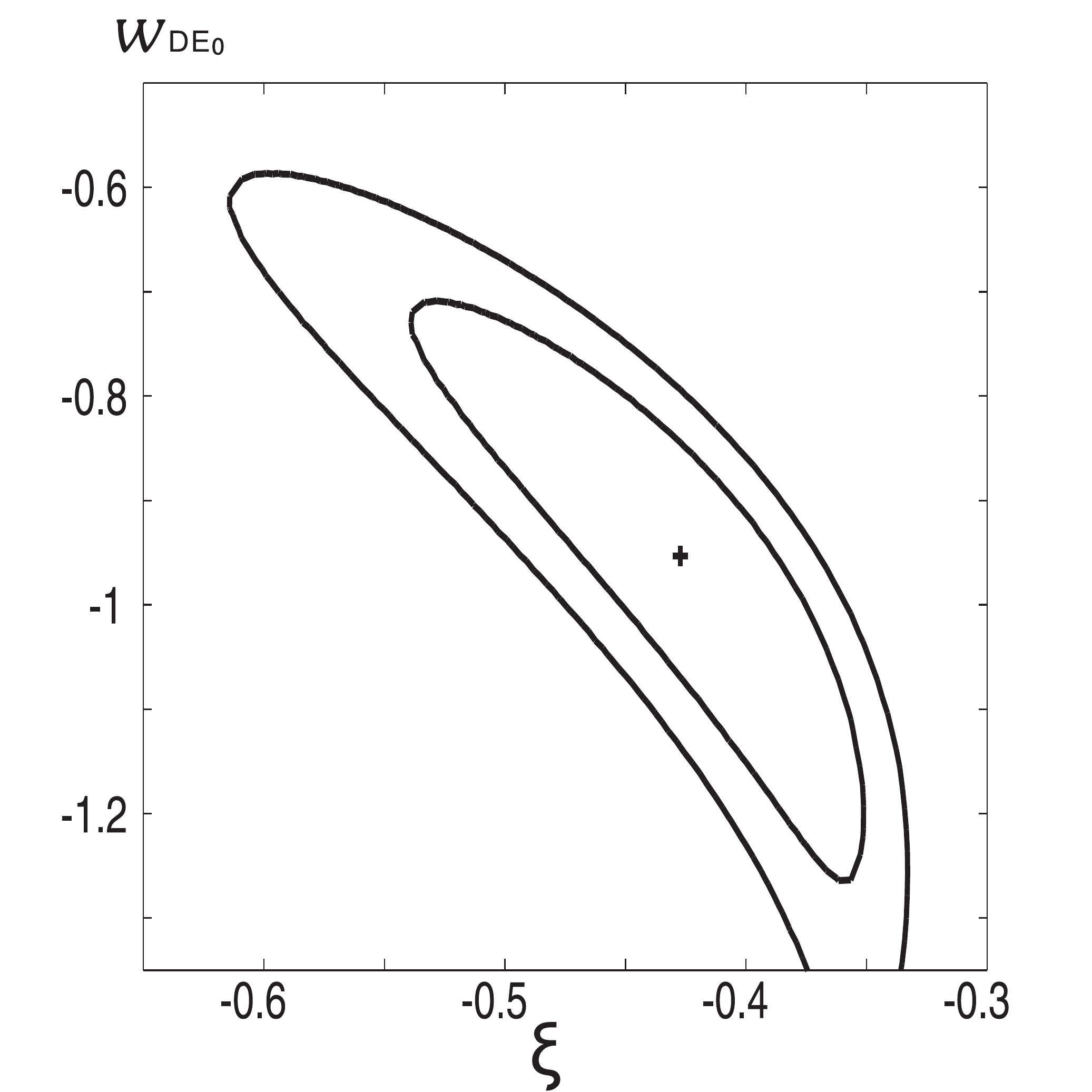}
\caption{{\it{ Contour plots of the present dark-energy
equation-of-state parameter $w_{DE_0}$ versus the non-minimal coupling
parameter $\xi$ under SNIa, BAO and CMB
observational data, in the teleparallel dark energy scenario with the
quartic potential $V(\phi)=V_0 \phi^4$. The curves correspond
to 1$\sigma$ and 2$\sigma$ confidence levels, respectively, and the cross
marks the best-fit point. From  \cite{Geng:2011ka}.}} }
\label{phi42}
\end{center}
\end{figure}
Interestingly enough we observe that the non-minimal coupling is favored
by the data, and in particular a small $\xi$ is responsible for the
quintessence regime, while a larger one leads to the phantom regime. Note that
the best-fit value of $w_{DE_0}|_{b.f}\approx-0.98$
is very close to the
cosmological constant.
 \begin{figure}[ht]
\begin{center}
\includegraphics[width=6.5cm]{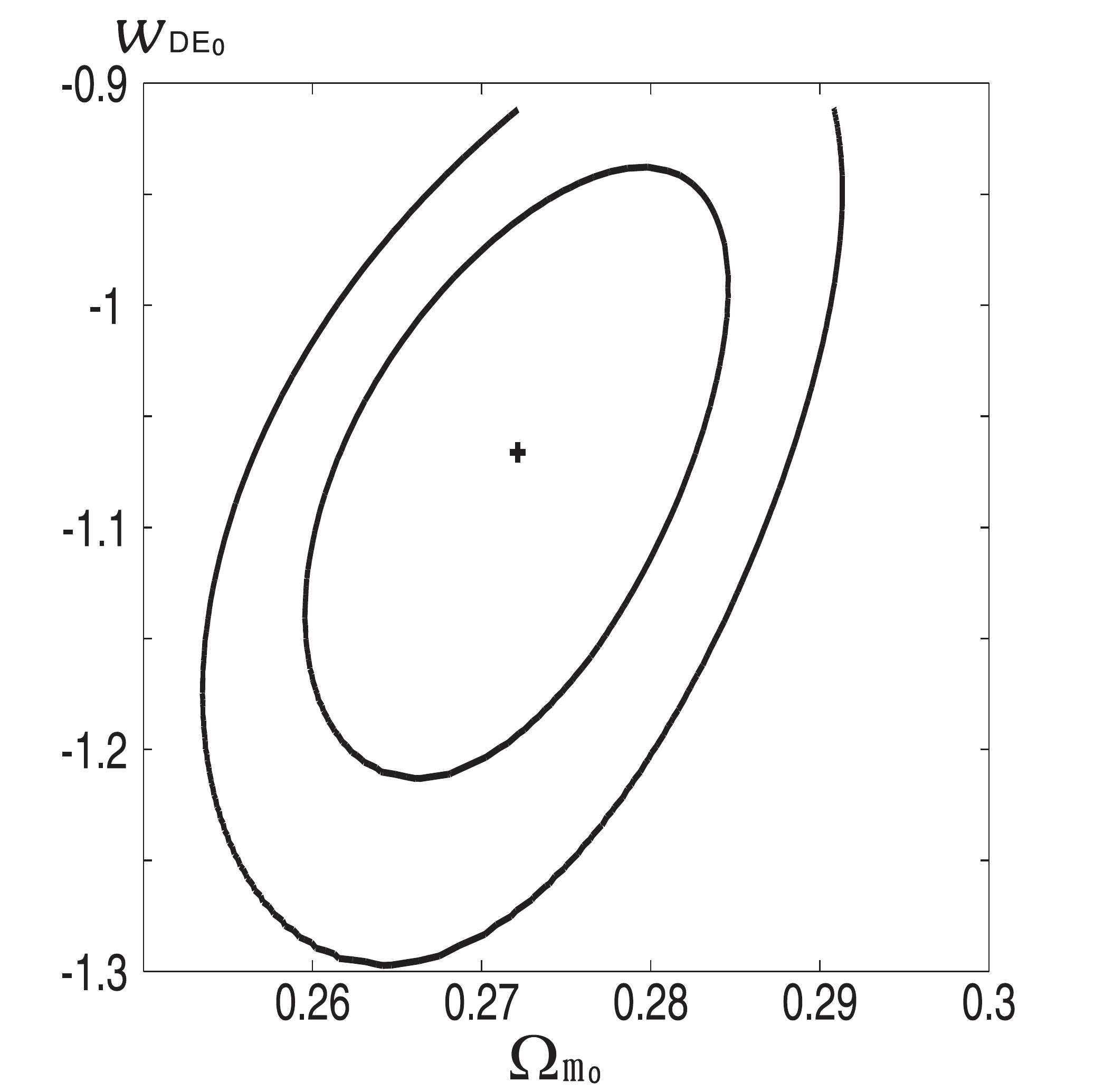}
\caption{{\it{Legend is the same as Fig.~\ref{phi41} but with
the exponential potential $V(\phi)=V_0 e^{-\sqrt{8\pi G}\lambda\phi}$.  From
 \cite{Geng:2011ka}.
}} }
\label{exp1teleDE}
\end{center}
\end{figure}

In Fig. \ref{exp1teleDE} we present the likelihood contours for $w_{DE_0}$  and
$\Omega_{m0}$, for the teleparallel dark energy scenario under the
exponential potential.
As we observe, this scenario  is consistent with observations, and it can describe both
the quintessence and phantom regimes, with the phantom regime favored by the data.
Furthermore, in Fig. \ref{exp2teleDE} we present the likelihood contours for $w_{DE_0}$
and $\xi$, for the exponential potential. From this graph we deduce that a non-minimal
coupling is favored by the data, and we observe that $w_{DE_0}$-values close to the
cosmological constant bound, either above or below it, can be induced by a relative
large $\xi$-interval, which is an advantage of this scenario.
\begin{figure}[ht]
\begin{center}
\includegraphics[width=6.5cm]{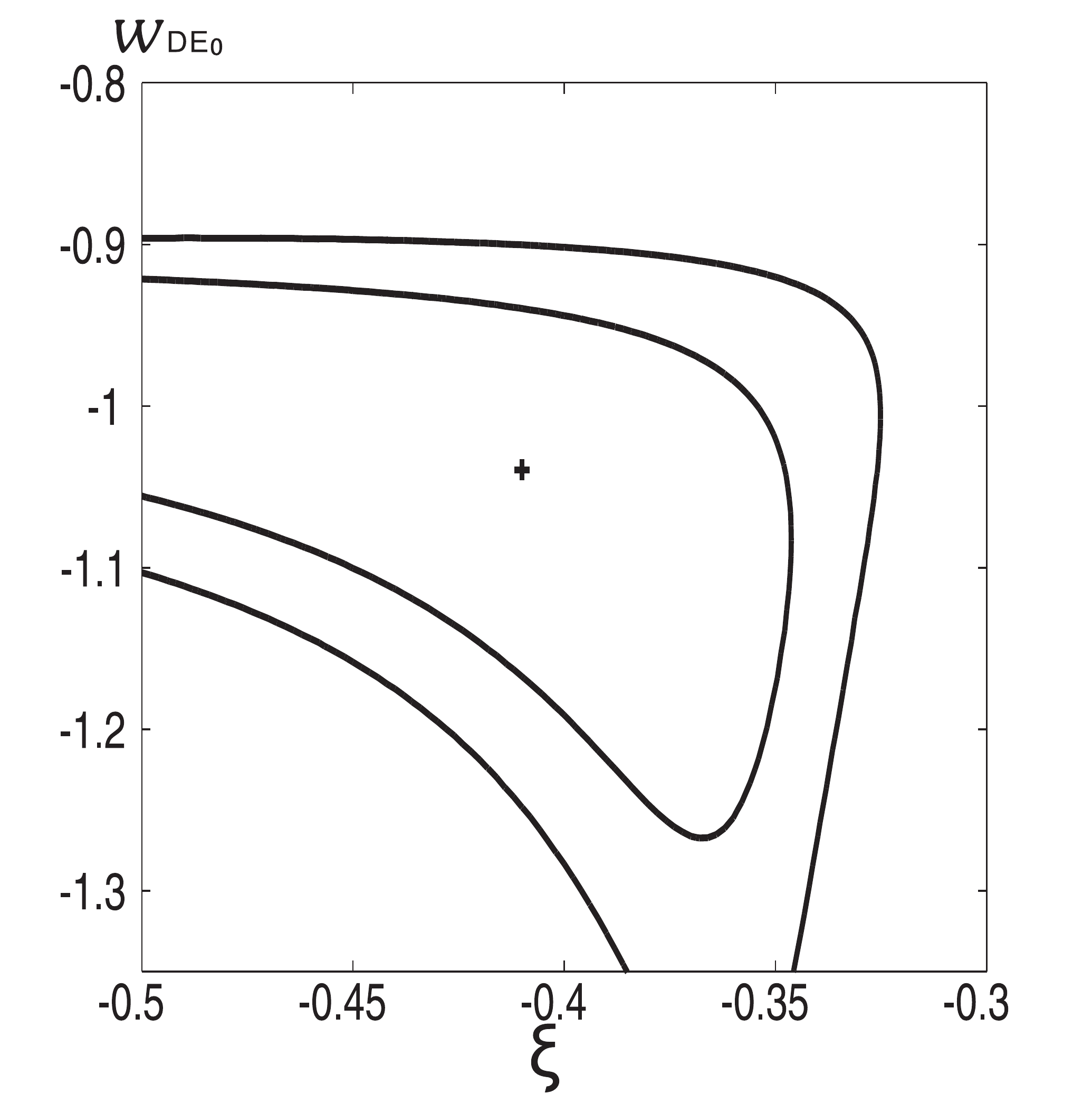}
\caption{{\it{
Legend is the same as Fig.~\ref{phi42} but with
the exponential potential $V(\phi)=V_0 e^{-\sqrt{8\pi G}\lambda\phi}$. From
 \cite{Geng:2011ka}.
}} }
\label{exp2teleDE}
\end{center}
\end{figure}

We close this section with a comment on the positive values of the non-minimal coupling
$\xi$. In particular, $\xi$ must always be bounded according to a physical constraint,
namely it must lead to positive $\rho_{DE}$ in (\ref{teleDErho}) and to positive
 $H^2$ in the Friedmann equation. In practice, $\xi$ is found to be mainly negative (in
our convention), and only a small window of positive values is theoretically
allowed. In the case of quartic and exponential potentials one finds the interesting
result that for the theoretically allowed positive $\xi$ ($0\leq\xi\lesssim 0.2$)
$w_{DE}$ is always very close to a constant $w_{DE_0}$ with
$|w_{DE_0}-w_{DE}|\lesssim 10^{-3}$  \cite{Geng:2011ka}. The reason is that the scenario
of teleparallel dark energy for positive $\xi$ (sufficiently small in order for the
positivity of $\rho_{DE}$ and $H^2$ not to be spoiled)  always results in the
stabilization of $w_{DE_0}$ close to the cosmological constant value, as was proven by
the detailed phase-space analysis presented above and in more details in
\cite{Xu:2012jf}. Such a behavior is an advantage from both observational and theoretical
point of view.

In summary, the scenario of teleparallel dark energy is compatible with observations, for
all the examined scalar-field potentials. Furthermore, although the scalar field is
canonical, the model can describe both the quintessence and phantom regimes. These
features are an advantage from both observational and theoretical point of view, and they
make the scenario at hand a good candidate for the description of nature. Finally, the
data favor a non-minimal coupling, and thus the model is distinguishable from standard
quintessence, since the two scenarios are equivalent only for the minimal coupling.

\subsection{$f(T,T_G)$ gravity}
\label{TTGgravity}

In this subsection we describe another extension of teleparallel and $f(T)$ gravity,
based on the Teleparallel Equivalent of the Gauss-Bonnet combination, following
\cite{Kofinas:2014owa,Kofinas:2014aka,Kofinas:2014daa} (see also \cite{Gonzalez:2015sha}
for the extension to the teleparallel equivalent of Lovelock gravity). The inspiration
for
this extension comes from the fact that in curvature gravity, apart from the simple
modifications such as the $f(R)$ models, one can construct more complicated actions
introducing higher-curvature corrections such as the Gauss-Bonnet combination $G$
\cite{Wheeler:1985nh,Antoniadis:1993jc} or arbitrary functions $f(G)$
\cite{Nojiri:2005jg,DeFelice:2008wz,Davis:2007ta}. Hence, one can follow the same
direction starting from the teleparallel formulation of gravity, and construct actions
involving higher-torsion corrections (see
\cite{MuellerHoissen:1983vc,Mardones:1990qc,Chandia:1997hu} for different constructions
of torsional actions).

As we mentioned in Section \ref{SecionTeleparallelgrav}, the basic strategy for the
construction of
Teleparallel Equivalent of GR is to express the curvature scalar
$R$ corresponding to a general connection,  as the curvature scalar $\bar{R}$
corresponding to Levi-Civita connection plus terms arising from the torsion
tensor. Then, by imposing the teleparallelism condition $R^{a}_{\,\,\,bcd}=0$,
we acquire that $\bar{R}$ is equal to a torsion scalar plus a total derivative,
namely the relation:
\begin{equation}
e\bar{R}=-eT+2(eT_{\nu}^{\,\,\,\nu\mu})_{,\mu}\,.
\label{TGricciscalar}
\end{equation}
Thus, this torsion scalar provides the Teleparallel Equivalent of GR, in
a sense that if one uses it as a Lagrangian, then exactly the same equations with
GR are obtained.

Hence, inspired by this strategy, in  \cite{Kofinas:2014owa} the authors followed the same
steps for the Gauss-Bonnet combination, which as it is well known reads as:
\begin{equation}
G=R^{2}-4R_{\mu\nu}R^{\mu\nu}+R_{\mu\nu\kappa\lambda}R^{\mu\nu\kappa\lambda}.
\label{TGGBdef}
\end{equation}
In particular, the idea is to extract a relation of the form
\begin{equation}
e\bar{G}
\!=\!eT_{G}\!+\!\text{total diverg.},
\label{TGTGbasic}
\end{equation}
where $\bar{G}$ is the Gauss-Bonnet combination calculated using the Levi-Civita
connection. Thus, from this expression one can read the $T_G$, i.e. the Teleparallel
Equivalent of the Gauss-Bonnet combination.

Indeed, after formulating the problem in the form language instead of the coordinate one,
one can extract relation (\ref{TGTGbasic}), with $T_G$ written as
\begin{eqnarray}
&&\!\!\!\!\!\!\!\!\!
T_G=(\mathcal{K}^{A_{1}}_{\,\,\,\,EA}\mathcal{K}^{EA_{2}}_{\,\,\,\,\,\,\,\,B}
\mathcal{K}^{A_{3}}_{\,\,\,\,FC}\mathcal{K}^{fA_{4}}_{\,\,\,\,\,\,\,D}
-2\mathcal{K}^{A_{1}\!A_{2}}_{\,\,\,\,\,\,\,\,\,\,\,A}\mathcal{K}^{A_{3}}_{
\,\,\,\,\,EB}\mathcal{K}^{E}_{\,\,FC}\mathcal{K}^{FA_{4}}_{\,\,\,\,\,\,\,\,\,D}
\nn\\
&& \ \ \ \
\,+2\mathcal{K}^{A_{1}\!A_{2}}_{\,\,\,\,\,\,\,\,\,\,\,A}\mathcal{K}^{A_{3}}_{
\,\,\,\,\,EB}\mathcal{K}^{EA_{4}}_{\,\,\,\,\,\,\,\,F}\mathcal{K}^{F}_{\,\,\,CD}
\nn\\
&& \ \ \ \
\,+2\mathcal{K}^{A_{1}\!A_{2}}_{\,\,\,\,\,\,\,\,\,\,\,A}\mathcal{K}^{A_{3}}_{
\,\,\,\,\,EB}\mathcal{K}^{EA_{4}}_{\,\,\,\,\,\,\,\,\,C,D})\delta^{\,A\,B\,C\,D}_{
A_{1}A_{2}A_{3}A_{4}}\,,
\label{TGTG}
\end{eqnarray}
where comma denotes differentiation. In this expression the generalized $\delta$
is the determinant of the Kronecker deltas, whereas $\mathcal{K}_{ABC}$ is the
contorsion tensor
\begin{equation}
\mathcal{K}_{ABC}=\frac{1}{2}(T_{CAB}-T_{BCA}-T_{ABC}
)=-\mathcal{K}_{BAC},
\end{equation}
with $T_{ABC}$ the torsion tensor (the conventions are the standard ones of this Review,
namely Greek indices span the space-time while capital Latin  indices span the tangent
space, assumed to be a Minkowski space with metric $\eta_{AB}=\mbox{diag}(+1,-1,-1,-1)$).
We refer the reader to \cite{Kofinas:2014owa} for the
lengthy calculations of this construction.
Thus, $T_G$ is the teleparallel equivalent of $\bar{G}$, in the sense
that the action
\begin{equation}
S_{T_G}
=\frac{1}{2\kappa_{D}^{2}}\int_{M}\!\!d^{D}\!x\,
e\,T_{G}\,,
\label{TGteleaction2}
\end{equation}
varied in terms of the vielbein gives exactly the same equations with the
action
\begin{equation}
S_{GB}
=\frac{1}{2\kappa_{D}^{2}}\int_{M}\!\!d^{D}\!x\,
e\,\bar{G}\,,
\label{TGGBaction2}
\end{equation}
varied in terms of the metric, and this equivalence holds for arbitrary dimensions
($\kappa_{D}^{2}$ is the $D$-dimensional gravitational constant).

Having constructed the teleparallel equivalent of the Gauss-Bonnet combination $T_G$, and
having also the usual teleparallel equivalent of the Ricci scalar $T$, one can be based
on
them in order to build modified gravitational theories. In particular, one can combine
both possible extensions and construct the $f(T,T_G)$ modified gravity
\cite{Kofinas:2014owa}
\begin{equation}
S=\frac{1}{2\kappa_{D}^{2}}\!\int d^{D}\!x\,e\,f(T,T_G)\,.
\label{TGFTTGgravity}
\end{equation}
Note that since $T_G$ is quartic in torsion tensor then $f(T_G)$ cannot arise from any
$f(T)$. Hence, clearly, $f(T,T_G)$ gravity is different from both $f(T)$ theory as well
as
from $f(R,G)$ gravity  \cite{Nojiri:2005jg,DeFelice:2008wz,Davis:2007ta}, and therefore it
is a novel gravitational modification.

Let us now extract the equations of motion, focusing in the most interesting case of
$D=4$. Varying the action (\ref{TGFTTGgravity}) in terms of the vierbein, after various
steps, one finally obtains  \cite{Kofinas:2014owa}
\begin{eqnarray}
&&\!\!\!\!\!\!\!\!\!\!\!\!\!\!\!\!\!\!\!\!\!
2(H^{[AC]B}\!+\!H^{[BA]C}\!-\!H^{[CB]A})_{,C}\nonumber\\
&&\!\!\!\!\!\!\!\!\!\!\!\!\!\!\!\!\!\!\!\!\!
+2(H^{[AC]B}\!+\!H^{[BA]C}
\!-\!H^{[CB]A})J^{D}_{\,\,\,DC}\nn\\
&&\!\!\!\!\!\!\!\!\!\!\!\!\!\!\!\!\!\!\!\!\!
+(2H^{[AC]D}\!+\!H^{DCA})J^{B}_{\,\,\,CD}
\!+\!4H^{[DB]C}J_{(DC)}^{\,\,\,\,\,\,\,\,A}\nonumber\\
&&\!\!\!\!\!\!\!\!\!\!\!\!\!\!\!\!\!\!\!\!\!
+T^{A}_{\,\,\,CD}H^{CDB}-h^{AB}
+(f\!-\!Tf_{T}\!-\!T_{G}f_{T_{G}})\eta^{AB}=0\,,
\label{TGgenequations}
\end{eqnarray}
where  $J^{C}_{\,\,\,AB}$ are the structure coefficients functions given by
\begin{equation}
J^{C}_{\,\,\,AB}=e_{A}^{\,\,\,\mu}
e_{B}^{\,\,\,\nu}(e^{C}_{\,\,\,\mu,\nu}-e^{C}_{\,\,\,\nu,\mu})
\label{structurefunKofinas}\,,
\end{equation}
and where we have defined
\begin{eqnarray}
&& \!\!\!\!\!\!\!
H^{ABC}=f_{T}(\eta^{AC}\mathcal{K}^{BD}_{\,\,\,\,\,\,D}-\mathcal{K}^{BCA})+f_{T_{G}}\Big[
\nn\\
&& \ \ \ \ \ \ \
\epsilon^{CPRT}\big(2\epsilon^{A}_{\,\,\,DKM}\mathcal{K}^
{BK}_{\,\,\,\,\,\,\,P}
\mathcal{K}^{D}_{\,\,\,QR}
+\epsilon_{QDKM}\mathcal{K}^{AK}_{\,\,\,\,\,\,P}\mathcal{K}^{BD}_{\,\,\,\,\,\,R}
\nonumber\\
&& \ \ \ \ \ \ \ \ \ \ \ \ \ \ \ \ \
+
\epsilon^{AB}_{\,\,\,\,\,\,KM}\mathcal{K}^{K}_{\,\,\,DP}\mathcal{K}^{D}_{\,\,
\,QR}\big)
\mathcal{K}^{QM}_{\,\,\,\,\,\,\,T}\nn\\
&& \ \ \ \ \ \ \
+\epsilon^{CPRT}\epsilon^{AB}_{\,\,\,\,\,\,KD}\mathcal{K}^{MD}_{\,\,\,\,\,\,P
}
\big(\mathcal{K}^{K}_{\,\,MR,T}-\frac{1}{2}\mathcal{K}^{K}_{\,\,MQ}J^{Q}_
{\,\,\,TR}\big)\nn\\
&& \ \ \ \ \ \ \
+\epsilon^{CPRT}\epsilon^{AK}_{\,\,\,\,\,\,\,DM}\mathcal{K}^{DM}_{\,\,\,\,\,\,P}
\big(\mathcal{K}^{B}_{\,\,KR,T}-\frac{1}{2}\mathcal{K}^{B}_{\,\,KQ}J^{Q}_
{\,\,\,TR}\big)\Big]
\nn\\
&& \ \ \ \ \ \ \ +\epsilon^{CPRT}\epsilon^{A}_{\,\,\,KDM}\Big[
\big(f_{T_{G}}\mathcal{K}^{BK}_{\,\,\,\,\,\,P}
\mathcal{K}^{DM}_{\,\,\,\,\,\,R}\big)_{,T}\nonumber\\
&& \ \ \ \ \ \ \  \ \ \ \ \ \ \  \ \ \ \ \ \ \  \ \ \ \ \ \ \ \,
+
f_{T_{G}}J^{Q}_{\,\,\,PT}\mathcal{K}^{BK}_{\,\,\,\,\,\,[Q}\mathcal{K}^{DM}_{\,\,\,\,\,\,R]
}
\Big]\label{TGHabc22}
\end{eqnarray}
and
\begin{equation}
h^{AB}=f_{T}\,\epsilon^{A}_{\,\,\,KCD}\epsilon^{BPQD}\mathcal{K}^{K}_{\,\,\,MP}
\mathcal{K}^{MC}_{\,\,\,\,\,\,\,\,Q}\,.
\label{TGhab22}
\end{equation}
We have used the notation $f_{T}=\partial f/\partial T$, $f_{T_{G}}=\partial f/\partial
T_{G}$,
the (anti)symmetrization symbol contains the factor $1/2$, while the antisymmetric symbol
$\epsilon_{ABCD}$
has $\epsilon_{1234}=1$, $\epsilon^{1234}=-1$.

Before proceeding to the cosmological application of $f(T,T_G)$ gravity, let us make some
comments. The first has to do with the Lorentz violation. In particular, as it was
discussed also in  \cite{Kofinas:2014owa}, under the use of the Weitzenb{\"{o}}ck
connection the torsion scalar $T$ remains diffeomorphism invariant, however the Lorentz
invariance has been lost since we have chosen a specific class of frames, namely the
autoparallel orthonormal frames. Nevertheless, the equations of motion of the Lagrangian
$eT$, being the Einstein equations, are still Lorentz covariant. On the contrary, when we
replace  $T$  by a general function $f(T)$ in the action, the new equations of motion
will not be covariant under Lorentz rotations of the vielbein, although they will indeed
be form-invariant, and the same features appear in the $f(T,T_G)$ extension. However,
this is not a deficit (it is a sort of analogue of gauge fixing in gauge theories), and
the theory, although not Lorentz covariant, is meaningful. Definitely, not all vielbeins
will be solutions of the equations of motion, but those which solve the equations will
determine the metric uniquely. Obviously, the above problem will be solved if one
formulates $f(T,T_G)$ gravity in a covariant way, i.e using both the vierbein and the
spin connection, as we did in subsection \ref{restoringLI} for $f(T)$ gravity. Such a
construction is still missing.

The second comment is related to possible acausalities  and problems with the Cauchy
development of a constant-time hypersurface. Indeed, there are works claiming that a
departure from TEGR, as for instance in $f(T)$ gravity, with the subsequent local Lorentz
violation, will lead to the above problems  \cite{Ong:2013qja,Izumi:2013dca}. In order to
examine whether one also has these problems in the present scenario of $f(T,T_G)$ gravity,
he would need to perform a very complicated analysis, extending the characteristics
method of \cite{Ong:2013qja,Izumi:2013dca} for this case, although at first sight one
does expect to indeed find them. Nevertheless, even if this proves to be the case, it
does
not mean that the theory has to be ruled out, since one could still handle $f(T)$ gravity
(and similarly $f(T,T_G)$ one) as an effective theory, in the regime of validity of
which the extra degrees of freedom can be removed or be excited in a healthy way
(alternatively one could re-formulate the theory using Lagrange multipliers)
\cite{Ong:2013qja,Izumi:2013dca}. However, there is a possibility that these problems
might be related to the restricting use of the  Weitzenb{\"{o}}ck connection, since the
formulation of TEGR and its modifications using other connections (still in the
``teleparallel class'') does not seem to be problematic, and thus, the general
formulation of $f(T,T_G)$ gravity that was presented in  \cite{Kofinas:2014owa} might be
free of the above disadvantages. These issues definitely need further investigation, and
the discussion is still open in the literature.

Let us now apply $f(T,T_G)$ gravity in a cosmological framework. Firstly, we add the
matter sector along the gravitational one, that is we start by the total action
\begin{eqnarray}
S_{tot} =\frac{1}{16 \pi G}\!\int d^{4}\!x\,e\,f(T,T_G)\,+S_m\,,
\label{TGfGBtelaction}
\end{eqnarray}
with $8\pi G=\kappa_4$ the four-dimensional Newton's constant, and where $S_m$
corresponds to a matter fluid of energy density $\rho_m$ and pressure $p_m$.
Secondly, in order to investigate the cosmological implications of the above
action, we consider as usual the spatially flat FRW geometry (\ref{weproudlyuse}), namely
\begin{equation}
\label{TGweproudlyuse2}
 e_{\mu}^A=\mathrm{diag}(1,a,a,a) ~,
\end{equation}
where $a(t)$ is the scale factor. Inserting this vierbein choice into (\ref{TGTG}) one
can easily find that
\begin{eqnarray}
\label{TGTGcosmo}
T_G=24\frac{\dot{a}^2}{a^2}
\,\frac{\ddot{a}}{a}=24H^2\big(\dot{H}+H^2\big),
\end{eqnarray}
while as usual $T=-6H^2$ (note that in the present Review we use the usual conventions,
which are slightly different than those of  \cite{Kofinas:2014owa}).
  Additionally, inserting
(\ref{TGweproudlyuse2}) into the general equations of motion
(\ref{TGgenequations}), after some algebra one obtains the Friedmann
equations  \cite{Kofinas:2014owa}
\begin{equation}
f+12H^{2}f_{T}-T_{G}f_{T_{G}}+24H^{3}\dot{f_{T_{G}}}=16 \pi G\rho_m
\label{TGeqmN}
\end{equation}
\begin{eqnarray}
&&\!\!\!\!\!\!\!\!\!\!\!\!\!f+4(\dot{H}+3H^{2})f_{T}+4H\dot{f_{T}}\nn\\
&& \!\!\!\!\!\!\!\!\!- T_{G}f_{T_{G}}+\frac{2}{3H}T_{G}
\dot{f_{T_{G}}}+8H^{2}\ddot{f_{T_{G}}}=-16 \pi G p_m\,,
\label{TGeqma}
\end{eqnarray}
where the right hand sides arise from the independent variation of the
matter action. In the above expressions
it is $\dot{f_{T}}=f_{TT}\dot{T}+f_{TT_{G}}\dot{T}_{G}$,
$\dot{f_{T_{G}}}=f_{TT_{G}}\dot{T}+f_{T_{G}T_{G}}\dot{T}_{G}$,
$\ddot{f_{T_{G}}}=f_{TTT_{G}}\dot{T}^{2}+2f_{TT_{G}T_{G}}\dot{T}
\dot{T}_{G}+f_{T_{G}T_{G}T_{G}}\dot{T}_{G}^{\,\,2}+
f_{TT_{G}}\ddot{T}+f_{T_{G}T_{G}}\ddot{T}_{G}$,
with $f_{TT}$, $f_{TT_{G}}$,\,... denoting multiple partial differentiations
of $f$ with respect to $T$, $T_{G}$.
Finally, $\dot{T}$, $\ddot{T}$ and $\dot{T}_{G}$, $\ddot{T}_{G}$ are obtained
by differentiating $T=-6H^2$ and (\ref{TGTGcosmo}) respectively with
respect to time.

We would like to mention that for the derivation of the above Friedmann equations we
presented the robust way following  \cite{Kofinas:2014owa}, that is one first
performs the general variation of the action resulting to the general
equations of motion (\ref{TGgenequations}), and then he inserts the cosmological ansatz
(\ref{TGweproudlyuse2}), obtaining (\ref{TGeqmN}) and (\ref{TGeqma}). However, as the
same authors showed in  \cite{Kofinas:2014daa}, one can obtain exactly the same result in
a faster way, that is following the shortcut procedure where one first inserts the
cosmological ansatz in the action and then performs variation. This shortcut method is a
sort of minisuperspace procedure since the (potential) additional degrees of freedom
other than those contained in the scale factor are frozen. The equivalence of the two
methods is not always guaranteed, since variation and ansatz-insertion do not commute in
general, especially in theories with higher-order derivatives
\cite{Deser:2004yh,Weinberg:2008zzc}. However, the equivalence of the two methods is
indeed the case for the scenario at hand   \cite{Kofinas:2014daa}, and an additional
cross-check for the calculation correctness.

The Friedmann equations (\ref{TGeqmN}), (\ref{TGeqma}) can be rewritten in the usual form
\begin{eqnarray}
\label{TGFr1a}
H^{2}&=&\frac{8\pi G}{3}(\rho_m+\rho_{DE})\\
\label{TGFr2s}
\dot{H}&=&-4\pi G(\rho_m+p_m+\rho_{DE}+p_{DE}),
\end{eqnarray}
where the energy density and pressure of the effective dark energy sector are
defined as
\begin{eqnarray}
&&\!\!\!\!\!\!\!\!\!\!\!\!\!\!\!\!\!\!\!\!\!\!\!
\rho_{DE}\!=\!\frac{1}{16\pi G}(6H^{2}\!-\!f\!-\!12H^{2}f_{T}\!+\!T_{G}f_
{T_{G}}
\!-\!24H^{3}\dot{f_{T_{G}}})
\label{TGrhode}\\
&&\!\!\!\!\!\!\!\!\!\!\!\!\!\!\!\!\!\!\!\!\!\!\!
p_{DE}\!=\!\frac{1}{16 \pi G}\Big[\!\!-\!2(2\dot{H}\!+\!3H^{2}
)\!+\!f\!+\!4(\dot{H}\!+\!3H^{2})f_{T}\nn\\
&& \ \ \ \ \
+4H\dot{f_{T}}\!-\!T_{G}f_{T_{G}}\!+\!\frac{2}{3H}T_{G}\dot{f_{T_{G}}}
\!+\!8H^{2}\ddot{f_{T_{G}}}\!\Big].
\label{TGpde}
\end{eqnarray}
Since the standard matter is conserved independently,
namely $\dot{\rho}_m+3H(\rho_m+p_m)=0$,
we obtain from (\ref{TGrhode}),(\ref{TGpde}) that the dark energy density and
pressure also satisfy the usual evolution equation
\begin{eqnarray}
\dot{\rho}_{DE}+3H(\rho_{DE}+p_{DE})=0\,.
\end{eqnarray}
Additionally, we can define the dark energy equation-of-state parameter as
\begin{eqnarray}
w_{DE}= \frac{p_{DE}}{\rho_{DE}}\,.
\end{eqnarray}

Finally, we mention that for $f(T,T_G)=T$ the Friedmann equations (\ref{TGeqmN}),
(\ref{TGeqma}) become the usual equations of Teleparallel Equivalent of General
Relativity (and thus of GR), while for $f(T,T_G)=f(T)$ one re-obtains
the equations of standard $f(T)$ gravity.

\subsubsection{Cosmological behavior}

Let us now investigate in detail the cosmological applications of  $f(T,T_G)$ gravity,
both at early and late times  following
\cite{Kofinas:2014owa,Kofinas:2014aka,Kofinas:2014daa} (see also
\cite{Jawad:2014vwa,Jawad:2015wea,Waheed:2015bga}). In particular, we proceed to the
investigation of some specific $f(T,T_{G})$ cases, focusing on the evolution of
observables such as the various density parameters $\Omega_i=8\pi G\rho_i/(3H^2)$ and the
dark energy equation-of-state parameter $w_{DE}$.

Since $T_G$ contains quartic torsion terms, it will in general and
approximately be of the same order with $T^2$. Therefore, $T$ and
$\sqrt{T^{2}+\beta_{2}T_{G}}$
are of the same order, and thus, if one of them contributes during the
evolution the other will contribute too. Therefore, it would be very
interesting to consider modifications of the form
$f(T,T_{G})=-T+\beta_{1}\sqrt{T^{2}+\beta_{2}T_{G}}$, which are expected to
play an important role at late times. Note that the couplings $\beta_{1},\beta_{2}$ are
dimensionless, and therefore no new mass scale enters at late times.
Nevertheless, in order to describe the early-times cosmology, one should
additionally include higher order corrections like
$T^{2}$. Since the scalar $T_{G}$ is of the same order with $T^{2}$, it
should be also included. However, since $T_{G}$ is topological
in four dimensions it cannot be included as it is, and therefore we use the
term $T\sqrt{|T_{G}|}$ which is also of the same order
with $T^{2}$ and non-trivial. Thus, the total function $f$ is taken to
be
\begin{equation}
f(T,T_{G})=-T+\beta_{1}\sqrt{T^{2}+\beta_{2}T_{G}}+
\alpha_{1} T^{2}+\alpha_{2}T\sqrt{|T_{G}|}\,.
\label{TGnhs}
\end{equation}
In summary, when the above function is used as an action, it gives rise
to a gravitational theory that can describe both inflation and late-times
acceleration in a unified way.
\begin{figure}[ht]
\includegraphics[scale=0.48]{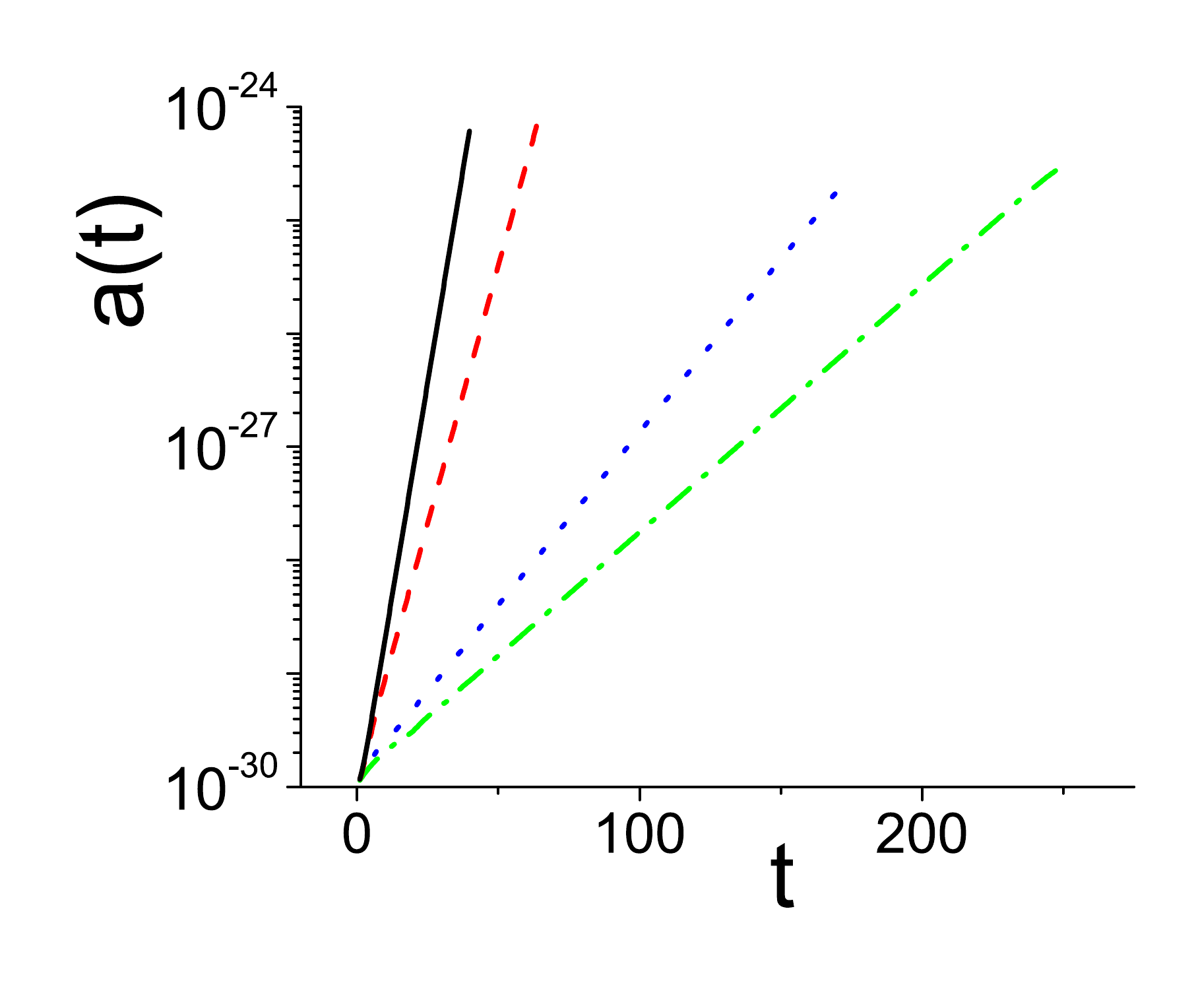}
\caption{
{\it{Four inflationary solutions for the   ansatz $f(T,T_{G})=\alpha_{1}
T^{2}+\alpha_{2}T\sqrt{|T_{G}|}-T+\beta_{1}\sqrt{T^{2}+\beta_{2}T_{G}}$,
corresponding to
a)  $\alpha_1=-2.8$, $\alpha_2=8$, $\beta_1=0.001$,  $\beta_2=1$
(black-solid),
b) $\alpha_1=-2$, $\alpha_2=8$, $\beta_1=0.001$, $\beta_2=1$ (red-dashed),
c) $\alpha_1=8$, $\alpha_2=8$, $\beta_1=0.001$, $\beta_2=1$ (blue-dotted),
d) $\alpha_1=20$, $\alpha_2=5$, $\beta_1=0.001$, $\beta_2=1$
(green-dashed-dotted).
All parameters are in Planck units. From  \cite{Kofinas:2014daa}.}} }
\label{TGInflation}
\end{figure}

In order to examine the cosmological evolution of a universe governed by the
above unified action, we perform a numerical elaboration of the Friedmann
equations (\ref{TGFr1a}), (\ref{TGFr2s}), with $\rho_{DE}$, $p_{DE}$ given by
relations (\ref{TGrhode}), (\ref{TGpde}), under
the ansatz (\ref{TGnhs}). In Fig. \ref{TGInflation} we present the
early-times, inflationary solutions for four parameter choices.
As we observe, inflationary, de-Sitter exponential expansions can be easily
obtained (with the exponent of the expansion determined by the model
parameters), although there is not an explicit cosmological constant term in
the action, which is an advantage of the scenario. This was expected, since one
can easily extract analytical solutions of the Friedmann equations
(\ref{TGFr1a}), (\ref{TGFr2s}) with
$H\approx$ const (in which case $T$ and $T_G$ as also constants).

Let us now focus on the late-times evolution. In Fig. \ref{TGDarkenergy}
we depict the evolution of the matter and effective dark energy density
parameters, as well as the behavior of the dark energy equation-of-state
parameter, for a specific choice of the model parameters. As we see, we can obtain the
observed behavior, where $\Omega_m$ decreases,
resulting to its current value of $\Omega_{m0}\approx0.3$, while
$\Omega_{DE}=1-\Omega_m$ increases. Concerning $w_{DE}$, we can see that in
this example it lies in the quintessence regime.
Finally, note that, as it is usual in modified gravity \cite{Nojiri:2013ru}, the model at
hand can describe the phantom regime too, for a region of the parameter space, which is
an
additional advantage \cite{Kofinas:2014daa,Chattopadhyay:2014xaa}.
\begin{figure}[ht]
\includegraphics[scale=0.48]{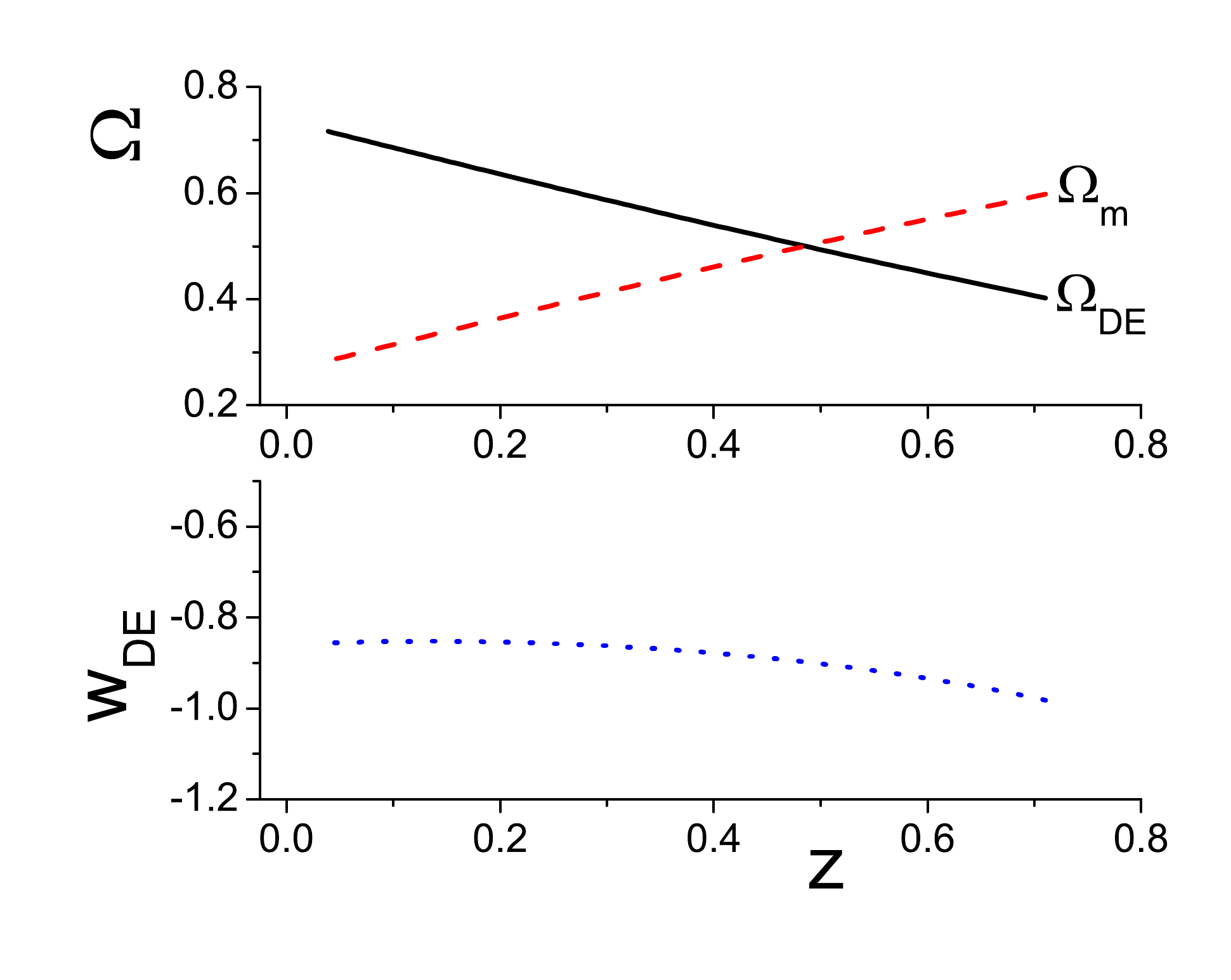}
\caption{ {\it{Upper graph: The evolution of the dark energy density
parameter $\Omega_{DE}$ (black-solid)  and the matter density
parameter $\Omega_{m}$ (red-dashed), as a function of the redshift $z$,
for the ansatz $f(T,T_{G})=\alpha_{1}
T^{2}+\alpha_{2}T\sqrt{|T_{G}|}-T+\beta_{1}\sqrt{T^{2}+\beta_{2}T_{G}}$
with $\alpha_1=0.001$, $\alpha_2=0.001$, $\beta_1=2.5$,
$\beta_2=1.5$.  Lower graph: The evolution of the corresponding dark
energy equation-of-state parameter $w_{DE}$. All parameters are in
units where the present Hubble parameter is $H_0=1$, and we have
imposed $\Omega_{m0}\approx0.3$, $\Omega_{DE0}\approx0.7$  at
present. From  \cite{Kofinas:2014daa}.}}}
\label{TGDarkenergy}
\end{figure}

One could proceed to the investigation of other $f(T,T_{G})$ ansatzen, going beyond the
simple model (\ref{TGnhs}). For instance
since  $T_G$ contains quartic torsion terms, it will
in general and approximately  be of the same order with $T^2$. Therefore,  it
would be interesting to consider modifications of the form
$f(T,T_G)=-T+F(T^2+\beta_{2}T_G)$   \cite{Kofinas:2014daa}. The involved building block is
an extension of the simple $T$, and thus, it can significantly  improve the
detailed cosmological behavior of a suitable reconstructed $f(T)$. Indeed, choices like
 $f(T,T_{G})=-T+\beta_1(T^2+\beta_{2}T_G)+\beta_3(T^2+\beta_{4}T_G)^2$, can lead to a
description of inflation and late-time acceleration in a unified way
 \cite{Kofinas:2014daa}.

We close this paragraph with an important point. In the above analysis we showed that
$f(T,T_G)$ cosmology can be very efficient in describing the evolution of the universe at
the background level. However, before considering any cosmological model as a candidate
for the description of nature it is necessary to perform a detailed investigation of its
perturbations, namely to examine whether the obtained solutions are stable. Furthermore,
especially in theories with local Lorentz invariance violation, new degrees of freedom
are introduced, the behavior of which is not guaranteed that is stable (for instance this
is the case in the initial version of Ho\v{r}ava-Lifshitz gravity  \cite{Bogdanos:2009uj},
in the initial version of de Rham-Gabadadze-Tolley massive gravity
 \cite{DeFelice:2012mx},
etc), and this makes the perturbation analysis of such theories even more imperative.
Thus, such a detailed and complete analysis of the cosmological perturbations of
$f(T,T_G)$ gravity is necessary. However, for the moment we would like to mention that in
the case of simple $f(T)$ gravity, the perturbations of which have been examined in
detail  \cite{Chen:2010va,Li:2011wu}, one does obtain instabilities, but there are many
classes of $f(T)$ ansatzen and/or parameter-space regions, where the perturbations are
well-behaved. This is a good indication that we could  expect to find a similar behavior
in $f(T,T_G)$ gravity too, although one needs to indeed verify this under the detailed
perturbation analysis.

\subsubsection{Dynamical analysis}

The interesting cosmological behavior of $f(T,T_G)$ gravity makes it necessary to
perform a detailed dynamical and stability analysis, examining in a systematic way the
possible cosmological behaviors, focusing on the late-time stable solutions. In
particular, as an example, following  \cite{Kofinas:2014aka} we will study the case where
\begin{equation}
f(T,T_G)=-T+\alpha_1\sqrt{T^2+\alpha_2 T_G}\,,
\label{TGansantzdynanal}
\end{equation}
 where the couplings $\alpha_{1},\alpha_{2}$ are dimensionless and the model is
expected to play an important role at late times. Indeed, this
model, although simple, can lead to interesting cosmological behavior,
revealing the advantages, the capabilities, and the new features of
$f(T,T_{G})$ cosmology. We mention here that when $\alpha_2=0$ this scenario
reduces to TEGR, that is to GR, with just a rescaled
Newton's constant, whose dynamical analysis has been performed in detail in
the literature  \cite{Copeland:1997et,Ferreira:1997au,Chen:2008ft}. Thus, in
the following we restrict our analysis to the case $\alpha_2\neq0$.

For the ansatz (\ref{TGansantzdynanal}), the effective energy density (\ref{TGrhode}) and
pressure (\ref{TGpde}), respectively become   \cite{Kofinas:2014aka}
\begin{eqnarray}
\label{TGdrhodeb}
 &&\!\!\!\!\!\!\!\!\!\!\!\!\!\!\!\!\!\!\!\!\!\!\!\!\!\!\!
 8\pi G \rho_{DE}=
 \frac{\sqrt{3} \alpha_1 H^2 \left\{\alpha_2^2 \ddot H+9
\alpha_2 H \dot H \right\}}{D^{3/2}}\nonumber\\
&&\ \
+
 \frac{\sqrt{3} \alpha_1 H^5  \big[(3-2 \alpha_2) \alpha_2+9\big]
}{D^{3/2}},
   \end{eqnarray}
   and
\begin{eqnarray}
\label{TGdpdeb}
&&\!\!\!\!\!\!
8\pi G p_{DE}
=\frac{\alpha_1    (2
\alpha_2 + 3) \big[\alpha_2 (10 \alpha_2 - 51) - 18\big] H^5\dot{H}}
{\sqrt{3} D^{5/2}}
   \nonumber\\
   &&
  + \frac{\alpha_1    \alpha_2 \big[4
   \alpha_2 (5 \alpha_2 - 21) - 90\big] H^3 \dot{H}^2  - 54 \alpha_1\alpha_2^2 H \dot H^3
  }
{\sqrt{3} D^{5/2}}
   \nonumber\\
     &&
 -
\frac{2
   \alpha_1 \alpha_2^2 \ddot H \left[2 (\alpha_2 - 3) H^2 \dot H+2 \alpha_2
\dot H^2+(6 \alpha_2 + 9) H^4\right]}{\sqrt{3} D^{5/2}}
   \nonumber\\
   &&
   + \frac{\sqrt{3}\alpha_1   (\alpha_2 - 3) (2
\alpha_2 + 3)^2 H^7 }
{ D^{5/2}}
\nonumber\\
 &&
   -\frac{\alpha_1 \alpha_{2}^{2} H \dddot H}
{\sqrt{3} D^{3/2}}
 +
 \frac{\sqrt{3} \alpha_1 \alpha_2^3 H \ddot H^2}{D^{5/2}}
\,,
\end{eqnarray}
where $D=3H^2+2\alpha_2(\dot{H}+H^2)$.

In order to transform the two Friedmann equation (\ref{TGFr1a}), (\ref{TGFr2s}) into an
autonomous dynamical system, we introduce the auxiliary variables
 \cite{Kofinas:2014aka}:
\begin{eqnarray}
\label{TGxdefin}
&&x=\sqrt{\frac{D}{3H^2}}=\sqrt{1+\frac{2 \alpha_2}{3}\Big(1+\frac{\dot
H}{H^2}\Big)}\\
&&\Omega_m=\frac{8\pi G \rho_m}{3 H^2}.
\end{eqnarray}
Thus, the cosmological system is transformed to the autonomous form
\begin{align}
&x'=-\frac{x \left[3 \alpha_1 x^2-6
(1 - \Omega_m)x+\alpha_1 (3 - 4 \alpha_2)\right]}{2 \alpha_1 \alpha_2}
\label{TGdeqx}\\
&\Omega_m'= -\frac{\Omega_m \left(3
x^2+\alpha_2+3 \alpha_2 w_m-3\right)}{\alpha_2}\,, \label{TGdeqm}
\end{align}
where primes denote differentiation with respect to $\ln a$, thus
$f'= H^{-1} \dot f$. The above dynamical system is defined in the phase
space $\left\{(x,\Omega_m)| x\in [0,\infty), \Omega_m\in
[0,\infty]\right\}$.
\begin{table*}
{\begin{tabular}
{|c| c| c| c| c| }
 \hline
Cr. P. &$x$&$\Omega_m$ &Existence &Stability\\
\hline\hline
$P_1$  &$\sqrt{1-\frac{\alpha_2}{3}}$ & $\Omega_{m1}$ &  $\frac{6}{5}<\alpha_2<3,
\alpha_1\geq -2\sqrt{\frac{3(3-\alpha_2)}{(-6+
5\alpha_2)^2}}$ or &
Stable spiral for $\alpha_2<3$ and  \\
             &&& $\alpha_2=\frac{6}{5} $ or
&$-32 \sqrt{3}
\sqrt{\frac{(3-\alpha_2)^3}{\left(71 \alpha_2^2-336
\alpha_2+288\right)^2}}<\alpha_1<0$ or  \\
             &&& $\alpha_2<\frac{6}{5}, \, \alpha_1 \leq 2
\sqrt{\frac{3(3-\alpha_2)}{(-6+ 5\alpha_2)^2}}$ & $\alpha_1<0,\alpha_2\leq
\frac{1}{71} \left(168-36 \sqrt{6}\right)\approx 1.124$.\\
&&&& Saddle otherwise
(hyperbolic cases). \\[0.2cm]
\hline
$P_2$ & $x_2$  & $0$ &
$\alpha_2<\frac{3}{4},\,0<\alpha_1\leq \sqrt{\frac{3}{3-4 \alpha_2}}$ or &
Stable node for $\alpha_2<0, \,0<\alpha_1<2 \sqrt{\frac{3(3-\alpha_2)}{(5
\alpha_2-6)^2}}$
   \\
       &&& $ \alpha_1\neq 0, \, \alpha_2=\frac{3}{4}$ or
& or $\frac{6}{5}<\alpha_2\leq
   3, \,\alpha_1<-2 \sqrt{\frac{3(3- \alpha_2)}{(5 \alpha_2-6)^2}}$
\\
             &&& $\alpha_2>\frac{3}{4}, \,\alpha_1<0$ &or
$\alpha_2>3, \,\alpha_1<0$.
\\
             &&&   &Unstable node for $0<\alpha_2<\frac{3}{4},\,
0<\alpha_1<\sqrt{\frac{3}{3-4 \alpha_2}}.$
 \\
             &&&   &Saddle otherwise (hyperbolic cases).
\\[0.2cm]
\hline
$P_3$  & $x_3$ &$0$&
$\alpha_2<\frac{3}{4},\, 0<\alpha_1\leq  \sqrt{\frac{3}{3-4 \alpha_2}}$ or &
 Stable node for
       $\alpha_1>0,\alpha_2\geq \frac{6}{5}.$
   \\
&&& $\alpha_2\geq \frac{3}{4},\,
   \alpha_1>0$ &  Unstable node for  $\alpha_2<0,\,
0<\alpha_1<\frac{\sqrt{3}}{\sqrt{3-4 \alpha_2}}.$
 \\
             &&&   &Saddle otherwise (hyperbolic cases).
\\[0.2cm]
\hline
$P_4$ & $0$ & $0$ & Always & Unstable node for $\frac{3}{4}<\alpha_2<3$.
\\
             &&&   &Saddle otherwise (hyperbolic cases).
\\[0.2cm]
\hline
\end{tabular}}
\caption[TGdcrit]{The real and physically interesting
critical points of the autonomous system  \eqref{TGdeqx}-\eqref{TGdeqm}, along with their
existence
and stability conditions. We use the notations
$\Omega_{m1}=\frac{\alpha_1 \sqrt{9-3 \alpha_2}
(6-5 \alpha_2)+6 (\alpha_2-3)}{6
   (\alpha_2-3)}$, $x_{2}=\frac{3-\sqrt{3 \alpha_1^2 (4 \alpha_2-3)+9}}{3
\alpha_1}$ and $x_{3}=\frac{3+\sqrt{3 \alpha_1^2 (4 \alpha_2-3)+9}}{3
\alpha_1}$. From  \cite{Kofinas:2014aka}.}
\label{TGdtab1}
%\\ [0.2cm]
\end{table*}
\begin{table*}
{\begin{tabular}
{|c| c|c| }
 \hline
Cr. P.       & $\nu_1$  & $\nu_2$  \\
\hline\hline
$P_1$     &
$-\frac{\sqrt{\alpha_1 \left[(336-71 \alpha_2)
\alpha_2-288\right]+32
\sqrt{3} (3-\alpha_2)^{3/2}}}{4 \sqrt{\alpha_1} \alpha_2}-\frac{3}{4}$     &
$\frac{\sqrt{\alpha_1
\left[(336-71 \alpha_2) \alpha_2-288\right]+32 \sqrt{3}
(3-\alpha_2)^{3/2}}}{4
\sqrt{\alpha_1}
   \alpha_2}-\frac{3}{4}$
\\[0.2cm]
\hline
$P_2$      &
$\frac{2 \sqrt{3 \alpha_1^2 (4 \alpha_2-3)+9}}{\alpha_1^2
\alpha_2}-\frac{6}{\alpha_1^2 \alpha_2}+\frac{6}{\alpha_2}-5$   &
$\frac{\sqrt{3 \alpha_1^2 (4 \alpha_2-3)+9}}{\alpha_1^2
\alpha_2}-\frac{3}{\alpha_1^2 \alpha_2}+\frac{3}{\alpha_2}-4$
  \\[0.2cm]
\hline
$P_3$       &
$-\frac{2 \sqrt{3} \sqrt{4 \alpha_1^2 \alpha_2-3
\alpha_1^2+3}}{\alpha_1^2 \alpha_2}-\frac{6}{\alpha_1^2
   \alpha_2}+\frac{6}{\alpha_2}-5$ &
$-\frac{\sqrt{3} \sqrt{4 \alpha_1^2 \alpha_2-3 \alpha_1^2+3}}{\alpha_1^2
   \alpha_2}-\frac{3}{\alpha_1^2 \alpha_2}+\frac{3}{\alpha_2}-4$
  \\[0.2cm]
\hline
$P_4$    & $2-\frac{3}{2
\alpha_2}$  &
$\frac{3}{\alpha_2}-1$
\\[0.1cm]
\hline
\end{tabular}}
\caption[TGdcrit]{The real and physically interesting
critical points at the finite region of the autonomous system
\eqref{TGdeqx}-\eqref{TGdeqm}, and the corresponding eigenvalues $\nu_{1},\nu_{2}$ of the
matrix of the perturbation equations. We denote
$\Omega_{m1}=\frac{\alpha_1 \sqrt{9-3 \alpha_2}
(6-5 \alpha_2)+6 (\alpha_2-3)}{6
   (\alpha_2-3)}$. From  \cite{Kofinas:2014aka}.}
\label{TGdtab4}
%\\ [0.2cm]
\end{table*}
\begin{table*}
{\begin{tabular}
{| c|c| c| c| c|}
\hline
Cr. P. &$\Omega_{DE}$ &$q$  & $w_{DE}$ & Properties of solutions\\
\hline\hline
$P_1$ & $1-\Omega_{m1}$  & $\frac{1}{2}$& $0$ & Dark Energy - Dark Matter
scaling solution
\\[0.2cm]
\hline
$P_2$   & $1$ &
$q_2$ & $w_{DE2}$     & Decelerating solution for \\
&&&& $\alpha_2<0,\; \frac{3}{3-2 \alpha_2}<\alpha_1\leq  \sqrt{\frac{3}{3-4 \alpha_2}}$
or
 \\
&&&& $0<\alpha_2<\frac{3}{4},\; 0<\alpha_1\leq  \sqrt{\frac{3}{3-4 \alpha_2}}$ or \\
&&&& $\alpha_1\neq 0,\; \alpha_2=\frac{3}{4}$ or\\
&&&& $\frac{3}{4}<\alpha_2\leq \frac{3}{2},\; \alpha_1<0$ or $\alpha_2>\frac{3}{2},\;
\frac{3}{3-2 \alpha_2}<\alpha_1<0$.\\
&&&& Quintessence
   solution for \\
&&&&
$\alpha_2\leq -\frac{3}{2},\; \,0<\alpha_1<\frac{3}{3-2 \alpha_2}$ or \\
&&&&$-\frac{3}{2}<\alpha_2<0,\; - \sqrt{\frac{3(2
\alpha_2+3)}{(\alpha_2-3)^2}}<\alpha_1<\frac{3}{3-
2 \alpha_2}$ or \\
&&&& $\frac{3}{2}<\alpha_2\leq 3,\; \alpha_1<\frac{3}{3-2 \alpha_2}$ or \\
&&&& $\alpha_2>3,\; - \sqrt{\frac{3(2 \alpha_2+3)}{(\alpha_2-3)^2}}<\alpha_1<\frac{3}{3-2
\alpha_2}
$. \\
&&&& De Sitter solution for \\
&&&&
$-\frac{3}{2}<\alpha_2<0,\;\alpha_1=\sqrt{\frac{3(2
\alpha_2+3)}{(\alpha_2-3)^2}}$ or $\alpha_2>3,\;\alpha_1=-\sqrt{\frac{3(2
\alpha_2+3)}{(\alpha_2-3)^2}}$.\\
&&&& Phantom solution for \\
&&&& $-\frac{3}{2}<\alpha_2<0,\;0<\alpha_1<
\sqrt{\frac{3(2\alpha_2+3)}{(\alpha_2-3)^2}}$  or  $\alpha_2>3,\;\alpha_1<-\sqrt{\frac{3(2
\alpha_2+3)}{(\alpha_2-3)^2}}$.
\\[0.2cm]
\hline
$P_3$  & $1$ &
$q_3$ & $w_{DE3}$ & Decelerating solution for  \\
&&&& $\alpha_2<0,\; 0<\alpha_1\leq  \sqrt{\frac{3}{3-4 \alpha_2}}$ or  \\
&&&& $0<\alpha_2<\frac{3}{4},\; \frac{3}{3-2 \alpha_2}<\alpha_1\leq  \sqrt{\frac{3}{3-4
\alpha_2}}$
or \\
&&&&
$\frac{3}{4}\leq \alpha_2<\frac{3}{2},\; \alpha_1>\frac{3}{3-2 \alpha_2}$.\\
&&&&
Quintessence
solution for
\\
&&&& $0<\alpha_2<\frac{3}{2},\;  \sqrt{\frac{3(2
\alpha_2+3)}{(\alpha_2-3)^2}}<\alpha_1<-\frac{3}{2
\alpha_2-3}$ or
\\
&&&&
$\frac{3}{2}\leq \alpha_2<3,\; \alpha_1> \sqrt{\frac{3(2 \alpha_2+3)}{(\alpha_2-3)^2}}$.
\\
&&&&   De Sitter
solution for $0<\alpha_2<3,\; \alpha_1=\sqrt{\frac{3(2\alpha_2+3)}{(\alpha_2-3)^2}}$.\\
&&&& Phantom solution for
\\
&&&& $0<\alpha_2<3,\; 0<\alpha_1<\sqrt{\frac{3(2\alpha_2+3)}{(\alpha_2-3)^2}}$ or
$\alpha_2\geq 3,\;
\alpha_1>0$.
            \\[0.2cm]
\hline
$P_4$  & $1$ &$\frac{3}{2 \alpha_2}$ &
$\frac{1}{\alpha_2}-\frac{1}{3}$  & Decelerating
solution for $\alpha_2> 0$. \\
&&&& Quintessence DE dominated
solution for $\alpha_2< -\frac{3}{2}$. \\
&&&& De Sitter solution for $\alpha_2=-\frac{3}{2}$.   \\
&&&&  Phantom
solution for
$-\frac{3}{2}<\alpha_2<0$. \\
[0.2cm]
\hline
\end{tabular}}
\caption[TGdcrit2]{
The real and physically interesting
critical points of the autonomous system \eqref{TGdeqx}-\eqref{TGdeqm}, and the
corresponding values of the dark energy density parameter $\Omega_{DE}$, the
deceleration parameter $q$, and the dark energy
equation-of-state parameter $w_{DE}$. We use the notation
$\Omega_{m1}=\frac{\alpha_1 \sqrt{9-3 \alpha_2}
(6-5 \alpha_2)+6 (\alpha_2-3)}{6
   (\alpha_2-3)}$,   $q_{2}= \frac{\sqrt{3 \alpha_1^2 (4
\alpha_2-3)+9}}{\alpha_1^2
\alpha_2}-\frac{3}{\alpha_1^2 \alpha_2}+\frac{3}{\alpha_2}-2$,  $q_{3}=-
\frac{\sqrt{3 \alpha_1^2 (4
\alpha_2-3)+9}}{\alpha_1^2
\alpha_2}-\frac{3}{\alpha_1^2 \alpha_2}+\frac{3}{\alpha_2}-2$, $w_{DE
2}= \frac{2
\sqrt{3 \alpha_1^2 (4 \alpha_2-3)+9}}{3 \alpha_1^2
\alpha_2}-\frac{2}{\alpha_1^2
   \alpha_2}+\frac{2}{\alpha_2}-\frac{5}{3}$  and $w_{DE
3}=- \frac{2
\sqrt{3 \alpha_1^2 (4 \alpha_2-3)+9}}{3 \alpha_1^2
\alpha_2}-\frac{2}{\alpha_1^2
   \alpha_2}+\frac{2}{\alpha_2}-\frac{5}{3}$.
In the last column we summarize
their physical description. From  \cite{Kofinas:2014aka}. }
\label{TGdtab2}
\end{table*}

One can now express the various observables in terms of the above auxiliary
variables $\Omega_m$ and $x$ (note that $\Omega_m$ is an observable
itself, that is the matter density parameter). In particular, the deceleration
parameter $q\equiv -1-\dot{H}/H^2$ is given by
\begin{equation}
q=\frac{3 \left(1-x^2\right)}{2 \alpha_2}.
\label{TGddecc}
\end{equation}
Similarly, the dark energy density parameter straightaway reads
\begin{equation}
\Omega_{DE}\equiv \frac{8\pi G\rho_{DE}}{3 H^2}= 1-\Omega_m.
\label{TGddOmegaDE}
\end{equation}
The dark energy equation-of-state parameter $w_{DE}$ is given by the relation
$2q=1+3(w_{m}\Omega_{m}+w_{DE}\Omega_{DE})$, and therefore
\begin{equation}
w_{DE}= \frac{3 x^2+\alpha_2+3 \alpha_2 w_m \Omega_m-3}{3 \alpha_2
(\Omega_m-1)}\,,
\label{TGdwdephasespace}
\end{equation}
where $w_m\equiv \frac{p_m}{\rho_m}$ is the matter equation-of-state parameter. In the
following, without loss of generality we assume dust matter ($w_m=0$), but the extension
to general $w_m$ is straightforward.

We now proceed to the detailed phase-space analysis following  \cite{Kofinas:2014aka}. The
real and physically interesting (that is corresponding to an expanding universe) critical
points of the autonomous system \eqref{TGdeqx}-\eqref{TGdeqm}, obtained by setting the
left hand sides of these equations to zero, are presented in Table \ref{TGdtab1}. In the
same table we provide their existence conditions. For each critical point of Table
\ref{TGdtab1} we examine the sign of the real part of the eigenvalues of the $2\times2$
matrix ${\bf {Q}}$ of the
corresponding linearized perturbation equations, in order to
determine the type and stability of the point. The various eigenvalues are presented in
Table \ref{TGdtab4}. Hence, in  Table \ref{TGdtab2} we summarize the results, and
additionally for each critical point we calculate the values  of the deceleration
parameter $q$, of the dark energy density  parameter $\Omega_{DE}$, and of the dark
energy equation-of-state parameter $w_{DE}$, given by
(\ref{TGddecc})-(\ref{TGdwdephasespace}), and we present the results
in Table \ref{TGdtab2}. Moreover, in the same Table we summarize the
physical description of the solutions, which we analyze below.

Finally, apart from the above finite critical points, the scenario at hand possesses
critical points at infinity too, due to the fact that the dynamical system
\eqref{TGdeqx}-\eqref{TGdeqm} is non-compact. In order to extract
and analyze them, one needs to apply the Poincar\'e central projection method
 \cite{PoincareProj, Leon:2014rra}. However, since the obtained three critical points at
infinity are usually non-stable, and since their stability regimes are determined only
through numerical elaboration, we do not describe the whole procedure in detail, and we
refer the reader to the original work  \cite{Kofinas:2014aka}.

Let us now discuss the corresponding cosmological behavior. As usual, the features of the
solutions can be easily deduced by the values of the observables. In particular, $q<0$
($q>0$) corresponds to acceleration (deceleration), $q=-1$ to de Sitter solution,
$w_{DE}>-1$ ($w_{DE}<-1$) corresponds to quintessence-like (phantom-like) behavior, and
$\Omega_{DE}=1$ implies a dark-energy dominated universe.

Point $P_1$ is conditionally stable, and thus it can attract the universe at late times.
Since the dark energy and matter density parameters are of the same order, this point
represents a dark energy - dark matter scaling solution, alleviating the coincidence
problem (note that in order to handle the coincidence problem one should provide an
explanation of why the present $\Omega_m$ and $\Omega_{DE}$ are of the same order,
although they follow different evolution behaviors). However, it has the disadvantage
that $w_{DE}$ is $0$ and the universe is not accelerating. Although this picture is not
favored by observations, it may simply  imply that the universe at present has not yet
reached its asymptotic regime.

Point $P_2$ is conditionally stable, and therefore, it can be the late-times state of the
universe. It corresponds to a dark-energy dominated universe that can be accelerating.
Interestingly enough, depending on the model parameters, the dark energy
equation-of-state parameter can lie in the quintessence regime, it can be equal to the
cosmological constant value $-1$, or it can even lie in the phantom regime. These
features
are  a great advantage of the scenario at hand, since they are compatible with
observations, and
moreover they are obtained only due to the novel features of $f(T,T_G)$ gravity, without
the explicit inclusion of a cosmological constant or a scalar field, either canonical or
phantom one.

Point $P_3$ is conditionally stable, and therefore it can attract the universe at late
times. It has similar features with $P_2$, but for different parameter regions. Namely,
it corresponds to a dark-energy dominated universe that can be accelerating, where the
dark-energy equation-of-state parameter can lie in the quintessence or phantom regime, or
it can be exactly $-1$. These features make this point also a good candidate for the
description of Nature.

Point $P_4$ corresponds to a dark-energy dominated  universe that can be accelerating,
where the dark-energy equation-of-state parameter can lie in the quintessence or phantom
regime, or it can be exactly $-1$.  However, $P_{4}$ is not stable and thus it cannot
attract the universe at late times.

In order to present the aforementioned behavior more transparently, one evolves the
autonomous system \eqref{TGdeqx}-\eqref{TGdeqm} numerically for the parameter choices
$\alpha_1=-\sqrt{33}$ and $\alpha_2=4$, assuming the matter to be dust ($w_m=0$). The
corresponding phase-space behavior is depicted in Fig. \ref{TGdfig1}. In this case the
universe at late times is attracted by the dark-energy dominated de Sitter attractor
$P_2$, where the effective dark energy behaves like a cosmological constant.
Additionally, in Fig. \ref{TGdfig5} we can see some phase-space orbits for the choice
$\alpha_1=3$ and $\alpha_2=\frac{3}{2}$, with $w_m=0$. In this case, the universe is
attracted by the quintessence solution $P_3$.
\begin{figure}[ht]
\begin{center}
\includegraphics[width=6.5cm]{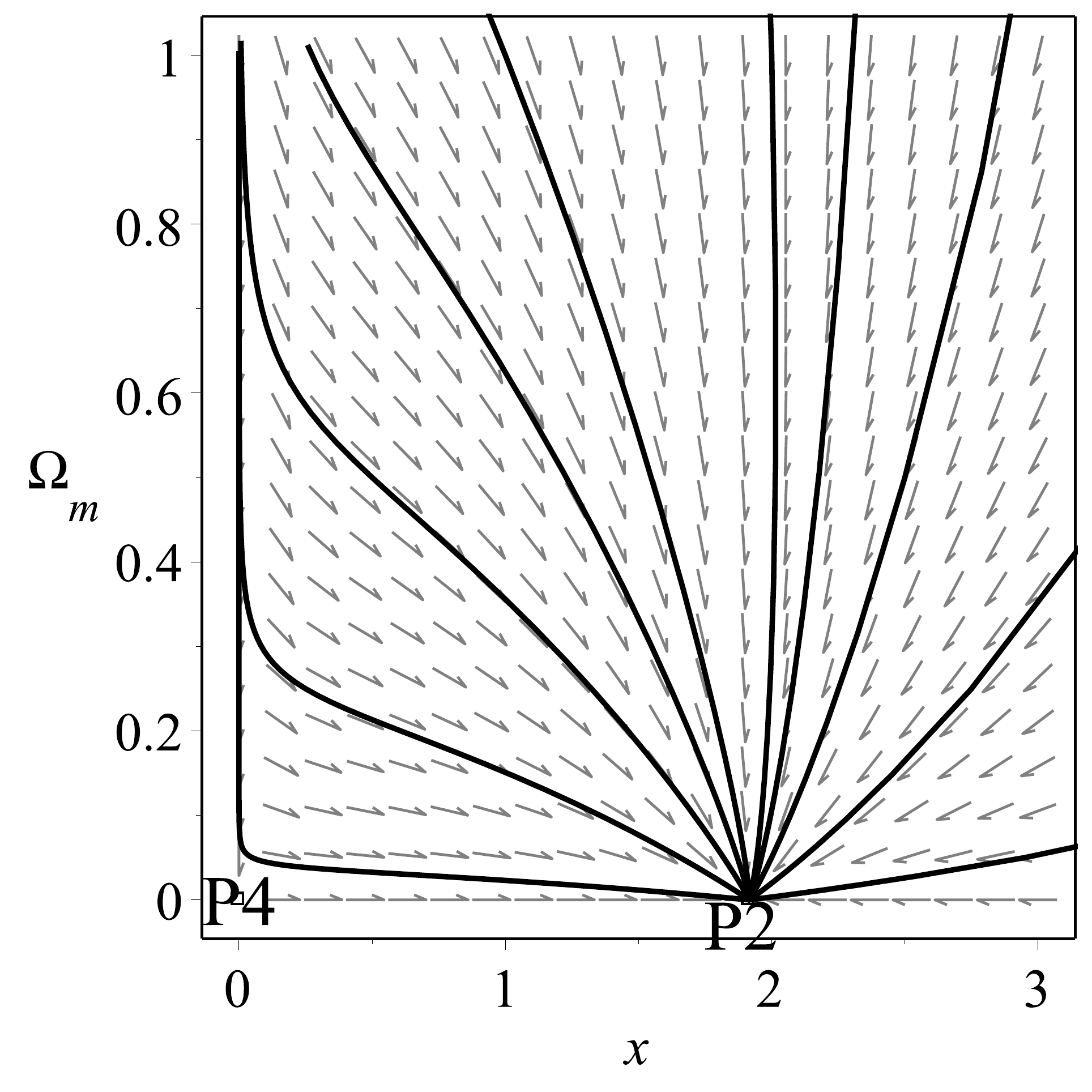}
\caption{{\it{Trajectories in the  phase space for the cosmological scenario
\eqref{TGdeqx}-\eqref{TGdeqm}, for the parameter choices $\alpha_1=-\sqrt{33}$ and $
\alpha_2=4$, and assuming the matter to be dust ($w_m=0$). In this specific example the
universe is led to the de Sitter attractor $P_2$, while $P_4$ is saddle. From
 \cite{Kofinas:2014aka}.}} }
\label{TGdfig1}
\end{center}
\end{figure}
\begin{figure}[ht]
\begin{center}
\includegraphics[width=6.5cm]{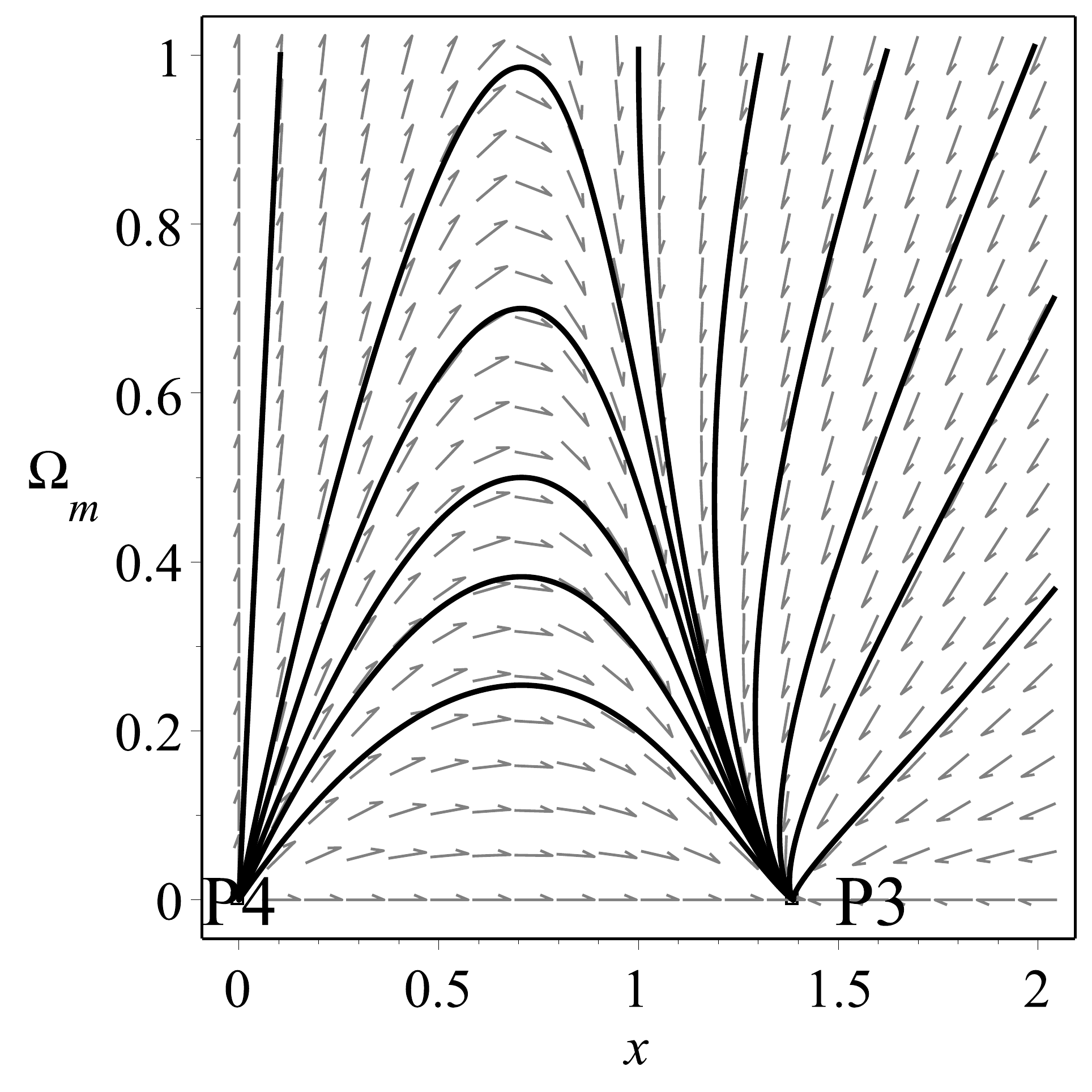}
\caption{{\it{Trajectories in the  phase space for the cosmological scenario
\eqref{TGdeqx}-\eqref{TGdeqm}, for the parameter choices  $\alpha_1=3$ and $
\alpha_2=\frac{3}{2}$, and assuming the matter to be dust ($w_m=0$).  In this specific
example the universe is led to the quintessence solution $P_3$. From
 \cite{Kofinas:2014aka}.}} }
\label{TGdfig5}
\end{center}
\end{figure}

In summary, the scenario of $f(T,T_G)$ cosmology exhibits interesting cosmological
behaviors. In particular, depending on the model parameters, the universe can result in a
dark energy dominated accelerating solution and the dark-energy equation-of-state
parameter can lie in the quintessence regime, it can be equal to the cosmological
constant value $-1$, or it can even lie in the phantom regime. Additionally, it can
result
in a dark energy - dark matter scaling solution, and thus it can alleviate the coincidence
problem. Definitely, before the scenario at hand can be considered as a good candidate
for the description of Nature, a detailed confrontation with observations should be
performed. In particular, one should use data from local gravity experiments (Solar System
observations), as well as SNIa, BAO, and CMB data, in order to impose constraints on
the model. These necessary investigations would be nice be performed in the near future.

\subsection{Non-minimal torsion-matter coupling extension of $f(T)$ gravity}

In this subsection we describe another extension of $f(T)$ gravity, in which one allows
for a non-minimal coupling between the torsion scalar $T$  and the matter Lagrangian,
first introduced in  \cite{Harko:2014sja}. The inspiration for this extension comes once
again from the corresponding models of curvature gravity.

In curvature gravity, indeed most of the modifications of the Einstein-Hilbert Lagrangian
involve a change in the geometric part of the action only,  and assume that the matter
Lagrangian plays a subordinate and passive role, which is implemented by the minimal
coupling of matter to geometry. However, a general theoretical principle forbidding an
arbitrary coupling between matter and geometry does not exist {\it a priori}, and indeed
if theoretical models in which matter is considered on an equal footing with geometry,
are allowed, then gravitational theories with many interesting and novel features can be
constructed.

A theory with an explicit coupling between an arbitrary function of the scalar curvature
and the Lagrangian density of matter was proposed in  
\cite{Nojiri:2004bi,Allemandi:2005qs,Bertolami:2007gv}. The
gravitational action of the latter model is of the form
$S=\int{\left\{f_1(R)+\left[1+\lambda f_2(R)\right]L_m\right\}\sqrt{-g}d^4x}$. In these
models an extra force acting on massive test particles arises, and the motion is no longer
geodesic. Moreover, in this framework, one can also explain dark matter
 \cite{Bertolami:2009ic}. The early ``linear''  geometry-matter coupling
 \cite{Bertolami:2007gv} was extended in  \cite{Harko:2008qz} and a maximal extension of
the Einstein-Hilbert action with geometry-matter coupling, of the form $S=\int d^{4}x
\sqrt{-g}f\left(R,L_m\right)$, was considered in  \cite{Harko:2010mv}.

Hence, inspired by the above non-minimal curvature-matter coupling scenarios, in
\cite{Harko:2014sja} the authors constructed similar models in the framework of
teleparallel and $f(T)$ gravity, that is allowing for a non-minimal
torsion-matter coupling. In particular, one considers the
action
\begin{equation}
S= \frac{1}{16\pi G}\,\int
d^{4}x\,e\,\left\{T+f_{1}(T)+\left[1+\lambda\,f_{2}
(T)\right]\,L_{m}\right\},
\label{fTLm1}
\end{equation}
where $f_{i}(T)$ (with $i=1,2$) are arbitrary functions of the
torsion scalar $T$ and $\lambda$ is a coupling constant with units of ${\rm
mass}^{-2}$. Varying the action with respect to the tetrad $e^{A}_{\rho}$
yields the field
equations
\begin{eqnarray}
&&\left(1+f_{1}'+\lambda f_{2}' L_{m}\right) \left[e^{-1}
\partial_{\mu}{(e e^{\alpha}_{A} S_{\alpha}{}^{\rho \mu})}-e^{\alpha}_{A}
T^{\mu}{}_{\nu \alpha} S_{\mu}{}^{\nu\rho}\right]\nonumber\\
&&\ \ \ \ \
+\left(f_{1}''+ \lambda f_{2}'' L_{m}
\right)  \partial_{\mu}{T} e^{\alpha}_{A} S_{\alpha}{}^{\rho\mu}+
e_{A}^{\rho} \left(\frac{f_{1}+T}{4}\right)
\nonumber\\
&&\ \ \ \ \
-\frac{1}{4} \lambda f_{2}' \,
\partial_{\mu}{T} e^{\alpha}_{A}  S^{(m)}_{\ \ \ \alpha}{}^{\rho \mu}
+ \lambda f_{2}'\, e^{\alpha}_{A} S_{\alpha}{}^{\rho\mu} \, \partial_{\mu}{L_{m}}
\nonumber\\
&&\ \ \ \ \
=4\pi G \left(1+\lambda f_{2}\right) e^{\alpha}_{A}
 {T}^{(m)}_{\ \ \ \alpha}{}^{\rho},
\label{fTLmgeneoms}
\end{eqnarray}
where we have defined
\begin{equation}
e^{\alpha}_{A}  S^{(m)}_{\ \ \ \alpha}{}^{\rho \mu}= S^{(m)}_{\ \ \ A}{}^{\rho
\mu}=\frac{\partial{L_{m}}}{\partial{\partial_{\mu}{e^{A}_{\rho}}}},
\label{fTLMStilde}
\end{equation}
and the prime denotes differentiation with respect to the torsion
scalar. As usual $S_{\rho}{}^{\mu\nu}\equiv\frac{1}{2}\Big(K^{\mu\nu}{}_{\rho}
+\delta^\mu_\rho
\:T^{\alpha\nu}{}_{\alpha}-\delta^\nu_\rho\:
T^{\alpha\mu}{}_{\alpha}\Big)$, with  the contorsion tensor given by
$K^{\mu\nu}{}_{\rho}\equiv-\frac{1}{2}\Big(T^{\mu\nu}{}_{\rho}
-T^{\nu\mu}{}_{\rho}-T_{\rho}{}^{\mu\nu}\Big)$. Note that, as expected, when $\lambda=0$
Eq.  (\ref{fTLmgeneoms}) reduces to Eq.  (\ref{eom_fT_general}) of simple $f(T)$ gravity.
Since the Lagrangian density of a perfect fluid is the energy scalar representing the
energy in a local rest frame for the fluid, a possible ``natural choice'' for the matter
Lagrangian density is $L_{m}/(16 \pi G)=-\rho_{m}$  \cite{Groen:2007zz,Bertolami:2008ab}.
In this case, we have $S^{(m)}_{\ \ \ A}{}^{\rho
\mu}=0$, and also the usual
form
of the energy momentum tensor for the perfect fluid
$ T^{(m)}_{\ \ \ \mu\nu}=(\rho_{m}+p_{m}) u_{\mu} u_{\nu}-p_{m} g_{\mu
\nu}$.

In summary, inserting the flat FRW vierbein choice (\ref{weproudlyuse}), namely
$e_{\mu}^A={\rm diag}(1,a(t),a(t),a(t))$, and
the above  matter Lagrangian density, into the field equations
\eqref{fTLmgeneoms}, we obtain the modified Friedmann equations
\begin{equation}\label{fTLmH0}
 H^{2}=\frac{8\pi G}{3} \left[1+\lambda
 \left(f_{2}+12 H^{2} f_{2}' \right)\right] \rho_m-\frac{1}{6} \left(f_{1}+12
H^{2} f_{1}'\right),
\end{equation}
\begin{equation}\label{fTLmH00}
 \dot{H}=-\frac{4\pi G\left( \rho _{m}+p_{m}\right) \left[ 1+\lambda
\left(
f_{2}+12H^{2}f_{2}^{\prime }\right) \right] }{1+f_{1}^{\prime
}-12H^{2}f_{1}^{\prime \prime }-16 \pi G \lambda \rho _{m}\left( f_{2}^{\prime
}-12H^{2}f_{2}^{\prime \prime }\right) }.
\end{equation}
In the limit $\lambda =0$, $f_1(T)\equiv f(T)$, and $f_2(T)\equiv 0$,
Eqs. ~(\ref{fTLmH0}) and (\ref{fTLmH00}) respectively reduce to Eqs. ~(\ref{background11})
and (\ref{background22}), of usual $f(T)$ gravity in FRW geometry.
The generalized Friedmann equations  can be rewritten  as
 \begin{eqnarray}
  3H^2&=& 8\pi G\left(\rho_{DE}+\rho_m  \right), \label{fTLmFr1} \\
  2\dot{H}+3H^2& =&-8\pi G\left(p_{DE}+p_m\right), \label{fTLmFr2}
\end{eqnarray}
where the effective energy density and effective pressure of the dark energy
sector are defined as
\begin{equation}
\label{fTLmrhode}
 \rho_{DE}:=-\frac{1}{16 \pi G} \left(f_{1}+12 H^{2}
f_{1}'\right)+ \lambda \rho_m\left(f_{2}+12 H^{2} f_{2}' \right) ,
\end{equation}
\begin{eqnarray}\label{fTLmpde}
&&p_{DE}:= \left(\rho _m+p_m\right)\times \nonumber\\
&&\left[\frac{  1+\lambda \left(
f_{2}+12H^{2}f_{2}^{\prime }\right)  }{1+f_{1}^{\prime
}-12H^{2}f_{1}^{\prime \prime }-16\pi G\lambda \rho _{m}\left( f_{2}^{\prime
}-12H^{2}f_{2}^{\prime \prime }\right) }-1\right]\nonumber\\
&&+\frac{1}{16 \pi G} \left(f_{1}+12 H^{2}
f_{1}'\right)- \lambda \rho_m\left(f_{2}+12 H^{2} f_{2}' \right).
\end{eqnarray}
Furthermore, we can define the dark-energy equation-of-state parameter
as usual
\begin{eqnarray}
w_{DE}\equiv \frac{p_{DE}}{\rho_{DE}}.
\label{fTLmwDE}
\end{eqnarray}
One can easily verify that the above affective dark energy density
and pressure satisfy the usual evolution equation
\begin{eqnarray}\label{fTLmcons}
 \dot{\rho}_{DE}+\dot{\rho} _m +3H\left(\rho_{DE}+\rho
_m+p_{DE}+p_m\right)=0.
\end{eqnarray}
Finally, we can introduce as usual the deceleration parameter $q=-1-\dot{H}/H^2$.

As we observe, the above equations of non-minimal coupled torsion-matter gravity are
different than those of non-minimal coupled curvature-matter gravity
 \cite{Bertolami:2007gv}, and obviously they are different from simple $f(T)$ gravity too.
Hence, the theory of non-minimal torsion-matter coupling is a novel class of
gravitational modification.

\subsubsection{Cosmological behavior}

Since we have extracted the basic background equations of motion for the $f(T)$ gravity
model with a non-minimal matter-torsion coupling, we are now able to investigate its
phenomenological implications, following  \cite{Harko:2014sja,Carloni:2015lsa}. Due to the 
usual relation
(\ref{TH2}), namely  $T = -6H^2$, for convenience in the following we will change the
$T$-dependence to $H$-dependence in the involved expressions, so that
$f_1(T)\equiv f_1(H)$, and $f_2(T)\equiv f_2(H)$. For the derivatives of the
functions $f_1(T)$ and $f_2(T)$ we obtain $f_i'(H)\equiv \left .f_i'(T)
\right |_{T\rightarrow -6H^2}$, and $f_i''(H)\equiv \left .f_i''(T) \right
|_{T\rightarrow -6H^2}$, $i=1,2$, respectively. In order to simplify the
notation, in this paragraph we set $8\pi G=1$.

The basic cosmological equations describing the time evolution of the
non-minimally matter-coupled  $f(T)$ gravity are given by Eqs. ~(\ref{fTLmH0}) and
(\ref{fTLmH00}). From Eq. ~(\ref{fTLmH0}) we can express the matter
density as
\be\label{fTLmc1}
\rho _m(t)=\frac{3H^2+\left[f_1(H)+12H^2\left. f_1'(T)\right  |_{T\rightarrow
-6H^2}\right]/2}{1+\lambda\left[f_2(H)+12H^2\left .f_2'(T)\right
|_{T\rightarrow -6H^2}\right]}.
\ee
By substituting the matter density $\rho _m$ into Eq. ~(\ref{fTLmH00}) we obtain
the basic as
\begin{widetext}
\begin{equation}
\label{fTLmc2}
2\dot{H}=-\frac{\left\{1+\lambda  \left[f_2(H)+12 H^2 \left .f_2'(T)\right
|_{T\rightarrow
-6H^2}\right]\right\}
   \left\{\frac{ \left[f_1(H)+12 H^2 \left .f_1'(T)\right |_{T\rightarrow
-6H^2}\right]/2+3   H^2}{1+\lambda  \left[f_2(H)+12 H^2
   \left .f_2'(T)\right |_{T\rightarrow -6H^2}\right]}+p_m\right\}}{1+\left .f_1'(T)
   \right|_{T\rightarrow -6H^2}-12 H^2
   \left .f_1''(T)\right |_{T\rightarrow -6H^2}-\frac{2\lambda  \left\{ \left[f_1(H)+12
H^2
   \left .f_1'(T)\right |_{T\rightarrow -6H^2}\right]/2+3 H^2\right\} \left[\left
.f_2'(T)\right |_{
T\rightarrow -6H^2}-12
   H^2 \left .f_2''(T)\right |_{T\rightarrow -6H^2}\right]}{1+\lambda  \left[f_2(H)+12 H^2
   \left .f_2'(T)\right |_{T\rightarrow -6H^2}\right]}}.
   \end{equation}
   \end{widetext}
Hence, once the functions $f_1(T)$ and $f_2(T)$ are fixed, Eqs. ~(\ref{fTLmc1}) and
(\ref{fTLmc2}) become a system of two ordinary differential equations for three
unknowns, $\left(H,\rho _m,p_m\right)$. In order to close the system of
equations, the matter equation of state $p_m=p_m\left(\rho _m\right)$ must
also be given. Finally, the dark energy equation-of-state parameter can be
expressed as
\be
\!w_{DE}=-\frac{2\dot{H}}{\rho _{DE}}-\frac{\rho _m+p_m}{\rho
_{DE}}-1= -\frac{2\dot{H}+3H^2+p_m}{3H^2-\rho_m}.
\label{fTLmwDE2}
\ee
In the following, we investigate the system of equations (\ref{fTLmc1}) and
(\ref{fTLmc2}), for different functional forms of $f_1(T)$ and $f_2(T)$.

\begin{itemize}
 \item{Model I}

Let us first consider the case where    \cite{Harko:2014sja}
 \begin{eqnarray}
&&f_1(T)=-\Lambda+\alpha _1T^2\nonumber\\
&&f_2(T)=\beta _1T^2,
\end{eqnarray}
with $\alpha _1$ and $\beta _1$ constants, since these are the first non-trivial
corrections to TEGR, that is to GR. As we mentioned above, it proves
convenient to
express the involved functions in terms of $H$. In particular, in
terms of $H$ the functional dependencies of $f_1$ and $f_2$ are given by
$f_1(H)=-\Lambda+\alpha H^4$ and $f_2(H)=\beta H^4$, respectively, with
$\alpha=36\alpha _1$, $\beta =36\beta _1$. For the derivatives of the
functions $f_1$ and $f_2$ we obtain $f_1'(H)=-\alpha H^2/3$, $f_2'(H)=-\beta
H^2/3$, $f_1''(H)=\alpha /18$, $f_2''(H)=\beta /18$. Moreover, we
restrict our analysis to the case of dust matter,  that is we take
$p_m=0$.
In this case the gravitational field equations (\ref{fTLmc1}) and (\ref{fTLmc2}) become
\be
\label{fTLmdem0}
\rho _m(t)=\frac{3 \alpha  H^4-6 H^2+\Lambda }{6 \beta  \lambda  H^4-2},
\ee
and
\be
\label{fTLmdem1}
\dot{H}(t)=\frac{\left(3 \alpha  H^4-6 H^2+\Lambda \right) \left(3 \beta  \lambda
H^4-1\right)}{4 H^2 \left(\alpha +\beta  \lambda  \Lambda -3 \beta  \lambda
H^2\right)-4},
\ee
respectively.

In order to investigate the behavior of the scenario at hand, we numerically elaborate
the
above equations for various parameter values, and in Fig. \ref{fTLmfig4} we depict the
evolution of the deceleration parameter as a function of time, while in Fig.
\ref{fTLmfig4a} we present the corresponding evolution of the dark-energy
equation-of-state parameter.
   \begin{figure}[ht]
   \centering
  \includegraphics[width=8cm]{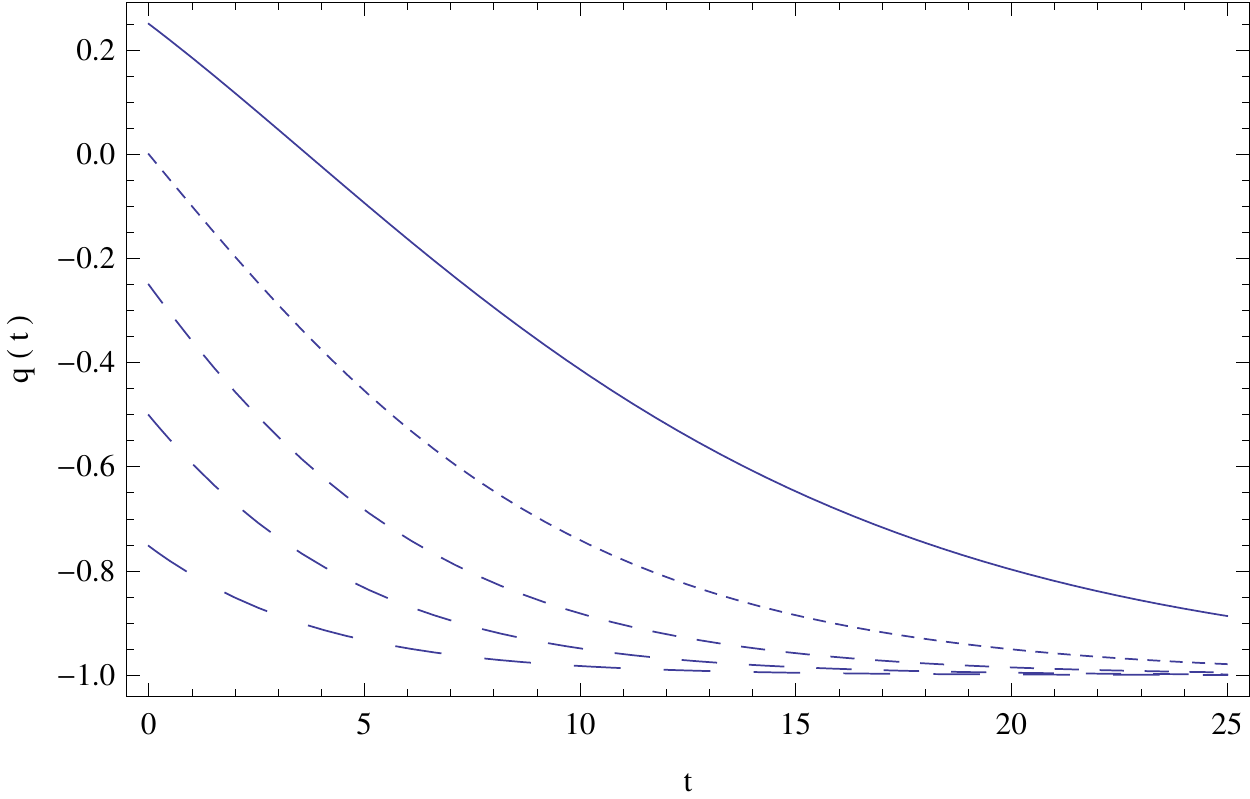}
  \caption{\it Time variation of the deceleration parameter $q(t)$ for
the non-minimally matter
coupled $f(T)$ gravity model with $f_1(T)=-\Lambda+\alpha _1T^2$ and
$f_2(T)=\beta _1T^2$, or equivalently   $f_1(H)=-\Lambda+\alpha H^4$ and
$f_2(H)=\beta H^4$ with $\alpha=36\alpha _1$, $\beta =36\beta _1$, for five
different choices of the parameters $\Lambda$, $\alpha $, $\beta $, and
$\lambda $: $\Lambda =0.01$, $\alpha =0.16$, $\beta =0.1$, and $\lambda =1$
(solid curve), $\Lambda =0.02$, $\alpha =0.18$, $\beta =0.3$, and $\lambda
=1.2$ (dotted curve), $\Lambda =0.03$, $\alpha =0.20$, $\beta =0.35$, and
$\lambda =1.4$ (short-dashed curve), $\Lambda =0.04$, $\alpha =0.30$, $\beta
=0.45$, and $\lambda =1.6$ (dashed curve), and $\Lambda =0.05$, $\alpha
=0.40$, $\beta =0.55$, and $\lambda =1.8$ (long-dashed curve), respectively.
The initial value for $H$ used to numerically integrate Eq. ~(\ref{fTLmdem1}) is
$H(0)=0.1$. From  \cite{Harko:2014sja}. }
\label{fTLmfig4}
   \end{figure}
\begin{figure}[ht]
   \centering
  \includegraphics[width=8.3cm]{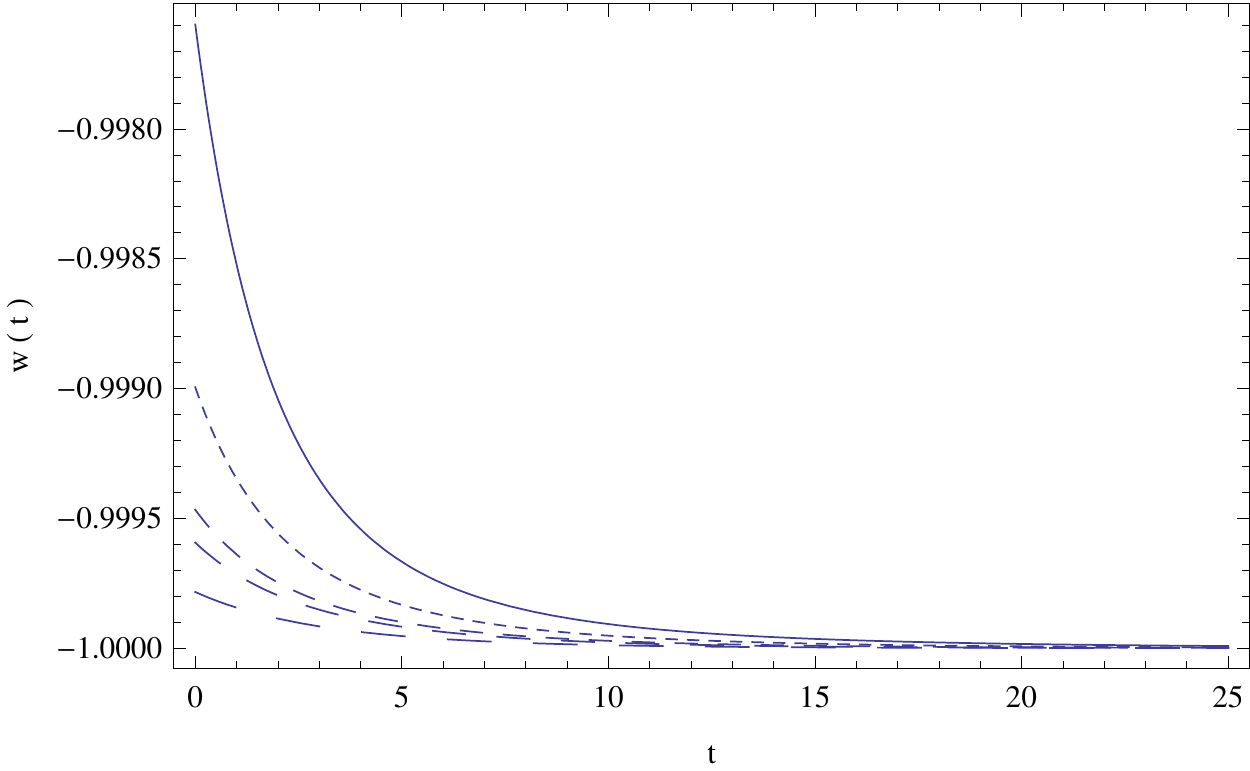}
  \caption{\it Time variation of the dark-energy equation-of-state parameter
$w_{DE}$ for the non-minimally
matter
coupled $f(T)$ gravity model with $f_1(T)=-\Lambda+\alpha _1T^2$ and
$f_2(T)=\beta _1T^2$, or equivalently   $f_1(H)=-\Lambda+\alpha H^4$ and
$f_2(H)=\beta H^4$ with $\alpha=36\alpha _1$, $\beta =36\beta _1$,
for five different choices of the parameters
$\Lambda$, $\alpha $, $\beta $, and $\lambda $:
$\Lambda =0.01$, $\alpha =0.16$, $\beta =0.1$, and $\lambda =1$ (solid
curve), $\Lambda =0.02$, $\alpha =0.18$, $\beta =0.3$, and $\lambda =1.2$
(dotted curve), $\Lambda =0.03$, $\alpha =0.20$, $\beta =0.35$, and $\lambda
=1.4$ (short dashed curve), $\Lambda =0.04$, $\alpha =0.30$, $\beta =0.45$,
and $\lambda =1.6$ (dashed curve), and $\Lambda =0.05$, $\alpha =0.40$,
$\beta =0.55$, and $\lambda =1.8$ (long dashed curve), respectively. The
initial value for $H$ used to numerically integrate Eq. ~(\ref{fTLmdem1}) is
$H(0)=0.1$. From   \cite{Harko:2014sja}.}
\label{fTLmfig4a}
\end{figure}
As we observe, for the considered range of values of the free parameters, in the $f(T)$
model with torsion-matter coupling, the Universe ends its evolution in an accelerating,
de Sitter-type phase. The deceleration parameter $q$ indicates a large variety of
dynamical behaviors of the $f(T)$ model with matter-torsion coupling. In particular, for
some values of the free parameters the Universe starts its evolution in the
matter-dominated phase from a decelerating phase, while some other values of the
parameters produce Universe models starting from a marginally accelerating phase ($q=0$),
and ending in a de Sitter state. Finally, for other parameter choices at the beginning of
the matter-dominated phase the Universe is already in an accelerating phase, that is with
$q<0$. Concerning the behavior of the dark-energy equation-of-state parameter presented
in
Fig. \ref{fTLmfig4a}, one can see that for these specific choices of the parameters
$w_{DE}$ is close to the value $-1$, to which it rigorously tends in the large-time
limits. This is an advantage, since in this model the effective torsion-matter coupling
can successfully mimic the cosmological constant, in agreement with observations.

After the above numerical elaboration, we analyze the properties of the equations in the
limit of small and large $H(t)$, extracting analytical expressions.

In the limit of small $H(t)$, that is at the late phases of the
cosmological evolution, Eq. ~(\ref{fTLmdem1}) becomes
   \be
 \dot{H}\approx \frac{1}{4} \left(\alpha  \Lambda +\beta  \lambda  \Lambda
   ^2-6\right)H^2 +\frac{\Lambda }{4}
   \ee
   yielding the following solution
 {\small{
 \begin{equation}
H(t)\approx\sqrt{H_{1}}\tan \left[ \tan ^{-1}\left(
\frac{H_{0}}{\sqrt{H_{1}}}\right) +\frac{%
\Lambda }{4\sqrt{H_{1}}}\left( t-t_{0}\right) \right],
\label{fTLmHapprox1}
\end{equation}
}}
with $H_{1}=\Lambda /\left[\Lambda (\alpha +\beta \lambda \Lambda )-6
\right]$, and  where we have used the initial condition
$H\left(t_0\right)=H_0$. Note that in the large-time limit the Hubble
function  becomes almost constant, implying that a de
Sitter-type evolution is possible in the framework of the present model.
Additionally, the matter energy-density (\ref{fTLmdem0}) can be approximated as
\be
   \rho _m(t)\approx3 H_1 \tan ^2\left[\tan
^{-1}\left(\frac{H_0}{\sqrt{H_1}}\right)+\frac{\Lambda  \left(t-t_0\right)}{4
\sqrt{H_1}}\right]-\frac{\Lambda }{2},
\ee
while the dark-energy equation-of-state parameter from (\ref{fTLmwDE2}),
becomes
 \bea
&&\!\!\!\!\!\!\!\!\!\!\!\!\! \!
w_{DE}(t)\approx -1-(\alpha+\beta\lambda
\Lambda)H_1    \nonumber\\
&& \ \ \ \ \ \ \times  \tan ^2
\left[\tan
^{-1}\left(\frac{H_0}{\sqrt{H_1}}\right)+\frac{\Lambda  \left(t-t_0\right)}{4
\sqrt{H_1}}\right].
  \eea
Interestingly enough, we
observe that according to the parameter values, $w_{DE}$ can be either above
or below $-1$, that is the effective dark-energy sector can be
quintessence-like or phantom-like. This feature, which is expected to happen
in modified gravity  \cite{Nojiri:2013ru}, is an additional advantage of the
scenario at hand.

In the limit of large $H$, corresponding to the early phases of the cosmological
evolution, in the first-order approximation the differential equation (\ref{fTLmdem1}),
describing the cosmological dynamics of the Hubble function, becomes
$\dot{H}=-3\alpha  H^4/4$,
   with the general solution given by
   \be
   H(t)\approx\frac{2^{2/3}H_0}{\left[4+9H_0^3\alpha \left(t-t_0\right)\right]^{1/3}}.
   \ee
   The behavior of the  matter energy density (\ref{fTLmdem0}) becomes
   \be
   \rho _m(t)\approx\frac{\alpha }{2 \beta  \lambda },
   \ee
   showing that during the time interval for which this approximation is
valid the energy density of the matter is approximately constant.
Finally, the dark-energy equation-of-state parameter from (\ref{fTLmwDE2}),
becomes
   \be
  w_{DE}(t)\approx-\frac{ \alpha}{2} H(t)^2=-\frac{
\alpha}{2}
\frac{2^{4/3}H_0^2}{\left[4+9H_0^3\alpha
\left(t-t_0\right)\right]^{2/3}}.
   \ee
 Again, we mention that according to the parameter choice, $w_{DE}$ can be
either above or below $-1$, that is the effective dark-energy sector can be
quintessence-like or phantom-like.

    \item{Model II}

Let us now consider the case where    \cite{Harko:2014sja}
 \begin{eqnarray}
&&f_1(T)=-\Lambda\nonumber\\
&&f_2(T)=\alpha _1T+\beta _1T^2,
\end{eqnarray}
where $\Lambda >0$, $\alpha _1$ and $\beta
_1$ are constants, since this scenario is also the first non-trivial
correction to TEGR, that is to GR. Equivalently, we impose
$f_1(H)=-\Lambda$  and $f_2(H)=\alpha H^2+\beta  H^4$, with $\alpha =-6\alpha
_1$ and $\beta =36\beta _1$. For the derivatives of the functions $f_1(T)$
and
$f_2(T)$ we obtain $f_1'(T)=f_1''(T)=0$, $f_2'(H)=-\alpha /6-\beta
H^2/3$, and $f_2''(H)=\beta /18$.  The basic evolution equations
 (\ref{fTLmc1}) and (\ref{fTLmc2}) in this case respectively become
\be
\label{fTLmmod21}
\rho _m(t)=\frac{\Lambda -6 H^2}{2 \alpha  \lambda  H^2+6 \beta  \lambda  H^4-2},
   \ee
   and
  \be\label{fTLmmod22}
   \dot{H}(t)=\frac{3 \left(\Lambda -6 H^2\right) \left(\alpha  \lambda  H^2+3 \beta
   \lambda  H^4-1\right)}{2 \left(\alpha  \lambda  \Lambda +6 \beta  \lambda
   H^2 \left(\Lambda -3 H^2\right)-6\right)}.
   \ee
One numerically evolves these equations  \cite{Harko:2014sja}, and in Fig. \ref{fTLmfig8a}
we depict the corresponding dark-energy equation-of-state parameter calculated from
(\ref{fTLmwDE2}),
  \begin{figure}[ht]
   \centering
  \includegraphics[width=8.5cm]{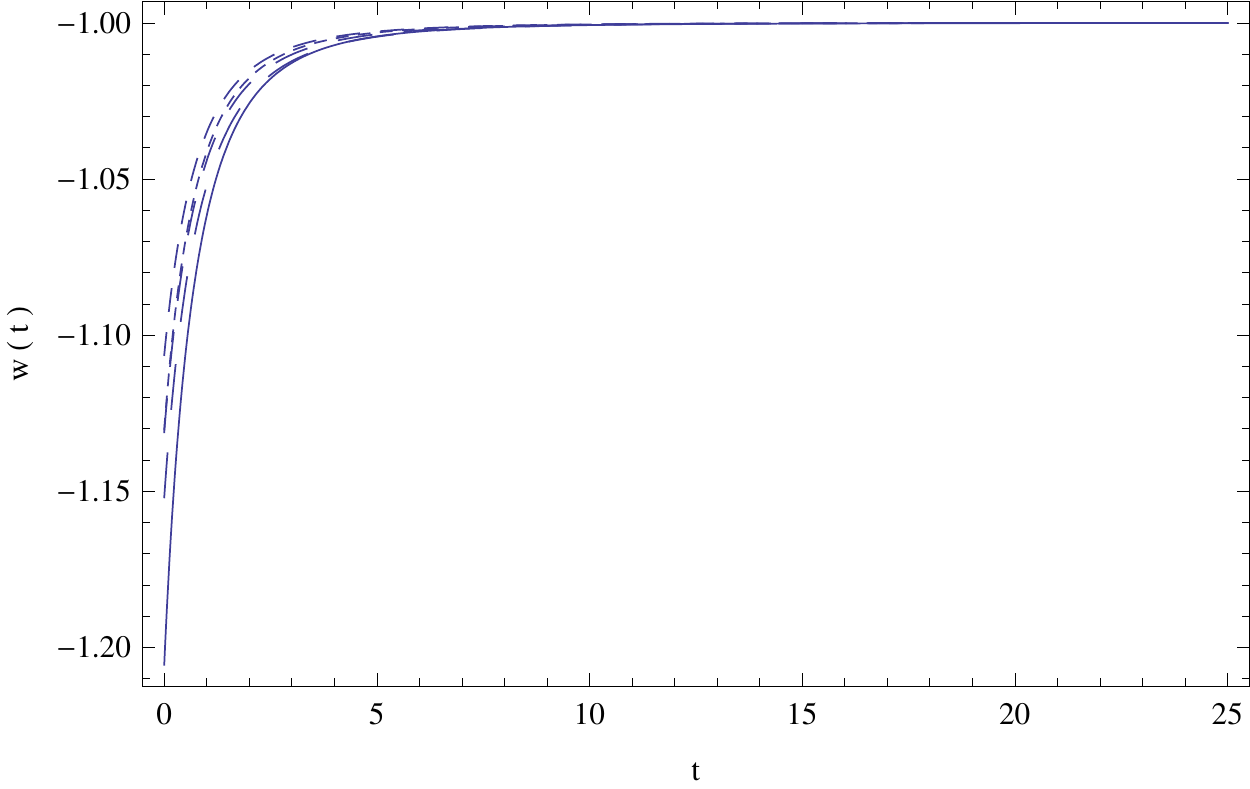}
  \caption{\it Time variation of the dark-energy equation-of-state parameter $w_{DE}$
for the non-minimally matter
coupled $f(T)$ gravity model with $f_1(T)=-\Lambda$ and $f_2(T)=\alpha
_1T+\beta _1T^2$, or equivalently  $f_1(H)=-\Lambda$ and $f_2(H)=\alpha
H^2+\beta H^4$ with $\alpha =-6\alpha_1$, $\beta =36\beta _1$,
for five different choices of the parameters $\Lambda$, $\alpha $, $\beta $,
and $\lambda $:
$\Lambda =0.01$, $\alpha =0.16$, $\beta =0.1$, and $\lambda =1$ (solid
curve), $\Lambda =0.02$, $\alpha =0.18$, $\beta =0.3$, and $\lambda =1.2$
(dotted curve), $\Lambda =0.03$, $\alpha =0.20$, $\beta =0.35$, and $\lambda
=1.4$ (short dashed curve), $\Lambda =0.04$, $\alpha =0.30$, $\beta =0.45$,
and $\lambda =1.6$ (dashed curve), and $\Lambda =0.05$, $\alpha =0.40$,
$\beta =0.55$, and $\lambda =1.8$ (long dashed curve), respectively. The
initial value for $H$ used to numerically integrate Eq. ~(\ref{fTLmmod22}) is
$H(0)=0.2$. From \cite{Harko:2014sja}. }\label{fTLmfig8a}
  \end{figure}
As we observe, for all parameter choices the Universe ends in a de Sitter phase. The
deceleration parameter indicates a dependence of the dynamical behavior of the Universe
on the model parameters. For the considered values, in all cases at the beginning of the
matter-dominated phase, the Universe is in a decelerating phase, with $q>0$, and after
a finite time it enters in the accelerated regime. Moreover, at large times the dark
energy equation-of-state parameter $w_{DE}$ tends to the value $-1$, thus showing that
this choice of the functions $f_1(T)$ and $f_2(T)$ can also successfully mimic an
effective cosmological constant. Note however, that for these specific parameter choices
$w_{DE}$ lies in the phantom regime, which is an advantage of the scenario at hand,
revealing its capabilities.

After the above numerical elaboration, we analyze the properties of the equations in the
limit of small and large $H(t)$, extracting analytical expressions.

In the limit of small $H(t)$, that is at the late phases of the cosmological evolution,
Eq. ~(\ref{fTLmmod22}) can be approximated
as
\be
\dot{H}\approx\frac{3 \Lambda }{2\left(6-\alpha  \lambda
\Lambda\right)}+\frac{3\left(\alpha^2  \lambda^2
\Lambda^2+6 \beta\lambda \Lambda^2-36\right)}{2\left(6-\alpha  \lambda
\Lambda\right)^2}H^2,
\ee
with the general solution given by
 \be
H(t)=\sqrt{H_{2}}\tan \left[ \tan ^{-1}\left(
\frac{H_{0}}{\sqrt{H_{2}}}\right) +    \frac{3 \Lambda  \left(
t-t_{0}\right)}{2\left(6-\alpha \lambda
\Lambda\right) \sqrt{H_{2}}  }      \right],
\label{fTLmHapprox22}
\ee
with $H_2=\Lambda(6-\alpha\lambda\Lambda)/\left(\alpha^2  \lambda^2
\Lambda^2+6 \beta\lambda \Lambda^2-36\right)$, and where we have used the
initial condition $H\left(t_0\right)=H_0$.
 Furthermore, the matter density (\ref{fTLmmod21}) becomes
\bea
 &&\!\!\!\!\!\!
 \rho _m(t)\approx  -\frac{\Lambda}{2}+\left(3-\frac{\alpha\lambda\Lambda}{2}\right)
H_2
  \nonumber\\
 && \!\!\!\!\! \!\times \tan^2 \left[ \tan ^{-1}\left(
\frac{H_{0}}{\sqrt{H_{2}}}\right) +    \frac{3 \Lambda  \left(
t-t_{0}\right)}{2\left(6-\alpha \lambda
\Lambda\right) \sqrt{H_{2}}  }      \right] .
\eea
Finally, the dark-energy equation-of-state parameter (\ref{fTLmwDE2}),
becomes
{\small{
  \bea
&&   w_{DE}(t)
\approx-1+\frac{\alpha\lambda\Lambda}{
\alpha\lambda\Lambda-6}  -\frac {
18\left(\alpha^2\lambda^2\Lambda+2\beta\lambda\Lambda-6\alpha\lambda
\right)}{ \left(6-\alpha\lambda\Lambda\right)^2}H_2
     \nonumber\\
 &&\ \ \ \ \ \ \ \ \
 \times   \tan^2 \left[ \tan ^{-1}\left(
\frac{H_{0}}{\sqrt{H_{2}}}\right) +    \frac{3 \Lambda  \left(
t-t_{0}\right)}{2\left(6-\alpha \lambda
\Lambda\right) \sqrt{H_{2}}  }      \right] ,
   \eea}}
which can lie both in the quintessence as well as in the phantom regime,
depending on the specific choices of the free parameters of the model,
namely on $\alpha $, $\beta $, $\lambda $ and $\Lambda $, respectively.

In the limit of large $H$, that is at early times, at first-order approximation Eq.
~(\ref{fTLmmod22}) becomes
$
\dot{H}-3 H^2/2=0,
$
with $H\left(t_0\right)=H_0$, and thus the general solution is given by
\be
H(t)= \frac{2 H_0}{2-3 H_0 \left(t-t_0\right)}.
\ee
Note that in this case the deceleration parameter is obtained as $q=-5/2$, that is the
universe at early times always starts with acceleration, which corresponds
to an inflationary stage. Finally, for the time variation of the matter
energy density in the large-$H$ regime we find $\rho_m(t)\approx
0$, which is consistent with the interpretation of this stage as inflationary.
We mention that the above expressions for $H$, $a$, $q$ and $\rho_m$ at
first-order approximation, are independent on the free parameters of the
model $\alpha $, $\beta $, $\lambda $ and $\Lambda $, respectively, and are
determined only by the initial value of $H$ at $t=t_0$.

\end{itemize}

In summary, as we saw, the theory of non-minimal torsion-matter coupling is a novel class
of gravitational modification, different from both non-minimal coupled curvature-matter
gravity, as well as from simple $f(T)$ gravity. From the physical point of view, in
this theory matter is not just a passive component in the space-time continuum, but it
plays an active role in the overall gravitational dynamics, which is strongly modified
due to the supplementary interaction between matter and geometry. Moreover, the major
advantage of the $f(T)$-type models, namely that the field equations are second order, is
not modified by the torsion-matter coupling.

We restricted our analysis in expanding evolutions, although contracting or bouncing
solutions can be easily obtained as well. In this case, one finds a universe
evolution in agreement with observations, that is a
matter-dominated era followed by an accelerating phase. Additionally, the effective
dark-energy equation-of-state parameter can lie in the quintessence or phantom regime,
which reveals the capabilities of the scenario. Finally, these models also accept
solutions with almost constant Hubble function, which can describe the inflationary
regime. Thus, the scenario of non-minimal torsion-matter coupling can offer a unified
description of the universe evolution, from its inflationary to the late-time accelerated
phases. Lastly, we mention that it is necessary to investigate the theory at hand in
more details, focusing on a thorough comparison with cosmological observations, and
performing the complete perturbation analysis.

\subsection{$f(T,\cal{T})$ gravity}
\label{FTTgravity}

In this subsection we describe another extension of $f(T)$ gravity, in which one allows
for a general coupling of the torsion scalar $T$ with the trace of the matter
energy-momentum tensor $\mathcal{T}$, following
 \cite{Harko:2014aja}. The inspiration for this extension comes once again from the
corresponding models of
curvature gravity.

As we mentioned in the previous subsection, in curvature gravity one can proceed to
modifications in which the geometric part of the action is coupled to the non-geometric
sector. The simplest models are those with non-minimally coupled and
non-minimal-derivatively coupled scalar fields, or more generally the Horndeski
 \cite{Horndeski:1974wa} and generalized Galileon theories  \cite{DeFelice:2010nf}.
However, since there is no theoretical reason against couplings between the gravitational
sector and the standard matter one, one can consider modified theories where the matter
Lagrangian is coupled to functions of the Ricci scalar  \cite{Bertolami:2007gv}, and
extend the theory to arbitrary functions $\left(R,\mathcal{L}_m\right)$
\cite{Harko:2008qz,Harko:2010mv}. Alternatively, one can consider models where the Ricci
scalar is coupled with the trace of the energy momentum tensor $\mathcal{T}$ and extend to
arbitrary functions, such as in $f(R,\mathcal{T})$ theory
\cite{Harko:2011kv,Momeni:2011am,Sharif:2012zzd,Alvarenga:2013syu,Shabani:2013djy}. We
stress that the above modifications, in which one handles the gravitational and matter
sectors on equal footing, do not present any problem at the theoretical level, and one
would only obtain observational constraints due to non-geodesic motion.

Hence, inspired by the  $f(R,\mathcal{T})$ theory, in  \cite{Harko:2014aja,Kiani:2013pba}
the authors constructed similar models in the framework of teleparallel and $f(T)$
gravity, that is allowing for a general coupling of the torsion scalar $T$ with the trace
of the matter energy-momentum tensor $\mathcal{T}$. In particular, one considers the
action
\begin{equation}
S= \frac{1}{16\,\pi\,G}\,\int d^{4}x\,e\,\left[T+f(T,\mathcal{T})\right]%
+\int d^{4}x\,e\,\mathcal{L}_{m},
\label{fTTaction1}
\end{equation}
where $f(T,\mathcal{T})$ is an arbitrary function of the torsion scalar $T$
and of the trace $\mathcal{T}$ of the matter energy-momentum tensor  $%
 T^{(m)}_{\ \ \ \rho}{}^{\nu}$, and  $\mathcal{L}_{m}$ is as usual the
matter Lagrangian density. Hereinafter, and following the standard
approach, we assume that $\mathcal{L}_{m}$ depends only on the
vierbein and not on its derivatives.

Varying the action, given by Eq. ~(\ref{fTTaction1}), with respect to the vierbeins yields
the field equations
\begin{eqnarray}
&&\left(1+f_{T}\right) \left[e^{-1} \partial_{\mu}{(e
e^{\alpha}_{A}
S_{\alpha}^{~\rho \mu})}-e^{\alpha}_{A} T^{\mu}_{~\nu \alpha} S_{\mu}^{~\nu
\rho}\right] \nonumber\\
&&
+\left(f_{TT} \partial_{\mu}{T}+f_{T\mathcal{T}} \partial_{\mu}{%
\mathcal{T}}\right) e^{\alpha}_{A} S_{\alpha}^{~\rho \mu}+ e_{A}^{\rho}
\left(\frac{f+T}{4}\right)
\notag \\
&&    -f_{\mathcal{T}} \left(\frac{e^{\alpha}_{A}
 \, T^{(m)}
{}_{\alpha}^{~~\rho}+p e^{\rho}_{A}}{2}\right)=4\pi G e^{\alpha}_{A}
\, T^{(m)}_{\ \ \ \alpha}{}^{\rho},
\label{fTTgeneoms}
\end{eqnarray}
where $f_{\mathcal{T}}=\partial{f}/\partial{\mathcal{T}}$ and
$f_{T\mathcal{T%
}}=\partial^2{f}/\partial{T} \partial{\mathcal{T}}$.

In order to apply the above theory in a cosmological framework, we insert
as usual the flat FRW vierbein ansatz (\ref{weproudlyuse}) into the
field equations \eqref{fTTgeneoms}, obtaining the modified Friedmann equations:
\begin{equation}
H^{2}=\frac{8\pi G}{3} \rho_m-\frac{1}{6}\left(f+12 H^{2} f_{T}\right)+f_{%
\mathcal{T}} \left(\frac{\rho_{m}+p_{m}}{3}\right),
\label{fTTFriedmann1}
\end{equation}
\begin{eqnarray}
\label{fTTF2}
 &&\!\!\!\!\!\!
 \dot{H}=-4\pi G \left(\rho_m+p_m\right)-\dot{H} \left(f_{T}-12 H^{2}
f_{TT}\right)   \nonumber\\
&&\ \ \ \   -H \left(\dot{\rho}_{m}-3\,\dot{p}_{m}\right)
f_{T\mathcal{T%
}}-f_{\mathcal{T}} \left(\frac{\rho_{m}+p_{m}}{2} \right).
\end{eqnarray}
We mention that in the above expressions we have used that
$\mathcal{T}=\rho_{m}-3\,p_{m}$, which holds in the case of a perfect matter
fluid.

Proceeding, we assume that the matter component of the Universe
satisfies a barotropic equation of state of the form $p_{m}=p_{m}\left(\rho
_{m}\right)$, with $w_m=p_m/\rho_m$ its equation-of-state parameter, and
$c_{s}^{2}=dp_{m}/d\rho _{m}$ the sound speed. Note that due to homogeneity
and isotropy, both $\rho_m$ and $p_m$ are function of $t$ only, and thus of
the Hubble parameter $H$. Thus, Eq. ~(\ref{fTTF2}) can be re-written
as
\begin{equation}
\dot{H}=-\frac{4\pi G\left( 1+f_{\mathcal{T}}/8\pi G\right) \left( \rho
_{m}+p_{m}\right) }{1+f_{T}-12H^{2}f_{TT}+H\left( d\rho _{m}/dH\right)
\left( 1-3c_{s}^{2}\right)f_{T\mathcal{T}} }.
\end{equation}

By defining the energy density and pressure of the effective dark energy
sector as
\begin{equation}
\rho _{DE}=-\frac{1}{16\pi G}\left[ f+12f_{T}H^{2}-2f_{\mathcal{T}}\left(
\rho _{m}+p_{m}\right) \right] ,  \label{fTTrhode}
\end{equation}%
{\small{
\begin{eqnarray}
&&\!\!\!\!\!\!\!\!\!\!\!\!\!\!\!p_{DE}\!= \! \left( \rho _{m}+p_{m}\right)
 \Big[ \frac{1+f_{\mathcal{T}}/8\pi G}{\!1+\!f_{T}\!-\!12H^{2}f_{TT}\!+\!H\frac{d\rho
_{m}}{dH} \left( 1-3c_{s}^{2}\right) f_{T\mathcal{T}}}
\nonumber\\
&&\ \ \ \ \
-1\Big]
   +\frac{1}{16\pi G}\left[ f+12H^{2}f_{T}-2f_{\mathcal{T}}\left( \rho
_{m}+p_{m}\right) \right] ,
 \label{fTTpde}
\end{eqnarray}}}
respectively,  the cosmological field equations of the $f(T,\mathcal{T})$
theory are rewritten in the usual form
\begin{eqnarray}
H^{2} &=&\frac{8\pi G}{3}\left( \rho _{DE}+\rho _{m}\right) ,  \label{fTTFr2} \\
\dot{H} &=&-4\pi G\left( \rho _{DE}+p_{DE}+\rho _{m}+p_{m}\right) .
\end{eqnarray}
Furthermore, we define the dark energy equation-of-state parameter as
\begin{equation}
w_{DE}=\frac{p_{DE}}{\rho _{DE}},
\label{fTTwDE1}
\end{equation}
and it proves convenient to introduce also the total equation-of-state parameter $w$,
given by
\be
w=\frac{p_{DE}+p_m}{\rho _{DE}+\rho _m}.
\ee
Note that in the case of the dust universe, with $p_m=0$, we have $w=w_{DE}/\left(1+\rho
_m/\rho _{DE}\right)$. Finally, as an indicator of the accelerating dynamics of the
Universe we use the deceleration parameter $q=-1-\dot{H}/H^2$.

As we can see from Eqs. ~(\ref{fTTFr2}), the matter energy density and pressure, and the
effective dark energy density and pressure, satisfy the
conservation equation
\begin{equation}
\dot{\rho} _{DE}+\dot{\rho }_m+3H\left(\rho _m+\rho
_{DE}+p_m+p_{DE}\right)=0.
\end{equation}
Thus, one obtains an effective interaction between the dark energy and
matter sectors, which is usual in modified matter coupling theories
 \cite{Harko:2008qz,Harko:2010mv,Harko:2012hm,Wang:2012rw,Harko:2014gwa,
Harko:2011kv}.  Therefore, in the present model the effective dark energy is not
conserved alone, and there is an effective coupling between dark energy and  normal
matter, with the possibility of energy transfer from one component to the other. The dark
energy alone satisfies the ``conservation'' equation
\begin{equation}
\dot{\rho} _{DE}+3H\left(\rho_{DE}+p_{DE}\right)=-Q\left(\rho _m,p_m\right),
\end{equation}
where the effective dark energy ``source'' function $Q \left(\rho _m,p_m\right)$ is
\be
Q \left(\rho _m,p_m\right)=\dot{\rho} _m+3H\left(\rho _m+p_m\right).
\ee
Hence, in the present model it is allowed to have an energy transfer from ordinary matter
to dark energy (which, even geometric in its origin, contains a matter contribution), and
this process may be interpreted in triggering the accelerating expansion of the universe.

As we observe, the above equations of $f(T,\cal{T})$ gravity are different than those of
$f(R,\cal{T})$ gravity  \cite{Harko:2011kv}, and obviously they are different from simple
$f(T)$ gravity too. Hence, the theory of $f(T,\cal{T})$ gravity is a novel class of
gravitational modification.

\subsubsection{Cosmological behavior}

Let us now study the cosmological applications of $f(T,\cal{T})$ gravity, both at early
and late times, following  \cite{Harko:2014aja} (see also
\cite{Kiani:2013pba,Momeni:2014jja,Junior:2015bva,Nassur:2015zba,Salako:2015xra}). In
particular, we proceed to the investigation of some specific $f(T,\mathcal{T})$ ansatzen,
focusing on the evolution of observables such as the various density parameters
$\Omega_i=8\pi G\rho_i/(3H^2)$ and the dark energy equation-of-state parameter $w_{DE}$.
For convenience, in this paragraph we use the natural system of units with $8\pi G=c=1$.

From the analysis above we saw that the basic equations
describing the cosmological dynamics are the two Friedmann equations
(\ref{fTTFriedmann1}) and (\ref{fTTF2}). These can be re-written respectively as
\begin{equation}
\rho _{m}=\frac{3H^{2}+\left( f+12H^{2}f_{T}|_{T\rightarrow -6H^{2}}\right)
/2-f_{\mathcal{T}}p_{m}}{1+f_{\mathcal{T}}},  \label{fTTeq1}
\end{equation}%
and
{\small{
\begin{equation}
\dot{H}= -\frac{\left( 1+f_{\mathcal{T}}\right) \left( \rho
_{m}+p_{m}\right) /2+H\left( \dot{\rho}_{m}-3\,\dot{p}_{m}\right) f_{T%
\mathcal{T}}|_{T\rightarrow -6H^{2}}}{1+f_{T}|_{T\rightarrow
-6H^{2}}-12H^{2}f_{TT}|_{T\rightarrow -6H^{2}}}.  \label{fTTeq2}
\end{equation}}}
Equations (\ref{fTTeq1}) and (\ref{fTTeq2}) compose a system of two
differential equations for three unknown functions, namely $\left(H,\rho
_{m},p_{m}\right) $. In order to close the system of equations one needs to
impose the matter equation of state   $p_{m}=p_{m}\left(\rho
_{m}\right)$. In the following, we restrict to the case of dust
matter, that is $p_{m}=0$, and thus $\mathcal{T}=\rho _{m}$.

We will investigate two specific  $f(T,\mathcal{T})$ models,
corresponding to simple non-trivial extensions of TEGR, that is of GR.
However, although simple, these models  reveal the new features and the
capabilities of the theory.

In order to relate the model with cosmological observations we will present the results
of the numerical computations for the Hubble function, matter energy density,
deceleration parameter and the parameter of the dark energy equation of state as
functions of the cosmological redshift $z$, defined as $z=a_0/a-1$, where $a_0$ is the
present day value of the scale factor, which we take as one, namely $a_0=1$. In terms of
the redshift the derivatives with respect to time are expressed as $
d/dt=-(1+z)H(z)d/dz$.
Finally, in order to numerically integrate the gravitational field equations one needs to
fix the value of the Hubble function at $z=0$, $H(0)=H_0$.

In the following, we investigate the  scenario at hand, for two different functional
forms of $f\left(T,\mathcal{T}\right)$.

\begin{itemize}
 \item{Model I}

Let us first consider the case where   \cite{Harko:2014aja}
 \begin{equation}
f\left(T,\mathcal{T}\right)=\protect\alpha \protect
T^n\,\mathcal{T}+\Lambda,
 \end{equation}
which describes a simple departure from GR, where $\alpha $, $n\neq
0$ and $\Lambda $ are arbitrary constants. In the case of a dust perfect fluid, this
ansatz becomes $f\left(T,\mathcal{T}\right)=\alpha T^n \rho _m+\Lambda$.
One thus  obtains straightforwardly
 $f=\alpha \left(-6H^2\right)^n\rho _m +\Lambda$,    $f_T=n\alpha
\rho_m\left(-6H^2\right)^{n-1}$, $f_{TT}=\alpha
n(n-1)\left(-6H^2\right)^{n-2}$, $f_{T\mathcal{T}}=\alpha n \left(-6H^2\right)^{n-1}$,
and
$f_{\mathcal{T}}=\alpha \left(-6 H^2\right)^n$.
Hence, inserting these into  (\ref{fTTeq1}) one can acquire the matter energy density as
a function of the Hubble function as
\begin{equation}
\label{fTTeqrA}
\rho _m=\frac{3H^2+\Lambda /2}{1+\alpha (n+1/2)\left(-6 H^2\right)^n}.
\end{equation}
Therefore, substituting the above expression into  (\ref{fTTeq2})  and (\ref{fTTwDE1}),
we extract the time-variation of the Hubble function, and of the dark-energy
equation-of-state parameter, as functions of $H$, namely
\begin{widetext}
\begin{eqnarray}
&\label{fTTeqHAb}
\dot{H}=-\frac{3H^{2}\left( 6H^{2}+\Lambda \right) \left[ \alpha 6^{n}\left(
-H^{2}\right) ^{n}+1\right] \left[ \alpha 6^{n}(2n+1)\left( -H^{2}\right)
^{n}+2\right] }{\alpha ^{2}36^{n}(2n+1)\left( -H^{2}\right) ^{2n}\left[
6(n+1)H^{2}+\Lambda n\right] -\alpha 2^{n+1}3^{n}\left( -H^{2}\right) ^{n}%
\left[ 6(n-2)(2n+1)H^{2}+\Lambda n(2n-1)\right] +24H^{2}},
\end{eqnarray}
and
\begin{eqnarray}
\label{fTTwn}
&w_{DE}=-
\frac{3H^{2}\left[ \alpha 6^{n}(2n+1)\left( -H^{2}\right) ^{n}+2\right]
\left\{ \alpha _{1}\alpha _{3}\left( -H^{2}\right) ^{n}H^{2}+\alpha
_{4}-\alpha _{2}\left( -H^{2}\right) ^{2n}\left[ 6(n-1)H^{2}+\Lambda (n-2)%
\right] +4\Lambda \right\} }{\left[ \alpha _{1}(2n+1)\left( -H^{2}\right)
^{n+1}+\Lambda \right] \left\{ \alpha _{2}\left( -H^{2}\right) ^{2n}\left[
6(n+1)H^{2}+\Lambda n\right] -\alpha _{1}\left( -H^{2}\right) ^{n}\left[
\alpha _{5}H^{2}+\alpha _{6}\right] +24H^{2}\right\} },
\end{eqnarray}
\end{widetext}
where for convenience we have defined the parameters $\alpha _{1}=\alpha
2^{n+1}3^{n}$, $\alpha _{2}=\alpha
^{2}36^{n}\left( 2n+1\right) $, $\alpha_{3}=6[n(2n-1)+1]$, $\alpha
_{4}=\Lambda \left( 2n^{2}+n+3\right) $, $\alpha _{5}=6(n-2)(2n+1)$, and $
\alpha _{6}=\Lambda n(2n-1)$. Note that relations (\ref{fTTeqrA}) and (\ref{fTTeqHAb})
hold for every $\alpha$, including $\alpha=0$ (in which case we obtain the GR
expressions), while  (\ref{fTTwn}) holds for $\alpha\neq0$, since for $\alpha=0$ the
effective dark energy sector does not exist at all (both $\rho_{DE}$ and
$p_{DE}$ are zero).

$\quad$ In the simplest case $n=1$, that is for $f\left(T,\mathcal{T}\right)= \alpha
T \mathcal{T}=\alpha T \rho _m+\Lambda$, from (\ref{fTTwn}) we can immediately see the
interesting feature that $w_{DE}$ can be quintessence-like or phantom-like, or even
experience the phantom-divide crossing during the evolution, depending on the choice of
the parameter range. This feature is an additional advantage of the scenario, since such
behaviors are difficult to be obtained in simple dark-energy constructions. In order to
present the above features in a more transparent way, one proceeds to a detailed
numerical elaboration for various parameter choices. In Fig. \ref{fTTfig4} one depicts the
corresponding evolution of the dark-energy equation-of-state parameter.
\begin{figure}[ht]
\centering
\includegraphics[width=8cm]{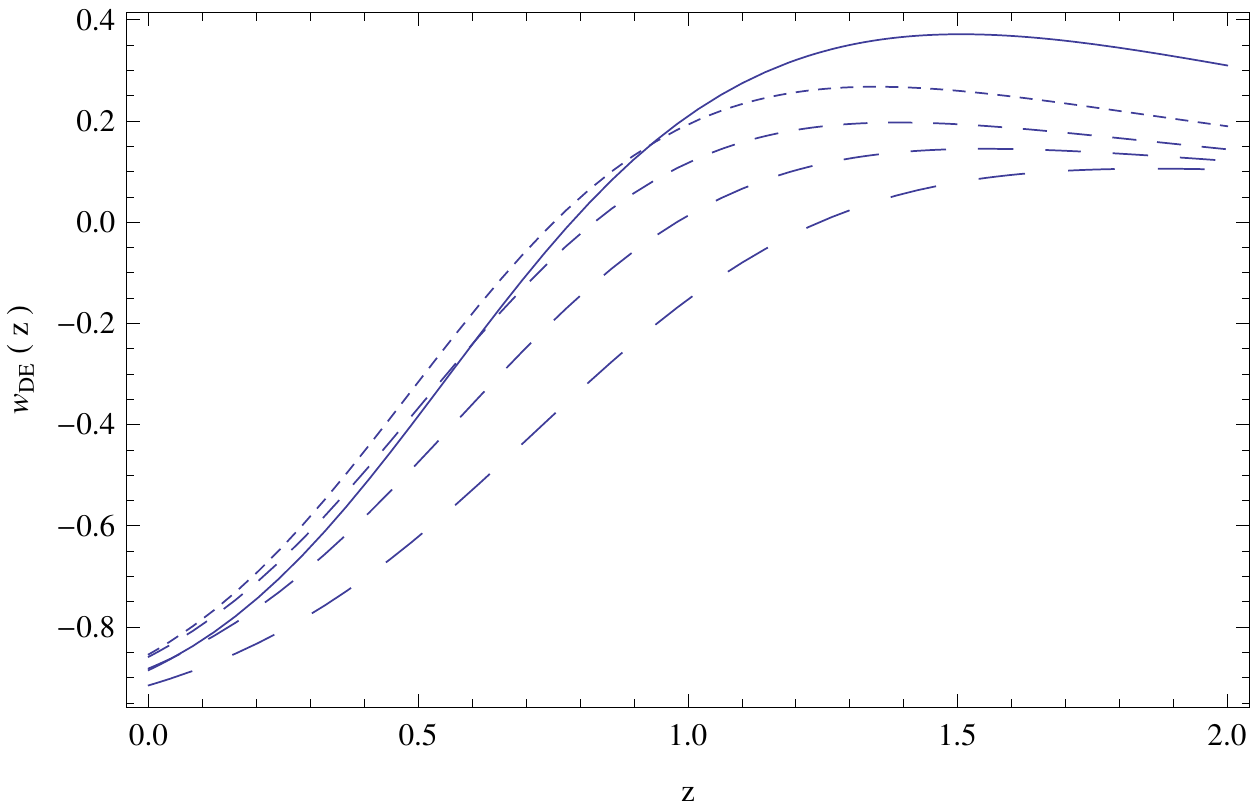}
\caption{\it Variation of the dark-energy equation-of-state parameter
$w_{DE}(z)$ as a function of the redshift $z$ for the model
$f(T,\mathcal{T})=\protect\alpha \protect\mathcal{T}T+\Lambda $, for five
different choices of the parameters values:
  $\protect\alpha  H_0^2=-0.01$, $\Lambda/H_0^2 =-3$ (solid curve), $\protect\alpha  H_0^2
=-0.02$, $\Lambda/H_0^2 =-3.5$ (dotted curve), $%
\protect\alpha  H_0^2=-0.03$, $\Lambda/H_0^2 =-4$ (short-dashed curve), $\protect\alpha
H_0^2
=-0.04$, $\Lambda/H_0^2 =-4.5$ (dashed
curve), and $\protect\alpha  H_0^2=-0.05$, $\Lambda/H_0^2 =-5$ (long-dashed curve),
respectively, with   $H_0\approx
2.3\times 10^{-18}$ s$^{-1}$ the present value of the Hubble parameter
 \cite{Ade:2013uln}. From  \cite{Harko:2014aja}.}
\label{fTTfig4}
\end{figure}
As we can see, depending on the values of the parameters $\alpha$ and $\Lambda$, the
Universe can exhibit a very interesting dynamics. In particular, the dust-filled Universe
starts its evolution from a decelerating state, entering in an accelerating phase at
around $z\approx 0.5$, a behavior which is in agreement with the observed behavior of the
recent Universe. Note that at asymptotically large times $w_{DE}$ tends towards negative
values, and the Universe ends in a de Sitter, dark-energy-dominated, expansion, with its
dynamics dominated by the effective dark energy component, mimicking a cosmological
constant. Indeed, one can verify that at large times, and for $\Lambda < 0$, relation
(\ref{fTTeqHAb}) leads to $q=-1$, $H=H_0=\sqrt{\Lambda /6}$ and $a\propto
\exp\left(H_0t\right)$ (however, for $\alpha >0$  the positivity of the
matter energy density constrains the $\alpha$-values in the region that leads to
$9\alpha H^2<1$  \cite{Harko:2014aja}).

$\quad$ In the general case $n\neq 1$, and in order to investigate the effects of
different $n$, one evolves numerically the cosmological equations
(\ref{fTTeqrA})-(\ref{fTTwn}), for fixed $\Lambda/H_0^2$ and $\alpha  H_0^2$, and varying
$n$. In Figs. \ref{fTTfig3a2} and \ref{fTTfig4a} one can see the evolution of the
deceleration parameter and of the dark-energy equation-of-state parameter, respectively,
for $n=1,2,3,4,5$.
  \begin{figure}[ht]
\centering
\includegraphics[width=8cm]{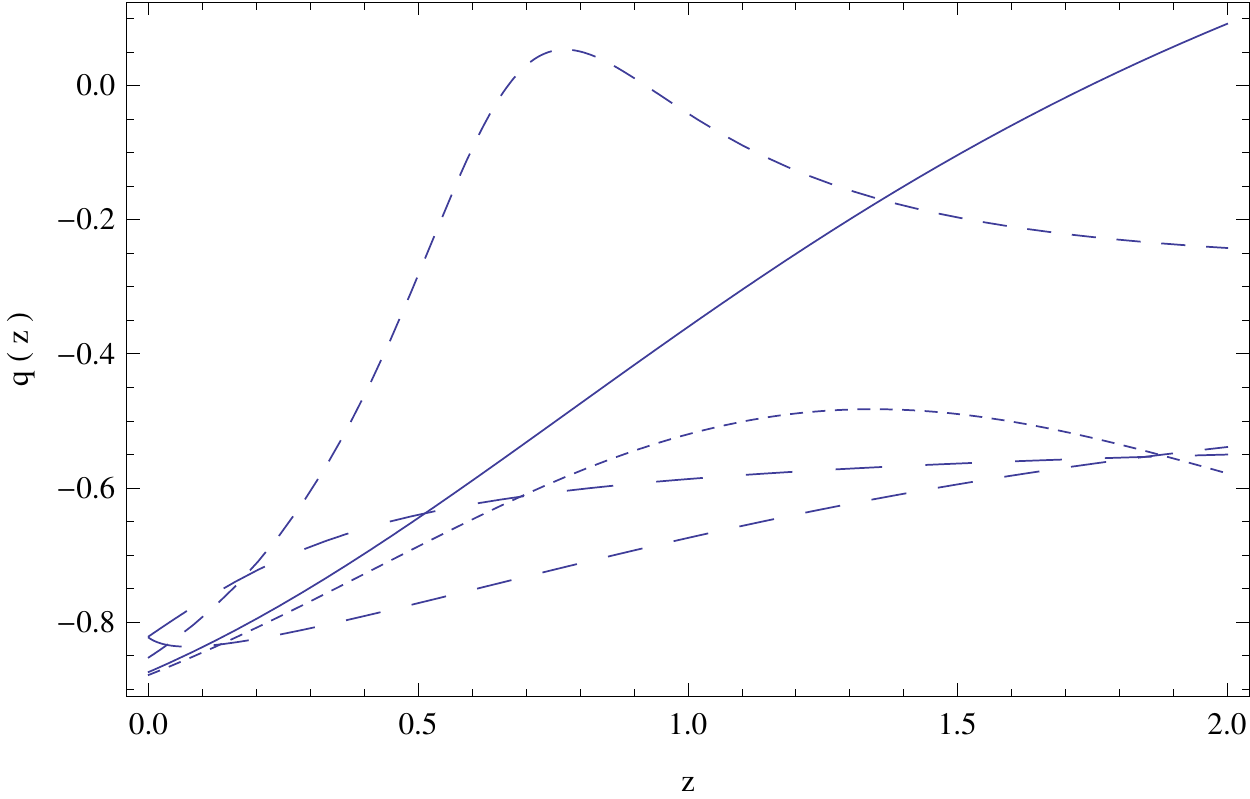}
\caption{\it Variation of the deceleration parameter $q(z)$ as a function of the redshift
$z$
in the
$f(T,\mathcal{T%
})$ gravity theory with $f(T,\mathcal{T})=\protect\alpha \protect\rho
_mT^n+\Lambda $,
for $\alpha  H_0^2=-0.0011$, $\Lambda/H_0^2 =-5.5$, and for five different values of
$n$:
$n=1$ (solid curve), $n=2$ (dotted curve),
 $n=3$ (short-dashed curve),  $n=4$ (dashed
curve), and $n=5$, respectively. From  \cite{Harko:2014aja}.}
\label{fTTfig3a2}
\end{figure}
\begin{figure}[!]
\centering
\includegraphics[width=8cm]{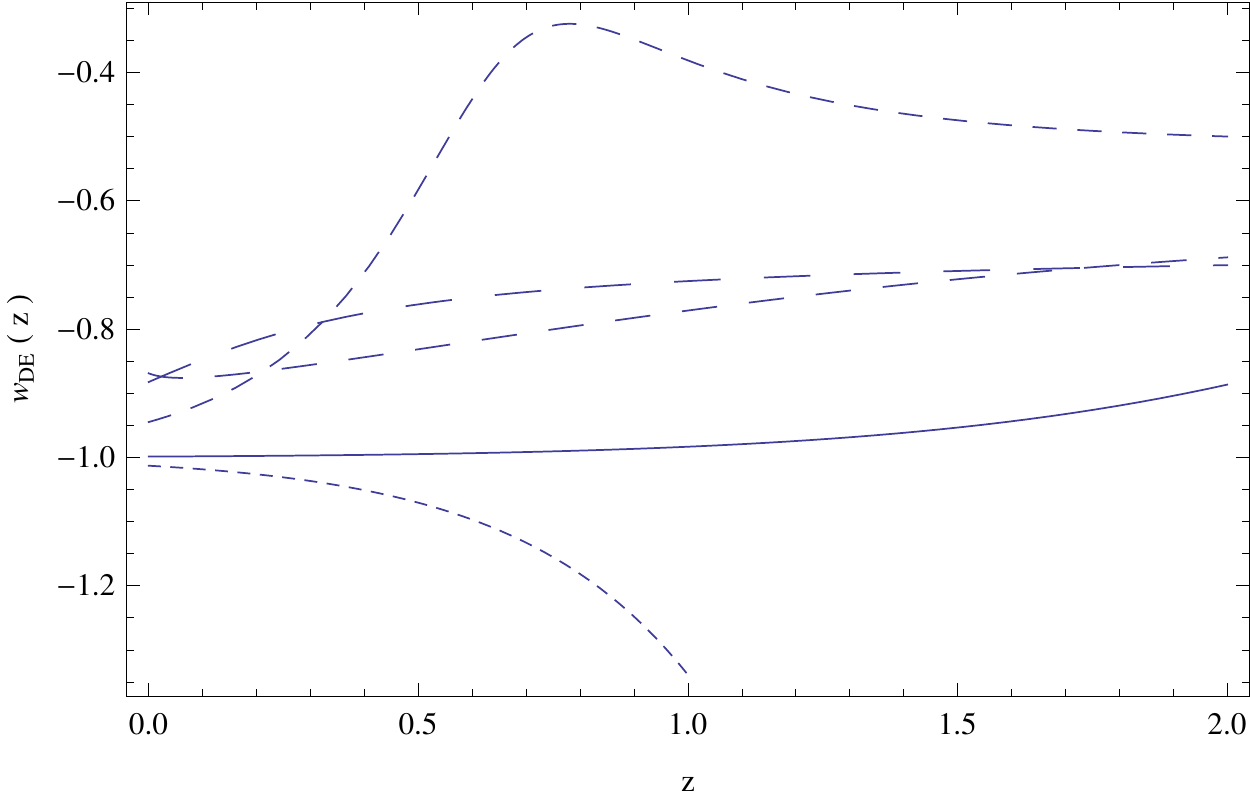}
\caption{\it Variation of  the dark-energy equation-of-state parameter
$w_{DE}(z)$ as a function of the redshift $z$ in the $f(T,\mathcal{T%
})$ gravity theory with $f(T,\mathcal{T})=\protect\alpha \protect\rho
_mT^n+\Lambda $,
for $\alpha  H_0^2=-0.0011$, $\Lambda/H_0^2 =-5.5$, and for five different values of
$n$:
$n=1$ (solid curve), $n=2$ (dotted curve),
 $n=3$ (short-dashed curve),  $n=4$ (dashed
curve), and $n=5$, respectively. From  \cite{Harko:2014aja}.}
\label{fTTfig4a}
\end{figure}
Interestingly enough, we observe that the dynamics of the Universe is very different for
different values of $n$, as can be revealed by the behavior of the deceleration
parameter. In particular, while for $n=1$ the Universe starts its evolution from a
decelerating phase, followed by an accelerating one, and ends in an eternally
accelerating de Sitter phase, for $n>1$ all cosmological models begin their evolution in
an accelerating  phase, with $q<0$ at $z=2$, before entering in a de Sitter exponential
expansion ($q=-1$) at $z=0$. However, the models with $n>1$ exhibit a radical difference
in the behavior of the dark energy sector, which is visible in the evolution of $w_{DE}$.
Specifically, $w_{DE}$ can lie in the quintessence or phantom regime, depending on the
value of $n$. Thus, models that present a similar behavior in the global dynamics, can be
distinguished by the behavior of the dark energy sector. Nevertheless, note that at late
times $w_{DE}\rightarrow-1$ independently of the value of $n$, and thus in order to
distinguish the various models one should use $w_{DE}$ at large redshifts.

We mention that, as can be deduced from Eqs. ~(\ref{fTTeqrA}) and (\ref{fTTeqHAb}),
independently of $n$, once the condition $\alpha (n+1/2)\left(-6 H^2\right)^n \ll 1$ is
satisfied, for $\Lambda \neq 0$ the Universe results in the de Sitter accelerating stage,
while for $\Lambda =0$
its evolution ends in the Einstein--de Sitter, matter-dominated decelerating phase.
Furthermore, from Fig.~(\ref{fTTfig4a}) notice the interesting behavior that $w_{DE}$
can be either quintessence-like or phantom-like. This is easily
explained by recalling that $w=w_{DE}/\left(1+\rho _m/\rho _{DE}\right)$, and thus
according to the signs of $\rho _{DE}$ and $p_{DE}$ all combinations are possible.
Finally, note the interesting evolution for $n=3$, where an initial accelerating phase
is followed by a decelerating one, with a subsequent transition to a final acceleration
at
late times, a behavior in agreement with the observed thermal history of the Universe,
namely a first inflationary stage, a transition to non-accelerating, matter-dominated
expansion, and then the transition to late-time accelerating phase. Thus,
the scenario at hand offers a unified description of the universe evolution.

\item{Model II}

Let us now consider the case where  \cite{Harko:2014aja}
 \begin{eqnarray}
f(T,\mathcal{T})=\protect\alpha \mathcal{T}+
\gamma T^{2},
\end{eqnarray}
  which describes another simple departure from GR, where $\alpha $ and
$\gamma$  are constants. In the case of dust perfect fluid, this ansatz
becomes
$f(T,\mathcal{T})=\alpha \rho _{m}+\gamma T^{2}=\alpha \rho
_{m}+\beta H^{4}$, with $\beta=36\gamma $.  In
this
case we obtain $f_{T}=\beta T/18=-\beta H^{2}/3$, $f_{TT}=\beta /18$, $f_{%
\mathcal{T}}=\alpha $, and $f_{T\mathcal{T}}=0$, respectively. Thus,
inserting these into  (\ref{fTTeq1}) one can acquire the matter energy density in terms
of the Hubble function as
\begin{equation}
\rho _{m}=\frac{3\left( 1-\beta H^{2}/2\right) H^{2}}{1+\alpha /2},
\label{fTT1B}
\end{equation}%
while the time variation of the Hubble function (\ref{fTTeq2}) yields
\begin{equation}
\dot{H}=-\frac{3\left( 1+\alpha \right) }{\alpha +2}\frac{\left( 1-\beta
H^{2}/2\right) H^{2}}{1-\beta H^{2}}.  \label{fTT2B}
\end{equation}
Additionally, the effective dark energy density (\ref{fTTrhode}) and pressure
(\ref{fTTpde}), become
\be
\label{fTT528}
\rho _{DE}=\frac{3 H^2 \left(\alpha +\beta  H^2\right)}{\alpha +2},
\ee
\be
p_{DE}=-\frac{3 H^2 \left(\alpha +\beta  H^2\right)}{(\alpha +2) \left(\beta
H^2-1\right)},
\ee
and therefore the dark-energy equation-of-state parameter (\ref{fTTwDE1}) writes as
\be
w_{DE}=\frac{1}{1-\beta  H^2}.
\ee
Finally, let us make a comment on the limiting behavior of the model at hand. First of
all, the positivity of the matter energy-density implies that for positive values of
$\alpha $ and $\beta $ we must have $\beta H^{2}/2<1$. Moreover, for $\alpha <-2$ no
negative values of $\beta $ are allowed, and the Hubble function must satisfy the
constraint $\beta H^{2}/2\geq 1$. For small $H$, that is at late times, and in particular
for the time interval of the cosmological evolution for which $\beta H^{2}/2 \ll 1$, the
Hubble function satisfies the equation $\dot{H}\approx -3\left( 1+\alpha \right)
H^{2}/\left( \alpha +2\right) $, giving $H=\left[ \left( \alpha +2\right) /3\left(
1+\alpha \right) \right] \left( 1/t\right) $, $a\propto t^{\left( \alpha +2\right)
/3\left( 1+\alpha \right) }$, and $q\approx \left( 1+2\alpha \right) /\left( \alpha
+2\right) $. Thus, the deceleration parameter is negative for $\alpha \in \left(
-2,-1/2\right) $, however the accelerating phase is not of a de Sitter type, but it is
described by a simple power-law expansion.

\end{itemize}

\subsubsection{Scalar perturbations and stability analysis}
\label{perturb0}

One of the most important tests in every gravitational theory is the investigation of the
perturbations  \cite{Mukhanov:1990me}. Firstly, such a study reveals the stability
behavior of the theory. Secondly, it allows the correlation of the gravitational
perturbations with the growth of matter overdensities, and thus one can use growth-index
data in order to constrain the parameters of the scenario. In this paragraph we examine
the scalar perturbations of $f(T,\mathcal{T})$ gravity at the linear level following
\cite{Harko:2014aja}. Specifically, one extracts the set of gravitational and
energy-momentum-tensor perturbations and using them he examines the stability.
Additionally, one constructs the equation for the growth of matter overdensities.

As usual in theories where the fundamental field is the vierbein, one imposes a vierbein
perturbation, which will then lead to the perturbed metric. Without loss of generality we
present the calculations in the Newtonian gauge. Denoting the perturbed vierbein with
${e}_{\mu}^A$ and the unperturbed one with $\bar{e}_{\mu}^A$, the scalar perturbations,
keeping up to first-order terms, write as
\begin{eqnarray}
e_{\mu}^A = \bar{e}_{\mu}^A + t_{\mu}^A,
\end{eqnarray}
with
\begin{eqnarray}
\label{fTTpert1a}
\!\!\!\!\!\!\!\!\!\!
&&\bar{e}_{\mu}^0 = \delta_{\mu}^0,\, \,\,\,\,
\bar{e}_{\mu}^a = \delta_{\mu}^aa,\,\,\,\,\,
\bar{e}^{\mu}_0 = \delta^{\mu}_0, \,\,\,\,\, \bar{e}^{\mu}_a =
\frac{\delta^{\mu}_a}{a},
\\ \!\!\!\!\!\!\!\!\!\!
&&t_{\mu}^0 = \delta_{\mu}^0\psi, \,\,\,\, t_{\mu}^a
=-\delta_{\mu}^a a\phi, \,\,\,\, t^{\mu}_0 =
-\delta_{0}^{\mu}\psi, \,\,\,\, t^{\mu}_a =
\frac{\delta^{\mu}_a}{a}\phi. \ \ \ \  \ \label{fTTpert2a}
\end{eqnarray}
Note the simplifying assumption that the scalar perturbations $t_{\mu}^A$ are diagonal,
which is sufficient in order to study the stability. Furthermore, subscripts zero and one
denote zeroth and linear order values respectively. In the above expressions the scalar
modes $\psi$ and $\phi$ were introduced, which depend on $\bx$ and $t$. The various
coefficients have been considered in a way that the induced metric perturbation to have
the usual form in the Newtonian gauge, that is
\begin{eqnarray}
\label{fTTpertmetric}
 ds^2 = (1 + 2\psi)dt^2
-a^2(1-2\phi)\delta_{ij}dx^idx^j.
\end{eqnarray}

Let us now calculate the various perturbed quantities under the
perturbations
(\ref{fTTpert1a}) and (\ref{fTTpert2a}). Firstly, the vierbein
determinant reads
\begin{eqnarray}
e = \textrm{det}(e_{\mu}^A) = a^3(1+\psi - 3\phi).
\end{eqnarray}

Similarly, the torsion tensor  $T^{\lambda}{}_{\mu\nu}$ from (\ref{torsiontensor}) and
the
auxiliary
tensor $S_{\lambda}{}^{\mu\nu}$ from (\ref{Stensor}) read (indices are not summed over):
\begin{eqnarray}
&&T^{0}{}_{\mu\nu} = \partial_{\mu}\psi\delta_{\nu}^0 -
\partial_{\nu}\psi\delta_{\mu}^0,\ \ \ \ T^{i}{}_{0i} = H -
\dot{\phi} \nonumber\\
&&S_{0}{}^{0i} = \frac{\partial_i \phi}{a^2},\ \ \ \
S_{i}{}^{0i} = -H + \dot{\phi} + 2H\psi \nonumber\\
&& T^{i}{}_{ij} = \partial_j\phi ,\ \ \ \  S_{i}{}^{ij} =
\frac{1}{2a^2}\partial_j (\phi - \psi).
\end{eqnarray}
Thus, the torsion scalar can be straightforwardly calculated using the
(\ref{telelag}), leading to
\begin{eqnarray}
T =T_0+\delta T,
\end{eqnarray}
where
\begin{eqnarray}
&& T_0=-6H^2\\
&& \delta T=12H(\dot{\phi}+H\psi)
\end{eqnarray}
are respectively the zeroth and first order results.

Having performed the perturbations of the gravitational sector we proceed to
the perturbations of the energy-momentum tensor. As usual they are expressed
as
\begin{eqnarray}
\label{fTTT00pert}
\delta  T^{(m)}_{\ \ \ 0}{}^0 &=& \delta\rho_m\\
\delta  T^{(m)}_{\ \ \ 0}{}^i &=& (\rho_m + p_m)\partial^i
\delta v\\
\delta  T^{(m)}_{\ \ \ i}{}^0 &=& -a^{2} (\rho_m +
p_m) \partial_i\delta v \label{fTTTa0pert}
\\
\delta  T^{(m)}_{\ \ \ i}{}^j &=& -\delta^{j}_{i}\delta
p_m-\partial_i\partial^j\pi^{S}, \label{fTTTabpert}
\end{eqnarray}
where $\delta{\rho_{m}}$, $\delta{p_{m}}$, $\delta{v}$ are respectively
the fluctuations of energy density, pressure and fluid velocity, while
$\pi^S$ is the scalar component of the anisotropic stress. Additionally,
since $\mathcal{T}\equiv
T^{(m)}_{\ \ \ \mu}{}^{\mu}=T^{(m)}_{\ \ \ 0}{}^{0}+T^{(m)}_{\ \ \ i}{}^{i}$,  one
concludes that
\begin{eqnarray}
\mathcal{T} =\mathcal{T}_0+\delta \mathcal{T},
\end{eqnarray}
with
\begin{eqnarray}
&& \mathcal{T}_0=\rho_{m}-3p_{m}\\
&& \delta \mathcal{T}=\delta\rho_{m}-3\delta p_{m}-\nabla^{2}{\pi^{S}},
\end{eqnarray}
where we have defined $\nabla^2 =\sum_i \partial_i\partial^i$.

Finally, one can express the variations of the various $f$-derivatives that
appear in the background equations of motion as:
\begin{eqnarray}
&&\delta f=f_T\delta
T+f_{\cal{T}}\delta \cal{T}\nonumber\\
 &&\delta f_T=f_{TT}\delta T+f_{T\cal{T}}\delta \cal{T}\nonumber\\
  &&\delta f_{TT}=f_{TTT}\delta T+f_{TT\cal{T}}\delta \cal{T}\nonumber\\
  &&\delta f_{\cal{T}}=f_{T\cal{T}}\delta T+f_{\cal{T}\cal{T}}\delta
\cal{T}\nonumber\\
  &&\delta f_{T\cal{T}}=f_{TT\cal{T}}\delta
T+f_{T\cal{T}\cal{T}}\delta \cal{T},
\end{eqnarray}
 where the various $f$-derivatives are calculated at the background values
$T_0$ and $\mathcal{T}_0$, for instance  $f_T\equiv
\frac{df}{dT}\Big|_{T=T_0, \mathcal{T}=\mathcal{T}_0}$.

Inserting everything in the equations of motion \eqref{fTTgeneoms},  we
acquire the scalar perturbation equations  \cite{Harko:2014aja}:
\begin{widetext}
\begin{eqnarray}
\label{fTTeqHA}
&&
\left(1+f_{T}\right)\left[\frac{\nabla^{2}{\phi}}{a^{2}}-6H\left(\dot{\phi}
+H\psi\right)\right]
%\nonumber\\
%&&
+\left[3H^{2}f_{TT}+\frac{1+f_{T}}{4}-\frac{
\left(\rho_ { m }
+p_{m}\right)f_{T\mathcal{T}}}{2}\right]
%\nonumber\\
%&&
%\ \
%\cdot
\left[12H(\dot{\phi}
+H\psi)\right ]
\notag \\
&&+
\left[3H^{2}f_{T\mathcal{T}}+\frac{f_{\mathcal{T}}}{4}-\frac{\left(\rho_{m}
+p_{m}\right)f_{\mathcal{T}\mathcal{T}}}{2}\right]
%\nonumber\\
%&&
%\ \
%\cdot
\left(\delta{\rho_{
m}}-3 \delta{p_{m}}-\nabla^{2}{\pi^{S}}\right)
\nonumber\\
&&
-\frac{f_{\mathcal{
T } } } { 2 } \left(\delta{\rho_{m}}+\delta{p_{m}}\right)=4\pi G
\,\delta{\rho_{m}},
\end{eqnarray}
\begin{eqnarray}
&&
-\left(1+f_{T}\right)
\partial^{i}{\left(\dot{\phi}+H\psi\right)}
%\nonumber\\
%&&
+\left[12H\dot{H}f_{TT}
-\left(\dot{\rho}_{m}-3\dot{p}_{m}\right)f_{T\mathcal{T}}\right]\partial^{i}{
\phi}\nonumber\\
&&
-\frac{a^{2}f_{\mathcal{T}}}{2}\left(\rho_{m}+p_{m}\right)
\partial^{i}{\delta{v}}
=4\pi G a^{2}\left(\rho_{m}+p_{m}\right) \partial^{i}{\delta{v}},
\end{eqnarray}
\begin{eqnarray}
&&
-\left(1+f_{T}\right)
\partial_{i}{\left(\dot{\phi}+H \psi\right)}
%\nonumber\\
%&&
+H \partial_{i}\left[12 H f_
{TT} \left(\dot{\phi}+H \psi\right)
%\right.
%\nonumber\\
%&&
%\left.
+f_{T\mathcal{T}} \left(\delta{\rho_{m}}-3 \delta{p_{m}}-\nabla^{2}{\pi^{S}}
\right)\right]
\notag \\
&&
- \frac{a^{2}
f_{\mathcal{T}}}{2} \left(\rho_{m}+p_{m}\right) \partial_{i}{\delta{v}}
=4 \pi G a^{2} \left(\rho_{m}+p_{m}\right) \partial_{i}{\delta{v}},
\end{eqnarray}
\begin{eqnarray}
&&
\left(1+f_{T}\right) \Big[-H \left(\dot{\psi}+6 \dot{\phi}\right)-2
\psi \left(3 H^{2}+\dot{H}\right)
%\nonumber\\
%&&
-\ddot{\phi}+\frac{\nabla^{2}{
\left(\phi-\psi\right)}}{3 a^{2}}\Big]
\notag \\
&&+
 12 H f_{TT} \left[\dot{H} \left(\dot{\phi}+H \psi\right)+H
\left(\ddot{\phi}+\dot{H} \psi+H \dot{\psi}\right)
\right]
%\nonumber\\
%&&
+H f_{T\mathcal{T}}
\left(\dot{\delta{\rho}}_{m}-3 \dot{\delta{p}}_{m}-\nabla^{2}{\dot{\pi}^{S}}
\right)
\notag \\
&&
+
\left[12 H \left(\dot{\phi}+H \psi\right)\right]
\left\{f_{TT} \left(3 H^{2}+\dot{H}\right)
%\right.\nonumber\\
%&&
%\left. \ \ \
-H \left[12 H
\dot{H} f_{TTT
}-f_{T\mathcal{T}T} \left(\dot{\rho}_{m}-3 \dot{p}_{m}\right)\right]+\frac{
1+f_{T}}{4}\right\}
\notag
\\
&&
+ \left(\delta{
\rho_{m}}-3 \delta{p_{m}}-\nabla^{2}{\pi^{S}}\right)
\Big\{f_{T\mathcal{T}} \left(3 H^{2}+\dot{H}\right)
%\nonumber\\
%&&  \ \ \
-H \left[12 H \dot{
H} f_{TT\mathcal{T}}-f_{T\mathcal{T}\mathcal{T}} \left(\dot{\rho}_{m}-3
\dot{p}_{m}\right)\right]+\frac{f_{\mathcal{T}}}{4}\Big\}
\notag
\\
&&
+\left(\dot{\phi}+2 H \psi\right) \left[12 H \dot{H} f_{TT}-f_{T \mathcal{T}}
\left(\dot{\rho}_{m}-3
\dot{p}_{m}\right)\right]
%\nonumber\\
%&&
+\frac{f_{\mathcal{T}}}{6}
\nabla^{2}{\pi^ {S}}=-4 \pi G \left(
\delta{p}_{m}+\frac{\nabla^{2}{\pi^{S}}}{3}\right),
\label{fTTeq33}
\end{eqnarray}
\end{widetext}
and
\begin{eqnarray}
\left(1+f_{T}\right)
\left(\psi-\phi\right)=-8 \pi G a^{2} \left(1+\frac{f_{\mathcal{T}}}{8
\pi G}\right) \pi^{S},
\label{fTTphipsi}
\end{eqnarray}
respectively.

Since we have presented the linear perturbation equations, one can examine the basic
stability requirement by extracting the dispersion relation for the gravitational
perturbations. As usual, for simplicity one considers zero anisotropic stress
($\pi^{S}=0$), and in this case equation (\ref{fTTphipsi}) allows us to replace $\psi$ by
$\phi$, and thus remaining with only one gravitational perturbative degree of freedom.
One transforms it in the Fourier space as
\begin{eqnarray}
 \phi(t,\bx)=\int \frac{d^3k}{(2\pi)^\frac{3}{2}}
~\tilde{\phi}_k(t)e^{i\bk\cdot\bx},
\label{fTTphiexpansion}
\end{eqnarray}
and therefore $\nabla^2\phi=-k^2 \tilde{\phi}_k$.

Inserting this decomposition into (\ref{fTTeq33}), and using the other perturbative
equations in order to eliminate variables, after some algebra one obtains the following
equation of motion for the modes of the gravitational potential $\phi$
 \cite{Harko:2014aja}:
\begin{eqnarray}
\label{fTTphiddk}
\ddot{\tilde{\phi}}_k+\Gamma \dot{\tilde{\phi}}_k+\mu^2
\tilde{\phi}_k+c_s^2\frac{k^2}{a^2}
\tilde{\phi}_k=D.
\end{eqnarray}
The functions $\Gamma$, $\mu^2$ and $c_s^2$ are respectively the frictional term, the
effective mass, and the sound speed parameter for the gravitational potential $\phi$, and
along with the term $D$ are given in Appendix \ref{fTTCoefficients}. Clearly, in order
for
$f(T,\mathcal{T})$ gravity to be stable at the linear scalar perturbation level, one
requires $\mu^2\geq0$ and $c_s^2\geq0$.

Due to the complexity of the coefficients $\mu^2$ and $c_s^2$, one cannot extract
analytical relations for the stability conditions. This is usual in complicated modified
gravity models, for instance in generalized Galileon theory  \cite{DeFelice:2011bh}, in
Ho\v{r}ava-Lifshitz gravity  \cite{Bogdanos:2009uj,Wang:2009yz}, in cosmology with
non-minimal derivative coupling  \cite{Dent:2013awa}, etc. Furthermore, although in
almost all modified gravity models one can, at first stage, perform the
perturbations neglecting the matter sector, in the scenario at hand this
cannot be done, and this is an additional complexity, since in that case one
would kill the extra information of the model (which comes from the matter
sector itself) remaining with the usual $f(T)$ gravity. A significant
simplification arises if we consider as usual the matter to be dust, that is
$p_m=\delta p_m=0$, but still one needs to resort to numerical elaboration
of equation (\ref{fTTphiddk}) in order to ensure if a given $f(T,\mathcal{T})$
cosmological model is free of instabilities. However, we mention that
since the simple $f(T)$ gravity is free of instabilities for a large
class of $f(T)$ ansatzen  \cite{Chen:2010va,Dent:2011zz}, as we described in detail in
subsection \ref{sec:pert_fT_gravity}, we deduce that at least for
$f(T,\mathcal{T})$ models that are small deviations from the corresponding
$f(T)$ ones, the stability requirements $\mu^2\geq0$ and $c_s^2\geq0$ are
expected to be satisfied.

In summary, as we discussed, $f(T,\mathcal{T})$ theory is a novel class of gravitational
modification, different from both $f(R,\mathcal{T})$ gravity, as well as from simple
$f(T)$ gravity. Due to the extra freedom in the imposed Lagrangian, $f(T,\mathcal{T})$
cosmology allows for a very wide class of scenarios and behaviors. In particular, one
finds evolutions experiencing a transition from a decelerating to an accelerating state,
capable of describing the late-time cosmic acceleration and the dark energy epoch.
Additionally, one finds evolutions where an initial accelerating phase is followed by a
decelerating one, with a subsequent transition to a final acceleration at late times, a
behavior in agreement with the observed thermal history of the Universe, namely a first
inflationary stage, a transition to non-accelerating, matter-dominated expansion, and
then the transition to late-time accelerating phase. Thus, $f(T,\mathcal{T})$ cosmology
offers a unified description of the universe evolution. An additional advantage of the
scenario at hand, revealing its capabilities, is that the dark energy equation-of-state
parameter can lie in the quintessence or phantom regime, or experience the phantom-divide
crossing. Finally, a detailed study of the scalar perturbations at the linear level
reveals that $f(T,\mathcal{T})$ cosmology can be free of ghosts and instabilities for a
wide class of ansatzen and model parameters.

Lastly, note that apart from the above basic investigation of  $f(T,\mathcal{T})$ gravity
and cosmology, many relevant studies are necessary in order to consider this theory a
candidate for the description of Nature. In particular one should perform a detailed
comparison with cosmological and Solar System observations, which could constrain the
allowed ansatzen and parameter ranges. Additionally, one could use the scalar
perturbation equations extracted above in order to perform a detailed
confrontation with the growth-index data. Moreover, one could extend the perturbation
analysis to the vector and tensor modes, and use them in order to predict the
inflationary
induced tensor-to-scalar ratio, especially under the recent BICEP2  \cite{Ade:2014xna}
and
Planck 2015 measurements  \cite{Planck:2015xua} that can exclude a large class of models.

%%%%%%%%%%%%%%%%%%%%
\subsection{$f(R,T)$ teleparallel gravity}
%%%%%%%%%%%%%%%%%%%%%

Recently much interest has also been given to the $f(R,T)$ modified theories of gravity,
where the gravitational Lagrangian is constituted  by an arbitrary function of the Ricci
scalar $R$ and the torsion scalar $T$
\cite{Myrzakulov:2012ug,Sharif:2012gz,Myrzakulov:2013hca} (see also
\cite{Bahamonde:2015zma} for a similar modification based on an arbitrary function of the
torsion scalar as well as of the divergence of the torsion vector). This issue is not
redundant since the information contained in $f(R)$ gravity is not the same with the one
contained in $f(T)$ gravity. In particular, differences emerge, amongst others, when one
studies the symmetries of the theories. In this subsection, using Noether symmetry
analysis, we desire to show that we can obtain interesting
cosmological solutions in $f(R,T)$ gravity, that cannot be obtain in simple $f(R)$ or
$f(T)$ modifications. Additionally, we will point out how $R$ and $T$ degrees of freedom
can be discussed under the same standard, comparing holonomic and anholonomic coordinate
systems \cite{Capozziello:2014bna}.

\subsubsection{$f(R,T)$ field equations}

The action of $f(R,T)$ gravity reads \cite{Myrzakulov:2012qp}
\begin{equation}\label{actionMYZ}
 {\cal S}=\frac{1}{16 \pi G}\int d^4 x\sqrt{-g}\left[f(R,T)+{\cal L}_m\right]\,,
 \end{equation}
where we define  $|e|\equiv$ det$\left(e^i_\mu\right)=\sqrt{-g}$ in order to
connect the two formalisms. Varying the action and expressing everything in terms of the
vierbeins gives the field equations \cite{Myrzakulov:2012qp}:
\begin{eqnarray}
\label{field}
&&\!\!\!\!\!
\frac{1}{4}e^\mu_A f(R,T)+e^{-1}\left(e e^\sigma_A
S_\sigma^{\mu\nu}\right)\frac{\partial
f(R,T)}{\partial T}\partial_\nu \nonumber\\
&&\!\!\!\!\!
+e_{A\nu}\left[\left(\nabla^\mu
\nabla^\nu-g^{\mu\nu}\nabla^\lambda\nabla_\lambda\right)\frac{\partial f(R,T)}{\partial
R}-\frac{\partial f(R,T)}{\partial R}R^{\mu\nu}\right]
\nonumber\\
&&\!\!\!\!\!
+e^\sigma_A S^{\mu\nu}_\sigma\left(\frac{\partial^2
f(R,T)}{\partial T^2}\partial_\nu T+\frac{\partial^2 f(R,T)}{\partial T\partial
R}\partial_\nu R\right)
\nonumber\\
&&\!\!\!\!\!
-\frac{\partial f(R,T)}{\partial T}e^\gamma_A
S^{\rho\beta\mu}
T_{\rho\beta\gamma}
=4\pi G\,
e_{A}^{\, \nu}\,
{T^{(m)}}_{\nu}{}^{\mu}\,.
\end{eqnarray}
It is easy to see that from  $f(R,T)$ gravity both $f(T)$ and  $f(R)$ gravities can be
immediately recovered.

\subsubsection{$f(R,T)$  cosmology}

Let us now investigate the cosmological applications of $f(R,T)$ gravity. As usual,
the  cosmological equations can be derived from the field Eqs. (\ref{field}), or
equivalently they can be deduced from a point-like canonical Lagrangian ${\cal L}(a,{\dot
a}, R,{\dot R},T,{\dot T})$ related to the action (\ref{actionMYZ}). Here  ${\mathbb
Q}\equiv\{a,R,T\}$ is the configuration space from which it is possible to derive
${\mathbb{TQ}}\equiv \{a,\dot{a}, R, \dot{R}, T,{\dot T}\}$, the corresponding tangent
space  on which ${\cal L}$ is defined.  The variables $a(t)$, $R(t)$ and $T(t)$ are,
respectively, the scale factor, the  Ricci scalar and the torsion scalar defined in the
FRW metric. The Euler-Lagrange equations are
\begin{eqnarray}
\frac{d}{dt}\frac{\partial {\cal L}}{\partial {\dot a}}=\frac{\partial {\cal L}}{\partial
 a}\,, \qquad
\frac{d}{dt}\frac{\partial {\cal L}}{\partial {\dot R}}=\frac{\partial {\cal L}}{\partial
 R}\,,\qquad
\frac{d}{dt}\frac{\partial {\cal L}}{\partial {\dot T}}=\frac{\partial {\cal L}}{\partial
 T}\,,\nonumber\\
\label{moto3}
\end{eqnarray}
with the energy condition
\begin{eqnarray}
E_{\cal L}= \frac{\partial {\cal L}}{\partial {\dot a}}{\dot a}+\frac{\partial {\cal
L}}{\partial {\dot R}} {\dot R}+\frac{\partial {\cal L}}{\partial {\dot T}} {\dot
T}-{\cal
L}=0\,.
\label{energy}
\end{eqnarray}
Here the dot indicates the derivatives with respect to the cosmic time $t$. One can use
the method of Lagrange multipliers to set  $R$  and $T$ as  constraints for the dynamics
\cite{Capozziello:2013xn}. In fact, choosing suitable Lagrange multipliers and
integrating by parts to eliminate higher order derivatives, the Lagrangian ${\cal L}$
becomes canonical. In physical units,  the action  is
\begin{eqnarray}
 {\cal S}&=&2\pi^2\int dt\,a^3 \left\{  f(R,T)-\lambda_1 \left[R+6\left( \frac{{\ddot
a}}{a}+\frac{{
\dot a}^2}{a^2}\right)\right]\right.\nonumber\\&&\left.-\lambda_2\left[T+6\left(
\frac{{\dot a}^2}{
a^2}\right)\right] \right\}\,.
\end{eqnarray}
Here the usual relations of the Ricci scalar and the torsion scalar in flat FRW
metric can be used, namely $R=-6(\dot{H}+2H^2)$ and $ T=-6H^2$. It is worthy to
stress that the two Lagrange multipliers are comparable but the order of derivative is
higher for $R$.
By varying the action with respect to $R$ and $T$, one obtains
\begin{eqnarray}
\lambda_1= \frac{\partial f(R,T)}{\partial  R}\,,\qquad \lambda_2= \frac{\partial
f(R,T)}{\partial
T}\,.
\end{eqnarray}
Hence, finally the point-like Lagrangian acquires the following form
\begin{eqnarray}\label{PointLagra}
 {\cal L}&=&a^3\left[ f(R,T)-R \frac{\partial f(R,T)}{\partial  R}-T \frac{\partial
f(R,T)}{\partial  T}\right]\nonumber\\&&+6\,a{\dot a}^2 \left[\frac{\partial
f(R,T)}{\partial  R}-\frac{\partial f(R,T)}{\partial
T}\right]+\nonumber\\&&6\,a^2\,{\dot
a}\,\left[{\dot R}\frac{\partial^2 f(R,T)}{\partial  R^2}+ {\dot T}\frac{\partial^2
f(R,T)}{\partial R \partial T}\right]\,,
\end{eqnarray}
which is a canonical function of 3 coupled fields, namely $a$, $R$ and $T$, depending on
time $t$. Note that the first term in square brackets has the role of an effective
potential. It is worth  stressing again  that the Lagrange multipliers have been  chosen
by considering the definition of the Ricci curvature scalar $R$ and the torsion scalar
$T$. This fact allows us to consider the constrained dynamics as canonical.

It is interesting to consider some important  subcases of the Lagrangian
(\ref{PointLagra}). For $f(R,T)=R$,  the  GR Lagrangian   is recovered. In this case,  we
have $ {\cal L}=6a{\dot a}^2 +a^3R$, which after developing $R$ easily reduces to ${\cal
L}=-3a{\dot a}^2$, namely the standard point-like Lagrangian of FRW cosmology.
In the case $f(R,T)=f(R)$, we recover the usual relation \cite{Capozziello:2011et}
\begin{equation}
 {\cal L}= 6 a {\dot a}^2 f'(R)+ 6 a^2  {\dot a}{\dot
R}f''(R)+a^3\left[f(R)-Rf'(R)\right]\,\label{
LfR}.
\end{equation}
Finally, $f(T)$ cosmology is recovered for  $f(R,T)=f(T)$, and
then \cite{Basilakos:2013rua}
\begin{eqnarray}
 {\cal L}= a^3[f(T)-Tf'(T)]-6a{\dot a}^2f'(T)
 \label{LfT}\,.
\end{eqnarray}
Clearly, these cases deserve a specific investigation.

Let us now derive the Euler-Lagrange equations from Eqs.  (\ref{moto3})- (\ref{energy}).
They write as
\begin{eqnarray}
 &&\!\!\!\!\!\!\!\!\!\!\!
 \left(12{\dot
a}^2-6a^2+12a{\ddot a}\right)
\left[ \frac{\partial f(R,T)}{\partial R}-\frac{\partial
f(R,T)}{\partial
T}\right]\nonumber\\
&&\!\!\!\!\!\!
-3a^2 \left[f(R,T)-T\frac{\partial
f(R,T)}{\partial
T}-R\frac{\partial
f(R,T)}{\partial R}\right]\nonumber\\
&&\!\!\!\!\!\!
-12 a {\dot a}\left[{\dot T}\frac{\partial^2
f(R,T)}{\partial T^2}-{\dot R}\frac{\partial^2 f(R,T)}{\partial R^2}\right]
\nonumber\\
&&\!\!\!\!\!\!
-12 a {\dot
a}\left[{\dot R}\frac{
\partial^2 f(R,T)}{\partial R\partial T}-{\dot T}\frac{\partial^2 f(R,T)}{\partial
R\partial T}\right]\nonumber
\\
&&\!\!\!\!\!\!
+6a^2\left[{\ddot T}\frac{\partial^2 f(R,T)}{\partial
R\partial T}+{\ddot R}\frac{\partial^2 f(R,T)}{\partial R^2}+{\dot T}^2\frac{\partial^3
f(R,T)}{\partial R\partial T^2}\right.
\nonumber\\
&&
\ \ \ \  \ \,
\left.
+2{\dot R}{\dot T}\frac{\partial^3 f(R,T)}{\partial
R^2\partial T}+{\dot R}^2\frac{\partial^3 f(R,T)}{\partial R^3}\right] =0\,,
\label{moto11}
\end{eqnarray}
\begin{eqnarray}
&&\!\!\!\!\!\!\!
a^3 \left[R \frac{\partial^2 f(R,T)}{\partial R^2}+T\frac{\partial^2
f(R,T)}{\partial
R\partial T}
\right]\nonumber\\
&&
+6a{\dot a }^2\left[\frac{\partial^2 f(R,T)}{\partial R^2}+\frac{\partial^2
f(R,T)}{\partial
R\partial T}\right]
\nonumber\\
&&+6a^2{\ddot a}\frac{\partial^2 f(R,T)}{\partial R^2}=0\,,
\label{moto22}
\end{eqnarray}
 \begin{eqnarray}
&&\!\!\!\!\!\!\!
a^3 \left[T \frac{\partial^2 f(R,T)}{\partial T^2}+R\frac{\partial^2
f(R,T)}{\partial
R\partial T}
\right]
\nonumber\\
&&
+6a{\dot a }^2\left[\frac{\partial^2 f(R,T)}{\partial T^2}+\frac{\partial^2
f(R,T)}{\partial
R\partial T}\right]
\nonumber\\
&&
+6a^2{\ddot a}\frac{\partial^2 f(R,T)}{\partial R\partial T}=0\,.
\label{moto33}
\end{eqnarray}
Additionally, the  energy condition  (\ref{energy}), corresponding to the $00$-Einstein
equation, gives
\begin{eqnarray}
&&\!\!\!\!\!\!\!\!\!\!\!\!\!\!\!
E_{\cal L}=6a{\dot a}^2\left[\frac{\partial f(R,T)}{\partial R}-\frac{\partial
f(R,T)}{\partial T}
\right]\nonumber\\
&&
+a^3\left[f(R,T)-T\frac{\partial f(R,T)}{\partial T}-R\frac{\partial
f(R,T)}{\partial R}\right]\nonumber\\\nonumber\\&&-6a^2{\dot a}\left[{\dot
T}\frac{\partial ^2f(R,T)}{\partial R\partial T}+{\dot R}\frac{\partial^2
f(R,T)}{\partial
R^2}\right]=0\,.
\label{energyMYZ}
\end{eqnarray}
As we have already mentioned, the above system of cosmological equations, namely Eqs.
(\ref{moto11})-(\ref{energyMYZ}), can be alternatively derived, as usual, from the
general field equations (\ref{field}) inserting the cosmological vierbein
(\ref{weproudlyuse}).

%%%%%%%%%%%%%%%%%%%%%%%%%%%%%%%%
\subsubsection{The Noether symmetry approach}
\label{quattro}
%%%%%%%%%%%%%%%%%%%%%%%%%%

The existence of Noether symmetries leads to the extraction of constants of motion, which
allow for a simplification of the dynamics. Often such a dynamics is exactly solvable by a
straightforward transformation to cyclic variables \cite{Capozziello:2013qha}.
A Noether symmetry for the Lagrangian (\ref{PointLagra}) exists if the condition
\begin{eqnarray}
L_X {\cal L}\,=\,0 \qquad \Rightarrow \qquad X{\cal L}\,=\,0\,
\label{LX}
\end{eqnarray}
is valid. Here $L_X$ is the Lie derivative with respect to the Noether vector $X$.
Eq. (\ref{LX}) is
nothing else but the contraction of the Noether vector $X$, defined on the tangent space
${\mathbb{TQ}}\equiv\{a,\dot{a}, R, \dot{R}, T,{\dot T}\}$ of the Lagrangian ${\cal
L}={\cal L}(a,{\dot a}, R,
{\dot R},T,{\dot T})$, with the Cartan one-form, generically defined as
$
\theta_{\cal L} \equiv \frac{\partial {\cal L}}{\partial {\dot q}_i}dq^i$.
Condition (\ref{LX}) gives
$
i_X \theta_{\cal L} = \Sigma_0$,
where $i_X$ is the inner derivative and $\Sigma_0$ is the conserved quantity
 \cite{Capozziello:1999xs,Capozziello:1998nd,Capozziello:1996ay,
Capozziello:2014ioa,Capozziello:2014qla}. In
other words, the existence of the symmetry is connected to the existence of a vector field
\begin{eqnarray}
X= \alpha^i (q)\frac{\partial}{\partial
q^i}+\frac{d\alpha^i(q)}{dt}\frac{\partial}{\partial {\dot
q}^i}\,,
\end{eqnarray}
where at least one of the components  $\alpha^i(q)$ have to be different from zero to
generate a symmetry.

In our case, the generator of symmetry is
\begin{eqnarray}
X=\alpha \frac{\partial}{\partial a}+ \beta\frac{\partial}{\partial R}+\gamma
\frac{\partial}{\partial T}+{\dot \alpha} \frac{\partial}{\partial  \dot a}+  {\dot
\beta}\frac{\partial}{\partial \dot R}+{\dot \gamma}  \frac{\partial}{\partial\dot
T}\,.\nonumber\\
\label{ourX}
\end{eqnarray}
The functions $\alpha, \beta, \gamma$ depend on the variables $a, R, T$ and then
\begin{eqnarray}
{\dot \alpha}\,&=&\, \frac{\partial \alpha}{\partial a}{\dot a}+\frac{\partial
\alpha}{\partial R}{\dot R}+\frac{\partial \alpha}{\partial T}{\dot T}\,,\nonumber\\
{\dot \beta}\,&=&\, \frac{\partial \beta}{\partial a}{\dot a}+\frac{\partial
\beta}{\partial R}{\dot R}+\frac{\partial \beta}{\partial T}{\dot T}\,,\nonumber\\
{\dot \gamma}\,&=&\, \frac{\partial \gamma}{\partial a}{\dot a}+\frac{\partial
\gamma}{\partial R}{\dot R}+\frac{\partial \gamma}{\partial T}{\dot T}\,.
\end{eqnarray}
As stated above, a Noether symmetry exists if at least one of them is different from
zero.
Their analytic forms can be found by making Eq. (\ref{LX}) explicit, which corresponds to
a set of partial differential equations given by equating to zero the terms in ${\dot
a}^2$,${\dot a}{\dot T}$, ${\dot a}{\dot R}$, ${\dot
T}^2$, ${\dot R}^2$,${\dot R}{\dot T}$  and so on. In our specific case, we acquire a
system of 7 partial differential equations, due to the fact that being the
minisuperpace 3-dim (i.e $n=3$), the equations are $1+n(n+1)/2$ as
shown in \cite{GSW,Hammond:1994ds,DeSabbata:1991th,deSabbata:1992cp,Murase:1993xd}. In
particular, these equations read:
\begin{eqnarray}
\label{dota2}
&&\alpha \left[\frac{\partial f(R,T)}{\partial R}-\frac{\partial f(R,T)}{\partial
T}\right]\nonumber\\
&&
+\beta
a  \left[\frac{\partial^2 f(R,T)}{\partial R^2}-\frac{\partial^2 f(R,T)}{\partial
R\partial T}\right]
\nonumber\\
&&
+ \gamma a\left[\frac{\partial^2 f(R,T)}{\partial R\partial
T}-\frac{\partial^2 f(R,T)}{\partial T^2}\right]
\nonumber\\
&&
+2 a \frac{\partial
\alpha}{\partial a}\frac{\partial f(R,T)}{\partial R}-2 a\frac{\partial \alpha}{\partial
a}\frac{\partial f(R,T)}{\partial T}
\nonumber\\
&&
+a^2 \frac{\partial \beta}{\partial
a}\frac{\partial^2 f(R,T)}{\partial R^2}+a^2 \frac{\partial \gamma}{\partial
a}\frac{\partial^2 f(R,T)}{\partial R\partial T}=0\,,
\end{eqnarray}
\begin{eqnarray}
\label{dotadotT}
&&
2\alpha a \frac{\partial^2 f(R,T)}{\partial R\partial T}+ \beta a^2 \frac{\partial^3
f(R,T)}{\partial R^2\partial T}
\nonumber\\
&&
+ \gamma a^2 \frac{\partial^3 f(R,T)}{\partial R\partial
T^2}+ a^2  \frac{\partial \alpha}{\partial a} \frac{\partial^2 f(R,T)}{\partial R\partial
T}\nonumber\\
&&
+2a\frac{\partial \alpha}{\partial T} \frac{\partial f(R,T)}{\partial
R}-2a\frac{\partial \alpha}{\partial T}
 \frac{\partial f(R,T)}{\partial T}
 \nonumber\\
&&
+a^2\frac{\partial \beta}{\partial T}
\frac{\partial^2
f(R,T)}{\partial R^2}  +a^2 \frac{\partial \gamma}{\partial T}  \frac{\partial^2
f(R,T)}{\partial R\partial
T}=0\,,
\end{eqnarray}
\begin{eqnarray}
\label{dotadotR}
  &&
  2 \alpha a \frac{\partial^2 f(R,T)}{\partial R^2} +\beta a^2  \frac{\partial^3
f(R,T)}{\partial R^3}
\nonumber\\
&&
+\gamma a^2 \frac{\partial^3 f(R,T)}{\partial R^2\partial T}+a^2
\frac{\partial\alpha}{\partial a}\frac{\partial^2 f(R,T)}{\partial R^2} \nonumber\\
&&
+2a
\frac{\partial\alpha}{\partial R}\frac{\partial f(R,T)}{\partial R}
-2 a
\frac{\partial\alpha}{\partial R} \frac{\partial f(R,T)}{\partial T}
\nonumber
\\
&&+a^2
\frac{\partial\beta}{\partial R} \frac{\partial^2 f(R,T)}{\partial R^2}+a^2
\frac{\partial\gamma}{\partial R}\frac{\partial^2 f(R,T)}{\partial R\partial T}=0\,,
\end{eqnarray}
\begin{eqnarray}
\label{dotT2}
   a    \frac{\partial\alpha}{\partial T} \frac{\partial^2f(R,T)}{\partial R\partial
T}=0\,,
\end{eqnarray}
\begin{eqnarray}
\label{dotR2}
  a^2       \frac{\partial\alpha}{\partial R}\frac{\partial^2f(R,T)}{\partial R^2}=0,
\end{eqnarray}
 \begin{equation}
 \label{dotRdotT}
a^2  \frac{\partial\alpha}{\partial R}  \frac{\partial^2f(R,T)}{\partial
R\partial T}
+a^2  \frac{\partial\alpha}{\partial R} \frac{\partial^2f(R,T)}{\partial R^2}=0\,,
\end{equation}
\begin{eqnarray}
\label{senzadot}
&&
3\alpha a^2 \left[f(R,T)-T \frac{\partial f(R,T)}{\partial T}-R \frac{\partial
f(R,T)}{\partial R}
\right]
\nonumber
\\
&&-\beta a^3 \left[T \frac{\partial^2 f(R,T)}{\partial R\partial T}+ R
\frac{\partial^2 f(R,T)}
{\partial R^2}\right]\nonumber\\
&&
-\gamma a^3 \left[T \frac{\partial^2 f(R,T)}{\partial
T^2}+R \frac{\partial^2 f(R,T)}{\partial R\partial T}\right]=0\,.
  \end{eqnarray}
The above system is overdetermined and, if solvable, enables one to assign
$\alpha,\beta,\gamma$
and $f(R,T)$.
The analytic form of $f(R,T)$ can be fixed by imposing, in the last equation of system
(\ref{senzadot}), the conditions
\begin{equation}\label{sistesenza}
\left\{\begin{array}{ll}
f(R,T)-T \frac{\partial f(R,T)}{\partial T}-R \frac{\partial f(R,T)}{\partial R}=0\\
T \frac{\partial^2 f(R,T)}{\partial R\partial T}+ R \frac{\partial^2 f(R,T)}{\partial
R^2}=0\\
T \frac{\partial^2 f(R,T)}{\partial T^2}+R \frac{\partial^2 f(R,T)}{\partial R\partial
T}=0\\
\end{array}\right. ,
\end{equation}
where the second and third equations  are  symmetric. However, it is clear that this is
an
arbitrary choice, since more general conditions are possible.  In particular, we can
choose the functional forms:
\begin{eqnarray}
f(R,T) = f(R)+ f(T)\,, \qquad f(R,T) = f(R) f(T)\,,\nonumber\\
\end{eqnarray}
from which it is easy to prove that the functional forms compatible with the system
(\ref{sistesenza}) are:
\begin{eqnarray}
f(R,T)= F_0 R+ F_1 T\,,\qquad
f(R,T)= F_0 R^n T^{1-n}\,.\nonumber\\
\end{eqnarray}
The first case is nothing else but GR, while the second gives interesting cases of
possible extended theories as soon as $n\neq1$.

For  $n=2$,  the canonical  Lagrangian (\ref{PointLagra}) obtains the form
\begin{eqnarray}
\label{lagrasoln2}
{\cal L}=6 a^2 {\dot a} \left(\frac{2 {\dot R}}{T}-\frac{2 R {\dot T}}{T^2}\right)+6 a
{\dot a}^2 \left(\frac{R^2}{T^2}+\frac{2
   R}{T}\right).
\end{eqnarray}
We can introduce the variable ${\displaystyle \frac{R}{T}= \zeta}$ in order to reduce the
dynamics of the system, and hence the above Lagrangian is transformed into
\begin{eqnarray}
\label{lagraredun2}
{\cal L}= 2 a^2 {\dot a} {\dot \zeta}+ 2 a {\dot a}^2\zeta+ a{\dot a}^2 \zeta^2.
\end{eqnarray}
The Euler-Lagrange equations are
\begin{eqnarray}
&& \!\!\!\!\!\!\!\!\!\!\!\!\!\!\!\!\!\!\!\!
{\ddot \zeta}
+\left(\frac{\dot a}{a}\right)^2 \zeta\left(1+\frac{\zeta}{2}\right)
+\left(\frac{\ddot a}{a} \right)\zeta
\left(\zeta+2\right)
\nonumber\\
&& \!\!\!\!\!\!\!\!\!\!\!\!\!\!\!\!
+2\left( \frac{\dot a
}{a}\right) {\dot
\zeta} \left(\zeta+1\right)=0\,,
\label{motoz1a}
\end{eqnarray}
\begin{eqnarray}
&& \frac{\ddot a}{a} + \left(\frac{\dot a}{a}\right)^2(1- \zeta)   =0\,,
%\label{motoz2z}\\
\end{eqnarray}
and the energy condition reads
\begin{eqnarray}
 \left(\frac{\dot a}{a}\right)^2 \zeta
\left(\zeta+2\right)+
2\left( \frac{\dot a}{a}\right) {\dot \zeta}=0\label{energyz}\,.
\end{eqnarray}
Clearly, we have lost one equation since the relation between the two
variables $R$ and $T$ is fixed by $\zeta$.  Immediately, an exact solution is found to be
\begin{equation}
a(t)=a_0 t^{1/2}\,,\qquad \zeta=0\,,
\end{equation}
which corresponds to a radiation solution. Another solution is achieved for $\zeta=1$,
but it is a trivial one since $a(t)=a_0$. These imply that these two solutions, in
the case $n=2$, are quite natural due to the fact that the asymptotic behavior of $R$ is
$\sim1/t^2$, similar to that of $T$ which is always $\sim 1/t^2$. Thus, $\zeta$ can be
either equal to zero or equal to a constant.

\section{Comparing torsion and curvature gravity}
\label{Secioncomparison}

As  discussed in detailed in the above sections, in the  construction of any theory  of
gravity, one has to  i) make some assumptions; ii) insert observational and/or
experimental information; iii) follow  the flow in order to obtain mathematical
self-consistency. For instance, in constructing GR,  Einstein imposed the following
considerations that arise from observational information: metricity (during a parallel
transportation nature seems to maintain a vecor's length, i.e the hydrogen redshift is
the
same in different directions of the universe);  Lorentz invariance (including the fact
that he used quadratic metric); Equivalence Principle; causality. We stress that all
these
considerations do not arise from some fundamental principle, but rather they are
indicated
by observations and/or experiments. Additionally, Einstein made an extra, pure,
assumption, without any theoretical or observational/experimental justification: he used
the symmetric, Levi-Civita connection, that is he {\it{assumed}} that torsion is set to
zero and that geometry, and therefore gravity, is described only by curvature. Finally,
Einstein desired to have an action up to second-order in the metric and its derivatives,
that is why he resulted in the Einstein-Hilbert action, that is linear in the  Ricci
scalar $R$.

Even if one agrees with Einstein's conjecture that gravity is described through geometry,
since this is an accepted principle by the community (note however that this is not what
happens with the other three interactions), still he faces the crucial question {\it{what
kind of geometry to consider}}, since there still exists a huge freedom in the
corresponding choice. Riemannian geometry was considered the simplest option at the
beginning of the twentieth century. It is just one possible  choice but definitely not
the only one. For instance, one could abandon the assumptions of metricity, Lorentz
invariance and Equivalence Principle, but obviously remaining inside the
observational/experimental bounds (although they are constrained in narrow bounds, they
are still assumptions in the sense that no fundamental principle lies behind them).
However, even without these radical changes, one can safely and without any problem
abandon the arbitrary assumption of a symmetric connection, and describe gravity through
both curvature and torsion, or only through torsion as Einstein himself did in his
approach to Teleparallel Equivalent of GR (TEGR). In this case, one obtains a
gravitational theory based on torsion, which is completely equivalent to  GR at the level
of equations.

Since GR and TEGR are  equivalent, one can choose one of them and use it safely. Actually
this was indeed the case, and, since GR was already used for 13 years before the
formulation of TEGR, physicists chose this theory as the paradigm for gravitational
interactions. Hence, in the following decades the investigation on TEGR had only academic
interest, faced as a different mathematical ``tool'' that could lead to the extraction of
solutions. In particular, one of its advantages, in  comparison with  the standard GR,
was
the successful calculation of the energy of a solution, which in  GR is known to face
problems like the use of pseudo-tensors
\cite{Maluf:2002zc,daRochaNeto:2002bw,Aldrovandi:2008xv,Lucas:2009nq}. Nevertheless, all
other solutions and obtained information were the same in both theories. A similar
situation occurred for the Palatini formulation of GR: metric and metric affine
approaches revealed exactly the same information for the Hilbert-Einstein formulation of
gravity.

However, the last fifteen years physicists started to seriously consider the approach of
modifying gravity, as a way to successfully describe the accelerating phases of the
universe, both at early (inflation) and late times, and moreover to improve the
renormalizability of the gravitational theories. If one agrees with this severe decision,
then immediately  the question {\it{what formulation of gravity to modify}} arises. In
fact, it can be  proved  that even if one has two equivalent at the level of equations
gravitational theories, their modifications in general will not be equivalent any more.
Clearly, a priori, and at the theoretical level, one cannot exclude some modifications
and
accept the others and vice versa, since only observations and experiments could do this.
Hence, it is at least interesting and worthy to try to investigate every possible
gravitational modification.

Up to now, almost all the works in modified gravity were starting from the usual
curvature formulation of gravity and were extending the Einstein-Hilbert action, namely
the Ricci scalar, with functions and combinations of various curvature invariants.
Such procedure is obviously not wrong, but as we described in detailed above, it is not
the only one. In particular, one has ``equal justification'' to start from a torsion-based
formulation of gravity, with TEGR being the simplest choice, and replace the torsional
Lagrangian, namely the torsion scalar, with functions and combinations of various
torsion quantities.

The simplest curvature-based modified gravity is the $f(R)$ paradigm, while the simplest
torsion-based modified gravity is the $f(T)$ paradigm. With the above considerations in
mind, the issue to compare  $f(R)$ and $f(T)$ gravity immediately emerges. A first and
crucial comment is that these two gravitational modifications are different, despite the
fact that GR is completely equivalent with TEGR at the level of equations. Definitely,
one
cannot argue that one of them is wrong and the other ones is correct, however since they
correspond to different gravitational modifications it is interesting and worthy to
analyze their properties and cosmological modifications in detail.

An advantage of $f(T)$ gravity, comparing to $f(R)$, is that the background field
equations are always second order differential equations. This profound advantage makes
the cosmologies in $f(T)$ theory much simpler than those in $f(R)$ gravity at background
level. However, note that the Palatini formulation of $f(R)$ gravity points out that $g$
and $\Gamma$ can be dealt as two fundamental objects so that most of the cosmological
dynamics could be addressed by this approach. In general, in this Review we showed in
detail the cosmological implications of $f(T)$ gravity, which can then be compared one by
one with those of $f(R)$ gravity. Obviously, the discussion is completely open at the
moment, and no final probe exists in pointing out what is the direction towards which it
is better to extend gravity.

%Beside cosmology,  other fundamental issues are related to quantum gravity. The question
%is ``What to quantize? Metric, Vierbeins or connections?'' Also in this case, the
%self-consistency of any quantization scheme could be related to some experimental probe
%discriminating among the competing theories.

%\pagebreak

%---------------------------------------
\section{Summary and Conclusions}
\label{Seciontconclusions}

\subsection{Summary}

In this Review, we started from the historical point of view that GR gives a geometrical
interpretation of gravity and that the presence of torsion can provide a gauging
description of gravity in restoring its force picture. The role of torsion becomes
significant in the translational gauge theories of gravity, namely the TEGR and the
Eistein-Cartan-Sciama-Kibble theory. Additionally, other gauge theories of gravity can
be investigated introducing torsion, which proves a key element as in Poincar\'{e} Gauge
Theory.

The simplest modification of gravitational theories based on torsion is the $f(T)$
paradigm, where the central quantity is the torsion scalar. Several basic issues
concerning the field equations, Lorentz invariance, degrees of freedom, perturbation
analysis, thermodynamics, etc., were discussed extensively.

With the basic picture of $f(T)$ gravity in mind, we then provided a comprehensive review
of its cosmological implications. As other modified gravity theories, $f(T)$ gravity,
when applied to late times, can provide an interpretation of the present observed
acceleration of the universe, driven by the torsion effects. Various ``effective dark
energy'' scenarios, based on  specific choices of $f(T)$ forms,  have been  discussed in
order to demonstrate that the observed universe can undergo a period  of late-time
acceleration without affecting the observed thermal expanding history. Moreover, we
performed a detailed dynamical analysis of these scenarios in order to extract the
stable late-time solutions. As we showed, in the framework of this theory the effective
dark energy component, related to $f(T)$, can realize the quintessence and phantom
regimes, and also the phantom-divide crossing, which is an advantage and reveals the
capabilities of the scenario. As a separate study, we addressed the possibility to find
Noether's symmetries and conserved quantities in the context of $f(T)$ gravity, which
lead
to the extraction of exact solutions that cannot be obtained using the usual methods.

Another important advantage of $f(T)$ cosmology is that a sufficiently long accelerating
phase of the universe at early times can be naturally achieved, without the need of
introducing an inflaton field. We reviewed this inflationary realization and we
commented on its perturbation analysis. Furthermore, we investigated bouncing solutions,
showing that they can easily arise for suitably designed $f(T)$ forms, and we performed
the associated perturbation analysis.

Using the obtain solutions, at the background and perturbations levels, we used detailed
data from SNIa, BAO, and CMB observations and we constrained several representative 
models. 
Moreover,
we applied $f(T)$ cosmography, which is a very useful tool to discriminate among
competing
models with respect to observations.

A crucial topic in every gravitational theory, especially concerning the forthcoming
observations, is the investigation of gravitational waves. We performed a detailed
analysis of gravitational waves in $f(T)$ gravity in the post-Minkowskian limit, and we
pointed out that their dynamics has significant differences with respect to $f(R)$
gravity.

Going beyond the cosmological applications, we investigated in detail spherically
symmetric and black hole solutions, charged black hole solutions, cylindrical solutions,
wormhole solutions, and other topics of astrophysical interest. This is an important
subject, since it reveals the deeper features of the theory and allows for a comparison
with other gravitational modifications.

Inspired by the study of $f(T)$ gravity, one may further modify the theory into various
extended versions, addressing other dynamical issues. In particular, we studied
non-minimal couplings of the torsion scalar with scalar fields, as well as the
incorporation of higher-order torsional invariants such as the teleparallel equivalent of
the Gauss-Bonnet combination. Additionally, we reviewed scenarios in which the torsion
scalar couples to the matter Lagrangian or with the trace of the matter energy-momentum
tensor. The additional degrees of freedom allow to improve the dynamics towards more
self-consistent cosmologies.

Finally, we compared the curvature and the torsion gravitational modifications, pointing
out that the discussion on what could be the geometrical description of gravity is still
completely open.

\subsection{Concluding remarks}

The goal of the present manuscript was to provide a comprehensive review of torsional,
teleparallel and $f(T)$ gravity, and its various cosmological and astronomical
implications, in order to act as a starting point for readers with less experience in the
field, as well as a reference guide for experts on the subject.

Modified torsional gravity, the simplest representative of which is $f(T)$ gravity, has
made a fast progress in the last five years and the whole discussion has helped
significantly towards the understanding of (classical) gravity, as well as its gauging
description, which may be connected to more fundamental aspects and indeed enlighten the
discussion towards its quantization. Definitely, there are still many unexplored
aspects in the framework of $f(T)$ teleparallel gravity, the investigation of which may
turn out to be fruitful. In the following, we list some of them:
\begin{itemize}

\item
Can a universe driven by this type of gravity accommodate with all cosmological and
astronomical observations, and furthermore solve or avoid the theoretical problems
existing in standard inflationary $\Lambda$CDM paradigm?

\item
Although simple curvature modified gravities, such as $f(R)$ gravity, are different from
simple torsional modified gravities, such as $f(T)$ gravity, will the most general
curvature modified gravity, including every possible curvature term up to infinite order,
coincide with the most general torsional modified gravity, including every possible
torsion term up to infinite order?

\item
How can the effective approach of $f(T)$ or in general of torsional modified gravity, be
combined to or arisen from fundamental theories, such as string theory or loop quantum
gravity?

 \item
If the gravitational interaction is gauged through torsion, is it possible to develop a
grand unified theory that combines the standard model of particle physics and the
gravitational sector in one single framework?

\item
 Since gauging a theory is the first step towards its quantization, will this framework
be
enlightening towards gravitational quantization?

\item
Since in torsional gravity one uses the vierbein instead of the metric as a fundamental
variable, is the above discussion enlightening on which classical variable (i.e. the
metric, the vierbein or the connection) would be the one to be quantized, namely which
one would correspond to the graviton? Or are all of them equivalent to capture the
degrees of freedom and dynamics of quantum gravity?

\end{itemize}

The above open questions and issues require thorough and systematic theoretical
investigation in the field. Additionally, the confrontation with high-accuracy
observational data, that are expected to appear in the relatively near future, would
assist in constraining the possible classes of theories, having in mind however that a
consistent cosmology is not a proof for the consistency of the underlying theory of
gravity, namely it is a necessary but not sufficient condition. We believe that this
Review will trigger the readers to work towards these issues, and consider torsional
modified gravity as a candidate for the description of Nature.

%---------------------------------------
\section*{Acknowledgments}

We are grateful to K.~Atazadeh, A. Aviles, K. Bamba, S. Basilakos, F. Becattini, A. 
Bravetti, V.F. 
Cardone, S. Carloni, S.-H. Chen, R. Cianci, F.~Darabi, J. B. Dent, S. Dutta, L. Fabbri, 
H. 
Farajollahi, V. Faraoni, P.~A.~Gonzalez, T.~Harko, G. Kofinas,
M.~Kr\v{s}\v{s}\'ak, G. Lambiase, G.~Leon, M. Li, F.~S.~N.~Lobo, O. Luongo, A. Marciano, 
R.-X. Miao,
 R. Myrzakulov, S.~Nojiri, S.~D.~Odintsov, G.~Otalora, A. Paliathanasis, E. 
Papantonopoulos, J.~G.
~Pereira, A. Ravanpak, J.~Saavedra, D. Saez-Gomez, L. Smolin, A. Starobinsky, P. 
Stavrinos, P. J. 
Steinhardt, C. Stornaiolo, N. ~Tamanini, M.~Tsamparlis, Y.~Vasquez, S. Vignolo, C. 
Wetterich, M.~L.
~Yan and many other colleagues for long-termed collaborations in the field of $f(T)$ 
gravity and/or 
valuable comments on the manuscript.
YFC is supported in part by the Chinese National Youth Thousand Talents Program, by the 
USTC start-
up funding (Grant No.~KY2030000049), by the NSFC (Grant No. 11421303) and by the 
Department of 
Physics at McGill.
SC and MDL are supported by INFN ({\it iniziative specifiche} TEONGRAV and QGSKY).
MDL is partially supported by the Ministry of Education and Science (Russia). This 
research was partially supported by ERC Synergy
Grant ``BlackHoleCam'' Imaging the Event Horizon of Black Holes awarded by the ERC in
2013 (Grant No. 610058).
The research of ENS is implemented within the framework of the Operational Program 
``Education and 
Lifelong Learning'' (Actions Beneficiary: General Secretariat for Research and 
Technology), and is 
co-financedby the European Social Fund (ESF) and the Greek State.
%---------------------------------------
\appendix

\section{Conventions}
\label{SecionConventions}

It is worth  stressing that the conventions of physical parameters and quantities, such
as
the metric, coordinates indices, etc., appearing in the literature, vary significantly.
In order to avoid possible confusions for the readers, we provide a unified convention
throughout the Review, which is summarized is the following:

$\bullet~$ Throughout the review we take the natural units by requiring $c = \hbar =1$.

$\bullet~$ We adopt the signature of the metric $g_{\mu\nu}$ to be ($+,-,-,-$).

$\bullet~$ Concerning the gravitational constant we use the symbol $G$, related to the
reduced Planck mass $M_p$ and the Planck mass $M_{\rm PL}$ through
\begin{eqnarray}
  &&
  M_p=1/\sqrt{8\pi G}=2.4357\times 10^{18}{\rm GeV} ~,\nonumber\\
  &&M_{\rm
PL}=1/\sqrt{G}=1.2211\times 10^{19} {\rm GeV}~.
\end{eqnarray}
In the literature one can also find the symbol $\kappa^2=8\pi G$.

$\bullet~$ The 4-dimensional space-time coordinates are depicted by Greek letters,
namely we use $\mu, ~\nu, ...~$ to run over the values $0,~1,~2,~3$.

$\bullet~$ The 3-dimensional spatial coordinates are depicted by lower case Latin
letters, namely we use $i, ~j, ...~$ to run over the values $1,~2,~3$.

$\bullet~$ The 4-dimensional tangent space-time coordinates are depicted by capital
Latin letters, namely we use $A, ~B, ...~$ to run over the values $0,~1,~2,~3$.

$\bullet~$ The 3-dimensional tangent spatial coordinates are depicted by the lower case
Latin letters, namely we use $a, ~b, ...~$ to run over the values $1,~2,~3$.

$\bullet~$ If not specified differently, the indices that are repeated in the same
formula ought to be summed.

%$\bullet~$ The subscript ``$_{,}$'' denotes the regular derivative with respect to some
%variable and
%``$_{;}$'' represents for the covariant derivative. For instance, ``$_{,\mu} =
%\partial_\mu$'' and
%``$_{;\mu} = \nabla_\mu$''.

$\bullet~$ In General Relativity we take the Christopher symbol to be:
$$ \Gamma^{\alpha}_{\lambda \mu}  = \frac{1}{2} g^{\alpha \nu}(g_{\mu \nu,\lambda}
+g_{\nu \lambda,\mu} -g_{\lambda \mu, \nu})~.$$
The Riemann tensor is given by
$$ R^{\lambda}_{\mu \nu \kappa} =\Gamma^{\lambda}_{\mu \nu, \kappa}-\Gamma^{\lambda}_{\mu
\kappa, \nu} +\Gamma^{\sigma}_{\mu \nu}\Gamma^{\lambda}_{\sigma \kappa}
-\Gamma^{\sigma}_{\mu \kappa}\Gamma^{\lambda}_{\sigma \nu}~.$$
The Ricci tensor takes the form
$$ R_{\mu\kappa} = R^{\lambda}_{\mu \lambda \kappa}~. $$
The Ricci scalar is expressed as,
$$ R  = g^{\mu \nu} R_{\nu \mu}~.$$
Einstein equations are,
$$ G_{\mu\nu} \equiv R_{\mu \nu}-\frac{1}{2}g_{\mu \nu} R= 8\pi GT_{\mu\nu}~.$$
The Einstein-Hilbert action is expressed as
$$ {\cal S}_{EH}=\int d^4x\sqrt{-g} \frac{R}{16\pi G}~.$$
The energy momentum tensor can be derived via
$$ \delta {\cal S}_m =\frac{1}{2} \int d^4x \sqrt{-g} T^{\mu\nu}\delta g_{\mu\nu}~.$$

$\bullet~$ For the cases of torsional, teleparallel, $f(T)$ gravity, the fundamental
variables that
describe the manifold geometry are the vierbein fields ${\bf e}_A(x^\mu)$. They form an
orthonormal
basis for the tangent space-time at each point $x^\mu$, i.e., ${\bf e}_A\cdot{\bf e}_B =
\eta_{AB}$
where $\eta_{AB} = {\rm diag}(+1,-1,-1,-1)$ is the Minkowski metric for the tangent
space-time. The
component form of the vierbein vector is given by
$$ {\bf e}_A = e_A^\mu\partial_\mu~. $$
The metric of physical space-time can be expressed as
$$ g_{\mu\nu}(x) = \eta_{AB} e^A_\mu(x)e^B_\nu(x) ~.$$
Moreover, the vierbein components follow the relations
$$ e^\mu_A e^A_\nu = \delta^\mu_\nu~,~~e^\mu_A e^B_\mu = \delta^B_A~.$$
The Weitzenb\"ock connection is defined as
$$ \hat\Gamma^\lambda_{\mu\nu} \equiv e^\lambda_A \partial_\nu e^A_\mu = -e_\mu^A
\partial_\nu e_A^\lambda~.$$
The torsion tensor is given by
$$ T^\lambda_{\mu\nu} \equiv \hat\Gamma^\lambda_{\mu\nu} -
\hat\Gamma^\lambda_{\nu\mu}
= e^\lambda_A ~ ( \partial_\mu e^A_\nu - \partial_\nu e^A_\mu) ~. $$
The difference between the Levi-Civita and Weitzenb\"ock connections is known as the
contortion
tensor, which takes the form of
$$ K^{\mu\nu}_{\:\:\:\:\rho} = - \frac{1}{2} \Big( T^{\mu\nu}_{\:\:\:\:\rho} -
T^{\nu\mu}_{\:\:\:\:\rho} - T_{\rho}^{\:\:\:\:\mu\nu} \Big) ~.$$
We define the ``superpotential''
$$ S_\rho^{\:\:\:\mu\nu} = \frac{1}{2}\Big(K^{\mu\nu}_{\:\:\:\:\rho} +\delta^\mu_\rho
T^{\alpha\nu}_{\:\:\:\:\alpha} -\delta^\nu_\rho {\cal
T}^{\alpha\mu}_{\:\:\:\:\alpha}\Big) ~. $$
The torsion scalar is defined as
$$ T \equiv S_\rho^{\:\:\:\mu\nu} ~ T^\rho_{\:\:\:\mu\nu} ~. $$

%---------------------------------------
%\subsection{The Noether Symmetry Approach}

%\subsection{The derivation of the quadratic action of cosmological perturbations}

\section{Coefficients of the stability equation of $f(T,\cal{T})$ gravity}
\label{fTTCoefficients}

In this appendix, we give the coefficients $\Gamma$, $\mu^2$, $c_s^2$ and $D$ of the
perturbation
equation (\ref{fTTphiddk}) of  $f(T,\cal{T})$ gravity of subsection \ref{FTTgravity}:
\begin{eqnarray}
\label{fTTphiddk22}
\ddot{\tilde{\phi}}_k+\Gamma \dot{\tilde{\phi}}_k+\mu^2
\tilde{\phi}_k+c_s^2\frac{k^2}{a^2}
\tilde{\phi}_k=D.
\end{eqnarray}
Concerning the effective mass we have
\begin{equation}
\label{fTTomega2}
 \mu^2=\mu^2_{(1)}+\mu^2_{(2)}+\mu^2_{(3)}+\mu^2_{(4)}+\mu^2_{(5)}+
\mu^2_{(6)},
\end{equation}
with
\begin{eqnarray}
 \mu^2_{(1)}&=&2\dot{H}\left(f_{T}+1\right)   \frac{
B}{E}
+2H   \left(\dot{\rho}_{m}-3 \dot{p}_{m}
\right) f_{T\mathcal{T}}\frac{B^2}{F}
\nonumber\\
&&
 +H^2     \Big[4 \dot{H} \left(3 A f_{T\mathcal{T}
}-5 B f_{TT}\right) \frac{3 B}{F}
\nonumber\\
&&
+A f_{\mathcal{T}}+B
f_{T}+B\Big] ,
\end{eqnarray}
\begin{eqnarray}
\!\!\!\!\!\!\mu^2_{(2)}&=&\frac{12 H^3}{F} \Big\{8
\dot{p}_{m}f_{T\mathcal{T}}\left(B f_{T\mathcal{T}}-A
f_{\mathcal{T}\mathcal{T}}\right)
\nonumber\\
&&
+ \left(\dot{\rho}_{m}- 3\dot{p}_{m}\right)
 \left[2 p_{m} f_{T\mathcal{T}}
\left(B f_{
T\mathcal{T}\mathcal{T}}-A f_{\mathcal{T}\mathcal{T}\mathcal{T}}
\right)\right.\notag \\
&&
+A B f_{T\mathcal{T}\mathcal{T}}-2 A \rho_{m} f_{\mathcal{T}\mathcal{T}
\mathcal{T}} f_{T\mathcal{T}}-3 A f_{\mathcal{T}\mathcal{T}} f_{T\mathcal
{T}}
\nonumber\\
&&
\left.
-B^2 f_{TT\mathcal{T}}+2 B \rho_{m} f_{T\mathcal{T}}
f_{T\mathcal{T}
\mathcal{T}}+3 B f_{T\mathcal{T}}^2\right]
\Big\},
\end{eqnarray}
\begin{eqnarray}
\!\!\!\!\!\!\!\!\!\!\!\!
\mu^2_{(3)}&\!\!=\!\!&\frac{12 H^4}{F} \Big\{4
f_{T\mathcal{T}}^2 \left[9 A
\dot{H}+B (p_{m}+\rho_{m})\right]
\nonumber\\
&&
+f_{T\mathcal{T}} \Big\{
 2 B( f_{T}+1)+A B
\notag \\
&&
 +24 \dot{H}
\Big[(p_{m}+\rho_{m}) (A f_{T\mathcal{T}\mathcal{T}}-B f_{TT\mathcal{T}}
)-3 B f_{TT}\Big]
\Big\}\notag \\
&&
 -3 B \Big[4 \dot{H} (A f_{TT\mathcal{T}}-B f_
{TTT})+f_{TT} (f_{\mathcal{T}}+B)\Big] \Big\},
\end{eqnarray}
\begin{eqnarray}
\mu^2_{(4)}&=&\frac{144 H^5}{F}
\Big\{8 \dot{p}_{m} f_{TT}   f_{T\mathcal{T}\mathcal{T}}
f_{T\mathcal{T}}
\nonumber\\
&&
+
\left(\dot{\rho}_{m}-3 \dot{p}_{m}  \right)
\big\{f_{T\mathcal{T}}
 \left\{A f_{T\mathcal{T}\mathcal{T}}-B
f_{TT\mathcal{T}}\right. \nonumber\\
 && \left.
 +f_{TT}
\left[2
(p_{m} +\rho_{m}) f_{\mathcal{T}\mathcal{T}\mathcal{T}}+3
f_{\mathcal{T}\mathcal{T }}\right]
\right\}
\notag \\
&&
-B f_{T\mathcal{T}\mathcal{T}} f_{TT}\big\}
\Big\},
\end{eqnarray}
\begin{eqnarray}
\mu^2_{(5)}&=& -\frac{432 H^6}{F} \Big\{f_{T\mathcal{T}}
\left\{4 \dot{H}
(A f_{TT\mathcal{T}}-B f_{TTT})
\right.
\nonumber\\
&&
\left.
+f_{TT} \left[8 \dot{H}
(p_{m}+\rho_{m}) f_{ T\mathcal{T}\mathcal{T}}+B\right]
\right\}\nonumber\\
&&
-4 B \dot{H} f_{TT} f_{TT\mathcal{T}}
+12 \dot{H} f_{TT} f_{T\mathcal{T}}^2\Big\},
\end{eqnarray}
and
\begin{eqnarray}
\mu^2_{(6)}&=&\frac{1728}{F}  H^7 f_{T\mathcal{T}} f_{TT}
\left[12 H \dot{H} f_{TT\mathcal{T}}
\right.\nonumber\\
&&\left.
+\left(3
\dot{p}_{m}-\dot{\rho}_{m} \right) f_{T\mathcal{T}\mathcal{T}}\right],
 \end{eqnarray}
respectively.

Concerning the sound speed, we have
\begin{eqnarray}
\label{fTTomega2b}
c_s^2=c^2_{s(1)}+c^2_{s(2)}+c^2_{s(3)}+c^2_{s(4)},
\end{eqnarray}
with the following relations
\begin{equation}
c^2_{s(1)}=\frac{\left(f_{T}+1\right)}{E}
\left(4 \dot{H} f_{T\mathcal{T}}+f_{\mathcal{T}}\right),
\end{equation}
\begin{eqnarray}
&&\!\!\!\!\!\!\!\!\!\!\!\!\!\!\!\!\!\!\!\!\!\!\!\!\!
c^2_{s(2)}=\frac{4 H}{F} \Big\{   B \left(\dot{\rho}_{m}-3 \dot{p}_{m}\right)
(f_{T}+1) f_{
T\mathcal{T}\mathcal{T}}
\nonumber\\
&&\ \ \ \
+f_{T\mathcal{T}}
\left\{
B
\left(\dot{\rho}_{m}-3 \dot{p}_{m}\right)
f_{T\mathcal{T}}
\right.  \notag \\
&&\ \ \ \ \ \ \ \
\left.
+\left(f_{T} +1\right)
\left[\left(\dot{p}_{m}-3 \dot{\rho}_{m}\right) f_{\mathcal{T}
\mathcal{T}}\right.
\right. \notag \\
&&\ \ \ \ \ \ \ \
\left.
\left.
+ 2 (p_{m}+\rho_{m}) (3 \dot{p}_{m}-\dot{\rho}_{m}) f_{\mathcal{T}\mathcal{
T}\mathcal{T}}\right]
\right\}
\Big\},
\end{eqnarray}
\begin{eqnarray}
c^2_{s(3)}&=&\frac{4H^2}{F}
\Big\{
-12 B \dot{H}\left[ f_{TT}f_{T\mathcal{T}}
+ (f_{T}+1) f_{TT\mathcal{T}}\right]
\notag
\\
&&
 +
  f_{T\mathcal{T}}(f_{T}+1) \big\{12 \dot{H} \left[2 (p_{m}+\rho_{m})
f_{T\mathcal{T}
\mathcal{T}}
\right.\nonumber\\
&& \left.
+3 f_{T\mathcal{T}}\right]
+B\big\} \Big\},
\end{eqnarray}
\begin{eqnarray}
\!\!\!  \!\!\!  \!\!\!
c^2_{s(4)}=-\frac{48 H^3}{F}     \left(3
\dot{p}_{m}-\dot{\rho}_{m}\right)
(f_{T}+1) f_{T\mathcal{T}} f_{T\mathcal{T}\mathcal{T}},
\end{eqnarray}
respectively.

Concerning the frictional coefficient we have
\begin{eqnarray}
\Gamma=\Gamma^{(1)}+\Gamma^{(2)}+\Gamma^{(3)}+\Gamma^{(4)}+\Gamma^{(5)}
+\Gamma^{(6)},
\end{eqnarray}
with
\begin{eqnarray}
&&\Gamma^{(1)}=
\frac{f_{T\mathcal{T}}}{F} \Big[20736\, H^7 \dot{H} f_{TT}
f_{TT\mathcal{T}}
\nonumber\\
&&
-\left(3 \dot{p}_{m}-\dot{\rho}_{m}\right) \left(B^2-1728
H^6 f_{T\mathcal{T}\mathcal{T}} f_{TT}\right)\Big],
\end{eqnarray}
\begin{eqnarray}
&&\Gamma^{(2)}=\frac{H}{E} \left\{4 \left[\dot{H} \left(6 A f_{T\mathcal{T}
}-9 B f_{TT}\right)+B f_{T}+B\right]
\right.\nonumber\\
&& \left.
+3 A f_{\mathcal{T}}+\frac{4
k^2}{
a^2} \left(f_{T}+1\right) f_{T\mathcal{T}}\right\},
\end{eqnarray}
\begin{eqnarray}
&&
\Gamma^{(3)}= \frac{\mu^2_{(2)}}{H},
 \end{eqnarray}
\begin{eqnarray}
 \Gamma^{(4)}&=&\frac{12 H^3}{F}
 \Big\{36 A \dot{H} f_{T\mathcal{T}}^2+4 A B  f_{T\mathcal{T}}
 \nonumber\\
&&
 - B \big[12 \dot{H} \left(A f_{TT\mathcal{T}}-B
f_{TTT}\right)
\nonumber\\
&&
\!\!\! \!
+f_{TT}  \left(3 f_{\mathcal{T}}+4 B\right)\big]
- 12 \dot{H} f_{T\mathcal{T}}   \left[5 B
f_{TT}
\right.\nonumber\\
&&  \left.
+2 (p_{m}+\rho_{m}) \left(B f_{
TT\mathcal{T}}   -A
f_{T\mathcal{T}\mathcal{T}}\right)\right]
\Big\},
 \end{eqnarray}
 \begin{eqnarray}
\Gamma^{(5)}=\frac{\mu^2_{(4)}}{H},
  \end{eqnarray}
  and
  \begin{eqnarray}
\!\!\!\!\!\!\!\!\Gamma^{(6)}&=&\frac{576 H^5}{F} \Big\{
3 B \dot{H} f_{TT} f_{TT\mathcal{T}}-9 \dot{H} f_{TT} f_{T\mathcal{T}}
^2
\nonumber\\
&&
-f_{T\mathcal{T}} \left[3 \dot{H}
(A f_{TT\mathcal{T}}-B f_{TTT})\right.
\nonumber\\
&&
\left.+ 6 \dot{H} (p_{m}+\rho_{m})
f_{TT} f_{ T\mathcal{T}\mathcal{T}}+Bf_{TT}  \right]
\Big\},
  \end{eqnarray}
  respectively.

The coefficient $D$ of the right-hand side of (\ref{fTTphiddk22}) is given by
\begin{eqnarray}
D=-D_{1} \delta{\dot{\tilde{p}}^{k}_{m}}-D_{2} \delta{\tilde{p}}^{k}_{m},
\end{eqnarray}
where
\begin{eqnarray}
 D_{1}=\frac{H  f_{T\mathcal{T}}}{E} (I+36 H^2 f_{T\mathcal{T}}),
\end{eqnarray}
and
\begin{eqnarray}
D_{2}=D_{2}^{(1)}+D_{2}^{(2)}+D_{2}^{(3)}+D_{2}^{(4)}+D_{2}^{(5)}+D_{2}^{(6)
},
\end{eqnarray}
with
\begin{equation}
D_{2}^{(1)}=\frac{1}{4 E} \left[(4 \dot{H} f_{T\mathcal{T}}+f_{\mathcal{T
}}) I-16 \pi G B\right],
\end{equation}
\begin{eqnarray}
\!\!\!\!\!\!\!\!\!
D_{2}^{(2)}&=&-\frac{H}{F}
\Big\{
\dot{\rho}_{m} \left\{f_{T\mathcal{T}} \left[f_{\mathcal{T}\mathcal{T}} (3
I-8 B)\right.
\right.\nonumber\\
&&\left.
\left.
 +2 I (p_{m}+\rho_{m})
f_{\mathcal{T}\mathcal{T}\mathcal{T}}\right]
-BI
f_{T\mathcal{T}\mathcal{T}}\right\}
\notag
\\
&&  -\dot{p}_{m} \left\{f_{T\mathcal{T}} \left[f_{
\mathcal{T}\mathcal{T}} (I-24 B)\right.
\right.\nonumber\\
&&\left.
\left.
+6 I (p_{m}+\rho_{m}) f_{\mathcal{T}
\mathcal{T}\mathcal{T}}\right]
-3 B I f_{T\mathcal{T}\mathcal{T}}\right\}
\Big\},
\end{eqnarray}
\begin{eqnarray}
D_{2}^{(3)}&=&\frac{3
H^2}{F}
\Big\{
4 \dot{H}\left[ f_{T\mathcal{T}}^2 (3 I-5
B)- I B f_{
TT\mathcal{T}}\right]
\nonumber\\
&&
+f_{T\mathcal{T}}
\big\{8 f_{\mathcal{T}} \big[5 \dot{H}
(p_{m}+\rho_{m})
f_{T\mathcal{T}\mathcal{T}}+B\big] \notag \\
&&
+3 \left[B-2 (p_{m}+\rho_{m})
f_{\mathcal{T}\mathcal{T}}\right]
\nonumber\\
&&
\big[8 \dot{H}
(p_{m}
+\rho_{m}) f_{T\mathcal{T}\mathcal{T}}+B\big] \big\}
\Big\},
\end{eqnarray}
\begin{eqnarray}
    D_{2}^{(4)}&=&\frac{12 H^3
f_{T\mathcal{T}}}{F}
\Big\{(\dot{\rho}_{m}-3
\dot{p}_{m}) f_{T\mathcal{T}\mathcal{T}} (I+3 B)
\notag \\
&&
+3 f_{T\mathcal{T}} \big[(\dot{p}_{m}-3 \dot{\rho}_{m})
f_{\mathcal{T}\mathcal{T}}
\nonumber\\
&& +
2 (p_{m}+\rho_{m}) (3 \dot{p}_{m}-\dot{\rho}_{m}) f_{\mathcal{T}\mathcal{
T}\mathcal{T}}\big]\Big\},
\end{eqnarray}
\begin{eqnarray}
&&\!\!\!\!\!\!\!\!\!\!\!\!\!\!\!\!\!\!\!\!
D_{2}^{(5)}=\frac{36 H^4 f_{T\mathcal{T}}}{F}
\Big\{ 3 f_{T\mathcal{T}}  B -4 \dot{H}
f_{TT\mathcal{T}} \left(I+3
B\right)\nonumber\\
&&
+3 f_{T\mathcal{T}}
\big\{4 \dot{H} \left[2 (p_{m}+\rho_{m})
f_{T\mathcal{T}\mathcal{T}}+3 f_
{T\mathcal{T}}\right]
\big\}
\Big\},
\end{eqnarray}
and
\begin{eqnarray}
  \!\!\!\!\!\!\!\!\!\!\!\!\!\!\!\!
  \!\!\!\!\!\!\!\!\!\!\!\!\!
  D_{2}^{(6)}&=&-\frac{432 H^5
f_{T\mathcal{T}}^2}{F} \Big[12 H \dot{H}
f_{TT\mathcal{T}}
\nonumber\\
&& \ \ \ \ \ \ \ \ \ \ \ \ \  \ \ \ \ \
+\left(3 \dot{p}_{m}-\dot{\rho}_{m}\right)
f_{T\mathcal{T}
\mathcal{T}}\Big],
 \end{eqnarray}
 respectively.

 Finally, in all the above expressions we have introduced the coefficients
\begin{eqnarray}
 &&A\equiv 2 \left(p_{m}+\rho_{m}\right) f_{T\mathcal{T}}+f_{T}+1,\nonumber\\
&&
 B\equiv
2 \left[8 \pi G+\left(p_{m}+\rho_{m}\right) f_{\mathcal{T
}\mathcal{T}}-6 H^2 f_{T\mathcal{T}}\right]+f_{\mathcal{T}},
\nonumber\\
&& E\equiv
12 H^2 \left[A f_{T\mathcal{T}}-f_{TT} \left(12 H^2 f_{T\mathcal{T}}
+B\right)\right] +B \left(f_{T}+1\right),\nonumber\\
&&
 I\equiv-6 (p_{m}+\rho_{m}) f_{\mathcal{T}\mathcal{T}}+5 f_{\mathcal{T}}
+3 B,\nonumber\\
&&
F\equiv B E.
\end{eqnarray}

\end{document}